\documentclass[12pt,letterpaper,openany,oneside]{book}  
\usepackage{placeins}  
\usepackage{amsmath}
\usepackage{nccmath}
\usepackage{siunitx}
\usepackage{adjustbox}
\usepackage{physics}
\usepackage{url}
\usepackage[doublespacing]{setspace}
\usepackage[top=1in, bottom=1in, left=1.5in, right=1in]{geometry}
\usepackage[T1]{fontenc}
\usepackage{tabularx,ragged2e,booktabs,caption}
\newcolumntype{C}[1]{>{\Centering}m{#1}}

\extrafloats{1500}
\usepackage{amssymb}
\usepackage{graphicx}
\DeclareGraphicsExtensions{%
    .png,.PNG,%
    .pdf,.PDF,%
    .jpg,.mps,.jpeg,.jbig2,.jb2,.JPG,.JPEG,.JBIG2,.JB2}
\usepackage{multirow}
\usepackage{subfig}
\usepackage{array}
\usepackage{tabu} 
\usepackage{hyperref}
\hypersetup{
    colorlinks,
    citecolor=red,
    filecolor=Darkgreen,
    linkcolor=blue,
    urlcolor=blue
}
\usepackage{mathtools}
\usepackage{etoolbox}
\def\ls{\mathrel{\lower4pt\vbox{\lineskip=0pt\baselineskip=0pt
           \hbox{$<$}\hbox{$\sim$}}}}
\def\gs{\mathrel{\lower4pt\vbox{\lineskip=0pt\baselineskip=0pt
           \hbox{$>$}\hbox{$\sim$}}}}

\begin{document}
\frontmatter

{\singlespacing \setlength{\parindent}{0pt} \large

\vspace*{1.5cm}
\underline{Jui-Jen Wang} \\
{\scriptsize \textit{Candidate}} 

\vspace{0.5cm}
\underline{Physics \& Astronomy} \\
{\scriptsize \textit{Department}} 

\vspace{1cm}
This dissertation is approved, and it is acceptable in quality \\
and form for publication:

\vspace{0.25cm}
\textit{Approved by the Dissertation Committee:}

\begin{centering}
\vspace{0.5cm}
\underline{Michael Gold, Chairperson}

\vspace{0.25cm}
\underline{Rouzbeh Allahverdi}

\vspace{0.25cm}
\underline{Dinesh Loomba}

\vspace{0.25cm}
\underline{John Matthews}

\vspace{0.25cm}
\underline{Keith Rielage}

\end{centering}

}


\clearpage
{\singlespacing \setlength{\parindent}{0pt} \centering \large

\vspace*{5.5cm}
{\LARGE \doublespacing \textbf{MiniCLEAN Dark Matter Experiment}}

\vspace{1cm}
by

\vspace{.5cm}
\textbf{Jui-Jen (Ryan) Wang}

\vspace{.25cm}
B.S., Physics , Tamkang University, 2004\\
M.S., Physics, University of New Mexico, 2013

\vspace{3cm}
DISSERTATION

\vspace{.25cm}
Submitted in Partial Fulfillment of the \\
Requirements for the Degree of \\
\textbf{Doctor of Philosophy}

\vspace{.25cm}
\textbf{Physics}

\vspace{.25cm}
The University of New Mexico \\
Albuquerque, New Mexico

\vspace{1cm}
\textbf{December, 2017}

}


\chapter{Acknowledgements}
First of all, I would like to thank Prof. Gold, he has been really kind and patient to me. In the early year of my graduate school, I am struggling with accommodating different culture and environment which took my focus away from the school. However, Prof. Gold never gave upon me, and gave me a lot of space to work on my issues. Throughout the time in my graduate school, he always be there for me and gives me freedom to do whatever I want to do but in the same time gives the guidance on my research topic. I am grateful to have him as my advisor. Franco, who is the postdoc at the time also provide me a lot of advise on experimental skills. I learned a lot from him and have a really good time to work with him. I also want to thank miles who is the undergraduate student at the time, he poses a excellent skill on experimental stuff, and help me to build up many tests. During the writing of this manuscript, Guy helps me to correct and polish my english, I am also appreciated for his help.\par
   I also would like to thanks my friends in MiniCLEAN collaboration. When I station at SNOLab, Kimberly Palladino, Steve Linden and Thomas Caldwell taught me many skills on operation in underground lab. Kim is the on-site manager at the time and she organize everything well, Steve who became on-site manager after Kim's departure also done a good job on operation of detector. I learned from them and also received lots advise from them. Tom taught me lots things of data analysis, which enable me to do the data analysis in the later time. Chris Benson share his experiences on the subsystem of MiniCLEAN, allows me to relate the data to the hardware even I was not on site. Chris Jackson help me on doing simulation and he also organize the analysis tasks well such that I just need to follow his plan. Josh takes over the spokesperson when MiniCLEAN collaboration in its darkest time, save the project and allows me to have a chance to finish my thesis. I am grateful for all they have done for the collaboration and myself.\par
   My friends in Albuquerque also helps me to go through some down time in my graduate school. Chih-feng and hung hung out with me from time to time to relief my stress in the period of time when I am not sure if I can keep going with my research. My friends in Taiwan also constantly encourage me, gave me energy to keep moving toward the goal. One of my best friends G.Y. Chen who now is the associate professor in prestigious university in Taiwan always listen to my complain and push me to pursue my dream. Finally, I would like to thanks my parents, they never told me what to do, just always supported me unconditionally. I could not have done this without their supports. They are the best parents to me in this world.


\clearpage 
\phantomsection
\addcontentsline{toc}{chapter}{Abstract}
\markright{}

{\singlespacing \setlength{\parindent}{0pt} \centering

\textbf{MiniCLEAN Dark Matter Experiment}

\vspace{.25cm}
by \\
\textbf{Jui-Jen (Ryan) Wang}

\vspace{.25cm}
B.S., Physics ,Tamkang University , 2004\\
M.S., Physics, University of New Mexico, 2013

\vspace{.25cm}
Ph.D., Physics, University of New Mexico, 2017

\vspace{1.5cm}
{\Large \textbf{ABSTRACT}}

}

Particle Dark Matter is a hypothesis accounting for a number of observed astrophysical phenomena such as the anomalous galactic rotation curves.  From these astronomical observation, about 23 \% of the Universe is made by dark matter. Among the possible candidates of dark matter, Weakly Interacting Massive Particle (WIMPs) seems to be most promising candidates. The hypothetical particle provides a mechanism of producing the dark matter and is in consistent with the results inferred by Cosmic Microwave Background (CMB), reproduce the correct relic density of dark matter. In particle physics, understanding dark matter may leads to new physics beyond standard model. \par
The MiniCLEAN dark matter experiment will exploit a single-phase liquid-argon detector instrumented with 92 photomultiplier tubes placed in the cryogen temperature with 4-$\pi$ coverage of a 500 kg (150 kg) target (fiducial) mass. The detector design strategy emphasizes scalability to target masses of order 10 tons or more. The detector is designed also for a liquid neon target that allows for an independent verification of signal and background and a test of the expected dependence of the WIMP-nucleus interaction rate.\par 
For MiniCLEAN, PMT stability and calibration are essential. The \textit{In-situ} optical calibration will be able monitor the PMT stability and maintain the calibration. In MiniCLEAN, we use a Light-Emitting Diode(LED)- based light injection system to provide single photons for calibration, the calibration can be performed in near real-time, providing a continuous monitor at the condition of the detector. The intrinsic $^{39}$Ar beta emitter provides another way to calibrate the detector thanks to well defined properties and uniformly distributed inside the detector volume. The energy scale can be determined by fitting the energy spectrum of experimental $^{39}$Ar data. Moreover, the preliminary results from cold gas run shows the best measurement on triplet lifetime ($\sim$ 3.5 $\mu$ s). The results confirms the high purity of argon is attained by MiniCLEAN's purification system. The long triplet lifetime in gaseous argon can be exploit to obtain better performance of pulse shape discrimination (PSD) for future dark matter detector, also the low density of gaseous argon reduced the multi-scattering neutron backgrounds. On the other hand, by injecting $^{39}$Ar spike, the electronic recoil events due to $^{39}$Ar beta decay can be used to test the limit of PSD in liquid argon. The results will be informative for future multi-tonne LAr detector.

\tableofcontents
\listoffigures
\listoftables


\mainmatter

\chapter{Introduction and Detection of Dark Matter}\label{ch:intro}
\section{\label{section:overview}Overview}
   Dark Matter is a long standing mystery of our universe. The energy of our universe today is a fossil of the progresses that took place as its early stages. About 5\% of our universe consists of visible matter , while 23\% consists of dark matter(DM). The relic abundance of either of these components requires physics beyond what has currently been established.\par
The standard model of particle physics gives an excellent description of physical processes at energies thus far probed by experiments. Unitarity of electroweak interaction, however, breaks down at energy scales $\lesssim$ O(Tev) in the absence of a mechanism to account for electroweak-symmetry breaking.  This imply that a new framework will be required at reduced Plank scale $M_{p} = (8\pi G_{Newton})^{-1/2} = 2.4\times10^{18}$ GeV, where the quantum gravitational effect becomes important. Moreover, the fact that $M_{p}/M_{W}$ is so huge which gives a strong indication that must be new physics beyond the Standard Model, and also because of the infamous "hierarchy problem".\cite{PhysRevD.13.974}  Supersymmetry theory provides a different approach to help physicist to resolve the problem. it can render a theory stable to the radiative corrections which would otherwise force a fine tuning of high-energy parameters.
In the following sections, the evidences from astrophysical observation are present. The possible candidates of dark matter will also be discussed. Finally, the efforts to dark matter direct detection will be described by the end of this chapter.\par
\section{Astronomical Evidence of Dark Matter}
\subsection{Coma Cluster}
The discrepancy  between the mass and the light produced by astronomical objects was found by Fritz Zwicky\cite{1933AcHPh...6..110Z}.  Zwicky studied the coma cluster which is about 99 Mpc from Earth, and observing doppler shifts in galactic spectra. With the observation, the velocity dispersion of galaxies in the Coma cluster can be computed. Zwicky then used the virial theorem to calculate the cluster's mass.
\begin{ceqn}\begin{align}
\langle T\rangle = -\frac{1}{2}\langle U\rangle,
\end{align}\end{ceqn}
where $\langle T\rangle$ is average kinetic energy which can be obtained from the velocity dispersion, and $\langle U\rangle$ is average potential energy. Once the $\langle T\rangle$ is found, using virial theorem can calculate the total mass of the whole system.  It was found by Zwicky that the total mass of the cluster was $M_c \simeq 4.5\times10^{13}M_{\odot}$. However, the mass of the cluster estimated from the standard M/L ratios just approximately 2\% of this value. Therefore the results from Zwicky imply that there are huge masses in the cluster which is ``missing'' for some reason or ``non-luminous''. Although Zwicky was not able to give a full explanation of the problem, his results inspired many researchers to continue probing the problem.  
\subsection{Rotation Curve}
\begin{figure}[tbp]
\centering
\graphicspath{{./fig/}}
\includegraphics[scale=0.3]{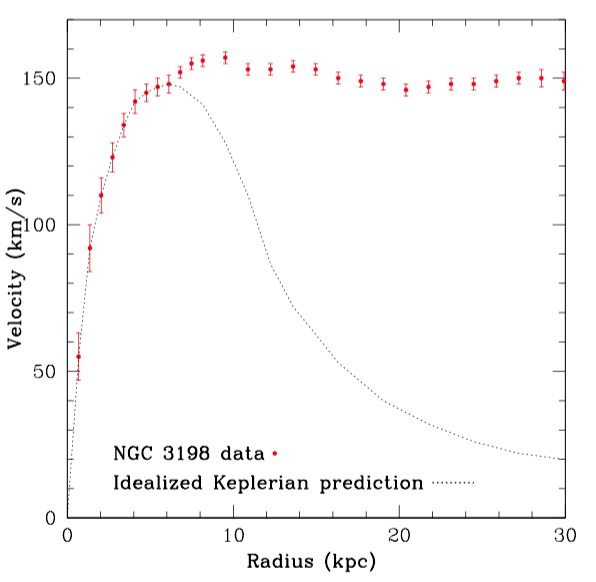}
\caption{ The measured rotation curve of HI regions in NGC 3198\cite{rotationcurve1} compare to the theoretical prediction(``Keplerian'').}
\label{fig:rotation_curve_1}
\end{figure}

\begin{figure}[tbp]
\centering
\graphicspath{{./fig/}}
\includegraphics[scale=0.25]{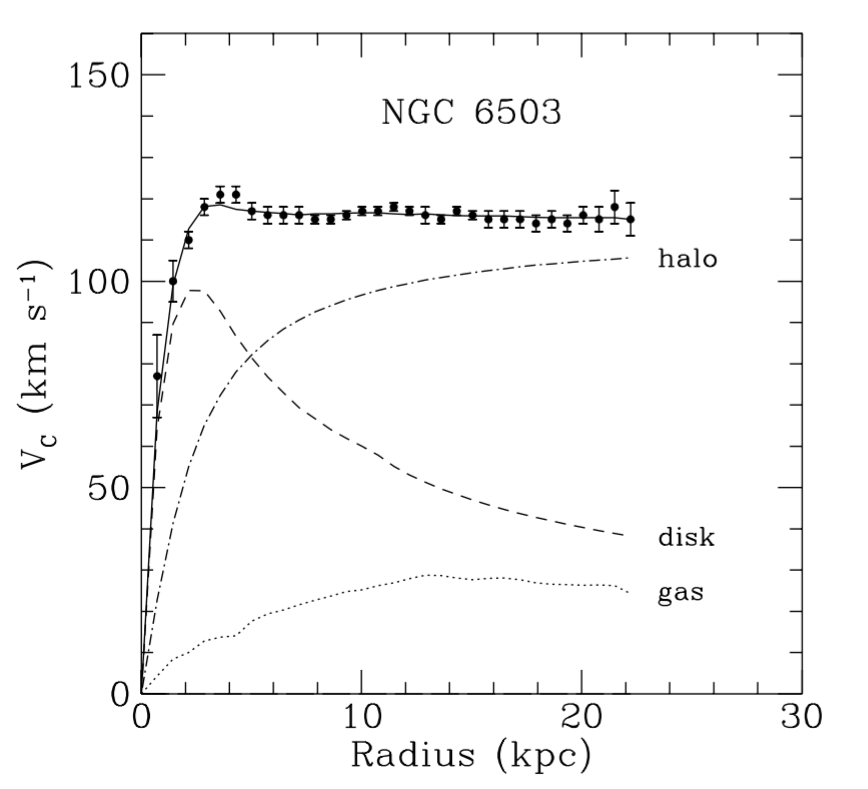}
\caption{ Measured rotation curve of NGC6503 with best fit and contributions from halo, disk and gas\cite{rotationcurve2}.}
\label{fig:rotation_curve_2}
\end{figure}

Observing the luminous object in the galaxy can gives a first order of estimation of the mass of the galaxy. Some interstellar gas, however, also contribute to the mass distribution of the galaxy. It was found that these gases also emit the electromagnetic wave and can be detected by the radio telescope. Most of the interstellar gas has hydrogen atom which first studied by van de Hulst \textit{et al.}\cite{1954BAN}. The hydrogen emits radio waves in 21 cm, which can be used to detect the interstellar gas and measure the velocity. From Newton's gravitational law, if the galaxy has mass $m(r)$ within the radius r, then the velocity as a function of radius r should be :
\begin{ceqn}\begin{align}
v(r) = \sqrt{G\cdot \frac{m(r)}{r}},
\end{align}\end{ceqn}
where G is the gravitational constant. This imply the velocity should decreased when the radius increased i.e. $v(r) \propto 1/\sqrt{r}$. This is generally referred to as ``Keplerian'' behavior. However, the experimental results on the rotation curve of spiral galaxy deviates the ``Keplerian'' behavior as shown in Fig. \ref{fig:rotation_curve_1}. The flat rotation curve was found in many galaxies indicating some huge mass is unaccounted for at large radius. With the technique to detector the velocity of interstellar gas, the contribution from dark matter can be estimated using the rotation curve as shown in Fig. \ref{fig:rotation_curve_2}. The ``dark matter halo'' envelope the galactic disc and give rises the local matter density. The experimental data of rotation curve shows the clear proof of existence of dark matter.
\subsection{CMB}
\begin{figure}[tbp]
\centering
\graphicspath{{./fig/}}
\includegraphics[scale=0.25]{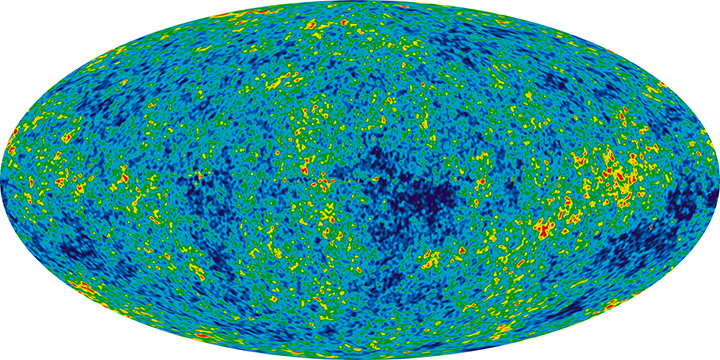}
\caption{CMB temperature fluctuations from WAMP. Image from \url{http://map.gsfc.nasa.gov/}.}
\label{fig:WMAP}
\end{figure}

Cosmic microwave background is the electromagnetic radiation from early universe. The photons scattering off last scattering surface (LSS) and redshifted due to the expansion of universe were observed in CMB. There are many cosmological parameters can be determined or constraint by CMB observations. The comprehensive review of CMB theory can be found in \cite{doi:10.1146/annurev.astro.40.060401.093926}. It is first discovered by Penzias and Wilson\cite{1965ApJ} in 1964 and the first map of CMB of universe is made by the Differential Microwave Radiometer (DMR) aboard NASA's Cosmic Background Explorer (COBE)\cite{COBE}. The first result from COBE has 7 degree of angular resolution, gives the snap shot of universe about 380,000 years after the big bang which is approximately 14 billion years ago from now. After years effort, the image released by Wilkinson Microwave Anisotropy (WMAP) with fraction-of-a-degree resolution as shown in FIg. \ref{fig:WMAP} shows that the temperature fluctuations of no more than  ~10$^{-5}$ and the spectrum follow precisely of a black body radiation with temperature T = 2.726 K.\par
\begin{figure}[tbp]
\centering
\graphicspath{{./fig/}}
\includegraphics[scale=0.25]{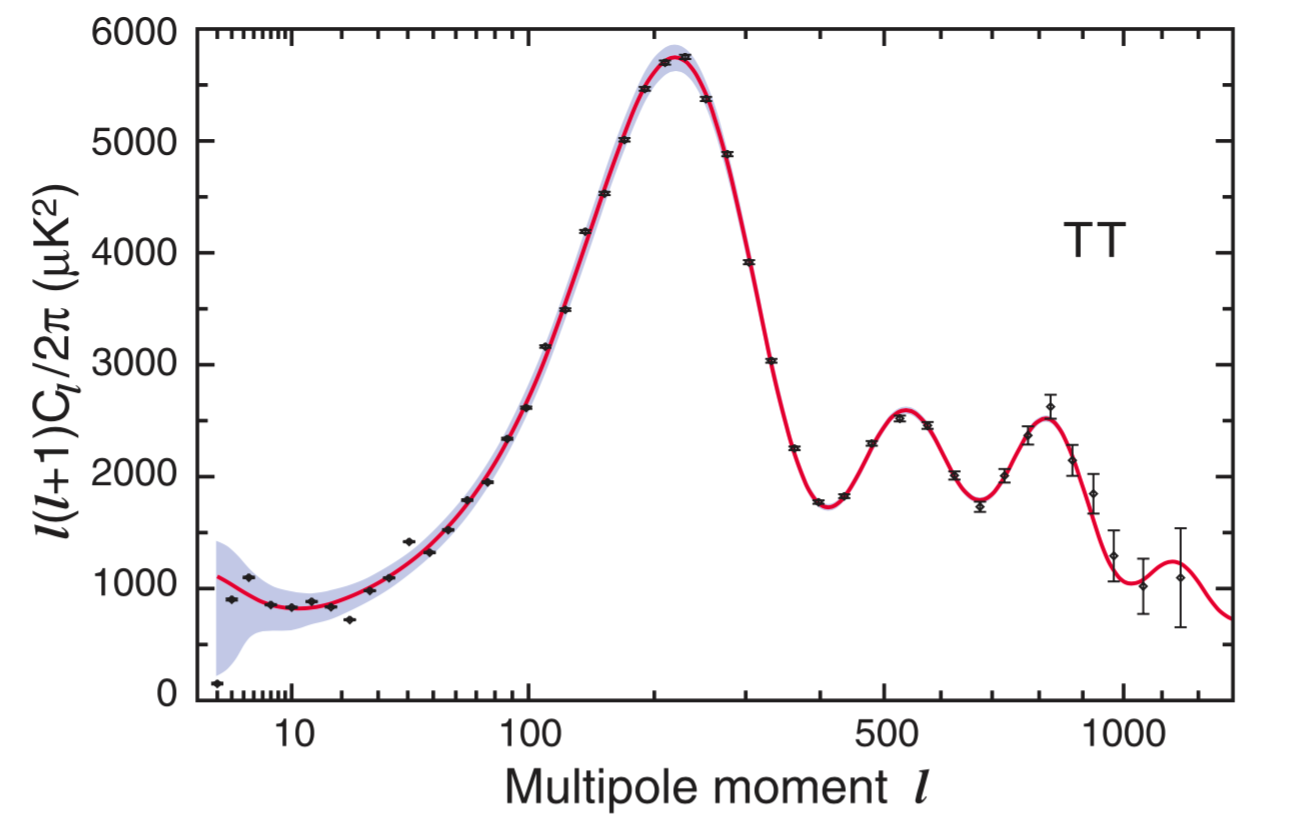}
\caption{Nine-year WMAP TT angular power spectrum. Plot is taken from\cite{0067-0049-208-2-20}.}
\label{fig:WMAP_power}
\end{figure}

The fluctuations of temperature observed by experiments can be expressed as 
\begin{ceqn}\begin{align}
\frac{\delta T}{T} (\theta,\phi)= \sum\limits_{l=2}^{+\infty}\sum\limits_{m=-l}^{+l} a_{lm}Y_{lm}(\theta,\phi),
\end{align}\end{ceqn}
where $Y_{lm}(\theta,\phi)$ is spherical harmonic function. The variance $C_l$ of $a_{lm}$ is given by 
\begin{ceqn}\begin{align}
C_l \equiv \langle |a_{lm}|^{2}\rangle\equiv\frac{1}{2l+1}\sum\limits_{m=-l}^{l}|a_{lm}|^2.
\end{align}\end{ceqn}
Assuming the temperature fluctuation is Gaussian, the power spectrum of CMB contains all the information.  Figure \ref{fig:WMAP_power} shows the power spectrum from WMAP's data. The abundance of baryon ($\Omega_bh^2$) and matter ($\Omega_mh^2$) in the universe can be calculated using the information extracted from CMB data and with fixed 6 parameters in cosmological model\cite{0067-0049-208-2-20}.  
\begin{ceqn}\begin{align}
\Omega_bh^2 = 0.02264 \pm 0.00050 \;\;\;\; \Omega_mh^2 = 0.01364\pm0.00440,
\end{align}\end{ceqn}
Various experiments\cite{0004-637X-591-2-556,0004-637X-600-1-32} dedicated to measure the power spectrum of CMB, and astronomical measurements of the power spectrum from large scale structure\cite{doi:10.1046/j.1365-8711.2001.04827.x}, gives the constraint of the abundance of baryon density in the universe :
\begin{ceqn}\begin{align}
0.018 < \Omega_bh^2 < 0.023
\end{align}\end{ceqn} 
which is in consistent with the predictions from Big Bang nucleosynthesis\cite{TASIdark}. These astronomical evidences all point to a large and opaque  massive object in our universe which is another strong evidence of dark matter. 
\subsection{Dark Matter Candidates}
\begin{figure}[tbp]
\centering
\graphicspath{{./fig/}}
\includegraphics[scale=0.3]{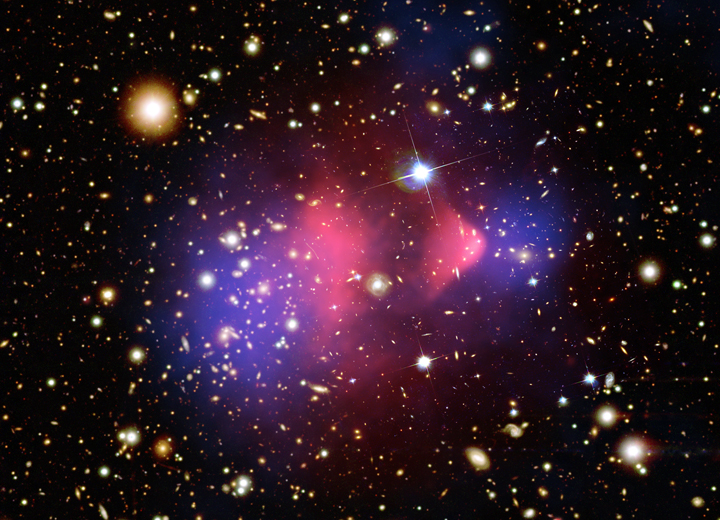}
\caption{Collision of two galaxy cluster to form bullet cluster. The red image is baryonic matter and the blue image is the dark matter. Image us taken from \url{http://chandra.harvard.edu/photo/2006/1e0657/}. }
\label{fig:bullet}
\end{figure}

\begin{figure}[tbp]
\centering
\graphicspath{{./fig/}}
\includegraphics[scale=0.3]{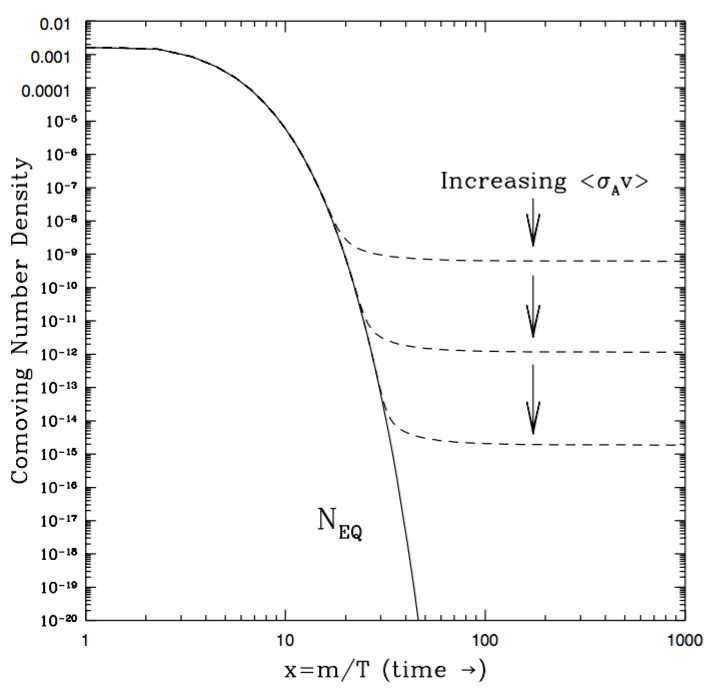}
\caption{Equilibrium (solid curve) and relic abundance (dashed curve) of WIMP particles. Figure is taken from \cite{JUNGMAN1996195}}
\label{fig:WIMP_miracle}
\end{figure}

The solid proof of existence of dark matter has been established in the previous sections. From the cosmological constraint, the dark matter should be non-baryonic. Therefore, the candidates of dark matter should satisfy several requirements.
\begin{itemize}
	\item Dark matter should have no or extremely weak interactions with photons, such that it is ``dark''.
	\item Self-interactions of dark matter should be small. If the interaction is not small, the dark matter halo should shrinks with time past due to the self-interactions. The observation of Bullet Cluster  provides a astronomical evidence(see Fig. \ref{fig:bullet}). The Bullet cluster was created by merger of two galaxy cluster. When the two cluster collides with each other, the two dark matter halos (Blue bulk in Fig. \ref{fig:bullet}) passed through. However, the baryonic gas (Red bulk in Fig. \ref{fig:bullet}) has shocked and is located between two halos.
	\item The interactions between dark matter and baryonic matter should also be weak. If not, the baryon-dark matter disk would be form which is in contradiction with the observed diffuse and extended dark matter halos.
	\item Dark matter can not be made up of Standard Model (SM) particles. The only suitable particles in SM is neutrino. However, a simple calculation shows the neutrinos can not responsible for all the dark matter in the universe. The relic density of neutrino is given by
	\begin{ceqn}\begin{align}
		\Omega_{\nu}h^2=\sum\limits_{i=1}^3\frac{m_i}{93\;eV}
	\end{align}\end{ceqn} 
	where $m_i$ is the mass of i-th neutrino. The current upper limit on sum of neutrino mass from cosmological observation is $m_{\nu}<0.13$ eV\cite{neutrinomass} with 95\% C.L. . This gives the upper bound on the total relic density of neutrino is 
	\begin{ceqn}\begin{align}
		\Omega_{\nu}h^2 \lesssim 0.0014
	\end{align}\end{ceqn}
	This means the neutrinos are simply not abundant enough to be responsible for all the dark matter in the universe. However, even if neutrino has much larger mass , they still can not responsible for the whole dark matter if they are traveling relativistically. They can free-stream and wash out the fluctuation which creates the large scale structure\cite{PhysRevLett.45.1980}. In the neutrino dominated universe, the galaxies only can form at the time z < 2\cite{Centrella1983}, which is in contrast to what has been observed.  
\end{itemize}
Some of the candidates which meet the requirement of above properties is summarized here.
\begin{itemize}
	\item Axions : Axions were introduced to attempt to solve the CP violation problem. The mass of axions are extremely small ($\lesssim$ 0.01 eV) which inferred from stellar cooling and the dynamics of supernova 1987A. Moreover, they have extremely weak interactions with ordinary matter which implies that they were not in thermal equilibrium in the early universe. The relic density of axions depends on the assumptions of the production mechanism, thus is uncertain. However, an acceptable range could be found for axions to satisfy all present day constraints, thus become a possible candidates of dark matter\cite{ROSENBERG20001}.  
	\item Sterile neutrinos : The hypothetical particles first proposed as a dark matter candidates in 1993 by Dodelson and Widrow\cite{PhysRevLett.72.17}. They are similar to the SM neutrino but without weak interactions. The results from WMAP reionization optical depth implies the massive stars were form prior to redshift $z$ > 20 while the dark matter structures were in place. Thus, it is not possible for dark matter particle mass is smaller than $\sim$ 10 keV\cite{1538-4357-591-1-L1}. The sterile neutrino provides a alternative explanation for the WMAP optical depth is reionization by decaying sterile neutrinos\cite{0004-637X-600-1-26}. 
	\item Standard Model neutrino :  The neutrinos in SM could also be a dark matter candidates but as mentioned previously, they can not responsible for all the dark matter in the universe.
	\item Weakly Interacting Massive Particles (WIMPs) :  They are most promising candidates to explain the dark matter. They were introduced by new physics at electroweak scale (i.e. supersymmetry ) as a new stable, weakly-interacting particles, with mass of order $M_{\chi} \sim $100 GeV. In supersymmetric theory (SUSY), the WIMP is the neutralino
	\begin{ceqn}\begin{align}
		\tilde{\chi} = \xi_{\gamma}\tilde{\gamma} +  \xi_{Z}\tilde{Z^0} +  \xi_{h_1}\tilde{h_1^0} +  \xi_{h_2}\tilde{h_2^0}, 
	\end{align}\end{ceqn}
	is a linear combination of the SUSY partners of the photon, $Z^0$ boson, and neutral Higgs bosons. WIMPs are stable and particle theory models suggest masses $M_{\chi} \sim$ 10 -10$^3$ GeV.
\end{itemize}
The WIMPs are assumed to be in thermal equilibrium at temperature $T \gtrsim M_{\chi}$ in the early universe. Using Boltzmann equation, the WIMP number density as a function of time $t$ is :
\begin{ceqn}\begin{align}
\frac{dn_{\chi}}{dt} = -3Hn_{\chi} - \langle\sigma_{ann}\cdot v\rangle(n_{\chi}^2 - n_{eq}^2).
\end{align}\end{ceqn}
where $n_{eq}$ is the desity of equilibrium, the Hubble constant in the early universe is  $H^2 = \rho_{rad}/3M_p^2$, and $ \langle\sigma_{ann}\cdot v\rangle$ is the e thermally averaged WIMP annihilation cross section times WIMP relative velocity. At the temperature cooled to $T_{fr}$ -- the freeze-out point, the annihilation rate of WIMPs overtaken by the expansion rate. Thus the WIMPs freeze-out and the number density in a co-moving volume becomes constant. Therefore, the present-day WIMP relic density can be approximated as :
\begin{ceqn}\begin{align}
	\Omega_{\chi}h^2\simeq \frac{s_0}{\rho_c/h^2}(\frac{45}{\pi^2g_*})^{1/2} \frac{1}{x_fM_p\langle\sigma_{ann}\cdot v\rangle},
\end{align}\end{ceqn}
the explanations and value of parameters can be found in \cite{1674-1137-38-9-090001}. The measured value of $\Omega_{\chi}h^2$ is $\simeq$ 0.12, thus 
\begin{ceqn}\begin{align}
\frac{\Omega_{\chi}h^2}{0.12} \simeq \frac{1}{\langle \frac{\sigma_{ann}}{10^{-36} cm^2}\frac{v/c}{0.1}\rangle}.
\end{align}\end{ceqn}
an annihilation cross section of weak strength of order $\sim$ 10$^{-36}$ cm$^2$ and WIMP freeze-out velocity give a correct present day relic density of dark matter, so-called ``WIMP-miracle'' (see Fig. \ref{fig:WIMP_miracle}). Therefore, most of direct search for dark matter experiments assumes the dark halo is made of WIMPs. In the following sections, WIMPs are considered as only candidates of dark matter.
\section{Direct Dark Matter Detection}
\begin{figure}[tbp]
\centering
\graphicspath{{./fig/}}
\includegraphics[scale=0.3]{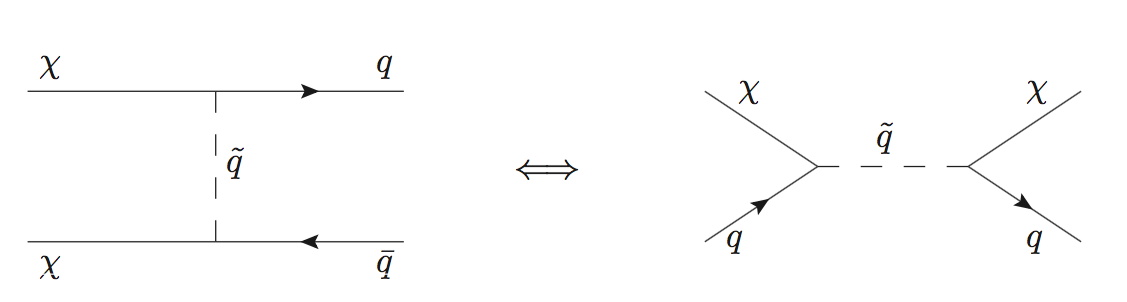}
\caption{Crossing symmetry between annihilation and scattering diagrams\cite{dark_matter_astrophysics}.}
\label{fig:feynman_WIMP}
\end{figure}

Assuming WIMPs make up the halo of the Milky Way, they will have a local spatial density $n_{\chi} \sim$ 0.004 ($M_{\chi}$/100GeV)$^{-1}$cm$^{-3}$. The velocity of WIMPs follows the Maxwellian velocity distribution with most probable velocities $v\sim$ 200 km sec$^{-1}$\cite{dark_matter_astrophysics}. WIMPs can interact with itself or baryonic matter with very small cross-section. The cross section  between annihilation $\chi\chi \rightarrow q\bar{q}$ and the elastic scattering $\chi q \rightarrow \chi q$ process are more or less the same (Fig. \ref{fig:feynman_WIMP} $\sim$ 10$^{-36}$ cm$^2$). Therefore, indirectly observation can be made through detecting the product of annihilation  ($\gamma$ rays, electrons, positrons) or directly observation through the interaction of WIMPs-target nucleus in low-background detector. MiniCLEAN is designed base on direct detection, thus I will focus on the technique required to directly detect WIMPs and compare the results between different experiments in the following sections.
\subsection{Basic Principle}
Due to extremely small cross-section between WIMPs and ordinary matter, when WIMPs incident on target atom, small energy will be deposited (1-100 keV). Moreover, the multiple scattering between WIMPs and target atom can be negligible for the same reason. A nuclear recoil is expected for interaction between WIMPs and target atom\cite{PhysRevD.31.3059}. The differential spectrum of dark matter interactions can be expressed  as \cite{LEWIN199687} :
\begin{ceqn}\begin{align}\label{eq:diffcross}
	\frac{dR}{dE}(E,t) = \frac{\rho_0}{m_{\chi}\cdot m_A}\cdot \int v\cdot f(\textbf{v},t)\cdot \frac{d\sigma}{dE}(E,v)d^3v,
\end{align}\end{ceqn}
where $\frac{d\sigma}{dE}(E,v)$ is the differential cross-section of WIMPs and nucleus interaction and $m_{\chi}$ is the mass of dark matter. The WIMP cross-section $\sigma$ and $m_{\chi}$ can be measured experimentally. The velocity of dark matter is defined as the velocity in the rest frame of the detector and $m_A$ is the nucleus mass. The local dark matter density $\rho_0$ and velocity distribution $f(\textbf{v},t)$ are the astrophysical parameters. Moreover, the velocity distribution will change  with time due to the revolution of the Earth around the Sun. In general, the energy produced by WIMPs-nucleus recoil is easier to be determined than the directional information. The Eq. \ref{eq:diffcross} can be approximated by
\begin{ceqn}\begin{align}
	\frac{dR}{dE}(E) \approx \left( \frac{dR}{dE} \right )_0 F^2(E) exp\left(-\frac{E}{E_c}\right),
\end{align}\end{ceqn}
where $\left( \frac{dR}{dE} \right )_0$ is the event rate at zero momentum transfer and $E_c$ is a constant parameterizing a characteristic energy scale which depends on the dark matter mass and target nucleus\cite{LEWIN199687}. $F^2(E)$ is the nuclear form factor which accounts for when the particle wavelength is not large compare to the nuclear radius, the cross-section decreases with increasing momentum transfer: $\sigma \propto \sigma_0\cdot F^2$, where $\sigma_0$ is the zero-momentum transfer cross-section. Therefore, at low recoil energy, the signal is dominated by exponential function.\par
Another signature of dark matter signal is ``annual modulation''.  Due to relative motion between Earth and dark matter halo in the Milky Way, the velocity of dark matter particle reaches maximum around June 2 and has minimum in December. This results in the events that produced by WIMP-nucleus recoil exceed detector's threshold also have maximum in June\cite{PhysRevD.33.3495}. For experiment which can observe multiple events in a year, the amplitude of the variation in event rates at different time of a year can be observed. The differential event rate for the modulation can be written as \cite{RevModPhys.85.1561}
\begin{ceqn}\begin{align}
	\frac{dR}{dE}(E,t) \approx S_0(E) + S_m(E)\cdot cos\left( \frac{2\pi(t-t_0)}{T} \right),
\end{align}\end{ceqn} 
where $T$ is the period of one year and $t_0$ is the phase which is expected at about 150 days. The modulation amplitude is given by $S_m$ while the time averaged events rate is $S_0$. The signature signal from modulation can help to discriminate the back ground signal and confirm the dark matter detection as well.\par
Additionally, the directionality is another desired capability of the detector for dark matter detection. As indicated in \cite{PhysRevD.37.1353}, the direction of WIMP-nucleus recoil has a strong angular dependance. The angle $\theta$ can be defined as the direction of the nuclear recoil relative to the mean direction of the solar motion, thus the differential rate equation gives :
\begin{ceqn}\begin{align}\label{eq:intedir}
	\frac{dR}{dE\;d\cos{\theta}} \propto exp\left[ \frac{-[(v_E+v_{\odot})\cos{\theta}-v_{min}]^2}{v_c^2}\right]. 
\end{align}\end{ceqn}
where $v_E$ is the Earth's motion, $v_{\odot}$ is the velocity of the Sun around the galactic centre, $v_{min}$ represents the minimum WIMP velocity that can produce a WIMP-nucleus recoil of energy E and $v_c$ is the circular velocity of the dark matter halo ($v_c = \sqrt{3/2}v_{\odot}$). The integrated rate from Eq. \ref{eq:intedir} shows that the event rates for forward scattering is a order of magnitude more than the backward scattering\cite{PhysRevD.37.1353}. The detector which can provide the directional information would be powerful to discriminate the background signal and also confirm the measurement of dark matter particles\cite{PhysRevD.61.101301}.\par
\begin{figure}[tbp]
\centering
\graphicspath{{./fig/}}
\includegraphics[scale=0.4]{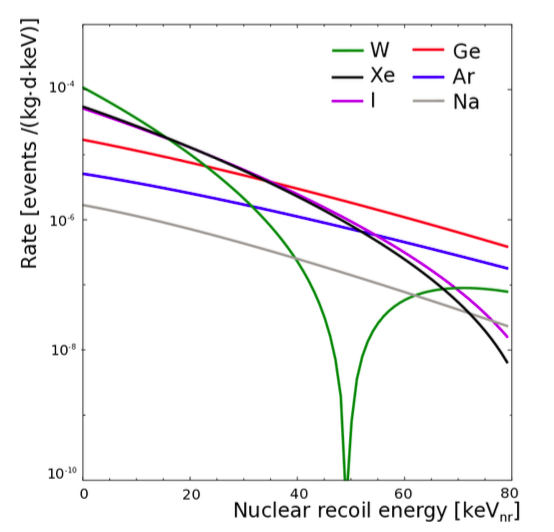}
\caption{Differential event rate for direct detection of a 100 GeV/c$^2$ WIMP with a cross-section of 10$^{-45}$ cm$^2$ for different material target. Plot is taken from \cite{0954-3899-43-1-013001}. }
\label{fig:event_rate_material}
\end{figure}

The WIMP-nucleus cross-section in Eq. \ref{eq:diffcross} can be written as the sum of a spin-independent (SI) contribution and spin-dependent (SD) contribution :
\begin{ceqn}\begin{align}
	\frac{d\sigma}{dE}=\frac{m_A}{2\mu_A^2v^2}\cdot \left(\sigma_0^{SI}\cdot F_{SI}^2(E) + \sigma_0^{SD}\cdot F_{SD}^2(E) \right).
\end{align}\end{ceqn}
The cross-section for spin-independent part can be expressed as
\begin{ceqn}\begin{align}
	\sigma_0^{SI} = \sigma_p\cdot \frac{\mu_A^2}{\mu_p^2}\cdot [Z\cdot f^p + (A-Z) \cdot f^n]^2
\end{align}\end{ceqn}
where $f^{p,n}$ denotes the contributions of protons and neutrons to the total coupling strength, respectively, and $\mu_p$ is the WIMP-nucleon reduced mass. A, Z are mass number and atomic number of target atom, respectively. In general, $f^p = F^n$ is assumed such that the cross-section is scaled according to $A^2$. Figure \ref{fig:event_rate_material} shows the event rate as a function of recoil energy and taking into account of form factor correction for different target material. For heavier elements will get higher events rates but also suffer from larger nucleus causing loss of coherence. On the other hand, the cross-section for spin-dependent can be determined from nuclear shell model\cite{PhysRevC.56.535,PhysRevC.79.044302} :
\begin{ceqn}\begin{align}
	\sigma_0^{SD} = \frac{32}{\pi}\mu^2_A\cdot G_F^2\cdot[a_p\cdot\langle S^p\rangle + a_n\cdot\langle S^n\rangle]^2\cdot \frac{J+1}{J}.
\end{align}\end{ceqn}  
where $G_F^2$ is the Fermi coupling constant, J is the total nuclear spin and $a_{p,n}$ is the effective proton (neutron) coupling, $\langle S^{p,n}\rangle$ is the expectation value of the nuclear spin content due to proton and neutron respectively. The chiral effective-field theory is used to carry out the couplings of WIMPs to nucleons\cite{PhysRevD.88.083516,PhysRevD.86.103511}. These calculations yield results in good agreement with the experimental data. 
\subsection{Detector Technologies}
Various detector materials can be exploit to produce dark matter signal in terms of phonon, charge or light signal.  Phonon signal comes from the incident particle induce the lattice vibrations. Typically, only few meV is needed to create  a photon in the solid target. For charged particles passing through a medium and ionize its atoms and produced charges  which can be collected by applying an electric field. In semi-conductor detector, few meV is required to create an electron-hole pair. While in the liquid noble gas scintillator, the photons are emitted by the relaxation of excited atom and the ionization energy is usually around 10-20 eV\cite{PhysRevA.9.1438,PhysRevA.12.1771}. For dark matter direct detection, several goals need to be achieved to successfully detect signals : 
\begin{itemize}
	\item Large detector mass : More detector mass will increase the probability to observe signal of WIMP-nucleus recoil.
	\item Low energy threshold of detector :  Low threshold allows the detector to observe low energy deposit in the detector.
	\item Low background : Some background will mimic the WIMP' signal while at low energy the electronic recoil will also resemble the WIMP's signal. Thus to eliminate the background will improve the signal significance.
	\item Stability : The detector need to be able to perform measurement continuously for several years to accumulate good statistics due to very low event rate. Thus a reliable and stable detector is needed.
\end{itemize}
\subsection{Scintillator Crystal}
\begin{figure}[tbp]
\centering
\graphicspath{{./fig/}}
\includegraphics[scale=0.3]{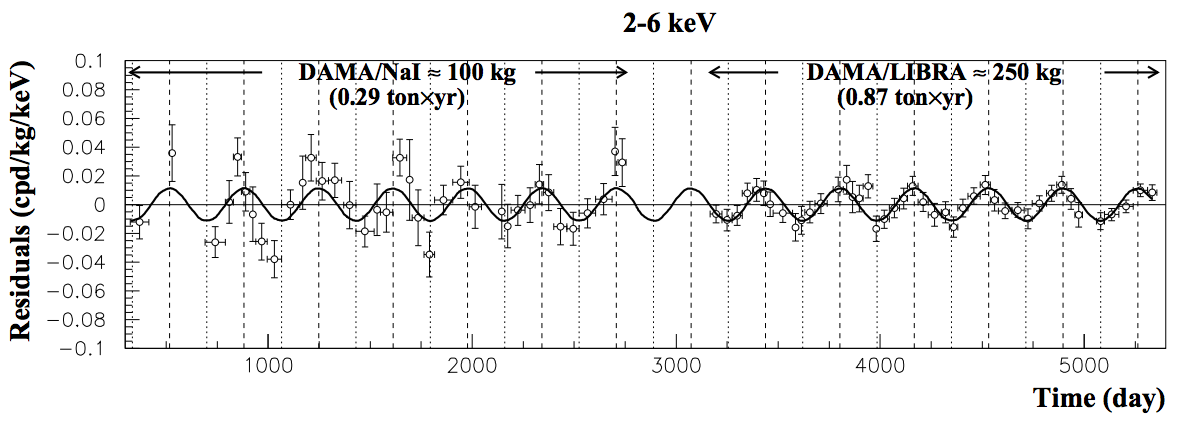}
\caption{Annual modulation of the measured residual single-hit event rate by the DAMA experiment in the (2-6) keV energy range. The superimposed curve is a sinusoidal function with a period of one year and a phase equal to 152.5 days (maximum on June 2 nd). The figure covers the period between 1996 to 2007. Figure from \cite{damaresults}. }
\label{fig:dama_mod}
\end{figure}

Scintillators are frequently used in particle physics. When particle pass through the crystal, the target atom will be excited and the subsequently de-excitation process will emit scintillation lights. NaI(Tl) and CsI(Tl) are mostly commonly used in the dark matter experiments. The scintillation lights are typically collected by Photomultiplier tube (PMT) provides the estimation of energy deposited by the incident particle. They have advantages with the high density (3.7 and 4.5 g/cm$^3$ for NaI and CsI) which gives larger probability for incident particle to deposit its energy in the detector target. In addition, they have good energy resolution (8\% for 1 MeV energy deposition) and lower energy threshold than other scintillator. However, no particle discrimination is possible, only the rejection of multiple hits in different crystal can be achieved. Therefore, the low background environment with active shielding are needed for crystal scintillator. The DAMA experiment at the LNGS underground laboratory using ultra low-radioactive NaI(Tl) crystal \cite{BERNABEI2008297}. With its successor DAMA/LIBRA, the combined 1.33 ton/year exposure shows the annual modulation signature of WIMPs-nucleus recoil\cite{Cerulli2017}. Figure \ref{fig:dama_mod} shows the modulation signal measured by DAMA. The maximum is agree with theoretical results at June 2nd within 2 $\sigma$. Moreover, the significance of dark matter signal reaches 9.3 $\sigma$ over a measurement of 14 annual cycles\cite{Bernabei2013}. DAMA experiment has demonstrated a stable long-term operation of dark matter direct detection. 
\subsection{Semi-conductor Detector}
\begin{figure}[tbp]
\centering
\graphicspath{{./fig/}}
\includegraphics[scale=0.7]{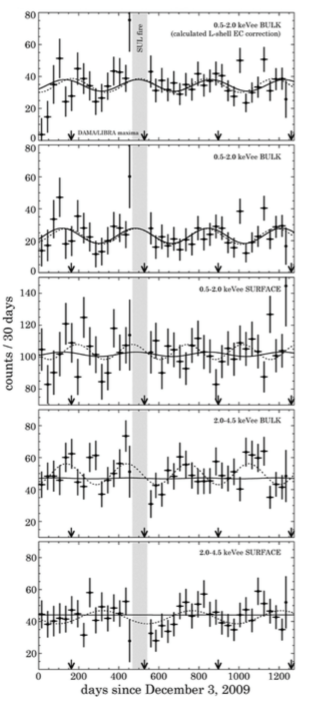}
\caption{Best-fit for four group of events. Vertical arrows point at the position of the DAMA/LIBRA modulation maxima\cite{Bernabei2010}. Figure from \cite{COGENT}. }
\label{fig:COGENT}
\end{figure}

Among semi-conductor detectors, the germanium detector is frequently choose to be used for direct detection. The germanium has a high radio-purity target material and a very low threshold ($\sim$ 0.5 keV$_{ee}$) allowing to search for WIMPs down to masses of a few GeV/c$^2$. With such low threshold, the germanium detector usually operate under the liquid nitrogen temperature to reduce the noise level. Moreover, the noise level scales up with increasing crystal size due to increased capacitance, thus the optimization of the detector is needed\cite{doi:10.1146/annurev.ns.38.120188.001245}. Germanium detector exhibit a excellent energy resolution (0.15\% at 1.3 MeV) which gives a ability to identify the background sources and can be used to reduce the background. Although for discriminating electronic recoil from nuclear recoil is not possible, the rise-time of the signal can be used to discriminate surface backgrounds. CoGeNT experiment\cite{COGENT} utilizing p-type point contact germanium detectors to do direct detection of dark matter. The total mass of 443 g and has energy threshold at 500 eV$_{ee}$ which has acquired 3.4 years of data in the Soudan Underground Laboratory. The annual modulation of dark matter signal has been reported by the group\cite{COGENT}, with a phase corresponding to the expectation for WIMPs at a level of 2.2 $\sigma$. However, from several independent analysis using different background model, no significant signal was found\cite{1475-7516-2014-08-014,COGENT_2}. The results from CoGeNT is shown in Fig. \ref{fig:COGENT}.
\subsection{Cryogenic Liquid noble gas detector}\label{sec:lquiddetector}
\begin{figure}[tbp]
\centering
\graphicspath{{./fig/}}
\includegraphics[scale=0.3]{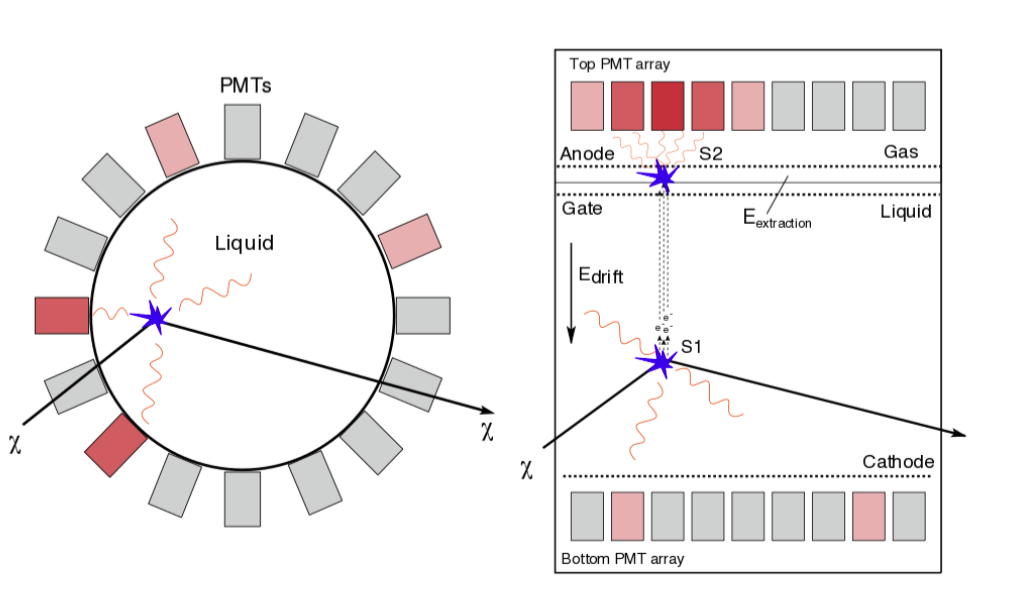}
\caption{Schematic of single phase (left) and dual phase (right) liquid noble gas detector. Figure from \cite{0954-3899-43-1-013001}. }
\label{fig:single_dual}
\end{figure}

Liquid noble gas detector provides high light yield and advantage of building large and homogenous detector. Currently, most experiments choose liquid argon (LAr) or liquid xenon (LXe) for building detector for direct dark matter detection. Two common design are used : single and dual phase. For single phase detector, the scintillation light from particle interact with atoms is the only signal will be observed. LAr is a good candidate for detector material due to the very different lifetime for two components in scintillation lights(see Chapter \ref{ch:scin}). This feature allows the LAr to perform the pulse shape discrimination (PSD) and can be used to discriminate electronic recoil from nuclear recoil. On the other hand, LXe is not suitable for constructing single phase detector due to small difference of lifetime between two components. However, with design of dual phase LXe detector, the ionization signal can be used to do the PSD. In LXe dual phase detector, the first signal comes from liquid phase. The scintillation light in liquid phase will produce light signal (S1), subsequently the electrons escape from ionization process will drift to the gaseous phase at the top of detector due to the external applied electric field ($\sim$ 10 kV/cm). When these electrons accelerated by external electric field reaches the gaseous phase will produce the scintillation light through electroluminescence and collect by the PMTs (S2). Using the ratio of S1/S2, the discrimination of electronic recoil against nuclear recoil can be achieved. Figure \ref{fig:single_dual} shows the basic design schematic of both single and dual phase liquid noble gas detector. Table \ref{tab:darkexperiment} summarize current dark matter experiments based on direct detection technique.

\begin{table}[htbp]
\caption{List of Dark matter direct detection experiment.} 
\centering 
\begin{adjustbox}{width=1\textwidth}
\begin{tabular}{c c c c} 
\toprule[1.5pt]
Experiment & Type & Signal type & Target Material  \\\midrule
DAMA/LIBRA\cite{Cerulli2017} & Solid scintillator & pulse shape discrimination & NaI(Tl)\\
SuperCDMS\cite{PhysRevD.95.082002} & Semi-conductor &  phonon/charge & Ge\\
EDELWEISS\cite{ARMENGAUD201751} &Semi-conductor &  phonon/charge & Ge\\
CRESST\cite{1475-7516-2014-05-018} &Crystal & phonon &CaWO$_4$ \\
ArDM\cite{Badertscher:2013soa} & Dual phase & light/ionization &LAr \\
WArP\cite{1742-6596-308-1-012005} & Dual phase& light/ionization & LAr\\
XENON100\cite{PhysRevD.96.022008} &Dual phase &light/ionization & LXe\\
LUX\cite{PhysRevLett.118.021303}&Dual phase &light/ionization & LXe\\
DarkSide\cite{Aalseth:2017fik}& Dual phase& light/ionization & LAr\\
PandaX\cite{PhysRevLett.117.121303} &Dual phase &light/ionization & LXe\\
DEAP3600\cite{Amaudruz:2017ekt}&Single Phase & light &LAr\\
MiniCLEAN\cite{Rielage:2014pfm} &Single Phase & light &LAr\\\bottomrule[1.25pt]
\end{tabular}
\end{adjustbox}

\label{tab:darkexperiment} 
\end{table}
\subsection{Current Results Comparison}
\begin{figure}[tbp]
\centering
\graphicspath{{./fig/}}
\includegraphics[scale=0.3]{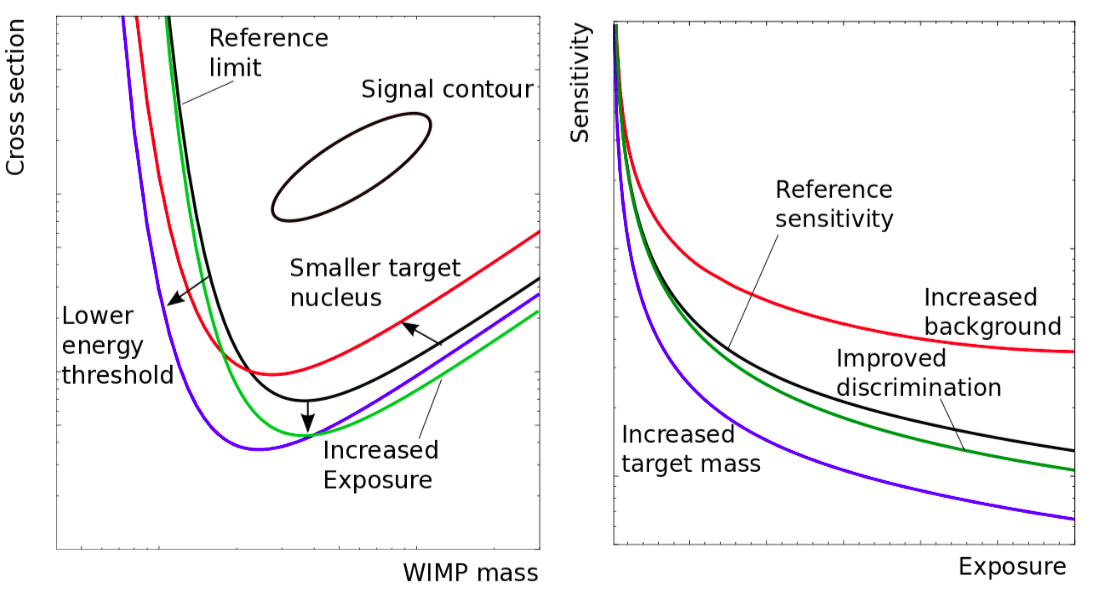}
\caption{Left: Illustration of a result from a direct dark-matter detector derived as a cross-section with matter as function of the WIMP mass. The black line shows a limit and signal for reference, while the coloured limits illustrate the variation of an upper limit due to changes in the detector design or properties. Right: Evolution of the sensitivity versus the exposure. Figure from \cite{0954-3899-43-1-013001} }
\label{fig:concept_exclusion}
\end{figure}

\begin{figure}[htbp]
\hfill
\subfloat[]{\includegraphics[width=7cm]{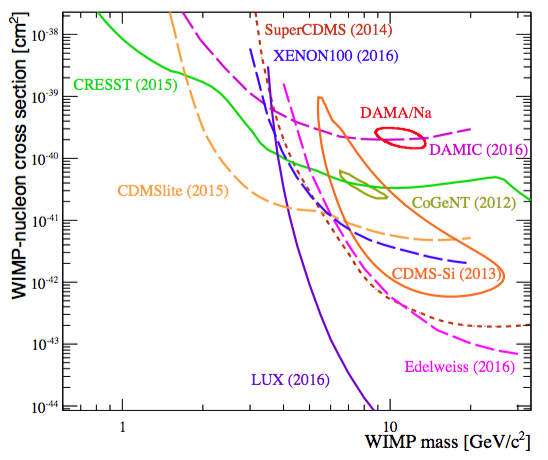}}
\hfill
\subfloat[]{\includegraphics[width=7cm]{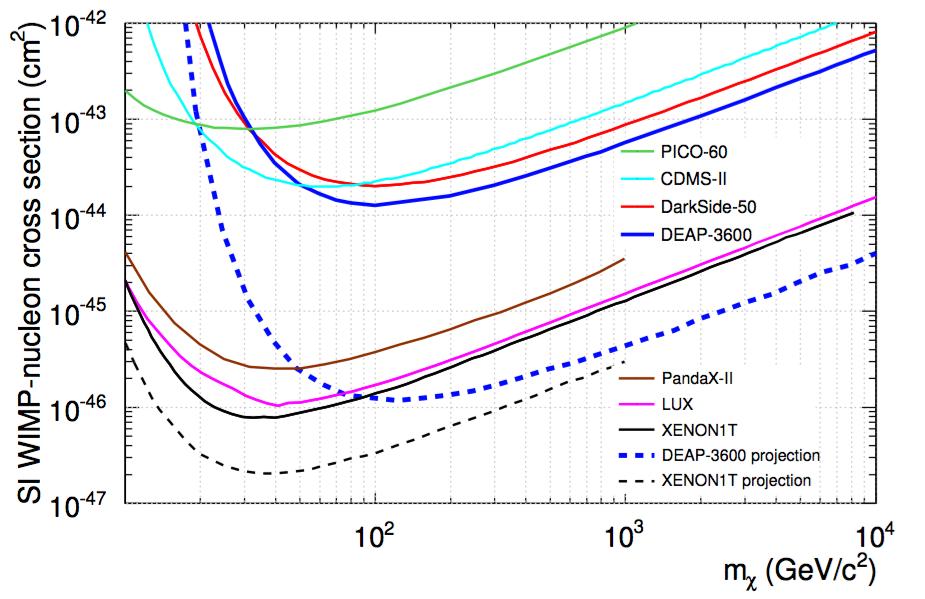}}
\hfill
\caption{(Overview of signal indications and exclusion limits from various experiments for spin-independent WIMP-nucleon cross-section for low WIMP masses (left)\cite{0954-3899-43-1-013001} and high WIMP masses (left)\cite{Amaudruz:2017ekt}.}
\label{fig:exclusion_low}
\end{figure}

Thus far, no WIMPs signal has been confirmed. Therefore, the exclusion plot of WIMPs-nucleus cross-section is the results from each experiment at the moment. For different type of detector will have different sensitive region on the exclusion plot. Figure \ref{fig:concept_exclusion} shows some basic principle for detector has different energy threshold and target nucleus. The differential rate for spin-independent interactions can be expressed as :
\begin{ceqn}\begin{align}
	\frac{dR}{dE}(E,t) = \frac{\rho_0}{2\mu_A^2\cdot m_{\chi}}\cdot\sigma_0\cdot A^2\cdot F^2 \int_{v_{min}}^{v_{esc}}\frac{f(\textbf{v},t)}{v}d^3v,
\end{align}\end{ceqn}
where the escape velocity is 544 km/s\cite{Smith:2006ym} and $v_{min}$ is 
\begin{ceqn}\begin{align}
	v_{min} = \sqrt{\frac{m_A\cdot E_{thr}}{2\mu_A^2}}.
\end{align}\end{ceqn}
where the $E_{thr}$ represents the energy threshold of the detector and $\mu_A$ is the reduced mass of WIMP-nucleus system. For experiments based on liquid noble gas detector, usually have large mass but higher energy threshold, is most sensitive to larger WIMP mass region. Conversely, the semi-conductor based detector has lower energy threshold, thus can probe to lower recoil energy and sensitive to lower WIMP mass region. However, they suffer from low mass of detector target, thus the probability of observing the WIMP-nucleus recoil is smaller, results in reduction of overall sensitivity. With improved discrimination also improves the overall sensitivity as shown in right panel of Fig. \ref{fig:concept_exclusion}. Figure \ref{fig:exclusion_low} shows the latest experimental results on exclusion plot. The results from each experiments is represent by color curve, and the parameter space above the curve is excluded. Nowadays, experiments are focus on specific region of parameter space to fully exploit the advantage of different type of detector. 


\chapter{Scintillation Process in Noble Liquid and Gas}\label{ch:scin}
   The scintillation light is the most important thing in single phase liquid noble gas detector. Especially in liquid argon, the very different decay time between fast and slow component provides excellent pulse shape discrimination, table \ref{tab:noblepro} summarize the basic properties of liquid noble gas. It is necessary to understand the scintillation process. Unfortunately, there has not yet a comprehensive theory to describe the energy  loss in liquid noble gas. However, many people have developed a working theoretical frame work that allow people to put in simulation and get reasonable results when compare to experimental data. To understand the process not only help to optimize detector setup but also crucial for detector response function, which will be useful when determine the energy scale of detector. 



\begin{table}[htbp]
\centering 
\begin{tabular}{ c c c c c c} 
\toprule[1.5pt]
& $^{2}_{4}$He & $^{10}_{20}$Ne &  $^{18}_{40}$Ar  & $^{36}_{84}$Kr & $^{54}_{132}$Xe \\\midrule
Liquid density (g/ml) &0.13 & 1.2 & 1.4 & 2.4 & 3.1\\
Boiling point (K) & 4.2 & 27.1 & 87.3 & 119.9 & 165.0\\
Electron yield ($e^{-}$/keV) & 39 & 46 & 42 & 49& 64\\
Photon yield ($\gamma$/keV) & 22 & 32 & 40 & 25 & 42\\
Singlet decay time (ns) & 10 & 10 & 7 & 7 & 5\\
Triplet decay time & 13 s &  15 $\mu$s & 1.5 $\mu$s & 85 ns & 27 ns\\
Scintillation wavelength (nm) & 80 & 78 & 128 & 148 & 175 \\
Radioactive isotope & None & None & $^{39}$Ar & $^{85}$Kr & $^{136}$Xe\\\bottomrule[1.25pt]
\end {tabular}
\captionof{table}{Relevant properties for scintillation of the noble liquid elements of interest for direct dark matter searches. } \label{tab:noblepro} 
\end{table}

\section{Particle energy transfer to liquid}
\begin{figure}[tbp]
\centering
\graphicspath{{./fig/}}
\includegraphics[scale=0.4]{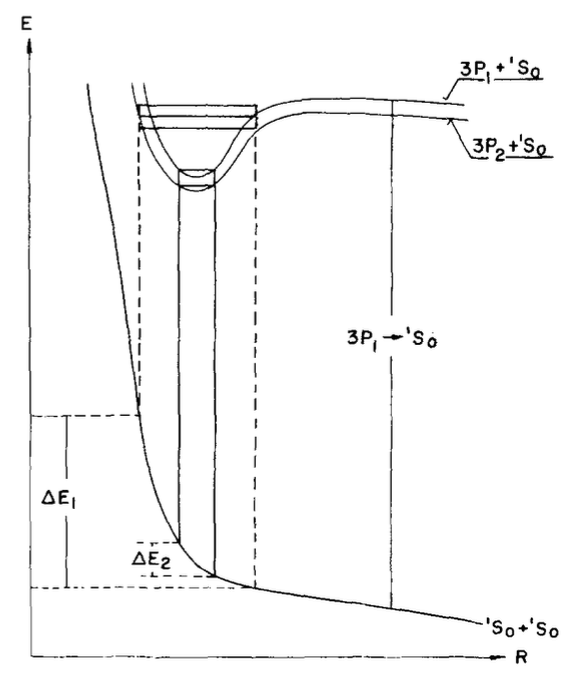}
\caption{ A schematic representation of the first and second continuum of the diatomic molecules of the rare gas. Figure is taken from \cite{doi:10.1063/1.1678779}.}
\label{fig:electronic_state}
\end{figure}

The dominant scintillation process in liquid noble gas due to the impinged particle is the scintillation light emitted by the lowest two vibrational states($^1\Sigma_u$, $^3\Sigma_u$)\cite{doi:10.1063/1.1672756} :  the bound excited $^1\Sigma_u$ molecular state (from $^3P_1 + ^1S_0$ atomic state) and of the $3\Sigma_u$ (from $^3P_2+^1S_0$) to the repulsive ground state (see Fig. \ref{fig:electronic_state}). The higher vibration states will also emit the scintillation light at range of 110 nm, however, is suppressed in the liquid(see Chapter\ref{sec:gaseous}). 
Although the transition directly from $^3\Sigma_u$ is forbidden, through the spin-orbital coupling\cite{0022-3719-11-12-024}, there's a small chance to transit to the ground state and emit a photon. This results in rather long lifetime for LAr ($\sim$ 1.5 $\mu$ s). However, as the coupling becomes stronger for molecules with higher atomic number, the triplet lifetime is significantly shorter for LXe ($\sim$ 27 ns).\par
These excited vibrational states are crated by either direct excitation from the ground state or radiative cascades from higher excited atomic states (or ionic states after recombination). The excitation process can be expressed :
\begin{center}
	$e^- + R \rightarrow R^* + e^-$ \\    
	$R^* + 2R \rightarrow R_2^{*} + R$\\
	$R_2^* \rightarrow 2R+\gamma$\\                 	
\end{center}
where R is noble element,  $R_2^*$ is the excited states and $\gamma $ is the VUV photon.  In addition, the VUV photon can be emitted through the ionization and recombine with electrons :
\begin{center}
	$e^- + R \rightarrow R^+ + 2e^-$ \\    
	$R^+ + 2R \rightarrow R_2^{*} + R$\\
	$R_2^+ + e^- \rightarrow R^{**} + R $ \\
	$R^** + 2R \rightarrow R_2^* + R$\\
	$R_2^* \rightarrow 2R + \gamma$\\	                 	
\end{center}
Note that the last step in recombination process is the same as the last step of excitation process, the ``dimer'' state is formed and emit the VUV photon. The electrons were ejected from atoms and undergo thermalization. The high-kinetic energy of the electrons is transfer to the surrounding medium and become non-thermal. After that, under the coulomb filed of the parent atom (now positive ion), the electrons perform a diffusive motion and may recombine with the positive ion or escape and becomes free electrons. For those free electrons, further recombination is possible. With external applied electric filed, these electrons may be collected to provide the ionization signal in dual phase detector (see Chapter \ref{sec:lquiddetector})\par
For different type of incident particles, the different degree of energy will transfer during the reaction and results in different ratio of the excitation/ionization process. For the electron recoil, assuming energy $E_0$ is transferred by incident particle, three processes will share the energy : ionization, excitation and non-radiative process (heat). The $E_0$ can be written as: 
\begin{ceqn}\begin{align}\label{eq:LET1}
E_0 = N_iE_i + N_{ex}E_{ex} + N_i\eta,
\end{align}\end{ceqn}  
where $E_i$ and $E_{ex}$ are the mean energy to create ionization and excitation; $N_i$ and $N_{ex}$ are the mean number of ionized and excited atoms, respectively. $\eta$ is the mean energy of secondary electrons which is generated in the first ionization and subsequently excite or ionize other atoms.  Below such energy, the electrons will just participate the elastic scattering and raises the temperature of the medium. It is found that the band structure of electronic states in solid noble gas also present in the liquified noble gas\cite{DOKE1976353}. Therefore , assuming the band gap $E_g$, the Eq. \ref{eq:LET1} can be rewritten as :
\begin{ceqn}\begin{align}\label{eq:LET2}
\frac{E_0}{E_g} = N_i\frac{E_i}{E_g} + N_{ex}\frac{E_{ex}}{E_g} + N_i\frac{\eta}{E_g},
\end{align}\end{ceqn}
The band gap was found to be $E_g$ = 14.2 eV for argon and 9.28 eV for xenon \cite{PhysRev.128.1562}. The W-value ($E_0/N_i$) is then defined as the average energy to produce an electron-ion pair. Eq. \ref{eq:LET2} can be further rewritten as :
\begin{ceqn}\begin{align}
\frac{W}{E_g} = \frac{E_i}{E_g} + \frac{E_{ex}}{E_g}\cdot \frac{N_{ex}}{N_i} + \frac{\eta}{E_g}.
\end{align}\end{ceqn}
The experimental W-value for argon is 23.6 $\pm$ 0.3 eV\cite{PhysRevA.9.1438} and 15.6 $\pm$0.3 for xenon\cite{PhysRevA.12.1771}. In addition, theoretical calculation\cite{PhysRevA.12.1771} of the ratio of $N_{ex}/N_{i}$ for argon (0.21) is in good agreement with experimental results (0.19\cite{PhysRevB.13.1649}). However, for liquid xenon, the discrepancy between experimental data (0.20) and calculation(0.06\cite{PhysRevA.12.1771}) is shown.\par
\begin{figure}[tbp]
\centering
\graphicspath{{./fig/}}
\includegraphics[scale=0.4]{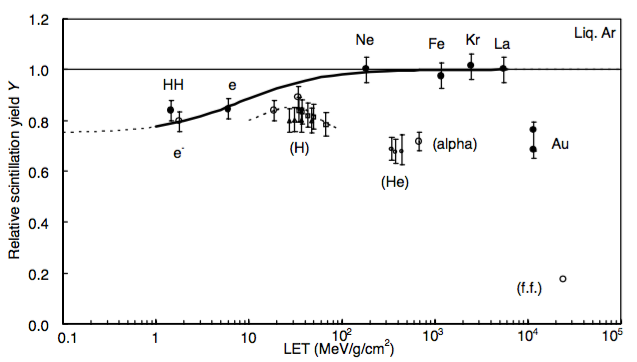}
\caption{ LET dependance of scintillation light yield in liquid argon. Solid circles show the yields for relativistic particles.Non-relativistic particles are represented by open circles. Open squares and triangles show the yields for non-relativistic protons whereas small open circles show those for non-relativistic heliumions\cite{1347-4065-41-3R-1538}.}
\label{fig:LET_low}
\end{figure}

The photon yield ($W_{ph}$) can be expressed in terms of the W-value and the ratio of number of excitations to the ionizations. 
\begin{ceqn}\begin{align}
W_ph = \frac{W}{1+\frac{N_{ex}}{N_i}}.
\end{align}\end{ceqn}
The different incident particle will transfer different amount of energy which results in different ratio of the excitations to the ionization, ultimately depends on the linear energy transfer (LET).  The relative light yield as a function of LET is shown in Fig. \ref{fig:LET_low}. The flat top response corresponds to the region where each of the excited and ionized species created by incident particle gives a photon. Notice that  at low LET region the light yield decreases. This can be explained by Onsager theory\cite{PhysRev.54.554}, in the ionization process, an electrons is slowed down to thermal energy within the Onsager radius from the parent ion. These electrons can not escape from the parent ion and the electron-ion pair recombination take place. The electron thermalization length in liquid argon is around 1500-1800 nm\cite{MOZUMDER1995143} and for xenon is around 4000-5000 nm\cite{MOZUMDER1995359}. These value are larger than Onsager radius (127 nm for liquid argon and 49 nm for liquid xenon). For high LET value created by nuclear recoil, a highly excited and ionized track is created by incident particle. Therefore, for electrons escape from the parent ion, there is a high probability that they will recombine with other ions in the track. However, for low LET, when these electrons escape from the parent ion, their lifetime in the liquid could be $>$ ms even without the external electric field applied, results in losing of light yield. In the high LET region, the process suffered from ``biexcitonic quenching''\cite{PhysRevB.46.11463}. 
\begin{center}
	$R^* + R^* \rightarrow R + R^+ + e^-$,
\end{center}
The two excitons collide with each other and results in non-radiative reaction. The excessive energy is carried away by the electrons which close to one excitation energy. These electrons will consume its energy before the further recombination. Conversely the $R^{+}$ can undergo the recombination with other electrons to produce a new excited state. This process will emit on photon, however, at the cost of two photons from original two excitons. This ``biexcitonic quenching'' effects mainly responsible for the reduction of light yield. \par
\begin{figure}[tbp]
\centering
\graphicspath{{./fig/}}
\includegraphics[scale=0.3]{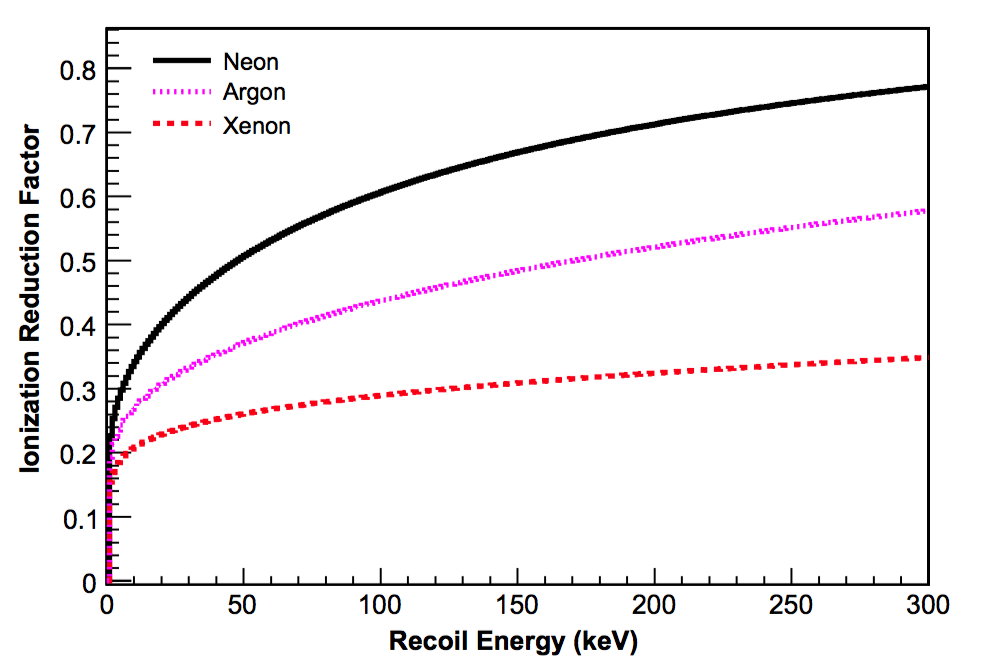}
\caption{ The reduction factor as a function of recoil energy for different noble elements. Plot is taken from \cite{MEI200812}.}
\label{fig:reduction_factor}
\end{figure}

\begin{figure}[tbp]
\centering
\graphicspath{{./fig/}}
\includegraphics[scale=0.4]{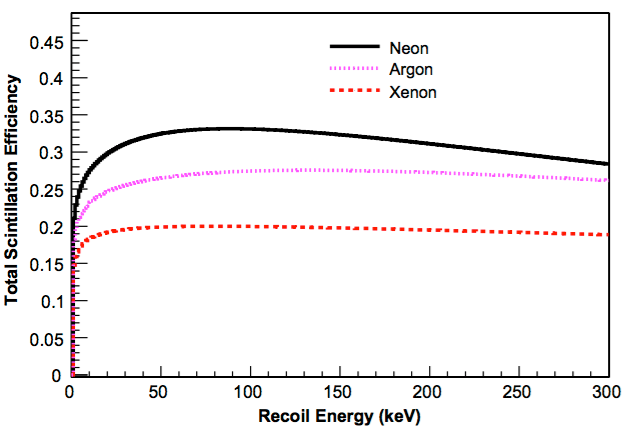}
\caption{ The scintillation efficiency for nuclear recoils as a function of recoil energy for different liquid noble elements. Plot is taken from \cite{MEI200812}}
\label{fig:scin_eff}
\end{figure}

\begin{figure}[tbp]
\centering
\graphicspath{{./fig/}}
\includegraphics[scale=0.25]{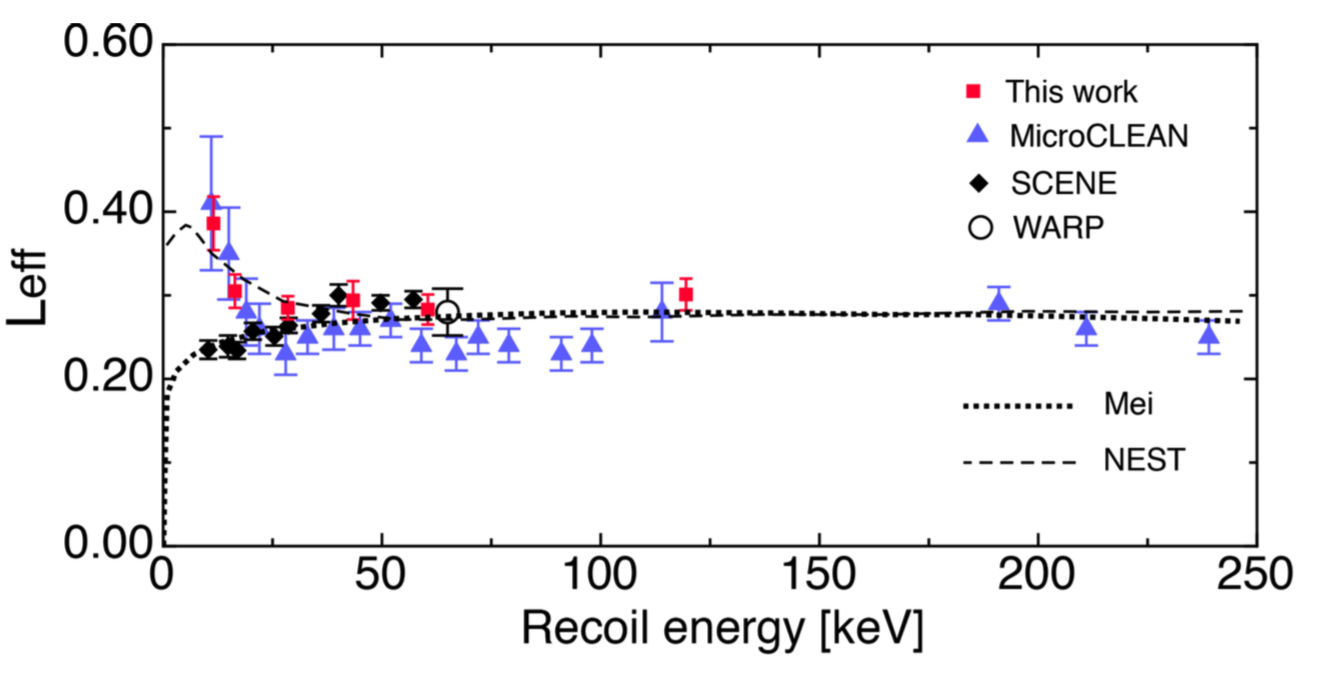}
\caption{ The relative scintillation efficiency ($L_{eff}$) as a function of recoil energy. Plot is taken from \cite{1748-0221-10-08-P08002}.}
\label{fig:Leff}
\end{figure}

In the nuclear recoil, not all the energy goes into producing excitation and ionization of the atoms. The nuclear stopping power defined as the amount of energy per unit length due to that transfer to recoiled atom in the form of kinetic energy. The energy reduction factor ($f_n$) from Lindhard \textit{et al.}\cite{Lindhard} can be written :
\begin{ceqn}\begin{align}
f_n = \frac{kg(\epsilon)}{1+kg(\epsilon)},
\end{align}\end{ceqn}
For nucleus  has atomic number A and Z protons, $\epsilon = 11.5E_R (keV)Z^{-7/3}$, $k=0.133Z^{2/3}A^{-1/2}$ and $g(\epsilon) = 3\epsilon^{0.15} + 0.7\epsilon^{0.6} + \epsilon$. The reduction factor as a function of recoil energy for different liquid noble element is shown in Fig. \ref{fig:reduction_factor}. In addition, the biexcitonic quenching introduce a extra quenching factor which can be defined using Birk's saturation\cite{MEI200812}:
\begin{ceqn}\begin{align}
f_l = \frac{1}{1+kB\frac{dE_e}{dx}}.
\end{align}\end{ceqn}
where $dE_e/dx$ represents the electronic stopping power, k is the collision probability at the core of the track, and A, kB are determined experimentally. The total scintillation efficiency($q_f$) can be obtained by 
\begin{ceqn}\begin{align}
q_f = f_n\times f_l.
\end{align}\end{ceqn}
The results from \cite{MEI200812} predicts the total scintillation efficiency for different liquid noble elements is shown in Fig. \ref{fig:scin_eff}. The scintillation efficiency of nuclear recoil relative to the electronic recoil can then be defined as :
\begin{ceqn}\begin{align}
L_{eff} = \frac{E_{er} \cdot n_{\gamma,nr}}{n_{\gamma,er} \cdot E_{nr}},
\end{align}\end{ceqn}
where the $E_{nr}$ and $E_{er}$ are the recoil energy of nuclear recoil and electronic recoil; $n_{\gamma,nr}$ and $n_{\gamma,er}$ are the number of photons detected in nuclear and electronic recoil respectively. In the recent experimental study, the experimental measured $L_{eff}$ seems disagree with the theoretical calculation using Lindhard theory and Birk's law as shown in Fig. \ref{fig:Leff}. Note that at low recoil energy, the theory predicts a down turn which means the light yield decreased, and is in agreement with earlier experimental results\cite{BRUNETTI2005265}. NEST \cite{BEZRUKOV2011119} developed a theoretical prediction to explain this phenomenon in liquid xenon\cite{1748-0221-6-10-P10002}. However, in the liquid argon, the experimental results from MicroCLEAN \cite{PhysRevC.85.065811} and W. Creus \textit{et al.}\cite{1748-0221-10-08-P08002} shows a up turn experimental data points, suggesting in low recoil energy, the light yield increased. No comprehensive theory has developed yet to explain the exact behavior in all energy region. According to W. Creus \textit{et al.}\cite{1748-0221-10-08-P08002}, this phenomenon could due to increasing ratio of exciton-ion in low recoil energy region. In that case, the light yield will increase since the exciton required less energy to emit the photons.
\section{Scintillation Process in Gaseous Argon}\label{sec:gaseous}
The gaseous argon has been widely investigated throughout the literature. The common application includes high pressure argon gas for UV laser, scintillation counter and dark matter detection etc. The scintillation light emit from gaseous argon by impinging charged particle, heavy ions or neutral particles is a well known phenomenon. When the particles pass through the gas, it creates so-called "excimer" states along the particle track.  These "excimer" states are either in the form of singlet excimer state ($^{1}\Sigma^{+}_{u}$) or triplet excimer state ($^{3}\Sigma^{+}_{u}$). Subsequently through the direct de-excitation or recombination the scintillation light are emitted in the process. The lifetime for singlet and triplet state are 6 ns and 3.2 $\mu$s respectively. The triplet lifetime has been measured by various groups(Table \ref{tab:title}), the main reason for the differences of measured lifetime is from impurity effect.\cite{doi:10.1063/1.452869} When the impurities present in the gaseous argon ($O_{2},N_{2},H_{2}O, etc.$) , there's a chance that argon excimer collide with impurities and going through a non-radiative collisional reaction. Thus this quenching process is in competition with the de-excitation process leading to VUV light emission. The singlet states are not affected by this process due to its very fast decay time. Therefore, the impurity effect mainly involved with triplet states, subsequently reduce the triplet lifetime. Thus the measured triplet lifetime is a good indicator for understanding the impurity level in GAr.\par
\begin{figure}[tbp]
  \centering
  \subfloat[]{\includegraphics[width=0.4\textwidth]{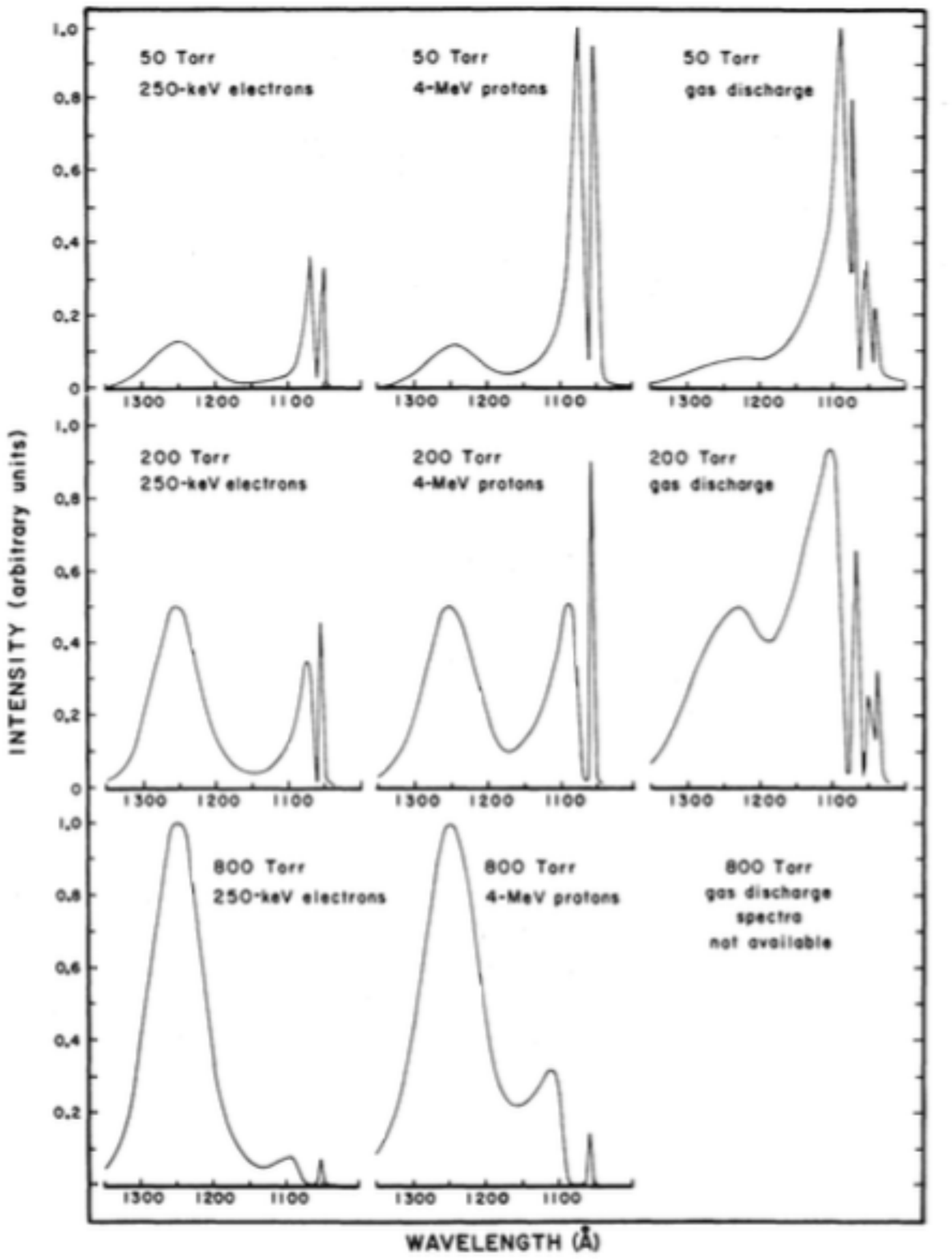}\label{fig:f1}}
  \hfill
  \subfloat[]{\includegraphics[width=0.6\textwidth]{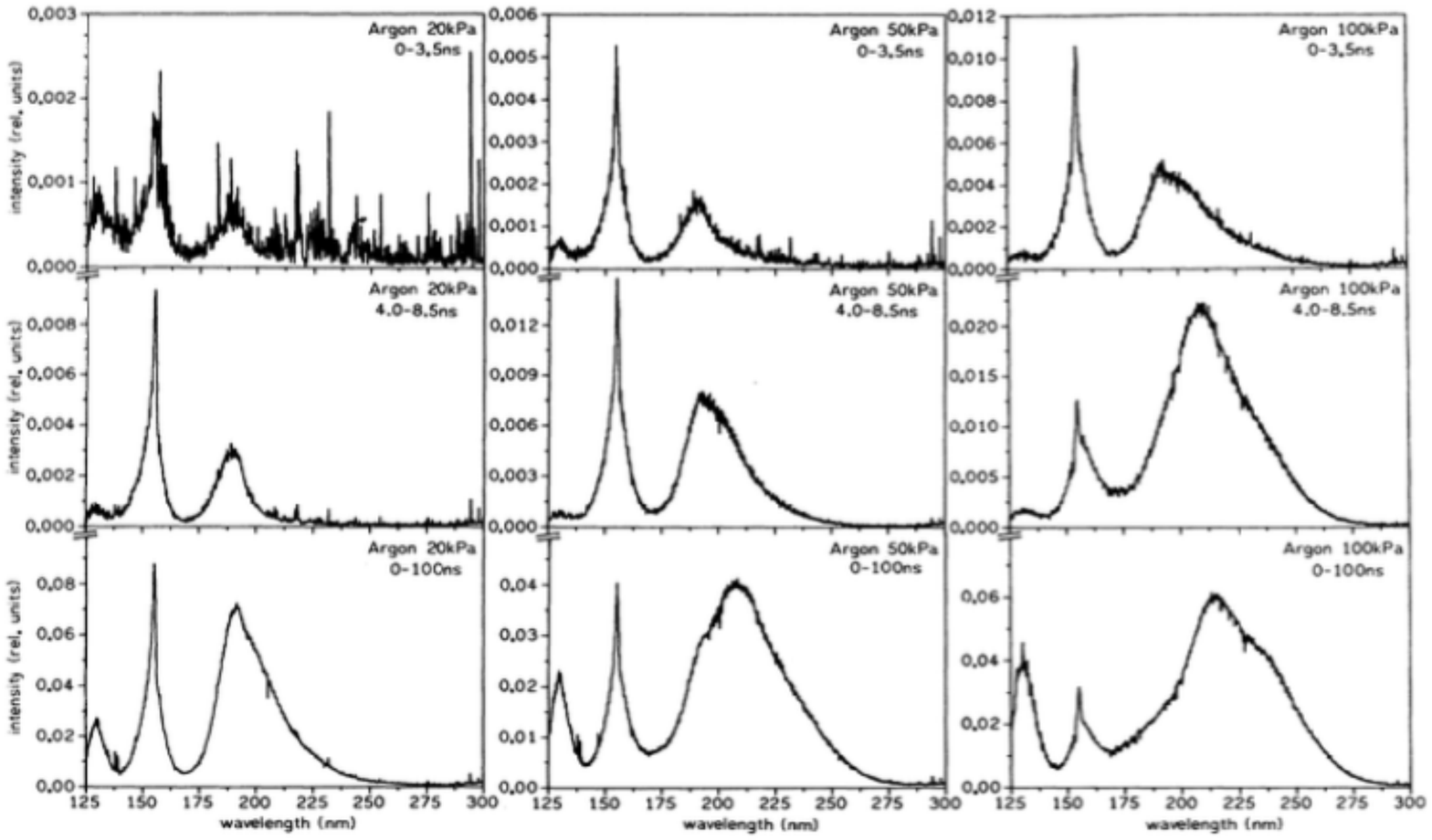}\label{fig:f12}}
  \caption{VUV/UV emission spectrum of argon. (a) Comparison of the vuv spectra of argon when excited by 250 keV electrons, 4 MeV photons, and gas discharges under the different pressure.\cite{PhysRevA.5.1110} (b) Time resolved emission spectra of argon under different pressures.\cite{PhysRevA.43.6089}}
\end{figure}

Traditionally, the scintillation light emitted from gaseous argon is considered from the de-excitation process of two lowest vibration state ($^{1}\Sigma^{+}_{u}$ , $^{3}\Sigma^{+}_{u}$) to the ground state ($^{1}\Sigma^{+}_{g}$) with characteristic wavelength peak at 128 nm. To detect the VUV scintillation light, the detector usually equipped with wavelength shifter (e.g. TPB) which can convert the VUV light to visible region where the PMT has highest detecting efficiency. The wavelength shifter will integrate all the scintillation light in VUV/UV region and convert them to visible photons.  The emitted scintillation light from 128 nm is dominant at typical operation pressure of detector (> 1bar),  however,there are actually some scintillation light from some other wavelength contribute to the total scintillation light. Fig \ref{fig:f1} and \ref{fig:f12} show the spectral of scintillation light in VUV/UV region.\par
\begin{figure}[tbp]
\centering
\graphicspath{{./fig/}}
\includegraphics[scale=0.4]{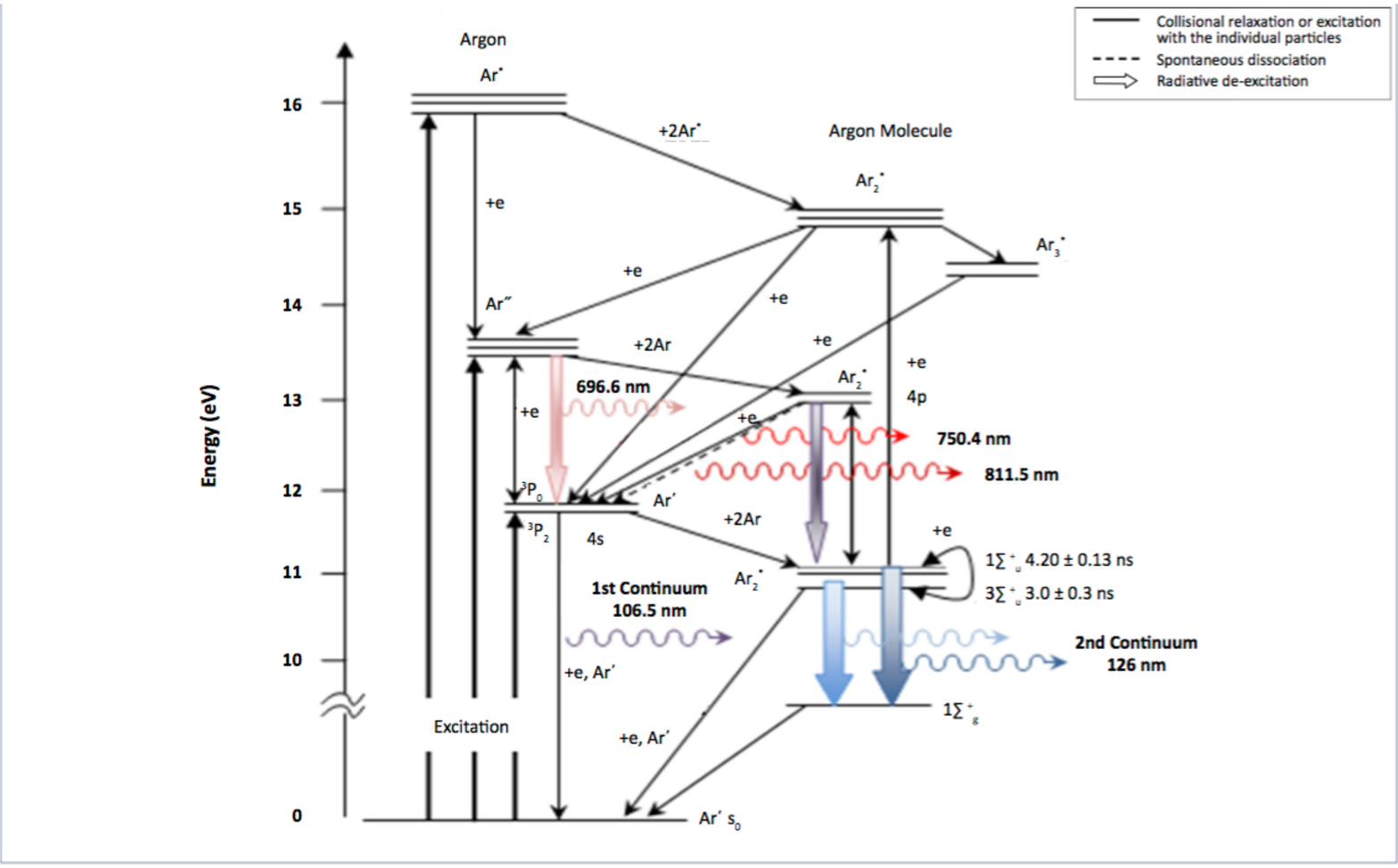}
\caption{ Energy level for atomic and molecular argon excited by a dielectric barrier discharge, showing transitions leading to the 126 nm, 750.4 nm and 811.5 nm emissions.}\label{fig:f2}
\end{figure}
\begin{figure}[!tbp]
  \centering
  \subfloat[]{\includegraphics[width=0.4\textwidth]{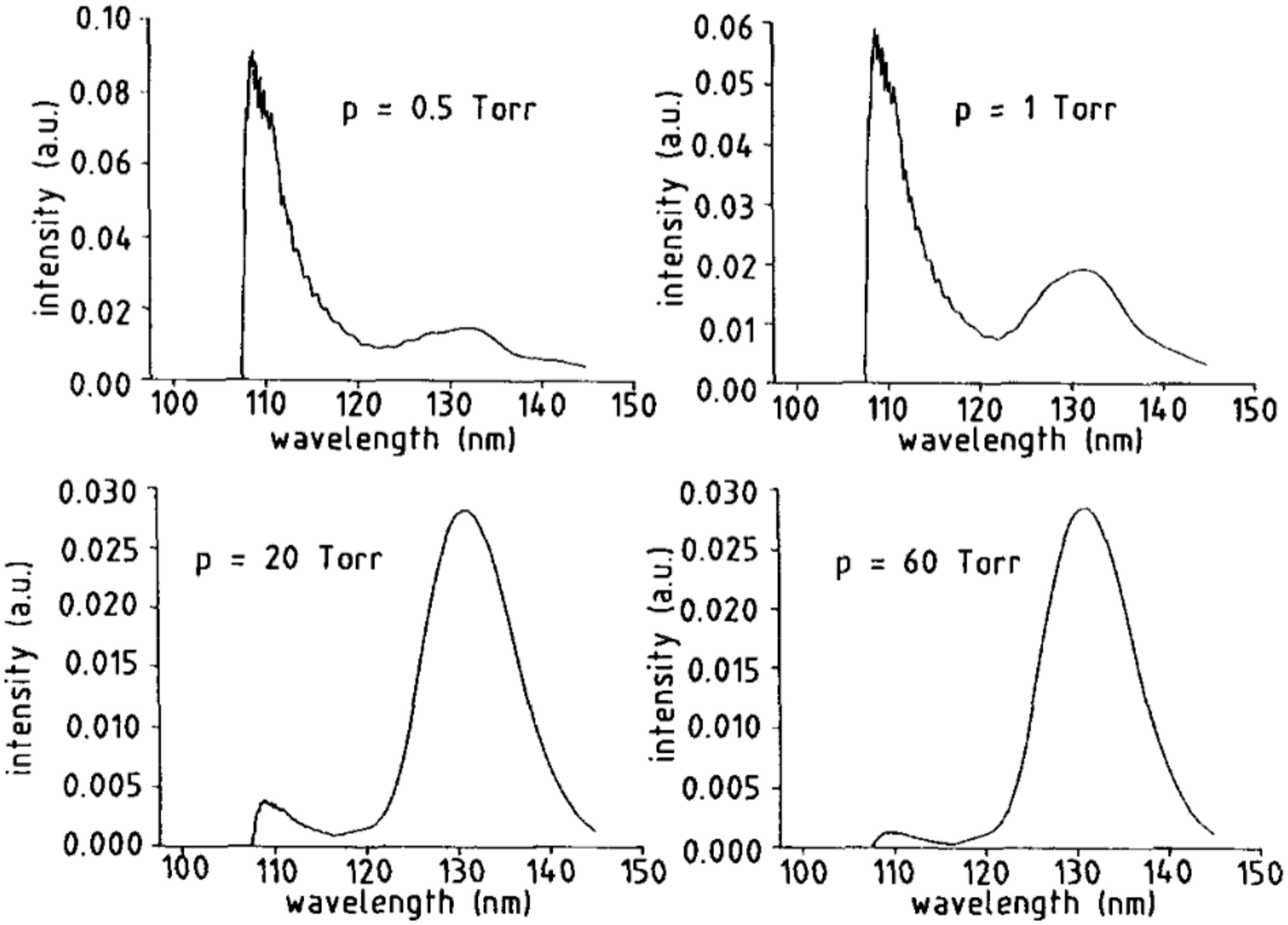}\label{fig:f3}}
  \hfill
  \subfloat[]{\includegraphics[width=0.4\textwidth]{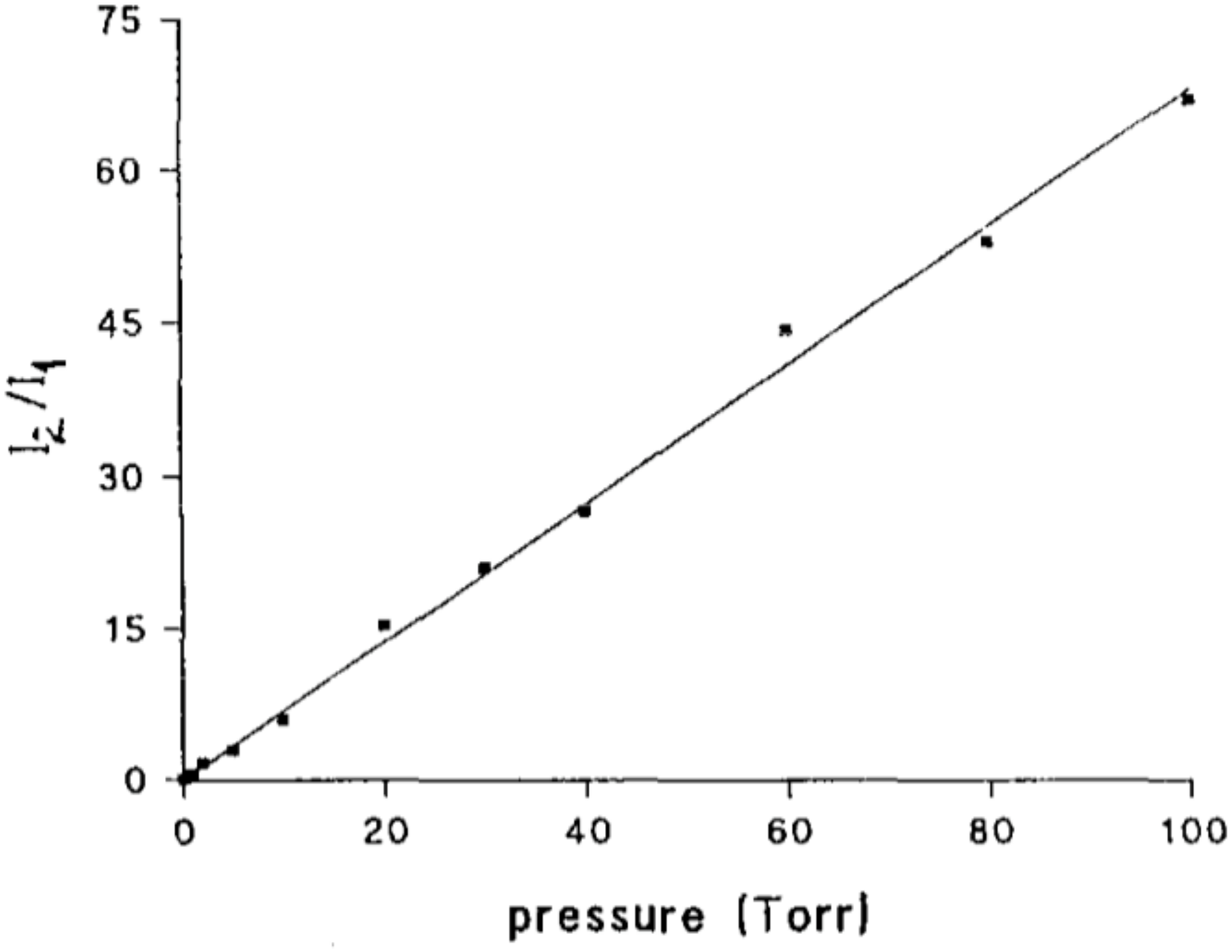}\label{fig:f32}}
  \caption{VUV emission spectrum of argon. (a) Calculated VUV spectra at pressure p=0.5,1,20,60 Torr (b) Linear dependence of $I_{2}/I_{1}$, upon the gas pressure}
\end{figure}

In atomic physics, these spectral line are referred to three continua. The first continuum ranges from 104 nm - 110 nm, second continuum peak at 128 nm and the third continuum ranges from 180 nm - 230 nm.
The origin of the second continuum is mentioned above, from the two lowest lying vibrational states of ($^{3}P_{1}$ and $^{3}P_{2}$) as shown in Fig \ref{fig:f2}.\cite{doi:10.1063/1.452869} The first continuum share the same origin with second continuum but from higher vibrational states.\cite{PhysRevA.5.1110}\cite{Goubert1994360} The origin of third continuum still under debate, several author have been trying to explain the kinetics behind it.\cite{doi:10.1063/1.436447}\cite{PhysRevA.43.6089}\cite{0953-4075-27-9-014}\cite{Grimm1987394}. From most recent research, it seems at least four different states are involved in the process and possibly from three body collision of Ar$^{++}$ (Ar$^{+*}$) with ground state argon atom. This process lead to the creation of Ar$_{2}^{++}$(Ar$^{+*}$) and then decay radiatively. In general, these three continua are pressure dependent. In fig.\ref{fig:f1}, the intensity of first continuum decreased with pressure increased. On the other hand, at low pressure the second continuum is not obvious, it becomes dominant when the pressure reaches 600 Torr while the first continuum is negligible.\cite{Goubert1994360}\cite{PhysRevA.5.1110} In the very low pressure (<10 mTorr), the lifetime of first continuum is around 160 ns\cite{PhysRevA.38.3456}, however with increased pressure it suffered from \textbf{radiation trapping} or \textbf{imprisonment}.\cite{PhysRev.83.1159}\cite{Boxall1975281} The simplified explanation is following : at low pressure the density of argon is also low (consider fixed T), so when the excited argon decay to the ground state, the released photons can easily escape from the gas without being absorbed. Conversely, when the pressure increased (density increased), the released photons can go through re-absorb and re-emit process many times such that we observer these photons late in the time. In the other words, the lifetime of these higher vibrational states are unchanged (natural lifetime) but the observed lifetime increased (apparent lifetime). The apparent lifetime for first continuum can be as long as 8 $\mu$s\cite{Boxall1975281}.  From the semiclassical calculation, the ratio of intensity from second continua and first continua($I_{2}/I_{1}$) increases with pressure increased(Fig. \ref{fig:f3}). For the second continuum, the formation of two lowest lying excited states are from three-body collision, so at low density(low pressure) the efficiency for forming these two states is low compare to the high density(high pressure). The difference of lifetime between singlet and triplet states is due to the forbidden transition from triplet to ground state. However, at short internuclear distance, spin-orbit coupling splits the triplet states, giving a slight oscillator strength to the ground state. \par

\begin{figure}[tbp]
\centering
\graphicspath{{./fig/}}
\includegraphics[scale=0.3]{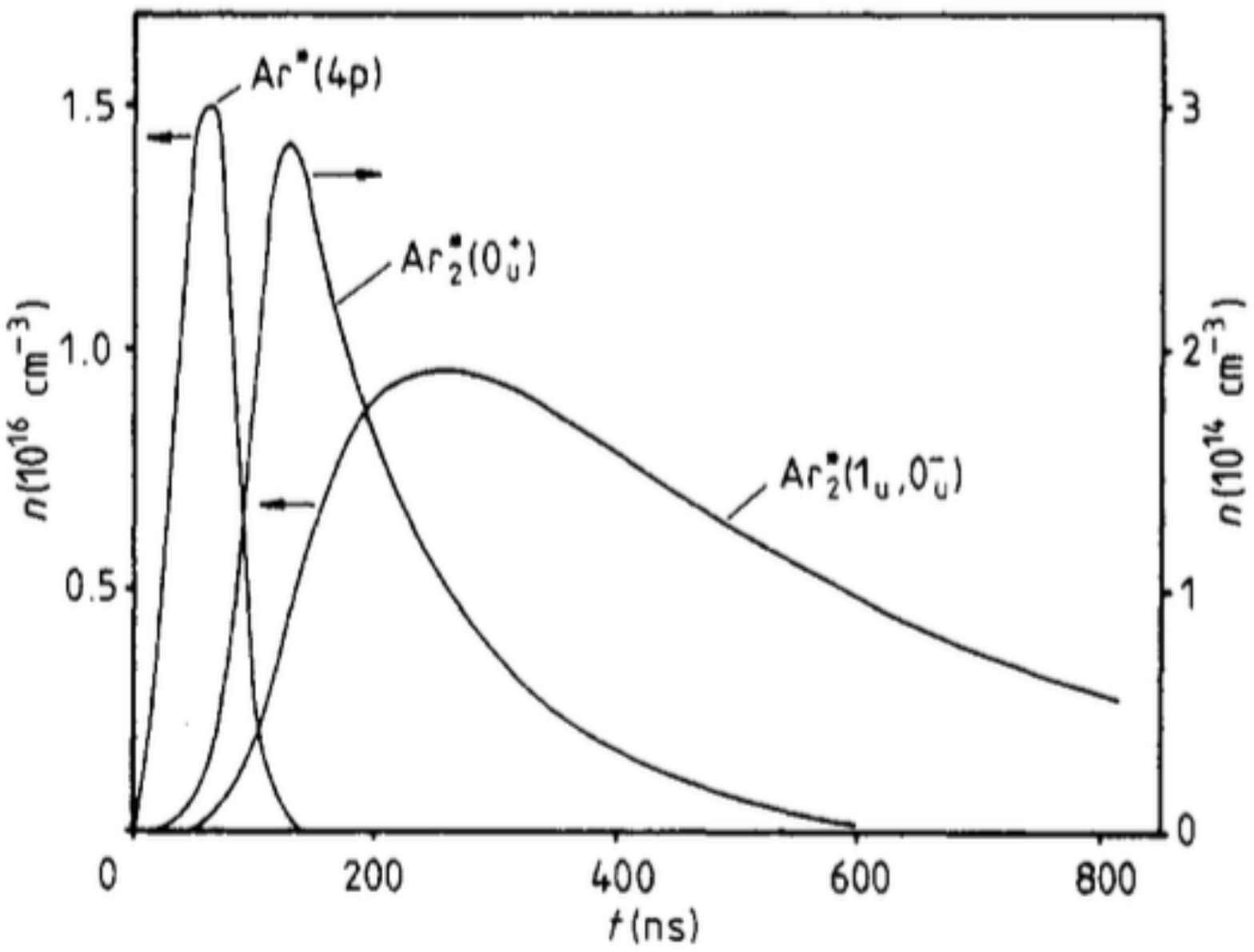}
\caption{ Results from the model calculations for an argon pressure of two bar. Where the Ar$^{*}$ contribute to the third continuum and the two lowest vibrational states of Ar$^{*}_{2}$}\label{fig:f4}
\end{figure}

It is also interesting to consider the timing analysis of these three continuum. The following discussion will be constraint to the pressure larger than 1 bar, since the first continua is suppressed  under this pressure region, I will focus on the second and third continuum.  In the early time, the scintillation light is dominant by third continuum and 155 nm peak, while the second continuum shows up later in the time\cite{Wieser2000233}. In the early stage of excitation and ionization process, there are several scintillation process competing with each other and results in the slowing down the production of Ar$_{2}^{*}$. The rate constant of different kinetic reactions can be found in \cite{0022-3700-17-10-011}. Fig.\ref{fig:f4} shows a decomposed  of the different components of scintillation light. This shows that the long time constant is mainly from the decay of the triplet states of Ar$^{*}_{2}$ (The first continuum has been suppressed under pressure > 1 bar).\par
The impurities in gaseous argon will lead to the quenching of the light yield and the decreased triplet lifetime through the Jesse effect\cite{10.2307/3573980}. In the MiniCLEAN cold gas run, the detector is operating at 1.5 bar and 120 K, the dominant  impurities are from N${_{2}}$ and O$_{2}$. When the impurities exist in the argon gas, the precursor of excimer states might have chance to collide with the impurities and then go through non-radiative collisional reaction.
\begin{ceqn}\begin{align}
	Ar^{*}_{2} + R_{2} \rightarrow 2Ar + R_{2}
\end{align}\end{ceqn}
Where R = O,N. These processes has been widely investigated in the following literature\cite{doi:10.1063/1.431621}\cite{doi:10.1063/1.452869}\cite{doi:10.106chen}\cite{doi:10.1063sheldon}, and the rate constant and cross-section for specific impurity can be found in \cite{doi:10.1063/1.436447}.  The detail study using MninCLEAN's cold gas data will be present in Chapter \ref{ch:cold}.

\chapter{MiniCLEAN Detector}\label{ch:detector}
The detail description of MiniCLEAN detector is provided in this chapter.
\section{Detector overview}
\begin{figure}[htbp]
\centering
\graphicspath{{./fig/Detector/}}
\includegraphics[scale=0.3]{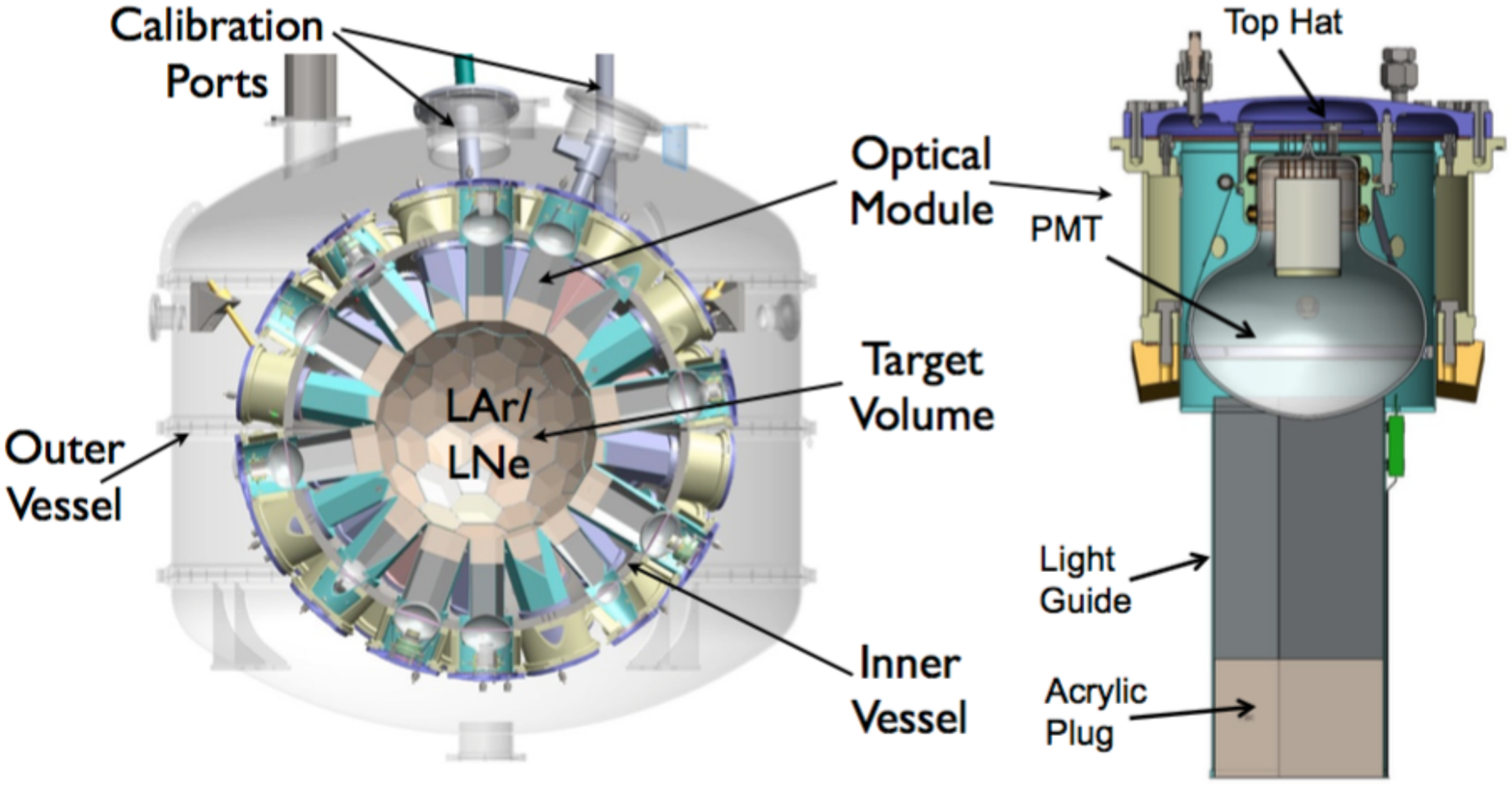}
\caption{ Left : The schematic of IV. Right : The schematic of optical cassette.  }
\label{fig:fivcas}
\end{figure}

\begin{figure}[htbp]
\centering
\graphicspath{{./fig/Detector/}}
\includegraphics[scale=0.3]{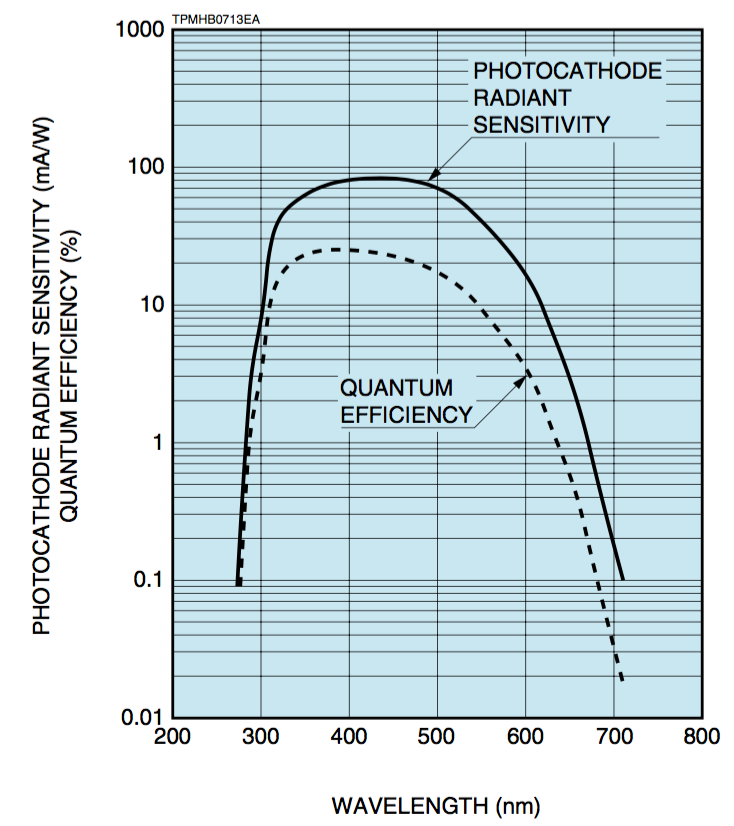}
\caption{ Typical spectral response characteristic of R5912-02. Figure from \cite{pmthama}. }
\label{fig:fpmtspec}
\end{figure}

\begin{figure}[htbp]
\centering
\graphicspath{{./fig/Detector/}}
\includegraphics[scale=0.3]{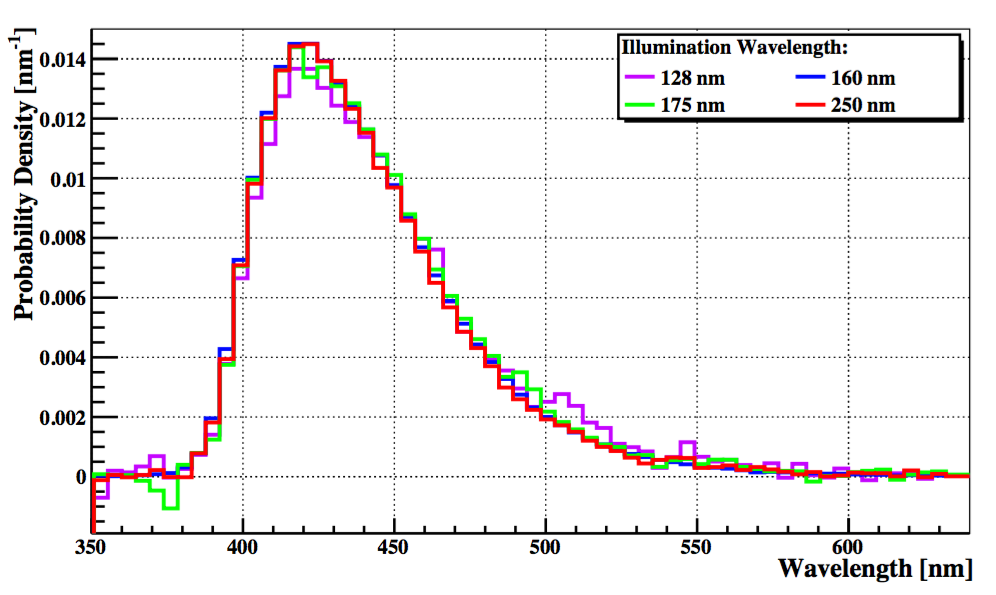}
\caption{ Visible re-emission spectrum for a TPB film illuminated with 128, 160, 175, and 250 nm light. Plot is taken from \cite{GEHMAN2011116}. }
\label{fig:TPB_reemit}
\end{figure}

\begin{figure}[htbp]
\centering
\graphicspath{{./fig/Detector/}}
\includegraphics[scale=0.3]{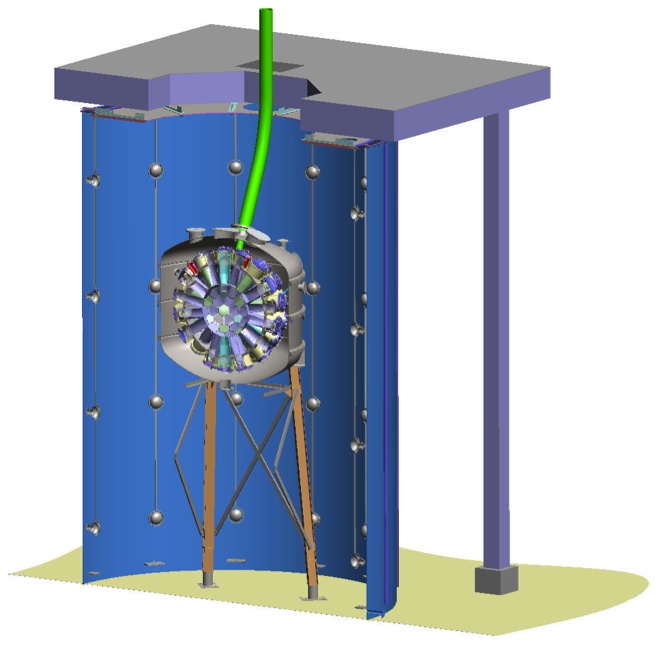}
\caption{ The MiniCLEAN water tank.  }
\label{fig:fwatertank}
\end{figure}

The \textbf{C}ryogenic \textbf{L}ow \textbf{E}nergy \textbf{A}strophysics with \textbf{N}oble Liquids (CLEAN) program utilize the liquid argon as the detector target. The detector is designed as a monolithic detector with maximum coverage of photomultiplier tubes viewing the active target. The liquid argon is held in a stainless steel Inner Vessel (IV) and surrounded by 92 optical cassettes (Fig. \ref{fig:fivcas}). The scintillation light in VUV range emitted from LAr is collected by 92 PMTs. The Hamamatsu R5912-02MOD photomultipliers has 8 inches diameters with borosilicate glass, the spectral response is shown in Fig. \ref{fig:fpmtspec}. The VUV scintillation light from argon is shifted to visible by Tetraphenyl butadiene (TPB) wavelength shifter (Fig. \ref{fig:TPB_reemit}). Each PMT is installed in a optical cassette as shown in Fig. \ref{fig:fivcas}. A 10 cm in thickness acrylic plug with TPB coated surface which is in contact of active volume is installed on the other end of cassette. In addition, the acrylic plug can moderate the neutron flux from the residual impurities in PMT glass which generated through U,Th ($\alpha$,n) decay chain. In order to improve the reflectivity of inside the optical cassette, the surface of light guide are lined with Vikuiti$^{TM}$ ESR foil by 3M. The light guides has different shape to maximize the coverage of the active volume. There are regular pentagon (12 ports), regular hexagon (20 ports) and irregular hexagon (60 ports) and is installed in different location of the port on IV sphere. The gaps between the optical cassettes are covered by ESR foil to minimize the photon leakage. The IV is held through three hangers inside the Outer Vessel (OV) which is pumped down to vacuum to serve as thermal insulation of IV.   The OV is sitting on a stand via a set of springs which dampen relative moment between the Cube Hall floor and the deck. The OV is inside an 18 ft diameter by 26 ft tall water tank (Fig. \ref{fig:fwatertank}) which provides shielding from external radiation to the detector. The water tank is located in the Cube Hall which is 6800 ft (6000 mwe) below surface in SNOLAB, sudbury, Canada.


\section{Data Acquisition System}
\begin{figure}[htbp]
\centering
\graphicspath{{./fig/Detector/}}
\includegraphics[scale=0.45]{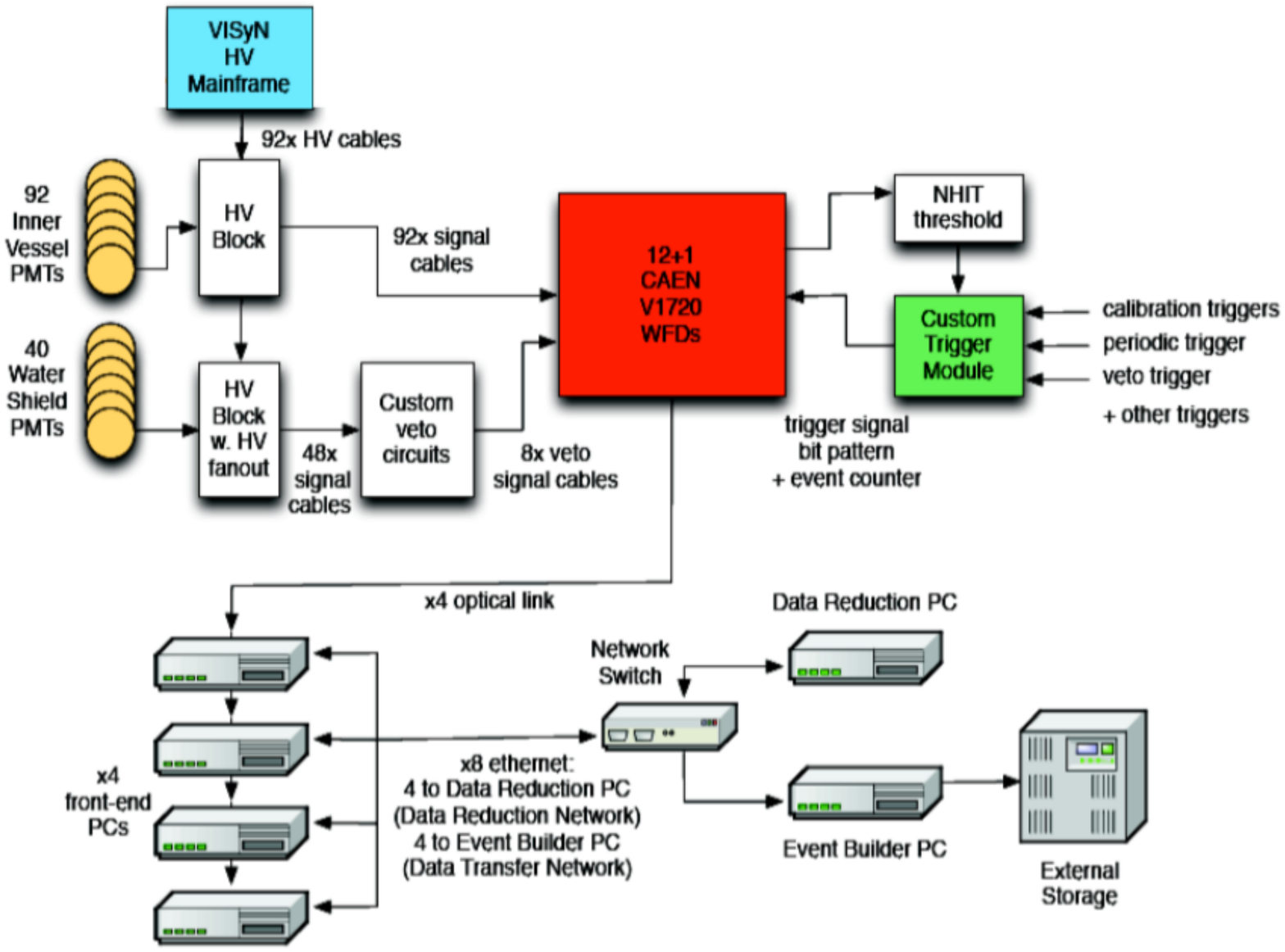}
\caption{ The diagram of DAQ system. }
\label{fig:fdaq}
\end{figure}

The DAQ system is build around 12 CAEN V1720 WFDs and each of WFD pick PMT pulses off of their HV lines, assembling pulses into events and saving to the disk.
Both the PMT signals and high voltage are transmit along a single cable (Gore 30 AWG in the OV vacuum and RG-68 in air).  A custom VME crete is housing the WFDs and fed signal from the PMTs via the HV-Block modules and are triggered by the VENTOR triggering system. Each WFD has 8 channel with 12-bit resolution over 2 V peak-to-peak input and 250 MHz sampling rate. The default length of the digitized waveform is set to 16 $\mu$s, which is approximately ten times of the slow scintillation time constant for LAr. The input signal from PMTs is digitized and output a signal if the number of channels above a programmable threshold. The hit sum (NHit) defined as five or more channels have crossed threshold within 16 ns coincidence window. The external triggers are allowed to trigger DAQ for calibration purposes. The diagram of the DAQ system is shown in Fig. \ref{fig:fdaq}.
The digitizer provides different waveform reduction to save the disk space. During the warm gas runs, the trigger rate is around 400 Hz, thus to store full waveform data for every events is impractical due to heavy traffic on data transferring. Therefore, the zero suppressed data (Zero Length Encoding) is stored. In ZLE mode, the region where the no samples cross the threshold will be discard. In addition, where the samples has crossed threshold, the programmable length pre- and post-sample regions are also stored in each ZLE block. Upon the trigger, the event summary was send to a PC which, in real time, determines wether the event could be of interest based on the amount of charge (prompt and late), and charge centroid. The event is categorized  depending on the event summary information, the DAQ system records them as either full ZLE waveform, summary information for each ZLE block, or summary data on channel level. The detail description can be found in Appendix \ref{app:reduction}.
\section{Purification and Cryogenic System}
\begin{figure}[htbp]
\centering
\graphicspath{{./fig/Detector/}}
\includegraphics[scale=0.35]{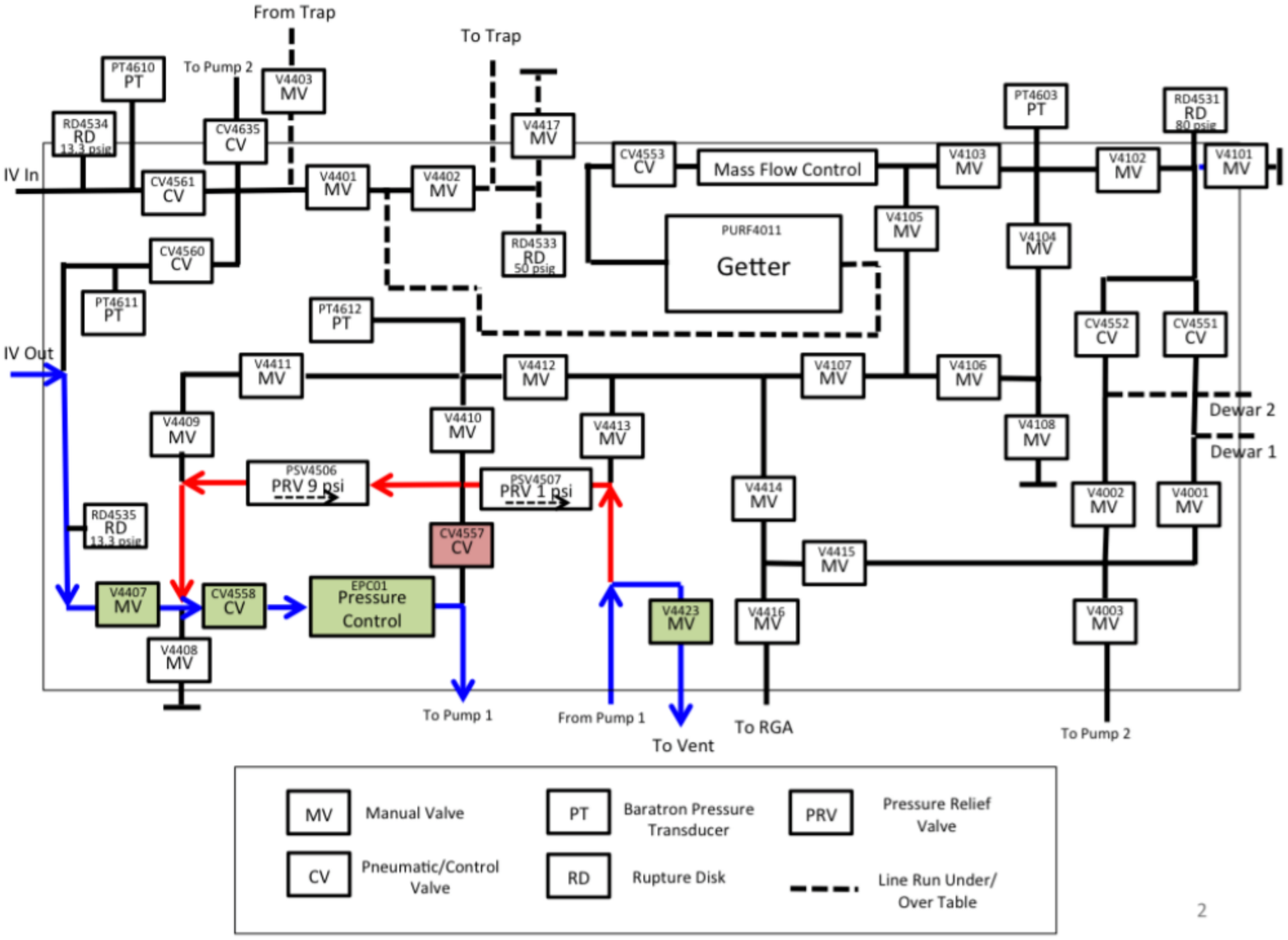}
\caption{ The schematic of purification system. }
\label{fig:Leak_source_purification_layup}
\end{figure}

\begin{figure}[htbp]
\centering
\graphicspath{{./fig/Detector/}}
\includegraphics[scale=0.25]{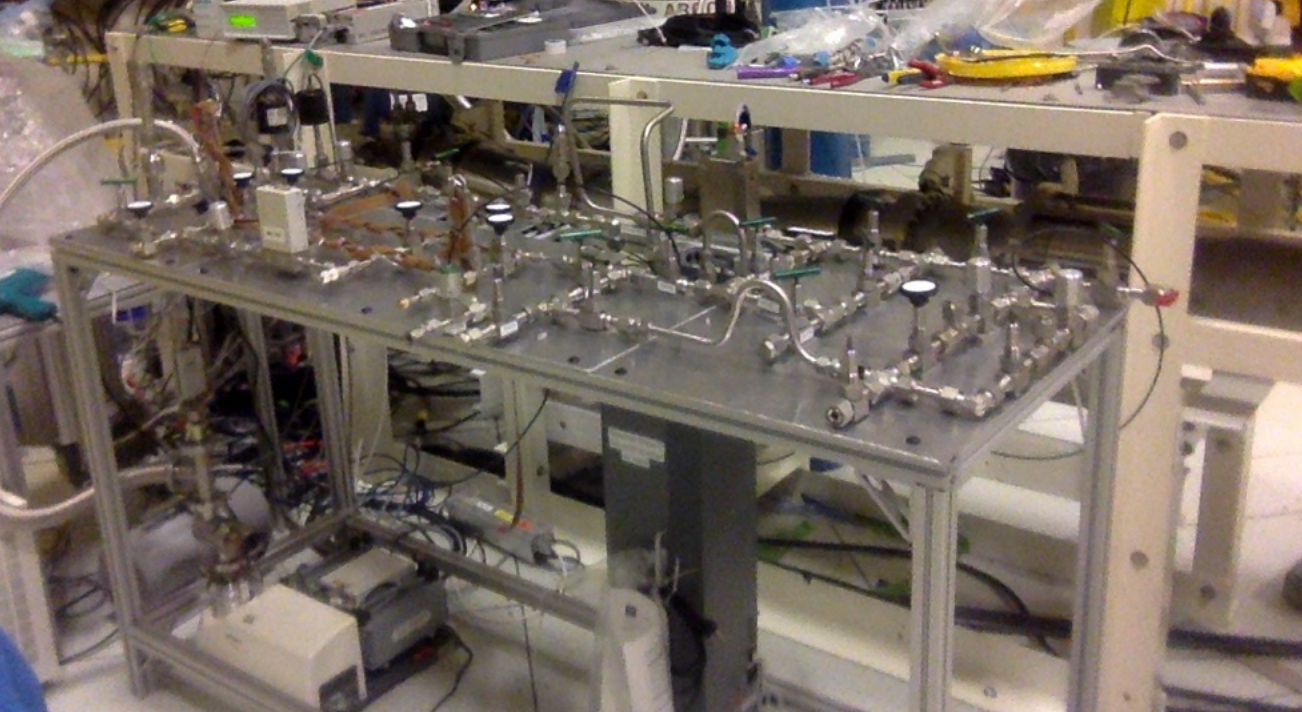}
\caption{ The photograph of purification system. }
\label{fig:purification_sys}
\end{figure}

The purpose of purification system is to further purify the argon and remove the radon particles in the argon. The purity of argon affects both triplet lifetime and the light yield. With high impurity level of argon (>10 ppb), the triplet lifetime is decreased which worsen the pulse shape discrimination and the energy resolution (See Chapter \ref{ch:cold}). Therefore to obtain pure argon is a key for successful measurement. The liquid argon purchased from AirLique has purity level as 99.999\%. For each dewar before connect to the purification system, is checked by Residual Gas Analyzer (RGA) to confirmed the required impurity level is met.  The boil-off argon from argon dewar is further purified by a SAES PS4-MT3-R-1 zirconium purifier.  The SAES getter requires 99.999\% inlet argon gas. In addition, the freezing point of radon (202 K) is well above the temperature of LAr, thus boil-off gas from LAr contains significant less radon than the output of gas cylinders. However, when the dewar begins to empty, the radon and other contaminants are readily boiled, thus the argon quality needs to be constantly monitor by RGA. The getter reduces most impurities ($H_2$,$H_2O$,$CO$,$CO_2$,$CH_4$,$N_2$,$O_2$, etc.) to below ppb concentration at flow rates of 5-20 SLPM. For higher flow rates (20-50 SLPM), the getter with lower electronegativity only reduces the impurities to below 10 ppb concentrations. Radon is not readily removed by the getter which requires additional purification through cryo-adsorption within an activated charcoal trap. The charcoal trap is cooled down to below radon freezing point such that with its large surface area, radon cryo-absorbs while allowing the purified argon to exit the trap. The schematic of argon flow is shown in Fig. \ref{fig:Leak_source_purification_layup} and the purification system is shown in \ref{fig:purification_sys}\par 

\begin{figure}[htbp]
\centering
\graphicspath{{./fig/Detector/}}
\includegraphics[scale=0.35]{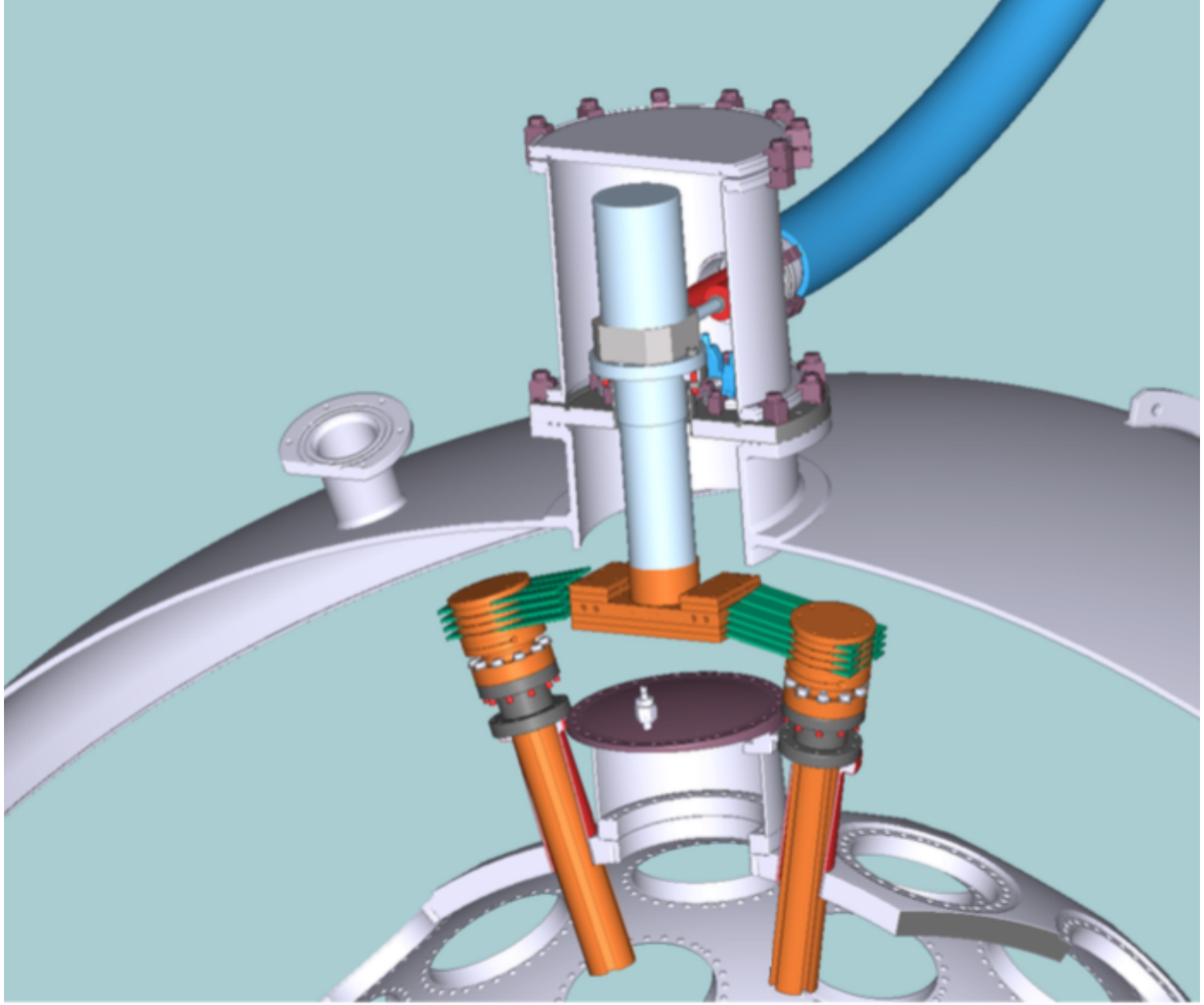}
\caption{ The cryocooler connection to cold fingers. }
\label{fig:cryocooler}
\end{figure}

\begin{figure}[htbp]
\centering
\graphicspath{{./fig/Detector/}}
\includegraphics[scale=0.35]{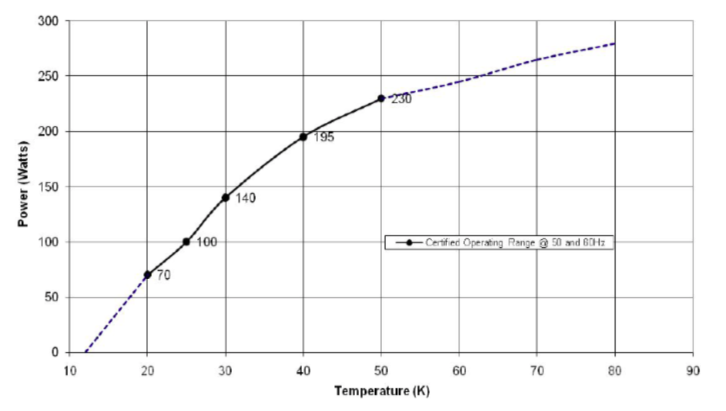}
\caption{ The cooling power of cryocooler as a function of the helium temperature\cite{tomphd}. }
\label{fig:cooling_power}
\end{figure}

In the original design, the cryogenic system consists of a Gifford-McMahon cryocooler mounted on the OV-D flange with 24 flexible OFHC copper braids extending to OFHC copper cold fingers mounted on two of the IV's 3 inches diameter ports as shown in Fig. \ref{fig:cryocooler}. The cooling power is from a pair of high pressure, vacuum jacketed helium lines which created a closed loop between cryocooler and the helium compressor. The calculated heat load of IV during normal  operation is in Table \ref{table:heatload}. The Multilayer insulation (MLI) installed on OV surface greatly reduced the thermal radiation. The cooling power of cryocooler is shown in Fig. \ref{fig:cooling_power}. The helium temperature at 40-50 K has sufficient cooling power to liquefy and maintain the LAr target. The IV temperature is monitored by 5 silicon diode temperature sensors mounted on the different locations of IV sphere. The temperature controller is linked to a set of four DC power supplies which are connected in parallel to a pair of 500 W Omegalux cartridge heaters which used to control the temperature of cold finger to maintain at suitable temperature and prevent the argon condensed on the cold finger such that reduces the cooling power.   
\begin{table}[htbp]
\centering 
\begin{tabular}{| c | c|} 
\hline
Component & Load (W)\\
\hline
Thermal radiation & 22.3\\
Free-molecular air conduction & 4.3\\
OV-IV supports & 11\\
Vent pipes & 9.4\\
PMT cables & 5.1\\
Other cables & 0.6\\
92 PMTs & 12.1\\
\hline 
Total & 64.8\\
\hline
\end{tabular}
\caption{Upper limit on the heat load to the Inner Vessel during normal operations. This
model assumes only 10 layers of MLI in the OV and uses twice the thermal emissivity of
MLI to account for any gaps, joints, etc.} 

\label{table:heatload} 
\end{table}

\section{$^{39}$Ar Spike}
\begin{figure}[htbp]
\centering
\graphicspath{{./fig/Detector/}}
\includegraphics[scale=0.35]{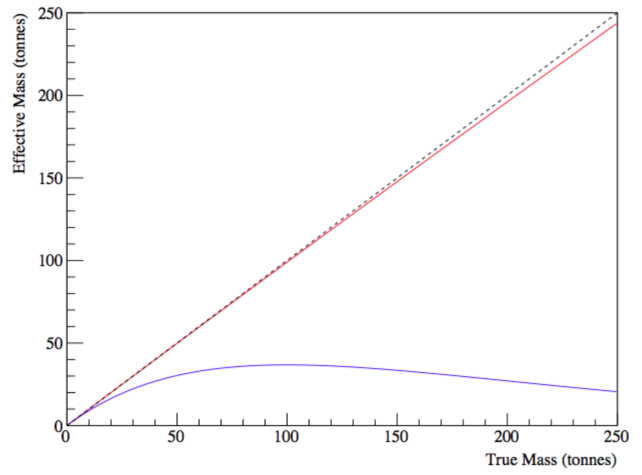}
\caption{  A lower bound on the effective mass of a large LAr detector given the true mass and the effective dead time induced by pileup of 39Ar events with WIMP candidates. A 10 $\mu$s event window is assumed. Atmospheric LAr is shown in blue while argon depleted by a factor of 100 in $^{39}$Ar is shown in red.
The dashed line indicates no loss of true mass due to $^{39}$Ar pileup\cite{tomphd}. }
\label{fig:mass_ar39}
\end{figure}

The primary goal of MiniCLEAN detector is to test the PSD background rejection ability. The intrinsic $^{39}$Ar beta decay produces the electronic recoil which could leak into the region of nuclear recoil. With larger mass of LAr, the $^{39}$Ar background limit energy threshold for single phase detector. To test the PSD background rejection ability, $^{39}$Ar spike is injected into the LAr to increase the concentration of $^{39}$Ar background. A concentrated sample of $^{39}$Ar will be mixed with natural argon and injected into the detector through the gas process system. The $^{39}$Ar spike samples has been produced at LANL by irradiating a potassium target with neutrons above 2 MeV and the $^{39}$Ar is produced through the $^{39}$K(n,p)$^{39}$Ar reaction. This allows the MiniCLEAN detector to test the ultimate PSD achievable in a single phase LAr detector as a function of energy. Moreover, the results can be informative for future detector design for the size and energy threshold needed. In addition, with increased concentration of $^{39}$Ar, the  pileup of electronic recoils with potential WIMP nuclear recoils induces an effective pileup dead time. In typical data taking with the $^{39}$ spike injected, PSD techniques assumes in the event window (16 $\mu$ s) only the recoil of interest happened. However, with increasing $^{39}$Ar rate. the process becomes non-trival, and the resulting ``dead time'' effectively reduces the mass of the detector as shown in Fig. \ref{fig:mass_ar39}. To test the $^{39}$Ar concentration in tens of tonnes detector with MiniCLEAN detector (0.5 tonnes), at least a 200 times spike would be required according to the Fig. \ref{fig:mass_ar39}. 
\section{Simulation and Analysis Software}
The data analysis and the simulation are done by analysis package \textit{RAT}. RAT incorporate the analysis software ROOT , simulation package GEANT4 and integrate with the DAQ system, originally developed by the Braidwood collaboration\cite{Bolton2005}. The aim of RAT is to provide a tool for photomultiplier-based detectors with scintillation targets. The particle propagation of the electromagnetic, hadronic physics process and the detector geometry implementation  are done by GEANT4. In addition, the add-on package (GLG4 scintillation) handles the scintillation and fluoresces process and generates photons according to the type and amount of energy deposited in the desired material. For both data taking and simulation, the ROOT framework handles the data processing and storage.  RAT simulates the following detector effects : 
\begin{itemize}
	\item GEANT4 handles the propagation of primary and secondary particles, including the electrons, gamma rays, nuclear recoils and neutrons through the detector materials.
	\item The UV(VUV) scintillation light produced by charged particles in the liquid argon. 
	\item The propagation of individual VUV photons and the optical properties of detector material including the wavelength shifted photons.
	\item PMT simulation including the realistic pulses shape, timing and charge, as well as the pre-pulsing, late-pulsing, double-pulsing and after-pulsing.
	\item Simulates the detector triggering, the digitized waveform, the waveform reduction of zero-suppressed mode. 
\end{itemize}
The simulation events and physical events from detector trigger are treated in the same way in RAT. For every iteration of events, a series of self-contained ``processor'' performed different task on the events such as extract raw data, calibration, pulse finding, event reconstruction, etc. User can defined their own processor to perform desired function on the events.
\subsection{Discriminant Parameters }
\begin{figure}[htbp]
\hfill
\subfloat[]{\includegraphics[width=7cm]{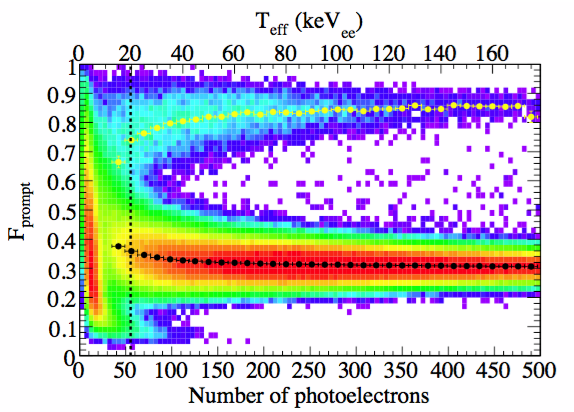}}
\hfill
\subfloat[]{\includegraphics[width=7cm]{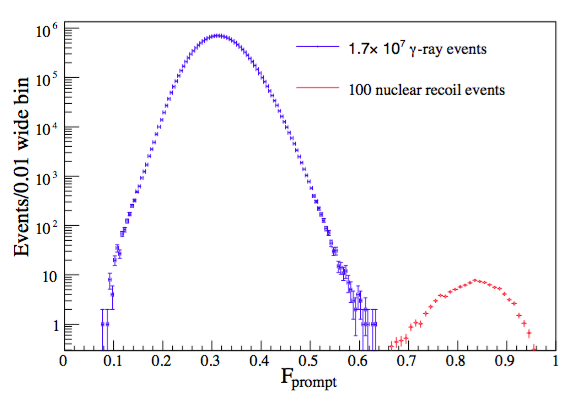}}
\hfill
\caption{(a) Fprompt versus energy distribution in LAr. The upper band is from nutron-induced nuclear recoils, the lower band is from gamma ray interactions. (b) Fprompt distribution for gaama ray events and the nuclear recoil events from the Am-Be calibration source. Figures are taken from \cite{AMAUDRUZ20161}.}
\label{fig:Fprompt_2deap1}
\end{figure}

Several basic parameters are defined to preserve the data quality. For the scintillation light produced by different particle incident on LAr target which described in detail in Chapter \ref{ch:scin} can be used to defined a discriminant parameters. The electronic recoil tend to produce more fraction of late light (triplet) than nuclear recoil. Thus the prompt-fraction, $F_p$ (Fprompt ) is defined as :
\begin{ceqn}\begin{align}
f_p = \frac{\int_{T_i}^{\eta} V(t) dt}{\int_{T_i}^{T_f} V(t) dt}
\end{align}\end{ceqn} 
where $V(t)$ is the voltage waveform, $T_i$ is some time before the maximum of the prompt peak which has been calibrated as time zero, $T_f$ is the end of acquisition window and $\eta$ is depending on the timing characteristic of the scintillator. In MiniCLEAN, the time window to acquire prompt charge has been optimized through simulation. The start time $T_i$ is 28 ns before the maximum of the prompt peak and $\eta$ is 80 ns after the maximum of the prompt peak. The example of using Fprompt as discriminant parameter to separate electronic and nuclear recoil in LAr is shown in Fig. \ref{fig:Fprompt_2deap1}.\par
\begin{figure}[htbp]
\centering
\graphicspath{{./fig/Detector/}}
\includegraphics[scale=0.35]{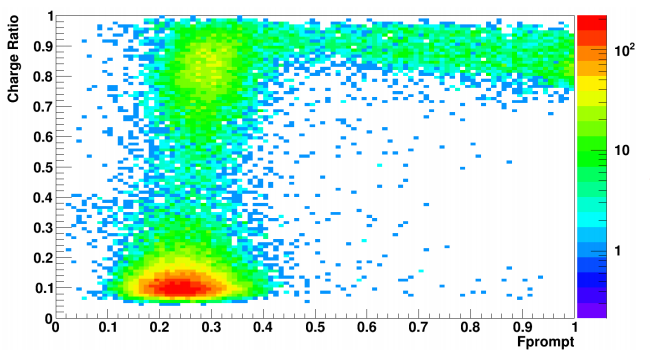}
\caption{$Q_R$ vs Fp in cold gas data with a cut (charge > 75 PE). }
\label{fig:charge_ratio_fp}
\end{figure}

The charge ratio ($Q_R$) can be used to determine the preliminary charge distribution within 92 PMTs. It is defined as the ratio of maximum charge in the PMTs to the total charge in the given event. In data, it is useful to discriminate against non-argon scintillation events. The example of the $Q_R$-Fp distribution is shown in Fig. \ref{fig:charge_ratio_fp}. In this figure, the group of events in the low $Q_R$ comes from electronic recoil induced by intrinsic $^{39}$Ar beta decay. The events in high $Q_R$ comes from instrument effects and Cherenkov light in the acrylic (see Ch. \ref{ch:gasrun}).

\subsection{Event Reconstruction}
\begin{figure}[htbp]
\centering
\graphicspath{{./fig/Detector/}}
\includegraphics[scale=0.25]{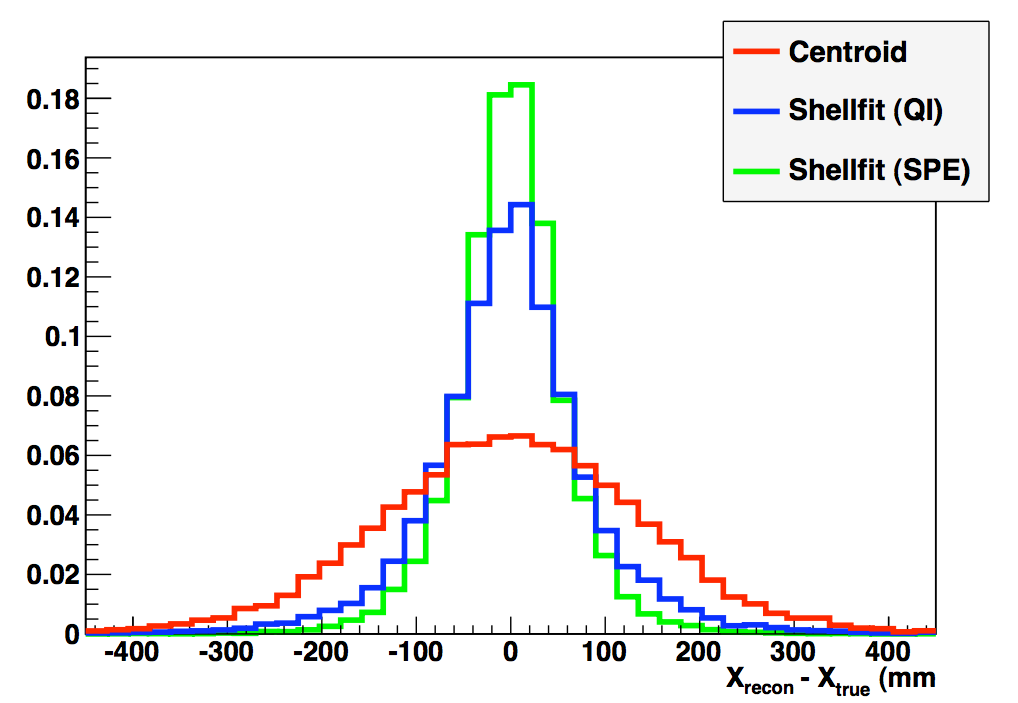}
\caption{The reconstruction resolution on X determined by simulation using different reconstructed methods\cite{shellfit_old}. The vertical axis is normalized counts.}
\label{fig:centroid_shellfit_compare}
\end{figure}

The charge centroid is used to reconstruct the event position, it is defined as :
\begin{ceqn}\begin{align}
\vec{R} = \frac{\sum\limits_{i} \vec{r_i}\cdot Q_i^2}{\sum\limits_{i}Q_i^2}
\end{align}\end{ceqn}
where $\vec{r_i}$ is the position of the $i$'th PMT and $Q_i$ is the total charge in that PMT. However, the charge centroid reconstruction is biased inward. The collaboration developed more sophisticated reconstruction method ``Shellfit'' to reconstruct energy and the position with better position resolution. The Shellfit is based on the maximum likelihood fit and incorporate with the optical properties of TPB. The four assumption about the detector configuration in the likelihood function :
\begin{enumerate}
	\item The detector is approximately spherically symmetric with respect to waveguide placement and the configuration.
	\item Scintillation light is isotropically emitted from the event vertex.
	\item TPB is applied to a spherical shell or fixed radius.
	\item TPB absorption and reemission is isotropic, with no directional information from the incoming UV photon passing to the reemitted visible photon(s).
\end{enumerate}
With these assumptions, the charge likelihood function for the events is 
\begin{ceqn}\begin{align}
L(N_{UV},\vec{r_{ev}}) = \prod\limits_{i=1}^M P(q_i | \vec{C}(\vec{r_i},N_{UV},\vec{r_{ev}})),
\end{align}\end{ceqn}
where $q_i$ is the charge in the $i$'th PMT, $\vec{r_i}$ is the position of $i'$th PMT, $\vec{C}(\vec{r_i},N_{UV},\vec{r_{ev}})$ is the mean number of photoelectrons detected by a PMT t position $\vec{r_i}$, given an event at $\vec{r_{ev}}$ which produces $N_{UV}$ scintillation photons. With observed charge $q_i$ in each PMT, the probability density $P(q_i | C)$ gives the mean charges $\vec{C}$ according to the Poisson distribution. The expected mean number of photoelectrons $\vec{C}$ at PMT $i$ can be computed by smapling the detector response at N points, $\vec{p_i}$ distributed over the TPB surface uniform in solid angle relative o the event position $\vec{r_{ev}}$ Therefore, the expected number of photoelectrons can be calculated by 
\begin{ceqn}\begin{align}
	\vec{C}(\vec{r_i},N_{UV},\vec{r_{ev}}) = N_{UV} \frac{1}{N}\sum\limits_{j=1}^{N} E(\theta_{ij})
\end{align}\end{ceqn}
where the $\theta_{ij}$ is the angle in radians between the PMT position vector $\vec{r_i}$ and the TPB sample point $\vec{p_j}$. The $E(\theta_{ij})$ is the detector response function which gives the number of detected photoelectrons at PMT $i$ given that the probability of a UV photon absorbed at $\vec{p_j}$. To convert the mean photoelectrons to a charge distribution, the single photoelectron distribution is used. The results from charge centroid fit, Shellfit using integral charge and shellfit using SPE distribution is shown in Fig. \ref{fig:centroid_shellfit_compare}. The results shows a dramatic improvement on the resolution of reconstructed radius. However, the shellfit required a scintillation target , and definite timing p.d.f. of scintillation timing profile. Therefore, the analysis in vacuum and gas run will use the charge centroid fit to perform the event reconstruction.\par
\subsection{Pulse Finding}\label{sec:pulsefinding}
\begin{figure}[htbp]
\centering
\graphicspath{{./fig/Detector/}}
\includegraphics[scale=0.35]{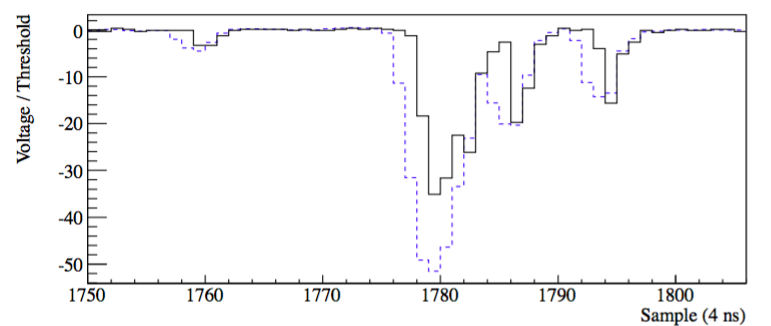}
\caption{A typical voltage waveform from a single PMT in MiniCLEAN Monte Carlo simulation. The top panel shows the waveform normalized by 5 times the RMS of the electronics noise profile (black, solid) compared to the sliding integral value normalized by the corresponding threshold (blue, dashed)\cite{tomphd}.}
\label{fig:pulse_finding}
\end{figure}

\begin{figure}[htbp]
\centering
\graphicspath{{./fig/Detector/}}
\includegraphics[scale=0.35]{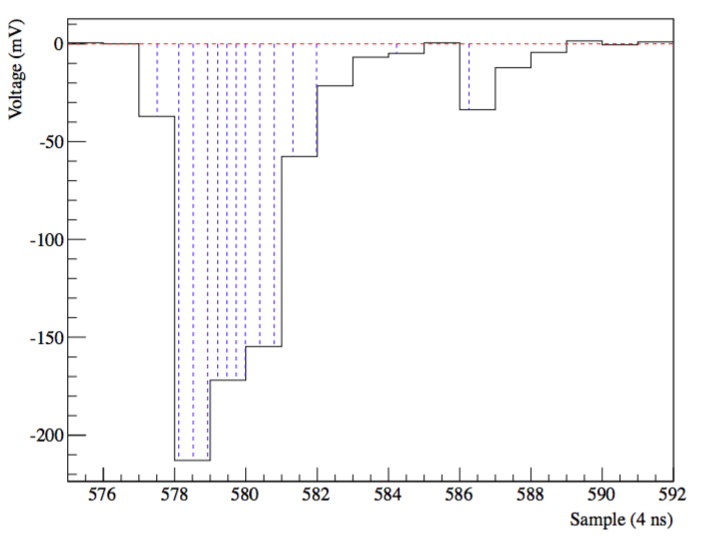}
\caption{ The assigned times using the waveform shape are shown by the vertical blue dashed lines.\cite{tomphd}.}
\label{fig:baye_example}
\end{figure}

A sliding integration window of 12 ns (3 samples) is used to scan the calibrated waveforms to identify the pulse. The pulse is extracted whenever the integral exceeds 5 times the RMS of noise samples times the square root of the number of samples in the window. The boundaries of the pulse region is defined when the sliding window integral drops o below the RMS divided by the square root of the number of samples. Figure \ref{fig:pulse_finding} shows a example pulse found by the pulse finding. A new algorithm developed by MiniCLEAN\cite{AkashiRonquest201540} utilizing the Baye's theorem to improve the estimation of single photoelectron arrival time. Using the bayesian technique and the characteristic scintillation timing profile for different type of recoil to estimate the single photoelectron arrival time. In addition, the energy reconstructed by Shellfit can provide more accurate information on photoelectron statistics, then as a prior to the Baye's theorem to get better results. The example using Bayesian technique to estimated the single photoelectron arrival time is shown in Fig. \ref{fig:baye_example}. 

\subsection{Particle Identification Using Likelihood Method}
\begin{figure}[htbp]
\centering
\graphicspath{{./fig/Detector/}}
\includegraphics[scale=0.35]{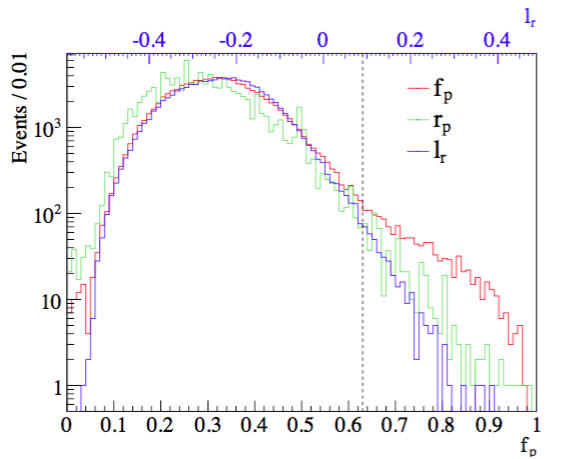}
\caption{Distribution of $f_p$, $r_p$, and $L_r$ test statistics for electronic recoils for 22Na calibration events in DEAP-1 with 30 PE. The vertical dashed line indicates 50\% nuclear recoil acceptance at 6.7 keVee. The $L_r$ values have been linearly transformed such that the median values for the electron and nuclear recoil distributions match those for $f_p$\cite{AkashiRonquest201540}.}
\label{fig:LR}
\end{figure}

The Fprompt parameters is useful for identify the electronic recoil from nuclear recoil. However, in low energy region, the Fprompt value for electronic recoil leaks into the nuclear recoil region. The MiniCLEAN collaboration developed new parameters to improve the discrimination power. With the single photoelectron time estimated by Bayesian technique (Chapter \ref{sec:pulsefinding}) for an event $\zeta$, a discrimination variable $L_r$ can be defined as a normalized log-likelihood difference, comparing the nuclear recoil hypothesis with the electronic recoil hypothesis : 
\begin{ceqn}\begin{align}
L_r = \frac{1}{m} \sum\limits_{t\in \zeta} (log P_n(t|E) - log P_e(t|E)),
\end{align}\end{ceqn}
where m is the number of photoelectrons in the event, $P_n$ ($P_e$)is the time probability density function for the nuclear recoil (electronic recoil) hypothesis given the energy E. Positive value of $L_r$ indicates the event is more nuclear recoil-like, and negative are more electronic recoil-like. Comparing $L_r$ to Fprompt and a simple statistic $r_p$ which is a discrete version of $f_p$ using the single photon arrival time estimated by Bayesian technique :
\begin{ceqn}\begin{align}
r_p = \frac{|\{t|t\in \zeta \wedge T_i < t < \epsilon \}|}{|\{t|t\in \zeta \wedge T_i < t < T_f \}|},
\end{align}\end{ceqn}
where $\epsilon$ defined the prompt window (same as $f_p$), $T_i$ and $T_f$ is the start time and the end time for counting window respectively. Figure \ref{fig:LR} shows the comparison between these discriminant parameters.  
\section{Background}
The summary of backgrounds of MiniCLEAN detector is described in the following sections.
\subsection{External Backgrounds}
\begin{figure}[htbp]
\centering
\graphicspath{{./fig/Detector/}}
\includegraphics[scale=0.35]{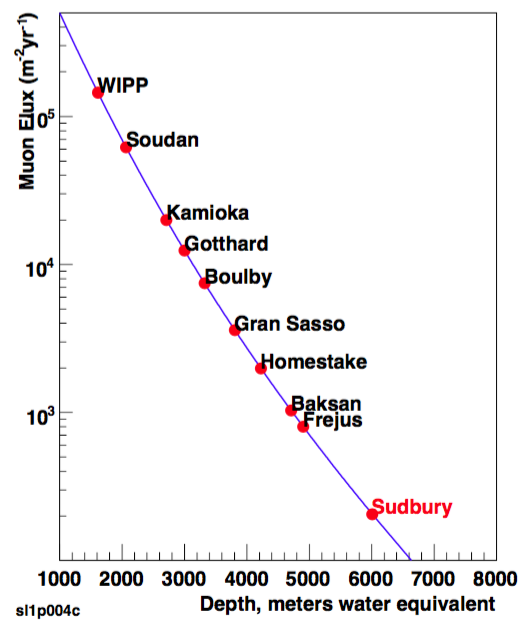}
\caption{Muon flux as a function of depth. Figure from \cite{snolabhandbook}.}
\label{fig:muon_flux}
\end{figure}

The MiniCLEAN detector locate at 6800 ft underground in SNOLAB. The muon flux is significant reduced in the underground laboratory. The muon flux as a function of depth is shown in Fig. \ref{fig:muon_flux}. The muon flux at SNOLAB underground laboratory is less than 0.27 $\mu$/m$^2$/day\cite{snolabhandbook}. The gamma ray flux from the rock is measured by SNO experiment and is tabulated in Table \ref{table:gammaflux}. The actual rate for MiniCLEAN detector will be lower due to the shielding of the water tank.The fast neutron flux from the rock is estimated to be 400 neutrons/m$^2$/day. The simulation shows the flux of fast neutron is reduced to much less than unity by the water shielding.

\begin{table}[htbp]
\centering 
\begin{tabular}{| c | c | c |} 
\hline
E$_{\gamma}$ (MeV) & Measured Flux ($\gamma m^{-2} d^{-1}$) & Calculated Flux ($\gamma m^{-2} d^{-1}$)\\
\hline
4.5-5 & 510$\pm$200 & \\
5-7 & 360$\pm$220 & 320\\
>7 & 180$\pm$90 &250\\
>8 & < 20 & 15\\
\hline
\end{tabular}
\caption{High energy $\gamma$-ray flues from rock. Gamma fluxes from norite, measured during the installation of SNO with a NaI(Tl) detector and various thicknesses of lead. ?The calculations are based on neutron capture in the elements of norite with neutron flux predicted from the mea- sured Th and U concentrations in the rock. From the SNOLAB User's Handbook\cite{snolabhandbook}} 
\label{table:gammaflux} 
\end{table}

\subsection{Internal Backgrounds}
The potential internal background source of MiniCLEAN are :
\begin{itemize}
	\item intrinsic $^{39}$Ar beta decay.
	\item Gamma from PMTs and IV/OV steel.
	\item fast neutrons from ($\alpha$,n) processes in the PMT bulb and IV/OV steel.
	\item Alpha decays from radon daughter.
\end{itemize}
The aim of MiniCLEAN is to eliminate these background in the fiducial volume ($\sim$ 150 kg) and an energy region of interest corresponding to 75-150 photoelectrons.
The radioactive isotope $^{39}$Ar has radioactivity of 1Bq/kg. It will produce the electrons through the beta decay with half-life 269 years and end point energy 565 keV. In order to obtain a background free fiducial argon volume over one year, the required PSD rejection ability needs to be better than parts per billion. However, it can be used as a calibration source to monitor the detector health and reconstruction bias. The detail description is in Chapter \ref{ch:ar39}.\par
The gamma ray can be produced by the relaxation of alpha-emitters in the $^{238}$U and $^{232}$Th  decay chains. The estimated rates is 808 mBq from $^{238}$U and 421 for $^{232}$Th. These gammas may produce the Compton electrons both in LAr and acrylic plug to create the Cherenkov light in the acrylic or electronic recoil in the liquid. The simulation of $^{238}$U and $^{232}$Th decay chain in OV/IV steel, PMT glass, and light guide steel/acrylic shows no events survived after all cuts. The results is summarized in Table \ref{table:gammasim}.\par
\begin{figure}[htbp]
\centering
\graphicspath{{./fig/Detector/}}
\includegraphics[scale=0.35]{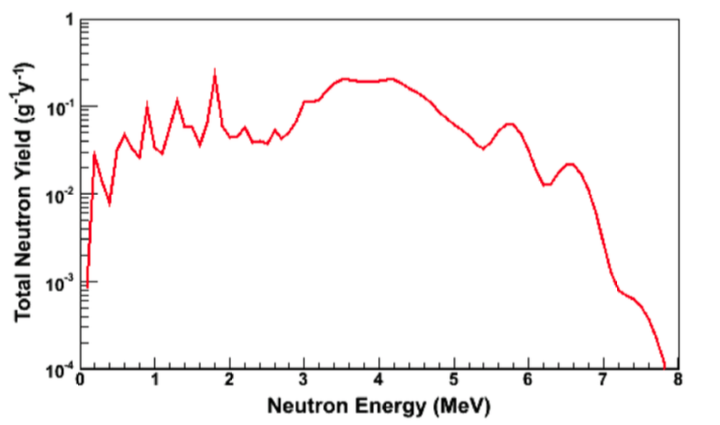}
\caption{($\alpha$,n) neutron energy spectrum for $^{238}$U and $^{232}$Th contamination in borosilicate glass\cite{sjaditz}.}
\label{fig:PMT_neutrons}
\end{figure}

The primary source of fast neutrons come from the ($\alpha$,n) interactions due to the $^{238}$U and $^{232}$Th both in the borosilicate glass of the PMTs and steel. Fast neutron will scatter elastically and inelastically from the target nuclei and produce a signal that is indistinguishable from a WIMP signal.  Table \ref{table:neutronflux} summarize the intrinsic radioactivity of the major components of the MiniCLEAN detector. The calculation from Mei \textit{et al}\cite{MEI2009651} predicts the neutrons from 66 kg of PMT glass are 42000 neutrons per year. A similar calculation for the steel IV and OV predicts a neutron yield of 1800 per year. Figure \ref{fig:PMT_neutrons} shows the energy spectrum for PMT neutrons. The PMT neutrons can be moderated by a 10-cm acrylic plug in the light guides. In addition, 20 cm LAr self-shielding also contribute to moderate the neutron from PMT or steel to get into the fiducial volume. The alpha decays of radon daughters deposit on the TPB will induces the alpha-TPB scintillation. In addition, the nucleus in the alpha decay will be injected into the LAr volume and create a signal in the region of interest. The detail description on identifying and discriminating against these events is in Chapter \ref{ch:vacuum}.

\begin{table}[htbp]
\caption{Summary of internal gamma background in the simulation at different stages of the cut\cite{chrisJgamma}.} 
\centering 
\begin{adjustbox}{width=1\textwidth}
\begin{tabular}{c c c c c c c} 
\hline\hline 
Material & Generated events & Triggered events & Energy (> 75 PE and < 150 PE) &Fiducial cut ( > 295 mm) & Fprompt ( > 0.681) & LRcoil (> 0.373) \\ [0.5ex] 
\hline 
OV Steel &1,000,000 & 1,100 & 13 & 6 & 0 &0\\
IV Steel & 1,000,000&7096 & 66 & 28 & 0 & 0\\
PMT Glass & 920,000 & 17,220 & 160 & 42 & 0 & 0\\
Light Guide Steel & 988,987 & 47,818 & 542 & 209 & 0 & 0\\
Light Guide Acrylic & 929,772 & 209,934 & 2,743 & 856 & 0 &0\\[0.5ex]
\hline\hline
\hline 
\end{tabular}
\end{adjustbox}

\label{table:gammasim} 
\end{table}

\begin{table}[htbp]
\centering 
\begin{tabular}{| c | c | c | c |} 
\hline
Component & Material & $^{238}$U/$^{232}$Th & Natural-K\\
\hline
Light guides & 480 kg Acrylic & 480/480 ng& 3ppb\\
\hline
PMT Sphere & 60 kg SiO$_2$ & 6.0/10.5 mg & 100 ppm\\
\hline
& 12 kg B$_2$O$_3$ & 1.2/2.1 mg & 100 ppm\\
\hline
& 1050 kg Steel & 1.05/1.05 mg& 2 ppm\\
\hline
Outer Cryostat & 1575 Steel & 1.58/1.58 mg & 2 ppm\\
\hline
& 150 kg Cu & 15/15$\mu$g & 10 ppb\\
\hline
\end{tabular}
\caption{Summary of the major components of the Mini-CLEAN detector and the projected intrinsic radioactivity used as input to the background model and simulations.} 
\label{table:neutronflux} 
\end{table}

\chapter{Construction and Cooling of MiniCLEAN detector }
The construction and cooling of MiniCLEAN detector is described in this chapter. The author maintained a full-time presence on-site at SNOLAB from February 2014 to May 2015, from the middle of detector assembly to the start of the detector cool down. 
\section{Inner Vessel Assembly}
\begin{figure}[htbp]
\centering
\graphicspath{{./fig/Detector/}}
\includegraphics[scale=0.35]{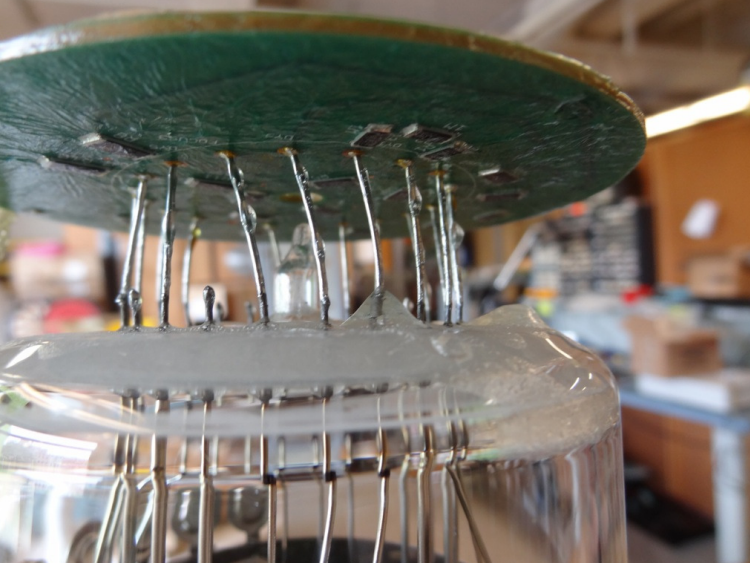}
\caption{Conformal coating on the base and neck of PMT.}
\label{fig:PMT_coating}
\end{figure}

The MiniCLEAN inner and outer vessels arrived on-site in Sudbury in Fall 2012. The construction of IV starts in early 2013. The PMTs and the DAQ system are tested in Boston University and shipped to SNOLAB in summer 2013. The author made several trips to SNOLAB in 2013 to install the LED pulser system and assist the PMT installation. The base and neck of each PMT are conformal coated to prevent the extra impedance through the Gas/Liquid as shown in Fig. \ref{fig:PMT_coating}. However, during the conformal coating, some material drip along the glass bulb which might flake off inside the IV during the cooling. Therefore, a thorough  examination of PMT was performed to remove these substance.\par
\begin{figure}[htbp]
\hfill
\subfloat[]{\includegraphics[width=7cm]{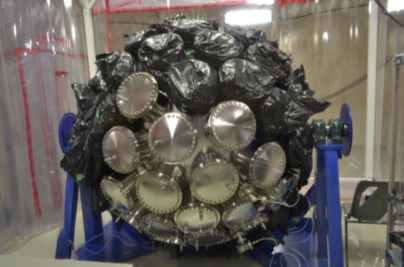}}
\hfill
\subfloat[]{\includegraphics[width=7cm]{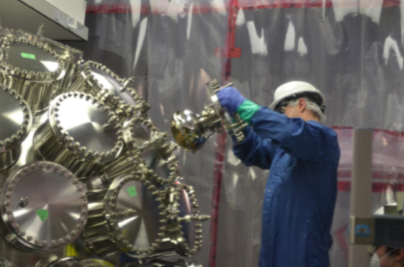}}
\hfill
\caption{(a) IV on the rotating stand. Plastic tubing connected to the nitrogen purge system is visible near the bottom right portion of the rotater stand (b) Installing the PMT in the inner vessel.}
\label{fig:SWCR}
\end{figure}
\begin{figure}[htbp]
\centering
\graphicspath{{./fig/Detector/}}
\includegraphics[scale=0.35]{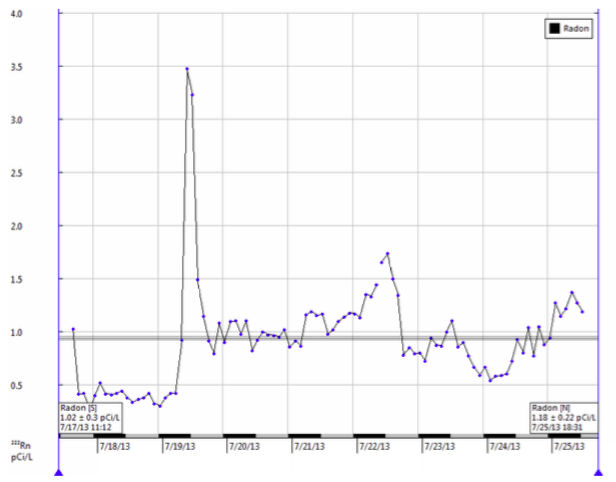}
\caption{The radon concentration in the radon-reduced clean room mea- sured by a RAD7 radon monitor over a week time period. The peak on July 19 is due to the introduction of mine air into the clean room by a portable air conditioner which was being tested. During assembly, radon levels in the clean room were typically 1-2 pCi/L while underground air typically contains 4-5 pCi/L of radon\cite{tomphd}.}
\label{fig:RAD7}
\end{figure}

The assembly of IV was performed in the softwall cleanroom (SWCR Fig. \ref{fig:SWCR}) modified to maintain a low radon atmosphere with provided compressed air. The radon level indie the SWCR is monitored with a RAD 7 radon monitor as shown in Fig. \ref{fig:RAD7}. After the assembly, each time the IV is open to the cleanroom space, the boil-of nitrogen gas is used to purge the cleanroom atmosphere to minimize the chance to leak radon into the IV. The optical cassettes house the PMT, and the top hat provides electrical feedthrough for connect the PMT HV/signal cable to the OV. The Vikuiti$^{TM}$ ESR foil lined the inner surface of the optical cassettes. When the ESR foil shipped to the SNOLAB, a thin layer of plastic to prevent the foil from scratch is removed outside the SWCR. However, when removing the plastic foil, the electrostatic force attract the radon particles to deposit onto the foil. This create excessive events in ESR foil scintillation which described in detail in Chapter \ref{sec:esrfoilscin}. After completion of IV assembly, the IV is filled with the argon gas. The DAQ system and the purification system (without charcoal trap) is then tested.  The IV has been tested throughly with DAQ and purification system by the summer of 2014. Subsequently, the preparation of moving IV into OV start while IV sit in the vacuum with continuously data taking to ensure the stability.

\section{Outer Vessel Assembly}
\begin{figure}[tbp]
\centering
\graphicspath{{./fig/}}
\includegraphics[scale=0.4]{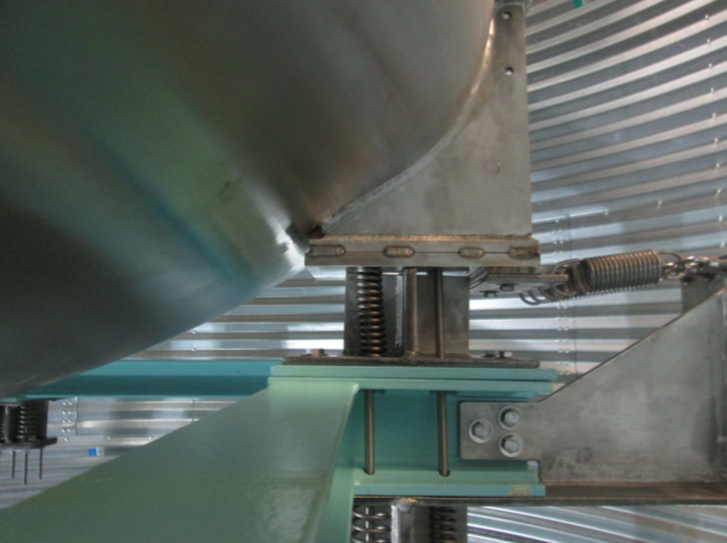}
\caption{ The spring support system for the OV.}
\label{fig:spring}
\end{figure}

The Outer vessel are assembled on the Cube Hall floor, and leak checked after assembly. The stand which support the OV was constructed inside the water tank. The seismic analysis indicates that the OV on a relatively rigid stand would not support the weight of IV in SNOLAB's design seismic event. Later , a spring support system was designed to mitigate the seismic hazard. Which reduce the movement of the OV and IV to approximately 1.5 mm in the 4.3 magnitude (on Nuttli scale) design seismic event. Figure \ref{fig:spring} shows the spring support system.\par
\begin{figure}[tbp]
\centering
\graphicspath{{./fig/}}
\includegraphics[scale=0.4]{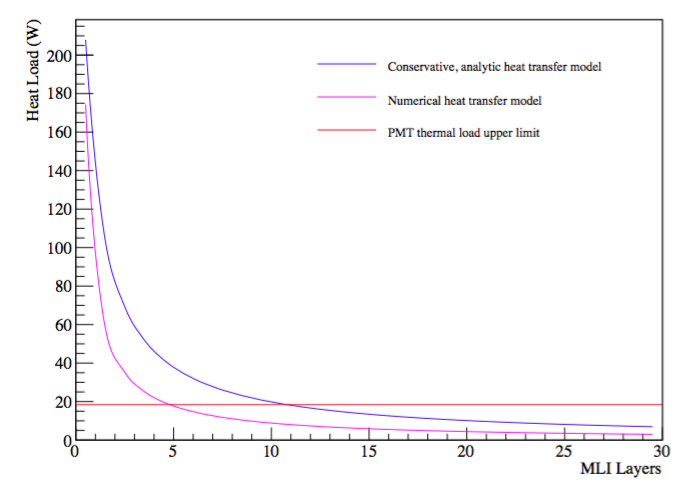}
\caption{ The estimated radiative heat load on the IV as a function of the number of layers of multi-layer insulation in a conservative analytic model (blue) and also a numerical model (magenta) which includes more geometrical detail. The horizontal line indicates an upper bound on the thermal load due to the PMT bases\cite{tomphd}.}
\label{fig:MLI_cal}
\end{figure}
\begin{figure}[tbp]
\centering
\graphicspath{{./fig/}}
\includegraphics[scale=0.4]{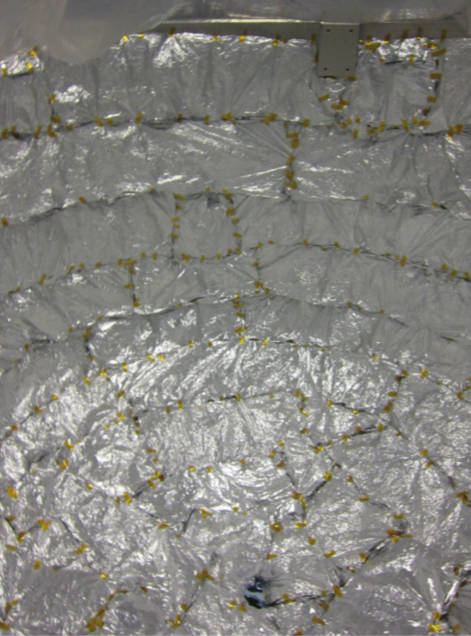}
\caption{ A photograph of ten MLI layer blanket attached to the inner surface of OV.}
\label{fig:MLI}
\end{figure}

When IV is contained in the OV vacuum space, the dominant heat transfer comes from the thermal radiation form the OV. The multi-layer insulation (MLI) is often used in the cryogenic system. It is consisting of alternating layers of poor thermal conductivity and high IR-reflectivity which can reduce the heat load on the cryogenic body. After the calculation which taking into account of the heat load of MiniCLEAN detector (Fig. \ref{fig:MLI_cal}), a 10 layers multi-layer insulation is applied to the inner surface of OV. The MLI layers are cut and prepared in the Cryo-pit, the krypton type was used to attached the MLI layers on to the inner surface of OV as shown in Fig. \ref{fig:MLI}. Each MLI layer is 400 angstroms thick with a thermal emissivity of 0.03.\par
\begin{figure}[htbp]
\hfill
\subfloat[]{\includegraphics[width=7cm]{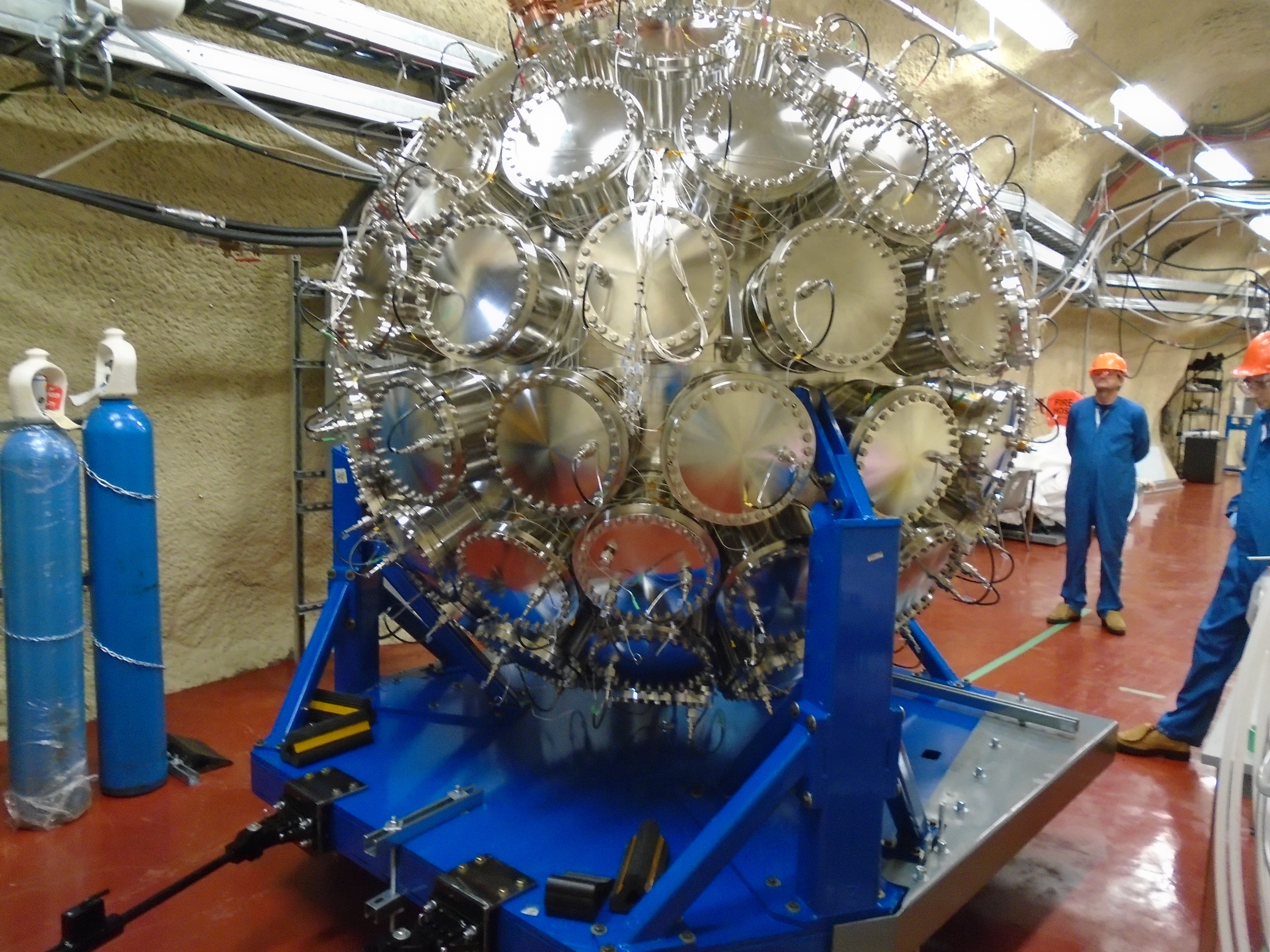}}
\hfill
\subfloat[]{\includegraphics[width=7cm]{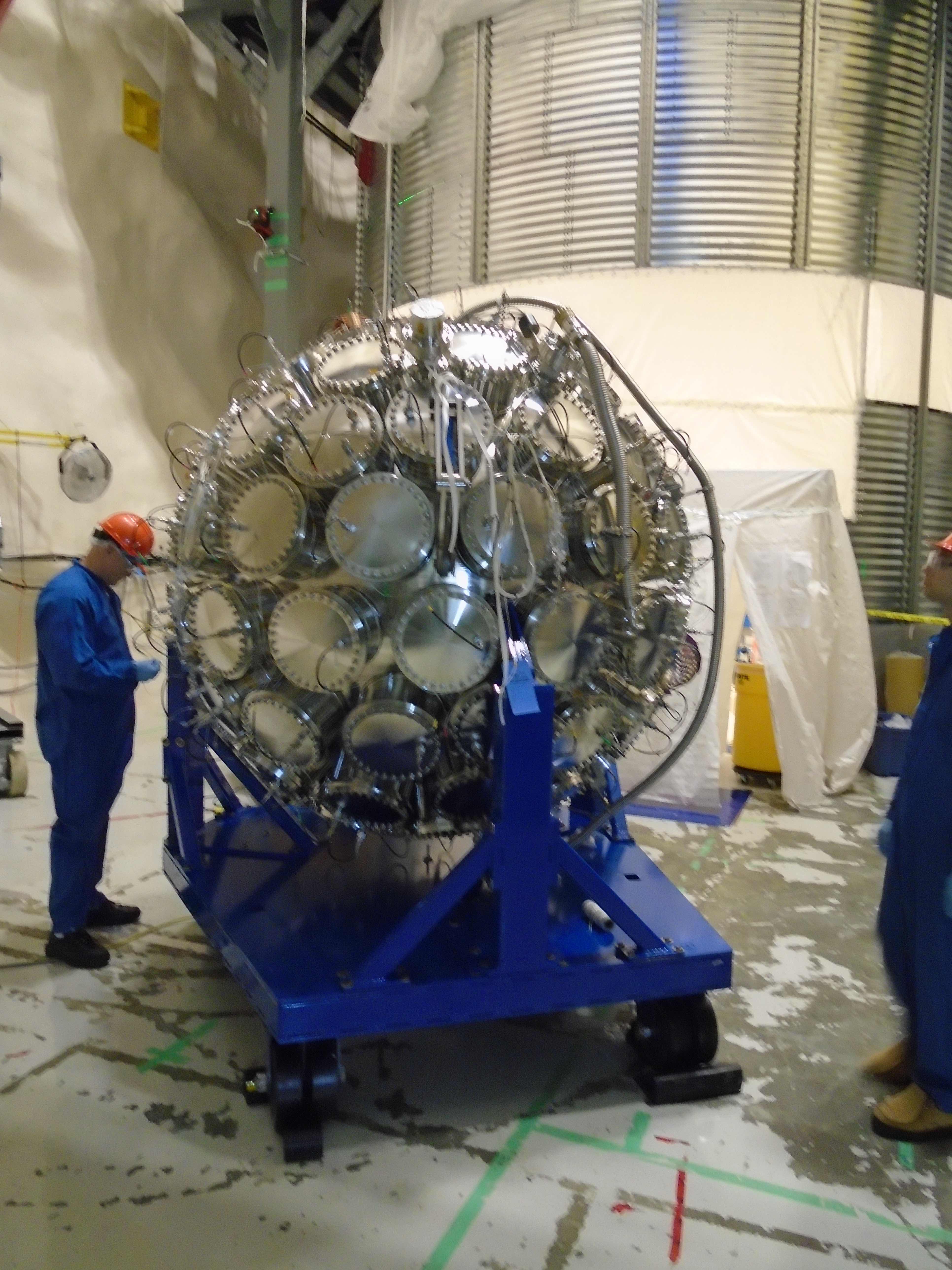}}
\hfill
\caption{(a) IV moving from Cryo-pit to CubeHall. (b) Final examination of IV in Cube Hall floor.}
\label{fig:IV_move}
\end{figure}
\begin{figure}[htbp]
\hfill
\subfloat[]{\includegraphics[width=7cm]{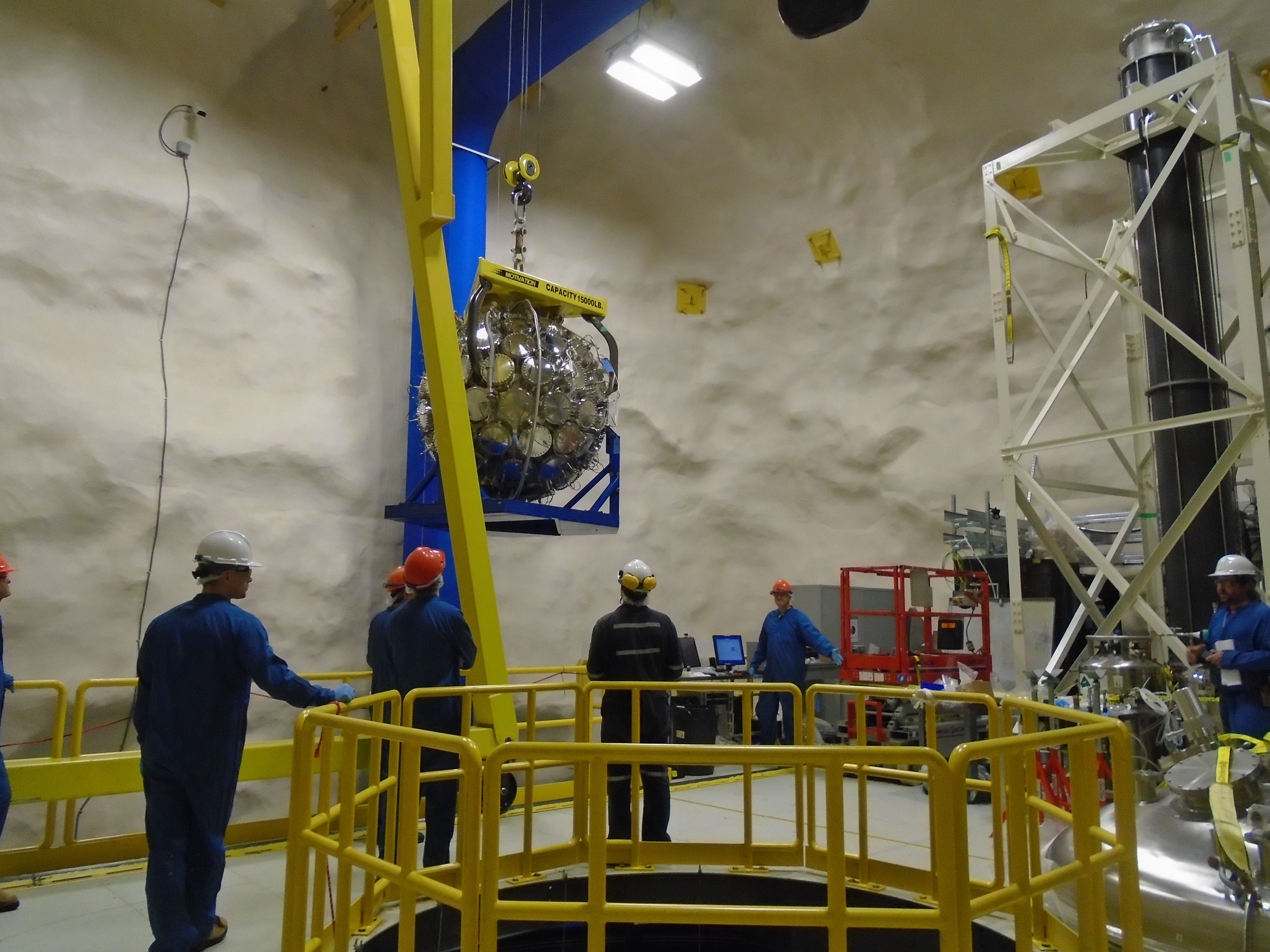}}
\hfill
\subfloat[]{\includegraphics[width=7cm]{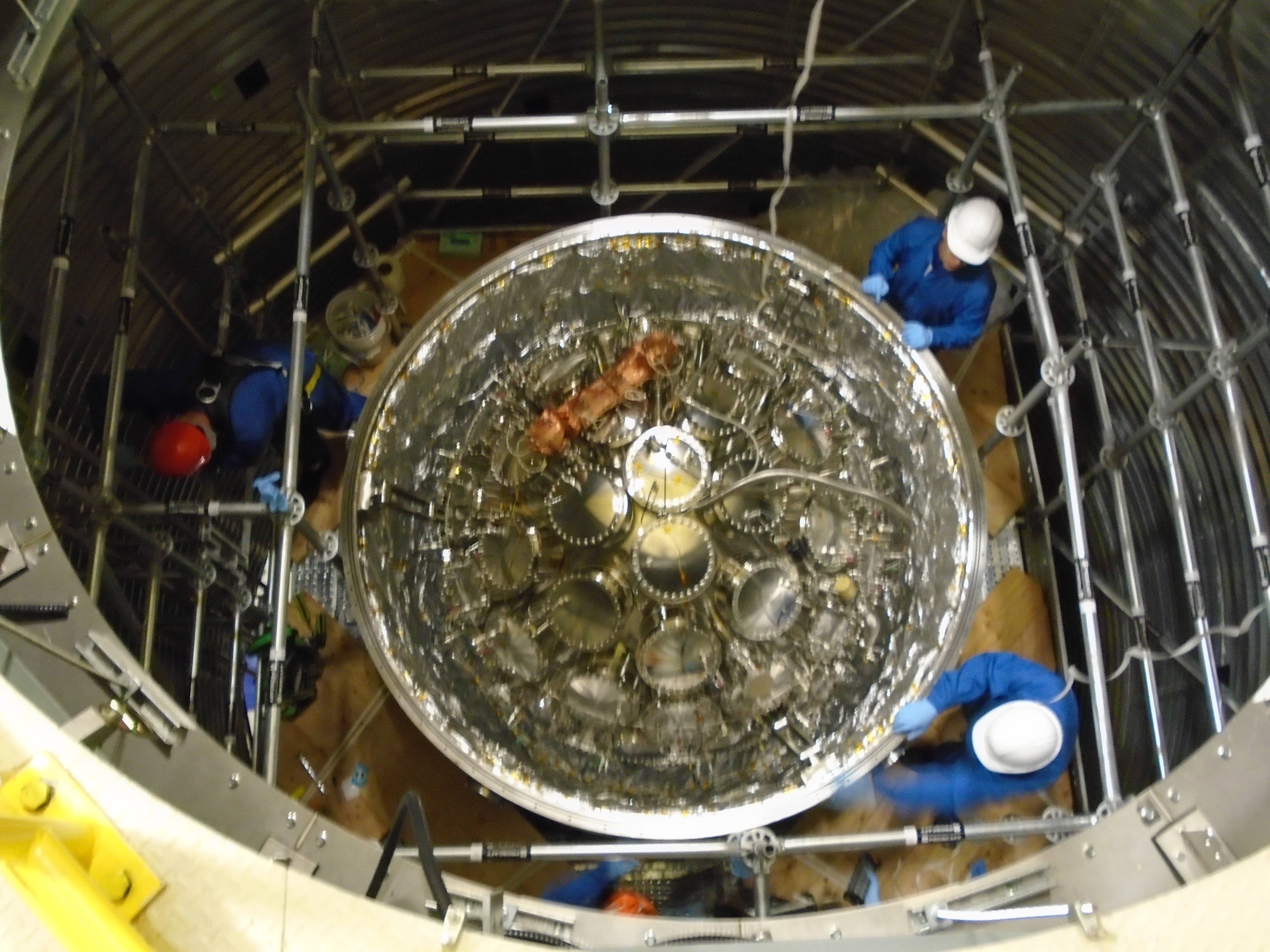}}
\hfill
\caption{(a) IV was on the way to the deck. (b) IV suspended inside the OV. }
\label{fig:IV_lift}
\end{figure}

The IV was moved into OV on November 17, 2014, as mentioned in last section, IV is prepared in the Cryo-pit and transported to the Cube Hall. During the transportation, the IV is filled with argon gas and kept the positive pressure in case a leak is created during the transportation.  Figure \ref{fig:IV_move} (a) shows the IV is on the move to the Cube Hall, and Fig. \ref{fig:IV_move} (b) shows the final examination of the IV before lifting. Subsequently, the IV was hoisted into the OV and suspended with three supporting arms as shown in Fig. \ref{fig:IV_lift}. A scaffolding was build around the OV for easy to access different elevation of OV to perform the final assembly. The copper components visible near the top of the IV make the connection to the cryogenic refrigerator which is mounted on the top dome of the OV. \par
A series of instrument cabling from IV to OV were carefully arranged in the following months. After the completion of connection of cable, the top dome of the OV was lowered to its position. The cables of  the air-side of the PMTs and instrument are housed in the water-proof hoses and extend to the deck to connect to the DAQ system and the corresponding equipment. The cooling lines are installed which extend to the top of water tank. The cryogenic and purification system connects to the OV from the deck and flow the gaseous argon into the IV to maintain the overpressure of the IV. 

\section{MiniCLEAN Detector in Cooling Phase}\label{sec:coolingphase}
\begin{figure}[tbp]
\centering
\graphicspath{{./fig/}}
\includegraphics[scale=0.4]{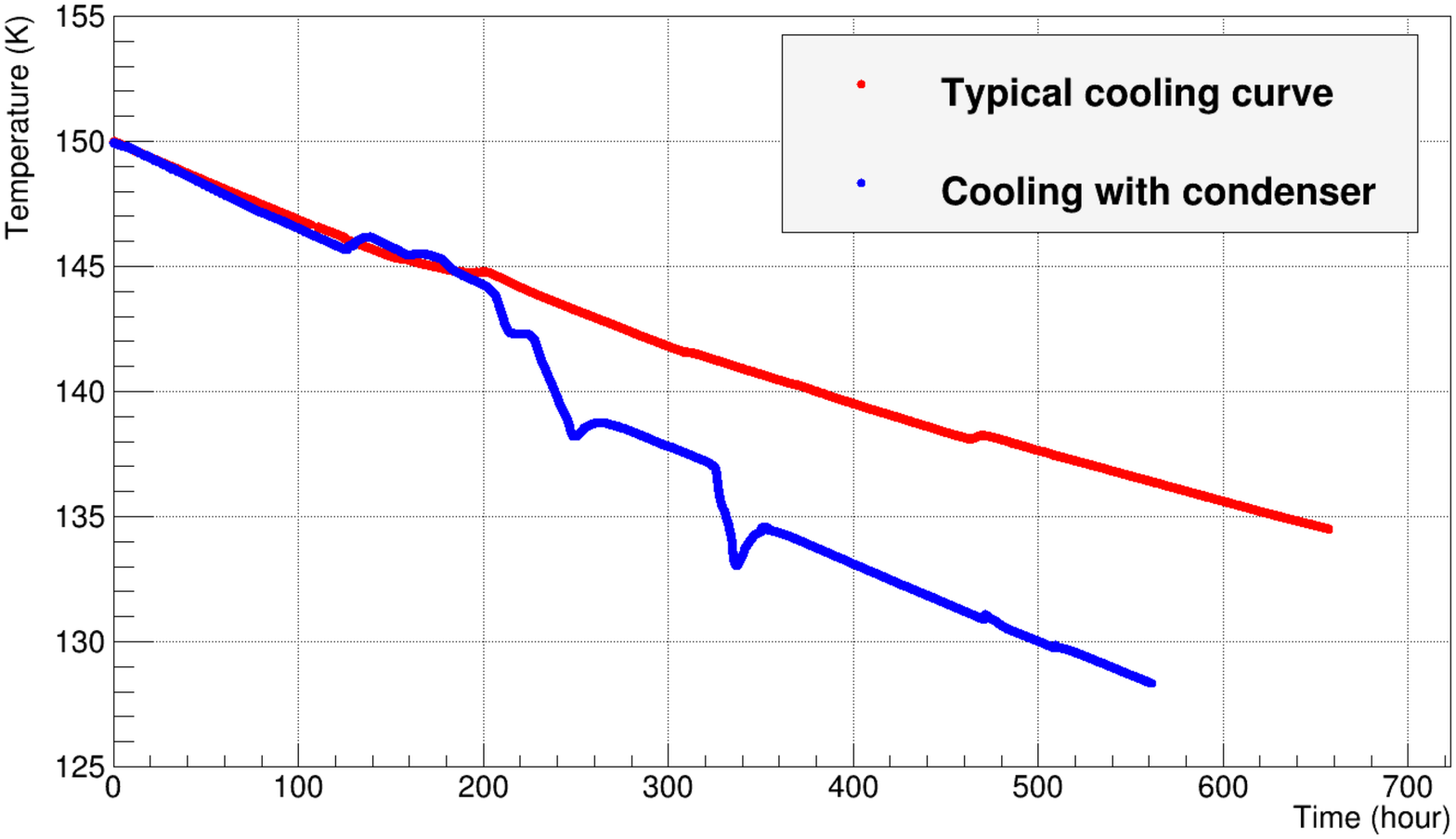}
\caption{ Comparison of cooling curve between typical cooling (red) and with condenser (blue). The condenser only running for short period of time between 200 hr and 350 hr, which effective expedite the cooling process.}
\label{fig:ftemp}
\end{figure}

\begin{figure}[tbp]
\centering
\graphicspath{{./fig/}}
\includegraphics[scale=0.4]{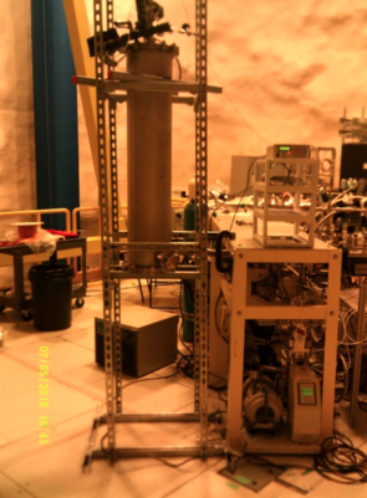}
\caption{ A photograph of condenser installed in support frame.}
\label{fig:condenser_photo}
\end{figure}

The construction of MiniCLEAN completed in June 2015 and the detector start cooling down in early August. The cooling rate was initially 6K per day but this rate gradually decreased.  Below 140 K, the cooling rate decreased to less than 1 K per day. We speculate that cooling rate deceased due to the decreased thermal conductivity of argon gas at low temperature. We experienced a leak incident in April 2016 resulting from over-bending the OV exhaust bellows while filling the water tank.  Adjusting the bellows required the opening of the OV vacuum. 
We expected that this operation would warm up the temperature of IV by 20 K, but in fact the temperature rose by over 100 K due due to an additional OV leak.
In order to improve the cooling rate we added an external condenser to our cryogenic system to drip liquid argon into the IV in addition to cooling the gas. 
We were thereby able to obtain a much larger cooling rate and the temperature of bottom of the  IV reached liquefaction point of Argon ($\sim$ 87K) on the 3rd of December 2016. Figure \ref{fig:ftemp} shows the changes in the cooling curve before and after the condenser was added. A condenser is a supplement cooling power designed to work with purification system. The condenser consists of two concentric cylinders of 3/8" steel with 1/2" plates closing both ends. The outer cylinder vacuum space provides insulation of inner cylinder which holds the liquid nitrogen. The purified argon gas enters the condenser and subsequently condensed inside the inner cylinder then flow into the IV  through 1/2" stainless steel pipe. The condenser is shown in Fig. \ref{fig:condenser_photo}\par
The condenser speed up the cooling process, however, it rely on the shipment of LN$_2$ dewar every week at SNOLAB. Therefore, the condenser running can not be operated continuously. In the first 6 hours, due to the long hose connect to the IV, only cold gas reaches the IV. When the long hose cold enough, the liquid flow into the IV which increase the rate of cooling. Failing to continuously filling from condenser causing the cooling process stalled.\par
A series of leak was found  after the bottom of the IV reached liquefaction points. The first leak was found in the exhaust vent line causing the reduction of triplet lifetime. In order to restore the purity of argon, a series of pump and purge cycle is performed to pump out the impurity efficiently. Later, various leak is identified in the condenser and the charcoal trap. Due to the pump and purge cycle and constantly checking the leak, the time to operate condenser is limited. Thus the temperature of IV is kept at around 130 -140 K. As of August 1st, 2017, the average temperature of IV is around 130 K. The comprehensive leak checking has performed on the all sub-system of MiniCLEAN. No source of leak is found, the MiniCLEAN detector will continue its cooling. In order to solve the problem of LN$_2$ supply for condenser. A plan to purchase a  cryocooler with 500 W cooling power at LN$_2$ temperature is made to operate the condenser 24/7. This new equipment should improve the cooling and filling rate. The estimated time to cool and fill up the IV with LAr is around 52 days\cite{fillprojection}. The detail description of the leak and the monitoring of triplet lifetime are in Chapter \ref{ch:gasrun} and \ref{ch:cold}


\chapter{\textit{In-Situ} Optical Calibration System}
The MiniCLEAN detector utilizes 92 PMTs to collect liquid/gaseous argon scintillation light.  Now a so-called Single PhotoElectron (SPE) calibration is important for ensuring the PMTs' stability, and in order to track PMT gain over time, an external, stable source is required.  Towards this, an LED light injection system was developed for this purpose.  An LED is extremely stable over short time scale ($\sim$minutes), its intensity can be changed (through software), and it can reach a high repetition rate ($\sim$100 MHz).  Therefore the LED serves as a stable, programable external light source.\par
The system consist of 6 blue and 6 UV LEDs.  Now at low intensity, the blue LED can be used to determine the PMT gain; and, at high repetition rate, the PMT stability gain be can tracked hourly.  On the other hand, UV LEDs can not only help check TPB stability, but also verify the integrity of the optical path of scintillation light in the detector.  The LED system will be described in this chapter and the \textit{In-Situ} optical calibration, in association with the preliminary LED data analysis, will be described in the next chapter.
\section{LEDs}
\begin{figure}[htbp]
\centering
\graphicspath{{./fig/LED_system/}}
\includegraphics[scale=0.3]{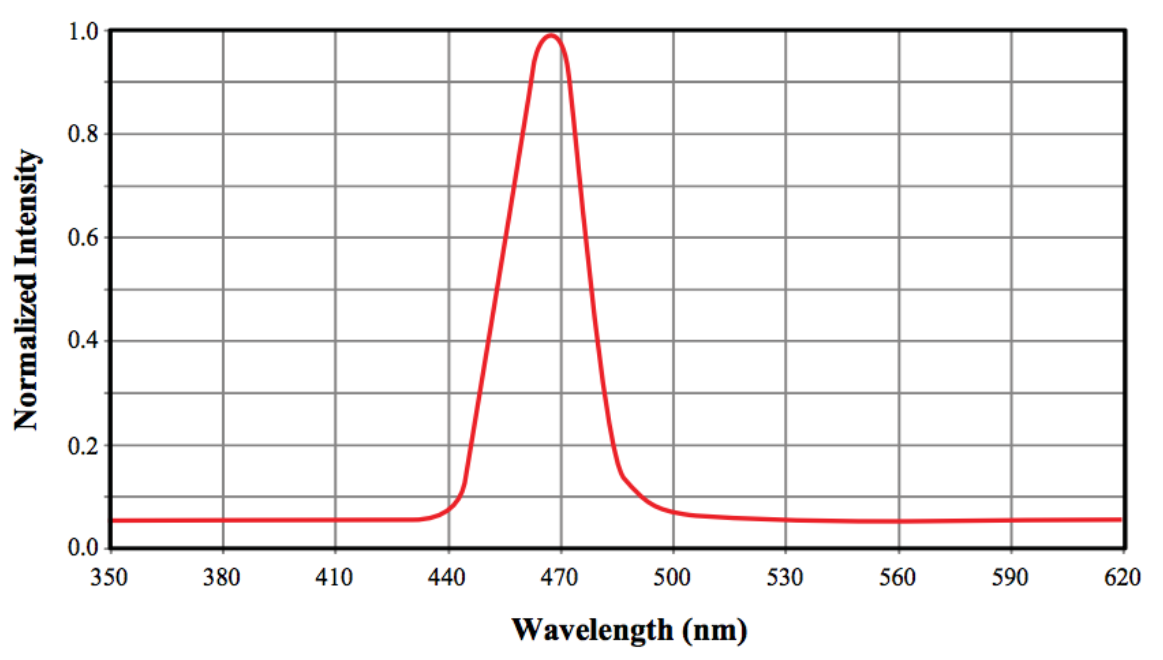}
\caption{ Typical spectral intensity distribution of blue LED. }
\label{fig:fbluespec}
\end{figure}

The blue LEDs are of type ThorLabs LED 465E\footnote{\url{https://www.thorlabs.com/thorproduct.cfm?partnumber=LED465E}}.  Their spectral intensity peaks at 465 nm as shown in Fig. \ref{fig:fbluespec}.  The VUV argon scintillation light is TPB shifted to the blue end of visible spectrum, and as such the blue LED was a natural choice.  The PMTs used by MiniCLEAN are described in section \ref{ch:detector}, and the Fig. \ref{fig:fpmtspec} shows the spectral response characteristic indicating that the most sensitive region is indeed at blue light range.  This ensures that when performing the SPE calibration, the very weak LED light will be detected with the highest possible collection efficiency.\par
\begin{figure}[htbp]
\centering
\graphicspath{{./fig/LED_system/}}
\includegraphics[scale=0.3]{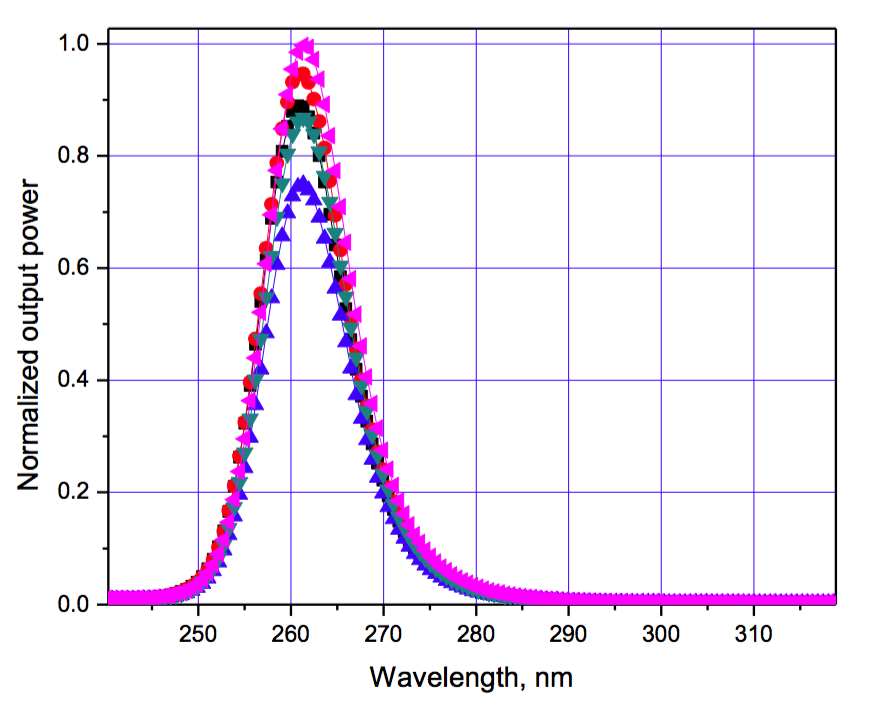}
\caption{ Typical spectral intensity distribution of UV LED. }
\label{fig:fuvspec}
\end{figure}
\begin{figure}[htbp]
\centering
\graphicspath{{./fig/LED_system/}}
\includegraphics[scale=0.5]{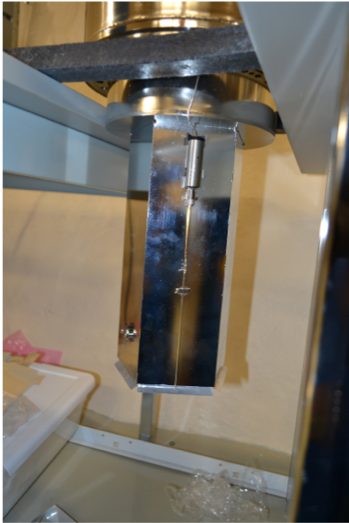}
\caption{  LED mounted on the side of optical cassettes. The upper part of cassettes is housing the PMT. The LED couples to a optical fiber and poke through the baffle then inject light into the active volume.}
\label{fig:fledcassettes}
\end{figure}
\begin{figure}[htbp]
\centering
\graphicspath{{./fig/LED_system/}}
\includegraphics[scale=0.2]{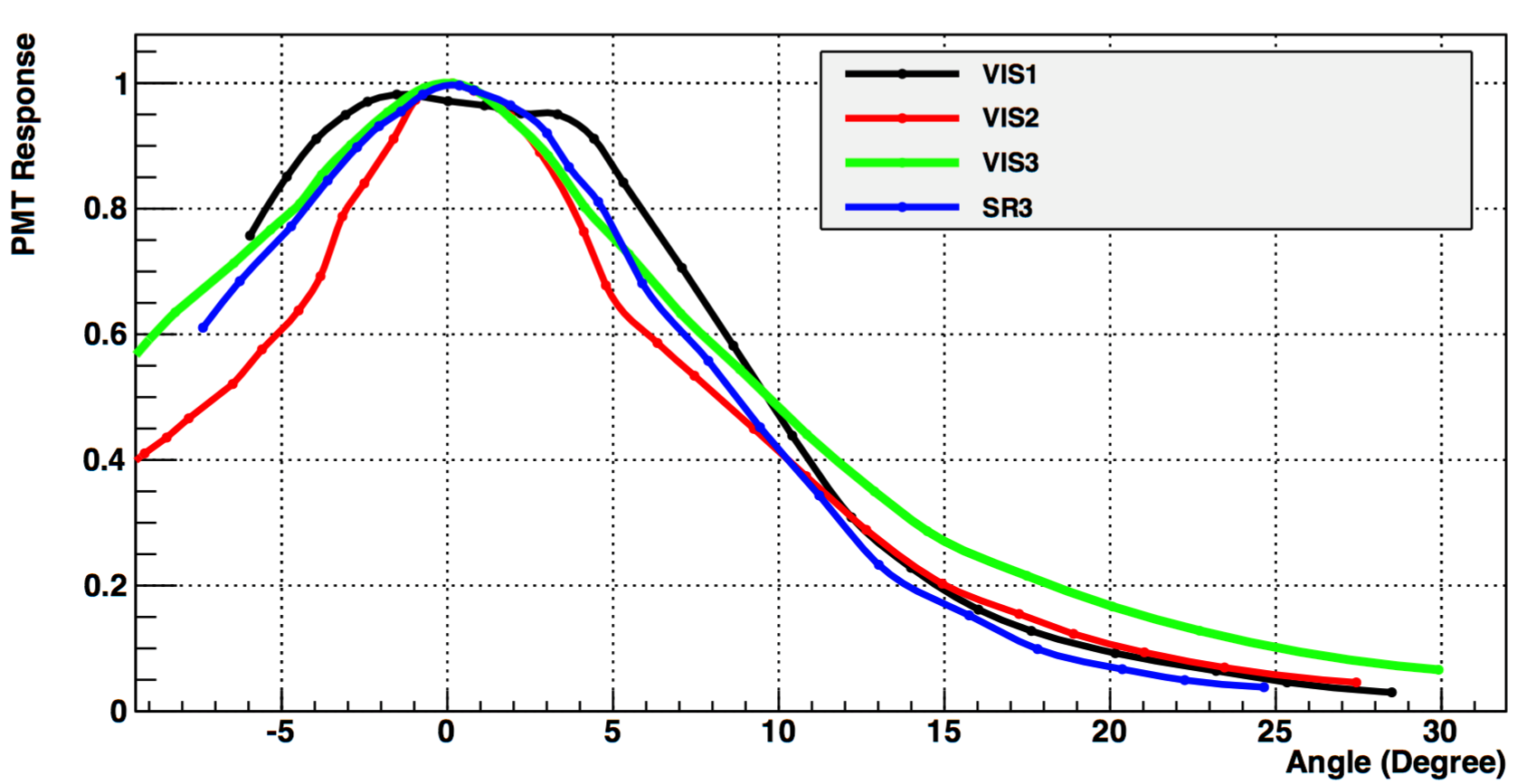}
\caption{ Angular distribution for different fibers.}
\label{fig:ffiberdis}
\end{figure}

The UV LEDs are from UVTOP 260 sensor electronic technology, inc., with peak intensity at 260 nm as shown in Fig \ref{fig:fuvspec}.  The LEDs are mounted on the side of optical cassettes and they are coupled to an optical fiber in order to inject the light into the detector active volume, as shown in Fig. \ref{fig:fledcassettes}.  Each LED is coupled to an optical fiber with a standard SMA connector on one end and bare on the other.  Both fiber ends were polished to a flat perpendicular surface by using the termination kit from Ocean Optics.  Fibers were tested to understand the characteristic, before making decision.  Figure \ref{fig:ffiberdis} shows the angular distribution of each tested fiber.\par
Because of its large angular coverage, fiber was chosen.  In the other words, the LED photons can reach more PMTs, giving an advantage towards performing the optical calibration.  Ocean Optics SR 600 nm\footnote{\url{https://oceanoptics.com/product-category/do-it-yourself-fibers/}} fiber was used with the UV LEDs, because of its increased transmittance in the UV range. For the blue LEDs, ThorLabs BFH48 600 nm\footnote{\url{https://www.thorlabs.com/catalogpages/Obsolete/2015/BFH48-600.pdf}} was used because of its higher numerical aperture.  Additionally, a convex lens (ThorLabs LB1157\footnote{\url{https://www.thorlabs.com/thorproduct.cfm?partnumber=LB1157-A}}) was placed between the fiber connection and the blue LEDs in order to focus the light onto the face of the fiber at an angle larger than the output angle of the LED.

\section{Angular distribution of LEDs}
\begin{figure}[htbp]
\centering
\graphicspath{{./fig/LED_system/}}
\includegraphics[scale=0.4]{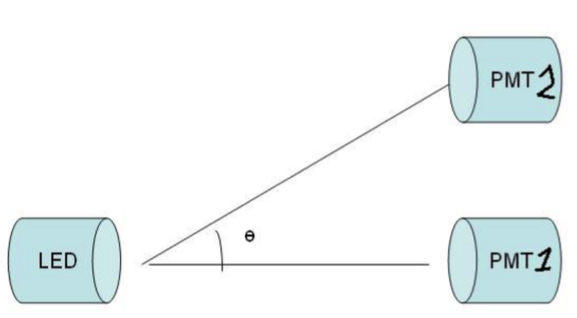}
\caption{ Experimental setup for measuring the angular distribution.}
\label{fig:fledangularsetup}
\end{figure}

In an attempt to increase the fiber emission light angular distribution, two modifications were tested.  The first approach used an irregular fiber tip shape to increase angular coverage, whereas the second directly coats the fiber tip with TPB in an attempt to further diffuse light.  Fig. \ref{fig:fledangularsetup} shows the experimental setup.  PMT1 received head-on photons from the fiber tip and was used to normalize the response from PMT2.  PMT2 was arranged such that different angles ($\theta$) could be attained.\par
\begin{figure}[htbp]
\hfill
\subfloat[]{\includegraphics[width=7cm]{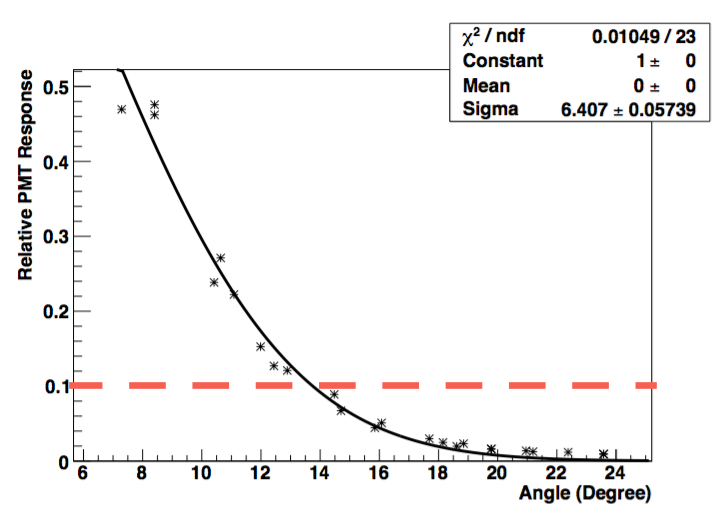}}
\hfill
\subfloat[]{\includegraphics[width=7cm]{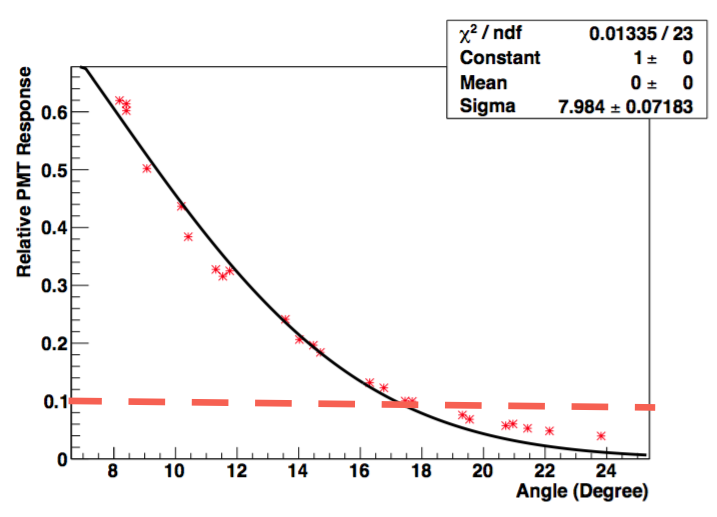}}
\hfill
\caption{The angular distribution for (a) Fiber with smooth tip without TPB coating (No. 1). (b) Fiber with irregular tip without TPB coating (No. 2).}
\label{fig:fiberangulartest1}
\end{figure}
\begin{figure}[htbp]
\hfill
\subfloat[]{\includegraphics[width=7cm]{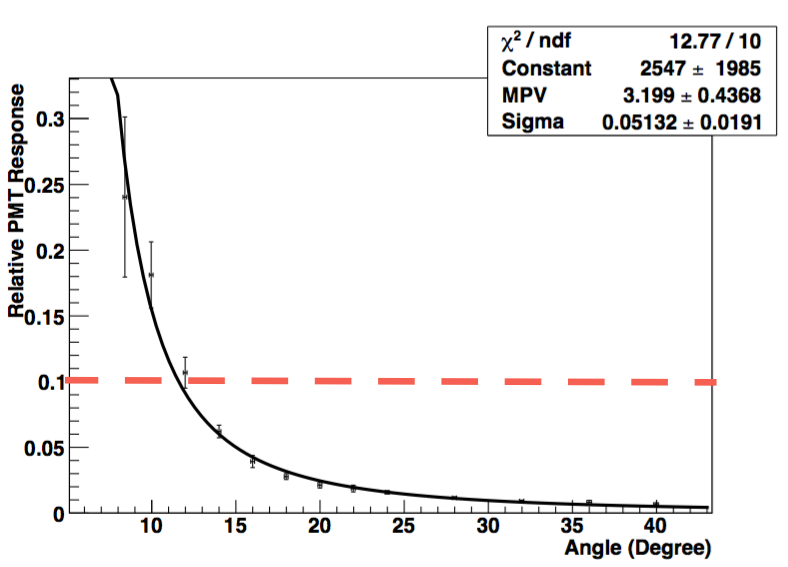}}
\hfill
\subfloat[]{\includegraphics[width=7cm]{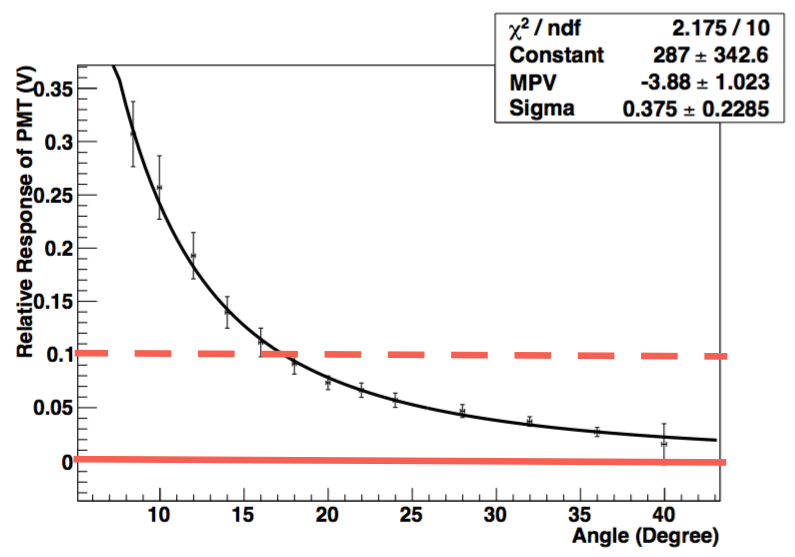}}
\hfill
\caption{The angular distribution for (a) Fiber with smooth tip with TPB(100 mg) coating (No. 1). (b) Fiber with irregular tip with TPB(150 mg) coating (No. 2).}
\label{fig:fiberangulartest2}
\end{figure}
\begin{figure}[htbp]
\centering
\graphicspath{{./fig/LED_system/}}
\includegraphics[scale=0.3]{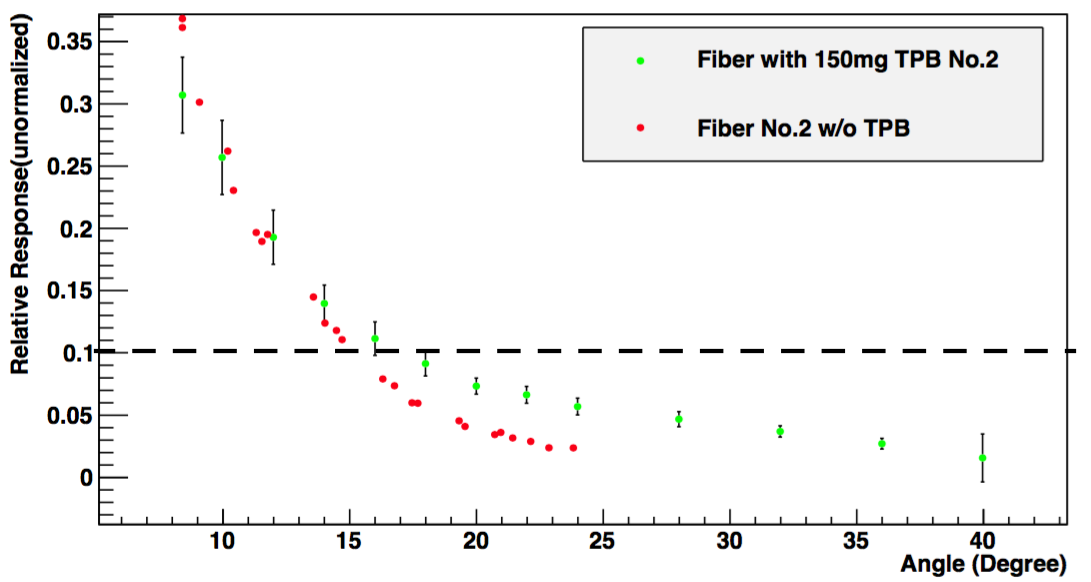}
\caption{Results of fiber No. 2, red dot shows the result of fiber tip without TPB, and green dot is the results of fiber tip with TPB(150 mg) coating.}
\label{fig:fiberangulartesttotal}
\end{figure}

Fibers No.1 had a smooth face, whereas No.2 had an irregular tip surface.  After one test, the two fibers were then tip coated with TPB: No.1 coated with 100 mg, and No.2 with 150 mg.  Figure \ref{fig:fiberangulartest1} shows the results from fibers without TPB.  The fiber with an irregular tip (No.2) shape reaches 10\% of intensity of PMT1 at around 17\si{\degree}, whereas the smooth-faced fiber (No. 1) has dropped to 2\%.  On the other hand, Fig. \ref{fig:fiberangulartest2} shows the results with both fibers TPB coated.  With first no TPB coating, both fibers reach the 10\% level at the same angle.  But with TPB coating a longer tail is obtained.
The results indicate a larger angular distribution, implying more PMTs illuminated by LED photons, for the irregular fiber tip shape.  Furthermore, with the TPB-coated fiber tip, a longer tail after 20\si{\degree} is obtained; however, the contribution from TPB is under 10\%, which is not al large improvement.  Figure \ref{fig:fiberangulartesttotal}, showing the angular distribution of fiber No.2 (no TPB coating), indicates that TPB doesn't make a pronounced improvement before 20\si{\degree}.  It was noticed that TPB can flake off the fiber tip, therefore if the LED light yield is lower than normal then the fiber tip coating integrity should be investigated.  Because of these observations, the TPB coating idea was discarded, but, because of the enhanced angular coverage, the irregular tips shape was retained.\par
Following these tests, a lens -- for the blue LED -- was added in order to improve light collection efficiency to the fiber, and towards the same goal, a spacer was added for the UV LED.  The angular distribution of emitted light from optical fiber was then measured for the LEDs.  The emitted light was projected onto a thin screen and a digital photograph was acquired.  Light emitted from the fiber tip was then projected onto a thin paper screen.  For the UV LED, the thin paper screen worked as a weak wavelength shifter, shifting invisible UV light into the blue end of the visible spectrum.  In order to minimize photon leakage, the whole operation was carried out in a dark box.  A 5 seconds default exposure time was used for the blue LED and the brightness was carefully reduced to avoid saturation near the peak.  The UV LED, however, did not produce enough light to be visible to the naked eye, and so a longer 15 seconds exposure time was employed.  A 10 seconds camera delay was used so that after set up, the dark box could be closed without rush.  Moreover, when the UV LED was in place, the room light was turned off in order to further reduce leaking photon into the dark box.\par
The 2848 x 2136 digital images were first scaled to 1/2 their original pixel count, to 1424 X 1068.  This resizing results in a local averaging that dampens extreme values, thereby improving peak finding.  Blue LED images were converted into gray-scale whereas the UV LED images had their blue component, of the RGB valued image, isolated and then converted to gray-scale.  The peak finder locates the maximum intensity and then checks values a few pixels around it.  If these surrounding pixels are less than 75\% of the peak the point is considered extreme (unusable) and the finder will then look for the next maximum value.  The intensity is then sampled at user-defined distances from the peak location.\par
\begin{figure}[htbp]
\centering
\graphicspath{{./fig/LED_system/}}
\includegraphics[scale=0.4]{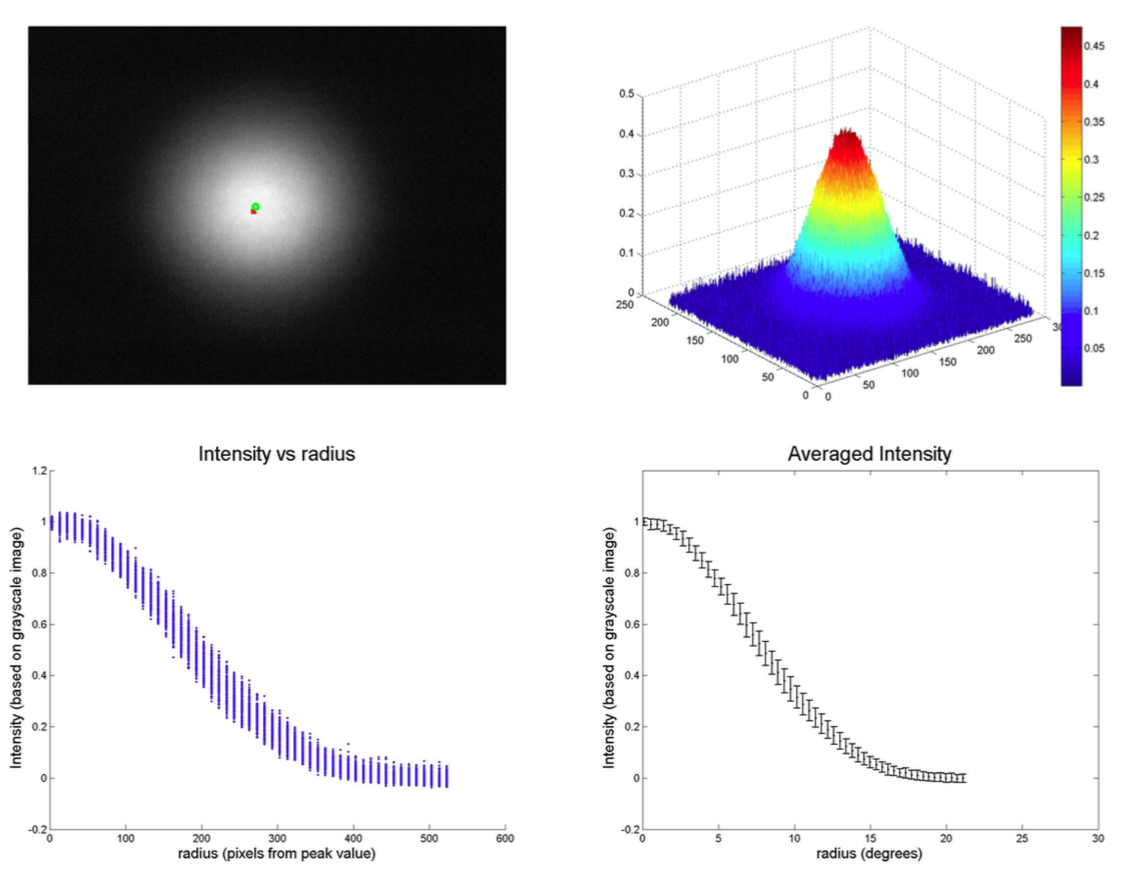}
\caption{Results from Blue LED. Upper left : Image after gray-scale conversion (see text). Upper right : 3-D representation of the gray-scale image. Lower left : Intensity as a function of distance. Lower right : Average intensity vs angular distance from the fiber axis.}
\label{fig:fblueana}
\end{figure}
\begin{figure}[htbp]
\centering
\graphicspath{{./fig/LED_system/}}
\includegraphics[scale=0.4]{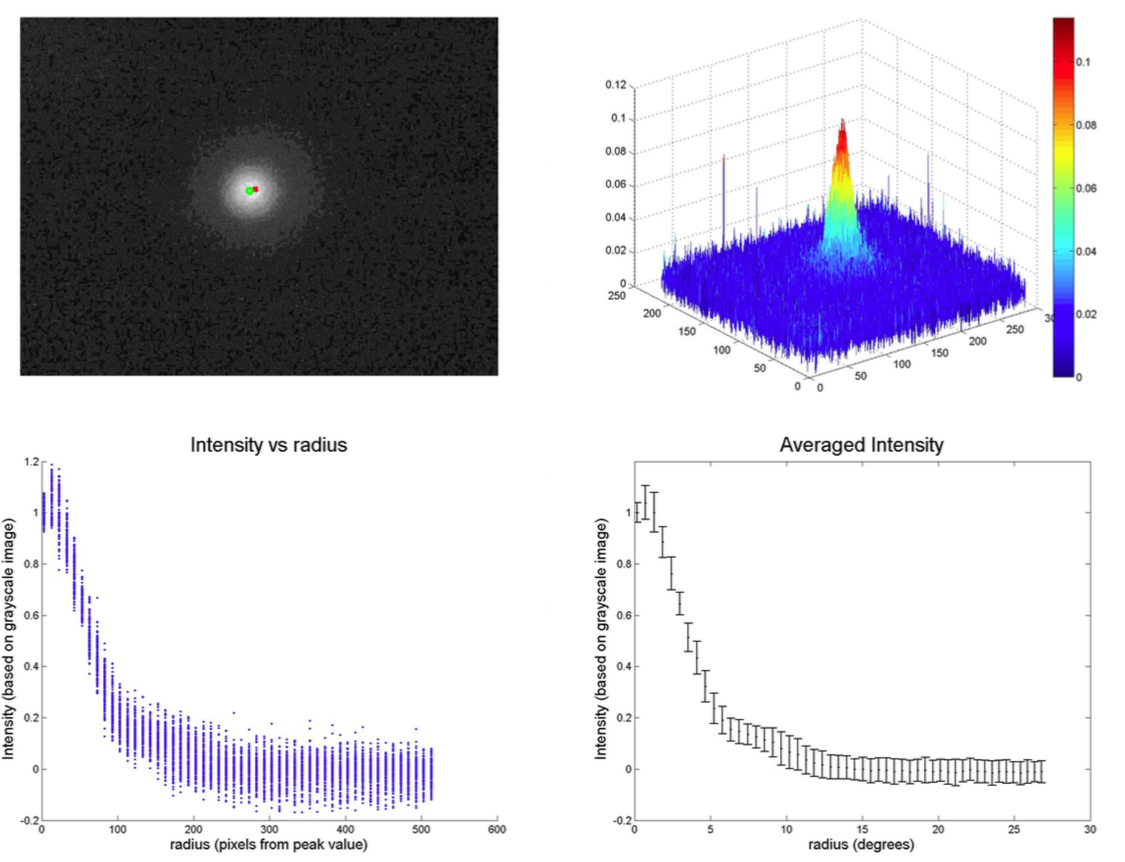}
\caption{ Results from UV LED. Upper left : Image after gray-scale conversion (see text). Upper right : 3-D representation of the gray-scale image. Lower left : Intensity as a function of distance. Lower right : Average intensity vs angular distance from the fiber axis.}
\label{fig:fuvana}
\end{figure}

To examine procedure reproducibility, two data sets for the blue LEDs were acquired.  Figures \ref{fig:fblueana} and \ref{fig:fuvana} show example plots for blue- and UV-LED tests, respectively.  The upper left displays the image after gray-scale conversion.  The red dot represents the last peak found with the peak finder, and the green circle is the peak location used in calculations.  If the two did not over lay then the peak location was inputted manually.  The upper right is a 3D representation of the gray-scale image (not normalized).  The lower left plots each intensity sampled (normalized) as a function of distance, in pixels, from the peak.  The lower right is the average intensity (normalized) as a function of angular distance from the fiber axis, which is assumed to coincide with the peak value.  The data analysis was carried out using MatLab. The final results are summarized in Tab. \ref{table:ledanadis}.\par

\begin{table}[htbp]
\caption{Average Angles at Select Relative Intensities} 
\centering 
\begin{adjustbox}{width=1\textwidth}
\begin{tabular}{c c c c c c c c c} 
\hline\hline 
&Blue 1 & Blue 2 & Blue 3 & Blue 4 & Blue 5 & Blue 6 & Blue Spare & Average \\ [0.5ex] 
\hline 
50\% & 8.3\si{\degree} & 7.7\si{\degree} & 7.7\si{\degree} & 6.9\si{\degree} I& 7.9\si{\degree} & 6.9\si{\degree} & 6.4\si{\degree} & 7.4\si{\degree}\\[0.5ex]
25\% & 11.2\si{\degree} & 12.0\si{\degree} & 11.4\si{\degree} & 10.6\si{\degree} & 11.0\si{\degree} & 11.0\si{\degree} & 9.8\si{\degree} & 11.0\si{\degree}\\[0.5ex]
10\% & 13.4\si{\degree} & 14.6\si{\degree} & 14.2\si{\degree} & 13.6\si{\degree} & 13.8\si{\degree} & 13.8\si{\degree} & 13.0\si{\degree} & 13.8\si{\degree}\\[0.5ex]
\hline
&Blue 1 redo& Blue 2 redo& Blue 3 redo& Blue 4 redo& Blue 5 redo& Blue 6 redo& Blue Spare redo& Average \\ [0.5ex] 
\hline
50\% & 8.1\si{\degree} & 7.3\si{\degree} & 8.3\si{\degree} & 8.1\si{\degree} & 8.1\si{\degree} & 6.9\si{\degree} & 6.9\si{\degree} & 7.7\si{\degree}\\[0.5ex]
25\% & 10.6\si{\degree} & 11.4\si{\degree} & 11.8\si{\degree} & 11.8\si{\degree} & 11.2\si{\degree} & 11.0\si{\degree} & 10.2\si{\degree} & 11.1\si{\degree}\\[0.5ex]
10\% & 13.0\si{\degree} & 13.8\si{\degree} & 14.6\si{\degree} & 14.2\si{\degree} & 13.8\si{\degree} & 13.8\si{\degree} & 13.4\si{\degree} & 13.8\si{\degree}\\[0.5ex]
\hline
&UV 1 & UV 2 & UV 3 & UV 4 & UV 5 & UV 6 & UV Spare & Average \\ [0.5ex] 
\hline
50\% & 4.7\si{\degree} & 5.2\si{\degree} & 9.1\si{\degree} & 4.7\si{\degree} & 3.5\si{\degree} & 4.7\si{\degree} & 5.2\si{\degree} & 5.3\si{\degree}\\[0.5ex]
25\% & 9.1\si{\degree} & 9.7\si{\degree} & 11.3\si{\degree} & 8.0\si{\degree} & 5.2\si{\degree} & 6.9\si{\degree} & 8.0\si{\degree} & 8.3\si{\degree}\\[0.5ex]
10\% & 12.4\si{\degree} & 12.9\si{\degree} & 12.9\si{\degree} & 12.4\si{\degree} & 9.1\si{\degree} & 11.3\si{\degree} & 12.4\si{\degree} & 11.9\si{\degree}\\[0.5ex]
\hline\hline
\hline 
\end{tabular}
\end{adjustbox}

\label{table:ledanadis} 
\end{table}
The blue LED results show a crisp 3D representation, with the average intensity error typically below 5\%.  Comparing the two sets of blue measurements in the above table, it is clear that they are consistent with each other with differences typically below 1 degree.  The average angles for the intensities are consistent, suggesting that the results are reproducible within the random error.  Blue 22 shows a distinct deformation near the peak, which is captured in both measurements sets.  Blue 4, on the other hand, shows a slight deformation at the peak that may be the remains of the double peak that is seen without the use of a lens.\par
The UV results provide a general trend of the distribution but are subject to large errors.  The error in the UV results were typically 10-15\%, with UV 1 and 3 having larger errors -- 15-20\% and 20-30\%, respectively -- but UV 5 much less than 10-15\%.  In an attempt to obtain clearer pictures and reduce the error UV 1 and 3 were repeated, but the results were similar thus the previous measurements were retained.  UV 5 is unusually crisp and narrow and was observed to be much brighter than the other LEDs.  Due to mechanical failures Blue spare and UV 6 were not installed in the final detector.

\section{UNM LED pulser}
\begin{figure}[htbp]
\hfill
\subfloat[]{\includegraphics[width=7cm]{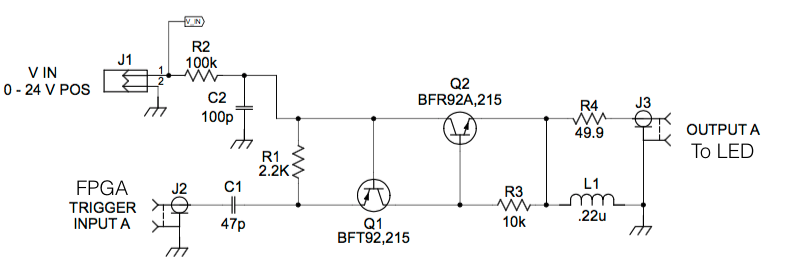}}
\hfill
\subfloat[]{\includegraphics[width=7cm]{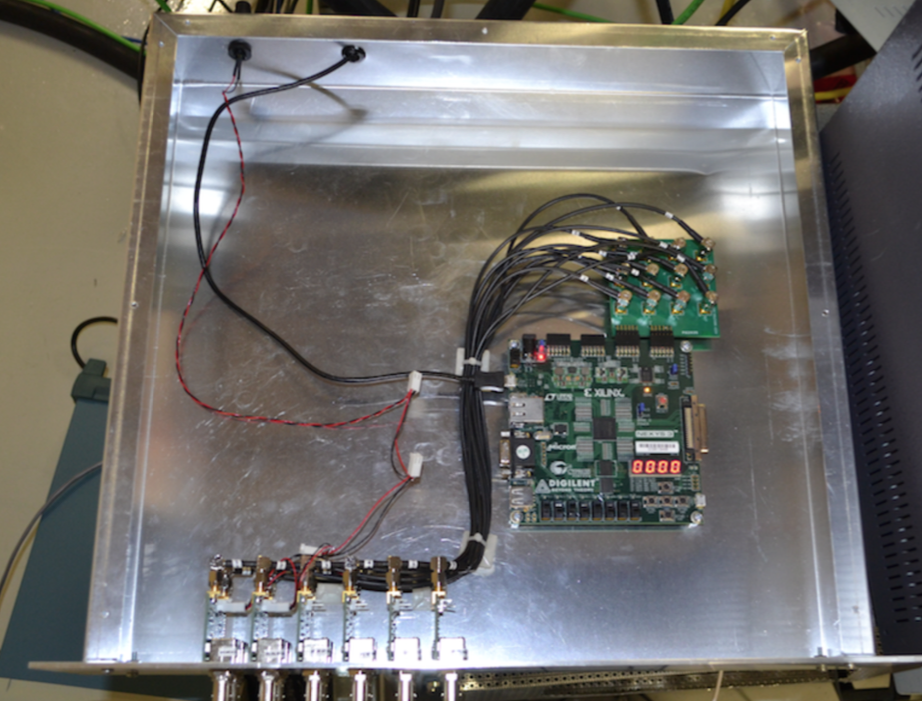}}
\hfill
\caption{ (a) UNM LED pulser circuit : FPGA and power supply are connected to the circuit as shown in the figure. When FPGA sends trigger signal the two transistors switches on and let the current from power supply flows through and illuminate LED. (b) LED pulser box (see text)}
\label{fig:fpulsercircuit}
\end{figure}

The prompt component of argon scintillation light has very fast 6 ns decay time, thus to best mimic this the LED pulses require a very high rate.  The LED pulse driven by a typical expensive, commercially-available pulser can achieve this goal.  Alternatively, Kapustinsky\cite{Kapustinsky1985612} offer a low cost pulser consisting of two fast transistors.  This pulser is based on the fast discharge of a small capacitor via a complementary pair of RF transistors.  We modified the original circuit to better accommodate both Blue and UV LEDs.  The resulting UNM LED pulser consist of two parts, a field-programmable gate array (FPGA) and the driver circuit.  The main circuit is shown in Fig. \ref{fig:fpulsercircuit}.  The FPGA sends a trigger pulse to the driver board through coaxial cables.  The trigger is a positive 3.5 V pulse that rides on the variable positive dc bias level, 0-24 V.  The dc component of the trigger charges a 100 pF capacitor.  The complementary-pair transistors are switched on by the trailing edge of the differentiated input pulse.\par
\begin{figure}[htbp]
\hfill
\subfloat[]{\includegraphics[width=7cm]{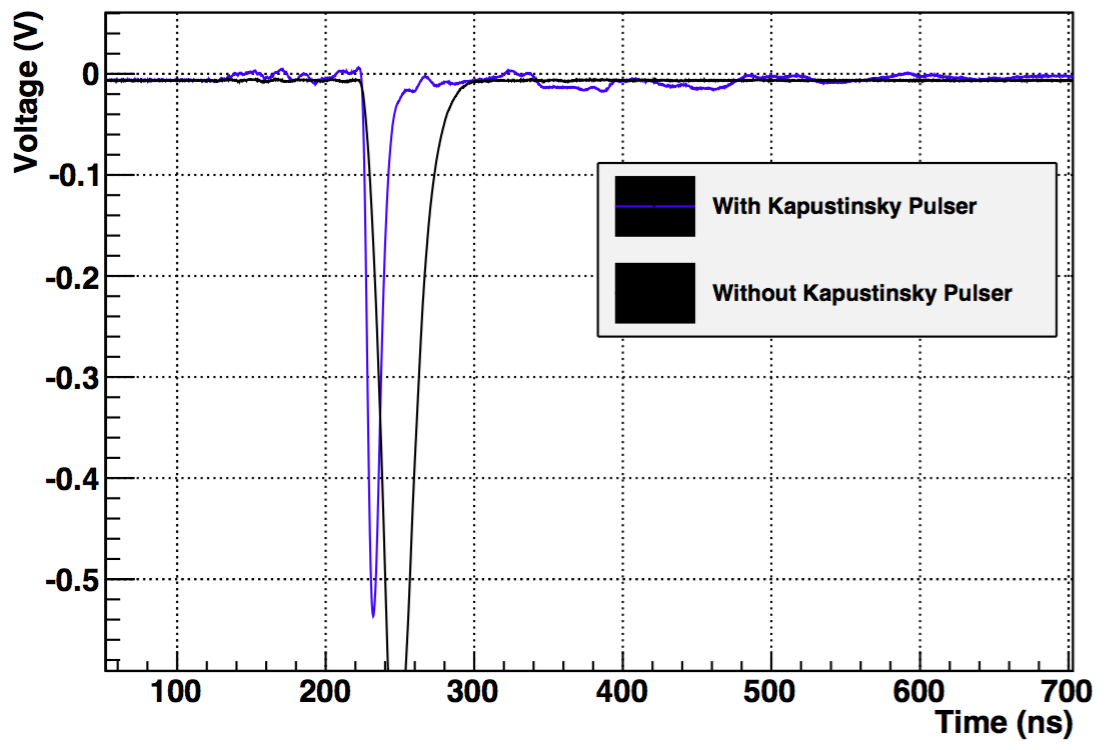}}
\hfill
\subfloat[]{\includegraphics[width=7cm]{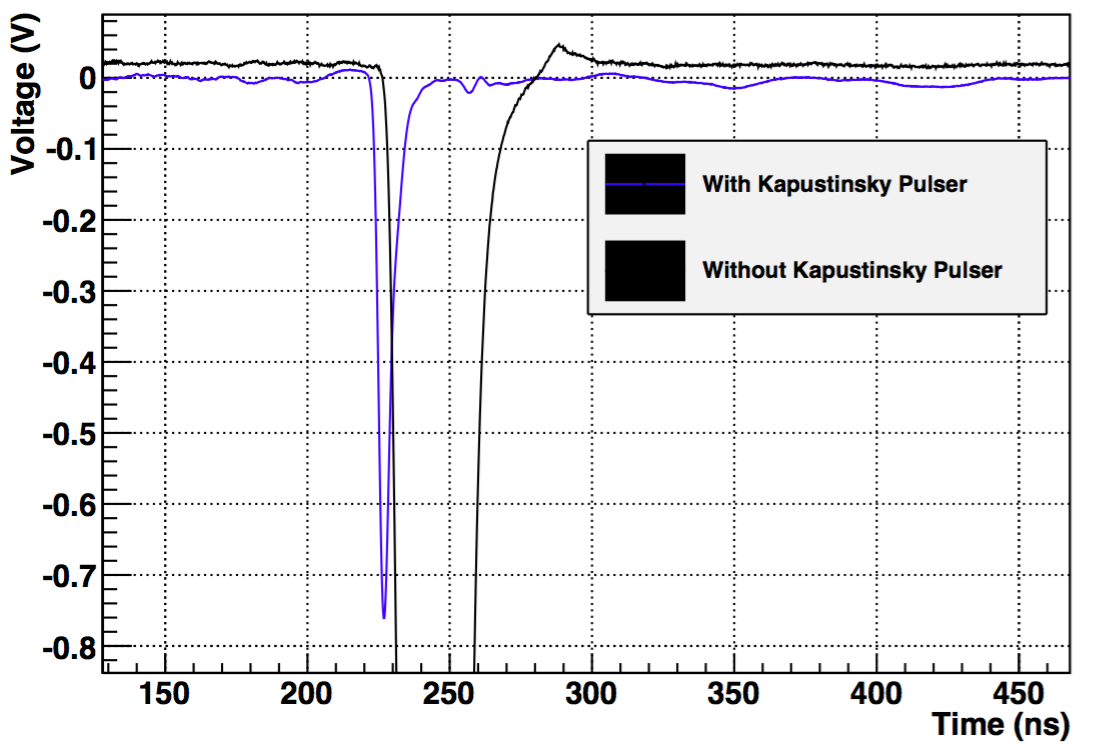}}
\hfill
\caption{ Pulse shape from LED on oscilloscope (a) The blue curve is from UNM LED pulser and the black curve is from normal pulser (500 MHz) for Blue LED. (b) The blue curve is from UNM LED pulser and the black curve is from normal pulser (500 MHz) for UV LED.}
\label{fig:fblueuvpulseshape}
\end{figure}
\begin{figure}[htbp]
\hfill
\subfloat[]{\includegraphics[width=7cm]{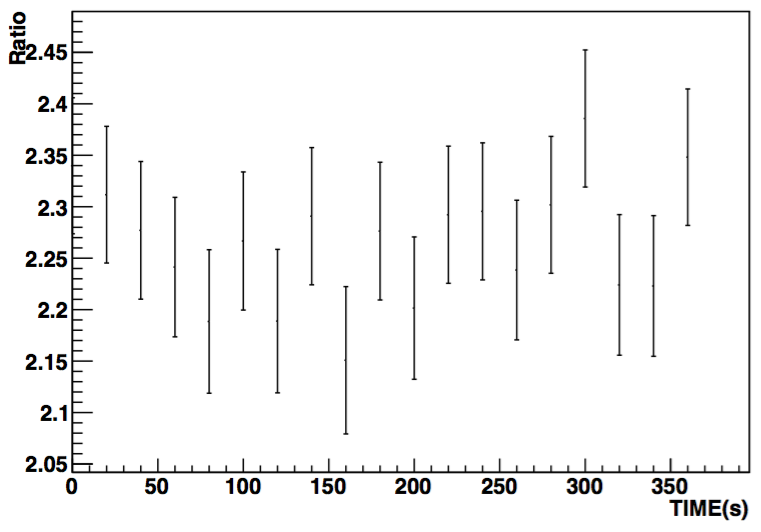}}
\hfill
\subfloat[]{\includegraphics[width=7cm]{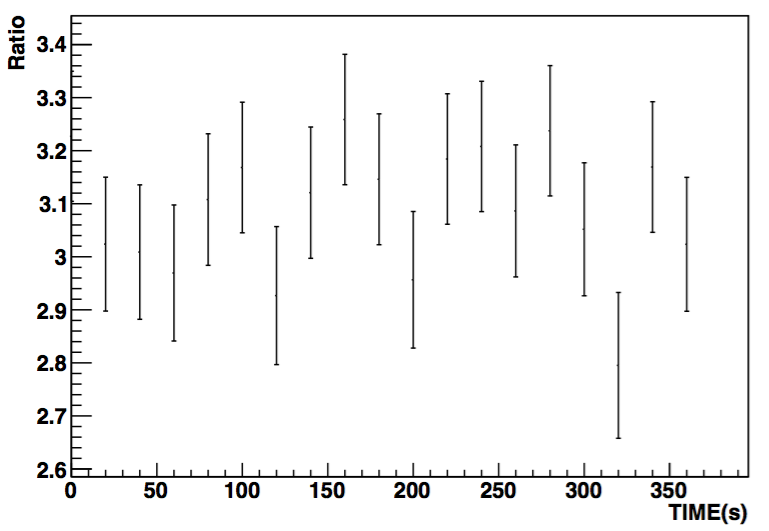}}
\hfill
\caption{Ratio of pulse width from normal pulser (500 MHz) and UNM LED pulser for (a) Blue LED. (b) UV LED.}
\label{fig:blueuvcompare}
\end{figure}
\begin{figure}[htbp]
\hfill
\subfloat[]{\includegraphics[width=7cm]{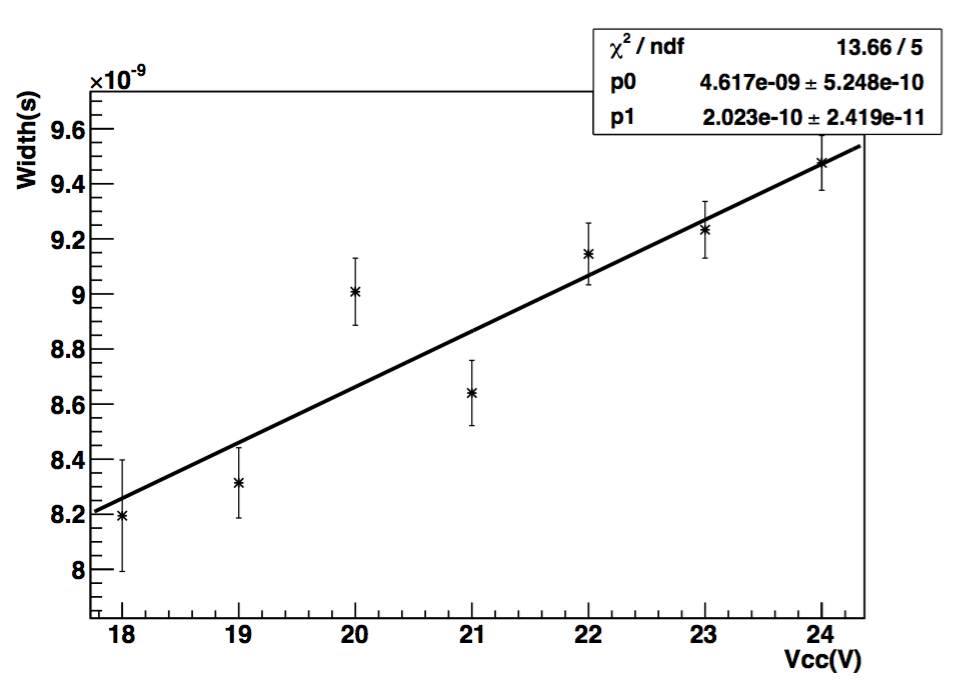}}
\hfill
\subfloat[]{\includegraphics[width=7cm]{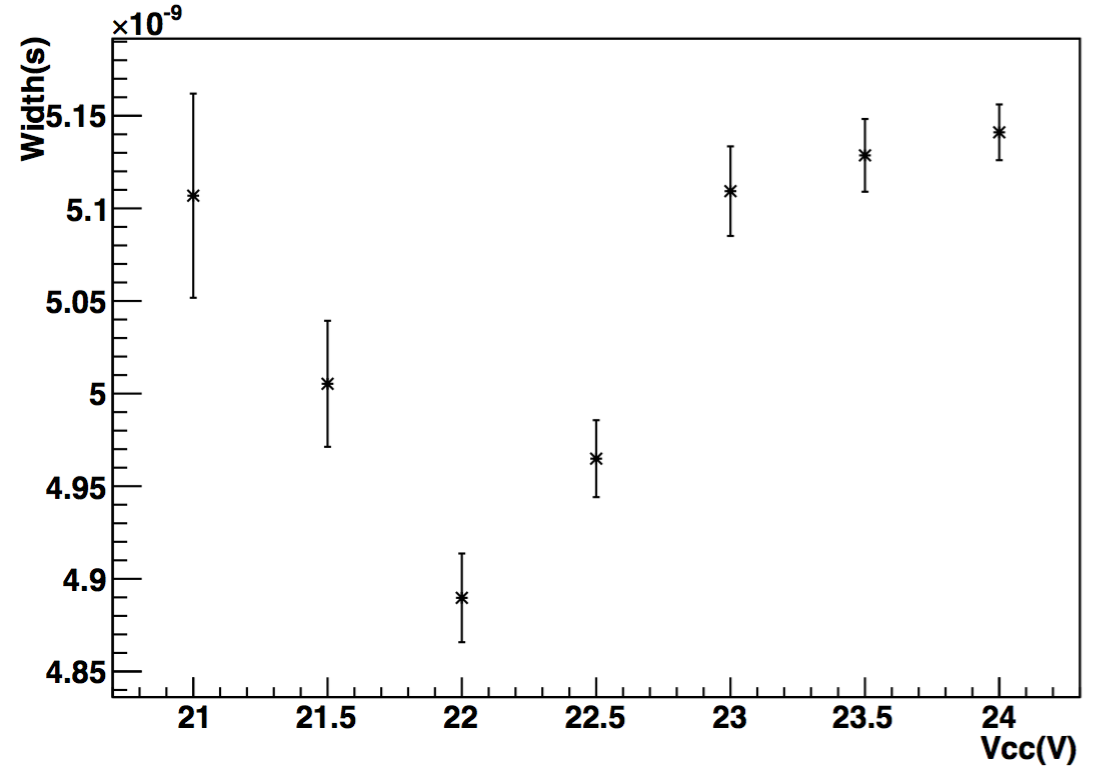}}
\hfill
\caption{Pulse width vs bias voltage (V$_{cc}$) for (a) Blue LED. (b) UV LED.}
\label{fig:blueuvcomparebias}
\end{figure}

The subsequent circuit path to ground provide a low impedance path for the capacitor to dump its charge through the LED.  The 220 nH inductor in the circuit develops charge in opposition to the discharging capacitor.  This action reduces the decay constant $\tau$ of the light pulse, which follows the time dependance e$^{-t/\tau}$.  The 220 nH inductor shortens the time constant to approximately 6 ns.  Figure \ref{fig:fblueuvpulseshape} compares the pulse shape driven by the UNM LED pulser and the normal pulser, and it is clearly seen that the UNM LED pulser improve the pulse width by factor of 2-3 as shown in Fig. \ref{fig:blueuvcompare}.  The output pulse width as a function of bias voltage (V$_cc$) is also investigated, and no apparent relationship between these two -- Fig. \ref{fig:blueuvcomparebias} -- is apparent.\par

\section{LED LN$_2$ dunk test }
\begin{figure}[htbp]
\hfill
\subfloat[]{\includegraphics[width=7cm]{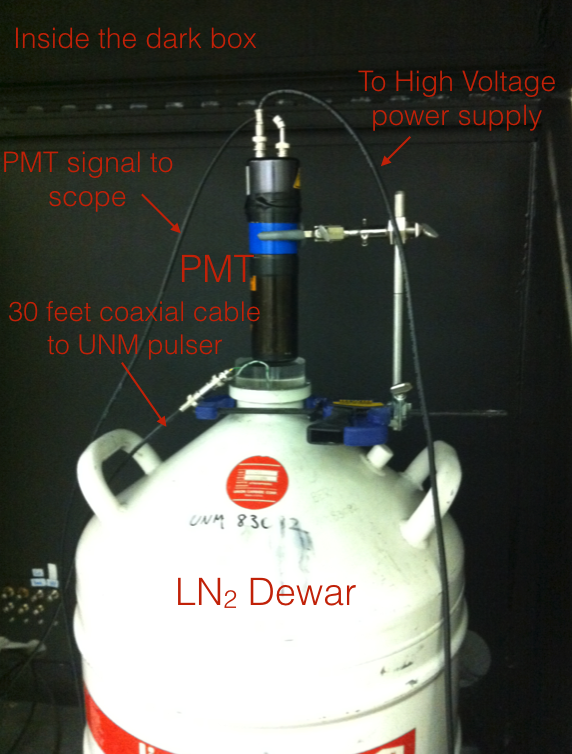}}
\hfill
\subfloat[]{\includegraphics[width=7cm]{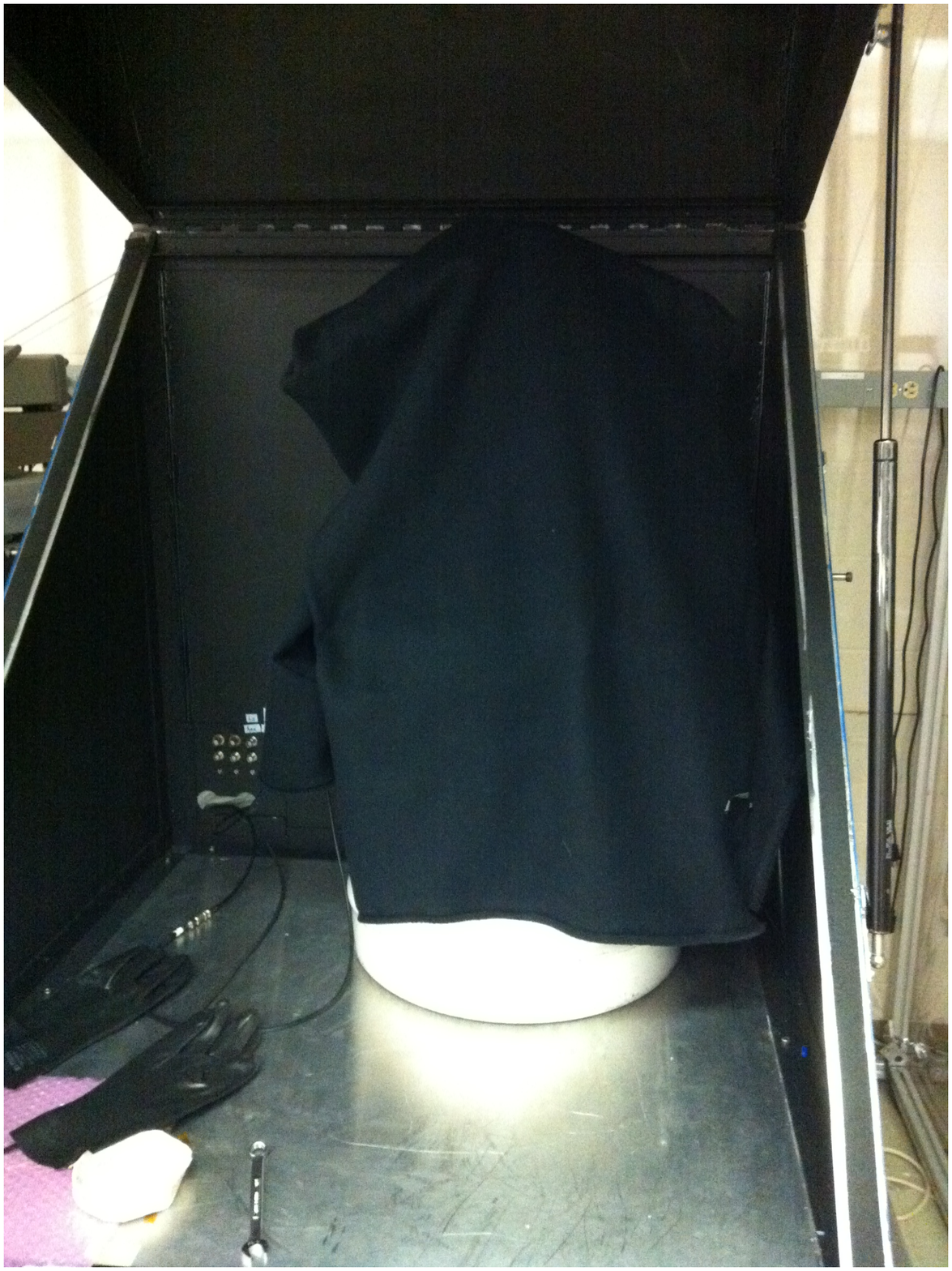}}
\hfill
\caption{Experimental setup for LED dunk test in liquid nitrogen. (a) Detail setup. (b) All the parts is covered with black blanket inside the dark box to reduce the photon leakage.}
\label{fig:dark_box}
\end{figure}

\begin{figure}[htbp]
\centering
\graphicspath{{./fig/LED_system/}}
\includegraphics[scale=0.4]{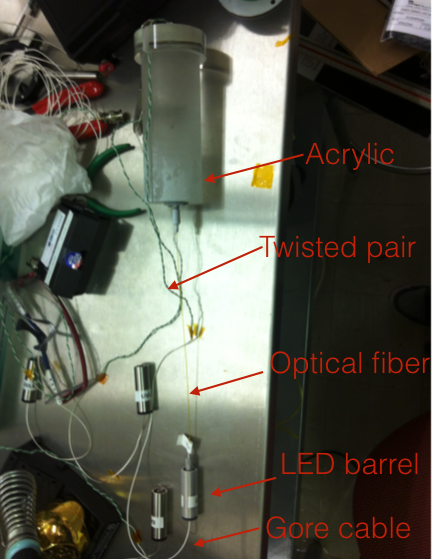}
\caption{ LED couples to acrylic.}
\label{fig:ledacrylic}
\end{figure}

In order to understand the stability of the LEDs under cryogenic temperature, the LN$_2$ dunk test was performed.  The experiment setup is shown in Fig. \ref{fig:dark_box}.  Each LED was assembled and coupled to optical fiber as described in previous sections.  The tip of the fiber coupled to the UV LED was coated with TPB to shift the UV light to visible.  In order to simulate the environment in the IV, a 20 cm cylindrical acrylic was used to transmit the light to the PMT.  Figure \ref{fig:ledacrylic} shows the LED coupled to the acrylic. An adaptor was installed on the bottom of the acrylic to couple to the optical fiber.  On the acrylic cylinder surface, a 1 cm groove was cut to accommodate the twisted pair of LEDs due to the limit space of the dewar bottle neck.  With this design, the acrylic cylinder can fully insert into the dewar through the opening.  A 3" PMT then couples to the top of acrylic cylinder in order to observe LED photons.  The whole system is positioned inside the dark box, and a black blanket was used to cover the dewar to further improve the system.\par

A Tektronix TDS 3054C oscilloscope was used to record both the PMT signal and the pulse driving the LED.  Data was then extracted from the oscilloscope -- and converted to text format -- using NI Express, after which ROOT analysis was performed.  Every LED was submerged in LN$_2$ for an intended 24 hours (light yield is collected by the PMT), but due to some accidental connection problems the immersion times of collected data is somewhat variable.  Nonetheless, each LED was continuously pulsed for at least 7 hours, giving ample opportunity to determine LED stability.  The raw waveform of each event during the test was read and the first 800 samples determined the baseline.  Subsequently, the baseline was subtracted from the raw waveform and the integrated charge was calculated.  The average integrated charge was obtained hourly and then plotted as a function of time.  The relative stability was defined as the standard deviation of integrated charge in a hour, divided by the average integrated charge.  This was used to track the hourly LED variation. \par
\begin{figure}[htbp]
\hfill
\subfloat[]{\includegraphics[width=7cm]{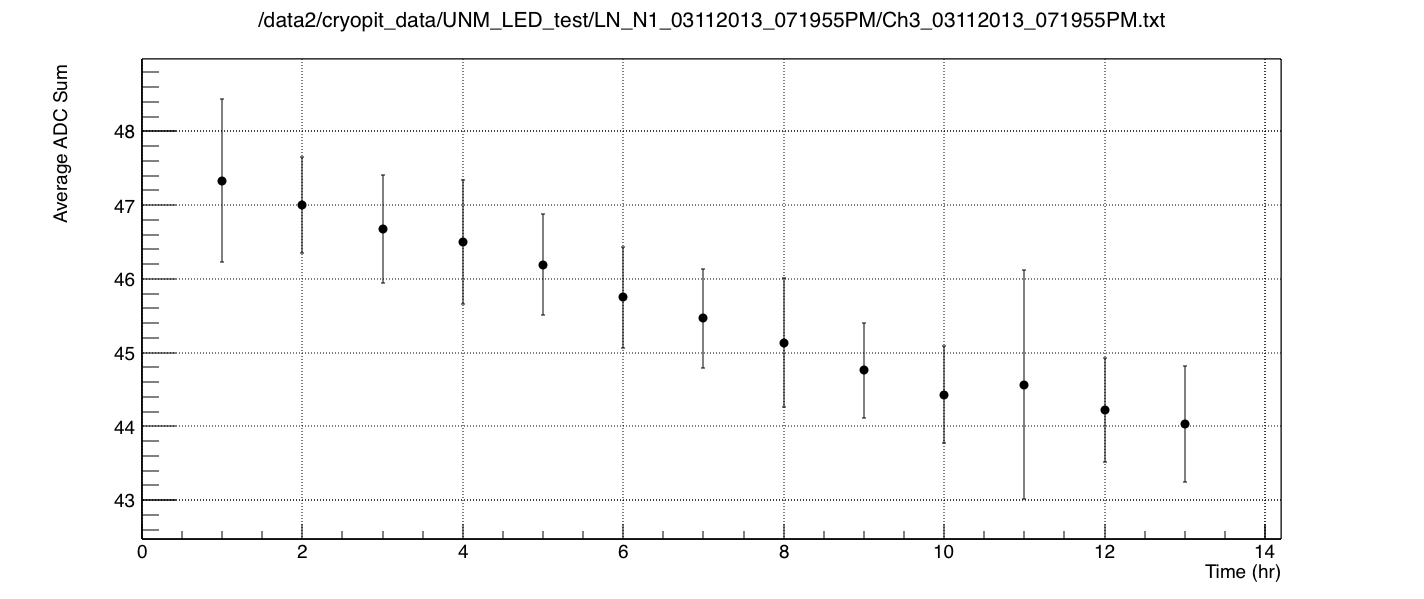}}
\hfill
\subfloat[]{\includegraphics[width=7cm]{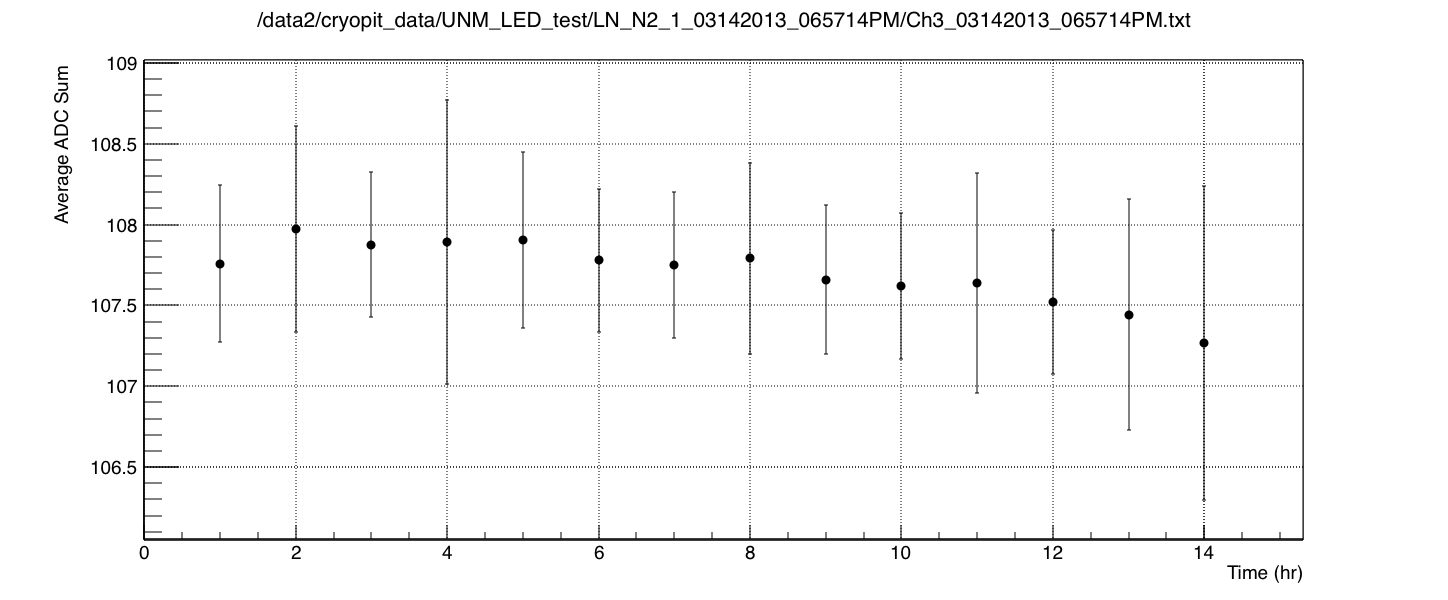}}
\hfill
\subfloat[]{\includegraphics[width=7cm]{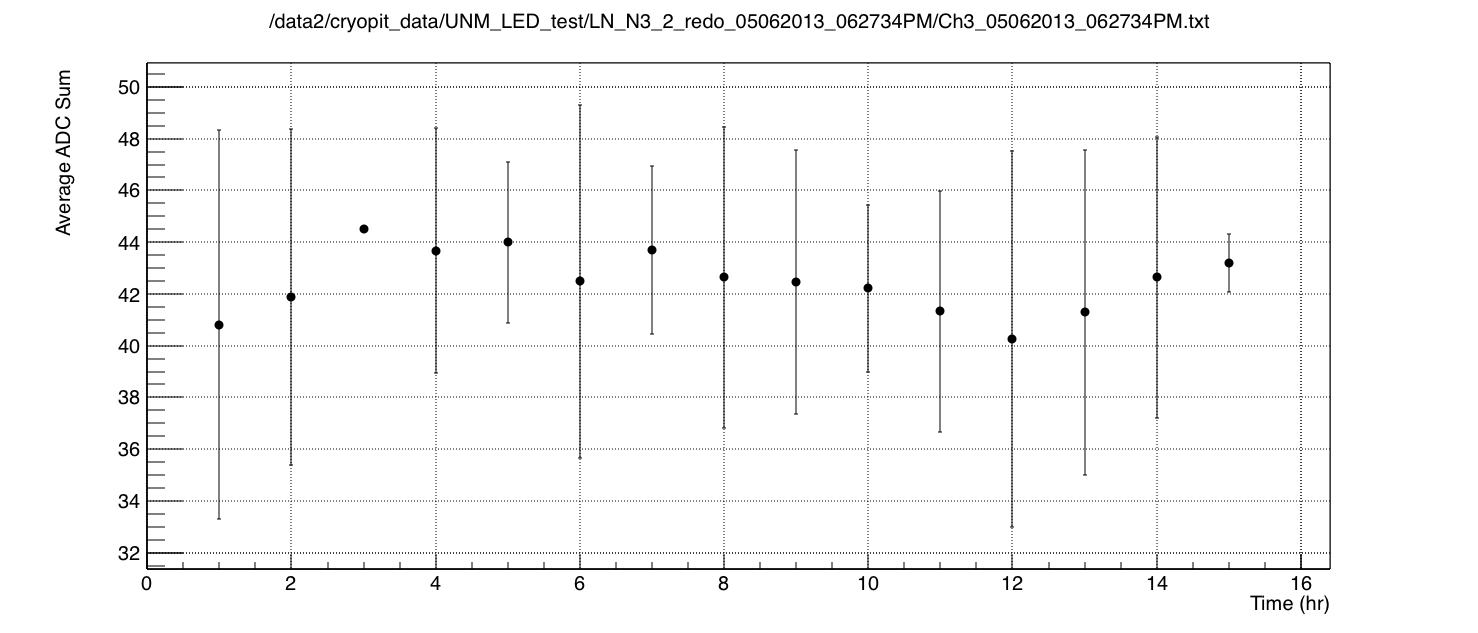}}
\hfill
\subfloat[]{\includegraphics[width=7cm]{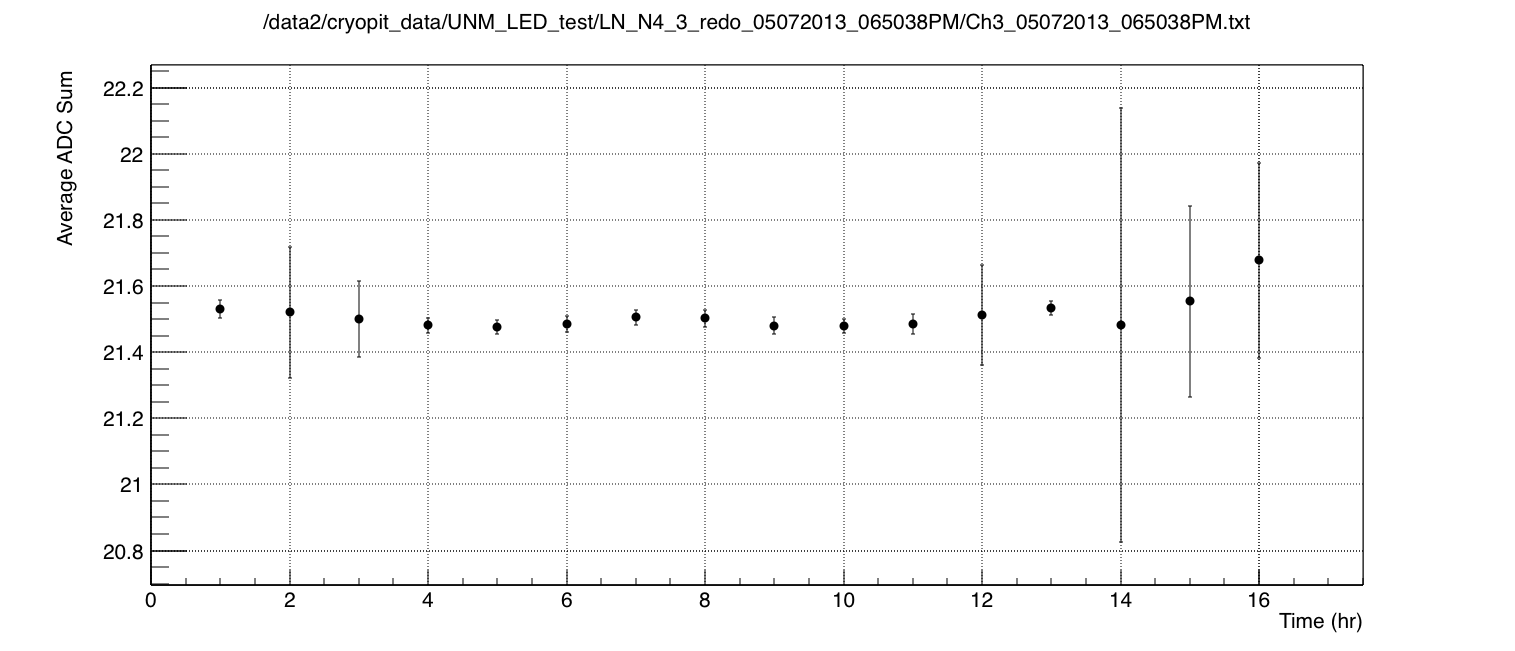}}
\hfill
\subfloat[]{\includegraphics[width=7cm]{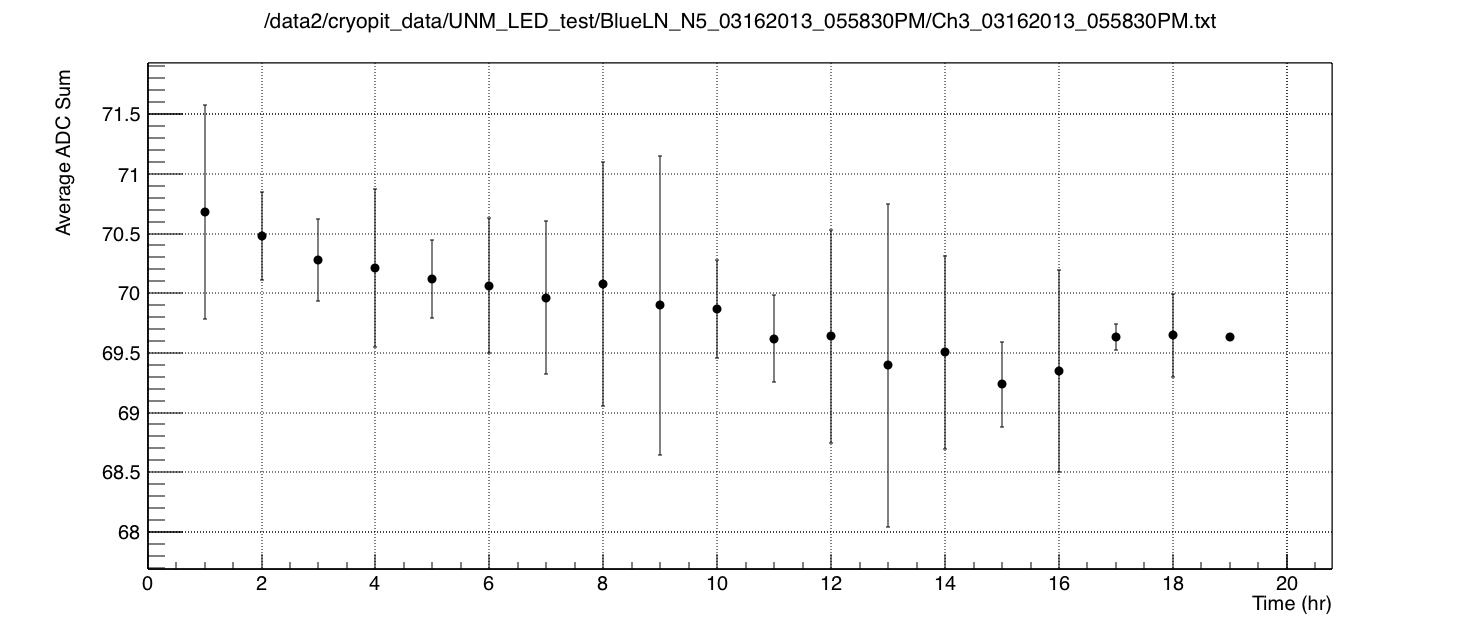}}
\hfill
\subfloat[]{\includegraphics[width=7cm]{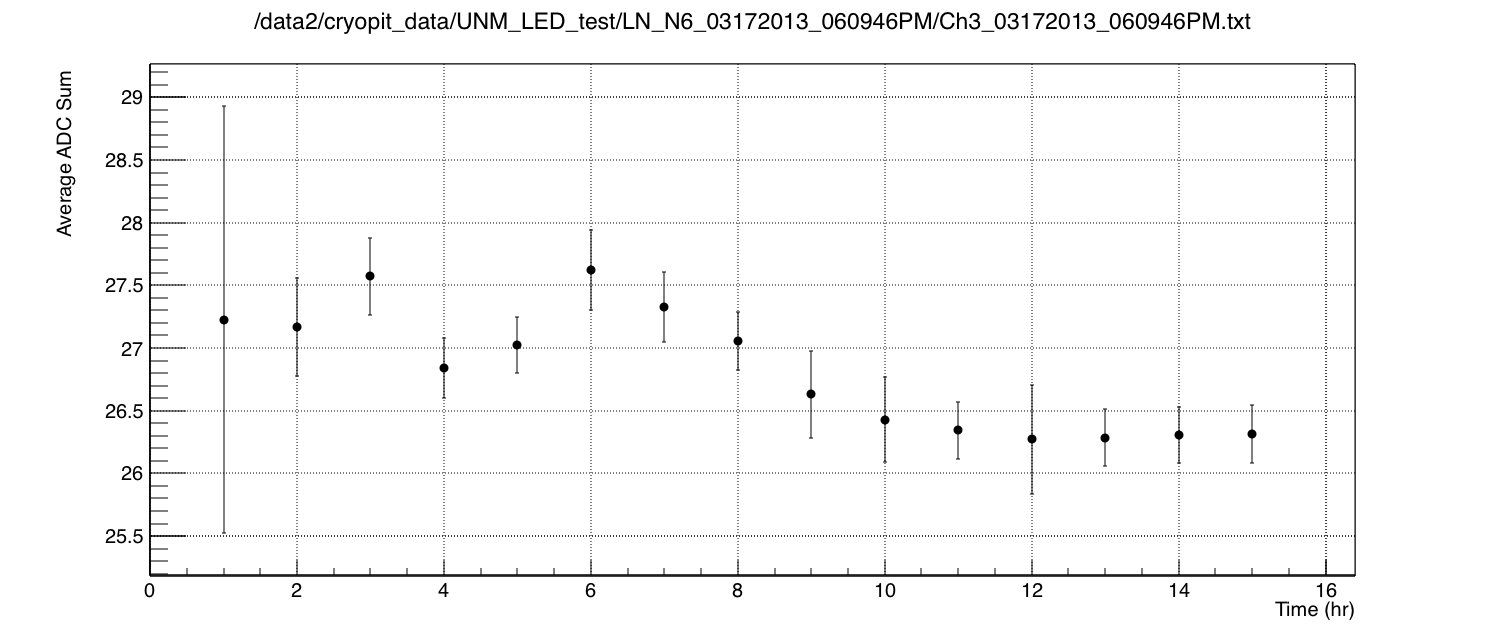}}
\hfill
\caption{Average integrated charge of the waveform vs dunk time for blue LEDs.}
\label{fig:bluestability}
\end{figure}
\begin{figure}[htbp]
\hfill
\subfloat[]{\includegraphics[width=7cm]{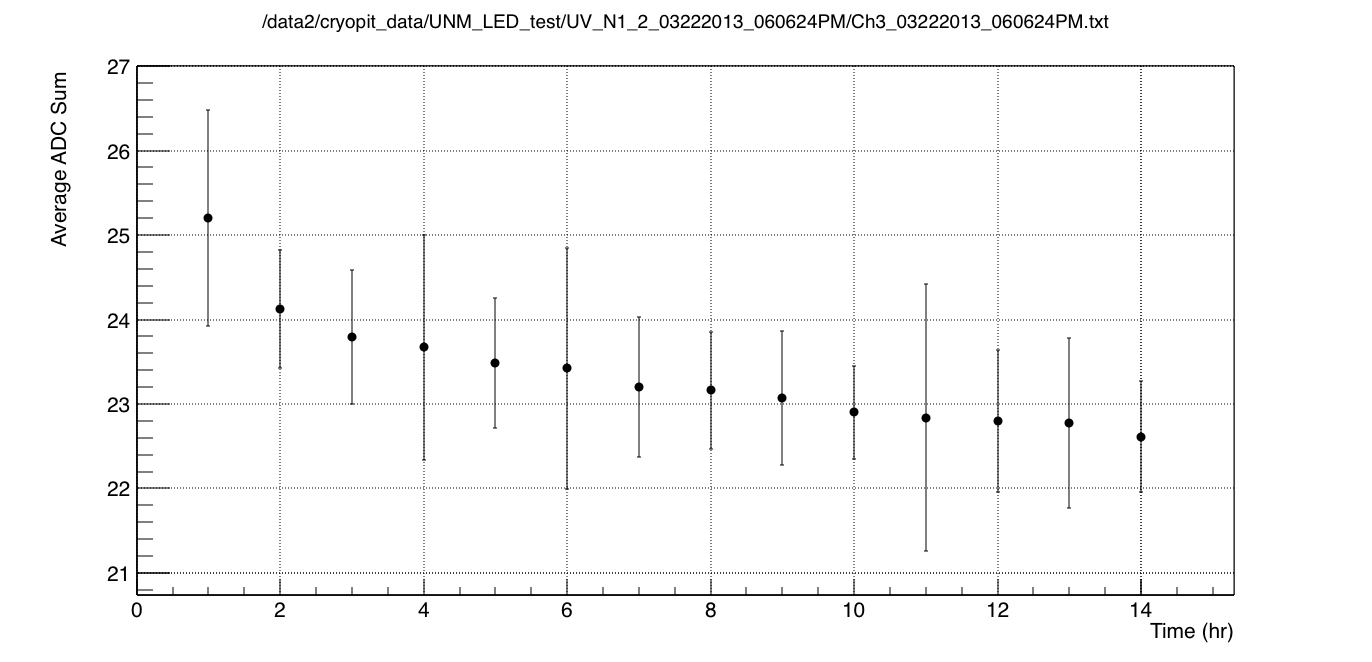}}
\hfill
\subfloat[]{\includegraphics[width=7cm]{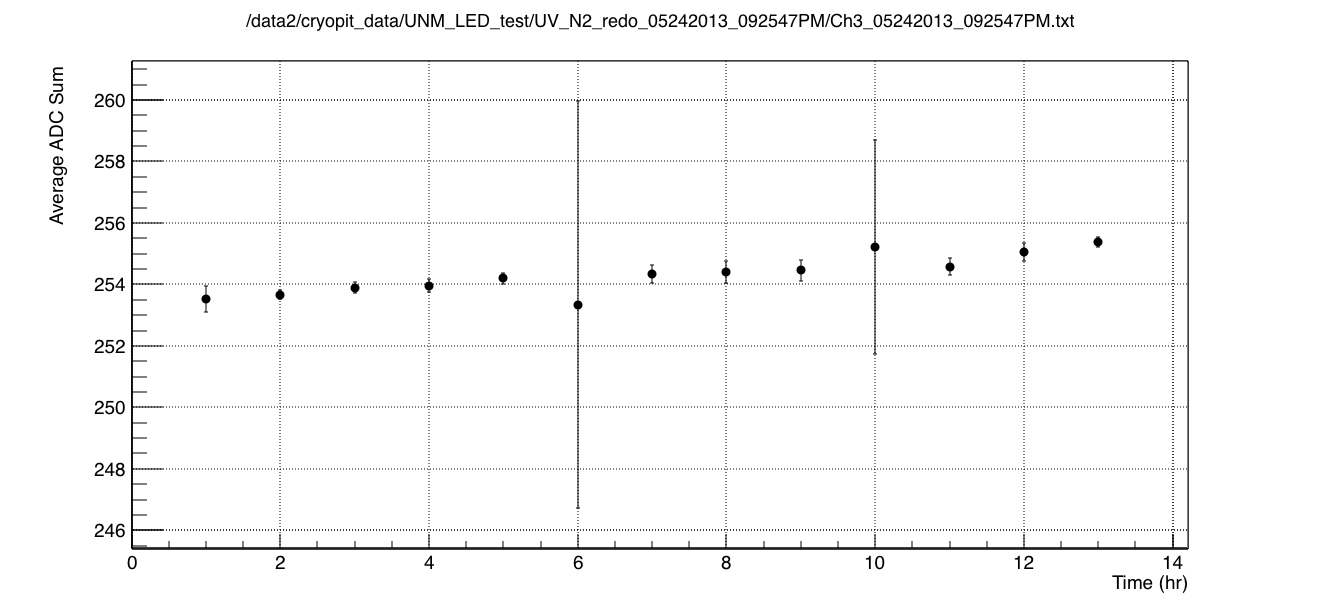}}
\hfill
\subfloat[]{\includegraphics[width=7cm]{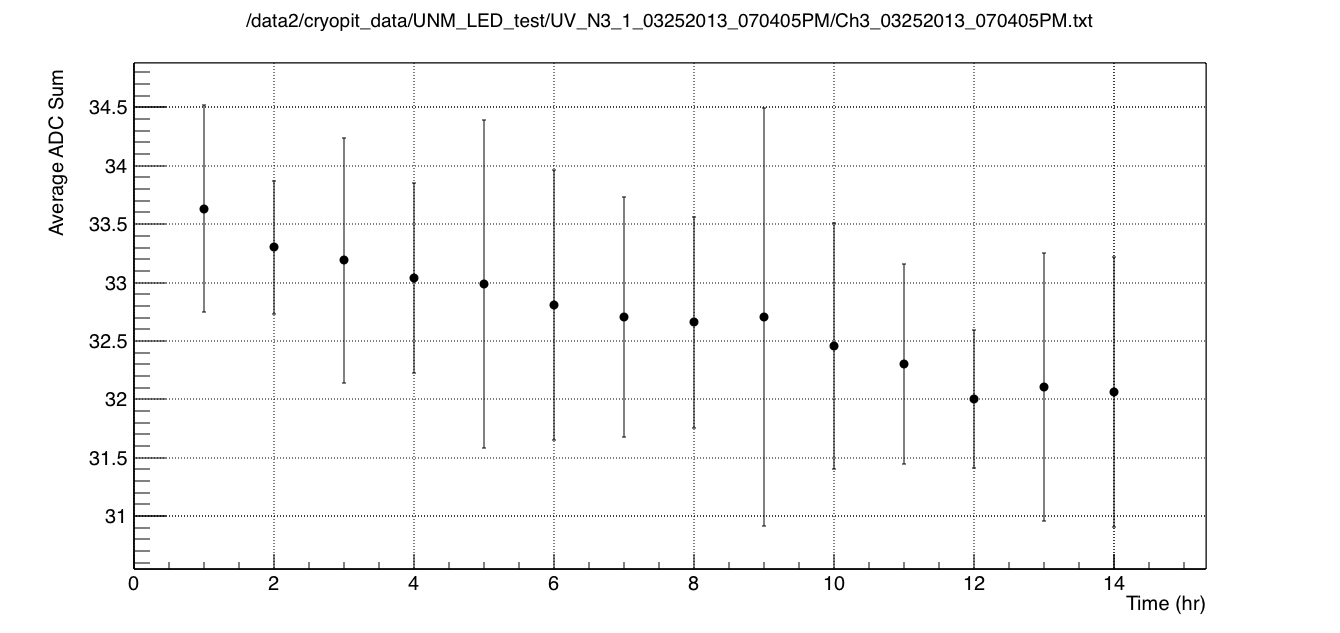}}
\hfill
\subfloat[]{\includegraphics[width=7cm]{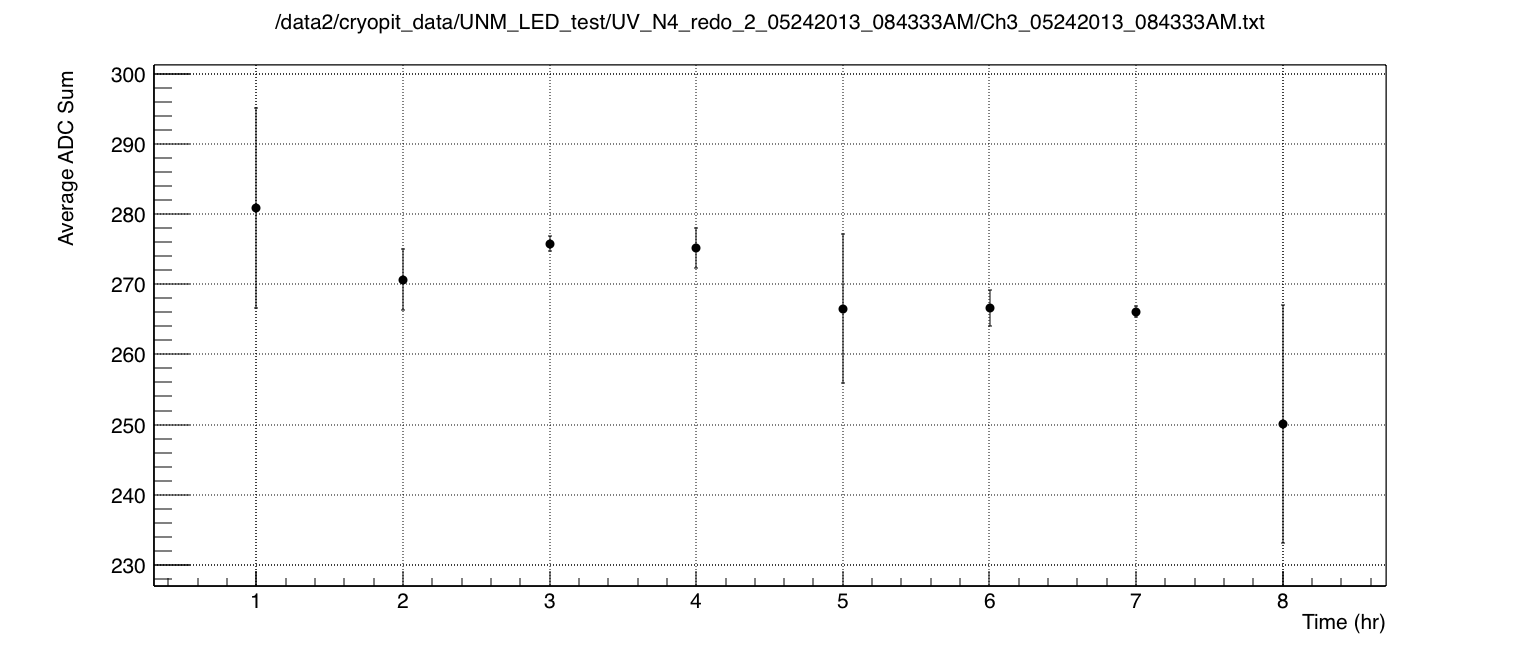}}
\hfill
\subfloat[]{\includegraphics[width=7cm]{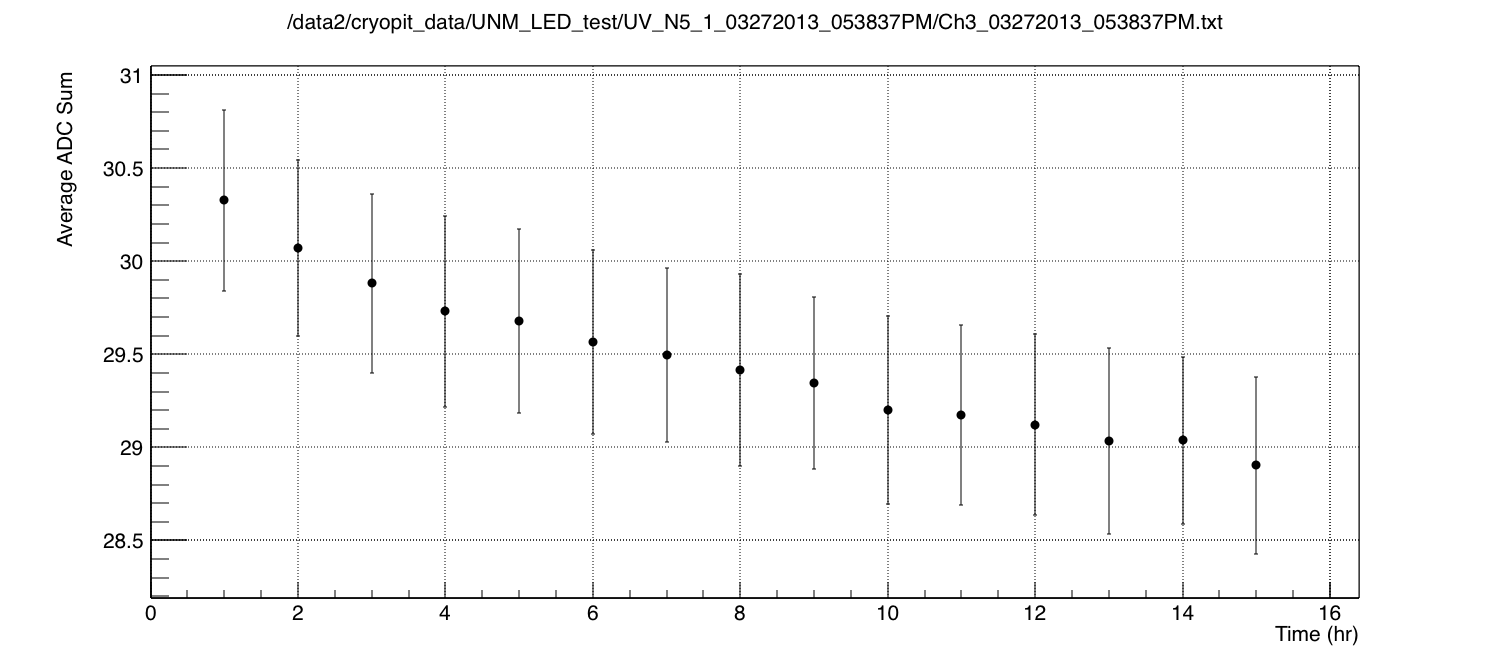}}
\hfill
\subfloat[]{\includegraphics[width=7cm]{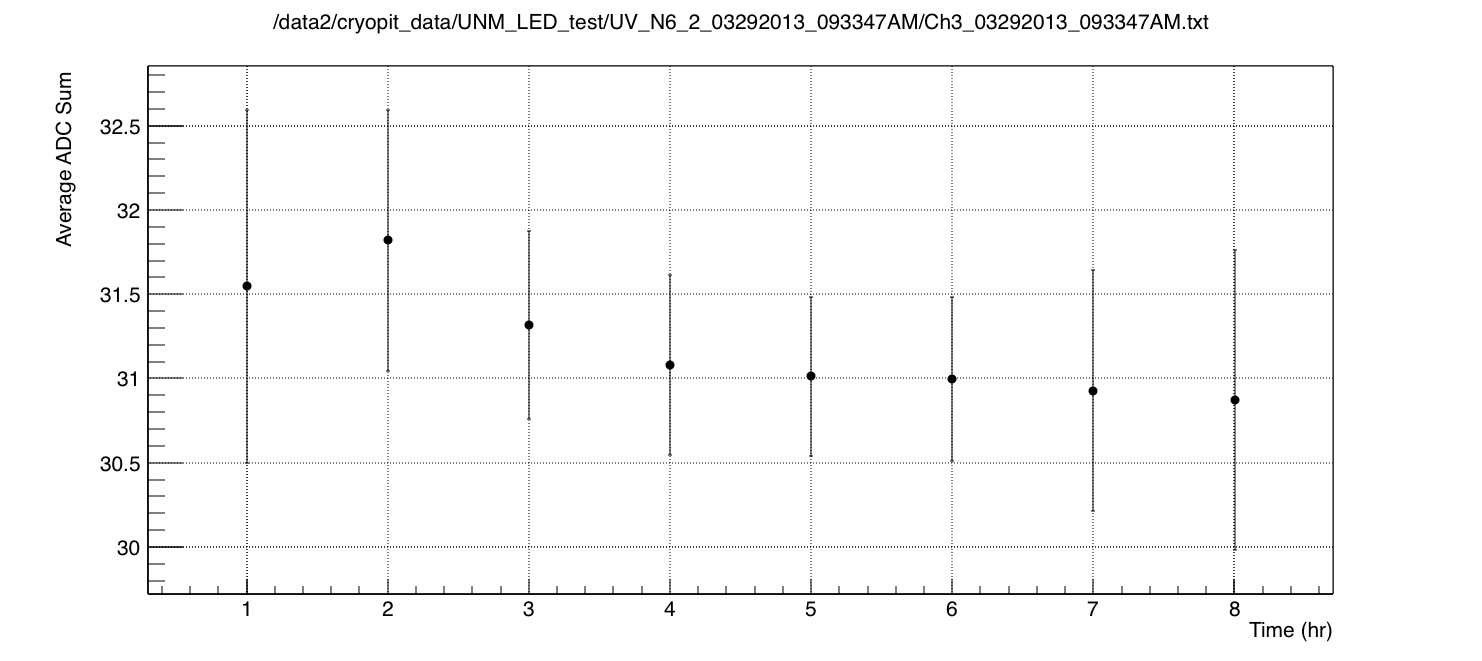}}
\hfill
\caption{Average integrated charge of the waveform vs dunk time for UV LEDs.}
\label{fig:uvstability}
\end{figure}
\begin{figure}[htbp]
\hfill
\subfloat[]{\includegraphics[width=7cm]{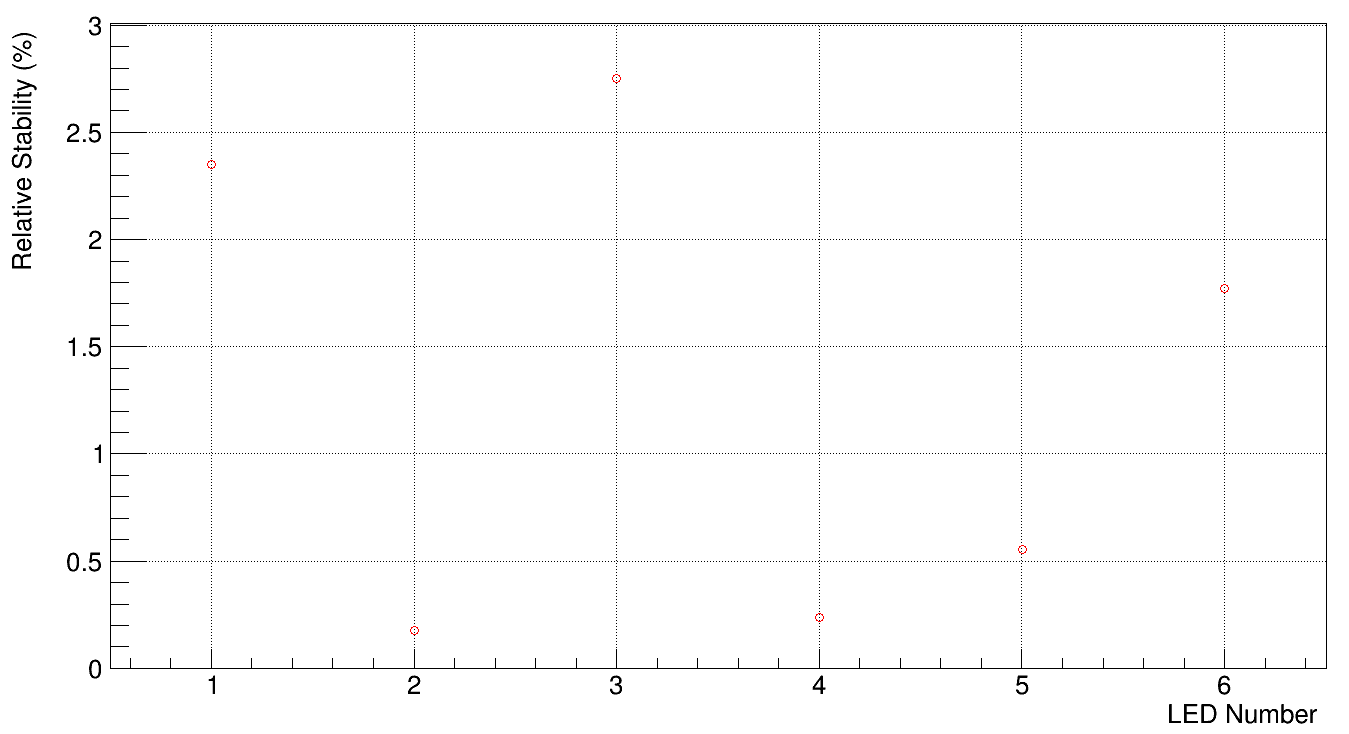}}
\hfill
\subfloat[]{\includegraphics[width=7cm]{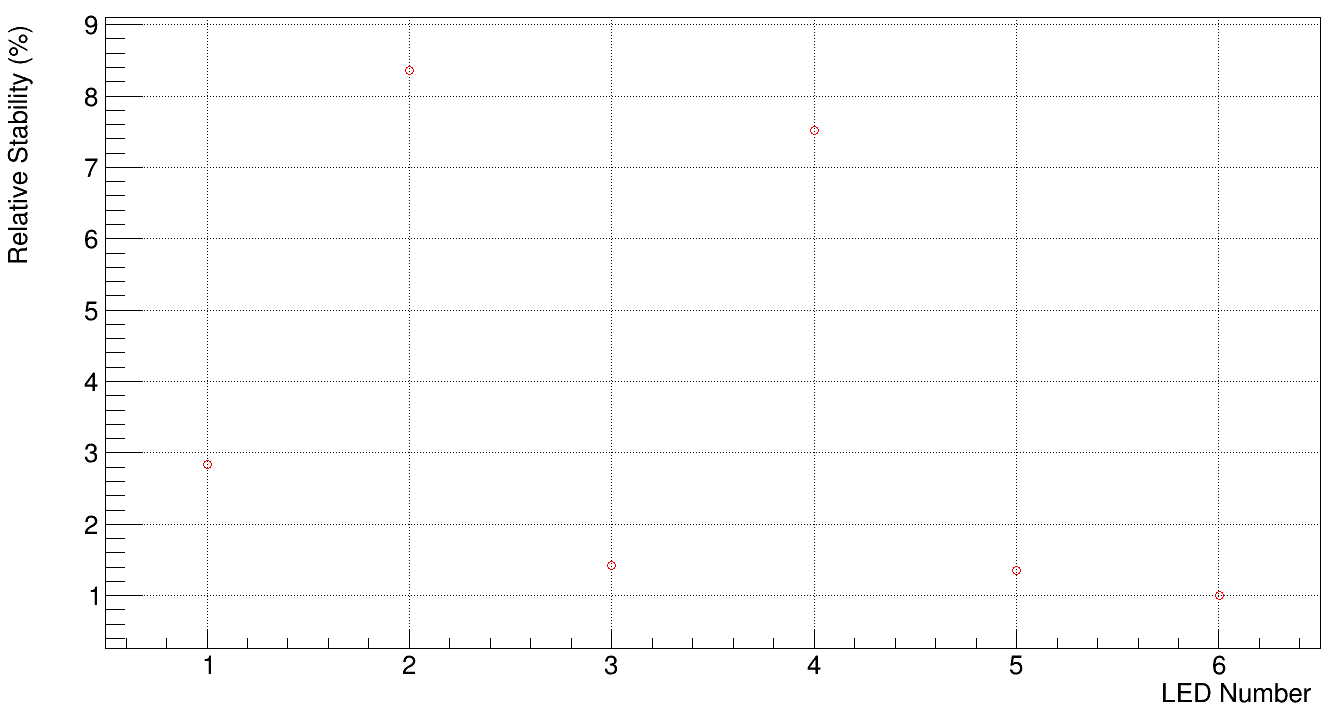}}
\hfill
\caption{Relative stability vs LED number for (a) Blue. (b) UV}
\label{fig:blueuvtotal}
\end{figure}

Figure \ref{fig:bluestability} and \ref{fig:uvstability} shows the average integrated charge as a function of time for the blue- and UV-LED, respectively.  Because every LED had a different response to the same voltage, the average integrated charge of each was in turn rather different.  The current is proportional to the light yield but exponentially proportional to the voltage, therefore a slight voltage change can cause a significant current change.  However, the relative stability over the course of a test is of most interest, not the light yield.  Figure \ref{fig:blueuvtotal} shows the relative stability of blue- and UV-LED, respectively.  The measured intensity drift is under 3\% for blue the LED and 10\% for UV LED.  Drift causes include the LED power supply, PMT high voltage supply, PMT gain variation, and so on.  Moreover, the LN$_2$ level in the dewar changed throughout the test and the resulting temperature changes could affect the optical properties inside the dewar, which in turn affected the collection efficiency of LED photons.  As for the UV LED test, any larger variation might be due to the additional TPB coating, which may flake off during the test.  The detailed estimation of uncertainties can not be obtained.  The tests proved that both LED types were functioning well under cryogenic temperature.  For the \textit{In-Situ} optical calibration, the LEDs were typically turned on for just tens of minutes, and a huge response variation was not expected in the daily calibration process.

\section{Preliminary PMT Gain calibration}
The \textbf{S}ingle \textbf{P}hoto \textbf{E}lectron (SPE) calibration runs with low intensity LED pulses and 1.7-2 kV PMT voltages, were performed at room temperature.  The data acquisition system was the same as the dunk test.  The digital oscilloscope read the PMT voltages at 0.2 ns intervals, each digitization giving a charge of $q_i$.  The baseline and the noise $\sigma_n$ were measured for each run, and a Gaussian fit to the central noise peak in the $q_i$ distribution was performed.  A simple pulse finding algorithm was used to scan each scope trigger for three sequential (baseline subtracted) voltages greater than 3$\sigma_n$, summing the pulse charge $q = \Sigma q_i$ until $q_i$ falls below $\sigma_n$.  The "noise" -- taken as an average over a random 200-ns long window normalized to the pulse length -- was subtracted on an event-by-event basis, shifting the charge gain by 11\% at 1.7 kV.\par
\begin{figure}[htbp]
\centering
\graphicspath{{./fig/LED_system/}}
\includegraphics[scale=0.4]{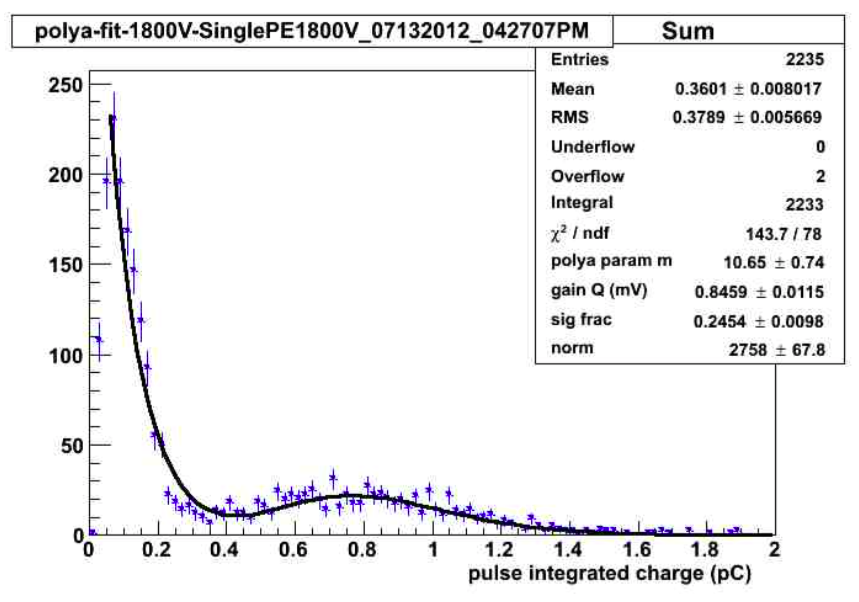}
\caption{ Example of fit for PMT at 1800 V.}
\label{fig:polyaexample}
\end{figure}
\begin{figure}[htbp]
\centering
\graphicspath{{./fig/LED_system/}}
\includegraphics[scale=0.4]{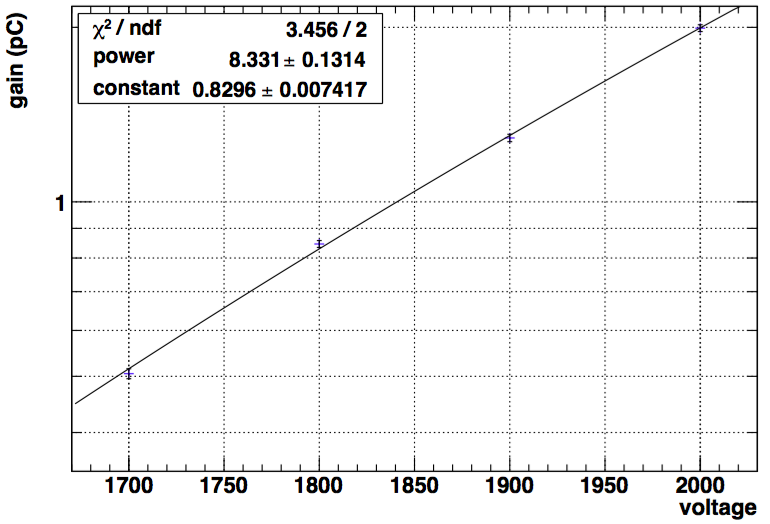}
\caption{ Fit to the power law g(v) = constant $\cdot$ (voltage/1800)$^{power}$ .}
\label{fig:polyaresults}
\end{figure}

The SPE charge distribution is fitted to a exponential plus Polya distribution with q in pC. The Polya "signal" piece is :
\begin{ceqn}\begin{align}
s(q) = \frac{m}{q}\cdot y^{(m-1)}\cdot \frac{e^{-y}}{\Gamma(m)}
\end{align}\end{ceqn}
with $y\equiv q\cdot m/Q$, where m is the Polya parameter which is an energy resolution parameter, a measure of the number of electrons from the first dynode given a single electron from the photo-cathode, and Q is the charge gain times the absolute value of the electron charge in pC. The exponential part of fitting function is : 
\begin{ceqn}\begin{align}
b(q) = N_{exp}\cdot \frac{ e^{(-q/Q_0)}}{Q_0}
\end{align}\end{ceqn}
where $Q_0$ is the exponential fit parameter and $N_{exp}$ normalizes $b(q)$ to one over the histogram range. Finally the total fitting function is :
\begin{ceqn}\begin{align}
PolyaE(q) = bw\cdot N\cdot (f\cdot s(q) + (1-f)\cdot b(q))
\end{align}\end{ceqn}
where $bw$ is the histogram bin width, $f$ is the signal fraction, and $N$ is the total number of events. Figure \ref{fig:polyaexample} shows the example of the fit and Fig. \ref{fig:polyaresults} shows the fit results as a function of PMT operating voltage. The latter plot fit well to the power law (see figure) indicating the gain varies with HV voltage exponentially. 

\section{Installation of LED light injection system}
\begin{figure}[htbp]
\hfill
\subfloat[]{\includegraphics[width=7cm]{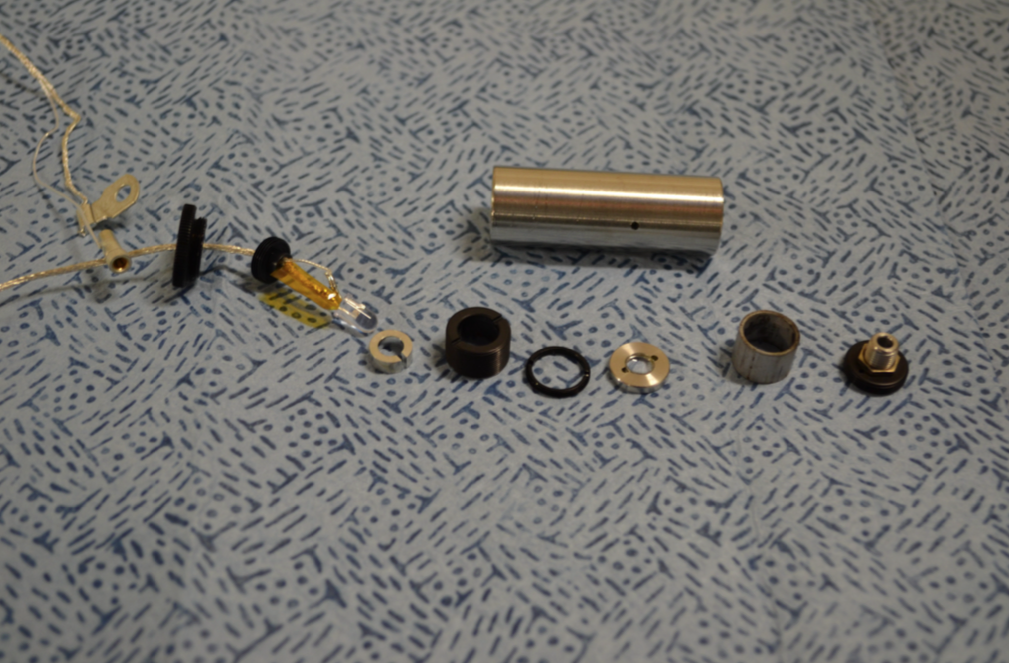}}
\hfill
\subfloat[]{\includegraphics[width=7cm]{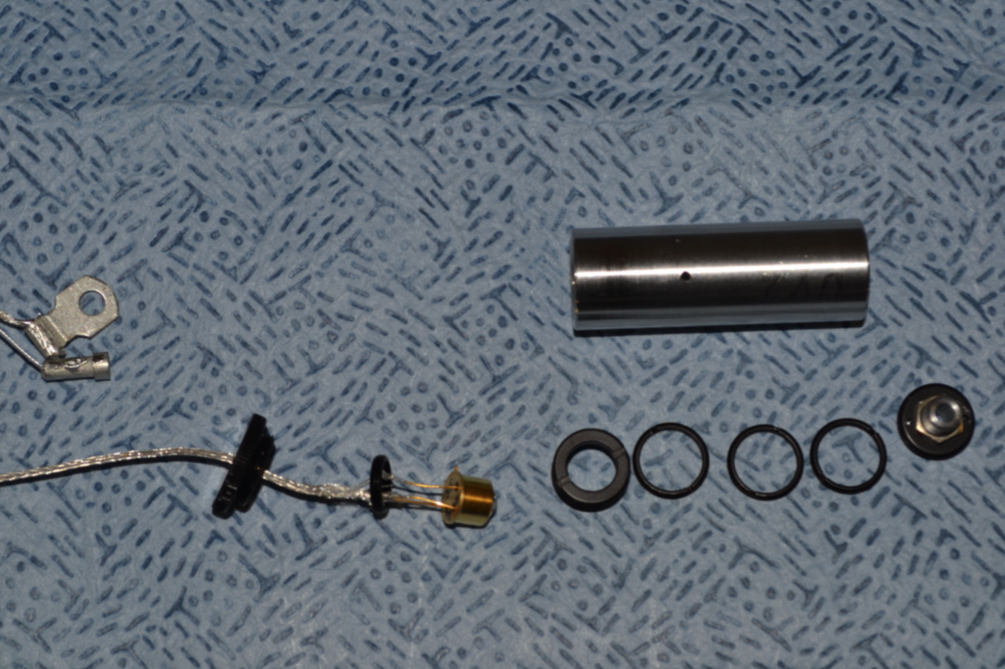}}
\hfill
\caption{(a) Blue LED assembly : (from left to right) Ground and lead connections, braided cable, end cap, LED mount back, BLUE LED, aluminum holder, LED mount, 4 lock rings (not all pictured), lens, spacer, connector. All components are installed in this order inside the LED barrel also pictured above. (b) UV LED assembly : (from left to right) Ground and lead connections, braided cable, end cap, LED lock ring, UV LED, LED mount, 3 lock rings, connector. All components are installed in this order inside the LED barrel also pictured above}
\label{fig:blueuvassemble}
\end{figure}

\begin{figure}[htbp]
\hfill
\subfloat[]{\includegraphics[width=7cm]{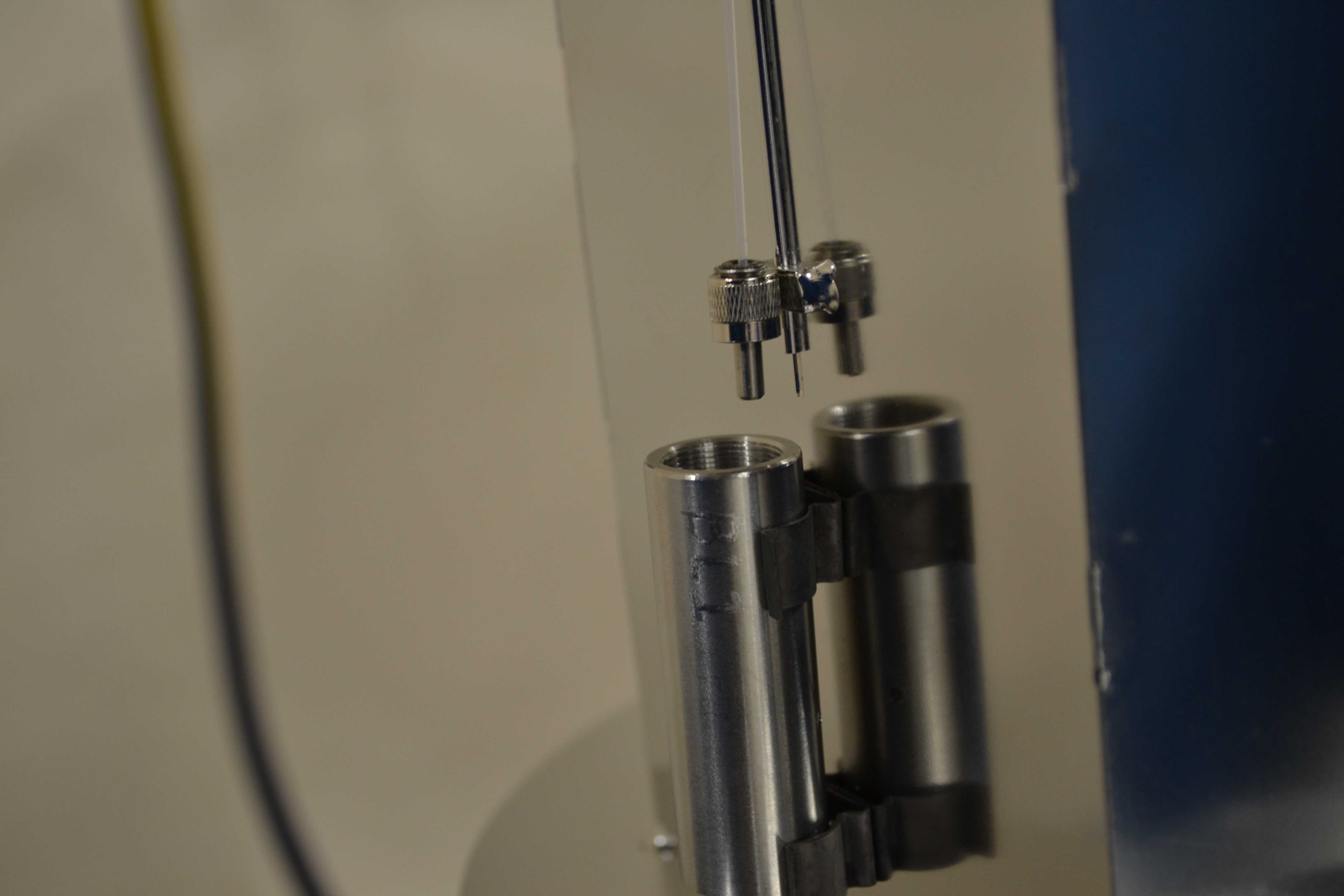}}
\hfill
\subfloat[]{\includegraphics[width=7cm]{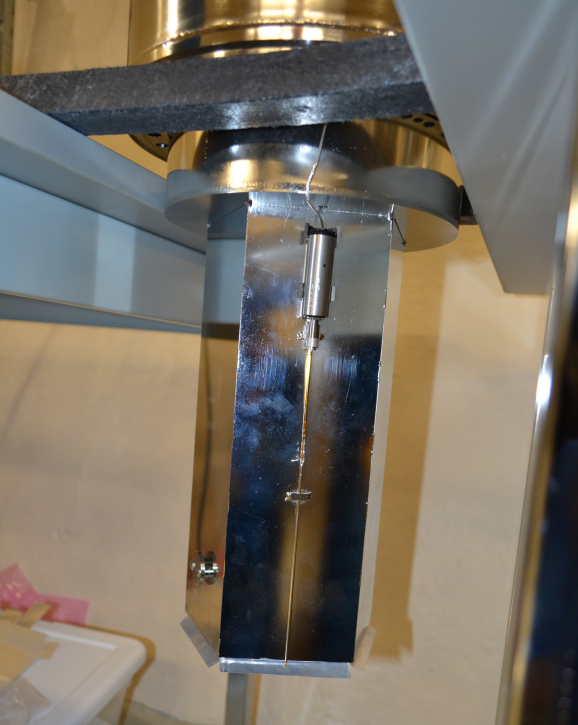}}
\hfill
\caption{(a) LED fixture before the modification, the position of the clip which fix the LED on the side of cassette is too far back than the design scheme. (b) After the adjustment, the LED can be accommodate in the clip and with proper length of the fiber. }
\label{fig:LED_fixure}
\end{figure}

\begin{figure}[htbp]
\centering
\graphicspath{{./fig/LED_system/}}
\includegraphics[scale=0.5]{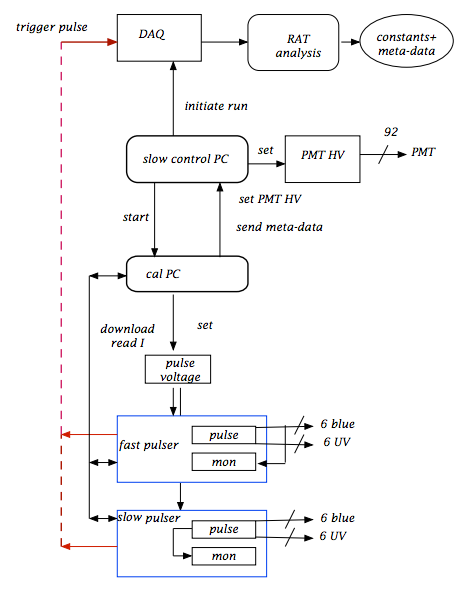}
\caption{ \textit{In-Situ} LED injection operational scheme.}
\label{fig:LED_scheme}
\end{figure}

In the SNOLAB underground laboratory, all parts -- LEDs, fibers, etc. -- were ultrasonically cleaned, thoroughly.  Subsequently, other components were hand cleaned with methanol, whereas the LEDs and fibers were cleaned with ultra pure water to remove any residual containment.  The LEDs were then assembled and mounted onto the optical cassettes (Fig. \ref{fig:blueuvassemble}).  The LED pulser box output channel connected to 50 foot 50 $\Omega$ coaxial cable (RG 58), and was routed through the top of the deck to the top hat.  Inside the OV, gore cable was used to connect from the top hat to the Optical cassettes. It was found the clip which used to fix the LED on the side of cassette is too small and the length of fiber need to be adjusted as shown in Fig. \ref{fig:LED_fixure}. After the adjustment, the LED can be fixed and the new fibers were made to extend far back to couple to the LED. The operational scheme is shown in Fig. \ref{fig:LED_scheme}, in which the operation of slow pulser is obsolete. \par



\chapter{LED Data Analysis }
The \textbf{S}ingle \textbf{P}hoto \textbf{E}lectron (SPE) calibration is important to the overall light detection experiment, the energy resolution of which depends on an understanding of the PMT response to a single photon.  The \textit{In-Situ} optical calibration provided a way to track the PMT gain hour by hour.  Therefore the LED light injection system can improve the energy resolution.  In this chapter a detailed description of data analysis from the \textit{In-Situ} optical calibration and the detector properties investigated by LED data, are presented.
\section{Preliminary test}
\begin{figure}[htbp]
\centering
\graphicspath{{./fig/LED_data/}}
\includegraphics[scale=0.25]{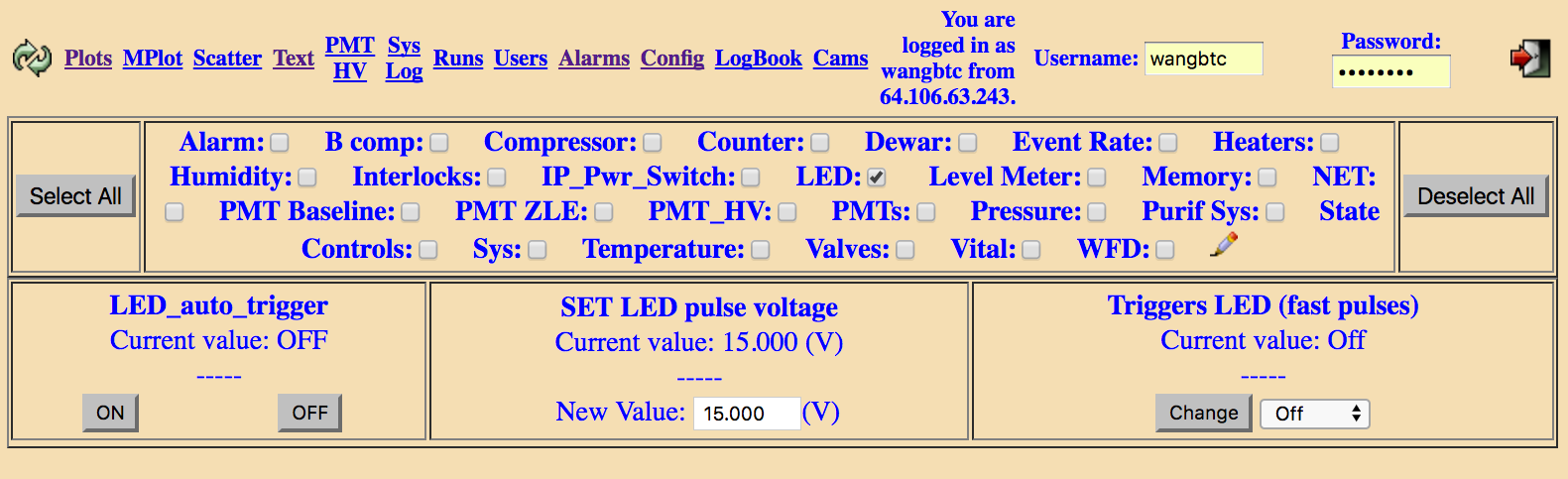}
\caption{ Slow control interface for LED light injection system.}
\label{fig:ledslowcontrol}
\end{figure}

The LED light injection system was tested to confirm its functionality.  Before the construction completion of the IV, the integration of the LED hardware was completed and the MiniCLEAN DAQ system was tested.  During regular operation, the slow control system was responsible for remotely monitoring and controlling the hardware.  The interface of the slow control is shown in Fig. \ref{fig:ledslowcontrol}.  The slow control system could manipulate the LED system in two ways: trigger any one LED with a desired voltage, or cycling the light injection.\par
\begin{figure}[htbp]
\centering
\graphicspath{{./fig/LED_data/}}
\includegraphics[scale=0.25]{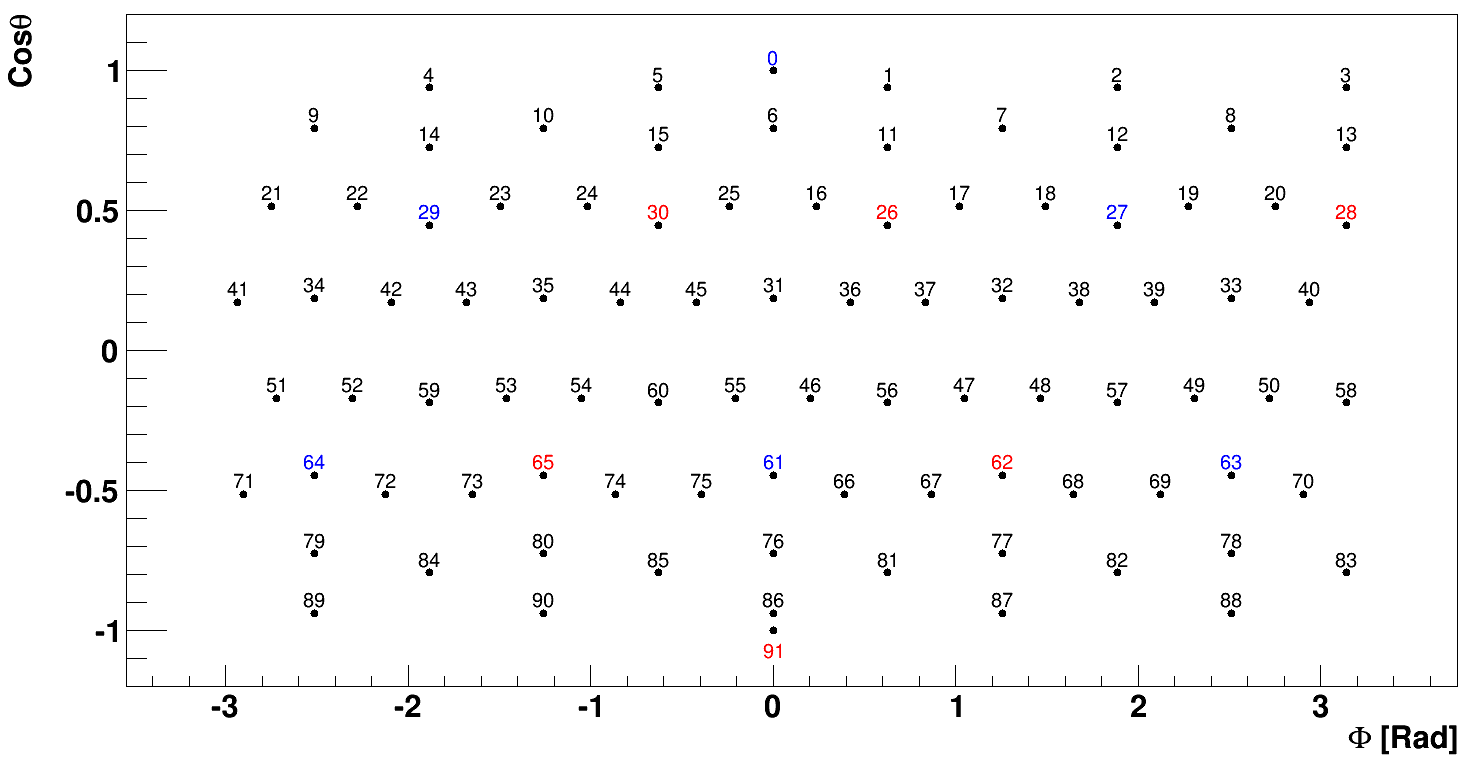}
\caption{ LED position. The red indicates  where the UV LEDs are mounted and blue are for blue LEDs.}
\label{fig:ledposition}
\end{figure}
\begin{figure}[htbp]
\hfill
\subfloat[]{\includegraphics[width=7cm]{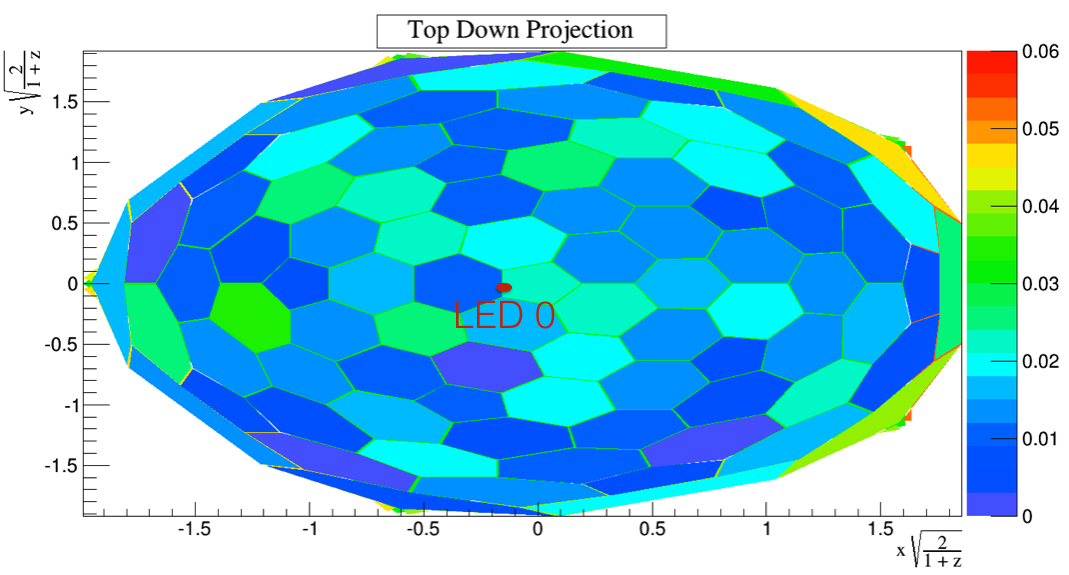}}
\hfill
\subfloat[]{\includegraphics[width=7cm]{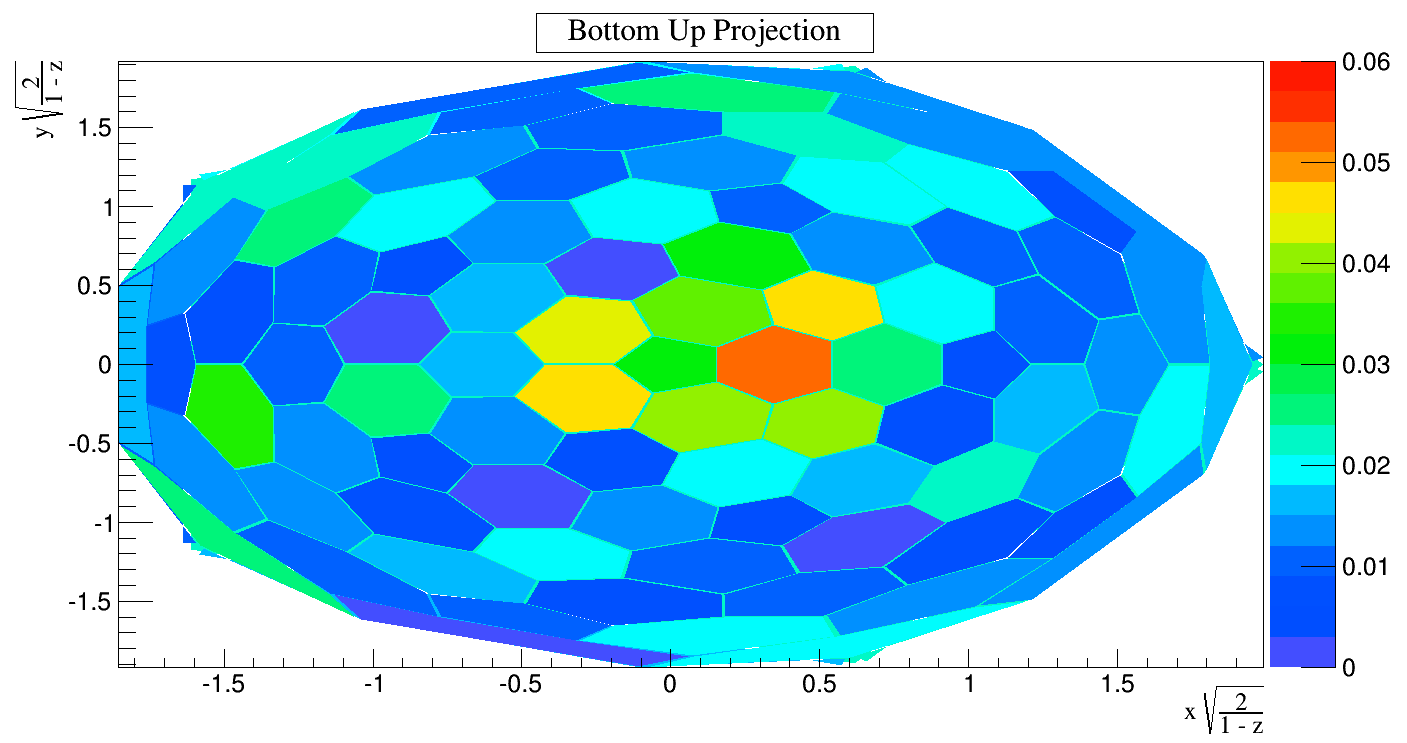}}
\hfill
\caption{Illumination plot from LED 0. The red dot in (a) indicate the approximate position of LED 0. (a) Top down view. (b) Bottom up view. }
\label{fig:illuminationled}
\end{figure}
\begin{figure}[htbp]
\hfill
\subfloat[]{\includegraphics[width=7cm]{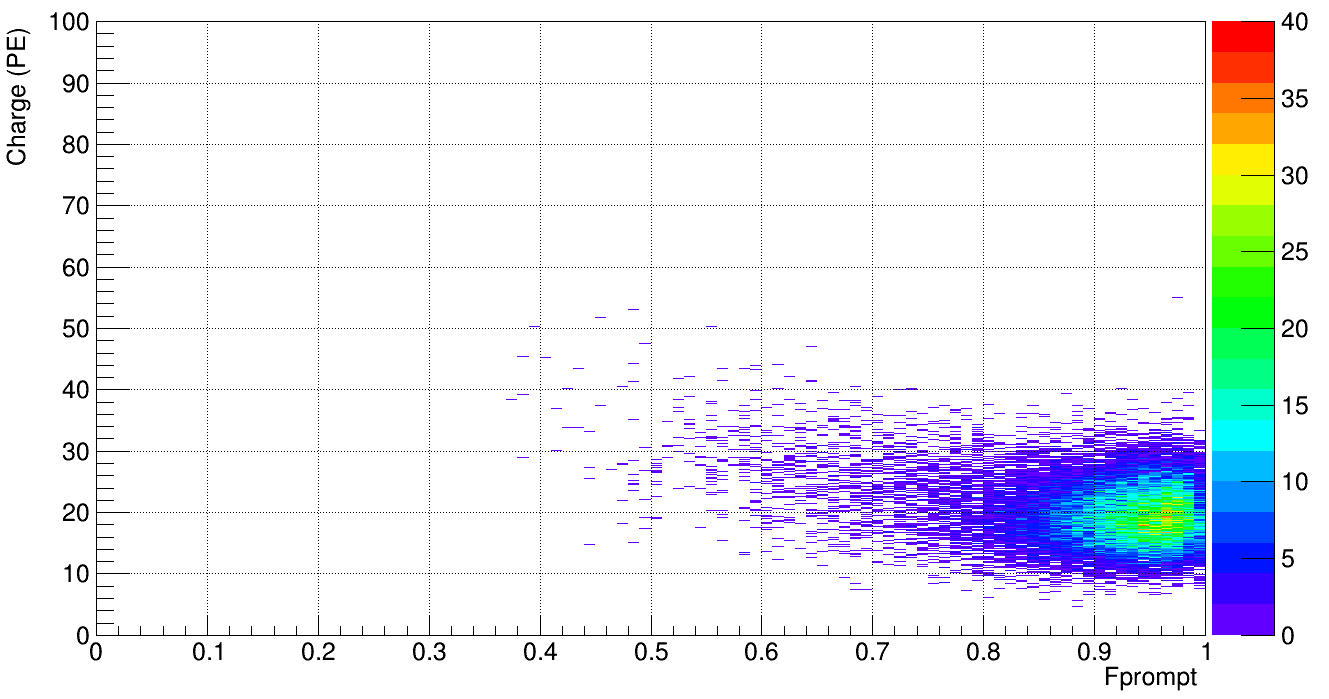}}
\hfill
\subfloat[]{\includegraphics[width=7cm]{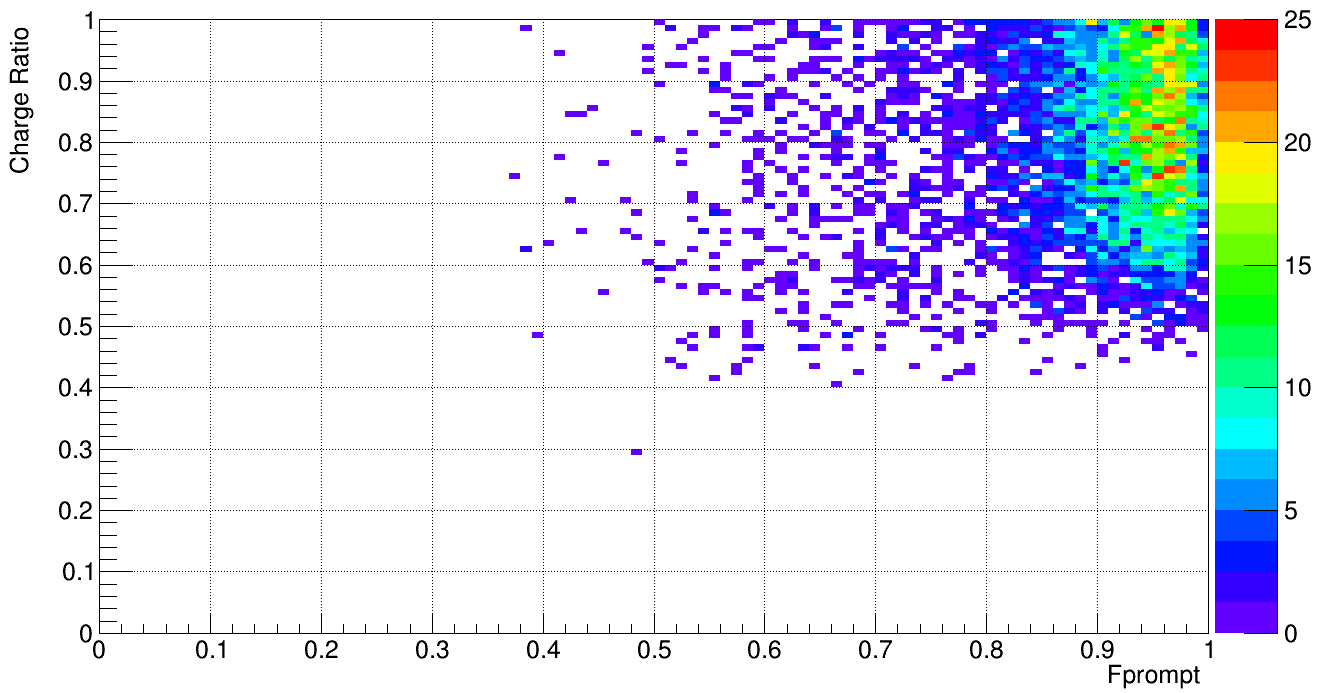}}
\hfill
\subfloat[]{\includegraphics[width=7cm]{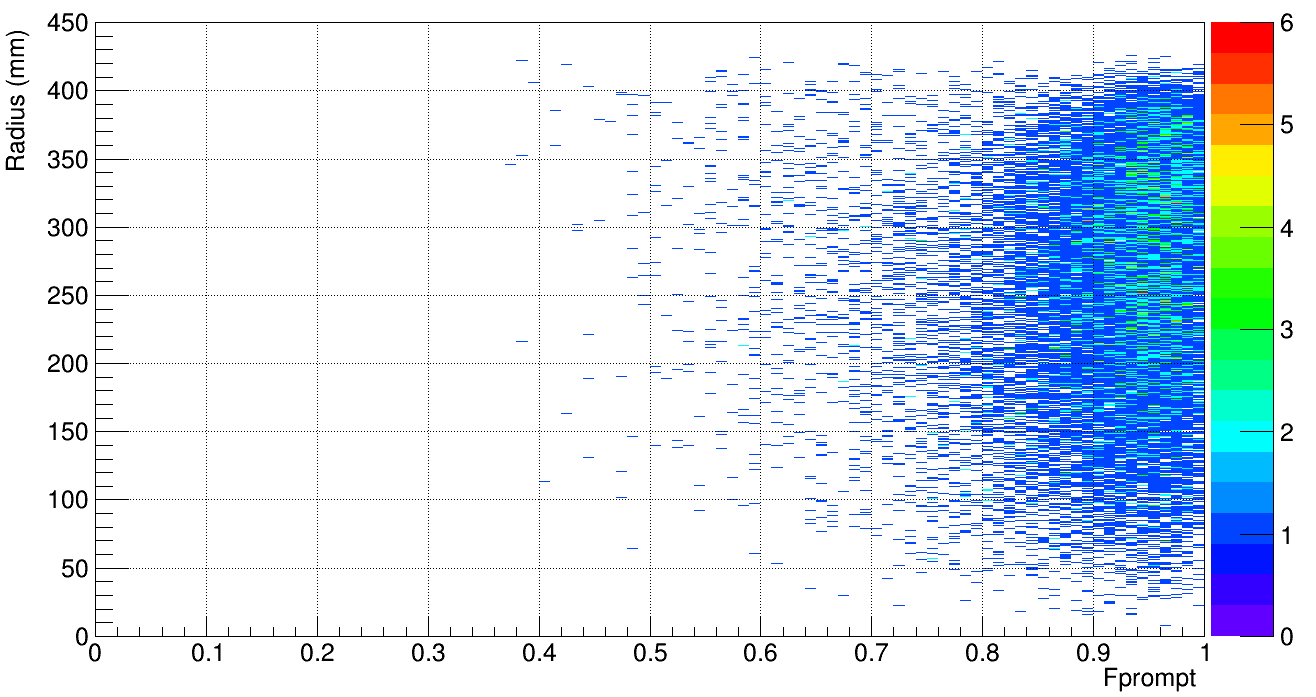}}
\hfill
\subfloat[]{\includegraphics[width=7cm]{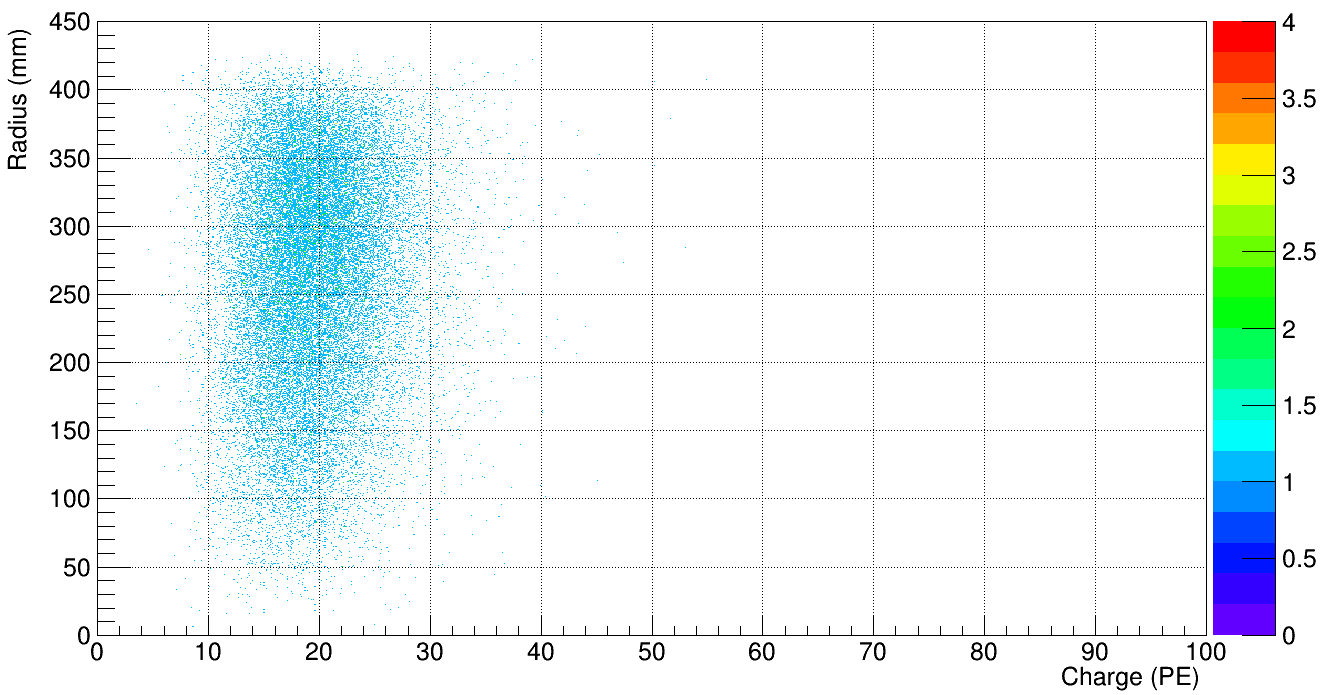}}
\hfill

\caption{(a) Fprompt vs Charge. (b) Fprompt vs $Q_r$. (c) Fprompt vs Radius. (d) Charge vs Radius. }
\label{fig:led2ds}
\end{figure}

For preliminary testing/debugging it would be convenient to pulse one LED at the time, but for the \textit{In-Situ} optical calibration, however, continuous pulsing through every LED is desired.  Indeed the ``\texttt{LED\_auto\_trigger}'' in Fig. \ref{fig:ledslowcontrol} provided this functionality for a daily calibration.  It fired both LED types, blue for 15 minutes then UV for 15 minutes alternatively.  Through hard-coding, each individual bias voltage pulsed each individual LED one by one, in order to obtain their optimal light emission values.  This gave good statistics towards tracking -- hourly -- the PMT or TPB stability.  The LED location map is shown in Fig. \ref{fig:ledposition}.
Figure \ref{fig:illuminationled} shows the illumination plot of the IV resulting from the pulsing of LED 0, where 0 indicates the LED was installed on the optical cassette housing for PMT 0.  The top down projection is viewing from the IV top (PMT 0) and the bottom up projection is viewing from the IV bottom (PMT 91).  These illumination plots show the fraction of photoelectrons (of the total) received by each PMT.
Additionally, some detector properties such as total charge (Q), charge ratio ($Q_R$), Fprompt (Fp) and the centroid reconstructed radius (R) are plotted as an aid towards helping check the integrity of both the detector and the data analysis.  Figure \ref{fig:led2ds} shows the 2-D plots of various parameters.  With low intensity LED emission the total charge from the LED events were usually lower than 50 PE, and the fluctuation depends on the optical properties in the detector such as ESR foil reflectivity and the UV-to-visible conversion efficiency of the TPB coating.  Light emitted from an LED should only be prompt light, making Fprompt equal to unity.  However, due to both after-pulsing (see section \ref{sec:afterpulsing}) in the late light region and random electronic noise, the Fprompt distribution is broadened; also, $Q_R$ and the centroid reconstruction are affected thereby causing some deviation from expectations.

\section{Single Photon Counting}
\begin{figure}[htbp]
\centering
\graphicspath{{./fig/LED_data/}}
\includegraphics[scale=0.4]{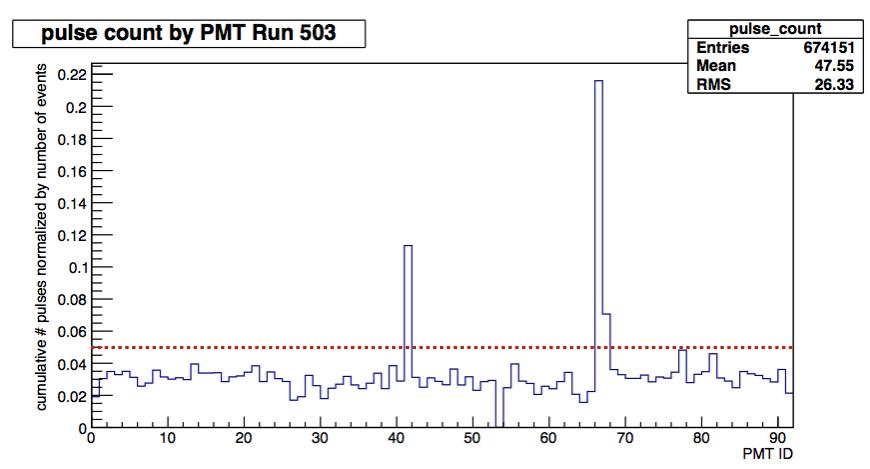}
\caption{ PMT occupancy rate. The red dashed line indicates the 5\% level.}
\label{fig:ledoccupancy}
\end{figure}

\begin{figure}[htbp]
\hfill
\subfloat[]{\includegraphics[width=7cm]{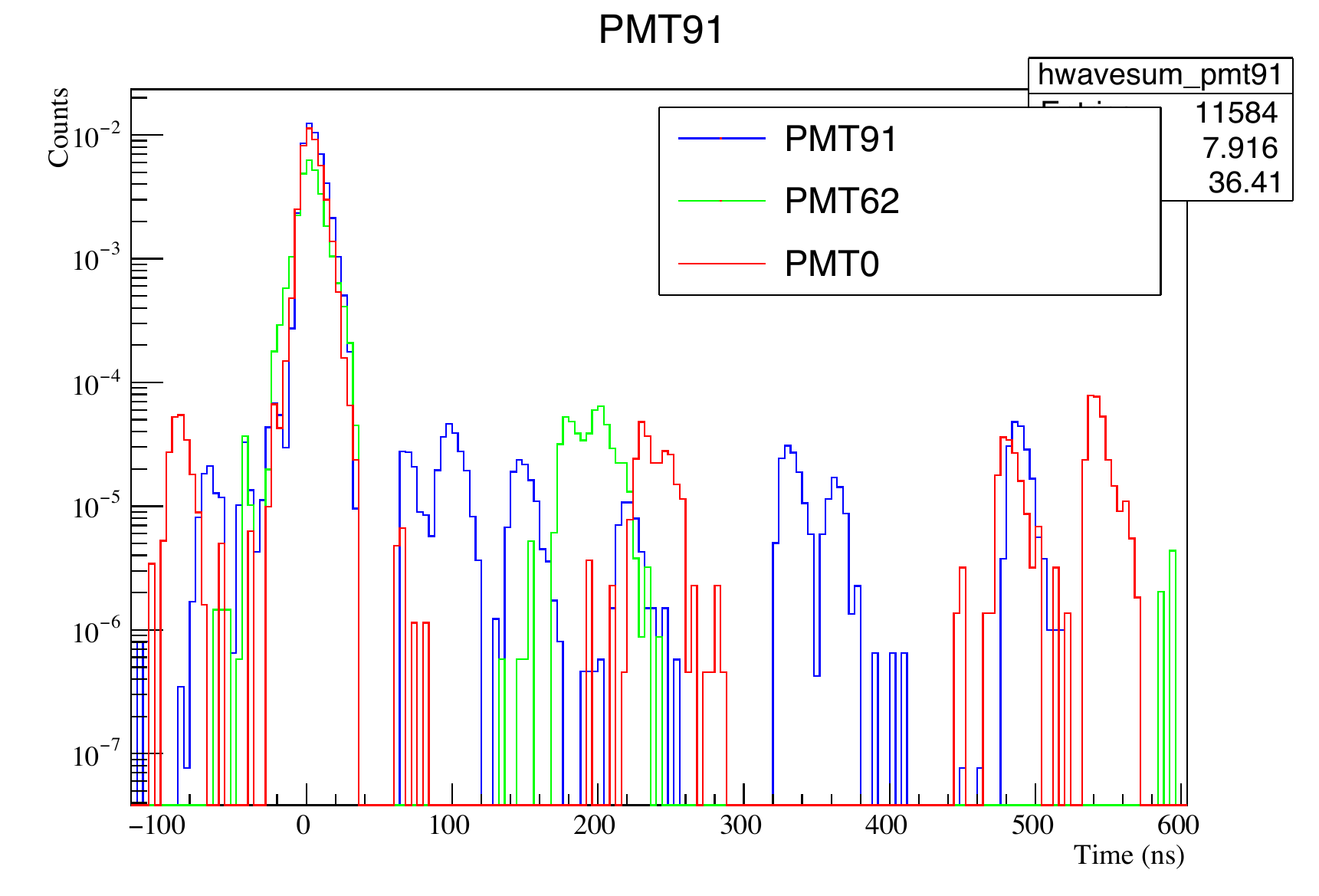}}
\hfill
\subfloat[]{\includegraphics[width=7cm]{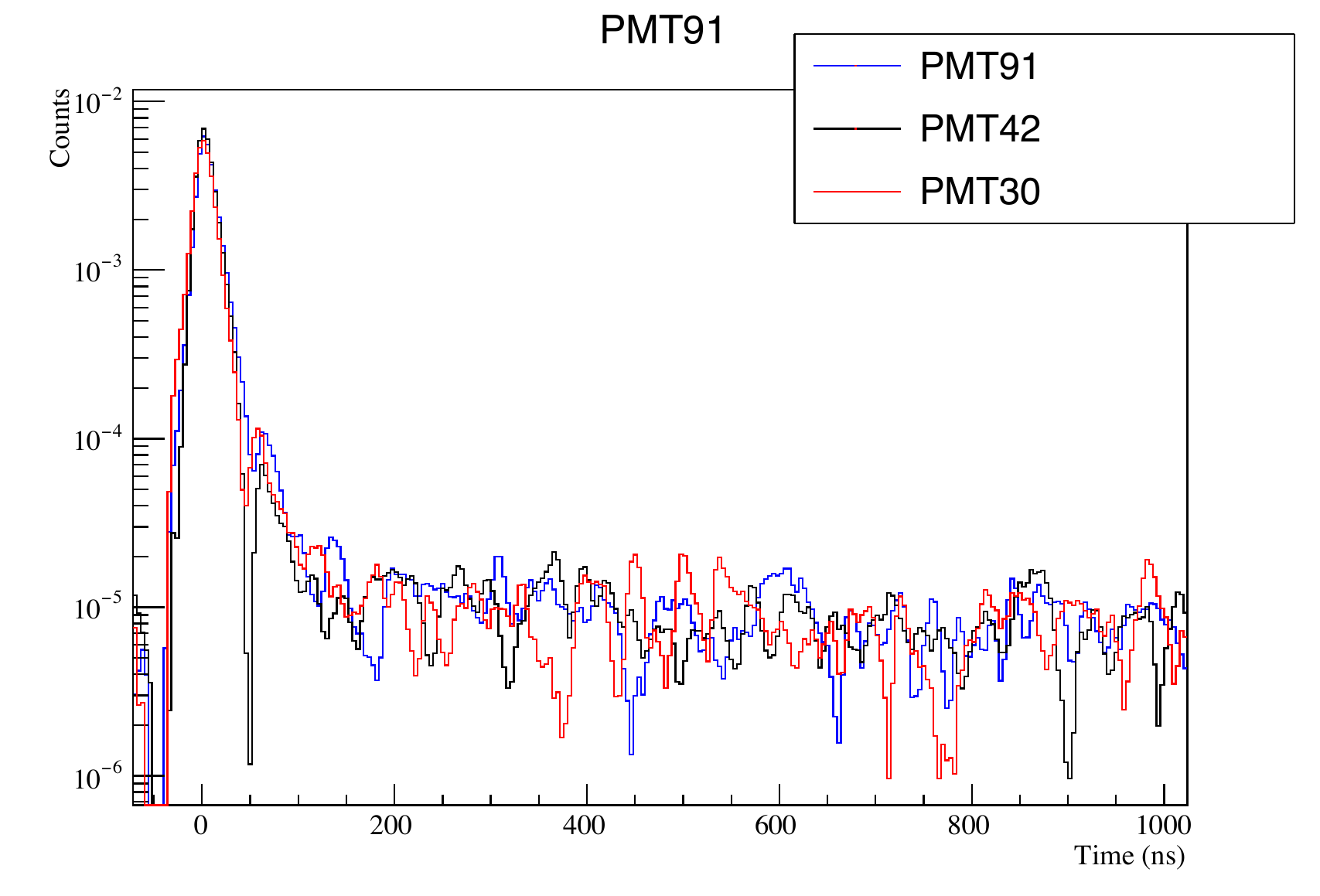}}
\hfill
\caption{(a) Pulse timing of blue LED. The position of blue LED is at PMT 0, the PMT 91 is directly opposite of LED and the PMT 62 is on the side of LED.  (b)  Pulse timing of UV LED. The position of blue LED is at PMT 91, the PMT 0 is directly opposite of LED and the PMT 42 is on the side of LED.}
\label{fig:ledpulsetiming}
\end{figure}

For low intensity, the distribution of the number of photons emitted by an LED follows the Poisson distribution, changing to the Gaussian distribution for higher intensities, as implied by the central limit theorem.  To check that the PMTs are, to high likelihood, indeed seeing only single photons from a low intensity LED, the Poisson Distribution was used to determine the specific probability of the photon being received by each PMT.  The Poisson distribution :
\begin{ceqn}\begin{align}
P(n;\mu) = \frac{\mu^n\cdot e^{-\mu}}{n!}
\end{align}\end{ceqn}
where $\mu$ is the mean number of photoelectrons observed by the PMT and $P(n;\mu)$ is the probability that $n$ photoelectrons are detected.  That is, occupancy rate for each PMT when the mean is $\mu$.  For the \textit{In-Situ} optical calibration, the required probability of observing multiple photons is set to <2.5\%. 
\begin{ceqn}\begin{align}\label{eq:multiple}
\frac{P(2)}{P(1)} = \frac{\mu}{2} = 0.025
\end{align}\end{ceqn}  
Eq. \ref{eq:multiple} prescribes the occupancy rate at 5 \%, to meet the requirement.  Note that the dark hits -- thermionic emission from photocathode -- could contribute to the LED charge distribution.  Now the probability of getting no photons (dark hits) can be expressed:
\begin{ceqn}\begin{align}\label{eq:noise}
P(0) = \frac{N_{noise}}{N_{trig}} = e^{-\mu}
\end{align}\end{ceqn}
where $N_{noise}$ is the number of events in the pedestal -- mostly from dark hits -- and $N_{trig}$ is the number of LED triggers.  For PMTs operating at LAr cryogenic temperature the dark hit rate is around 600 Hz, which is greatly reduced due to less thermionic motion at photocathode.  From \cite{DOSSI2000623}, the random coincidence rates of LED can be expressed :
\begin{ceqn}\begin{align}
f_{random} = f_{dark}\cdot f_{trig} \cdot \tau_{window}
\end{align}\end{ceqn}
where $f_{random}$ is the random coincidence rate, $f_{dark}$ is the dark hits rate, and $\tau_{window}$ is the acquisition window of LED events.  Furthermore, for small $\mu$ the events rate is (using Eq. \ref{eq:noise})
\begin{ceqn}\begin{align}
f_{events} = (1 - P(0))f_{trig} \simeq \mu f_{trig}
\end{align}\end{ceqn}
where the $f_{events}$ is the event rate of LED events. Thus, for the random coincidences' contribution at the level of 1\%, it is necessary to keep 
\begin{ceqn}\begin{align}\label{eq:occupancyrequirement}
\mu \geq \frac{f_{dark}\cdot \tau_{window}}{0.01}
\end{align}\end{ceqn}
For a 600 Hz dark hits rate and $\tau_{window}= 100$ ns, Eq. \ref{eq:occupancyrequirement} gives $\mu \geq 0.006$. Therefore, for $\mu = 0.05$ the LED charge distribution has both  negligible contribution from dark hits and multiple photoelectrons. Figure \ref{fig:ledoccupancy} shows the typical occupancy rate of each PMT while doing the optical calibration. Most PMT received a single photon with the probability less than 5\% except three PMTs which are at direct opposite of LED. The LED are mounted uniformly throughout the sphere, thus these PMTs received more than 5\% of single photon will have less than 5\% while some other LEDs are firing.
While occupancy rate ensure the PMT sees single photon, the timing of the pulse from LED can further ensure the PMT are seeing the photon from LED instead of random noise.
Figure \ref{fig:ledpulsetiming} shows the timing of pulse from LED in different relative position to the LED. For every trigger from LED, the timing of the pulse in different position are consistent with each other. \par


\section{Single Photoelectron Calibration}\label{sec:SPE}
The pulse shape of PMT pulses -- which has been throughly studied by the MiniCLEAN collaboration\cite{Caldwell2012} -- exhibits multiple timing components each of which are good fits to a lognormal distribution as shown in Fig. \ref{fig:pmtpulseshape}.  The fitting function is:
\begin{ceqn}\begin{align}\label{eq:pulseshape}
I(t) = \sum\limits_{i=1}^{n} \frac{Q_i}{(t+t_0)\sqrt{2\pi\sigma_i^2}}e^{-ln^2(\frac{t+t_0}{\tau_i})/2\sigma_i^2}
\end{align}\end{ceqn}
where $\tau_i$ is the geometric mean of the electron arrival time, $\sigma_i$ is the geometric RMS, $Q_i$ is the total charge in the component, and $t_0$ is a time offset that is left fixed for each time component.  For the double and triple lognormal distributions cases, $n$ = 2 and 3, respectively.  The charge of the pulse is taken to be the sum of $Q_i$ in Eq. \ref{eq:pulseshape}.  The charge distribution was fitted using two gamma distributions:
\begin{ceqn}\begin{align}\label{eq:doublegamma}
p(q) = \frac{p_1}{\Gamma(k_1)q_1^{k_1}}\cdot q^{k_1-1}\cdot e^{-q/q_1} + \frac{1-p_1}{\Gamma(k_2)}\cdot q^{k_2-1}\cdot e^{-q/q_2}
\end{align}\end{ceqn}
where the mean charge is determined from:
\begin{ceqn}\begin{align}\label{eq:doublegammamean}
\bar{q} = p_1k_1q_1 +  (1-p_1)k_2q_2
\end{align}\end{ceqn}
{Fig. \ref{fig:fitexled} shows an example fitted charge distribution, where no pedestal from noise is seen because the data was taken in ZLE mode.  As such, the baseline subtraction is performed before the waveform was recorded.  The fitting procedure cycles through the charge distribution of all 92 PMTs, thus the gain of each PMT channel can be determined, as shown in Fig. \ref{fig:pmtgainled}.  The PMT gains are determined approximately every 5 minutes, depending on the charge distribution statistics.  Therefore the PMT gain relative stability can be defined as the standard deviation of the gain in the given period of time divided by the average gain in the given time.  Figure \ref{fig:pmtgainledrel} shows the PMT gain relative stability for each PMT over the course of approximately a 7-hour data acquisition period, indicating that the PMTs' gain variation over 7 hours is less than 2.5\%.\par
\begin{figure}[htbp]
\hfill
\subfloat[]{\includegraphics[width=7cm]{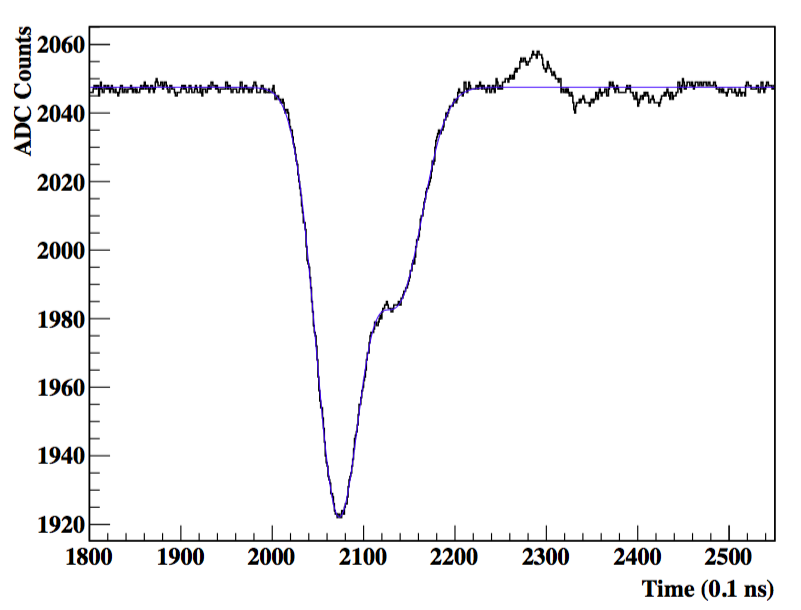}}
\hfill
\subfloat[]{\includegraphics[width=7cm]{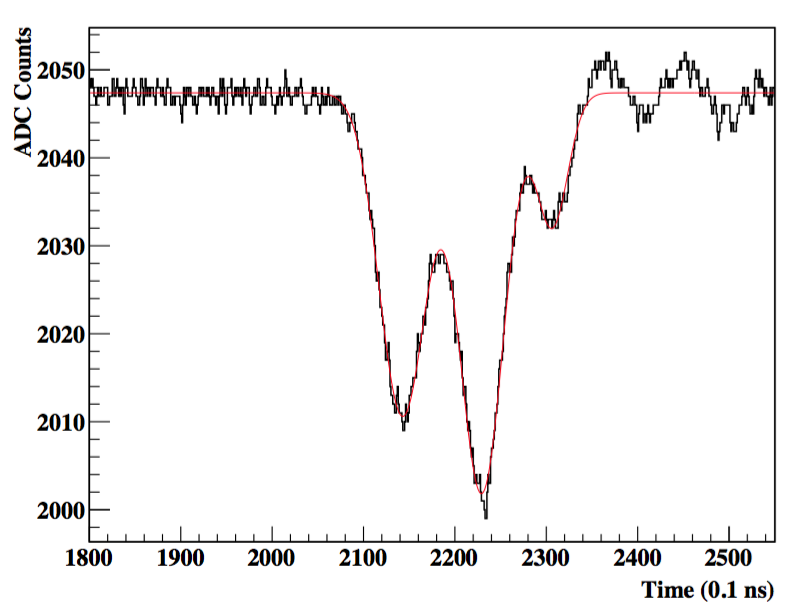}}
\hfill
\caption{(a) Two timing components was used to fit the typical R5912-02 pulse. (b) Three timing components was used to fit the typical R5912-02 pulse. Figures from \cite{Caldwell2012}.}
\label{fig:pmtpulseshape}
\end{figure}

\begin{figure}[htbp]
\centering
\graphicspath{{./fig/LED_data/}}
\includegraphics[scale=0.25]{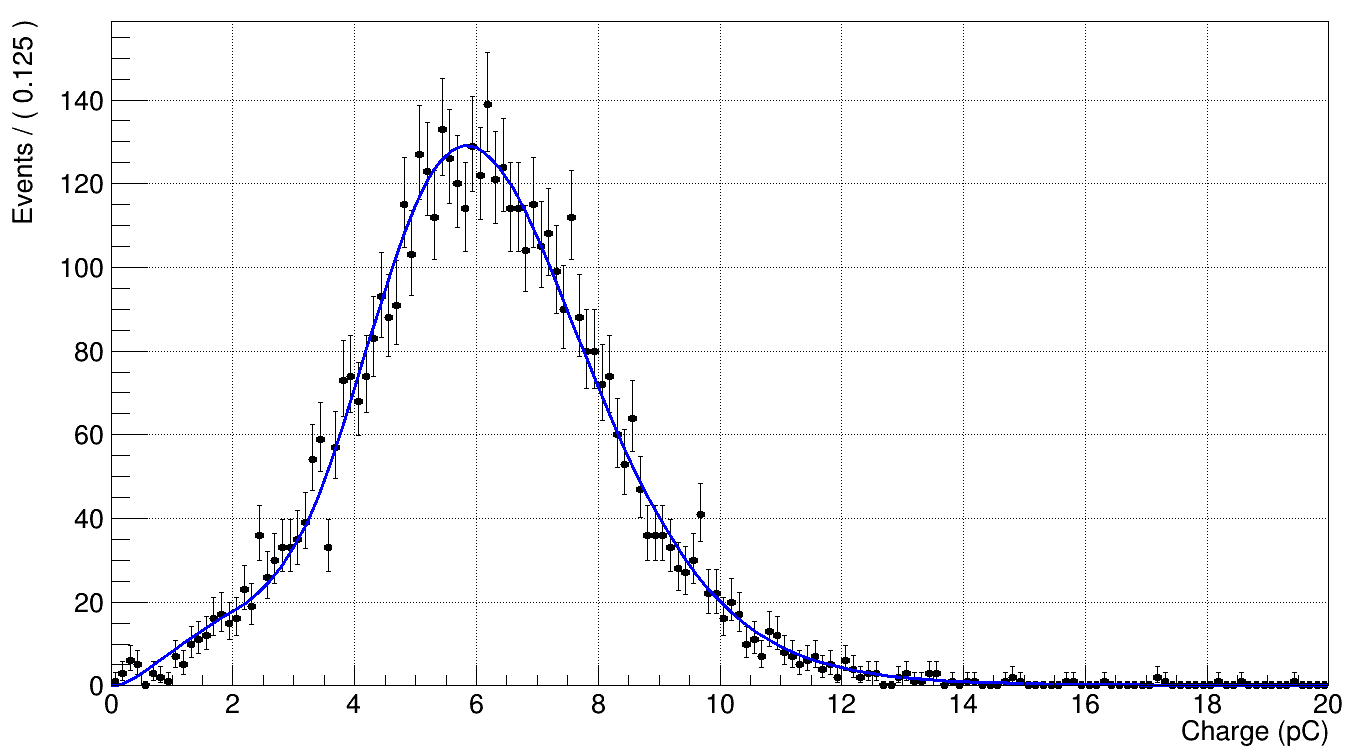}
\caption{Fitting example of charge distribution with two gamma function. The mean charge given by the fitting function is 6.15 pC.}
\label{fig:fitexled}
\end{figure}
\begin{figure}[htbp]
\centering
\graphicspath{{./fig/LED_data/}}
\includegraphics[scale=0.4]{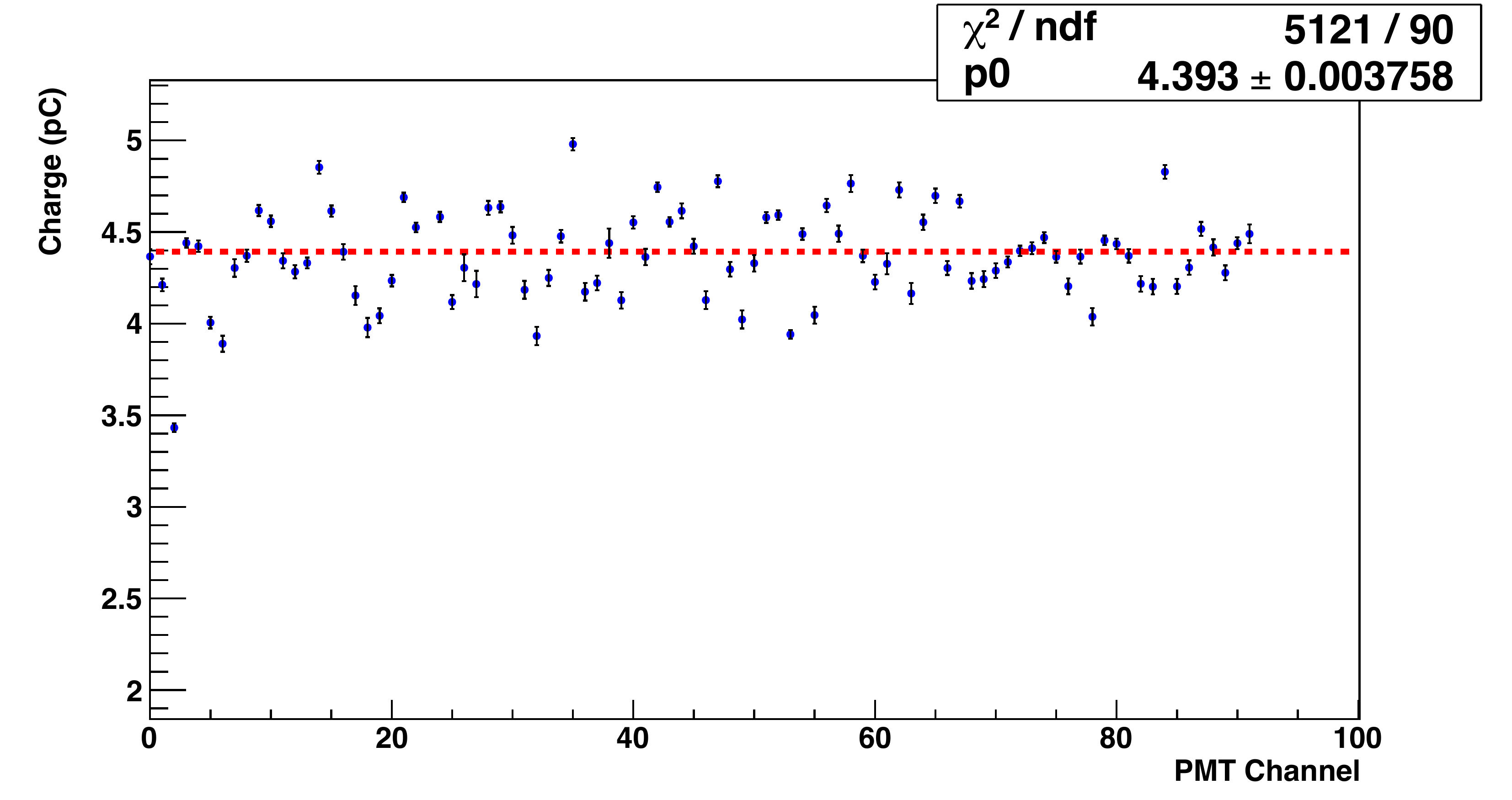}
\caption{PMT gain for PMTs determined by \textit{In-situ} optical calibration.}
\label{fig:pmtgainled}
\end{figure}

\begin{figure}[htbp]
\centering
\graphicspath{{./fig/LED_data/}}
\includegraphics[scale=0.4]{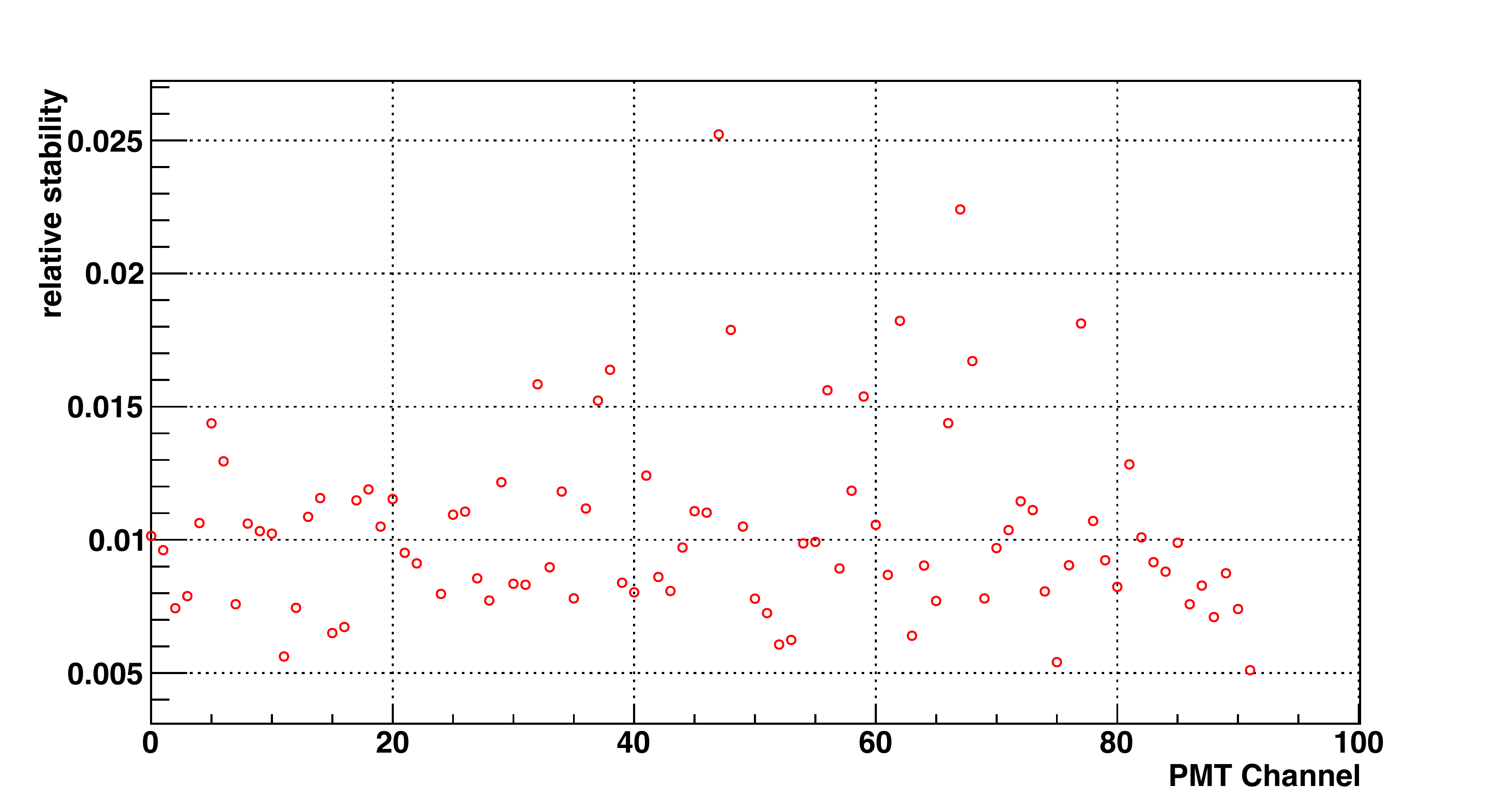}
\caption{Relative stability of PMT gain (see text). }
\label{fig:pmtgainledrel}
\end{figure}
The angular distribution of LED events can be helpful towards determining if the LED worked as expect and allowing a check of the detector optics, from the UV LED.  Figure \ref{fig:angularled} shows the intensity of PMTs as a function of the angle between each PMT and the firing LED.  For most LEDs, the highest intensity appears at 2.7 Rad because the LEDs were in fact mounted on the side of the cassettes, and therefore not aimed directly to the opposite PMT ($\sim$3.14 Rad.).  Notice that LED 61 (blue) has highest intensity at a 0.3 Rad angle.  This is because when the cassette of PMT 61 was adjusted the baffle became tilted by accident such that the fiber is blocked (by the baffle) instead of poking through it.  Therefore the photons emitted by LED 61 were mostly reflected to PMT 66 which is near PMT 61, thus resulting in the strange angular distribution. Nonetheless, for the sake of doing SPE calibration, by increasing the current through the LED, the rest of PMTs still be able to received enough photons. Therefore no further adjustment was performed in order to minimized the time of exposing the IV to the environment which is at risk to increase the radon level of the IV. On the other hand, the UV LED 62 seems not emitting any photons. After increasing the voltage to the maximum allowed voltage of LED pulser, very weak intensity still observed. The phenomenon was seen for all LEDs that the LED required more bias voltage to drive the LED at the same current with the test results obtained at UNM.  This is due to the fact that to reproduce the exact the same environment at UNM is impossible, thus the full setup of IV could introduce unexpected impedance to the LED pulser system. Nevertheless most LEDs still functioning well  by increasing the bias voltage by 1 or 2 V except the UV LED 62. \par
\begin{figure}[htbp]
\hfill
\subfloat[]{\includegraphics[width=7cm]{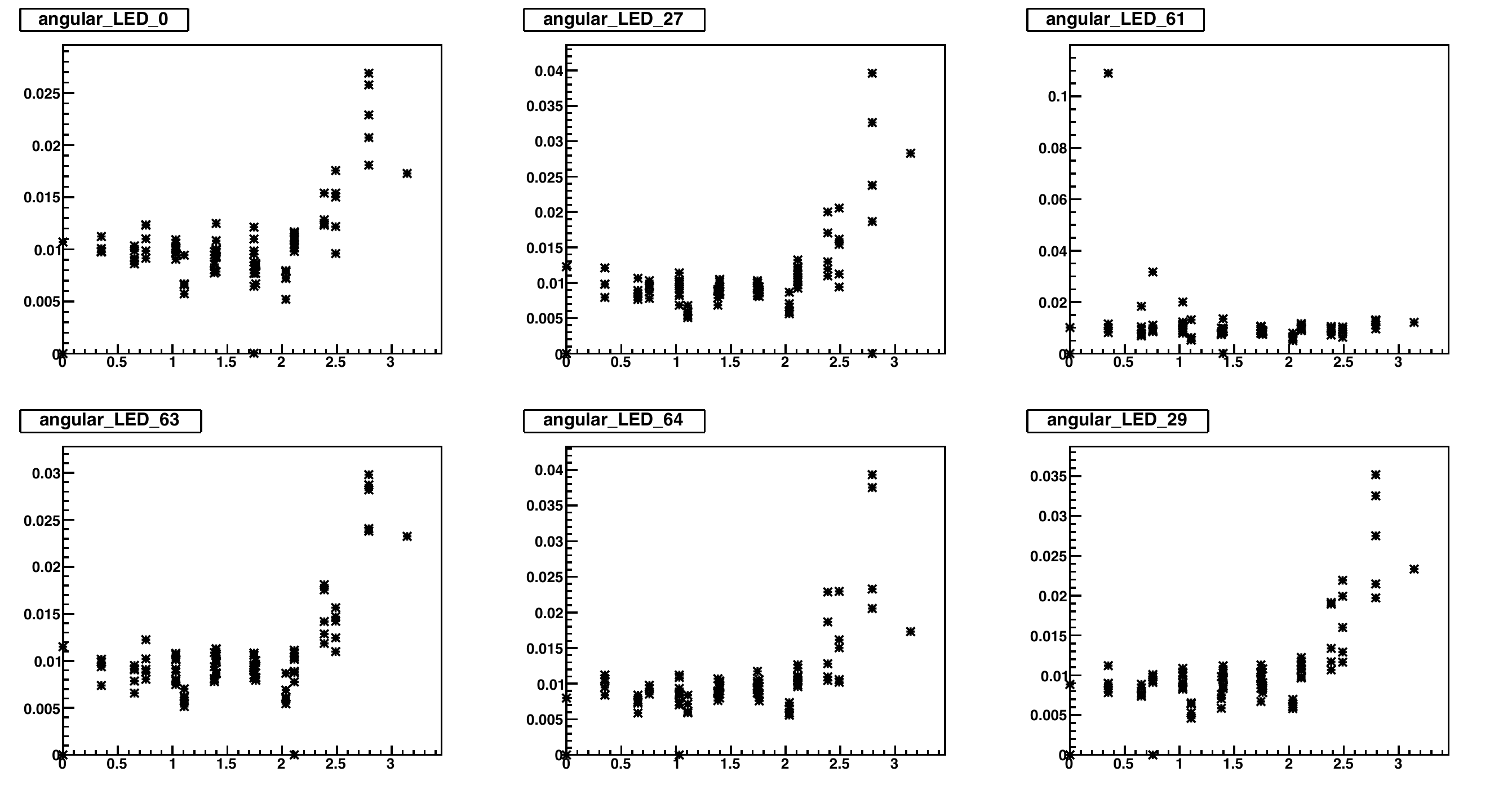}}
\hfill
\subfloat[]{\includegraphics[width=7cm]{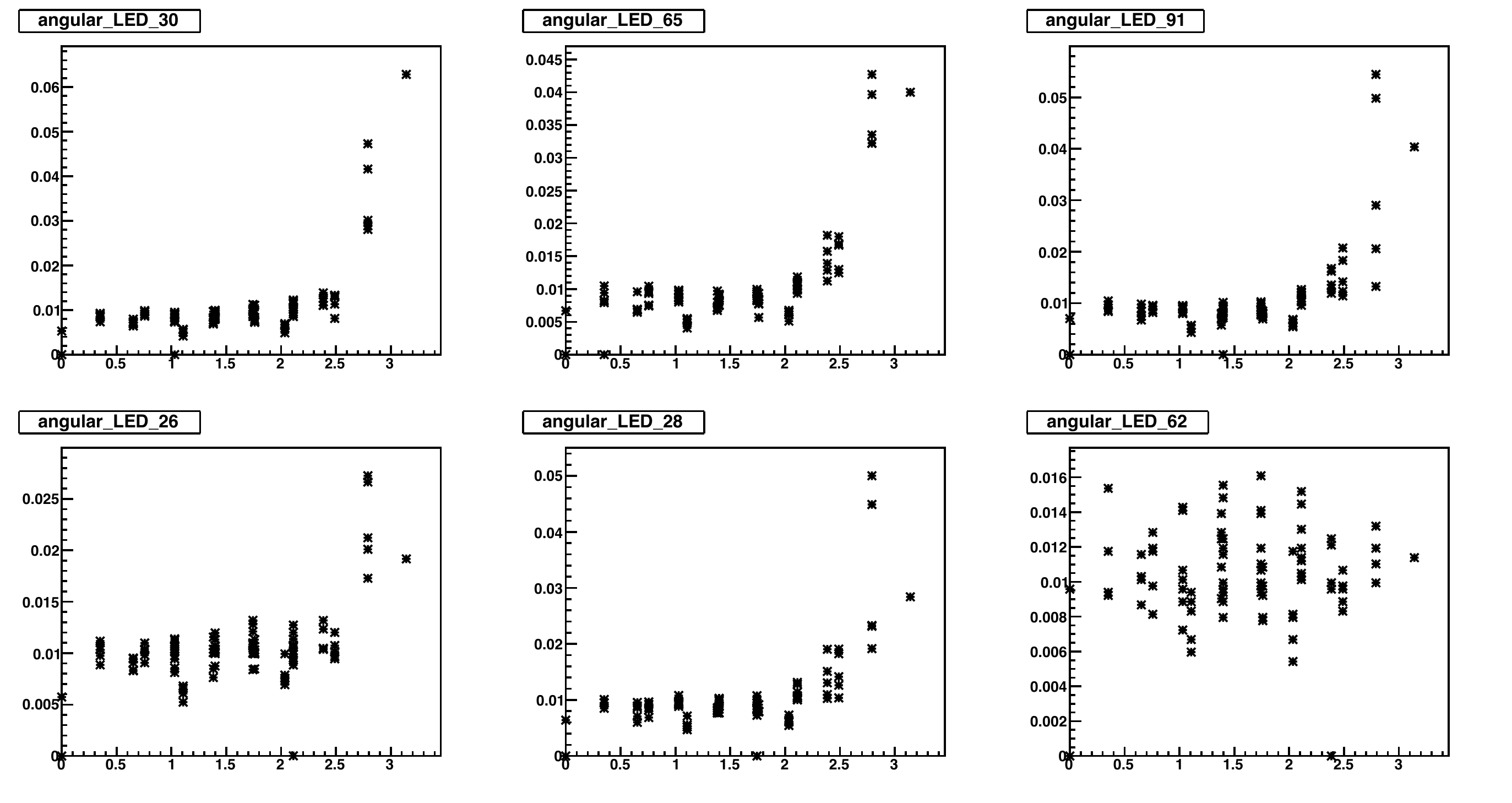}}
\hfill
\caption{(a) Angular distribution of blue LEDs. The y-axis is the intensity normalized by the total photoelectrons and x-axis is the angle between PMT and the firing LED in radian. (b) Angular distribution of UV LEDs. The y-axis is the intensity normalized by the total photoelectrons and x-axis is the angle between PMT and the firing LED in radian.}
\label{fig:angularled}
\end{figure}
Before moving the IV into the OV (with full shielding from water tank), it was sitting in the clean room without any further shielding.  Gamma rays, which are induced by the interaction between cosmic rays and underground rock, forms the major background source.  Such gamma photons passing through the acrylic creates Cherenkov light, which is detected by the PMTs.  This Cherenkov light -- which is blue and produces a very fast pulse in the PMT -- is suitable in helping to determine the PMT gain, or rather it is a method to help check the LED results.  In this regard, Figure \ref{fig:ledcherenkov} shows a comparison of SPE charge distribution between LED and Cherenkov light methods.\par
Notice that for Fig. \ref{fig:ledcherenkov}(b) there is a difference in the low charge part of the distribution, and several sources are reasons attributed to this difference.  Although each PMT uses the same model there are in reality some slightly difference in light collection efficiency.  The ``shoulder'' in the low charge might results from the under-amplified photoelectrons.  For Cherenkov light, the incident photons in a PMT could have large angles and might skip the first dynode, resulting in under-amplified photoelectrons.  Nonetheless, the PMT gain determined by the LED and Cherenkov methods are in good agreement, as shown in Fig. \ref{fig:ledcherenkovcom}.\par
\begin{figure}[htbp]
\hfill
\subfloat[]{\includegraphics[width=7cm]{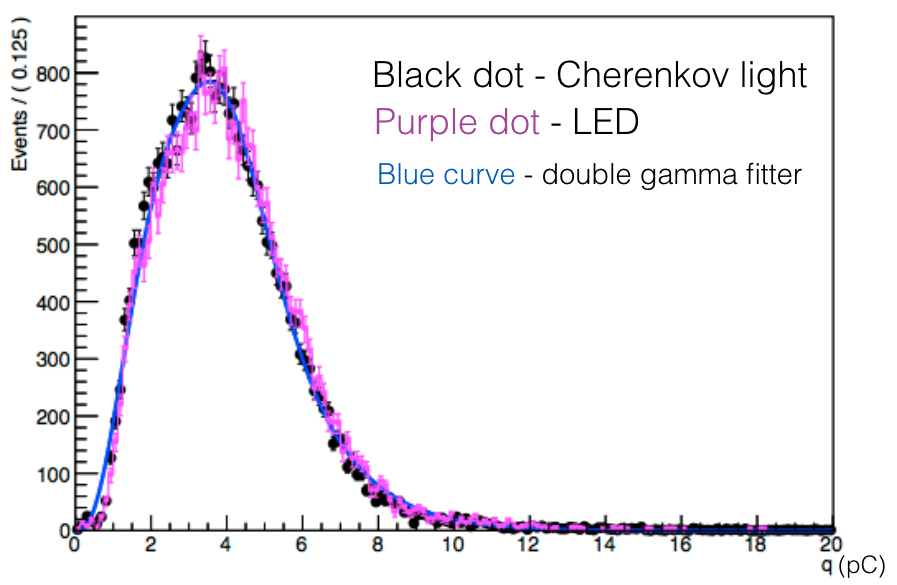}}
\hfill
\subfloat[]{\includegraphics[width=7cm]{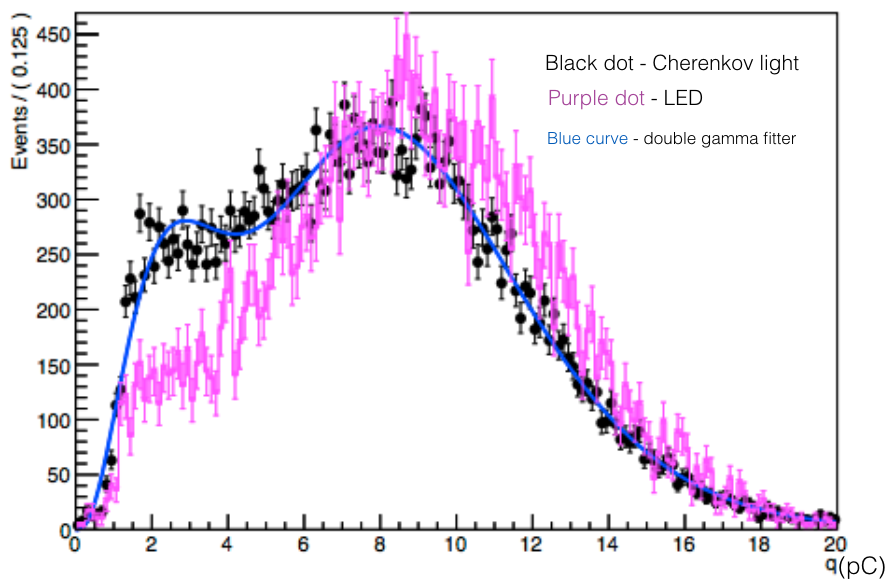}}
\hfill
\caption{(a) Two timing components was used to fit the typical R5912-02 pulse. (b) Three timing components was used to fit the typical R5912-02 pulse}
\label{fig:ledcherenkov}
\end{figure}
\begin{figure}[htbp]
\centering
\graphicspath{{./fig/LED_data/}}
\includegraphics[scale=0.4]{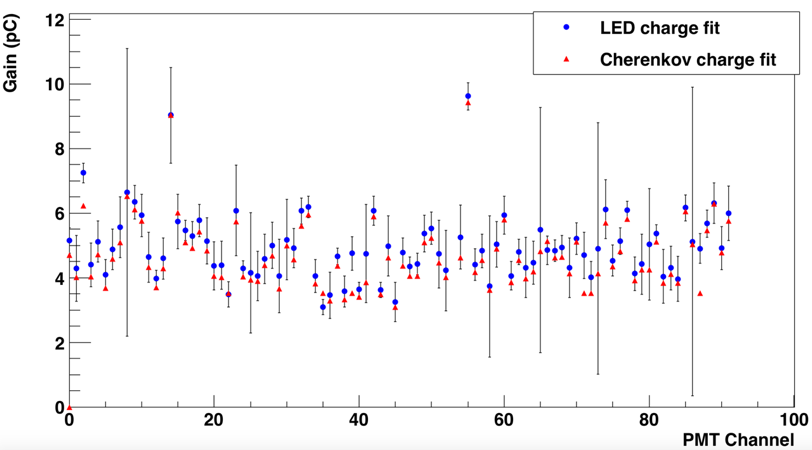}
\caption{PMT gain determined by LED and Cherenkov events. The blue dot is the results from LED and red dot is from Cherenkov events. }
\label{fig:ledcherenkovcom}
\end{figure}
The stability of the TPB coating can be determined by comparing the SPE calibrations (appropriately scaled) from blue and UV LEDs; for instance, Figure \ref{fig:leduvbluecompare} shows gains determined from blue and UV LED where differences are attributed to TPB efficiencies.  The TPB relative stability can be determined by comparing two measurements and is defined as the differences between two measurements in the given PMT, divided by the mean charge over the course of measurement determined by the UV LED.\par
A typical result is shown in Fig. \ref{fig:leduvrelative}.  The large change in relative stability indicates that there might be some other factors affecting the measurement. For example, the behavior of blue photons bouncing off the TPB surface might be very different from the UV-to-visible TPB reemission photons.  Moreover, the study from \cite{gerdaproposal} indicates that the UV photon can be shifted by the baffle (ESR foil, Fig. \ref{fig:ESR_spectral}) resulting in the systematic error on PMT gain measurement using UV LED.\par
\begin{figure}[htbp]
\centering
\graphicspath{{./fig/LED_data/}}
\includegraphics[scale=0.25]{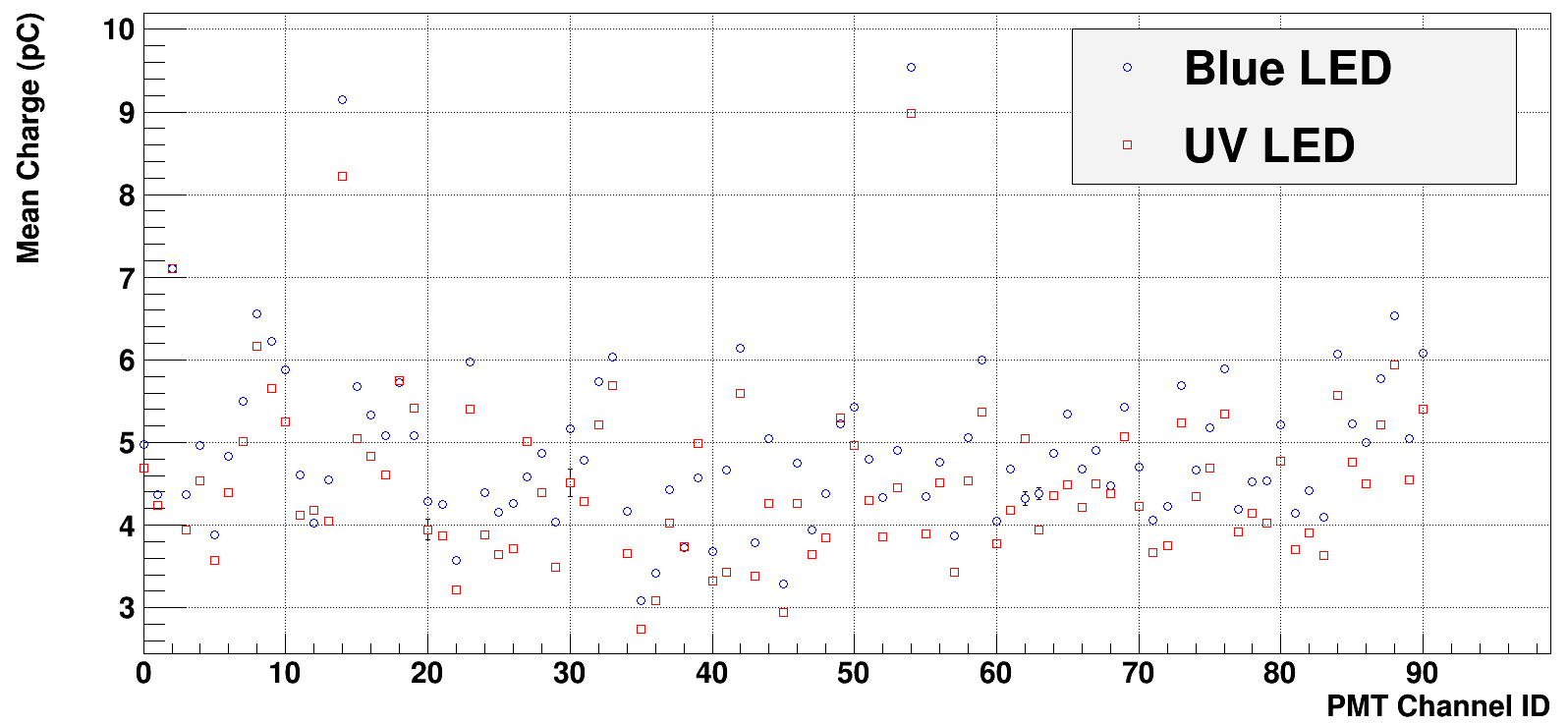}
\caption{PMT gain determined by blue and UV LED. The blue dot is the results from blue LED and red dot is from UV LEDs. }
\label{fig:leduvbluecompare}
\end{figure}

\begin{figure}[htbp]
\centering
\graphicspath{{./fig/LED_data/}}
\includegraphics[scale=0.25]{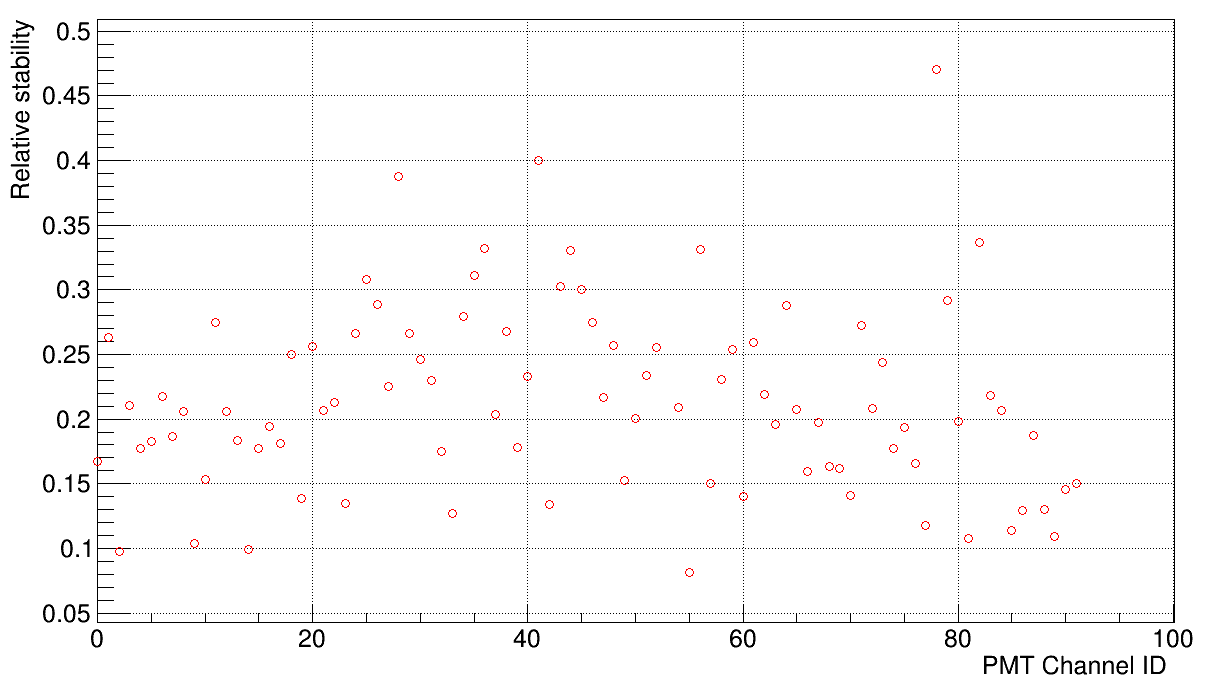}
\caption{Relative stability of TPB (see text). }
\label{fig:leduvrelative}
\end{figure}
\begin{figure}[htbp]
\centering
\graphicspath{{./fig/LED_data/}}
\includegraphics[scale=0.25]{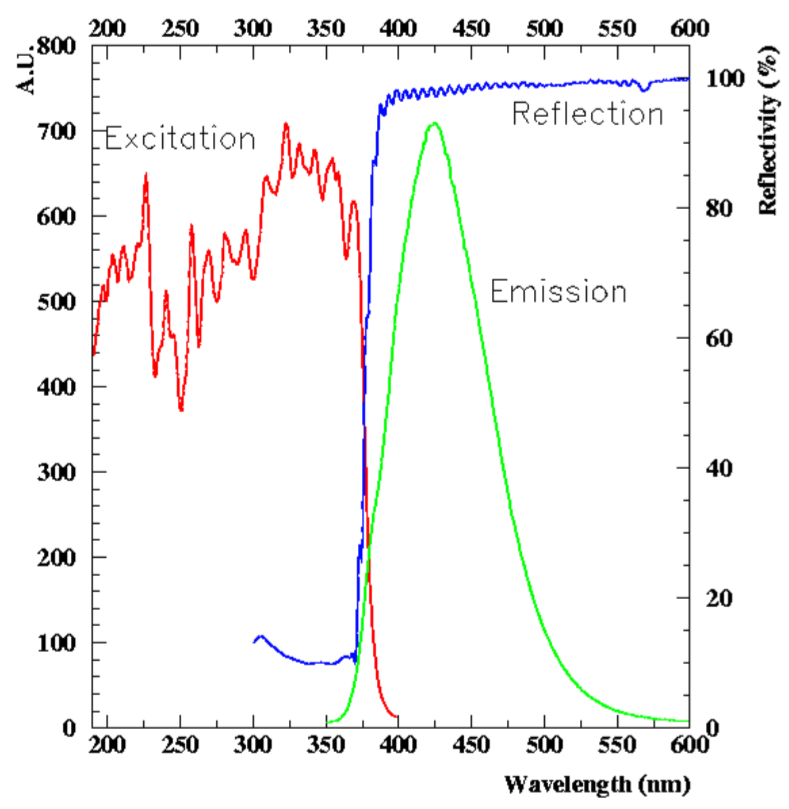}
\caption{Spectral response of ESR foil. }
\label{fig:ESR_spectral}
\end{figure}
\section{Pulse Timing Analysis}
The LED is a stable external light source which can be used to understand the pulse timing for different pulse types: pre-pulsing, prompt-pulsing, double pulsing, late pulsing and after pulsing.  These pulse types could affect the Fprompt and energy resolution, thus a thorough study is needed in order to understand the fraction of these pulses appearing in the data.  The definition and explanation of each pulse type follows :
\begin{itemize}
	\item \textbf{Pre-pulsing} : The incident photon did not convert to a photoelectron at the photocathode, but instead passes through the photocathode and hits the first dynode, create photoelectrons there.  This type of pulse usually arrives earlier than the prompt-pulsing (conversion to photoelectrons at photocathode).  The timing is around 30 ns earlier than prompt-pulsing which is equivalent to the electron transit time of the PMT (Fig. \ref{fig:typepulsing}(a)).
	\item \textbf{Prompt-pulsing} : The incident photon converts to a photoelectron at the photocathode and follows the series of dynode multiplications.  The trigger time is defined as timing of the peak of summed pulses (Fig. \ref{fig:typepulsing}(a)).
	\item \textbf{Double-pulsing} : The incident photon converts to photoelectrons at the photocathode with subsequent inelastic scattering off the first dynode.  This causes some photoelectrons to scatter backward and create the current at anode at a later time.  The charge of individual pulses is smaller than the pulse from prompt-pulsing, but the total charge of the pulses is equal to the prompt-pulsing (Fig. \ref{fig:typepulsing}(b)).
	\item \textbf{Late-pulsing} : The incident photon converts to photoelectrons at the photocathode and subsequently elastically scatters off the first dynode.  The charge of the pulse is the same with the pulse from prompt-pulsing, but arrives later in time.  The timing difference between this type of pulse and the prompt-pulsing is approximately twice the electron transit time of the PMT (Fig. \ref{fig:typepulsing}(c)).
	\item \textbf{After-pulsing} : The photoelectron could have some possibility to ionize the residual gases inside the PMT.  The resulting positive ion drift back to the photocathode, creating a cascade of photoelectrons going through the normal dynode chain.  The timing of this type of pulse depends on the type of ion (Fig. \ref{fig:typepulsing}(d)).
\end{itemize}
Figure \ref{fig:allpulsetiming} shows the pulse timing for all the pulses described above except after-pulsing.\par

\begin{figure}[htbp]
\hfill
\subfloat[]{\includegraphics[width=7cm]{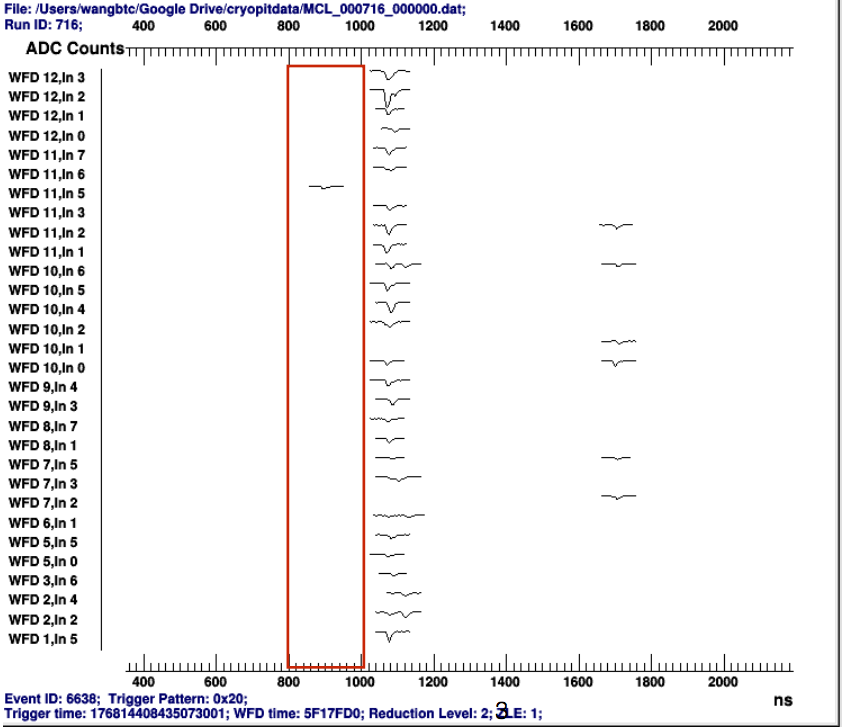}}
\hfill
\subfloat[]{\includegraphics[width=7cm]{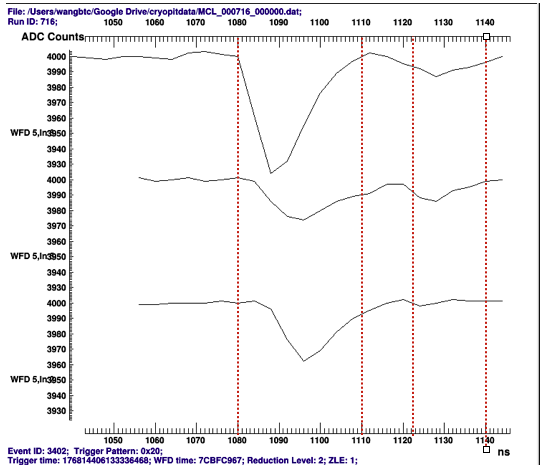}}
\hfill
\subfloat[]{\includegraphics[width=7cm]{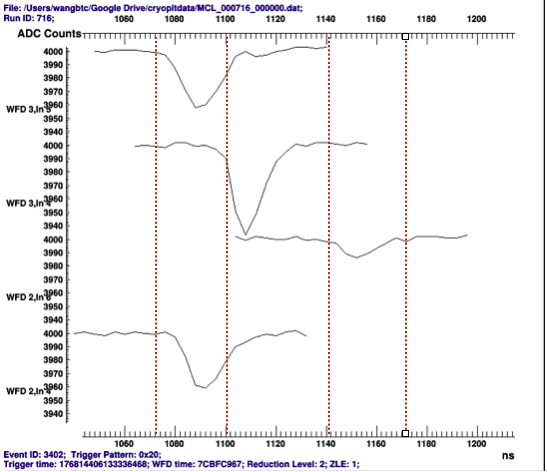}}
\hfill
\subfloat[]{\includegraphics[width=7cm]{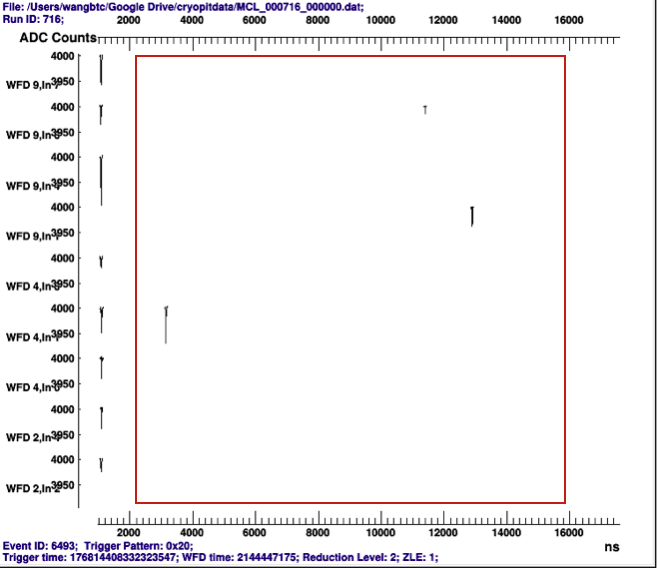}}
\hfill
\caption{(a) Pre-pulsing is indicated in the red box. It comes in before the Prompt-pulsing (a series of pulses line up on the left hand side of red box). (b) Double-pulsing. The dashed line indicates the first and second pulses. (c) Late-pulsing. The two red dashed line on the left indicates the Prompt-pulsing and the two red dashed line on the right indicates the Late-pulsing. (d) After-pulsing. The red box indicates the After-pulsing. }
\label{fig:typepulsing}
\end{figure}
\begin{figure}[htbp]
\centering
\graphicspath{{./fig/LED_data/}}
\includegraphics[scale=0.4]{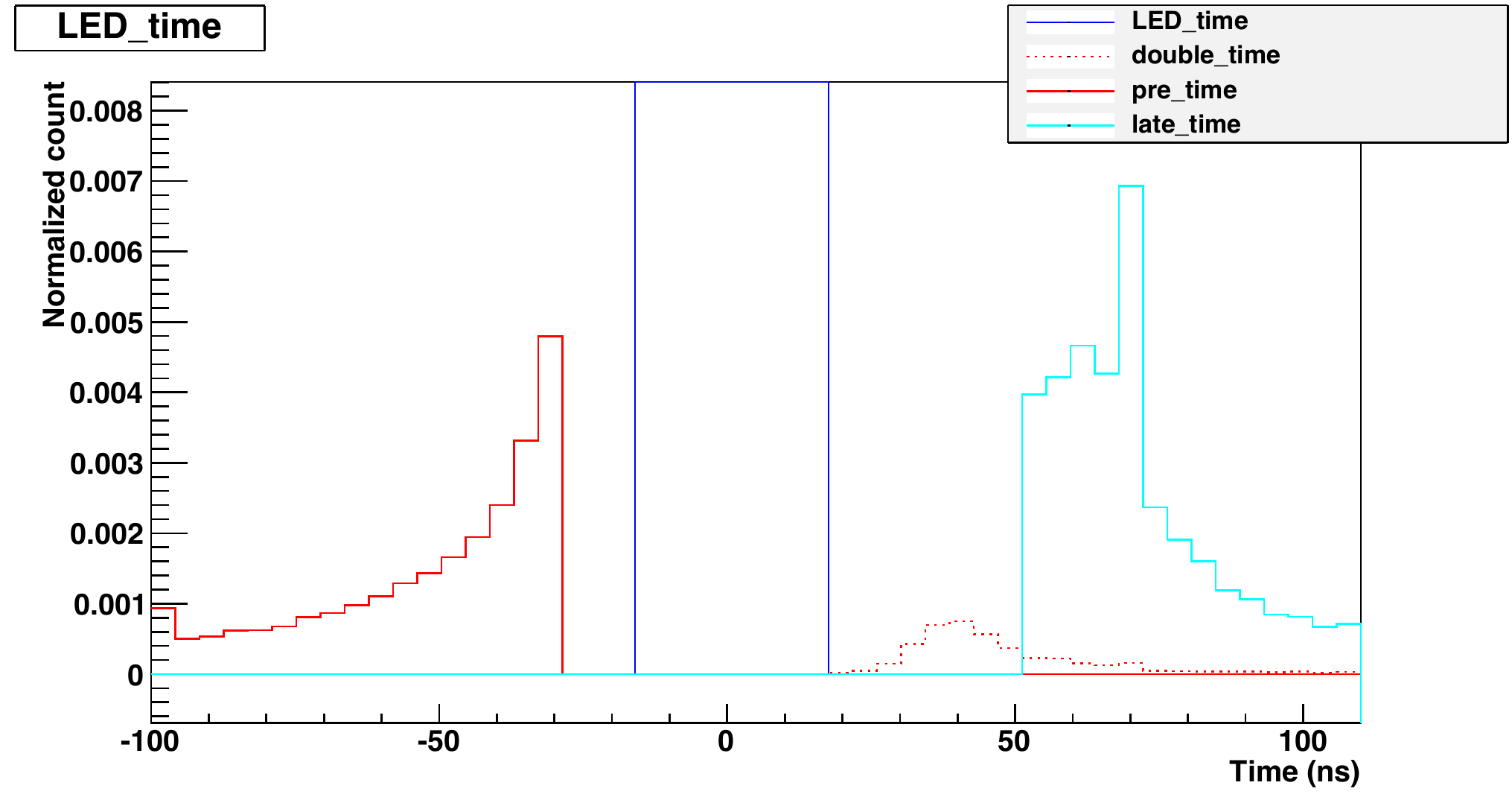}
\caption{Pulse timing for different type of pulse.}
\label{fig:allpulsetiming}
\end{figure}
Using the LED pulse as reference pulse (prompt-pulsing), the fraction of each type of pulse can be determined. Figure \ref{fig:ratepulsing} shows the rate of each type of pulse found in the data for each PMT. Among these different type of pulses, the pre-pulsing has lowest rate due to high conversion rate at photocathode. The double-pulsing also has low rate because if the photoelectron inelastic scattered from first dynode not far away depends on the kinetic energy it carried, therefore the pulse would be very close with the prior pulse, the pulse finding algorithm may not be able to distinguish it as double pulse.  On the other hand, with elastic scattering, the photoelectrons usually have larger kinetic energy which can travel in the opposite direction until it lose the kinetic energy and drift back to the first dynode. Therefore the timing is approximately twice of the electron transit time of the PMT. These three type of pulse, however, are well inside the prompt window and is coming from the real scintillation events. Therefore it will not affect the total charge and the Fprompt analysis. Conversely, the after-pulsing is from the interaction of residual gas and the photoelectrons, this additional pulse and charge will affect the energy resolution and the Fprompt analysis. Therefore more carefully treatment is needed to understand the impact of the after-pulsing.\par
\begin{figure}[htbp]
\hfill
\subfloat[]{\includegraphics[width=7cm]{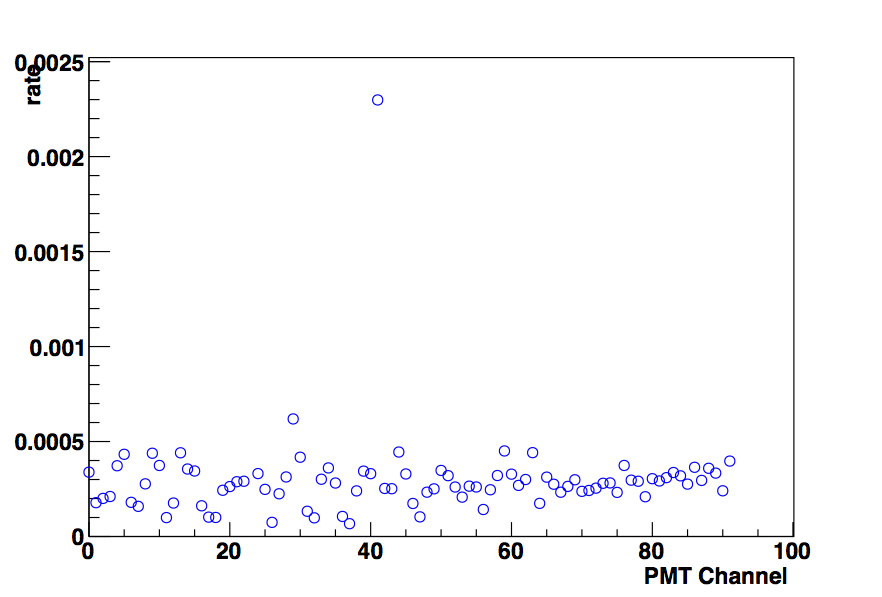}}
\hfill
\subfloat[]{\includegraphics[width=7cm]{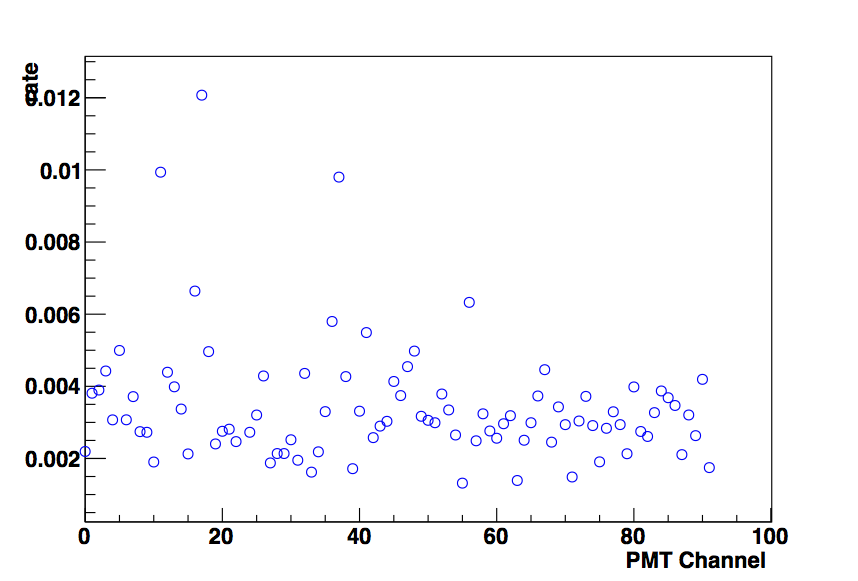}}
\hfill
\subfloat[]{\includegraphics[width=7cm]{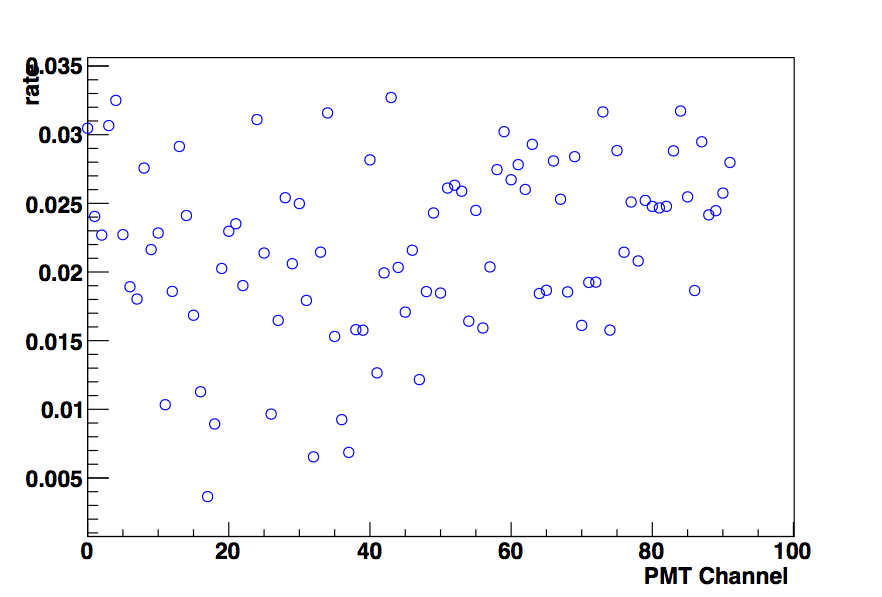}}
\hfill
\subfloat[]{\includegraphics[width=7cm]{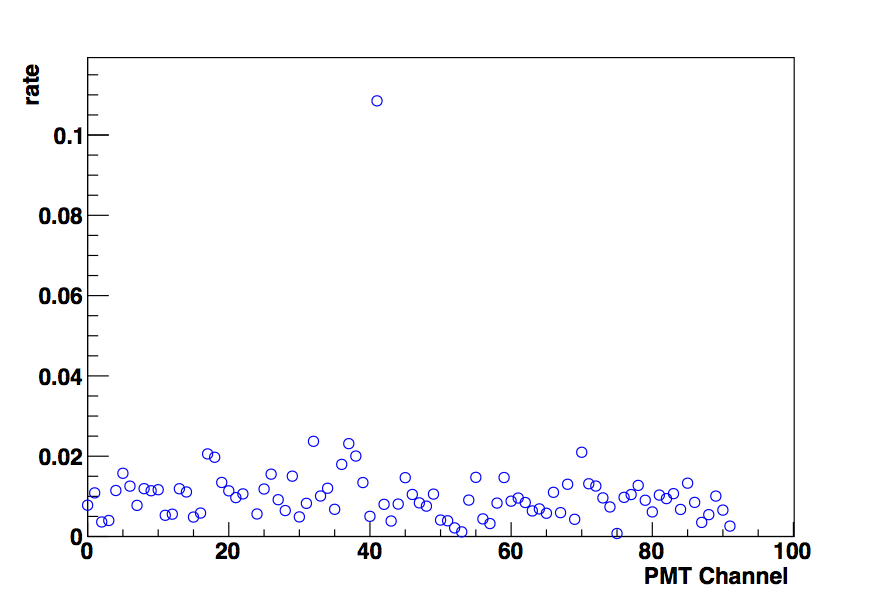}}
\hfill
\caption{ Rate of different type of pulse : (a) Pre-pulsing. (b) Double-pulsing. (c) Late-pulsing. (d) After-pulsing.  }
\label{fig:ratepulsing}
\end{figure}
\subsection{After-Pulsing}\label{sec:afterpulsing}
The timing of the after-pulsing depends on the type of ion and the supply voltage of PMT ( the voltage between first dynode and the photocathode). Figure \ref{fig:radiuspmt} shows the dimensionality of R5912-02MOD. The radius (0.131 m) is taken to be the distance between photocathode and the first dynode. The supply voltage (320 V) is determined by average the high voltage of 92 PMTs and calculate the voltage according to the voltage distribution ratio in the specification of PMT. If assuming the electric potential distribution between the first dynode and the photocathode is quadratic\cite{MA201193} :
\begin{ceqn}\begin{align}\label{eq:afterpulsingeq1}
V(s) = V_0\cdot(1-\frac{s}{L})^2
\end{align}\end{ceqn}
where s is the position inside the PMT, V(s) is the potential at $s$, $V_0$ is the potential of origin and $L$ is the distance between the first dynode and photocathode. Using the Eq. \ref{eq:afterpulsingeq1}, the delayed time of after-pulsing is :
\begin{ceqn}\begin{align}\label{eq:afterpulsingeq2}
t = \sqrt{\frac{m}{2qV_0}}\cdot L \int^L_{s_0} \frac{1}{(L-s_0)^2 - (L-s)^2}ds = \frac{4}{\pi}\sqrt{\frac{2m}{qV_0}}\cdot L
\end{align}\end{ceqn}
where $m$ is the mass of the ion, $q$ is the charge of ion and $s_0$ is the position of origin. With this formula, the expected delayed time of different ion can be calculated. Table \ref{table:afterpulsing} summarize the delayed time for usual residual gas ion in the PMT.\par
\begin{figure}[htbp]
\centering
\graphicspath{{./fig/LED_data/}}
\includegraphics[scale=0.4]{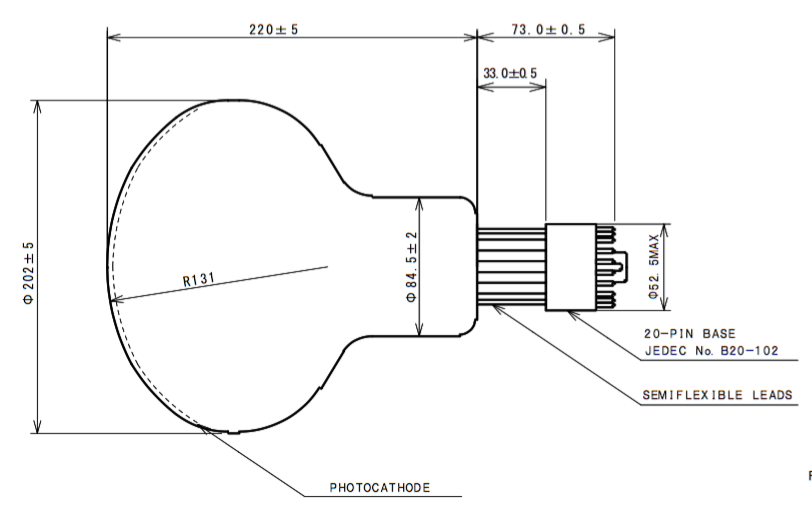}
\caption{Schematic of Hamamatsu R5912-02MOD PMT.}
\label{fig:radiuspmt}
\end{figure}

\begin{table}[htbp]
\caption{Calculated delayed time of after-pulsing induced by different ion.} 
\centering 
\begin{adjustbox}{width=1\textwidth}
\begin{tabular}{c c c c c c c c c} 
\hline\hline 
Type of ions& $H^+$& $He^+$ & $CO_2^+$ & $O^+$ & $CH_4^+$ & $N_2^+$ & $O_2^+$ & $Ar^+$ \\ [0.5ex] 
\hline 
Time ($\mu$s) & 1.43 & 2.86 & 9.45 & 5.71 & 5.71 & 7.55 & 8.08 & 9.01\\[0.5ex]
\hline 
\end{tabular}
\end{adjustbox}

\label{table:afterpulsing} 
\end{table}
Using the LED data taken with IV at room temperature, the peak from individual ion can be identified. Pulse time determined by the pulse finder algorithm is plotted against the charge of the pulse as shown in Fig \ref{fig:after2d}. Projecting the scatter plot onto X-axis (pulse time) with the charge larger than 6 PE to reject some background noise as shown in Fig. \ref{fig:afterfit}. This one dimension plot is fitted with three gaussian distributions to identify the peak position and the $\sigma$ of the peak.  The 6 PE charge cut is added after a abnormal background was found.  A peak was found at pulse time around 200 ns. However, according to the Eq. \ref{eq:afterpulsingeq2}, no ion can have such fast delayed time. After thorough study, it was found that this is due to the impedance mismatch along the LED coaxial cable as shown in Fig. \ref{fig:LEDreflection}. LED pulser is connected to a 60 ft. coaxial cable to the IV, the typical transit time per ft. is 1.5 ns results in 180 ns delay time for reflection signal to trigger the LED again. 
In addition, three peak are clear to be seen in the projection plot, the peak time is 1400 ns, 3157 ns and 7335 ns respectively. Compare this value to Table \ref{table:afterpulsing}, the first peak is from the $H^+$, second peak from the $He^+$ and third peak is from the $N_2^+$. The calculated delayed time is slightly larger than the fitted time. Possible systematic error could include the larger $\chi^2$ which indicates the fit is less accurate. Moreover, the constructing model of the voltage distribution in the PMT might be over simplified. In addition, the peak at 600 ns was found to have negligible contribution. 
This peak may due to the residual gas was ionized by photoelectron between the dynodes, thus faster delayed time was observed. The MiniCLEAN PMT is designed to be operated under cryogenic temperature. Therefore, the intensity of the peak and the rate of after-pulsing is expected to be lower under the cryogenic temperature due to some residual gas freezing onto the inner surface of PMT (e.g. $CO_2$, $O_2$, etc.).\par
\begin{figure}[htbp]
\centering
\graphicspath{{./fig/LED_data/}}
\includegraphics[scale=0.4]{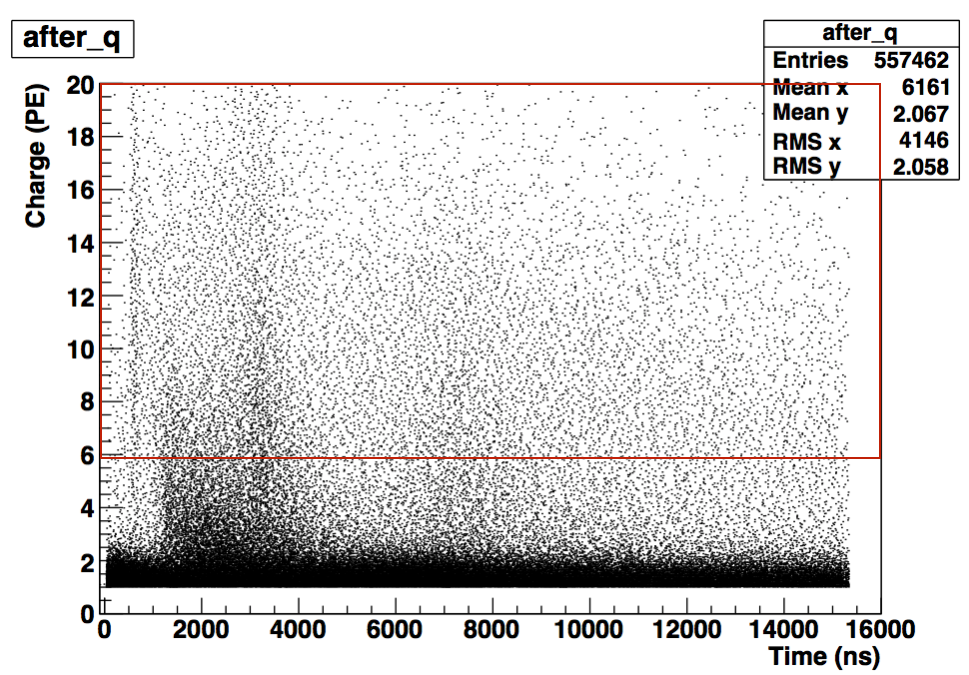}
\caption{Scatter plot of timing and charge of the pulse. The area in the red rectangular box indicates the events were selected for identify the after-pulsing peak (Fig. \ref{fig:afterfit}). }
\label{fig:after2d}
\end{figure}

\begin{figure}[htbp]
\centering
\graphicspath{{./fig/LED_data/}}
\includegraphics[scale=0.6]{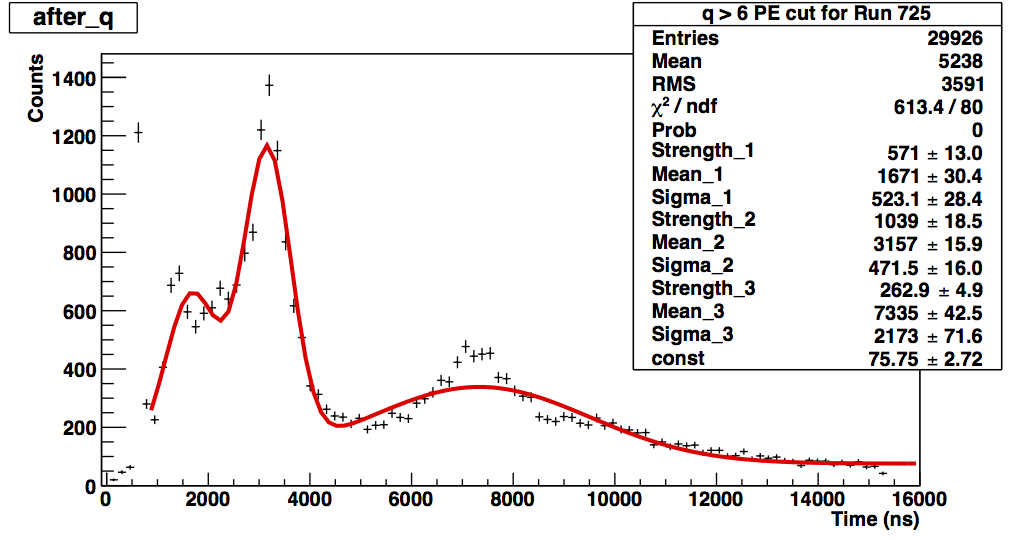}
\caption{Scatter plot of timing and charge of the pulse.}
\label{fig:afterfit}
\end{figure}
\begin{figure}[htbp]
\centering
\graphicspath{{./fig/LED_data/}}
\includegraphics[scale=0.6]{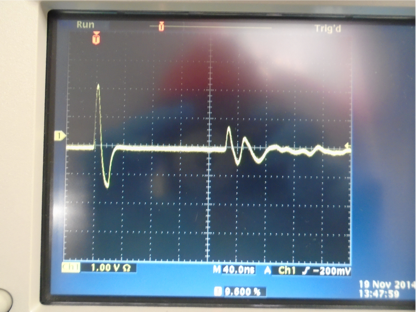}
\caption{Snap shot of the scope. The first sharp pulse is the driven signal for LED followed by the reflection and delay in the coaxial cable.}
\label{fig:LEDreflection}
\end{figure}
\subsection{TPB delayed light}
It has been found that the TPB response function to scintillation light has more complex structure with a delayed component(Segreto \cite{PhysRevC.91.035503}).  The functional form of no-exponential decay can be expressed\cite{doi:10.1080/15421406808082905}:
\begin{ceqn}\begin{align}\label{eq:tpb}
I(t)_{delayed} \simeq \eta_s \frac{N}{[1+A\cdot ln(1+t/t_a)]^2\cdot(1+t/t_a)}
\end{align}\end{ceqn}
where $N$ and $A$ are constants depending on the nature of the scintillator, $t_a$ is a relaxation time that is linked to the diffusion coefficient of triplet states in the scintillator, $\eta_s$ is the fluorescence yield.\par
In the LED data, the summed waveform  of blue and UV LED exhibit different shapes as shown in Fig. \ref{fig:LED_blue_uv_delayed}. This shows that the UV LED has extra component after the prompt peak. The UV photons are shifted by the TPB while it is transparent for the blue photons. Therefore, using the UV LED, the characteristic timing structure of TPB fluorescence can be determined by UV LED. To exclude any electronic effect, the pulse-time distribution of blue LED is fitted with  gaussian distribution  plus a exponential as shown in Fig. \ref{fig:blue_fit}. This shows no sign of extra components other than the exponential decay which exclude the possibility that any decay components in UV LED comes from electronic effect. The pulse-time distribution of UV LED is then fitted to Eq. \ref{eq:tpb} with the prompt peak fitted to a gaussian distribution as shown in Fig. \ref{fig:uv_fit}. The results shows the delayed light is about 5\% of the peak light. The surface alpha induced TPB scintillation(see \ref{sec:sa} ) in the vacuum can be used to cross check the results from UV LED. Figure \ref{fig:uv_alpha_fit} shows the fitting results using Eq. \ref{eq:tpb} of the summed waveform from UV LED and alpha-TPB scintillation. The fitting results seems agree with each other although the $\chi^2$/ndf seems too small which is because the wrong error bar is assigned to each bins. The parameter A (0.225) is in agreement with the results from Segreto(0.22), but the $t_a$ (11 ns) which is the relaxation time of the triplet states in the scintillator is almost factor of 5 smaller than the results from Segreto(50 ns). Moreover, the contribution of TPB-delayed light determined by UV LED is about 5\% which is also lower than Segreto's result (40\%). However, rather than using quenched scintillation light which used by Segreto, the UV LED provide more direct source to measure the TPB-delayed light and has less systematic errors.
\begin{figure}[htbp]
\centering
\graphicspath{{./fig/LED_data/}}
\includegraphics[scale=0.25]{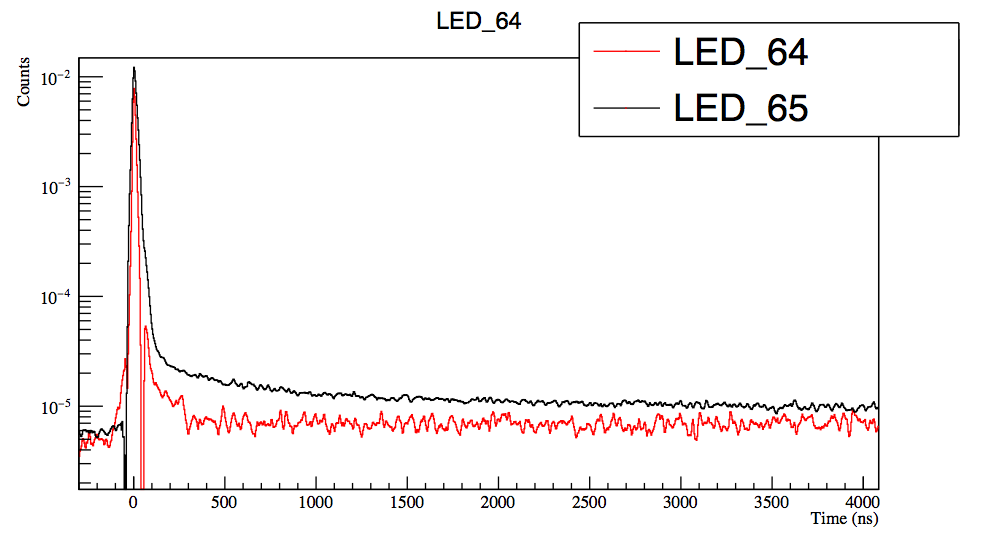}
\caption{Summed waveform of blue and UV LED. The LED 64 is blue LED and LED 65 is UV LED. }
\label{fig:LED_blue_uv_delayed}
\end{figure}
\begin{figure}[htbp]
\centering
\graphicspath{{./fig/LED_data/}}
\includegraphics[scale=0.25]{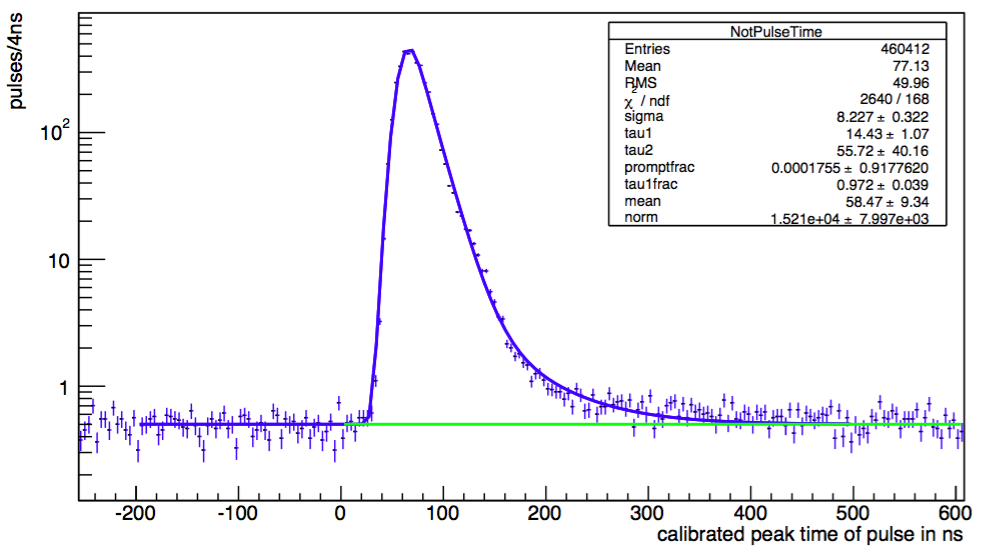}
\caption{The pulse-time distribution of UV LED. The prompt peak is fitted to a gaussian distribution and an exponential. The green line is the baseline taken from very late-time (14000 ns - 15000 ns) }
\label{fig:blue_fit}
\end{figure}
\begin{figure}[htbp]
\centering
\graphicspath{{./fig/LED_data/}}
\includegraphics[scale=0.25]{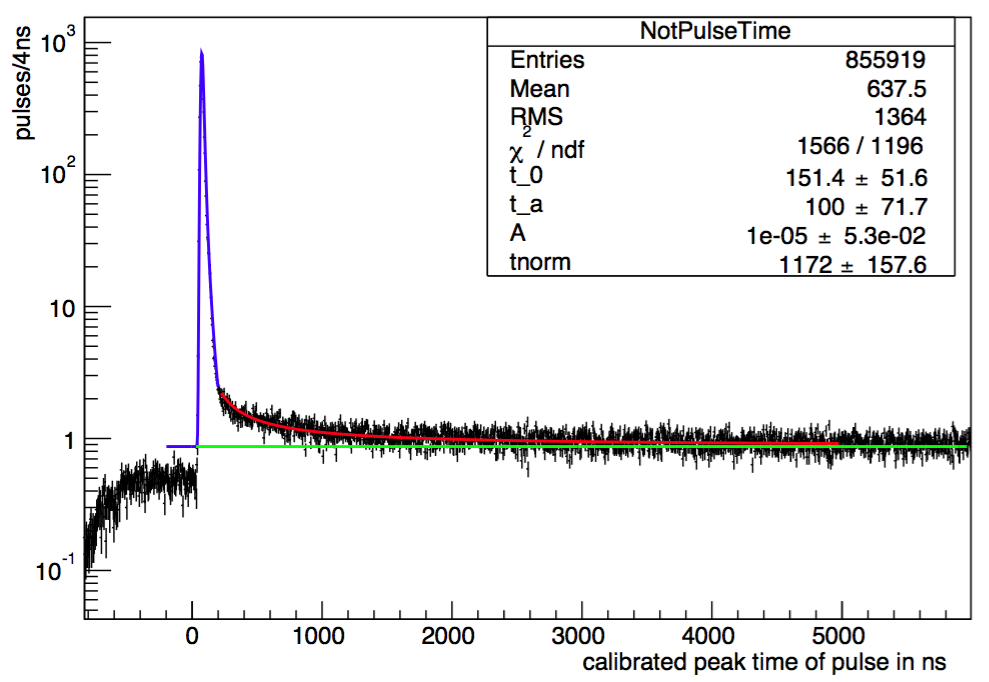}
\caption{The pulse-time distribution of UV LED. The prompt peak is fitted to a gaussian distribution and the decay component is fitted to Eq. \ref{eq:tpb}. The green line is the baseline taken from very late-time (14000 ns - 15000 ns) }
\label{fig:uv_fit}
\end{figure}

\begin{figure}[htbp]
\centering
\graphicspath{{./fig/LED_data/}}
\includegraphics[scale=0.25]{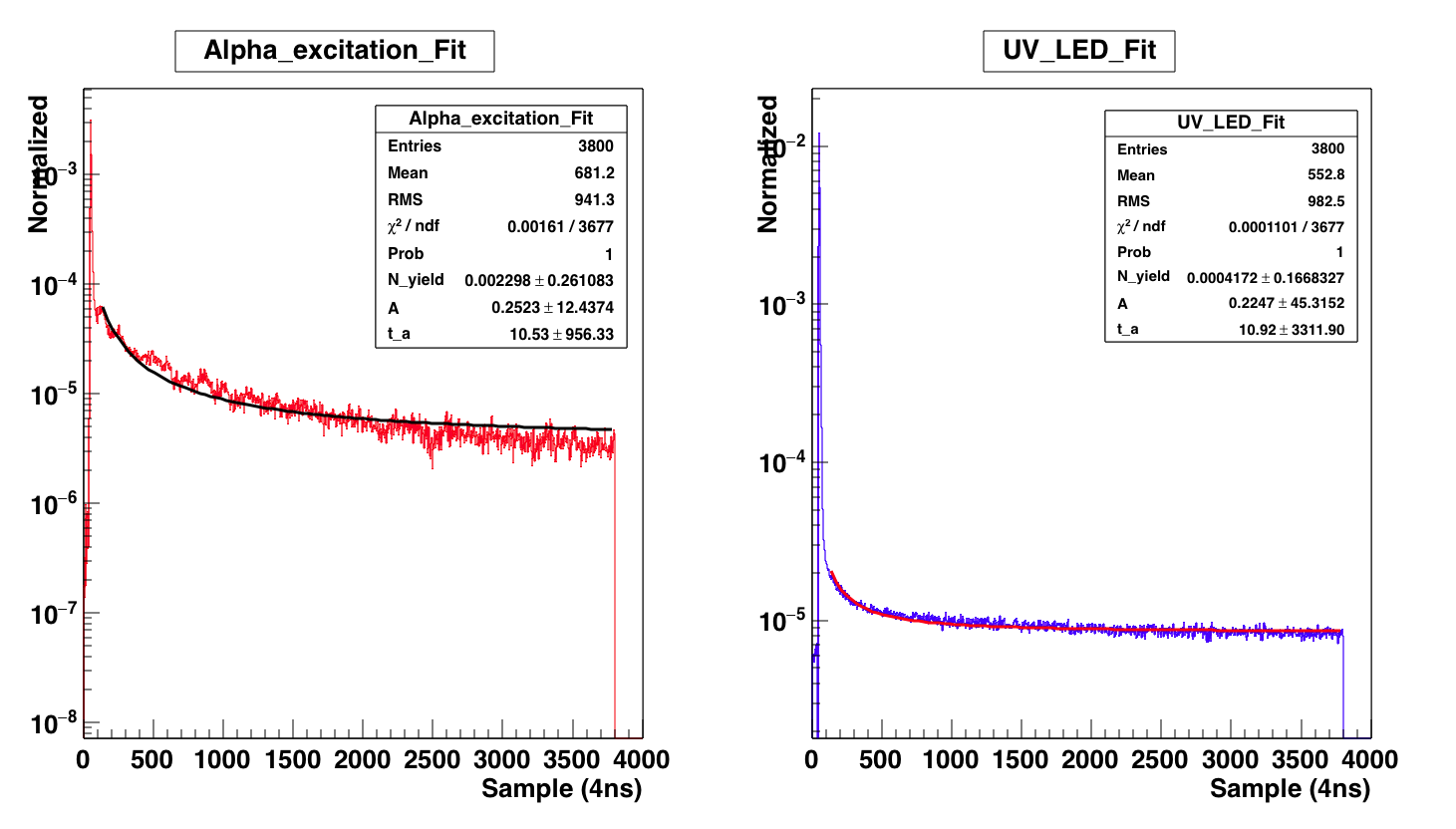}
\caption{The summed waveform of UV LED (Right) and Alpha-TPB scintillation (Left). The prompt peak is fitted to a gaussian distribution and the decay component is fitted to Eq. \ref{eq:tpb}.} 
\label{fig:uv_alpha_fit}
\end{figure}
\section{LED in Cold Gas}\label{sec:ledcoldgas}
The MiniCLEAN detector reaches 120 K in Oct. 2016. The preliminary data taking was performed to check the status of each detector components. The detail description of MiniCLEAN cold gas run will be described in Chapter \ref{ch:gasrun}. The LEDs were turned on to test the functionality. Unfortunately, from the connectivity test, only three LEDs has measured the reasonable impedance. All LEDs are still turned on to see if the pulse is seen in the data. However, no LED pulse is seen in the data. Figure \ref{fig:centroidcoldled} shows the reconstruction position from centroid fit in cold gas and compare with the results from simulation. The reconstructed LED events should concentrate at the opposite position of LED. Instead in the cold gas data, the reconstructed position is deviated from where it should be. Moreover, Fig. \ref{fig:LEDradius} shows the normalized counts as a function of reconstructed radius normalized to the TPB radius which is the maximum radius of active volume and compare with the results from previous vacuum data. The expected curve should concentrate at maximum radius of active volume, in vacuum data, due to some photons bounce off TPB surface or the baffles, broaden the peak. However, the results from cold gas seems more flat throughout the active volume, this indicates some sort of random noise was triggered over the LED events.\par
Looking into the individual waveform, the triggered pulse is shown in Fig. \ref{fig:LEDwaveformcold}. This shows that no physical photons were seen by PMT at trigger time which confirmed the results from connectivity test. Plotting the events as function of the angle between the LED and PMT, the flat distribution were seen for LED events in cold gas as shown in Fig. \ref{fig:LEDcoldspatial}. For LED events in vacuum data, a clear peak was seen at angle close to 0 degrees which indicates the direct opposite PMT received most of the LED photons. Possible reason for losing the connection with LEDs may be due to the incident which happened while cooling the detector. The incident causing the detector warm up more than 120 K in couple hours and might results in the loosen connector. The LED based \textit{In-situ} optical calibration might not be available after the commissioning of the detector, but the other method using the late scintillation light can be used to determine the PMT gain on daily basis. The method will be described in Chapter \ref{ch:gasrun}.

\begin{figure}[htbp]
\hfill
\subfloat[]{\includegraphics[width=7cm]{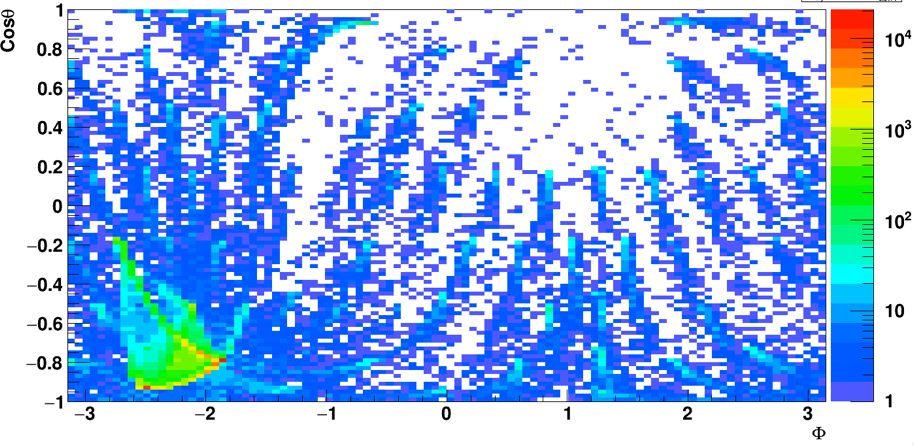}}
\hfill
\subfloat[]{\includegraphics[width=7cm]{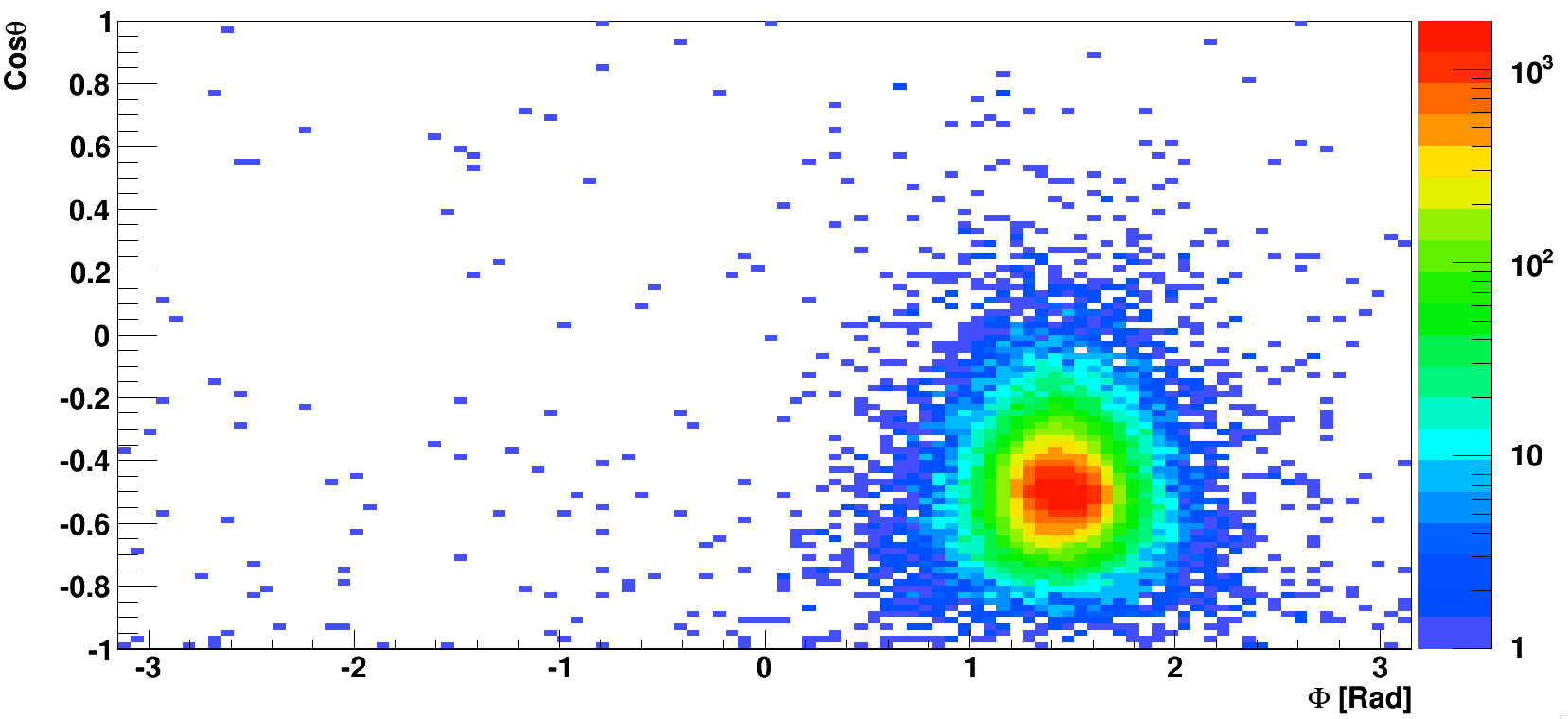}}
\hfill
\subfloat[]{\includegraphics[width=7cm]{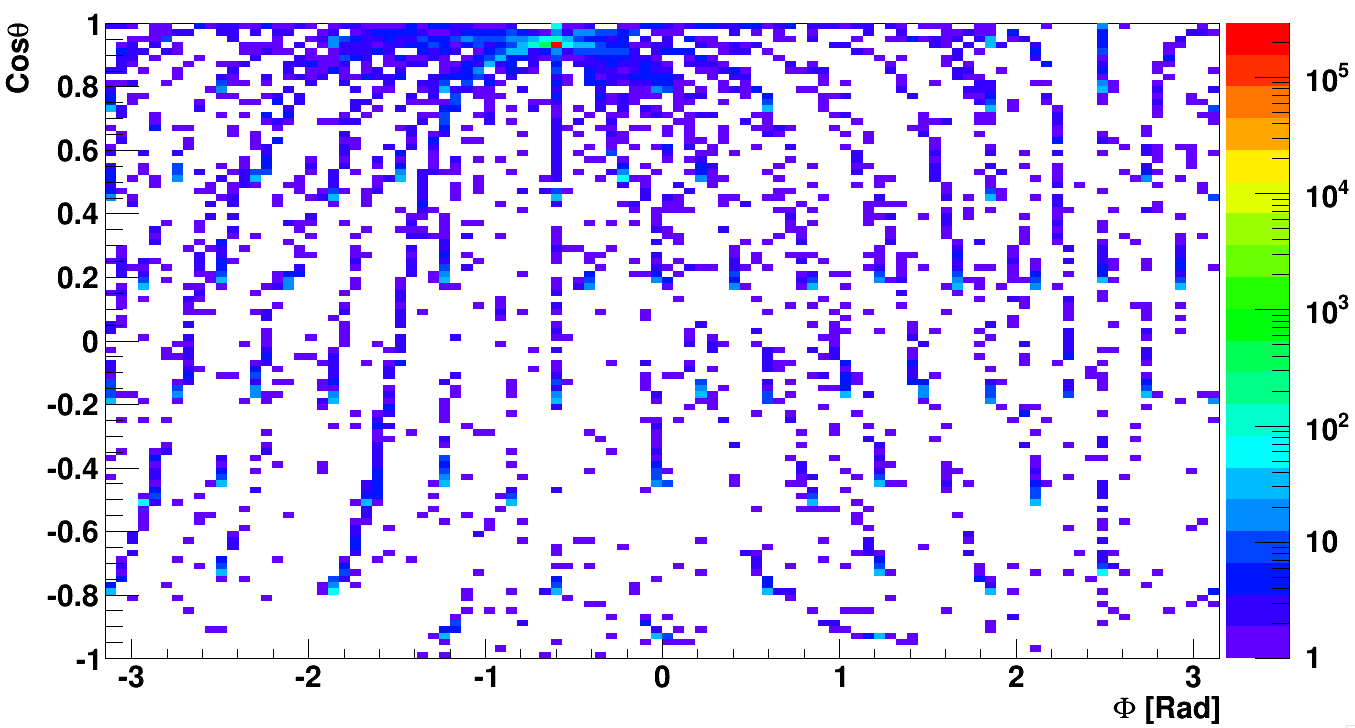}}
\hfill
\subfloat[]{\includegraphics[width=7cm]{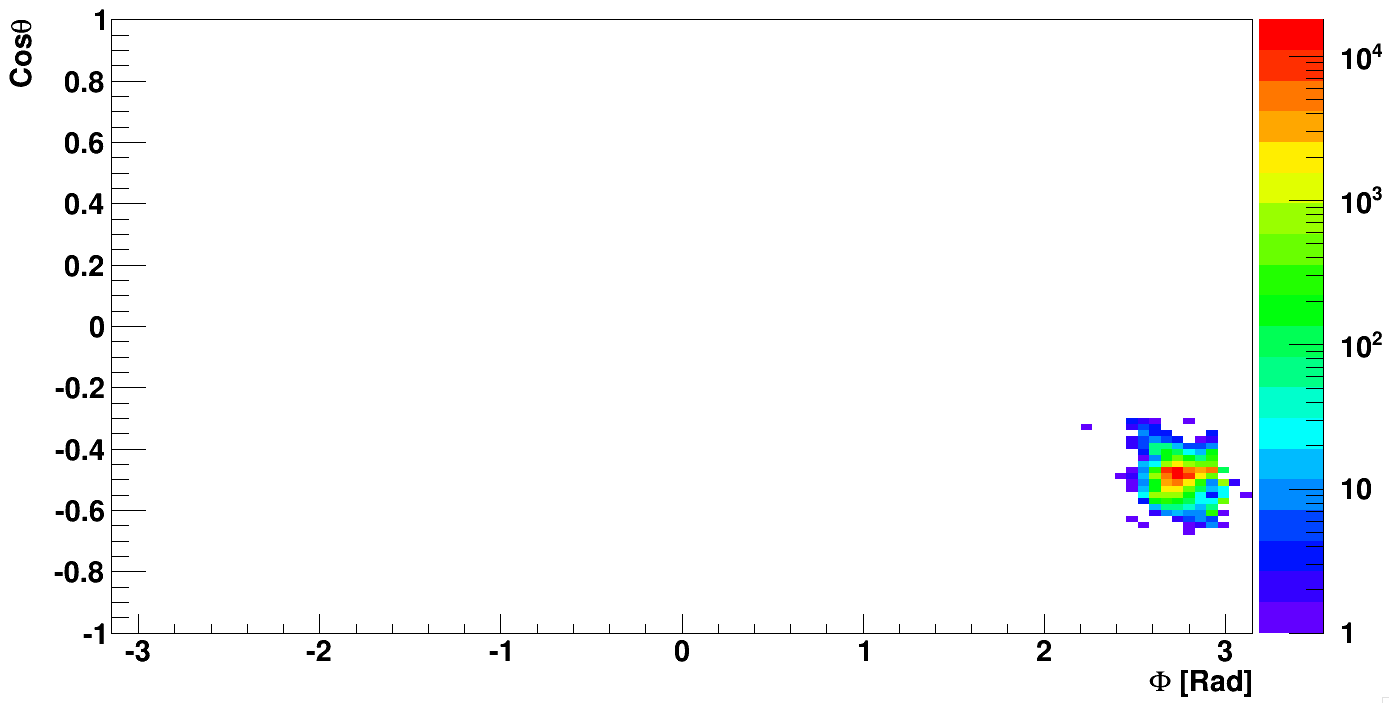}}
\hfill
\caption{ Reconstructed angular position of events (a) Blue (cold gas data) (b) Blue (simulation)  (c) UV (cold gas data)  (d) UV (simulation)  }
\label{fig:centroidcoldled}
\end{figure}
\begin{figure}[htbp]
\centering
\graphicspath{{./fig/LED_data/}}
\includegraphics[scale=0.4]{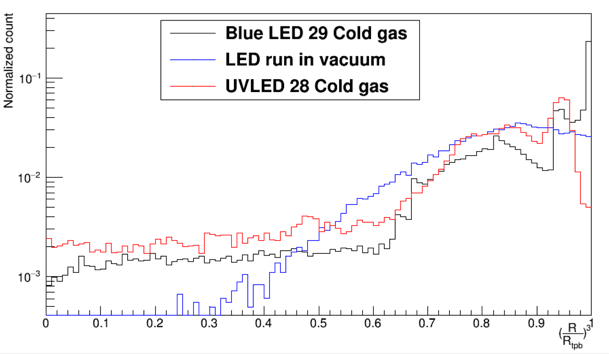}
\caption{ The normalized count as a function of reconstructed radius. Note that the X-axis is $(\frac{R}{R_{tpb}})^3$, for random isotropic events, the curve should be flat.}
\label{fig:LEDradius}
\end{figure}
\begin{figure}[htbp]
\centering
\graphicspath{{./fig/LED_data/}}
\includegraphics[scale=0.4]{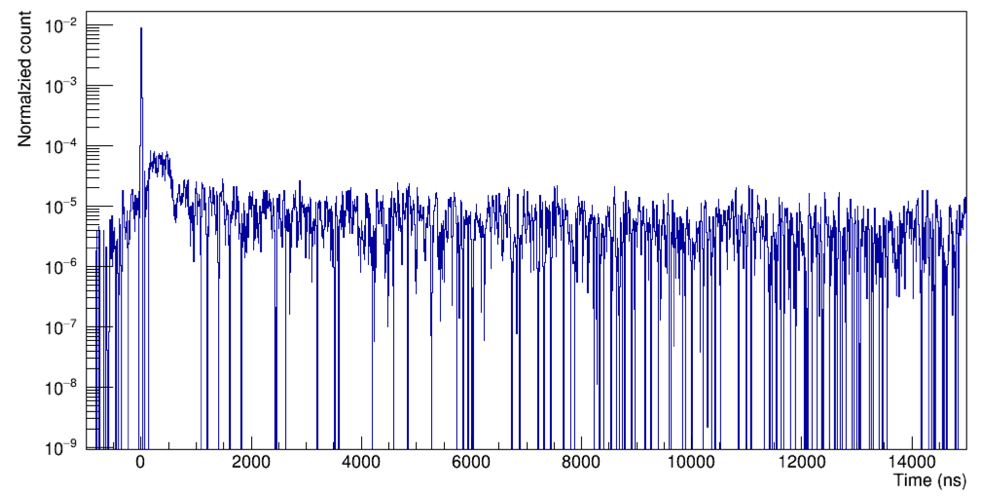}
\caption{ The normalized summed waveform as a function of time of LED events.}
\label{fig:LEDwaveformcold}
\end{figure}
\begin{figure}[htbp]
\centering
\graphicspath{{./fig/LED_data/}}
\includegraphics[scale=0.4]{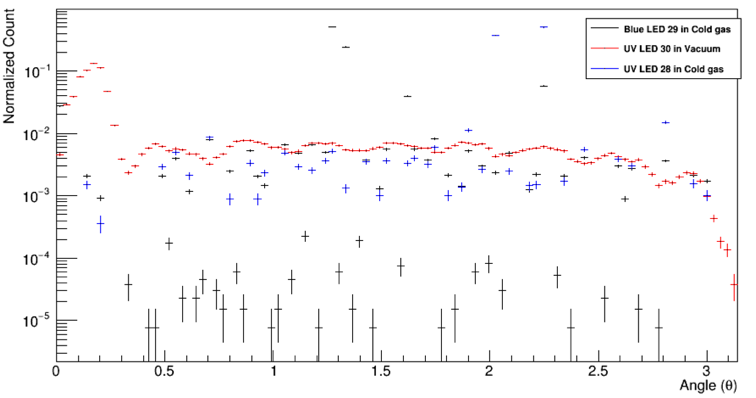}
\caption{ Spatial distribution of LED events in cold gas.}
\label{fig:LEDcoldspatial}
\end{figure}

\chapter{Vacuum Background Study}\label{ch:vacuum}
Minimizing the effects from background events is important for WIMPs search and demonstration of the PSD rejection ability. For background events which can fake the WIMP's signal the discrimination method is needed. Gamma and beta particles can create the indistinguishable signal of electron recoil and affect the energy resolution determined by $^{39}$Ar energy spectrum fitting. Therefore, it is important to understand the origin of the background such that it can be either eliminate from the data or quantified to be include in the data analysis. The detail of background events of MiniCLEAN has been discussed in \cite{sjaditz}. In this chapter, the strategy to eliminate the background events in the MiniCLEAN's preliminary vacuum data is discussed in detail.
\section{Surface alpha}\label{sec:sa}
The vacuum ``Golden run'' was taken after the fine tuning of the system to test the functionality of each system and before moving into OV. This data run was taken while IV is under the vacuum. The major backgrounds are from the external gamma which is the product of cosmic muons interact with the underground rock. In addition, the radon daughter  deposit on the TPB release the alpha particle in the process to decay to more stable elements. Table \ref{table:sumbkg} summarize the source and the estimated surviving rate after the cuts in LAr of the $\gamma/\alpha$ particles.\par

\begin{table}[htbp]
\caption{Summary of background sources for MiniCLEAN and their reduction via energy, fiducial volume, Fp, and $F_{\alpha}$ cuts, derived from simulation and tabulated in the internal document\cite{backgroundsum}} 
\centering 
\begin{adjustbox}{width=1\textwidth}
\begin{tabular}{c c c c c } 
\hline\hline 
Background Source&Raw Rate & 12.5-25 keV$_{ee}$ & R<29.5 cm& $F_p,F_{\alpha},L_{r}$\\ [0.5ex] 
\hline 
Intrinsic $^39$Ar & 1 Bq/kg & 4.2 $\times$ 10$^8$ & 1.2 $\times$10$^8$ & <1\\
$\alpha$ in Acrylic & 24,000/year & 284$\pm$4 & 0.02$\pm$0.01 & < 1$\times$ 10$^{-4}$\\
$\alpha$ at Acrylic-TPB interface & 10,000/year & 1.0$\pm$0.5 & < < 1 & 1$\times$ 10$^{-4}$\\
$\alpha$ in TPB & 1,000/year & 75$\pm$3 & 0.007$\pm$0.003 & 1$\times$ 10$^{-4}$\\
$\alpha$ at TPB-Ar interface & 10,000/year & 3000$\pm$5 & 0.82$\pm$0.09 & 0.24$\pm$0.05\\
PMT($\alpha$,n) & 42,000/year & 352.2$\pm$2.1 & 91.6$\pm$1.1 & 3.8$\pm$0.02\\
Steel ($alpha$,n) & 1,840/year & & & <0.2\\
$\gamma$ from PMTs & 20 $\times$ 10$^9$/year & 6$\times$10$^6$ & 3$\times$10$^5$ & < 0.08\\
$gamma$ from steel & 9$\times$10$^9$/year & 2$\times$10$^6$ & 1$\times$10$^5$ & < 0.02\\
$\gamma$-e Cherenkov & 29$\times$10$^9$/year & 3,500 & < 0.1 & < 0.1\\
Cosmogenic and wall n &3,650/year & 0.08$\pm$0.01& & \\ 
\hline\hline
\end{tabular}
\end{adjustbox}

\label{table:sumbkg} 
\end{table}
The radon daughter plate out onto the TPB surface during TPB-deposition, acrylic polishing or detector assembly creates alpha decay in the detector volume. Depending on the location that the radon daughter plated out, will have different contribution to the background. Figure. \ref{fig:radon_decay_cartoon} shows a cartoon describes the decay route of alpha particle at different locations. When alpha particle interact with the TPB, the scintillation is emitted as the results of the $\alpha$-TPB interaction\cite{POLLMANN2011127}. Figure \ref{fig:radon_decay_chain} shows the decay-chain of radon, the major source of the surface alpha comes from 
\begin{ceqn}\begin{align}
^{210}Po \rightarrow ^{206}Pb + \alpha (5.33 MeV)
\end{align}\end{ceqn}
As indicated by Fig. \ref{fig:radon_decay_cartoon}, this process can either inject the alpha particle into the acrylic and nucleus of polonium to the argon volume or vise versa. In vacuum data, the only interaction can be observed is when alpha particle pass through the TPB and producing the scintillation light. The surface alpha and the nucleus of polonium interact with argon gas will be described in Chapter \ref{ch:gasrun}.\par

\begin{figure}[htbp]
\centering
\graphicspath{{./fig/Vacuum/}}
\includegraphics[scale=0.4]{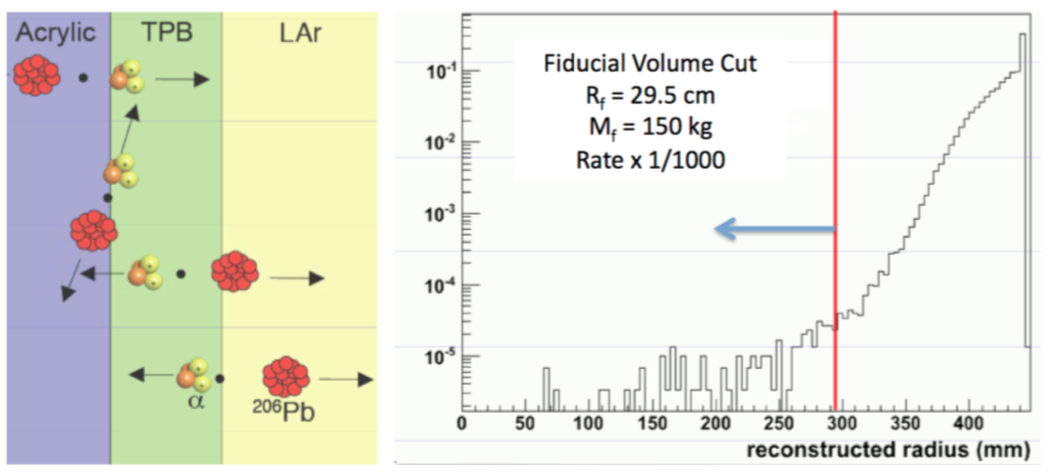}
\caption{ (Left) The decay process of $^{210}$Po, the black dot represent the alpha particle along with the decay product $^{206}$Pb. (Right) The fiducial volume cut on surface events.}
\label{fig:radon_decay_cartoon}
\end{figure}

\begin{figure}[htbp]
\centering
\graphicspath{{./fig/Vacuum/}}
\includegraphics[scale=0.5]{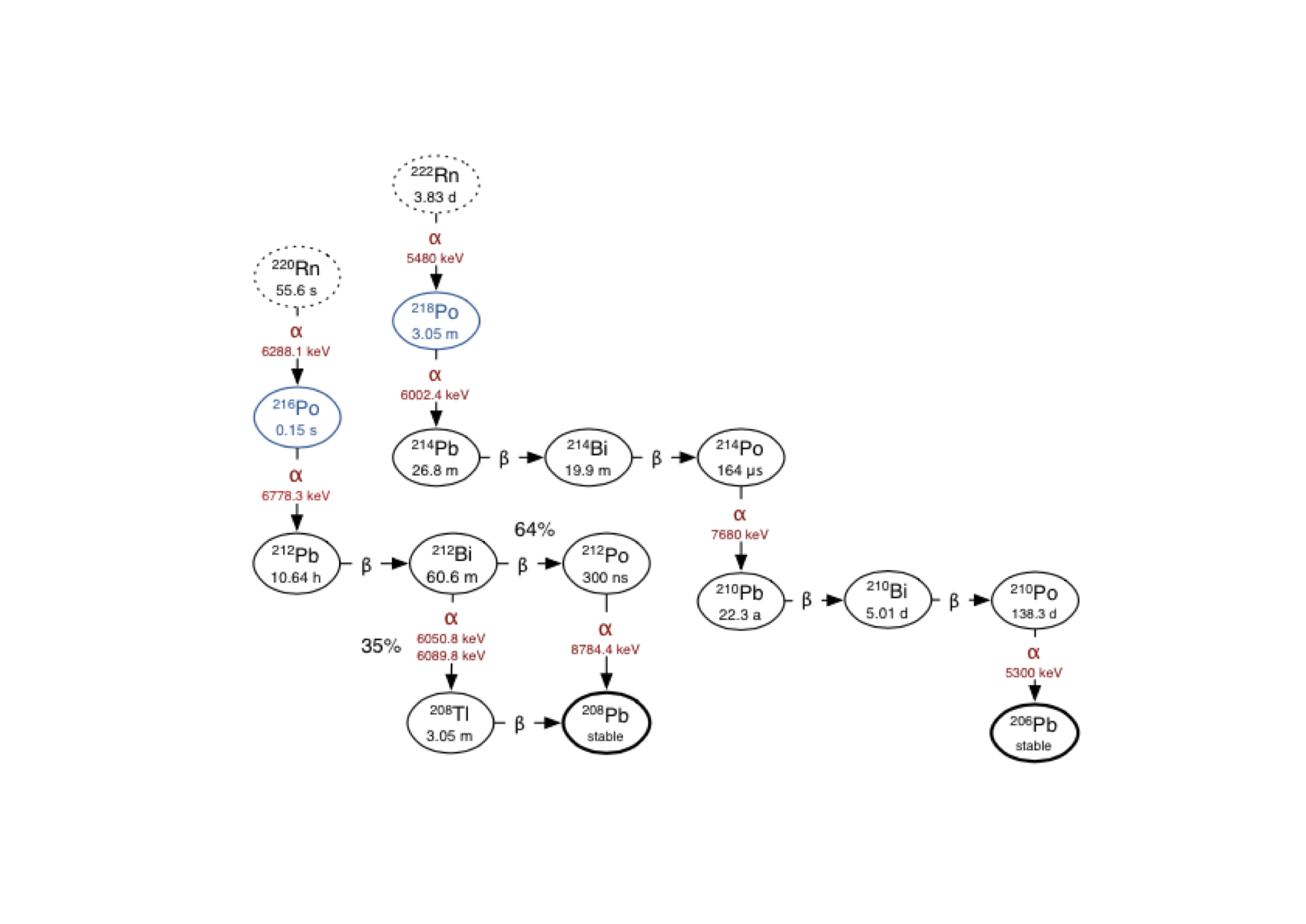}
\caption{ Radon decay chain.}
\label{fig:radon_decay_chain}
\end{figure}
%
The alpha-TPB interaction takes place in front of the PMT, thus ideally one PMT will get the major fraction of the charge of the event. However, the scintillation light could be shifted further by TPB and re-emitted isotropically. Thus the charge ratio is not expected to be high. Moreover, the characteristic decay time ($\sim$ 280 ns at room temperature and $\sim$ 800 ns at LAr temperature\cite{1748-0221-11-06-P06003}) push the Fprompt to lower region. Figure \ref{fig:alpha2d} shows the Fprompt-charge scatter plot, notice that the band at around fp $\sim$ 0.4 extend to high energy is from the alpha-TPB scintillation. 
\begin{figure}[htbp]
\centering
\graphicspath{{./fig/Vacuum/}}
\includegraphics[scale=0.4]{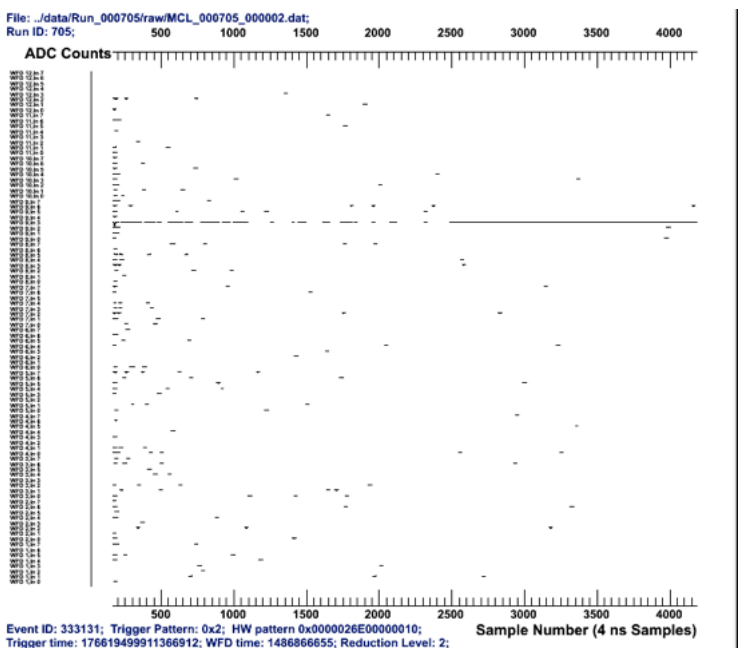}
\caption{ Alpha-TPB scintillation events in CLEANViewer.}
\label{fig:alphaclean}
\end{figure}

\begin{figure}[htbp]
\centering
\graphicspath{{./fig/Vacuum/}}
\includegraphics[scale=0.3]{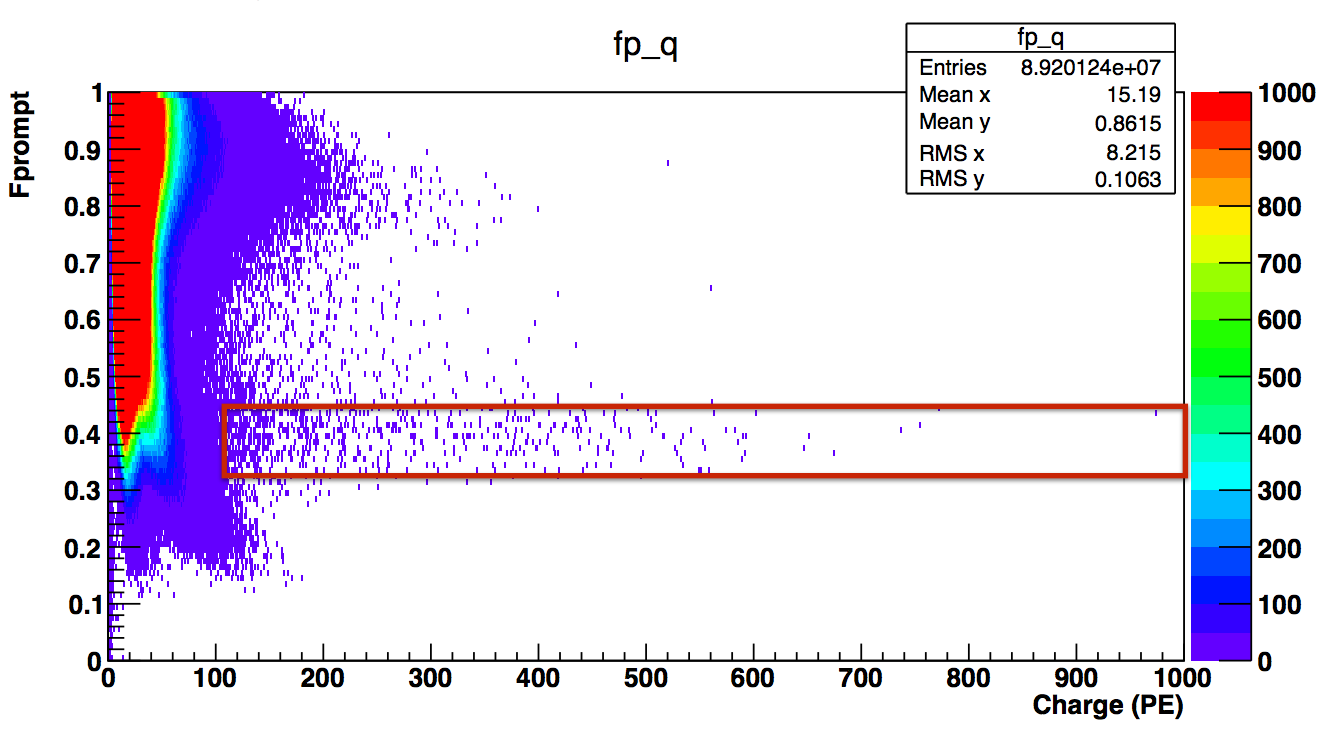}
\caption{ Fprompt vs charge distribution. Noticed that the band circled by red box is from alpha-TPB scintillation.}
\label{fig:alpha2d}
\end{figure}
To determine the rate of alpha-TPB interaction a cut need to be made to select the events. Projecting the Fig. \ref{fig:alpha2d} to Y(Fprompt) axis and fitted with Gaussian, the mean and $\sigma$ is obtained. The cut is defined as fp = mean $\pm$ 5$\sigma$ and charge > 100 PE. The expected rate of alpha-TPB interaction from Table \ref{table:sumbkg} is 6.8 event/hour which is consistent with the results from vacuum data (6.86 $\pm$ 2.80 events/hour with $\sim$ 55 hours of live data) . This indicates that the MiniCLEAN detector did not introduce unexpected radon daughters into the IV.


\section{Cherenkov Light in Acrylic}
When the charged particle travel in the medium with speed faster than the speed of light in the medium, the electromagnetic waves might be created which is called Cherenkov radiation. In MiniCLEAN detector, when the gamma particle moving through the acrylic the electrons will be produced through the Compton scattering.  The acrylic is polarized by the electromagnetic filed of the moving electrons. The Cherenkov light is emitted whenever the velocity of the moving electrons exceeds the speed of light in the medium or :
\begin{ceqn}\begin{align}\label{eq:cherenkovn}
\beta\cdot n > 1
\end{align}\end{ceqn}
where n is the refractive index of acrylic ($\sim$ 1.49) and $\beta$ is the ratio of the velocity of electrons to that of light in the vacuum. As indicated by Eq. \ref{eq:cherenkovn}, there is a minimum required velocity of electrons to produce the Cherenkov light. In terms of energy of moving electrons, a  threshold energy is given by 
\begin{ceqn}\begin{align}
E_{th}=m_0c^2\left(-1+\sqrt{1+\frac{1}{n^2-1}} \right)
\end{align}\end{ceqn}
where $m_0c^2$ is the electron rest-mass energy (0.511 MeV). Figure \ref{fig:chrenkovthreshold} (a) plots the threshold energy of electron and the minimum energy of gamma which can produce the Compton electrons as a function of refractive index. The time takes the electrons to slow from its initial velocity to below the threshold velocity is very short, typically of the order of picoseconds. Therefore the Fprompt of Cherenkov events should be near 1 (some electric noise and after-pulsing increases the late charge) as shown in Fig. \ref{fig:chrenkovfpq}. The photon yield from Cherenkov radiation is low therefore in the MiniCLEAN vacuum data, it can be used to do the SPE calibration (see \ref{sec:SPE}). The calculated yield of Cherenkov photons in the 300 to 600 nm wavelength region is shown in Fig.\ref{fig:chrenkovthreshold} (b). Consider the detector efficiency, each PMT mostly received single photon from Cherenkov events. Moreover, the scintillation light from LAr is emitted isotropically, for Cherenkov photons, however, are emitted preferentially along the direction of the electron velocity. Figure \ref{fig:chrenkovtravel} shows the direction of Cherenkov photons relative to the direction of moving electrons.\par
\begin{figure}[htbp]
\hfill
\subfloat[]{\includegraphics[width=5cm]{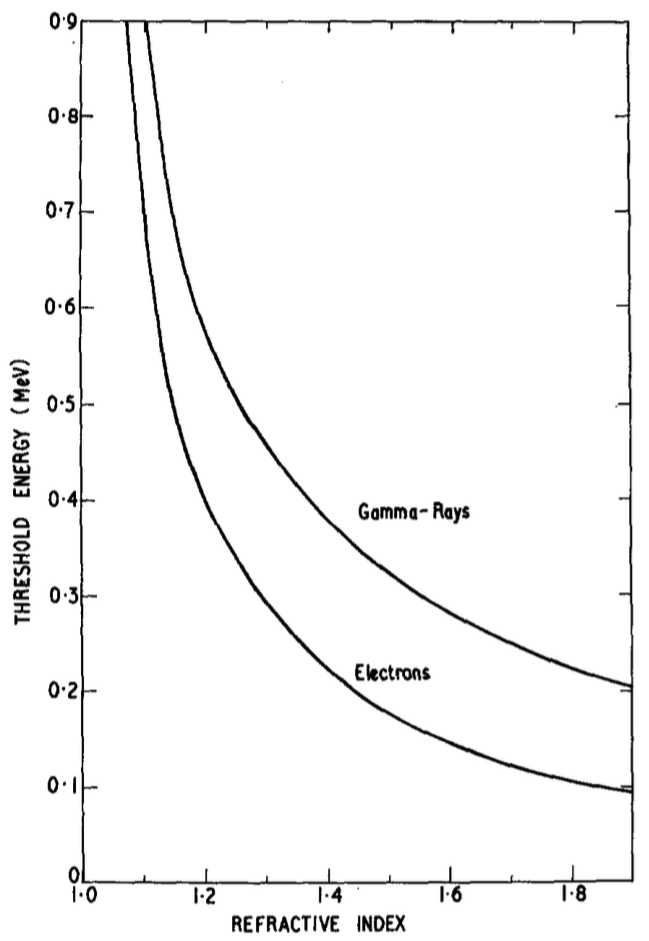}}
\hfill
\subfloat[]{\includegraphics[width=5cm]{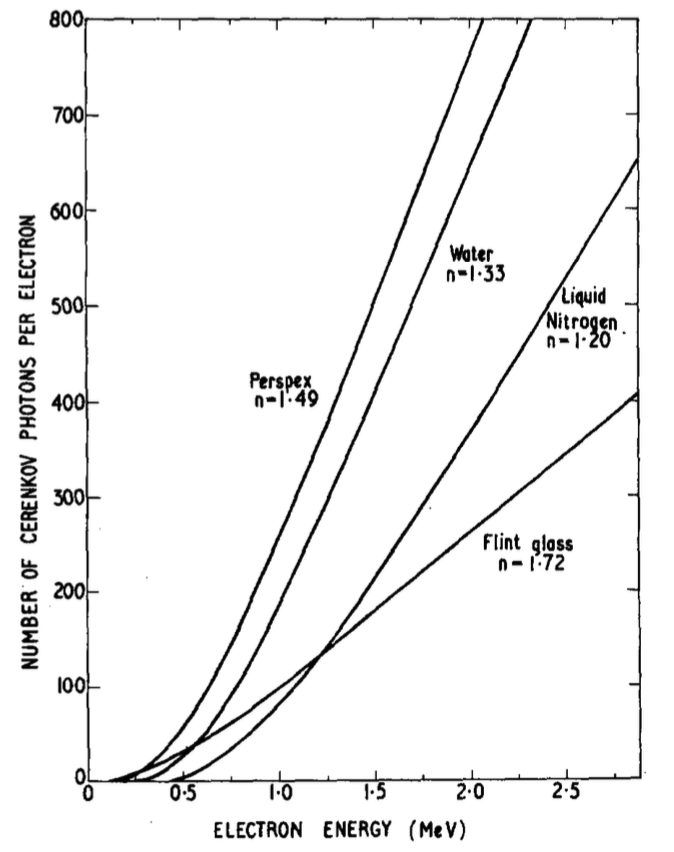}}
\hfill
\caption{ (A) The threshold energy can produce the Cherenkov light as a function of refractive index. Curves shows both for electrons and gamma particle that can produce the Compton electrons through 180\si{\degree} Compton scattering\cite{SOWERBY1971145}. (b)Calculated yield of Cherenkov photons in the 300-600 nm wavelength region for different meida\cite{SOWERBY1971145}.}
\label{fig:chrenkovthreshold}
\end{figure}

\begin{figure}[htbp]
\centering
\graphicspath{{./fig/Vacuum/}}
\includegraphics[scale=0.25]{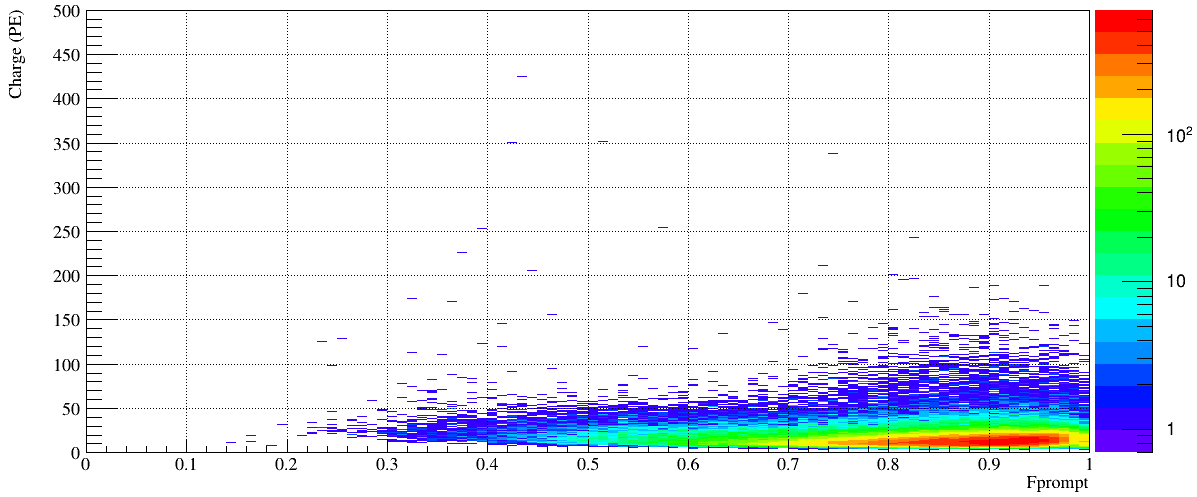}
\caption{ Charge vs Fprompt. Noticed the group of events near the high Fprompt and low charge region is from Cherenkov events.}
\label{fig:chrenkovfpq}
\end{figure}

\begin{figure}[htbp]
\centering
\graphicspath{{./fig/Vacuum/}}
\includegraphics[scale=0.3]{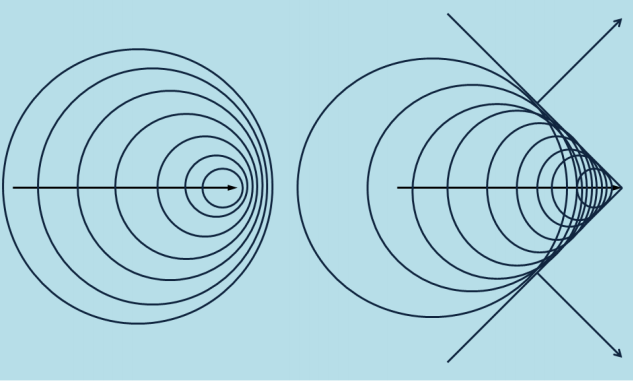}
\caption{ Left : The energy of moving electron does not pass the threshold, thus no Cherenkov radiation is emitted. Right : The energy of moving electrons pass the threshold and emit the Cherenkov radiation. The direction is indicated by two arrows from wave front.}
\label{fig:chrenkovtravel}
\end{figure}
Several cut can be made to eliminate the Cherenkov events from the data. A Monte Carlo simulation of WIMPs events (3 M) in the MiniCLEN detector is carried out to understand the cut efficiency. The first cut is the charge cut, from previous discussion the light yield of Cherenkov events is low and the detector threshold is 75 PE assuming the target light yield (6 PE/keV) is achieved with MiniCLEAN detector. The second cut is from the Fprompt parameters, the WIMPs-nuclear recoil produce the Fprompt centered at 0.7, while for Cherenkov events the Fprompt is near 1 due to the very fast light emission. The third cut is from the charge ratio, the Cherenkov events usually produced inside the acrylic thus the PMT which just behind the acrylic will received most of the photons. On the other hand, the scintillation events are created inside the active volume and emitted light isotropically. Moreover, the TPB re-emission is also isotropic in both forward and backward direction. Thus no one PMT should get large fraction of the scintillation light. The fourth has been defined is the polarization of Cherenkov light. As mentioned in the last paragraph, the Cherenkov light has preferential direction in contrast of scintillation light. The concept from moment of inertia of classic dynamics can be used to calculate the spreads of the light of the Cherenkov events. Assuming the position vector  of PMT is 
\begin{ceqn}\begin{align}
\textbf{r}_i = x_i\textbf{i} + y_i\textbf{j} + z_i\textbf{k}
\end{align}\end{ceqn}
The charge ($Q_i$ where i is the i-th PMT) of each PMT received in the Cherenkov event is known, the moment of inertia can be written as 
\begin{ceqn}\begin{align}
I_{xQy} =  \sum\limits_{i} Q_i z_i^2, \;\; I_{yQz} =  \sum\limits_{i} Q_i x_i^2, \;\; I_{zQx} =  \sum\limits_{i} Q_i y_i^2,
\end{align}\end{ceqn}
The moment of inertia of the system about $x$, $y$ and $z$ axes are
\begin{align}
I_{xx} = \sum\limits_{i} Q_i \left( y_i^2 + z_i^2 \right), &\\
I_{yy} = \sum\limits_{i} Q_i \left( z_i^2 + x_i^2 \right), &\\
I_{zz} = \sum\limits_{i} Q_i \left( x_i^2 + y_i^2 \right), 
\end{align}
Solving the above equations, the moment of inertia for both Cherenkov events and WIMPs-nuclei recoil events can be obtained with respect to $x$, $y$ and $z$. Figure \ref{fig:chrenkovmoment} shows the moment of inertia of $x$, $y$ and $z$ for both type of events. As can be seen in this plot, the moment of inertia of WIMPs-nuclei recoil are around 0.6 to 0.8, which means the light is emitted isotropically and no preferential direction was found. On the other hand, the moment of inertia of Cherenkov events exhibit the expected behavior. For x and y direction, the moment of inertia is large but for z direction, the moment of inertia is small. This shows the Cherenkov events emit the photons in the preferential direction and the shape looks like the ``pen'' such that one direction has smaller moment of inertia. The last cut is the fiducial radius cut. The fiducial volume determined from Monte Carlo simulation indicates within radius of 295 mm, the volume is free of backgrounds. Figure \ref{fig:cutplot} shows the plots of charge distribution, Fprompt distribution and charge ratio distribution of both Cherenkov events and WIMP-nuclei recoil events. The cut efficiency at various stage of the cut process is summarized in Table \ref{table:cutsum}
\begin{figure}[htbp]
\centering
\graphicspath{{./fig/Vacuum/}}
\includegraphics[scale=0.2]{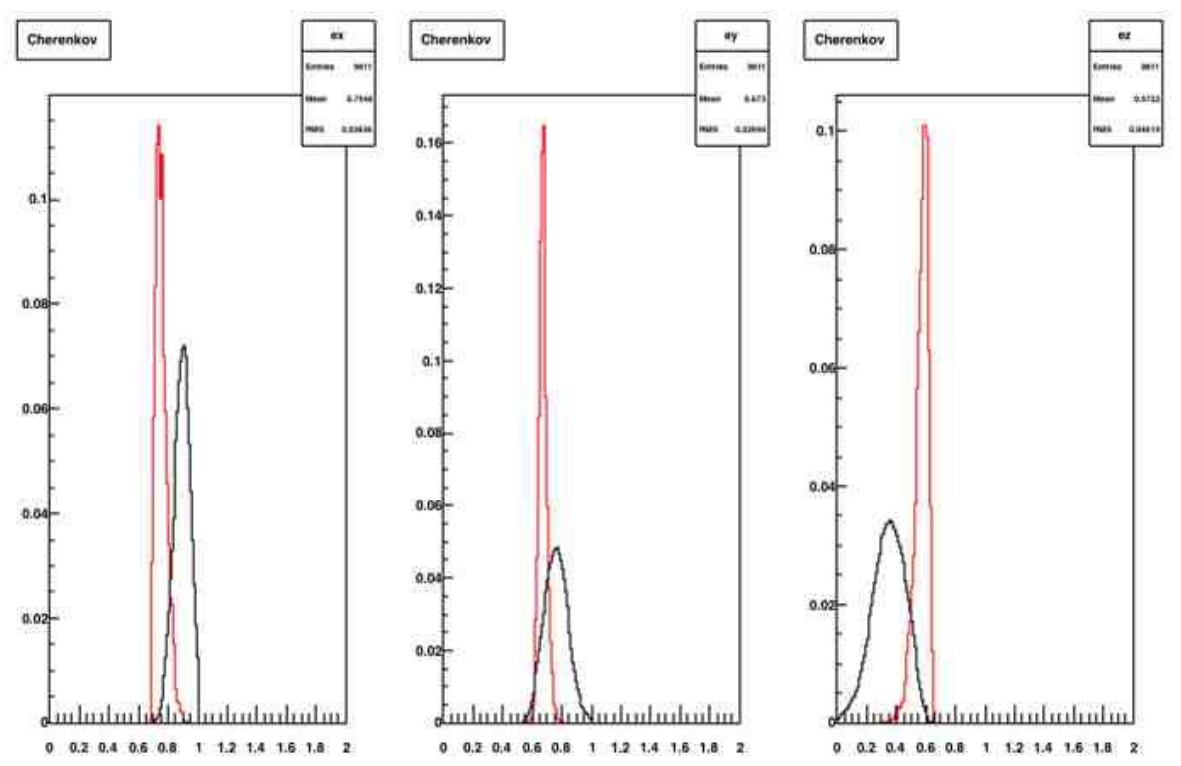}
\caption{The moment of inertia for x, y and z (from left to right). The red curve is from the WIMPs-nuclei recoil and black curve is from Cherenkov events.}
\label{fig:chrenkovmoment}
\end{figure}
\begin{figure}[htbp]
\hfill
\subfloat[]{\includegraphics[width=7cm]{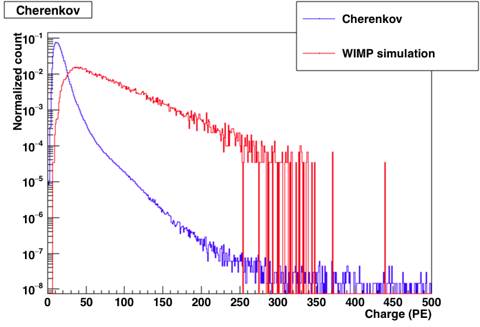}}
\hfill
\subfloat[]{\includegraphics[width=7cm]{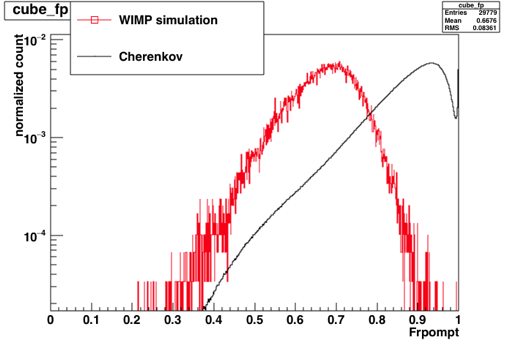}}
\hfill
\subfloat[]{\includegraphics[width=7cm]{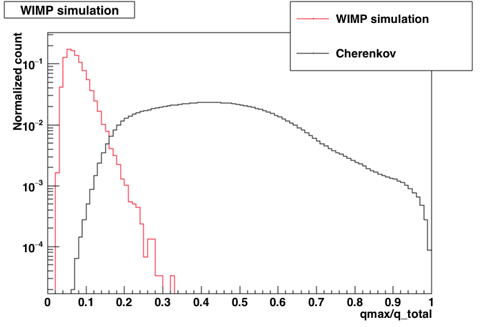}}
\hfill
\subfloat[]{\includegraphics[width=7cm]{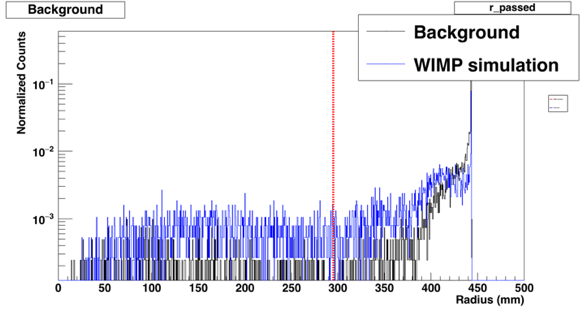}}
\hfill
\caption{ Comparing various distribution between Cherenkov events and WIMPs-nuclei recoil. (a) Charge distribution. (b) Fprompt distribution. (c) Charge ratio distribution. (d) Radius distribution.}
\label{fig:cutplot}
\end{figure}

\begin{table}[htbp]
\caption{Summary of background and signal events survived after the cuts} 
\centering 
\begin{adjustbox}{width=1\textwidth}
\begin{tabular}{c c c } 
\hline\hline 
Cut& Number of background events (fraction)& Number of signal MC events (fraction)\\ [0.5ex] 
\hline 
Before cuts & 2.19 $\times$ 10$^8$ & 264465\\
Charge > 75 PE & 312199 (0.0014) & 81290 (0.31)\\
Frprompt < 0.92 & 221129 (10$^{-3}$) & 81290 (0.31)\\
$Q_R$ < 0.44 & 19999 (10$^{-4}$) & 81290 (0.31)\\
Moment of inertia > 0.3 & 19920 (10$^{-4}$) & 81290 (0.31)\\
R < 295 mm & 2017 (9 $\times$10$^{-6}$) & 27392 (0.1)\\
\hline\hline
\end{tabular}
\end{adjustbox}

\label{table:cutsum} 
\end{table}
Prior to fill water in water tank (full shielding), the background rate drops due to the partial shielding. Table \ref{table:cutsumstage} summarize the background rejection efficiency at different stage of status of IV ( before and after PMT gain matching\footnote{ Gain matching is performed while IV under vacuum, all PMT gains are calibrated to 5 pC}, and IV moved into OV). After IV moving into OV (partial shielding), the background rejection efficiency rose due to the decreasing background from external gamma particles. The estimated background rate after IV moved into OV is 2.36 events/hour.\par

\begin{table}[htbp]
\caption{Summary of background and signal events survived after the cuts at different stage of status of IV.} 
\centering 
\begin{adjustbox}{width=1\textwidth}
\begin{tabular}{c c c c} 
\hline\hline 
Cut& Before gain matching (fraction)& After gain matching & After IV moved into OV\\ [0.5ex] 
\hline 
Raw counts & 1.9 $\times$10$^7$ (330.1 Hz)& 6.9 $\times$10$^7$ (424.8 Hz)& 5.2 $\times$10$^5$ (171.3 Hz)\\
Charge > 75 PE & 26671(0.0014) & 101743 (0.0015) & 1520 (0.003)\\
Fprompt < 0.9 & 15244 (0.0008) & 65569 (0.00095) & 886 (0.0017)\\
Charge ratio < 0.4 & 1491 (7.8$\times$10$^{-5}$) & 5474 (7.9$\times$10$^{-5}$) & 44 (8.4 $\times$10$^{-5}$)\\
R <295 mm & 224 (1.2$\times$10$^{-5}$) & 847 (1.2$\times$10$^{-5}$) & 2 (3.8$\times$10$^{-6}$)\\
\hline\hline
\end{tabular}
\end{adjustbox}

\label{table:cutsumstage} 
\end{table}

In summary, the expected Cherenkov event rate of MiniCLEAN detector is 9.7 $\pm$ 0.6 events/hour in the vacuum data without any shielding. After the full shielding of IV is in place, the event rate decreased from 400 Hz to 17 Hz. While IV under vacuum without any shielding, the main source that produce the Cherenkov events is from rock gamma and internal gamma emitter. With full shielding the only dominant source of gamma is from the internal gamma emitter. Thus the expected Cherenkov rate should decrease by the factor of 110 ($^{39}$Ar is a beta emitter which account for 2 Hz of trigger rate, see Chapter \ref{ch:gasrun}). Therefore the expected Cherenkov event rate with full shielding is 0.364 $\pm$ 0.023 events/hour.

\section{Foil scintillation}\label{sec:esrfoilscin}
In the vacuum data, a type of unexpected events was found. There are excessive events near the edge of the detector in the reconstructed radius distribution as shown in Fig. \ref{fig:esrradius}. In the vacuum data, these events can be seen by plotting the charge ratio as a function of reconstructed radius and project to the radius axis as shown in Fig. \ref{fig:ESR2d}. Noticed that if apply a charge cut ($Q_R < 0.8 $) and compare to the result without any cut, a clear peak at near the TPB radius is shown in the figure. This peak in the data confirmed that the results from simulation in Fig. \ref{fig:esrradius}. The origin of these ESR foil scintillation could be from high energy events (alpha, beta and gamma) bombard on the ESR foil and create the scintillation\cite{cresst}. The ESR foil which mentioned in Chapter \ref{ch:detector} was used to be the liner of the optical cassettes to increase the reflectivity inside it and to cover the gap between the optical cassettes as well. Therefore the gamma ray from the detector material, the alpha particle from the decay chain of radon daughter and the external high energy gamma ray are the major sources of the ESR foil scintillation.\par
\begin{figure}[htbp]
\centering
\graphicspath{{./fig/Vacuum/}}
\includegraphics[scale=0.4]{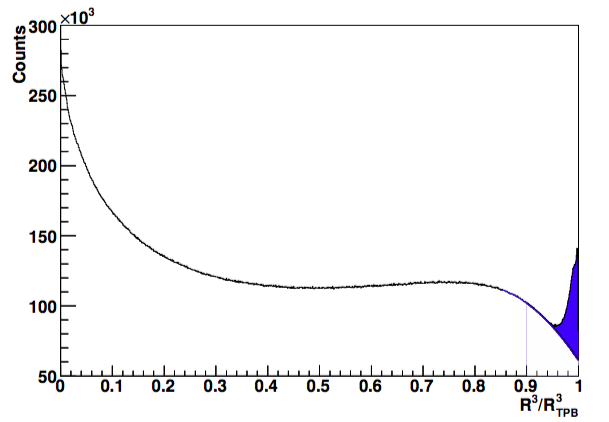}
\caption{The reconstructed radius distribution. The black curve is from the MC simulation and the blue color filled region is from the ESR foil scintillation\cite{sjaditz}.}
\label{fig:esrradius}
\end{figure}
\begin{figure}[htbp]
\hfill
\subfloat[]{\includegraphics[width=7cm]{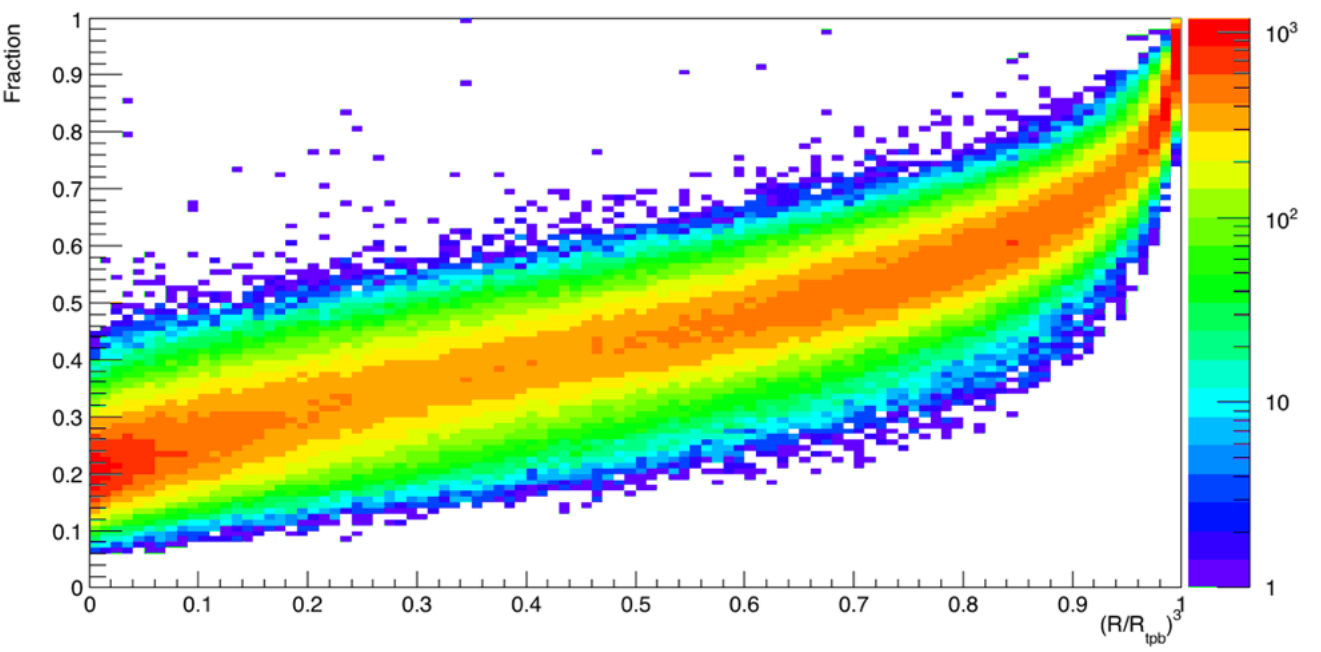}}
\hfill
\subfloat[]{\includegraphics[width=7cm]{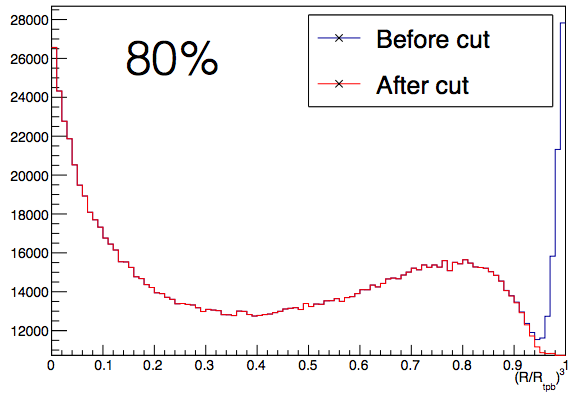}}
\hfill
\caption{(a) Charge ratio vs reconstructed radius normalized by radius of TPB. (b) Project the (a) to x-axis. Blue curve is before the charge ratio cut ($Q_R < 0.8$) and the red curve is the results after the cut.}
\label{fig:ESR2d}
\end{figure}
In order to eliminate the foil scintillation events from the data, the detail properties of the events needs to be understood. Figure \ref{fig:esrscatter} shows the scatter plot of charge and Fprompt distribution in the vacuum data. It is clear to see there are four different group of events in this plot. The group of events at high Fprompt and high charge ratio region is mainly from the Cherenkov events which has been carefully treated in the last section. The group of events at low Fprompt and low charge ratio is from the surface alpha-TPB scintillation which is described in detail in section \ref{sec:sa}. The events around the low Fprompt and low charge ratio is from the instrument effect which will be described in detail in the next section. The events around high Fprompt and low charge ratio is possibly from the ESR foil scintillation. A study from DEAP collaboration\cite{deapesr} found that the scintillation has four characteristic decay time as shown in Fig. \ref{fig:esrdecay}. The $^{241}$AM source which can emit the alpha particle at energy $\sim$ 5.4 MeV is placed before the ESR foil to induce the scintillation light. This confirmed what has been seen in the Fig. \ref{fig:esrscatter}, due to the longer decay component , the Fprompt is pushed to the lower Fprompt region.\par

\begin{figure}[htbp]
\centering
\graphicspath{{./fig/Vacuum/}}
\includegraphics[scale=0.25]{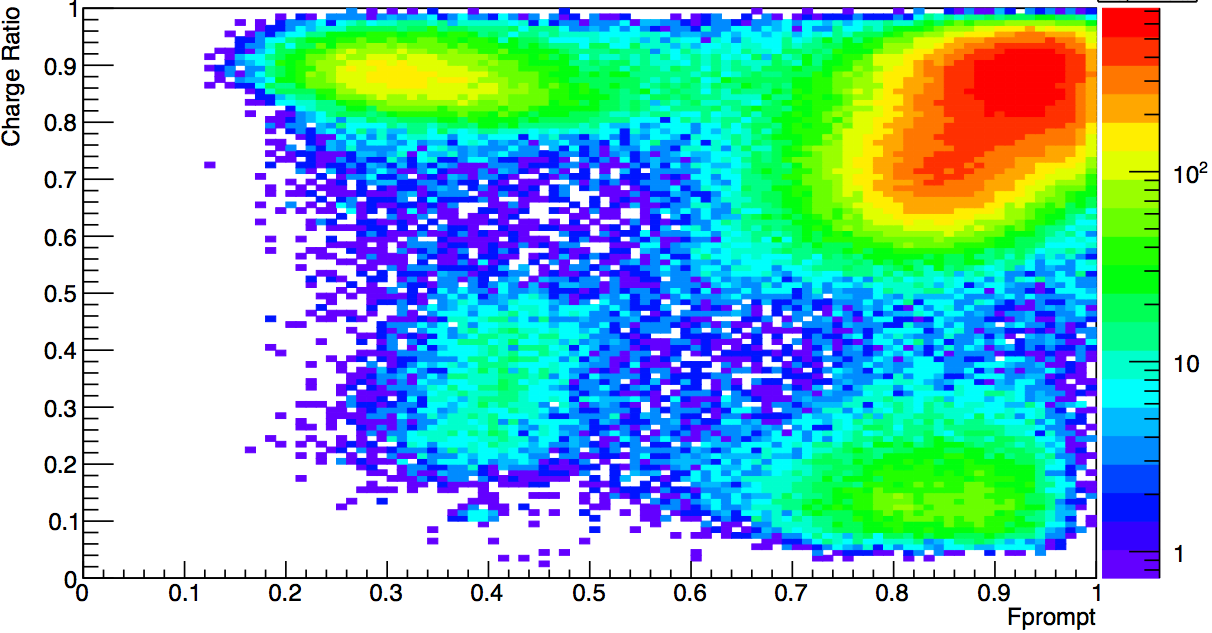}
\caption{The scatter plot of charge ratio and Fprompt distribution.}
\label{fig:esrscatter}
\end{figure}
\begin{figure}[htbp]
\centering
\graphicspath{{./fig/Vacuum/}}
\includegraphics[scale=0.25]{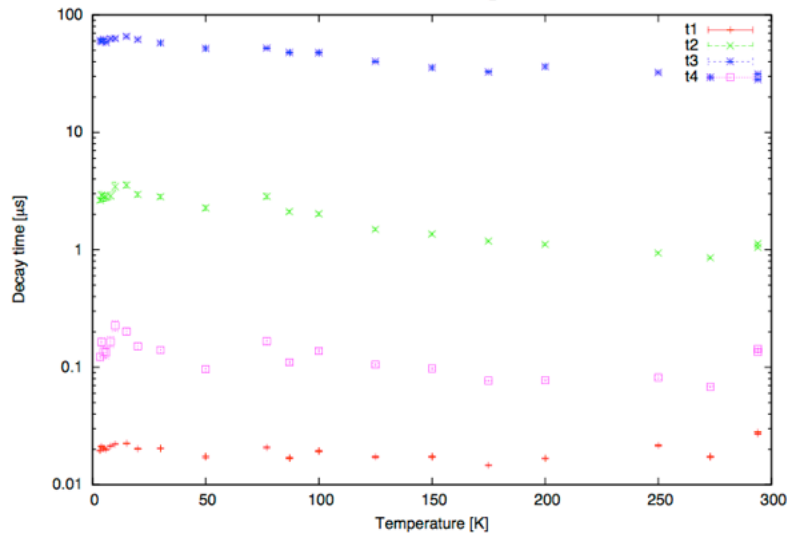}
\caption{The decay time of different components of ESR foil scintillation as a unction of temperature\cite{deapesr}. }
\label{fig:esrdecay}
\end{figure}
Selecting the possible ESR foil scintillation events from data according to the Fig. \ref{fig:esrscatter}, the scintillation timing structure can be fitted with three exponential  functions convoluted with Gaussian response function :
\begin{ceqn}\begin{align}
F(t) = A\cdot e^{-\frac{t}{\tau_1}} \otimes G(\mu_1;\sigma_1) +  B\cdot e^{-\frac{t}{\tau_2}} \otimes G(\mu_2;\sigma_2) +  (1-A-B)\cdot e^{-\frac{t}{\tau_3}} \otimes G(\mu_3;\sigma_3)
\end{align}\end{ceqn}
where the A, B are the fraction of the components and $\tau_i$, $\mu_i$, $\sigma_i$ are the decay time , mean, $\sigma$ of the Gaussian distribution for i-th component. The fitting results is summarized in the Table \ref{table:decayconstant} along with the results from DEAP collaboration. The fitting example is shown in Fig. \ref{fig:esrdecayfit}. The first two time constants are roughly agree with DEAP's result, due to the constrain on data acquisition window in MiniCLEAN (16 $\mu$s), the fourth time constant can not be measured. Therefore, the third time constant measured by MiniCELAN could be the combination of the third and fourth time constant, results in deviation from DEAP's result.\par

\begin{table}[htbp]
\caption{Decay time constant of ESR foil scintillation events.} 
\centering 
\begin{adjustbox}{width=1\textwidth}
\begin{tabular}{c c c c c} 
\hline\hline 
& Time constant 1 (ns) & Time constant 2 (ns)& Time constant 3 (ns)& Time constant 4(ns)\\ [0.5ex] 
\hline 
MiniCLEAN & 10 &252 &1886\\
DEAP & 20& 200 & 1200 & 30000\\
\hline\hline
\end{tabular}
\end{adjustbox}

\label{table:decayconstant} 
\end{table}
\begin{figure}[htbp]
\centering
\graphicspath{{./fig/Vacuum/}}
\includegraphics[scale=0.4]{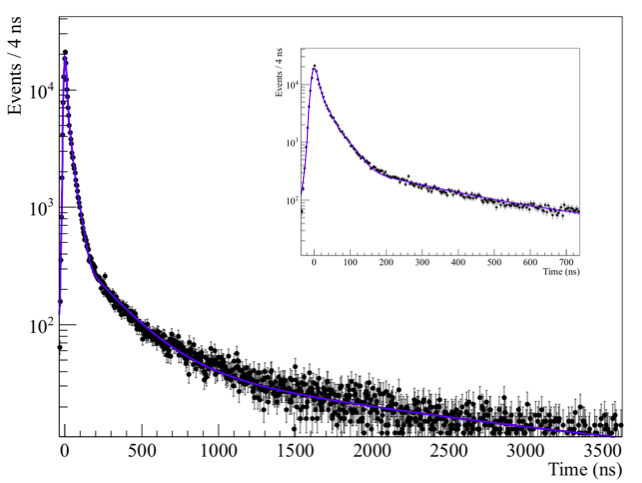}
\caption{The fitting example of  ESR foil scintillation events.}
\label{fig:esrdecayfit}
\end{figure}
With the timing P.D.F. of scintillation time structure, the log-likelihood ratio can be constructed and used to discriminate the ESR foil scintillation events from nuclear recoil events.
Figure. \ref{fig:esrdpdf} shows the timing P.D.F. of ESR foil scintillation events overlap with the nuclear recoil events. The test statistic is defined as : 
\begin{ceqn}\begin{align}
l_{ESR} = \frac{1}{n_{ph}}\sum\left ( logP_{ESR}(t) - logP_n(t) \right ) 
\end{align}\end{ceqn}
where $P_{ESR}(t)$ is the timing P.D.F. of ESR scintillation events, $P_n(t)$ is the timing P.D.F. of nuclear recoil events and the $n_{ph}$ is the total photoelectrons in the given events. The result is shown in Fig. \ref{fig:esrlesrplot}. For events passed the fiducial volume cut (R < 295 mm), the cut efficiency of $l_{ESR}$ as a function of cut value is shown in Fig. \ref{fig:esrcuteff}.
\begin{figure}[htbp]
\centering
\graphicspath{{./fig/Vacuum/}}
\includegraphics[scale=0.4]{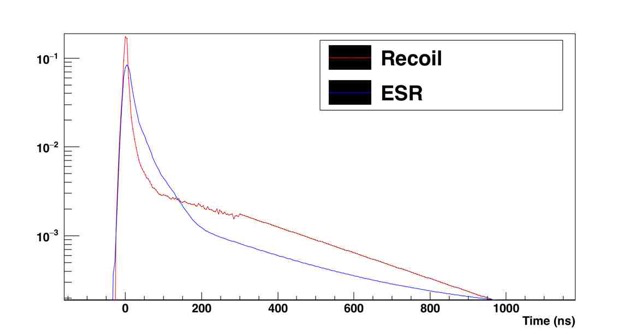}
\caption{The blue curve is the timing P.D.F. of the ESR foil scintillation events and the red curve is the timing P.D.F. from nuclear recoil.}
\label{fig:esrdpdf}
\end{figure}
\begin{figure}[htbp]
\centering
\graphicspath{{./fig/Vacuum/}}
\includegraphics[scale=0.4]{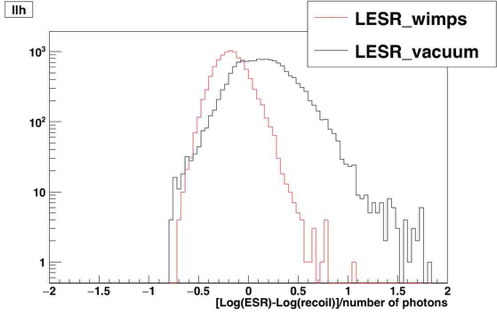}
\caption{$l_{ESR}$ value for both ESR scintillation events and nuclear recoil events.}
\label{fig:esrlesrplot}
\end{figure}
\begin{figure}[htbp]
\centering
\graphicspath{{./fig/Vacuum/}}
\includegraphics[scale=0.4]{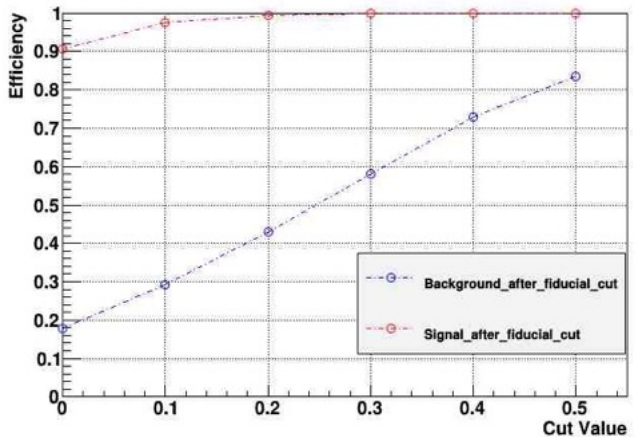}
\caption{Cut efficiency after pass the fiducial volume cut.}
\label{fig:esrcuteff}
\end{figure}
\section{Miscellaneous events}\label{sec:miscvacuum}
In Fig. \ref{fig:esrscatter}, the group of events at high charge ratio and low Fprompt are produced from instrument effects. Moreover, a ``arm'' structure can be seen in the charge-Fprompt scatter plot as shown in Fig. \ref{fig:fpqall1}. After the thorough study, PMT 41 seems to have significant amount of events in that area and responsible for the ``arm'' structure. Figure \ref{fig:fpqevents} shows the different charg-Fprompt scatter plot with different events are removed. Figure \ref{fig:fpqevents} (a) shows the result with no cut, Fig. \ref{fig:fpqevents}(b) shows the result with ESR-foil events are removed, Fig. \ref{fig:fpqevents} (c) shows the results excluded the PMT 41 and the results from PMT 41 is shown in Fig. \ref{fig:fpqevents} (d). It is clear to see that the arm stucture is mainly from PMT 41. In vacuum data, the events of high energy and low Fprompt is mostly from alpha-TPB scintillation, thus these events from PMT 41 can not be physical events unless some unknown source is contributing only to PMT 41.\par
 It becomes more clear by seeing the maximum charge of the pulses as a function of total charge of the event as shown in Fig. \ref{fig:maxqq}. Typically, the maximum charge is linearly corresponds to the total charge. However, for PMT 41, the distribution is relatively flat compare to the rest of events. This indicates some artificial effects are responsible for this behavior. Moreover, Fig. \ref{fig:maxqt} shows the maximum charge of the pulses vs time. It was found that for PMT 41, a significant portion of the maximum charge comes in later in time ($\sim$ 600 ns). At first glance, these pulses seems from the after-pulsing. However, from the discussion in Chapter \ref{sec:afterpulsing}, there are no source for the after-pulsing to be happened in such short delayed time. In addition, these pulses are not seen by the rest of PMTs, thus the origin of the pulses might not be from any physical events of IV.  Another piece of evidence shows that the PMT 41 is behaving very different from the rest of PMTs is from the profile plot of Fig. \ref{fig:fpqall1} as shown in Fig. \ref{fig:fpqallprofile}. The profile plot shows the mean value and associated error for charge (Y) and Fprompt (X).  The behavior of all PMTs except PMT 41 has relative flat curve, this is expected in the vacuum data since the Cherenkov events has low energy and no events with low Fprompt except noise events. PMT 41 shows the strange behavior in low Fprompt region, this also confirmed the observation from  previous plot. The exact origin of the events in PMT 41 is unknown. These events are probably due to some electrical malfunction such that the timing of the photoelectrons has a offset relative to the rest of the PMTs.\footnote{The timing mis-alignment is rejected for two reason : There are 8 PMTs plugging into the same WFD board and the channel has been confirmed with no issue.}
\begin{figure}[htbp]
\centering
\graphicspath{{./fig/Vacuum/}}
\includegraphics[scale=0.3]{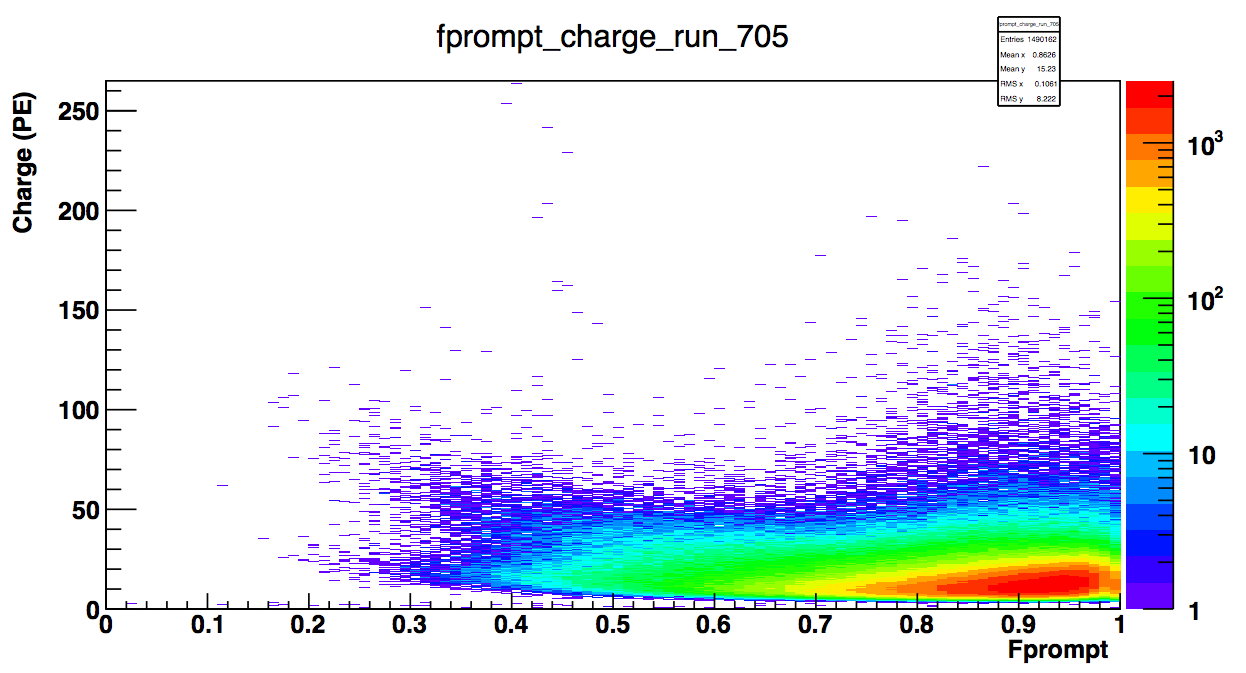}
\caption{Charge vs Fprompt in the vacuum data.}
\label{fig:fpqall1}
\end{figure}
\begin{figure}[htbp]
\hfill
\subfloat[]{\includegraphics[width=7cm]{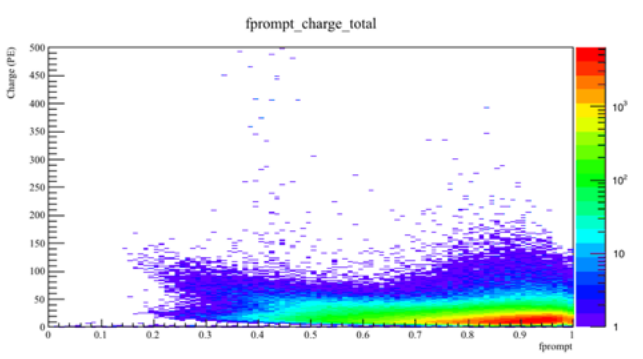}}
\hfill
\subfloat[]{\includegraphics[width=7cm]{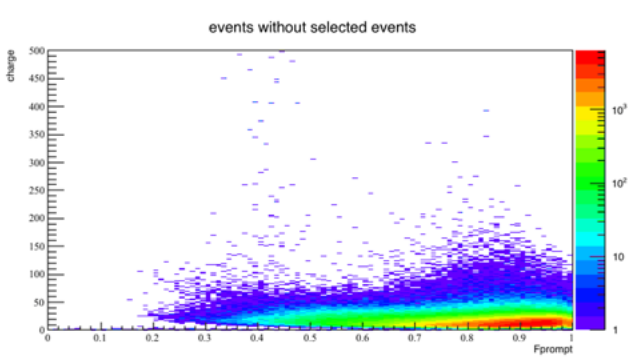}}
\hfill
\subfloat[]{\includegraphics[width=7cm]{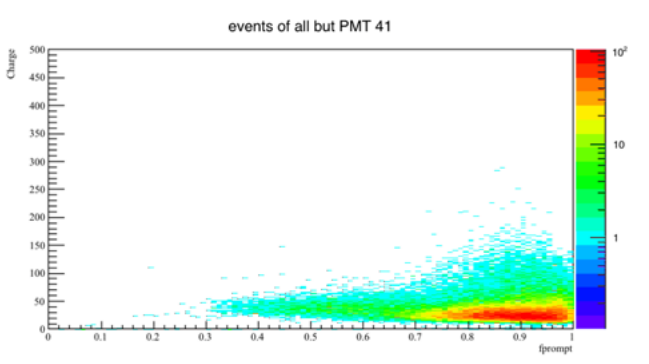}}
\hfill
\subfloat[]{\includegraphics[width=7cm]{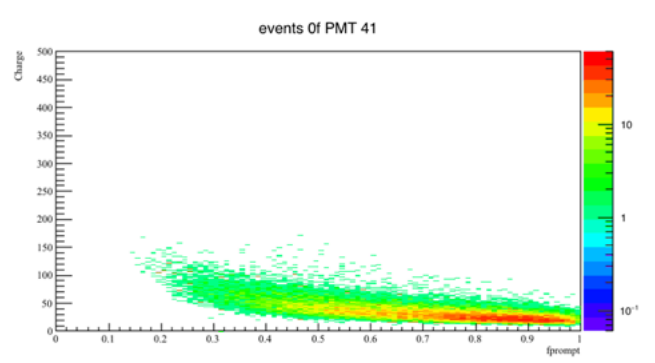}}
\hfill
\caption{Charge vs Fprompt in the vacuum data for different events are removed (a) No cut. (b) No ESR foil scintillation events. (c) All PMT except PMT 41. (d) Only PMT 41.}
\label{fig:fpqevents}
\end{figure}
\begin{figure}[htbp]
\hfill
\subfloat[]{\includegraphics[width=7cm]{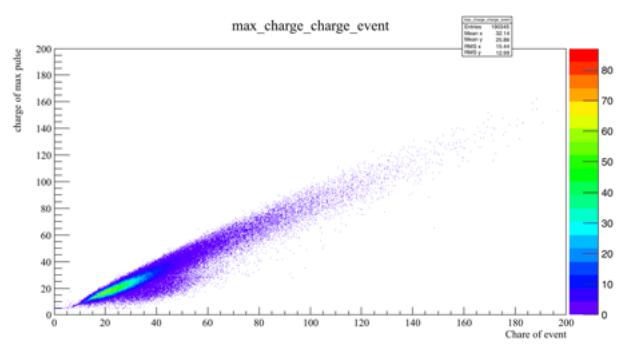}}
\hfill
\subfloat[]{\includegraphics[width=7cm]{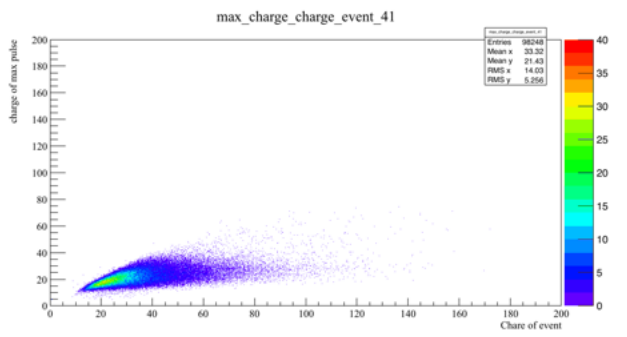}}
\hfill
\caption{ Maximum charge of the pulses in each PMT vs total charge of the event for (a) All PMT. (b) PMT 41.}
\label{fig:maxqq}
\end{figure}

\begin{figure}[htbp]
\hfill
\subfloat[]{\includegraphics[width=7cm]{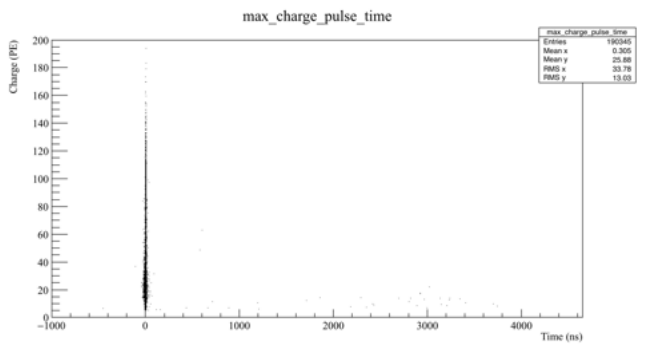}}
\hfill
\subfloat[]{\includegraphics[width=7cm]{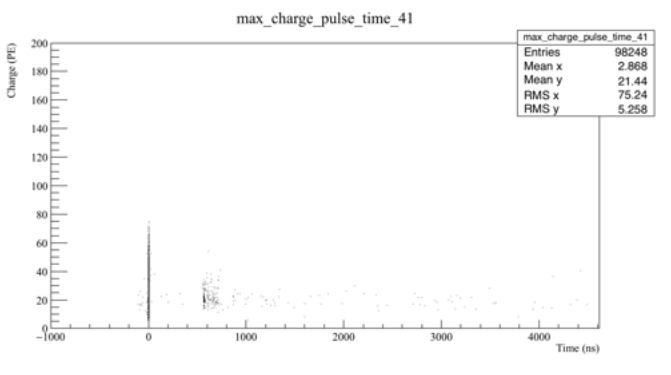}}
\hfill
\caption{ Maximum charge of the pulses in each PMT vs time for (a) All PMT. (b) PMT 41.}
\label{fig:maxqt}
\end{figure}
\begin{figure}[htbp]
\centering
\graphicspath{{./fig/Vacuum/}}
\includegraphics[scale=0.3]{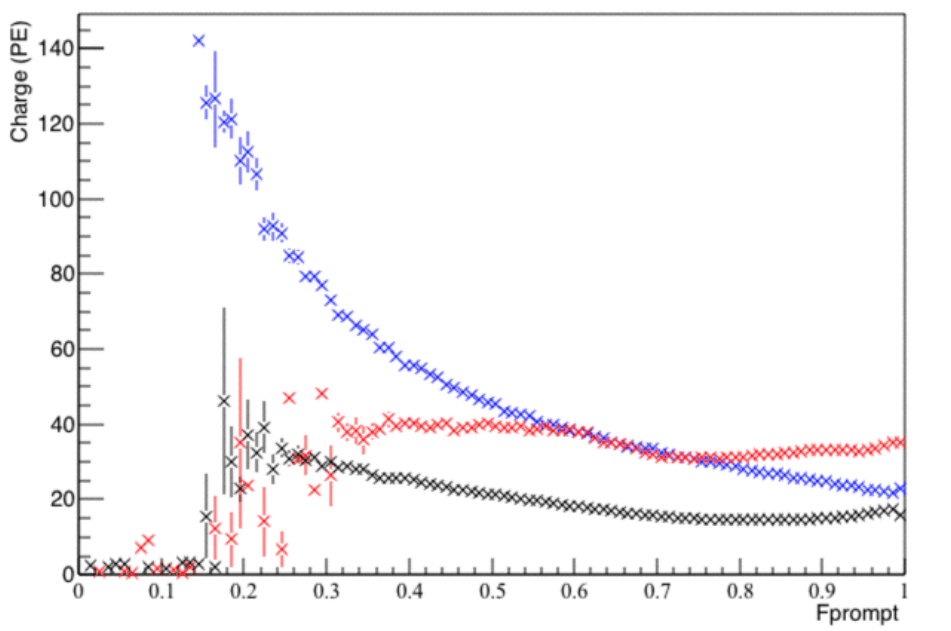}
\caption{Profile plot of Fig. \ref{fig:fpqall1}. Blue dot represent PMT 41 only, red dot represent all the rest of PMT and black dot shows all events.}
\label{fig:fpqallprofile}
\end{figure}
The other events fall into low Fprompt and high charge ratio is the cross talk induced by large pulse. Fro each WFD, there are 8 PMTs plugged into it. If any one of the PMT received a large pulse for some reason, it will induce a small bipolar pulse in the vicinity channels as shown in Fig. \ref{fig:crosstalk}.  These events happen more frequently to some PMTs than others (Fig. \ref{fig:crosstalkcounts}) due to the fact that the NHit trigger of MiniCLEAN requires 5 PMTs triggered in the same time. Therefore, for the PMTs which plugged into the middle channel will have larger chance to pass the NHit trigger. The more careful treatment and the cut will be described in Chapter \ref{ch:gasrun}.
\begin{figure}[htbp]
\centering
\graphicspath{{./fig/Vacuum/}}
\includegraphics[scale=0.3]{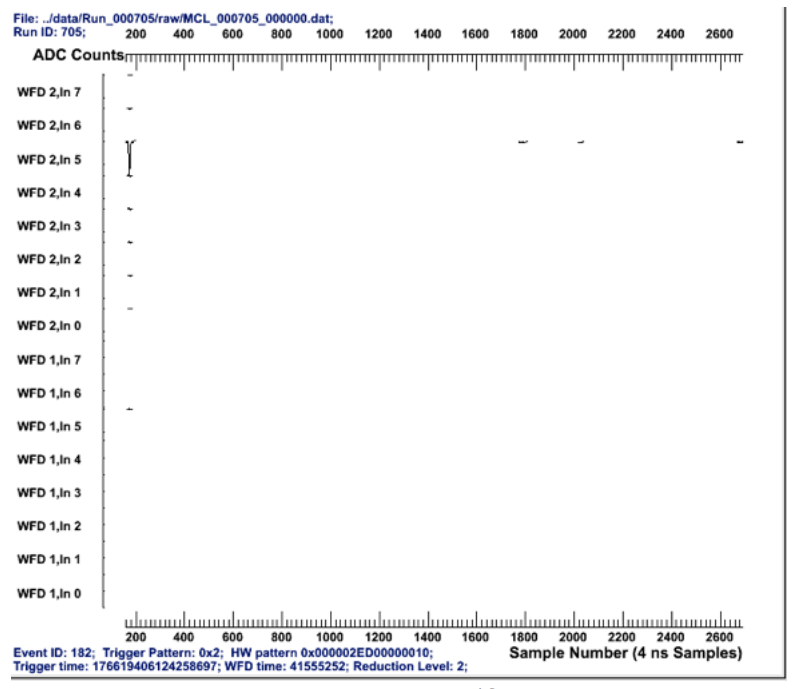}
\caption{Raw waveform of cross-talk events. The channel 5 on WFD 2 received large pulse and induce the cross talk in the vicinity channel.}
\label{fig:crosstalk}
\end{figure}
\begin{figure}[htbp]
\centering
\graphicspath{{./fig/Vacuum/}}
\includegraphics[scale=0.3]{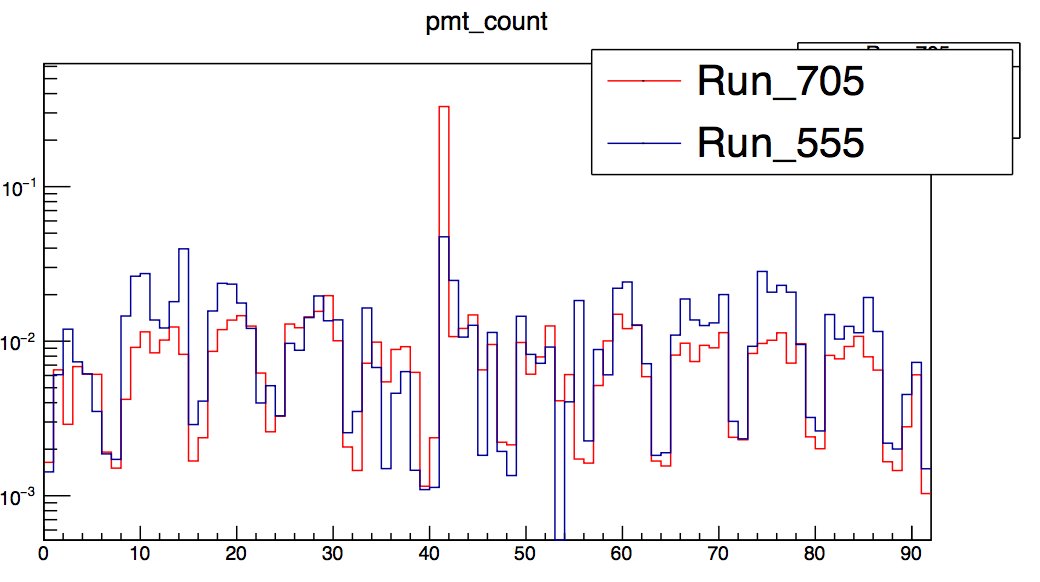}
\caption{Normalized count (X-axis) as a function of PMT channel ID.}
\label{fig:crosstalkcounts}
\end{figure}
\chapter{MiniCLEAN Gas Run}\label{ch:gasrun}
Before the final construction of MiniCLEAN is complete, the IV was filled with gas at room temperature to take data for testing the system.  The MiniCLEAN detector start cooling on Aug. 2015 and the data taking of the cold gas run started on Oct. 2016. The cold gas data taking is continuous throughout the cooling and filling phase of the MiniCLEAN detector. The main purpose of cold gas run is to monitor the detector health using the triplet lifetime. 

\section{Warm Gas run}
Warm gas data taking starts with IV filled with argon gas under the room temperature and without any shielding. Using the RGA to monitor the gas quality as shown in Fig. \ref{fig:impurity_warm},  which indicates the main source of the impurity of the IV is from the outgassing of acrylic (250 mtorr/l/min).  The water vapor is known to heavily quenched the triplet state of scintillation light. Figure \ref{fig:triplet_warm_hour} shows the triplet lifetime as a function of time in the first hour of filling.  It shows that the triplet lifetime drops nearly 15 \% in just one hour of the data.  The quenching effect can also be seen in the charge ratio-Fprompt distribution of surface alpha particle as shown in Fig. \ref{fig:surface_alpha_a}\par
\begin{figure}[htbp]
\centering
\graphicspath{{./fig/Warm_gas/}}
\includegraphics[scale=0.4]{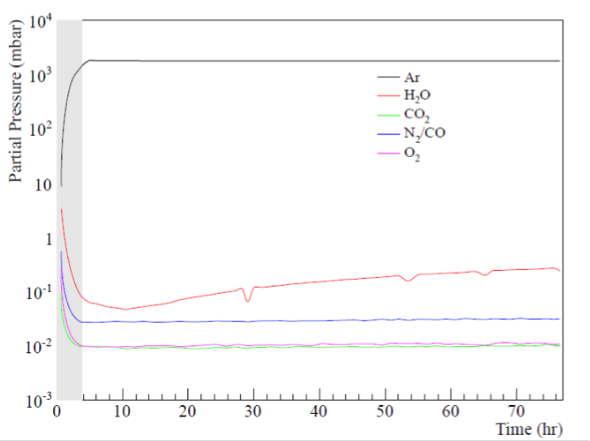}
\caption{Impurity level measured from RGA vs time for various impurities}
\label{fig:impurity_warm}
\end{figure}
\begin{figure}[htbp]
\centering
\graphicspath{{./fig/Warm_gas/}}
\includegraphics[scale=0.3]{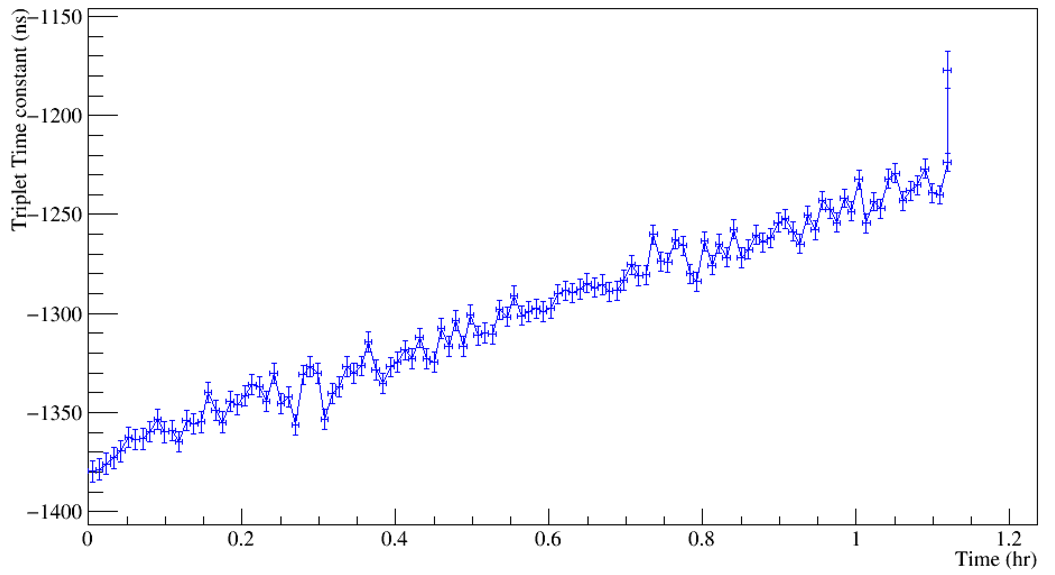}
\caption{Triplet lifetime as a function of time. The negative sign is from directly convert from the fitted rate.}
\label{fig:triplet_warm_hour}
\end{figure}
\begin{figure}[htbp]
\hfill
\subfloat[]{\includegraphics[width=7cm]{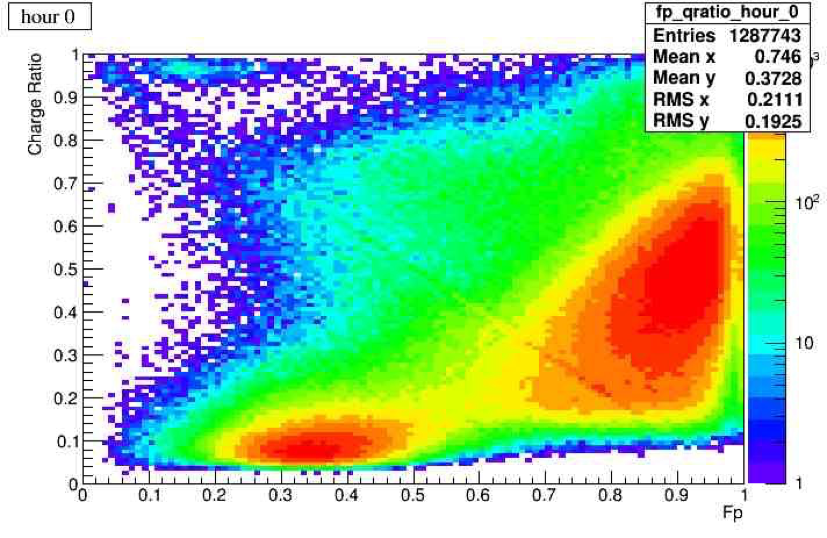}}
\hfill
\subfloat[]{\includegraphics[width=7cm]{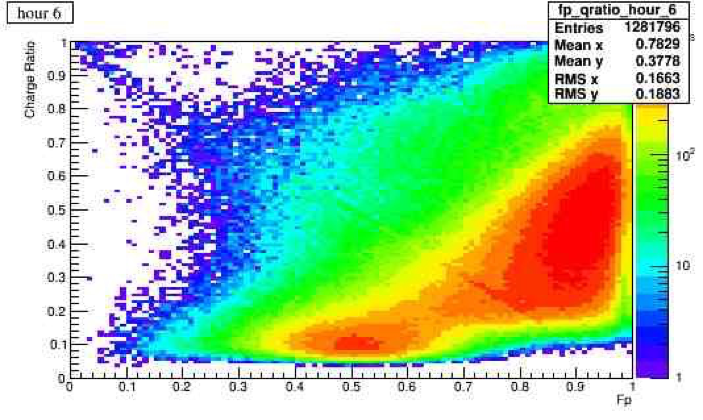}}
\hfill
\subfloat[]{\includegraphics[width=7cm]{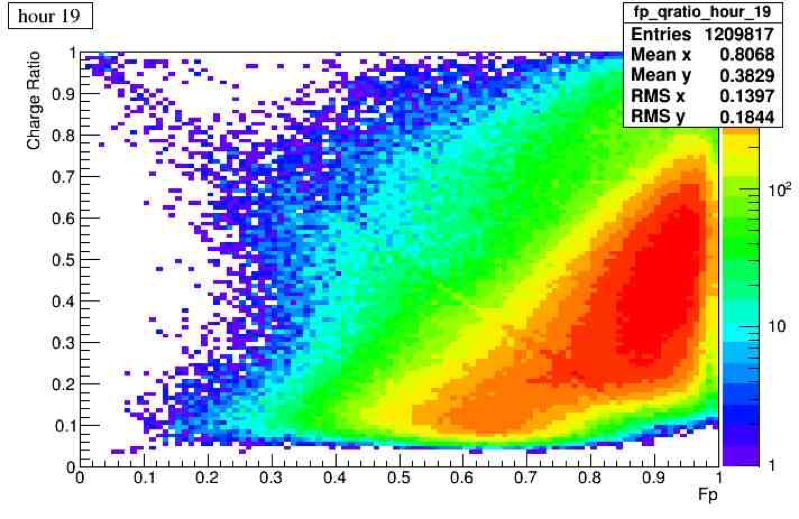}}
\hfill
\subfloat[]{\includegraphics[width=7cm]{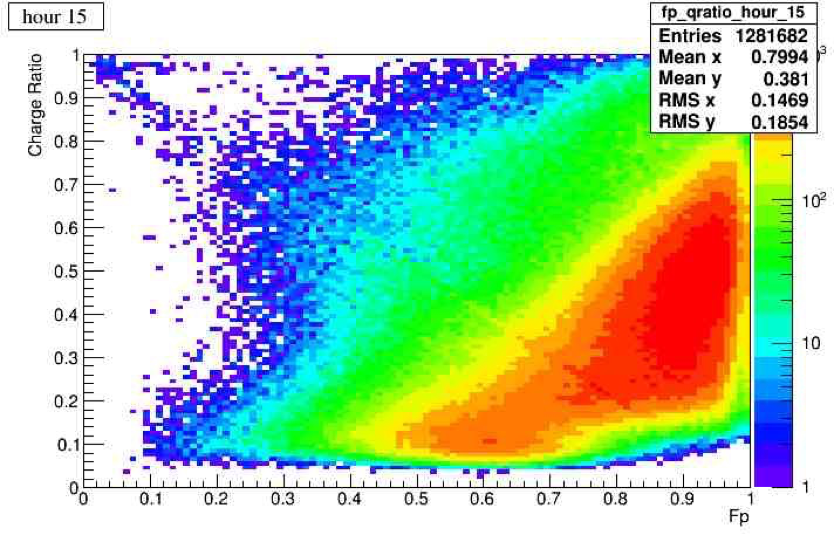}}
\hfill
\caption{Charge ratio - Fprompt distribution for (a) hour 0 after the fill. (b) hour 6 after the fill. (c) hour 15 after the fill. (d) hour 19 after the fill. Noticed that the group of events in low fprompt and low charge ratio move to the higher fprompt region, indicates the triplet states are quenched. }
\label{fig:surface_alpha_a}
\end{figure}
The tagged $^{22}$Na calibration run is performed after initial runs.  A radioactive $^{22}$Na emits a positron which annihilate to two back-to-back 511 keV gammas with radioactivity of 30 kBq. The source is in a NaI scintillator crystal and optically coupled to a PMT. The whole thing is placed on the top of IV. When the gamma scintillates in the crystal, the PMT attached to the scintillator fired. This assured the other gamma is emitted in the direction of IV. Therefore the back-to-back 511 keV gamma can be tagged in the data. In addition, a 1.275 MeV gamma is produced upon prompt relaxation. Figure \ref{fig:Na22_events} shows the Fprompt-charge distribution in warm gas run with (without) the $^{22}$Na presented. Figure \ref{fig:tag_fail} shows the Fprompt distribution between tag $^{22}$Na and no tag data. These two distribution are very similar indicating there is a problem with tagging. Later the problem was found that the NIM coincidence board was failed to record the tagging PMT with the DAQ properly.  
\begin{figure}[htbp]
\centering
\graphicspath{{./fig/Warm_gas/}}
\includegraphics[scale=0.3]{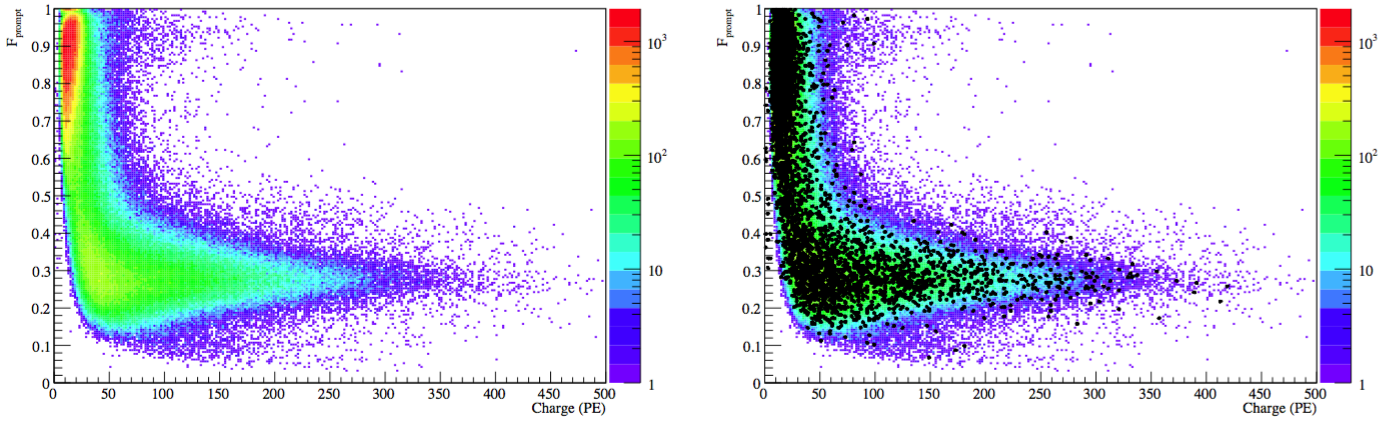}
\caption{Left : Fprompt as afunction of the charge for data acquired with the IV filled to 1800 mbar of argon gas. Right : The same distribution plotted with overlaid tagged $^{22}$Na events\cite{tomphd}.}
\label{fig:Na22_events}
\end{figure}
\begin{figure}[htbp]
\centering
\graphicspath{{./fig/Warm_gas/}}
\includegraphics[scale=0.4]{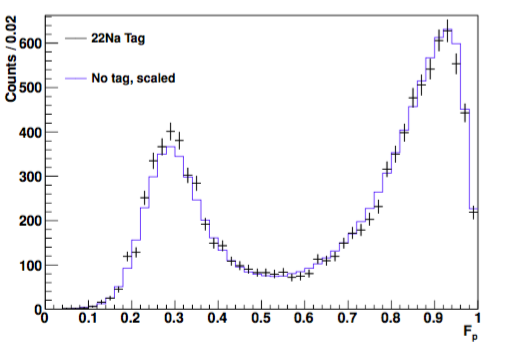}
\caption{The Fprompt distribution for tagged and untagged data. The two distribution are very similar indicating there was a problem with the tag. Figure from \cite{sjaditz}.}
\label{fig:tag_fail}
\end{figure}
The surface alpha rate is measured in the warm gas run. As mentioned in the previous Chapter, the surface alpha can be ejected into the Ar volume through the radon daughter decay chain. Figure \ref{fig:surface_alpha_gas} shows the alpha induced argon scintillation. Compare to the alpha-TPB scintillation in the vacuum, the alpha events in gaseous argon has larger energy. This is due to the scintillation efficiency for alpha-TPB scintillation ($\sim$ 880 photons/MeV\cite{POLLMANN2011127}) is smaller than in the gaseous argon (13000 photons/MeV). The measured surface alpha rate in the gaseous argon is 44.6 $\pm$ 0.7 events/hr\cite{sjaditz}.
\begin{figure}[htbp]
\hfill
\subfloat[]{\includegraphics[width=7cm]{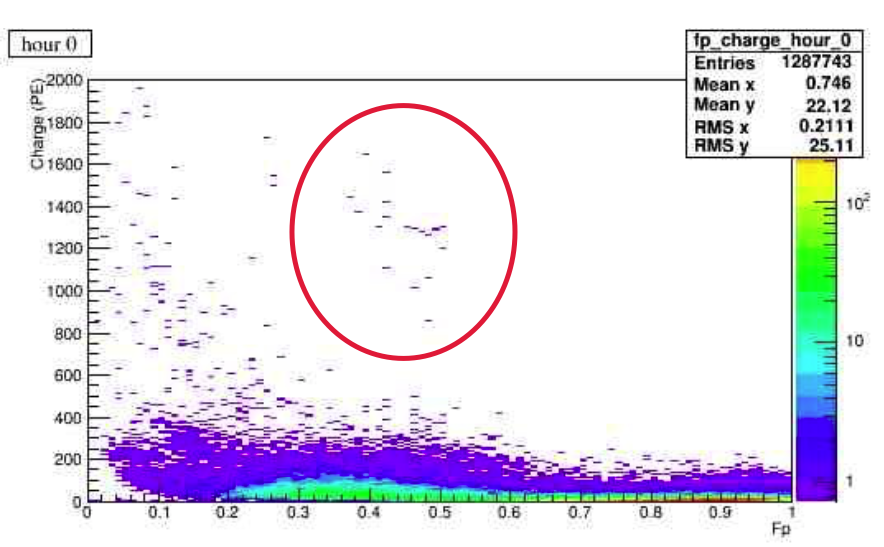}}
\hfill
\subfloat[]{\includegraphics[width=7cm]{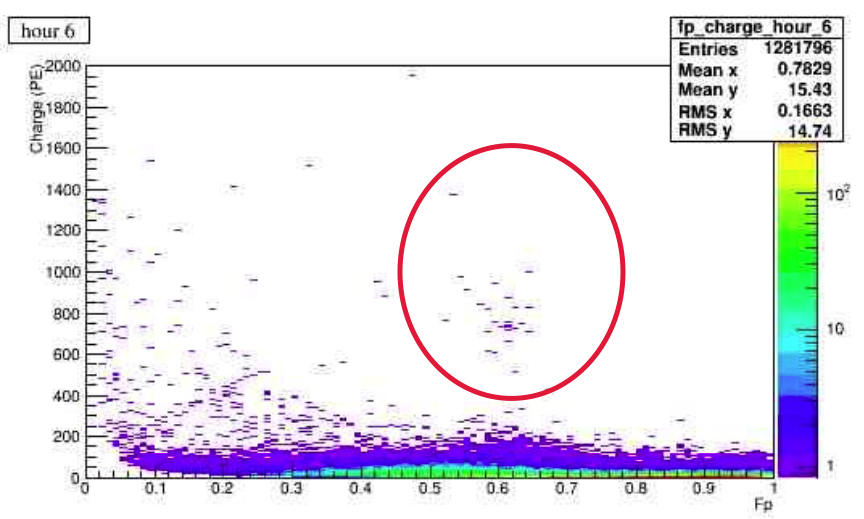}}
\hfill
\subfloat[]{\includegraphics[width=7cm]{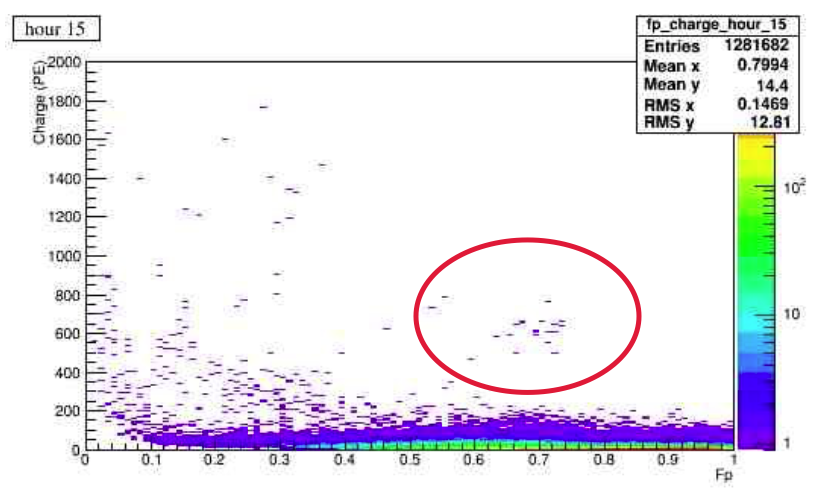}}
\hfill
\subfloat[]{\includegraphics[width=7cm]{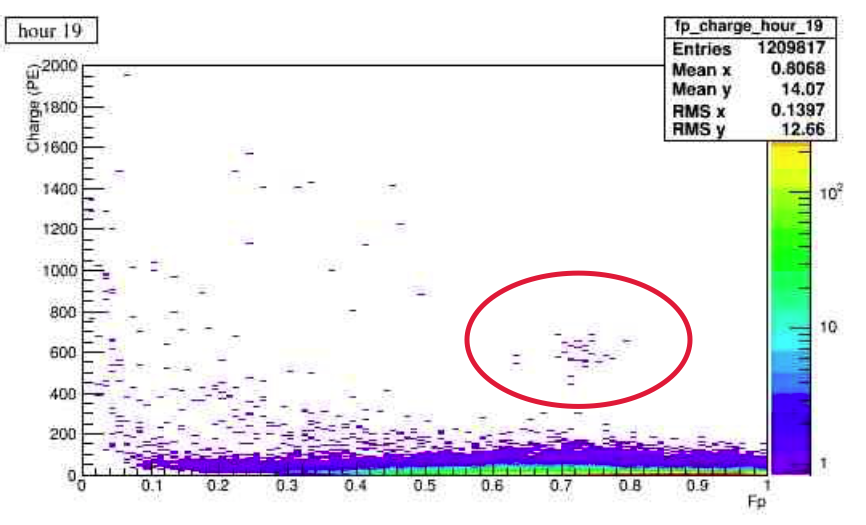}}
\hfill
\caption{Charge - Fprompt distribution for (a) hour 0 after the fill. (b) hour 6 after the fill. (c) hour 15 after the fill. (d) hour 19 after the fill. The circled events are from surface alpha scintillation in gaseous argon. Noticed that the alpha events move to higher Fprompt region due to the quenching of triplet component. }
\label{fig:surface_alpha_gas}
\end{figure}
\section{Cold gas run}
\subsection{Overview}
The data taking of cold gas run started from Oct. 2016. The motivation is to check the detector health and the functionality of each component. During the data taking, the temperature of IV is well under 140 K with pressure $\sim$ 1.5 bar. The cooling process still going on with lower cooling rate due to the extra heat load from operating PMTs. In the beginning of data taking, 18 PMTs initially deactivated. For 11 of these 18 deactivated PMTs we measured open circuits. We speculate that the connections became loose during the incident of rapid temperature rise described in Chapter \ref{sec:ledcoldgas}. The 7 other PMTs (less electrically insulating)
were intentionally deactivated due to observed high voltage tripping while when operating in warm gas. In the first two runs, we observe some PMTs with low gain and we increased the HV for these by 50 V. However, for 10 PMTs the gain was not improved.  Seven more PMTs that were turned off during the early pump and purge runs due to observing excessive ``flasher" events.\footnote{These events have very large fast pulses due primarily to electrical discharge.}  These PMTs are turned back on approximately 124 hours since beginning of pump and purge. In addition, PMT 41 is known to be noisy and PMT 35 was found to have large background during the data taking. Therefore in the following analysis, we will exclude these 30 (37 for early pump and purge run) PMTs which bring the number of total well-functioned PMTs to 62 (55 for early pump and purge run). A summary of PMT status by  angular distribution is shown in Figure \ref{fig:fmap}.\par
The main scintillation events are from the intrinsic $^{39}$Ar beta decay with full shielding of IV. There are some scintillation events are from the internal gamma particle induced electronic recoil. However, due to the low density of argon gas, only low energy gamma can produce the electronic recoil. Figure \ref{fig:attenuationargongamma} shows the attenuation coefficient as a function of density of argon. With the diameter of active volume of IV, the energy of gamma particle under 40 keV can leave the trace inside the IV. Figure \ref{fig:fp_q_cold_all} (a) shows the scatter plot of Fprompt and charge distribution. A band of electronic recoil can be seen at Fprompt $\sim$ 0.2. In the scatter plot of charge ratio and Fprompt as shown in Fig. \ref{fig:fp_q_cold_all} (b), a group of events at low Fprompt and low charge ratio can be seen. These events are corresponds to the band of events in the Fprompt-charge scatter plot. The Fprompt value of the electronic recoil is lower than the value in the $^{22}$Na warm gas run ($\sim 0.3$) due to the better purity level of the argon and thus longer triplet lifetime is attained.
\begin{figure}[htbp]
\centering
\graphicspath{{./fig/Triplet/}}
\includegraphics[scale=0.35]{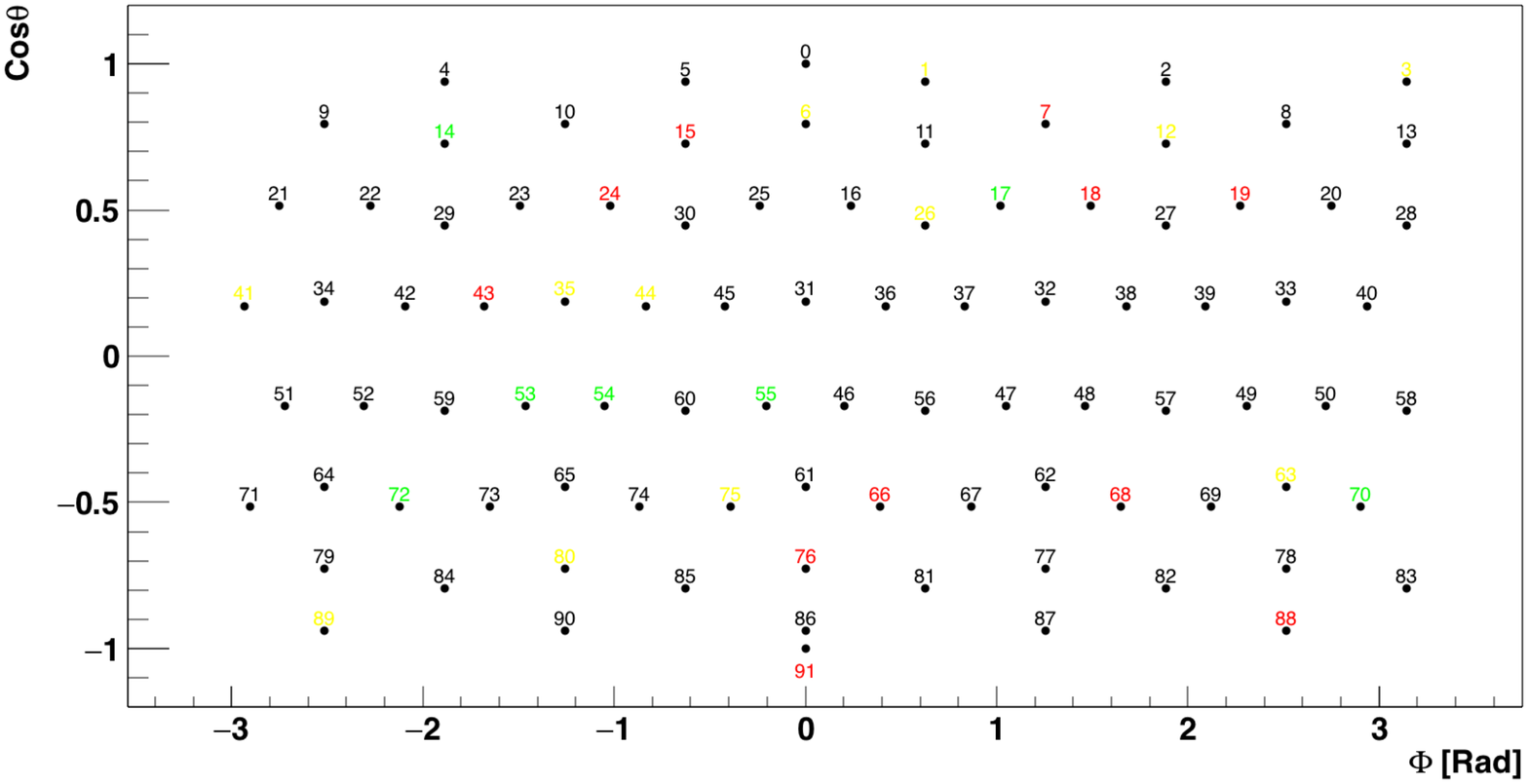}
\caption{ Red : PMTs are off (no connection). Green : PMTs without conformal coating (off in the gas run). Yellow : PMTs with very low gain except PMT 35 which has excessive background and PMT 41 is known as noisy PMT, these two PMTs are removed from analysis. }
\label{fig:fmap}
\end{figure}

\begin{figure}[htbp]
\centering
\graphicspath{{./fig/Warm_gas/}}
\includegraphics[scale=0.3]{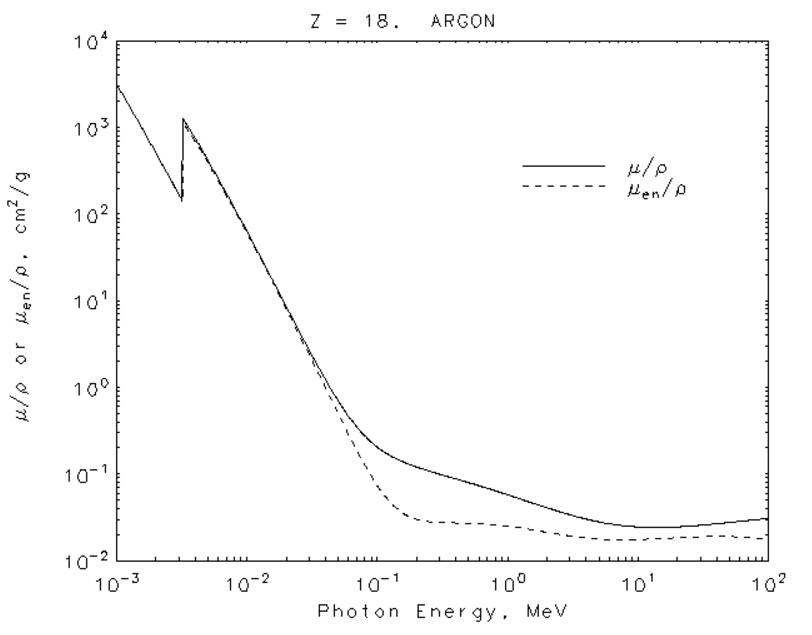}
\caption{Attenuation coefficient as a function of density of argon.}
\label{fig:attenuationargongamma}
\end{figure}
\begin{figure}[htbp]
\hfill
\subfloat[]{\includegraphics[width=7cm]{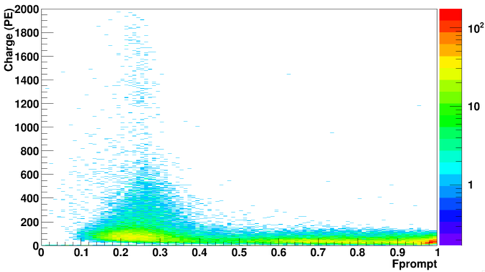}}
\hfill
\subfloat[]{\includegraphics[width=7cm]{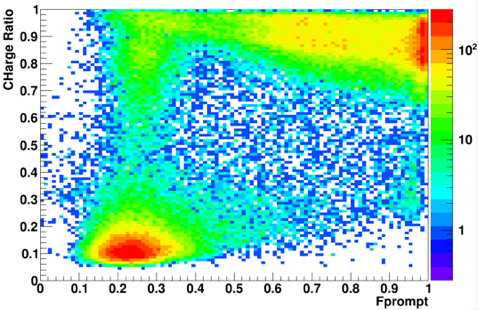}}
\hfill
\caption{ Maximum charge of the pulses in each PMT vs time for (a) All PMT. (b) PMT 41.}
\label{fig:fp_q_cold_all}
\end{figure}
\subsection{Triplet lifetime monitoring}
As mentioned in the previous section, the detector health can be monitored by observing the triplet lifetime. In the beginning of cold gas data taking, the triplet lifetime is stable at around 3.5 $\mu$s until the temperature of the bottom of IV reaches the liquefaction point of argon as shown in Fig. \ref{fig:triplet_temp_old}. At first, this behavior was explained due to some liquid build up in the bottom of IV such that when fitting the pulse time distribution from all PMTs, the average triplet lifetime is reduced due to the fact that triplet lifetime in the liquid is smaller ($\sim$ 1.6 $\mu$s). However, fitting the triplet lifetime for each active PMTs, no significant variations is observed as shown in Fig. \ref{fig:tripletpmtsold}. After thorough examination of the detector, a leak was found to be responsible foe the reduction of triplet lifetime.\par

\begin{figure}[htbp]
\hfill
\subfloat[]{\includegraphics[width=7cm]{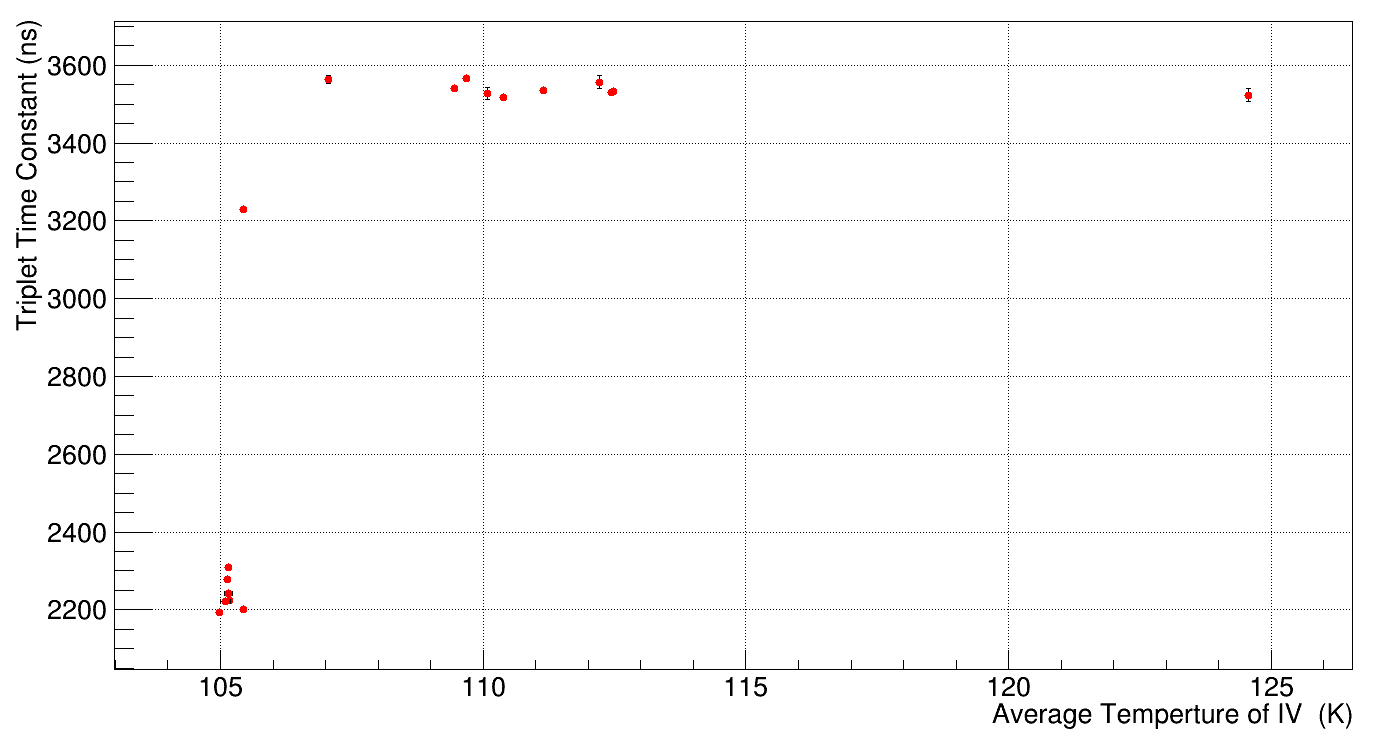}}
\hfill
\subfloat[]{\includegraphics[width=7cm]{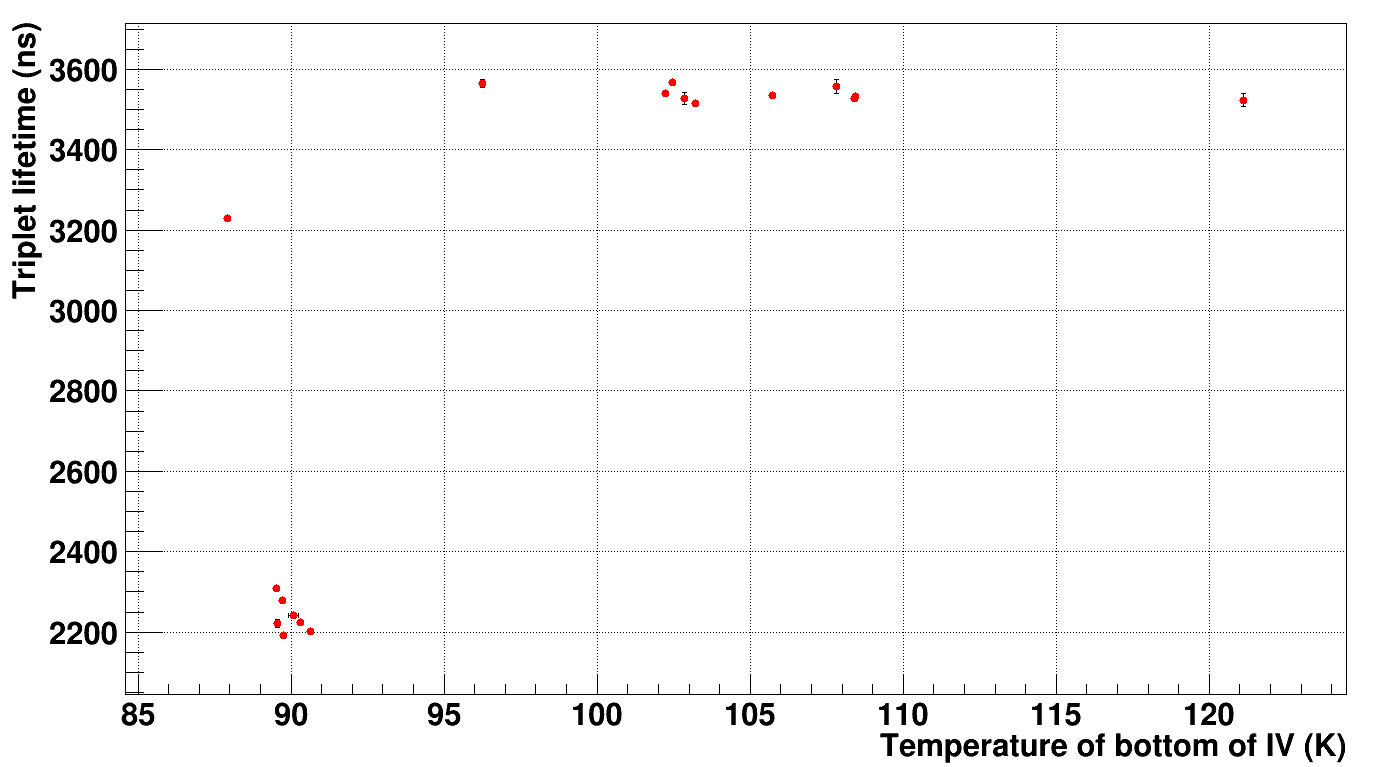}}
\hfill
\caption{ Triplet lifetime vs (a) Average temperature of IV. (b) Temperature of bottom of IV.}
\label{fig:triplet_temp_old}
\end{figure}
\begin{figure}[htbp]
\centering
\graphicspath{{./fig/Warm_gas/}}
\includegraphics[scale=0.25]{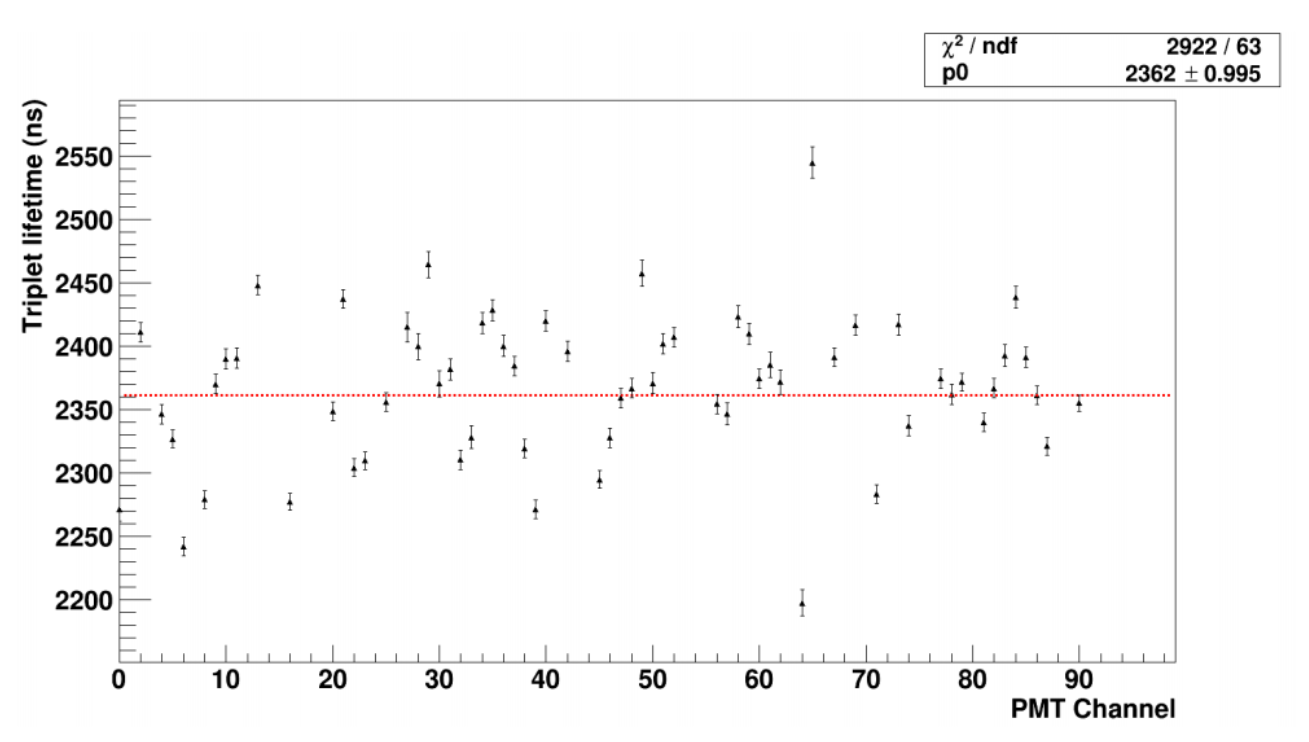}
\caption{Triplet lifetime for each PMT. The red dashed line indicates the weighted mean of triplet lifetime of all PMTs.}
\label{fig:tripletpmtsold}
\end{figure}

Figure \ref{fig:ftriptime} shows the triplet lifetime as a function of time. The triplet lifetime starts decreasing after the bottom temperature sensor reaches the liquefaction point of argon.  While the exact timing of degradation of triplet lifetime is unknown, leaks appear to have been developed sometime around mid-December 2016. The source of leak was found later in the exhaust from the IV pump. Shown in Fig. \ref{fig:fleak}, the source was likely back-flow of an argon/air mixture in the vent line through two check valves placed in series (opposite the allowed flow direction V4506 and PSV4507) leading to the IV, indicated by the red arrows\cite{Leak_fix}. Subsequently, the impurities were pump out of IV through a series pump and purge cycle(Detail description is in Chapter \ref{ch:cold}). The triplet lifetime is successfully restore to the values before the leak  after nearly 200 hours pump and purge cycle as shown in Fig. \ref{fig:fpumptime}. This shows the effectiveness of monitoring the triplet lifetime. The triplet lifetime is kept tracking whenever there's flow into IV to ensure the gas quality and detector health as well.\par

\begin{figure}[htbp]
\centering
\graphicspath{{./fig/Triplet/}}
\includegraphics[scale=0.4]{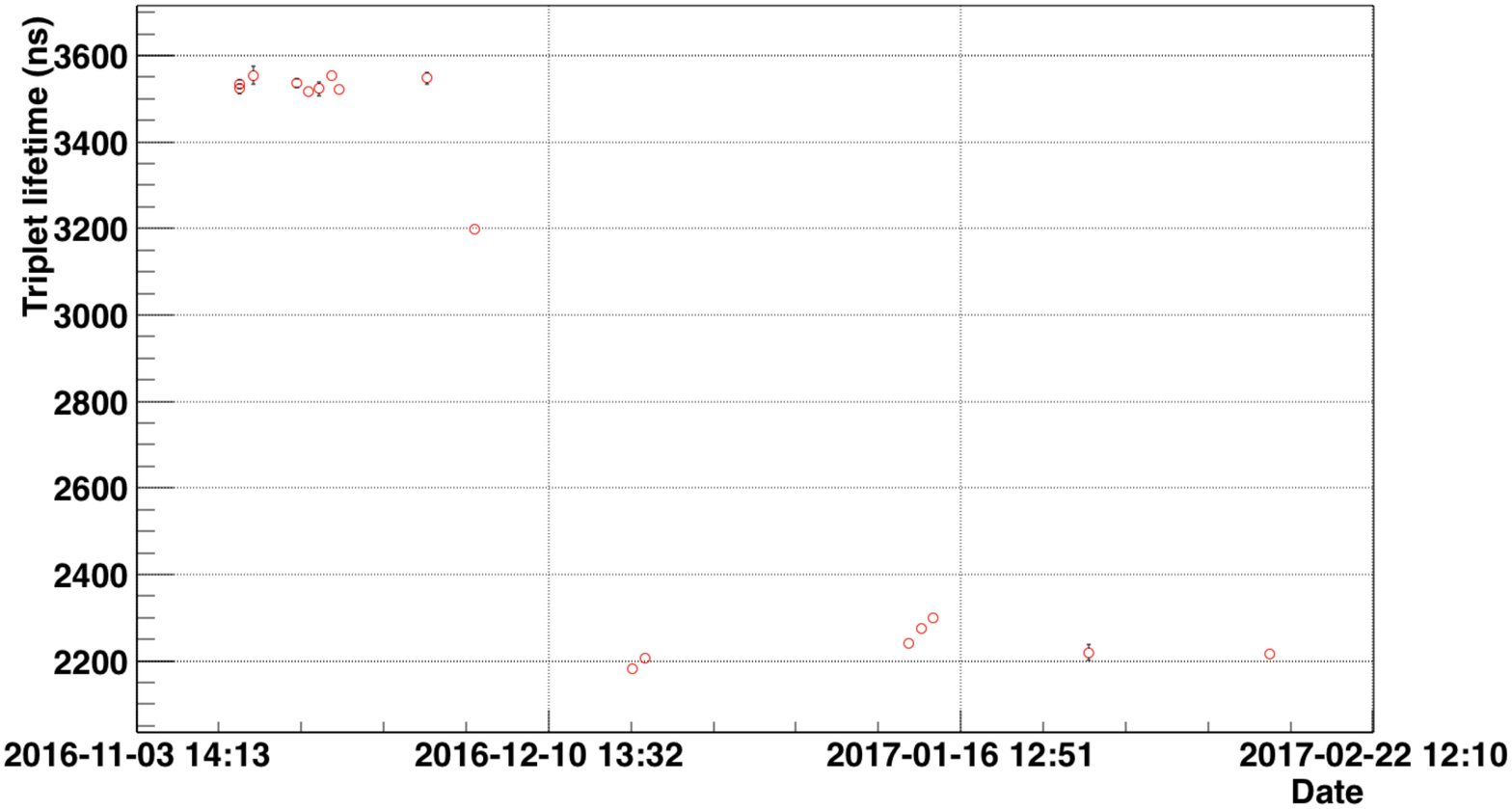}
\caption{ Triplet lifetime monitoring showing the period of time around the leak. }
\label{fig:ftriptime}
\end{figure}

\begin{figure}[htbp]
\centering
\graphicspath{{./fig/Triplet/}}
\includegraphics[scale=0.4]{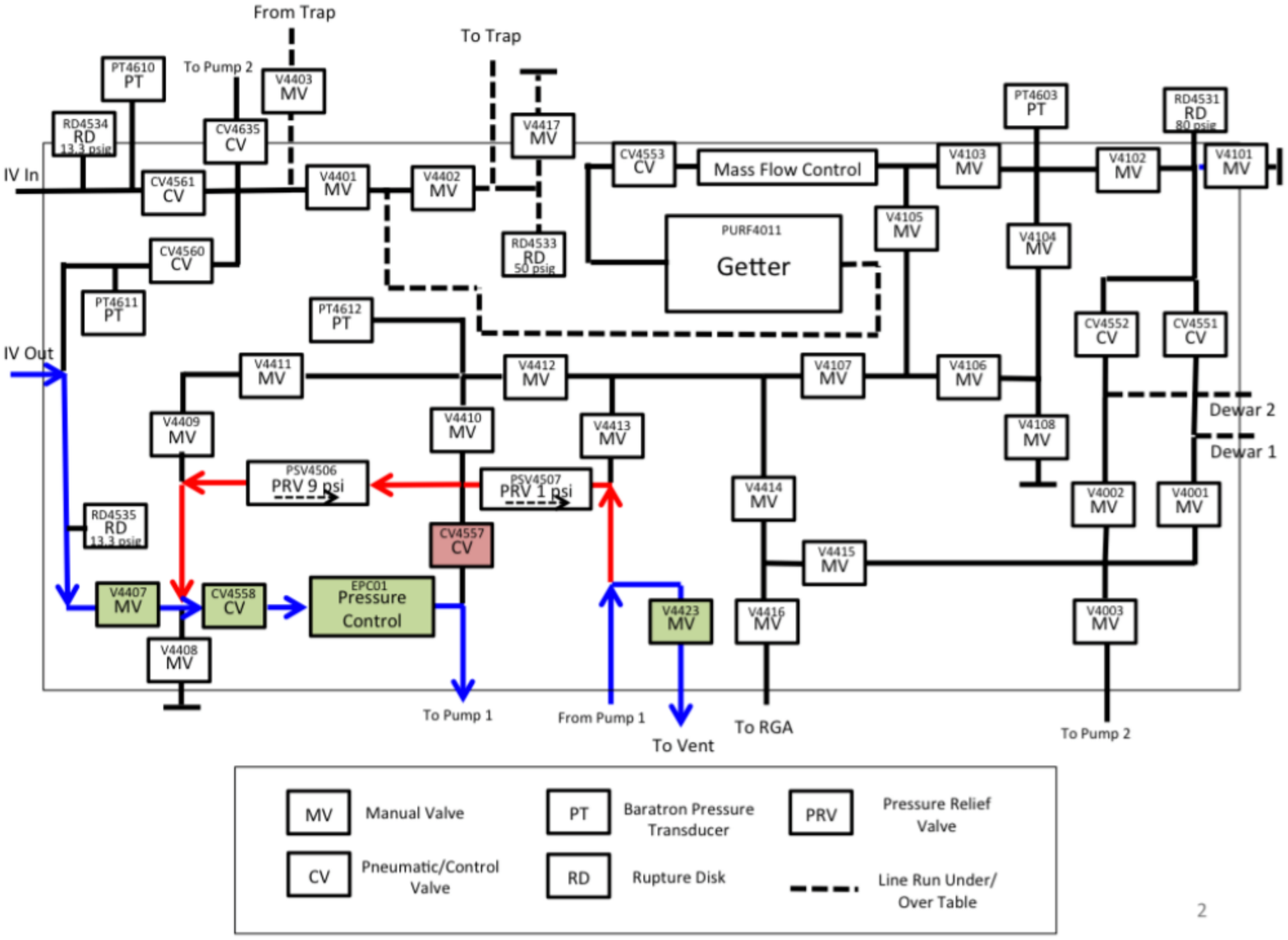}
\caption{ A cartoon schematic showing the cause of the air leak. }
\label{fig:fleak}
\end{figure}
\section{Trigger rate}
Trigger rate is another important indicator to understand the functionality of the whole system. The types of event can be roughly categorized as follows : 
\begin{itemize}
	\item \textbf{Raw trigger rate :} Total tigger rate calculated from events pass NHit trigger (at least 5 PMTs has hit in the coincident window).
	\item \textbf{Pass all cuts trigger rate :} Trigger rate of events passed all cuts (baseline, trigger time, saturation, trigger pileup).
	\item \textbf{Cherenkov trigger rate :} Events with charge ratio > 0.5 and Fprompt > 0.5.
	\item \textbf{Instrument event trigger rate :} Events with charge ratio >0.5 and Fprompt < 0.5.
	\item \textbf{ESR trigger rate :} Events with charge ratio < 0.5 and Fprompt > 0.5.
	\item \textbf{$^{39}$Ar trigger rate :} Events with charge ratio < 0.5 and Fprompt <0.5.
\end{itemize} 
The naming convention inherits from the vacuum data, the name indicates roughly the type of events appears in the vacuum data except $^{39}$Ar events which just appears in the gas runs. The trigger rate from each category are plotted against run number for clarity  as shown in Fig. \ref{fig:triggerrateall}.\par
\begin{figure}[htbp]
\centering
\graphicspath{{./fig/Warm_gas/}}
\includegraphics[scale=0.4]{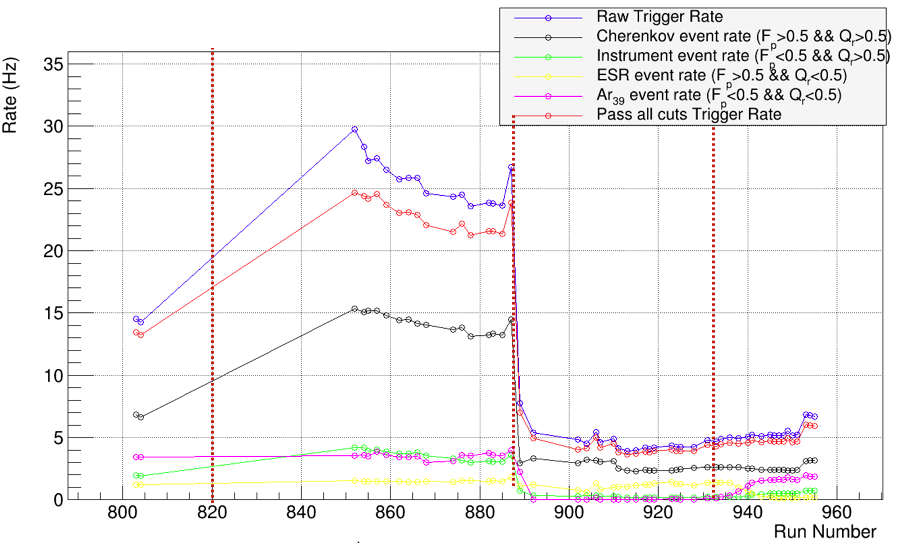}
\caption{ Trigger rates for different type of data in the cold gas runs. The rede dashed line around run 820 indicates the HV of PMTs increased by 50V. The dashed line around run 890 indicates the temperature of bottom of IV reaches liquefaction point. The red dashed line around run 930 indicates the beginning of pump and purge cycle. }
\label{fig:triggerrateall}
\end{figure}
The trigger rate starts off 15 Hz in the beginning of cold gas data taking. In the first two runs, the PMT gains were low, thus 50 V of HV was added to each PMT.  After gain adjustment, the raw trigger rate rises to near 30 Hz. The drops around run 850 due to observed very low gain of extra PMTs (1,3,12,26,41,44,63,75,80,89). This bring down the raw trigger rate to around 25 Hz. The raw trigger rate (charge per trigger) for selected PMTs is plotted against the run number and zoom in to the vicinity runs as shown in Fig. \ref{fig:triggerratepmts} (Fig. \ref{fig:triggerratechargepmts}). The PMTs plotted here except PMT 2,5,73, are having very low gain. Noticed that there's drop near the middle of the plot. The exact reason for this drop is unknown due to no data taking during that period of time. However, these malfunctioned  PMTs has way lower noise rate ($\sim$ 60 Hz) than the rest of PMTs ($\sim$ 400 Hz), indicating either they still have very low gain or some other unforeseeable reasons.\par
\begin{figure}[htbp]
\centering
\graphicspath{{./fig/Warm_gas/}}
\includegraphics[scale=0.32]{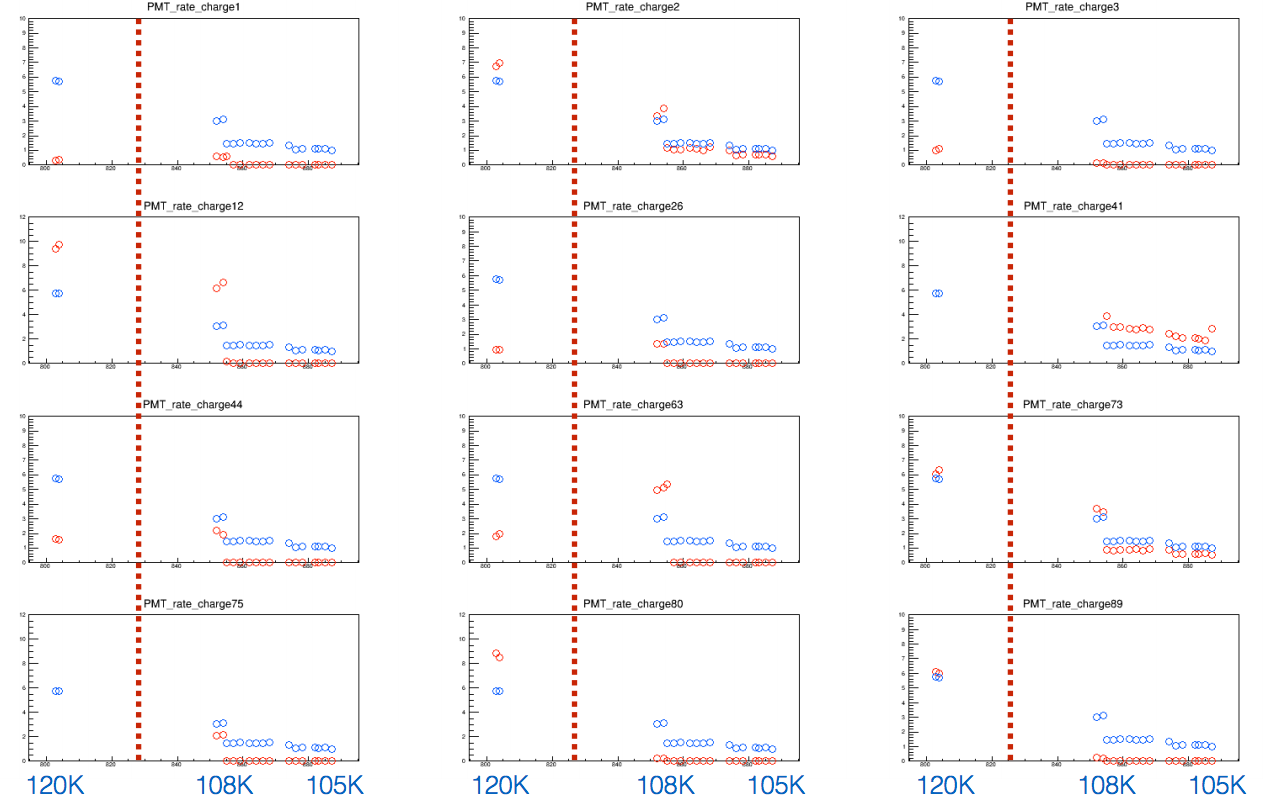}
\caption{ Trigger rates for selected PMTs (see text) vs run number. The blue circle is from PMT 5 (well functioned) serve as a reference. The red dashed line indicates the timing of rising PMT gain. PMT 2 and 73 are well functioned PMTs.}
\label{fig:triggerratepmts}
\end{figure}
The PMT gain seems decreasing before the leak. This may be due to the fact that temperature of the IV is getting colder and colder, thus the efficiency of  the multiplication through the dynode chain decreased. Therefore, the observed trigger rate decreased gradually according to the PMT gains. On the other hand, the trigger rate for different type of event are stable before the leak. The average temperature of IV is varying between 140 to 100 K and the average pressure is around 1500 mbar. The density of argon can be calculate according to the ideal gas law : 
\begin{ceqn}\begin{align}
\rho = \frac{P}{R_s T}
\end{align}\end{ceqn}
where $P$ is the pressure of IV and $R_s$ is the specific gas constant (for argon : 0.208 kJ/($kg\cdot K$)), and $T$ is the temperature of the IV. The beta decay rate of $^{39}$Ar is 1 bq/kg and IV is roughly sphere with 90 cm diameter. The expected trigger rate can be estimated from 1.985 to 2.748 Hz depends on the actual temperature and pressure of IV. The observed trigger rate for $^{39}$Ar is at reasonable range considering the detector efficiency. \par
\begin{figure}[htbp]
\centering
\graphicspath{{./fig/Warm_gas/}}
\includegraphics[scale=0.35]{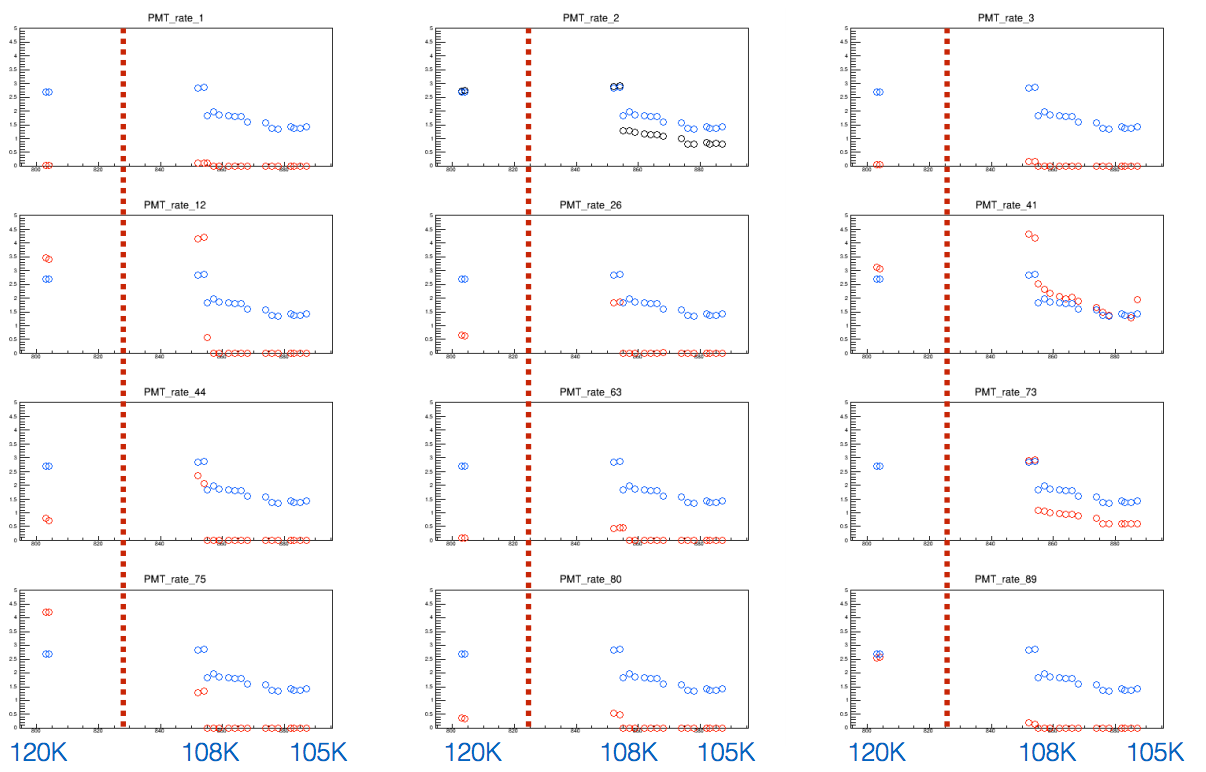}
\caption{  Charge pr trigger for selected PMTs (see text) vs run number. The blue circle is from PMT 5 (well functioned) serve as a reference. The red dashed line indicates the timing of rising PMT gain. PMT 2 and 73 are well functioned PMTs.}
\label{fig:triggerratechargepmts}
\end{figure}
After the leak, the trigger rate of all types of event decreased. It is reasonable that the $^{39}$Ar event rate decreased to zero since the impurity quenching the triplet states results in no or very little late light. The same for the instrument event rate which mostly come from the PMT discharging events. While the late light of scintillation events is gone, there is no source of event with low Fprompt to produce the PMT discharging. Therefore the discharging events are only exist in high Fprompt region. However, the raw trigger rate after the leak is less than third of the rate before the leak. This might be due to the decreasing PMT gain. Figure \ref{fig:triggerratechargepmtsdrop} shows the PMT charge per trigger for selected PMTs, noticed that since the beginning of leak, the PMT gain is decreasing gradually. Unlike the previous case, the temperature of IV is not decreasing and instead it is increasing due to the extra heat load from operating PMT continuously.
Therefore, the reason that the PMT gain is decreasing gradually might due to operating the PMT continuously for more than 400 hours during the pump and purge runs. This is confirmed by comparing the SPE value before and after the leak as shown in Fig. \ref{fig:triggerratechargepmtsdropspe}. Almost all PMT gain decreased by as much as factor of two.\par
\begin{figure}[htbp]
\centering
\graphicspath{{./fig/Warm_gas/}}
\includegraphics[scale=0.3]{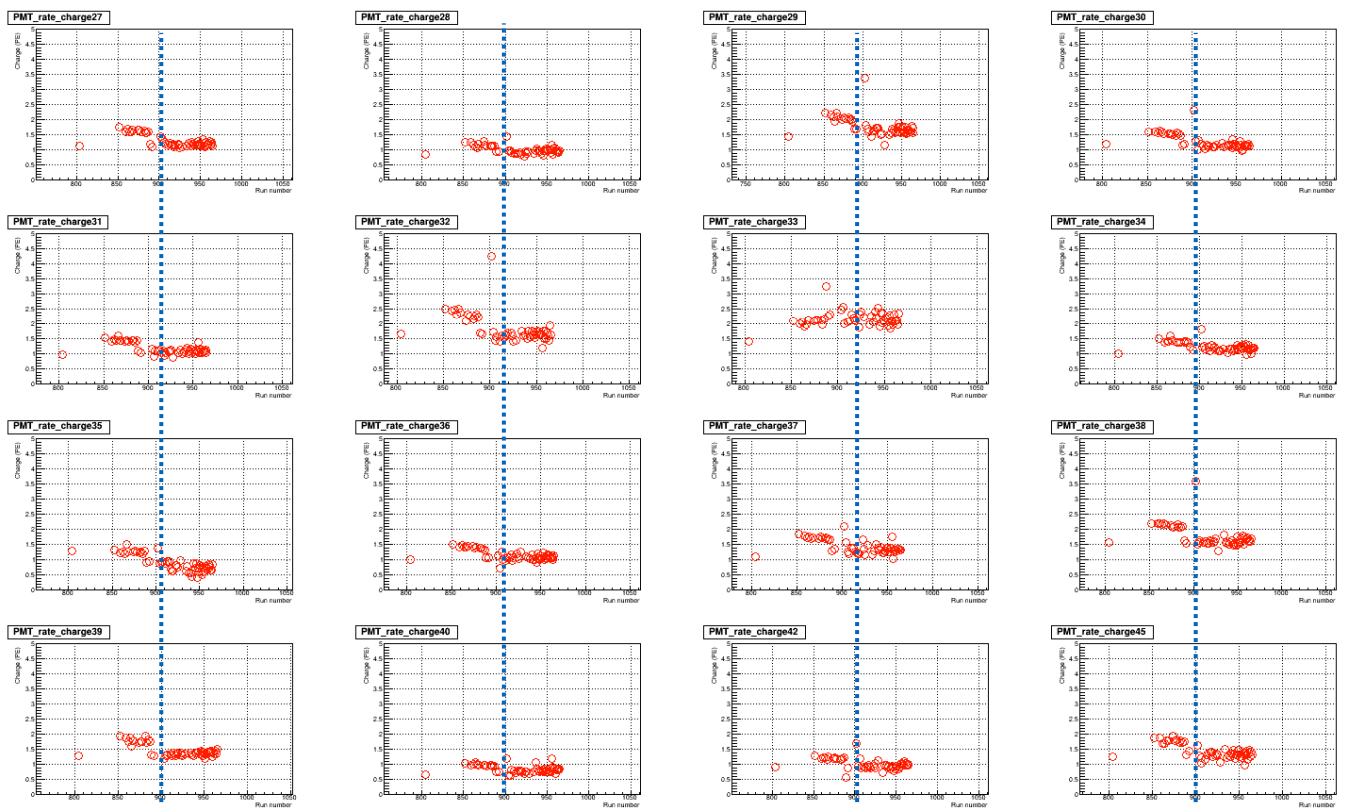}
\caption{  Charge pr trigger for selected PMTs (see text) vs run number. The blue dashed line separate the runs by the approximate timing of the leak. }
\label{fig:triggerratechargepmtsdrop}
\end{figure}
\begin{figure}[htbp]
\centering
\graphicspath{{./fig/Warm_gas/}}
\includegraphics[scale=0.3]{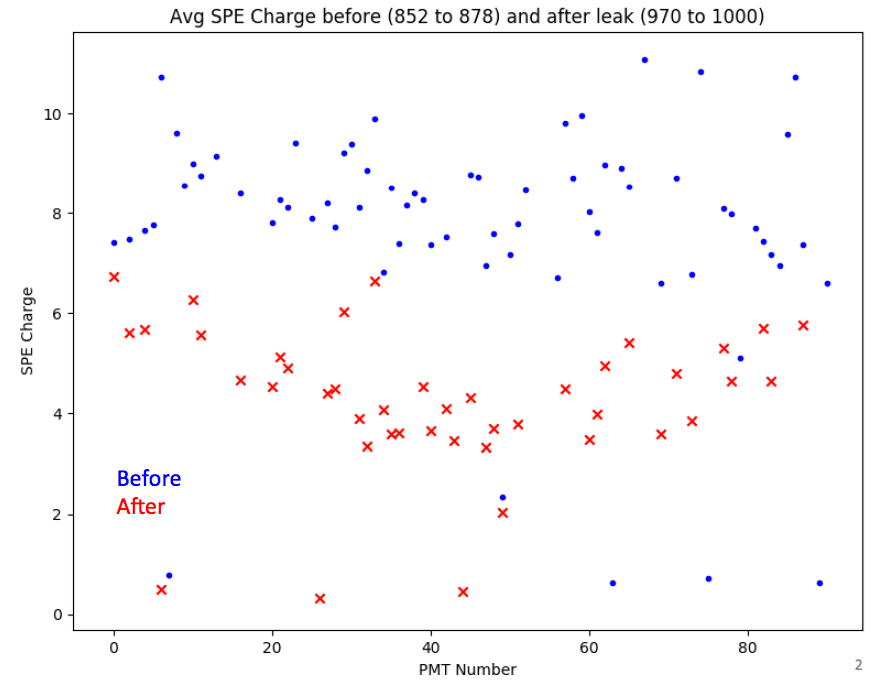}
\caption{  SPE value for runs before and after the leak. Figure from \cite{spevalue}. }
\label{fig:triggerratechargepmtsdropspe}
\end{figure}

\section{Instrument effect}
\subsection{WFD single fired}
As mentioned in Chapter \ref{sec:miscvacuum}, the huge pulse will induce the cross-talk within the same WFD board. These events also appears in the gas data and needed to be removed in order to improve the data quality. The event rate for single WFD fired events is increasing as the raw trigger rate increased. Figure \ref{fig:WFDsinglerate} shows the event rate for WFD single fired events before and after the leak. The cut is defined as in given events, if only the PMTs in the same WFD are fired, then the cut is true. Due to the PMTs relative position, the PMTs in the same WFD are on the same side of IV. Therefore, scintillation events in active volume can not trigger just one side of the IV due to the homogenous re-emission of TPB. This ensure that the events triggered just one WFD will highly unlikely from scintillation events. In addition, for the suspicious events, the maximum charge of the pulse in the prompt window (-20 to 80 ns) divided by total charge in the prompt window can further reduce the chance to cut on the scintillation events. Figure \ref{fig:WFDscatter} shows the scatter plot of above mentioned parameters. These two cuts can identify the single WFD fired events efficiently and preserved data quality. \par
The fraction of single WFD fired events before and after the leak is very similar. Figure \ref{fig:WFDabscatter} shows the scatter plot of charge-Fprompt distribution before and after the cut for events before and after the leak. The charge-Fprompt distribution for events removed by this cut is shown in \ref{fig:WFDcutevent}. This shows most single WFD fired events are with small charge. The Fprompt of these events are affect by random noise, after-pulsing and some strange events with unknown source mostly in PMT 41 as shown in Fig. \ref{fig:WFDstrange}.   Table \ref{table:WFDcut} summarize the number of events removed from the data and the fraction in each data sets. 
\begin{figure}[htbp]
\hfill
\subfloat[]{\includegraphics[width=7cm]{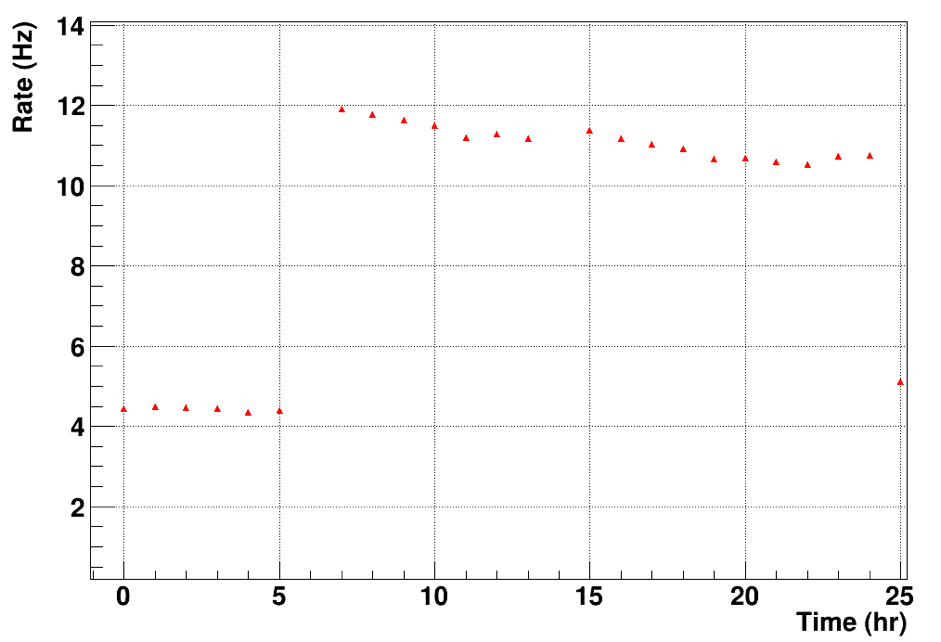}}
\hfill
\subfloat[]{\includegraphics[width=7cm]{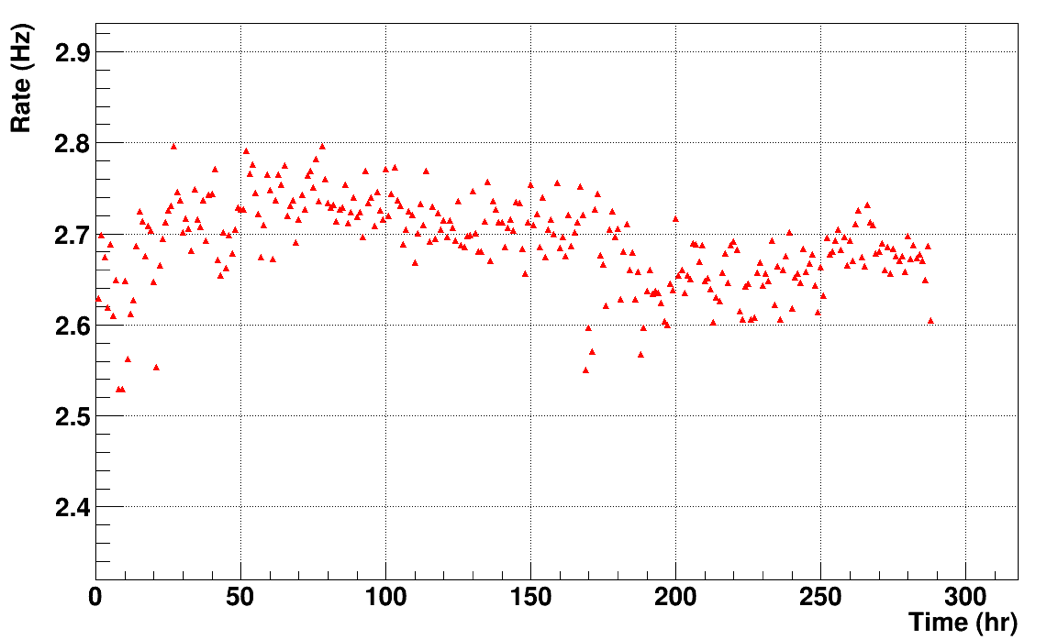}}
\hfill
\caption{ The trigger rate for WFD single fired events (a) Before the leak. Noticed that before hour 5, the raw trigger rate is 15 Hz and 25 Hz after hour 5. (b) After the leak. The raw trigger rate is around 6 Hz. }
\label{fig:WFDsinglerate}
\end{figure}

\begin{figure}[htbp]
\centering
\graphicspath{{./fig/Warm_gas/}}
\includegraphics[scale=0.35]{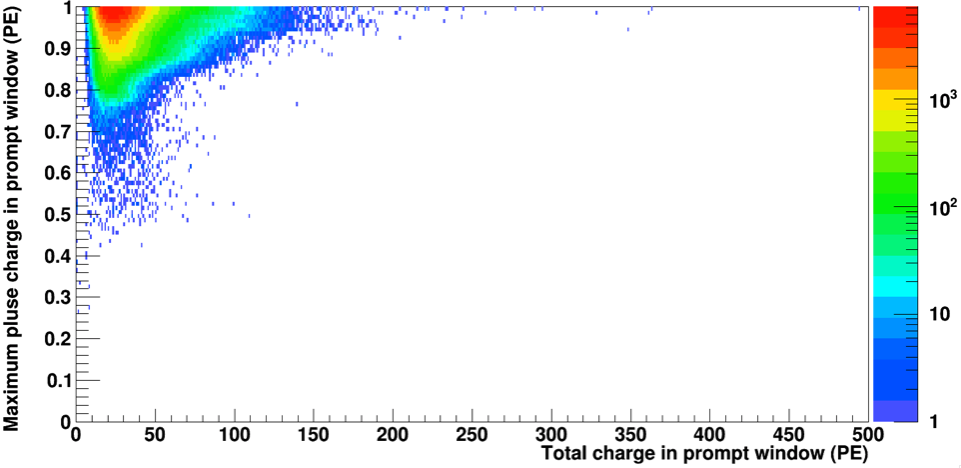}
\caption{  Maximum charge of the pulse in the prompt window divided by total charge in the prompt window vs the total charge in the prompt window. }
\label{fig:WFDscatter}
\end{figure}

\begin{figure}[htbp]
\hfill
\subfloat[]{\includegraphics[width=7cm]{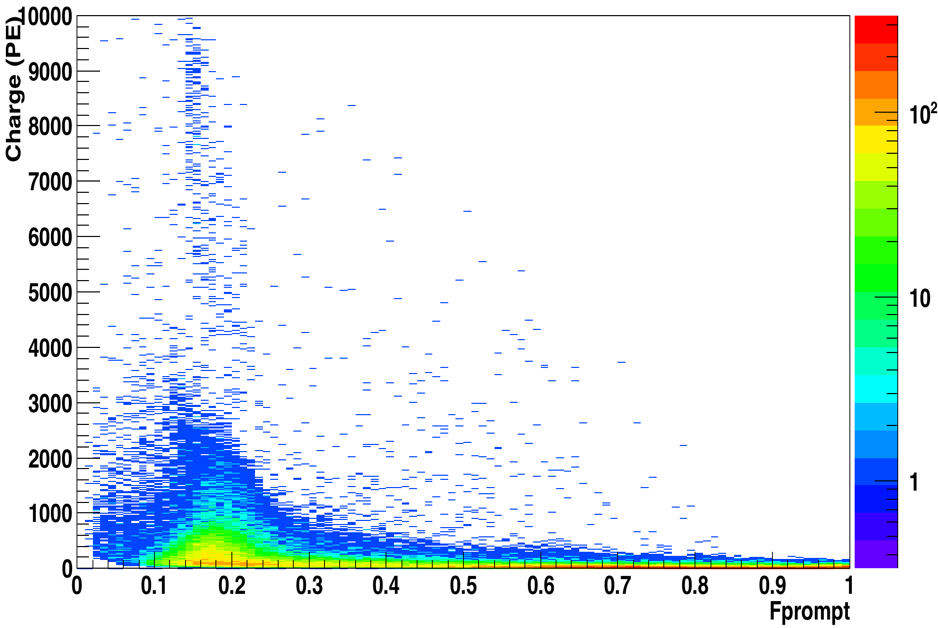}}
\hfill
\subfloat[]{\includegraphics[width=7cm]{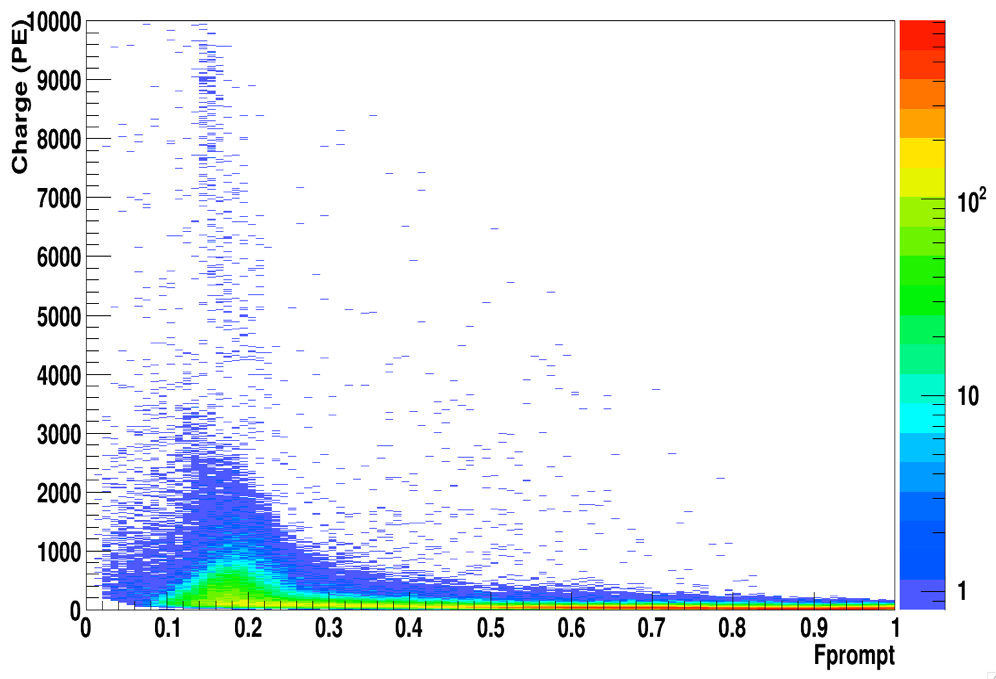}}
\hfill
\subfloat[]{\includegraphics[width=7cm]{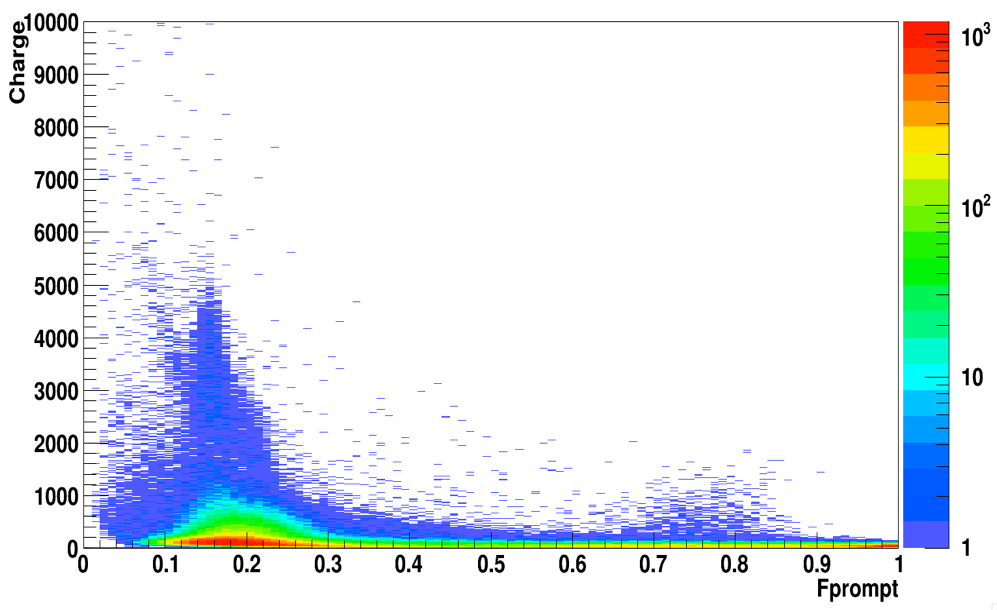}}
\hfill
\subfloat[]{\includegraphics[width=7cm]{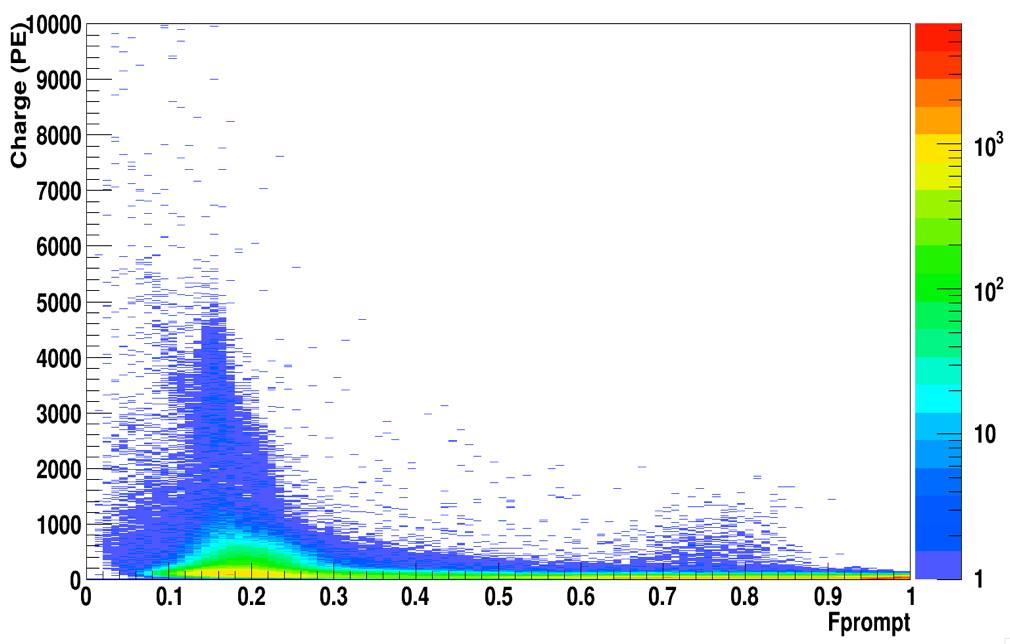}}
\hfill
\caption{ Charge-Fprompt distribution for (a) Events before the leak and before the cut. (b) Events before the leak and after the cut.  (c) Events after the leak and before the cut. (d) Events after the leak and after the cut. }
\label{fig:WFDabscatter}
\end{figure}

\begin{figure}[htbp]
\hfill
\subfloat[]{\includegraphics[width=7cm]{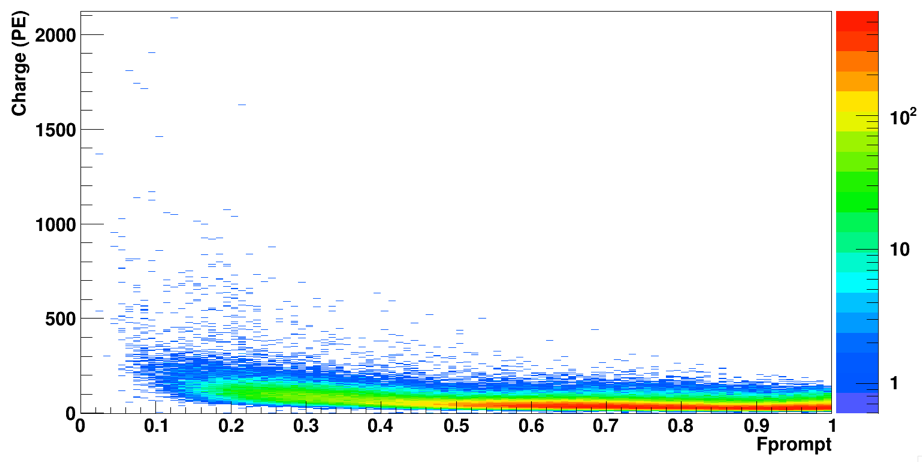}}
\hfill
\subfloat[]{\includegraphics[width=7cm]{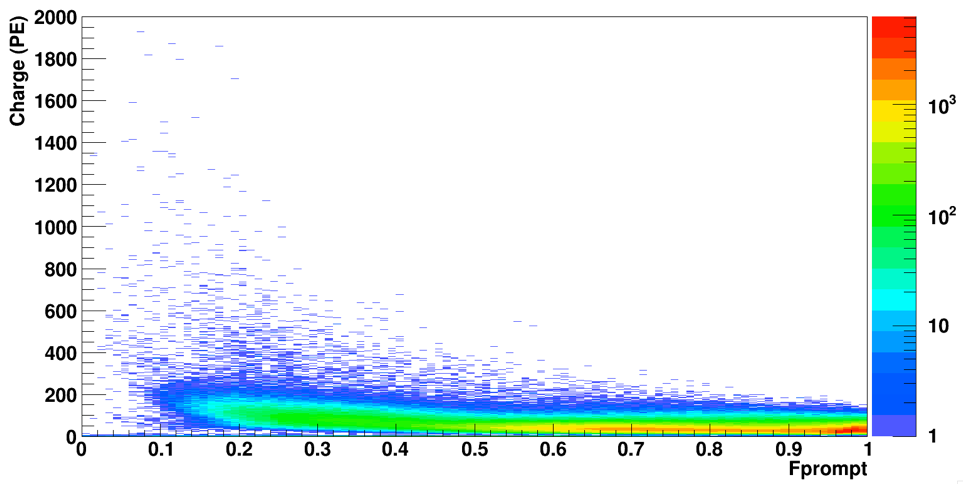}}
\hfill
\caption{ Charge-Fprompt distribution for events removed by the cut. (a) Before the leak. (b) After the leak. }
\label{fig:WFDcutevent}
\end{figure}

\begin{figure}[htbp]
\centering
\graphicspath{{./fig/Warm_gas/}}
\includegraphics[scale=0.35]{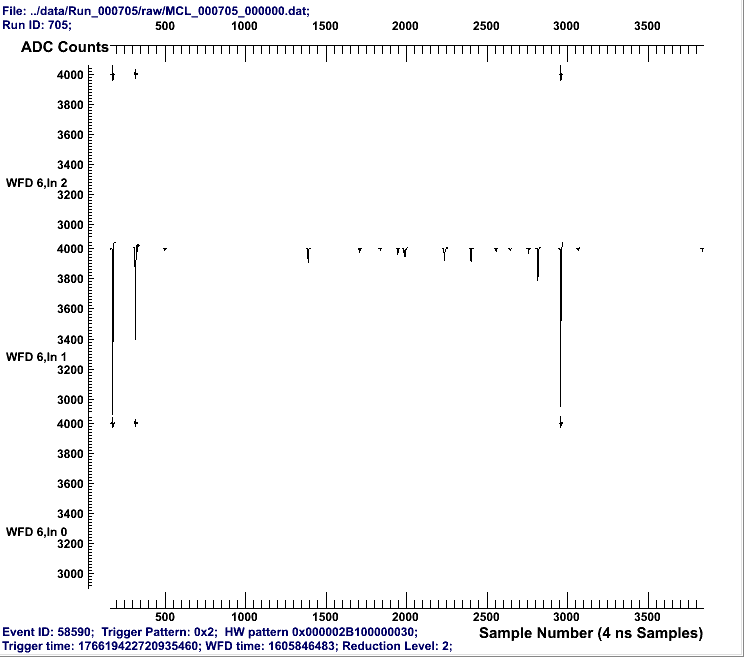}
\caption{  Strange pulses in the PMT. }
\label{fig:WFDstrange}
\end{figure}

\begin{table}[htbp]
\caption{Summary of  WFD single fired cut statistics} 
\centering 
\begin{adjustbox}{width=1\textwidth}
\begin{tabular}{c c c c c} 
\hline\hline 
&Total number of events before the cut& Total number of events after the cut & Total number of events removed& Fraction (\%)\\ [0.5ex] 
\hline 
Before the leak &2,061,316 & 1,134,626 & 926,690 & 45.00\\
After the leak &6,426,065 & 3,636,925 & 2,789,140 & 43.40\\
\hline\hline
\end{tabular}
\end{adjustbox}

\label{table:WFDcut} 
\end{table}
\subsection{Pulse-cut algorithm} \label{app:pulsecut}
When operating PMT in the gaseous argon, the PMT discharging through GAr is observed\cite{gerda}. The GAr provided a preferential route for the current at anode of PMT to discharge due to the low dielectric strength of GAr. This results in continuously register pulses throughout the trigger window and subsequently fake the long triplet lifetime. Figure \ref{fig:fdisexample}. shows the example events with one PMT channel is discharging through the GAr. These events usually has PMT discharging in one channel and with excessive charge such that bias the centroid reconstructed radius toward the given PMT. Therefore the radius cut is needed to remove these events from fitting the triplet lifetime. The pulse-cut is developed to identify these discharging event. Plotting the ratio of pulse height and pulse area against pulse area as shown in Fig. \ref{fig:fpcut}. Projecting these peak onto Y axis, we can see there's three distinct peaks.  The pulse shape of first peak (0 -0.05 in \ref{fig:fpcut}(b)) usually has wide base line and small amplitude as shown in Fig. \ref{fig:fpcut1}. These pulse shapes could results from baseline shift to produce the ringing. Second peak (0.05 -0.1 in \ref{fig:fpcut}(b)) includes the events has huge double pulse as shown in Fig. \ref{fig:fpcut2}, these might results from the scintillation events discharging through GAr. The third peak (0.15 -0.25 in \ref{fig:fpcut}(b)) contains pulse which has sharp pulse shape as shown in Fig. \ref{fig:fpcut3}, these pulse might results from the events with fast pulse discharging through GAr.\par
\begin{figure}[tbp]
\centering
\graphicspath{{./fig/}}
\includegraphics[scale=0.4]{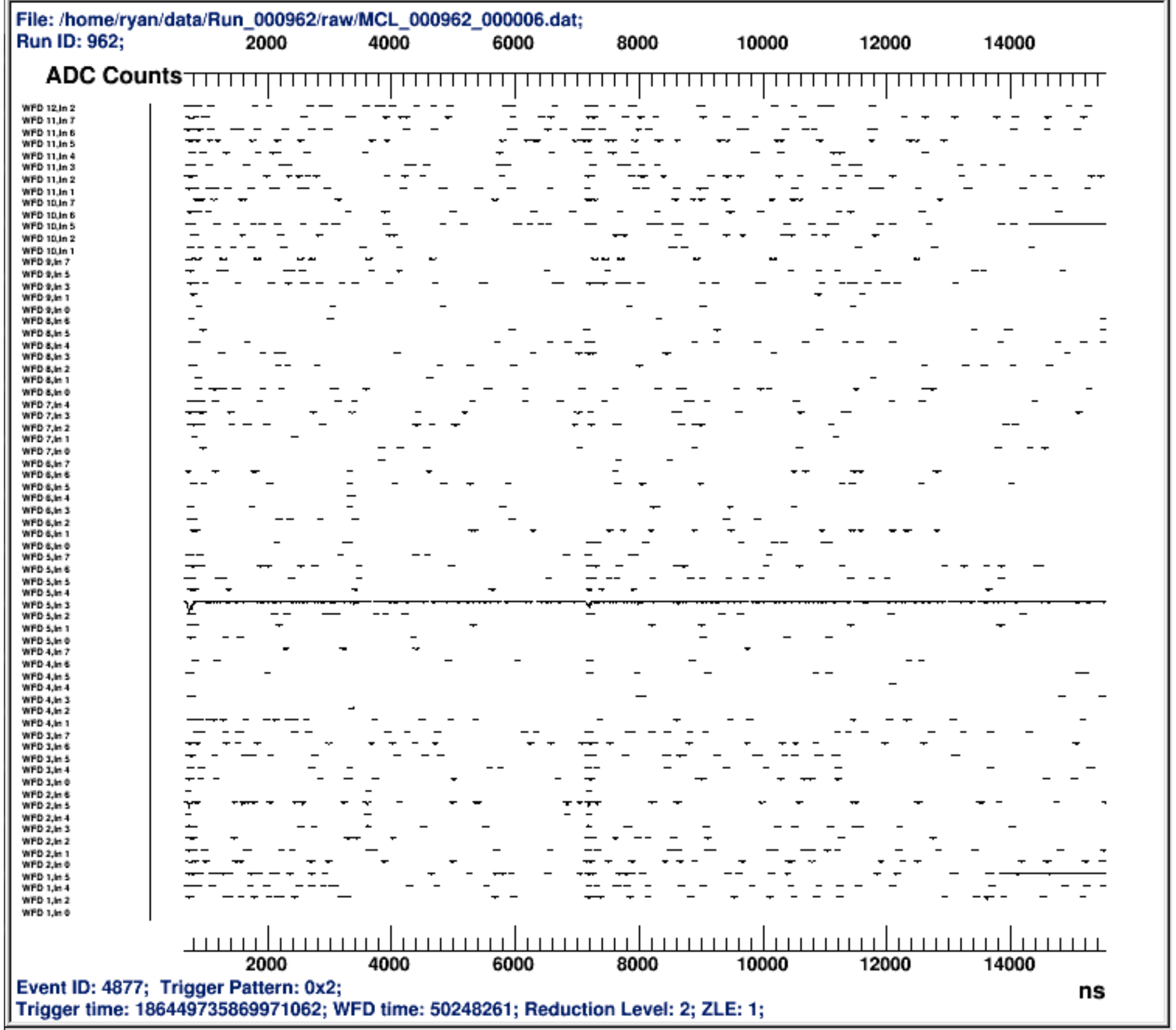}
\caption{ Example of discharging pulse viewing in  CLEANViewer. }
\label{fig:fdisexample}
\end{figure}
\begin{figure}
\hfill
\subfloat[]{\includegraphics[width=7cm]{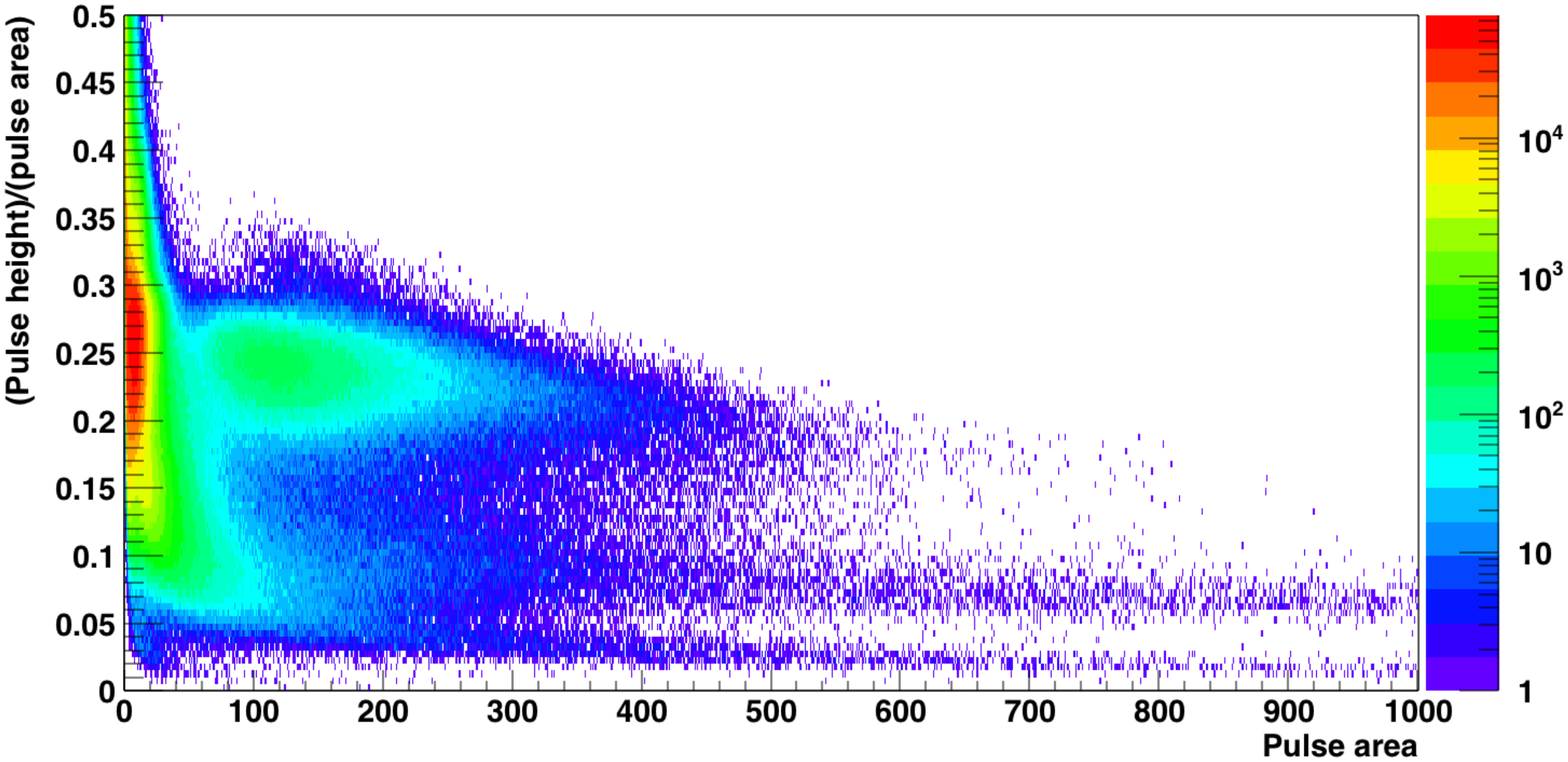}}
\hfill
\subfloat[]{\includegraphics[width=7cm]{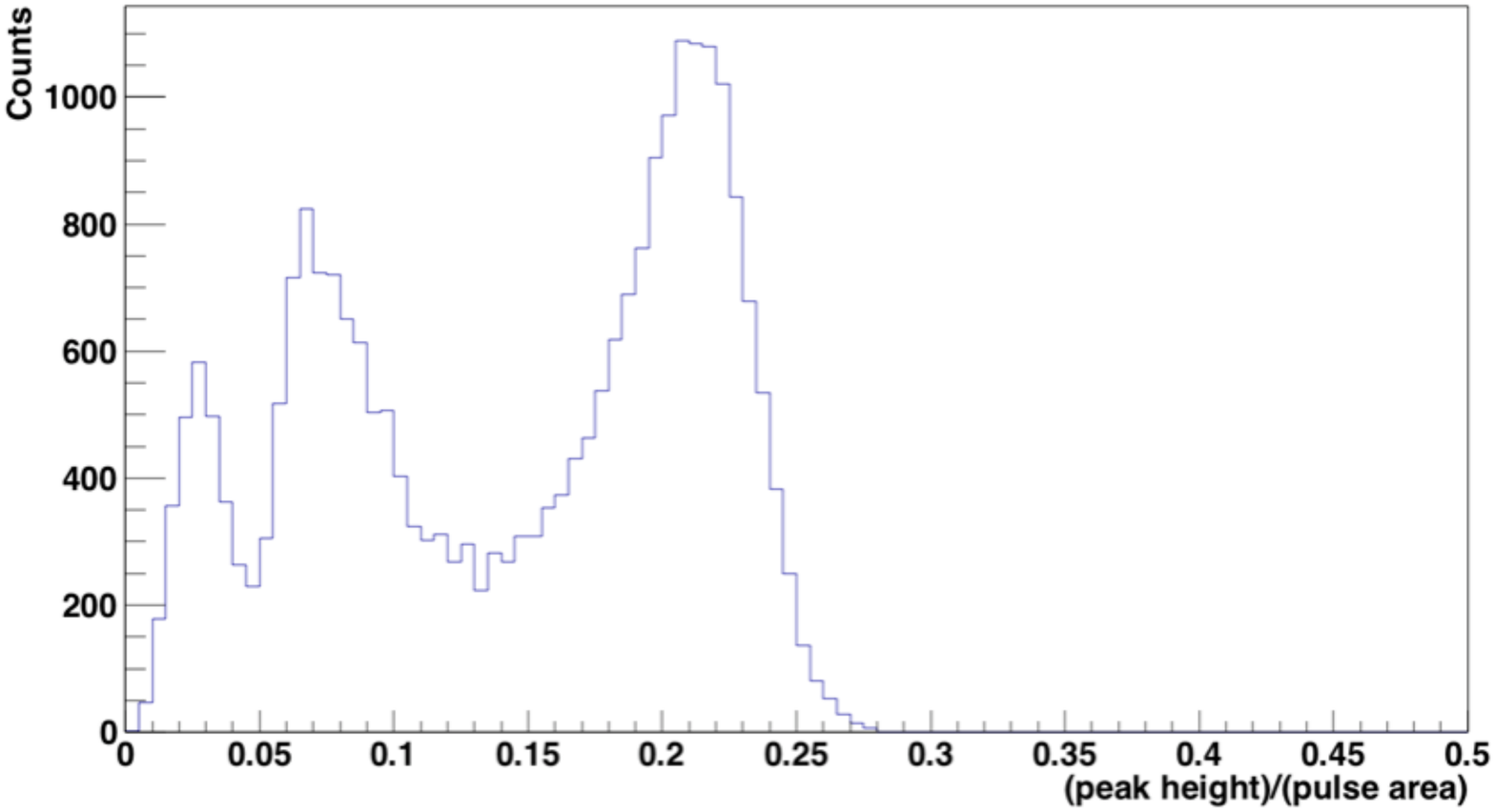}}
\hfill
\caption{(a) Ratio of pulse height divided by pulse area and plot against pulse area.  (b) Project the scatter plot on to Y axis. }
\label{fig:fpcut}
\end{figure}

\begin{figure}[tbp]
\centering
\graphicspath{{./fig/}}
\includegraphics[scale=0.4]{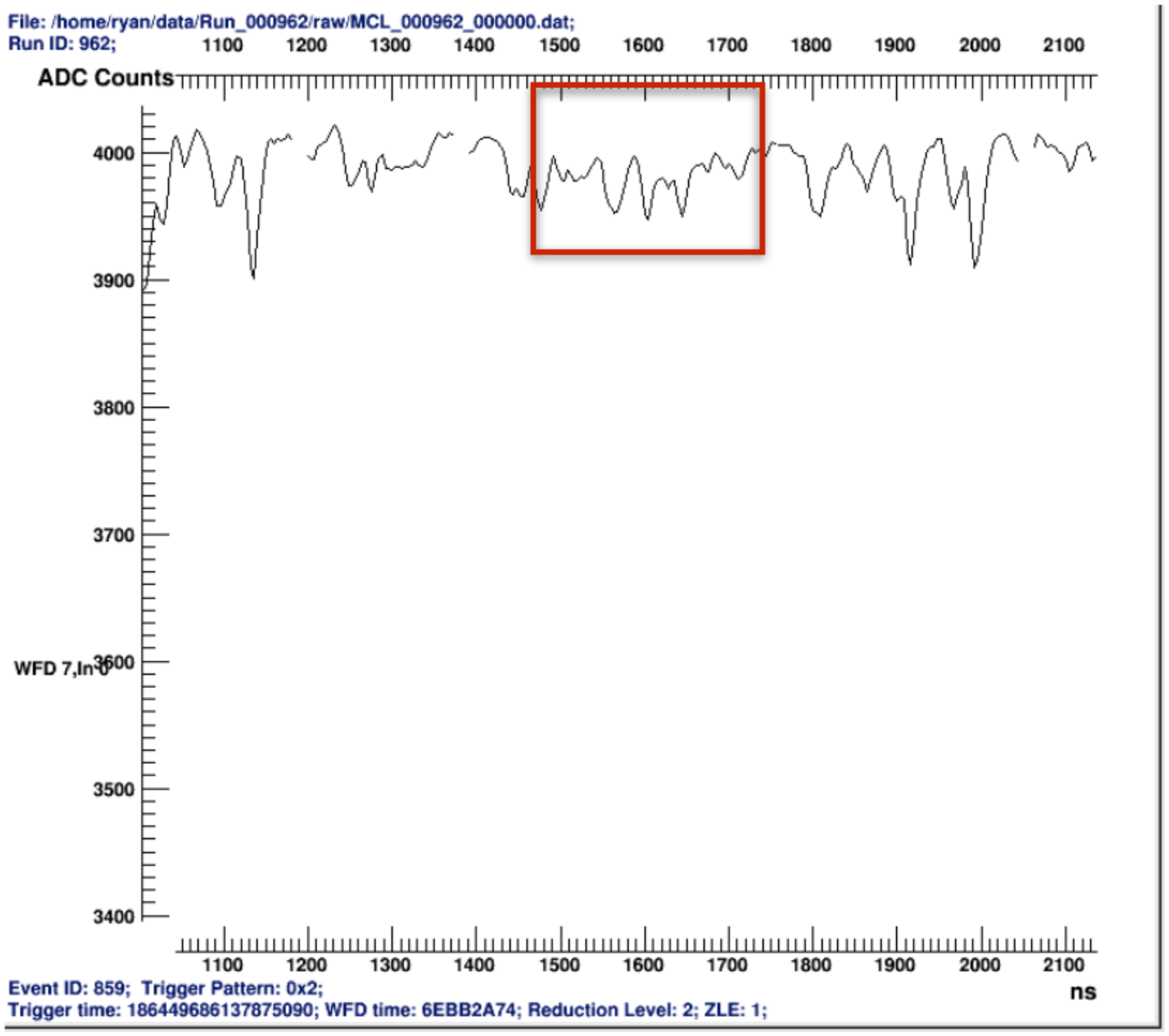}
\caption{ Example pulse shape of peak 1. The pulse in the red box is identify by the pulse cut. These pulse might results from baseline shift.  }
\label{fig:fpcut1}
\end{figure}

\begin{figure}[tbp]
\centering
\graphicspath{{./fig/}}
\includegraphics[scale=0.4]{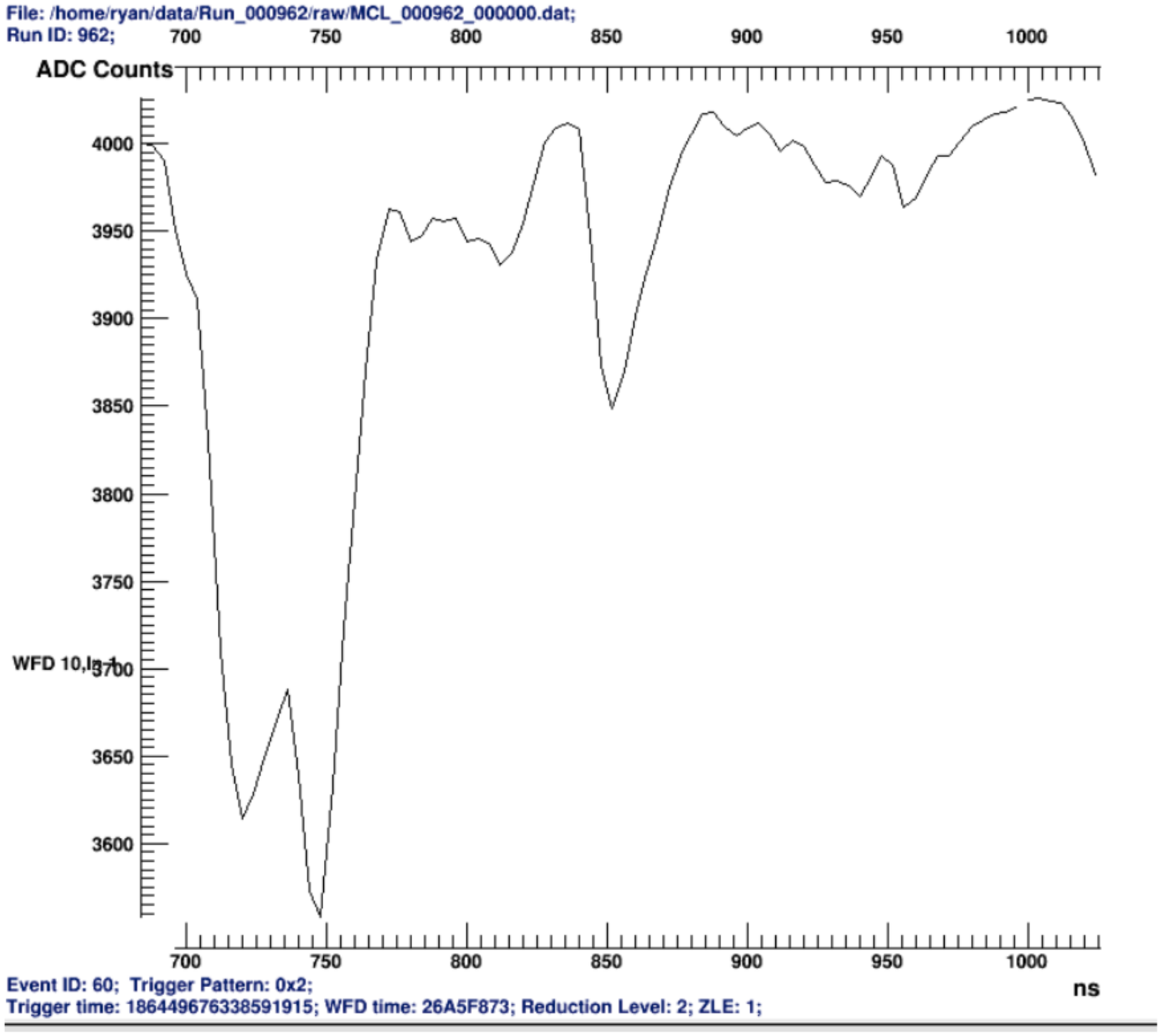}
\caption{ Example pulse shape of peak 2. These pulse might results from large current at anode discharging through the GAr.}
\label{fig:fpcut2}
\end{figure}
\begin{figure}[tbp]
\centering
\graphicspath{{./fig/}}
\includegraphics[scale=0.4]{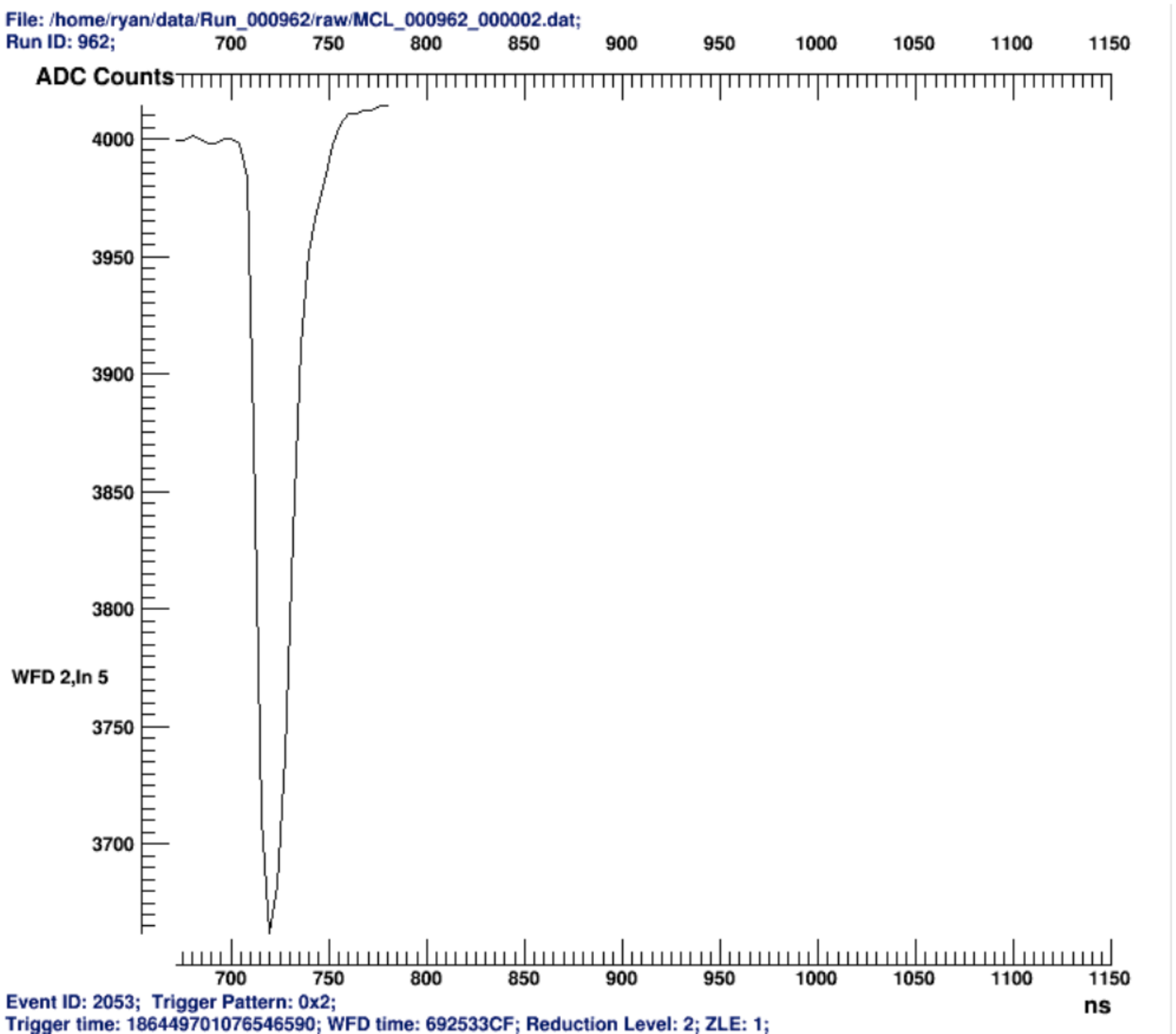}
\caption{ Example pulse shape of peak 3.}
\label{fig:fpcut3}
\end{figure}
The pulse cut is defined as following 
\begin{itemize}
	\item peak1 : Events fall in peak 1 region.
	\item peak2 : Events fall in peak 2 region and hits > 2 
	\item peak3 : Events fall in peak 3 region with maximum charge of the pulse > 100 pC.
\end{itemize}
A MC simulation of $^{39}$Ar is performed to estimate the cut efficiency. Figure \ref{fig:fpcutall} (a) shows the maximum charge in the PMTs plotted against the total charge of the event from MC simulation. Comparing to the same plot from the data as shown in Fig. \ref{fig:fpcutall} (b), there's a band of events which is not seen in the MC simulation. This shows the abnormal huge pulse results from the PMT discharging. After applied the pulse cut, these events are removed from the data as shown in Fig. \ref{fig:fpcutafter}. Although some huge pulse still remain in the data, but this shows the pulse cut removed most unwanted events and preserved good efficiency on the scintillation events. Table \ref{table:pulsecut} summarize the cut efficiency determined by the MC simulation. 

\begin{figure}
\hfill
\subfloat[]{\includegraphics[width=7cm]{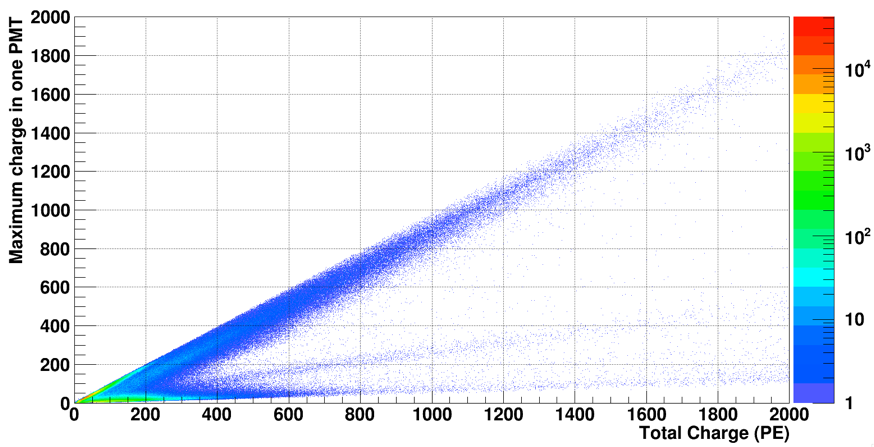}}
\hfill
\subfloat[]{\includegraphics[width=7cm]{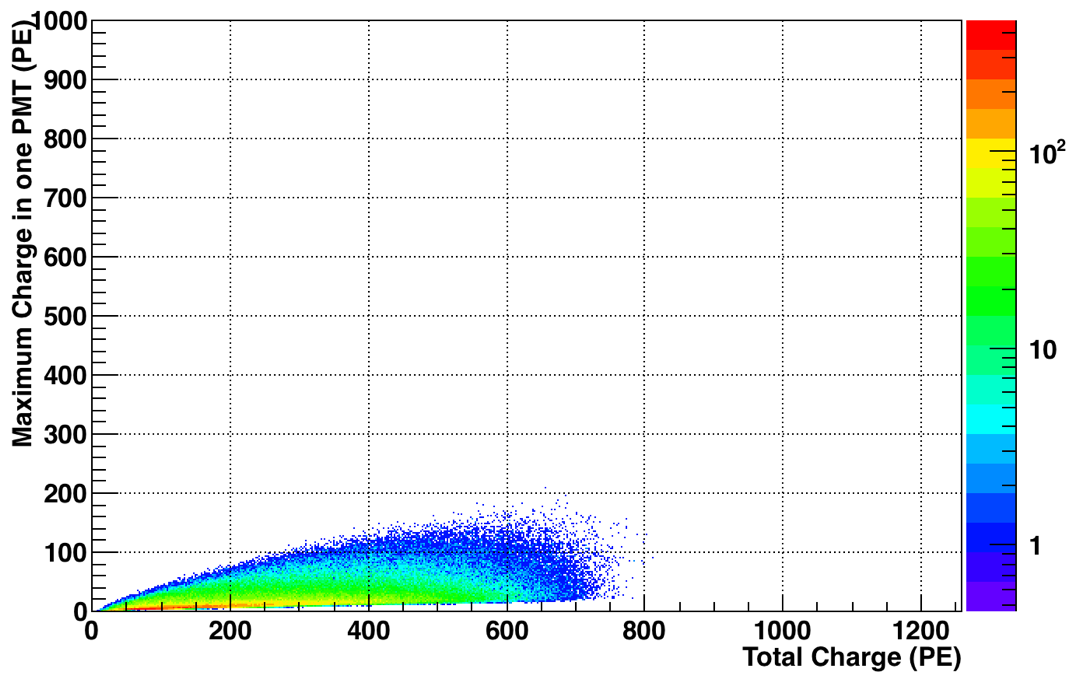}}
\hfill
\caption{Maximum charge in PMTs vs total charge in the event for (a) Data before the pulse cut.   (b) $^{39}$Ar events from MC simulation. }
\label{fig:fpcutall}
\end{figure}
\begin{figure}[tbp]
\centering
\graphicspath{{./fig/}}
\includegraphics[scale=0.4]{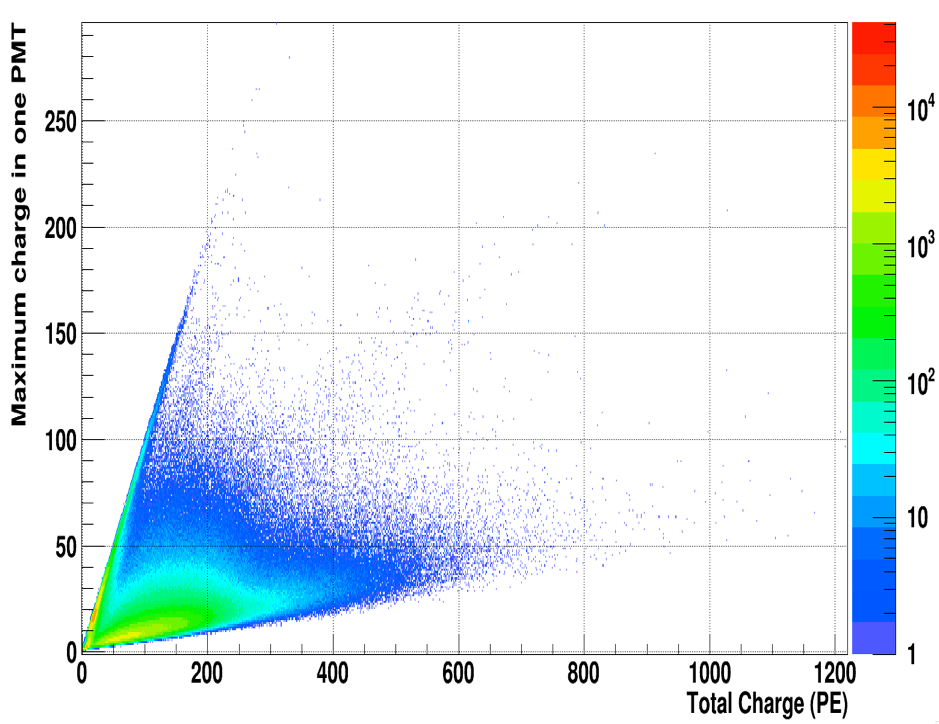}
\caption{Maximum charge in PMTs  vs total charge in the event after the pulse cut.}
\label{fig:fpcutafter}
\end{figure}


\begin{table}[htbp]
\caption{Cut efficiency of the pulse cut.} 
\centering 
\begin{adjustbox}{width=1\textwidth}
\begin{tabular}{c c c c} 
\hline\hline 
Cut & After the leak & before the leak & Simulation\\ [0.5ex] 
\hline 
Total number & 6,426,065 &2,061,316 &905,923\\
$Q_R$<0.6 and Fp<0.5 & 2,056,596 (32.004\%) & 352,208 (17.006\%) & 905,811 (99.988\%)\\
peak3 & 1,991,906 (30.997\%) & 326,029 (15.817\%) & 899,473 (99.288\%)\\
peak2 & 1,970,479 (30.664\%) & 312,625 (15,166\%) & 899,461 (99.287\%)\\
peak1 &1,970,445 (30.663\%) & 312,610 (15.165\%) &899,461 (99.287\%)\\
\hline\hline
\end{tabular}
\end{adjustbox}

\label{table:pulsecut} 
\end{table}
\subsection{Baseline Sag}
In the study of the pulse cut, 99\% of the PMT discharging events has maximum charge of the event in the PMT discharging channel as shown in \ref{fig:fsagchannel}. The rest of the events come from PMT baseline sag or multiple PMT discharging. When PMT gets larger amount of photons the photocathode is heavily depleted. During the time that photocathode restoring back to original potential, the low frequency baseline oscillation is presented. The digitizer's dynamic range is set to accept negtive voltage PMT pulse such that the positive voltage baseline oscillation will not be digitized.  This cause the charge integral value for channel has baseline sag artificially high as shown in Fig. \ref{fig:fsag}.
\begin{figure}[tbp]
\centering
\graphicspath{{./fig/}}
\includegraphics[scale=0.25]{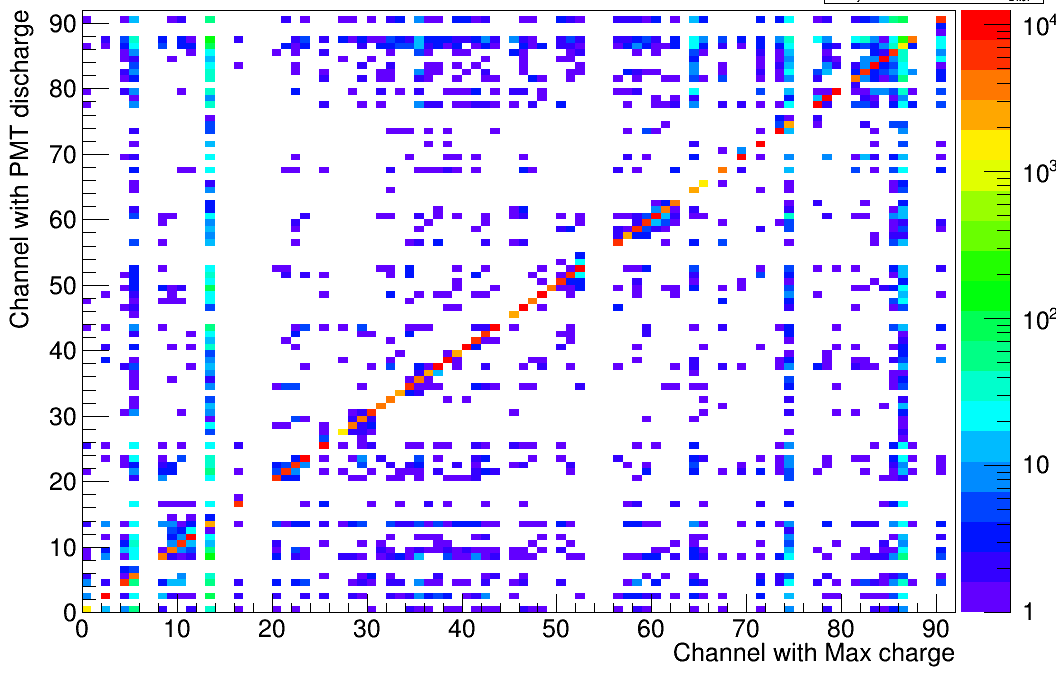}
\caption{Channel identified as PMT discharging vs PMT channel with maximum charge.}
\label{fig:fsagchannel}
\end{figure}
\begin{figure}[tbp]
\centering
\graphicspath{{./fig/}}
\includegraphics[scale=0.25]{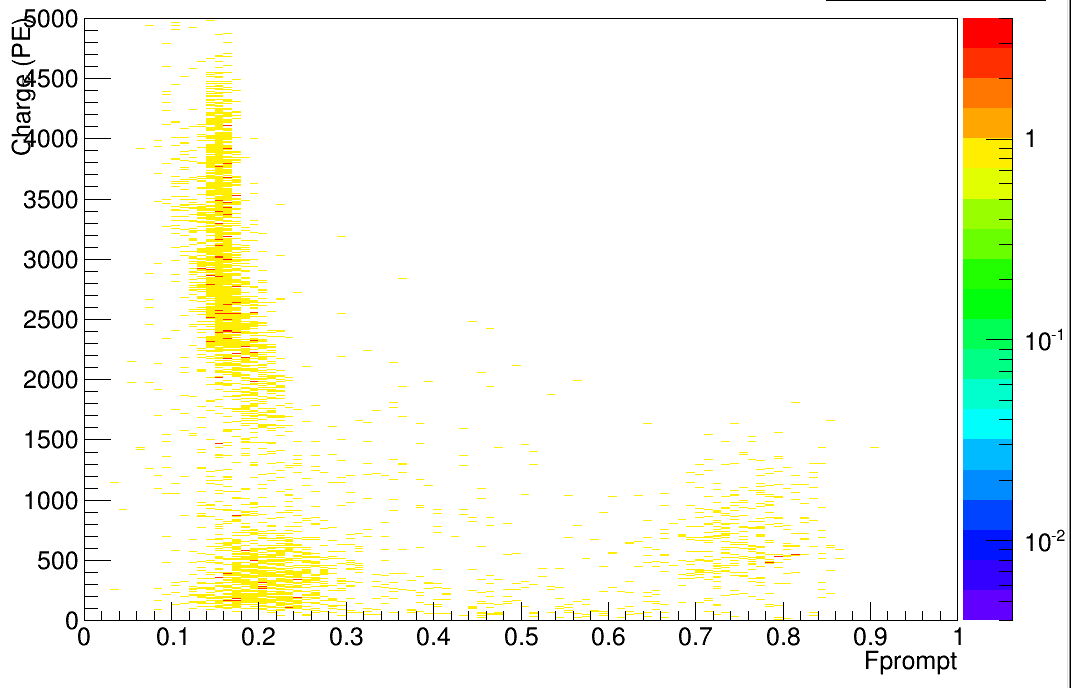}
\caption{Charge-Fprompt distribution for baseline sag events. Notice that the band in low fpormpt and high charge region are the results from the baseline sag which creates artificially high charge of the event.}
\label{fig:fsag}
\end{figure}
\subsection{Switching noise}
The DAQ system is continuously taking data after the leak was found. A type of noise event was found in the data. These pulses constantly present in some PMTs at around 2600 ns after the trigger time. Figure \ref{fig:fpulsetimeswitching} shows the pulse-time distribution from all PMTs, in the Run 918 no excessive pulses at around 2600 ns, but it emerged in Run 919. It was found these events are very similar to the high voltage switching noise which is studied previously\cite{hvswitchingnoise}. These bimodal pulses (Fig. \ref{fig:frawswitching}) origin from the power supply and picks up by the pulse finding algorithm. These events were found in the data and appear randomly in all PMTs. However, after the leak, these pulses are found just appear to some of the PMTs (PMT 26,35,41,42,43,44,45,46). Nonetheless, these can be easily removed from the data\cite{solutionswitching}.
\begin{figure}
\hfill
\subfloat[]{\includegraphics[width=7cm]{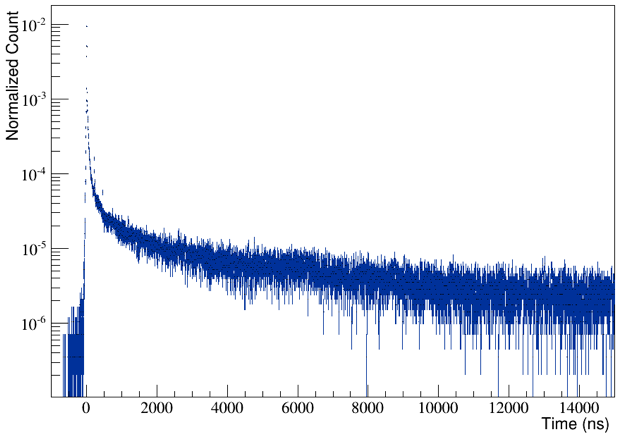}}
\hfill
\subfloat[]{\includegraphics[width=7cm]{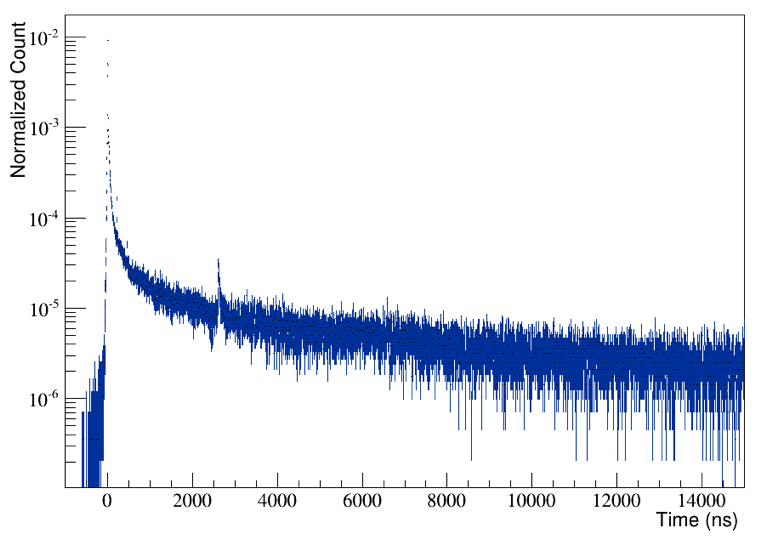}}
\hfill
\caption{ Pulse-time distribution for all PMTs. (a) Run 918. No excessive events at 2600 ns. (b) Run 919. Excessive events at 2600 ns.}
\label{fig:fpulsetimeswitching}
\end{figure}
\begin{figure}[tbp]
\centering
\graphicspath{{./fig/}}
\includegraphics[scale=0.3]{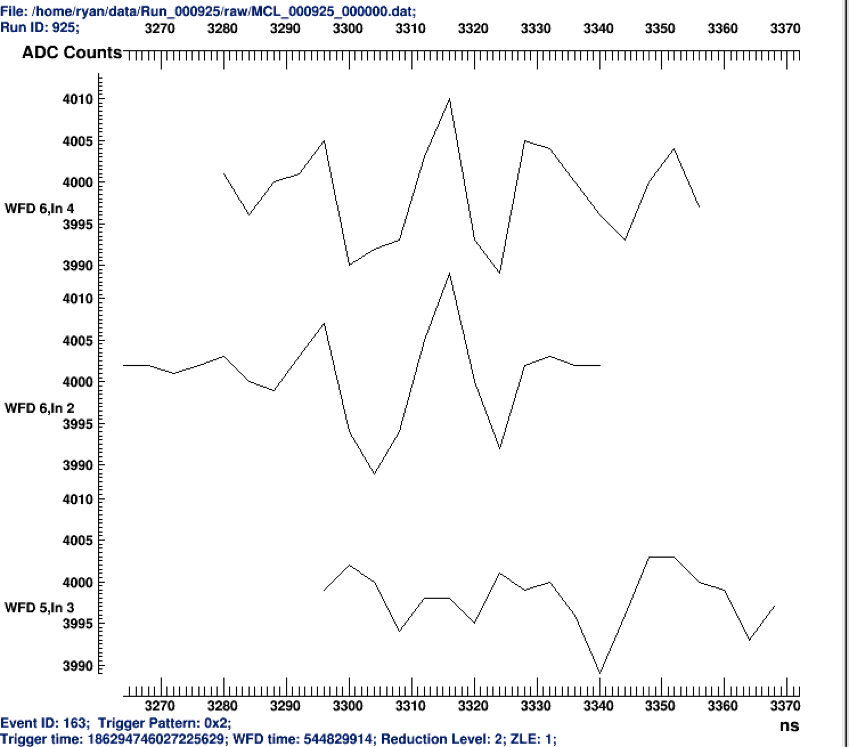}
\caption{Raw waveform of switching noise.}
\label{fig:frawswitching}
\end{figure}
\section{Relative light yield}
The PMTs are calibrated using single photoelectrons (SPE) from late scintillation light which is low intensity and therefore dominated by single photons (see Figure  \ref{fig:fexampleraw}). Using the timing p.d.f. from Figure \ref{fig:fscin}, we can calculate the pile-up probability and identify the region dominated by single photons and get the SPE charge distribution for each PMT. The SPE charge distribution contains two components, the contribution from real single photons and background (dark hits, electronic noise etc.). The background component is included in the fit procedure. The background model is obtained using a random, periodic trigger. The SPE distribution is fitted using double-gamma function\cite{Caldwell2012} as shown in Figure \ref{fig:fexampleSPEfit}.\par
\begin{figure}[tbp]
\centering
\graphicspath{{./fig/}}
\includegraphics[scale=0.5]{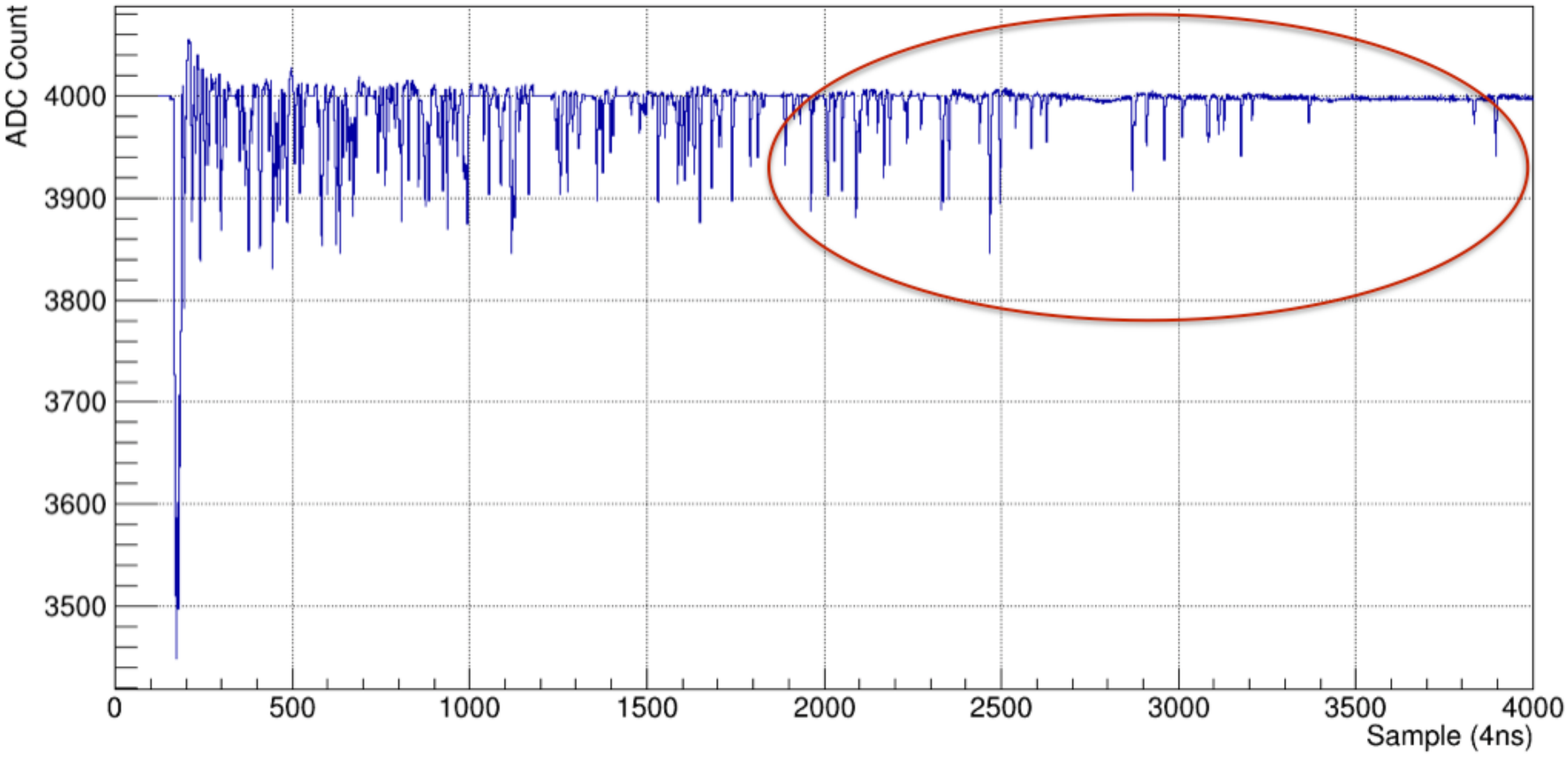}
\caption{Raw waveform of scintillation event. The region circled is dominated by single photon pulses.}
\label{fig:fexampleraw}
\end{figure}

\begin{figure}[tbp]
\centering
\graphicspath{{./fig/}}
\includegraphics[scale=0.5]{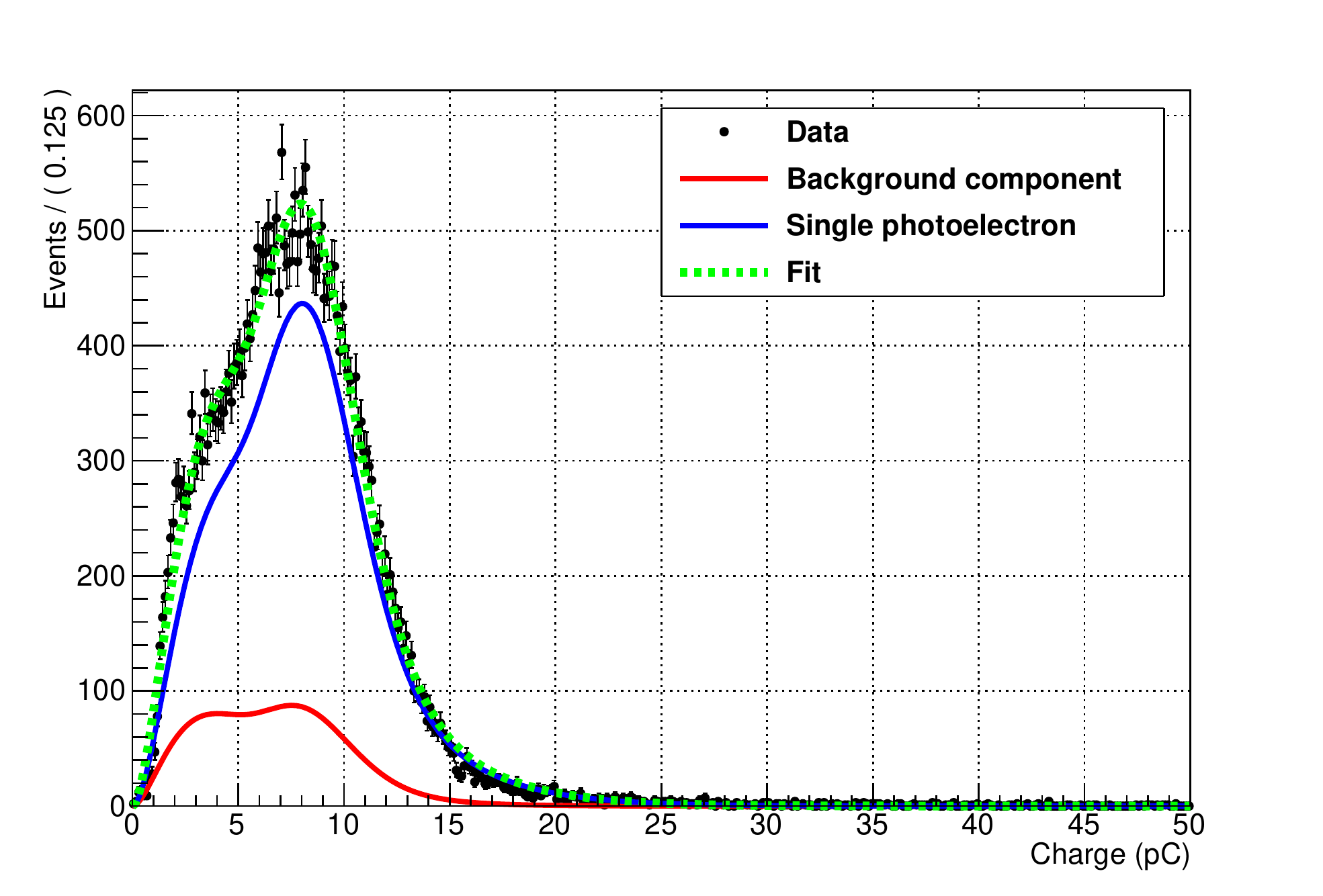}
\caption{Example of single photoelectron charge distribution fitted by double gamma distribution. Red curve represent the background component , blue curve is the contribution from single photoelectron and the green dashed line is from the fitting. The estimated SPE value from the fit for this distribution is 7.96 pC. }
\label{fig:fexampleSPEfit}
\end{figure}
The average charge of argon scintillation (Fp<0.5 and $Q_R$<0.6) events for each run can be calculated with calibrated charge distribution. Figure \ref{fig:chargediffruns} shows the charge distribution from different run at different stage of detector. For the run before the leak (Run 852), more events with higher charge in the charge distribution. When the triplet component quenched by impurity (Run 889), the high energy event disappeared.  For runs in the pump and purge cycle, with recovering of triplet lifetime which means the recovery of triplet component, the high energy events are increased. Noticed that the Run 852 and 955 both has triplet lifetime at around 3500 ns, but Run 955 has less high energy events. This can be seen in the plot of average charge as a function of run number as shown in Fig. \ref{fig:averagechargepmtgas}. Around run 820, the HV of PMTs were rose by 50 V, results in increased average charge. When the triplet state started quenched by impurity, the average charge decreased. With high impurity of the gas, huge fraction of light yield is suppressed and the detector starts to recover the light yield and triplet lifetime during the pump and purge. Noticed that even when the triplet lifetime is recovered to the previous value ($\sim$ 3500 ns), the light yield just barely back to the half as much as previous value. This difference might be explained by the difference of trigger rate. From previous section, the trigger rate before and after the leak is very different. The total raw trigger rate dropped from 25 Hz (before the leak) to 7 Hz indicating the contribution from some events is missing.\par

\begin{figure}[htbp]
\centering
\graphicspath{{./fig/Warm_gas/}}
\includegraphics[scale=0.3]{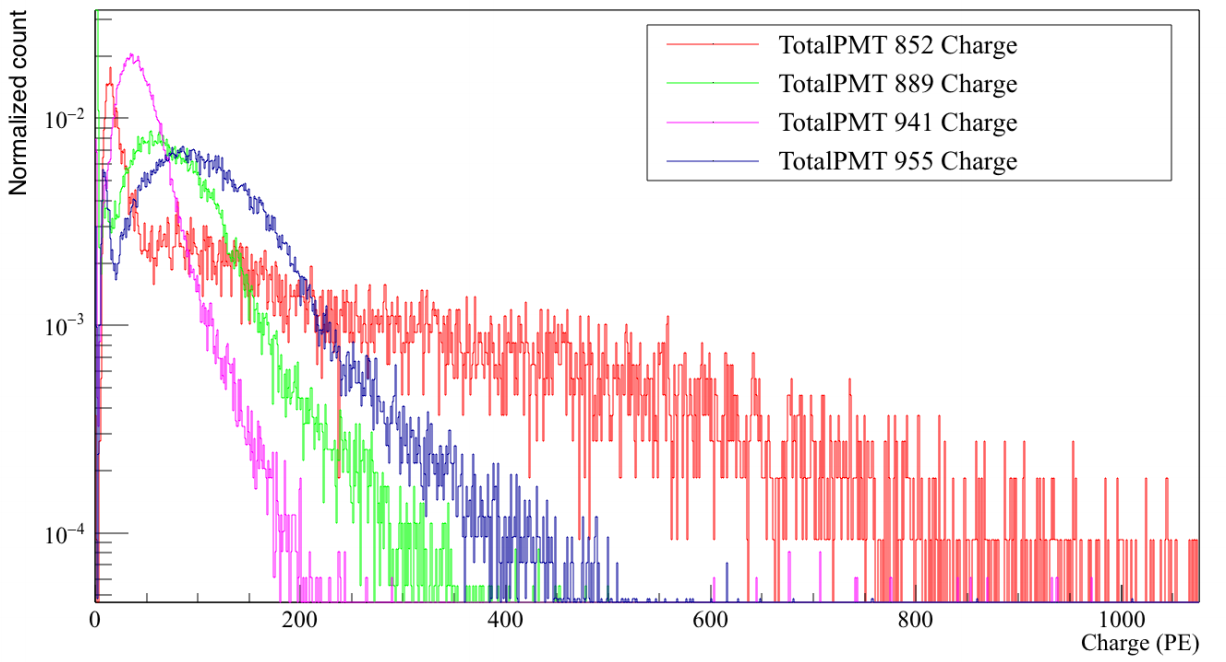}
\caption{  Charge distribution for different run. Run 852 : Before the leak with triplet lifetime at 3500 ns. Run 889 : Last run before triplet component disappeared with triplet lifetime at 2400 ns. Run 941 : pump and purge run with triplet lifetime at 1300 ns. Run 955 : pump and purge run with triplet lifetime at 3500 ns. }
\label{fig:chargediffruns}
\end{figure}
\begin{figure}[htbp]
\centering
\graphicspath{{./fig/Warm_gas/}}
\includegraphics[scale=0.4]{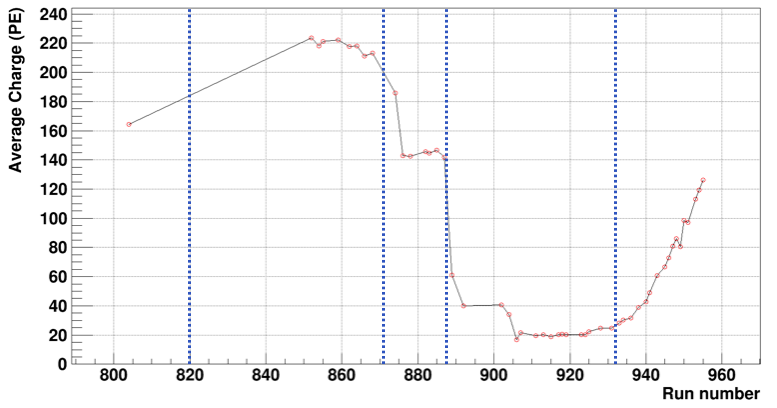}
\caption{  Average charge as a function of run number. The blue dashed lines denote the important changes of the detector. Run 820 : Raising HV by 50 V to each PMT. Run 870 : Triplet lifetime decreased to 3300 ns. Run 890 : First day of filling LAr. Run 930 : Beginning of pump and purge.}
\label{fig:averagechargepmtgas}
\end{figure}
After the a thorough study, there's a events from $^{39}$Ar but not fall in the cut region are unaccounted for the charge distribution. Figure \ref{fig:fpumppurgeruns} shows the charge ratio and Fprompt distribution for runs in the pump and purge cycle. Figure \ref{fig:fpumppurgeruns} (a) is from the earlier pump and purge run and (d) is the latter pump and purge run. Noticed that the events in low $Q_R$ and high Fprompt  in (a) is the $^{39}$Ar events with quenched late component. With the impurity pumped out of IV, $^{39}$Ar events recovering its late component and move gradually to the low Fprompt region throughout the course of pump and purge cycle. Finally when the triplet lifetime restore to the previous value, the group of events from $^{39}$Ar scintillation settles at low Fprompt region. A interesting phenomenon is observed, when the group of events in low $Q_R$ moving toward the low Fprompt, another group of events with high $Q_R$ and almost the same Fprompt move along with the $^{39}$Ar events toward the low Fprompt region. It is latter found that these events are mostly from $^{39}$Ar scintillation with one PMT channel with discharging such that it pushes the events to high $Q_R$ due to the fact that the discharging channel have artificially high charge. This also happens to the data before the leak but less frequent. Figure \ref{fig:dischargingexample} (a) shows the raw waveform of all PMTs from the discharging events. It is clear to see one channel has continuous pulses throughout the 16 $\mu$s window. Figure \ref{fig:dischargingexample} (b) shows the pulse-time distribution from discharging events with simple exponential fit. The fit result shows it is compatible with the lifetime acquired from $^{39}$Ar events and proved its origin.\par
\begin{figure}
\hfill
\subfloat[]{\includegraphics[width=7cm]{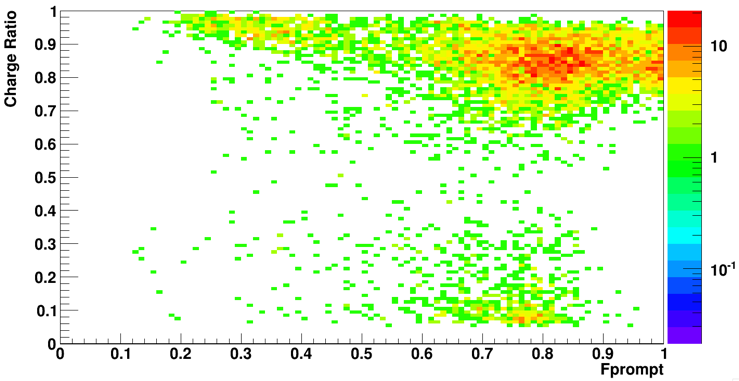}}
\hfill
\subfloat[]{\includegraphics[width=7cm]{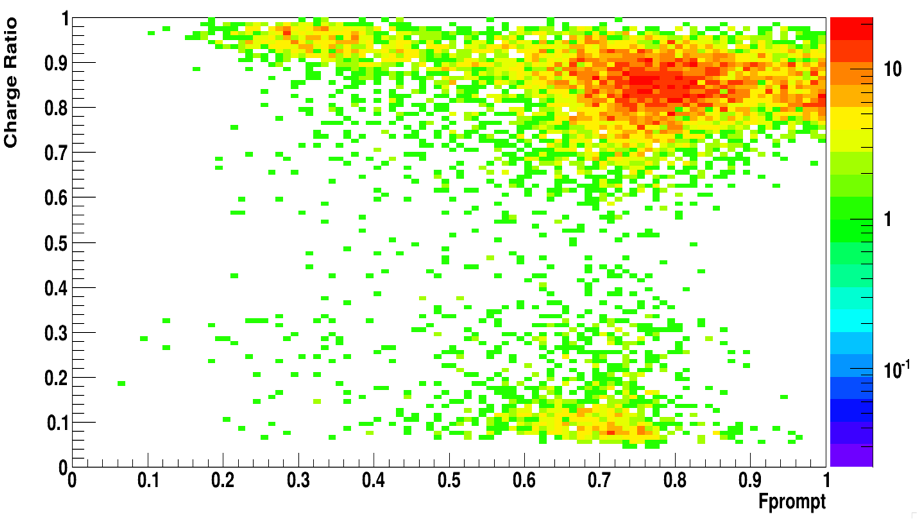}}
\hfill
\subfloat[]{\includegraphics[width=7cm]{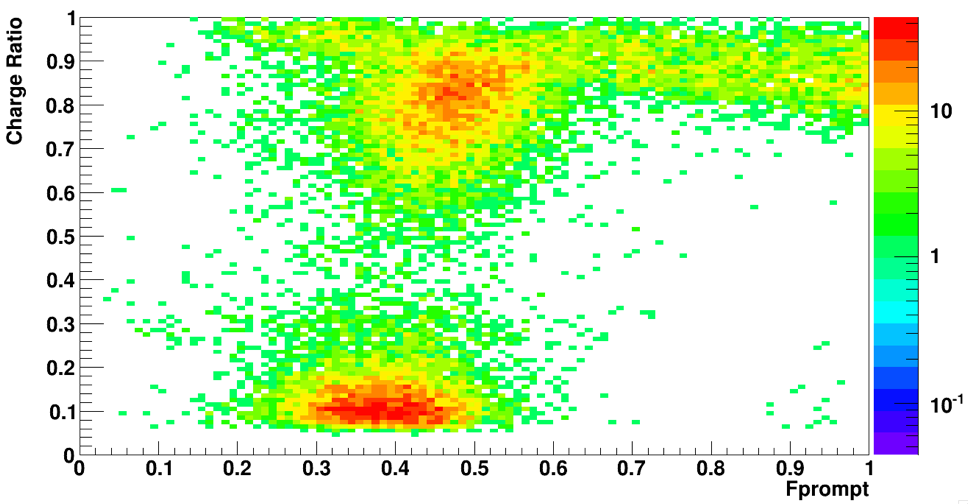}}
\hfill
\subfloat[]{\includegraphics[width=7cm]{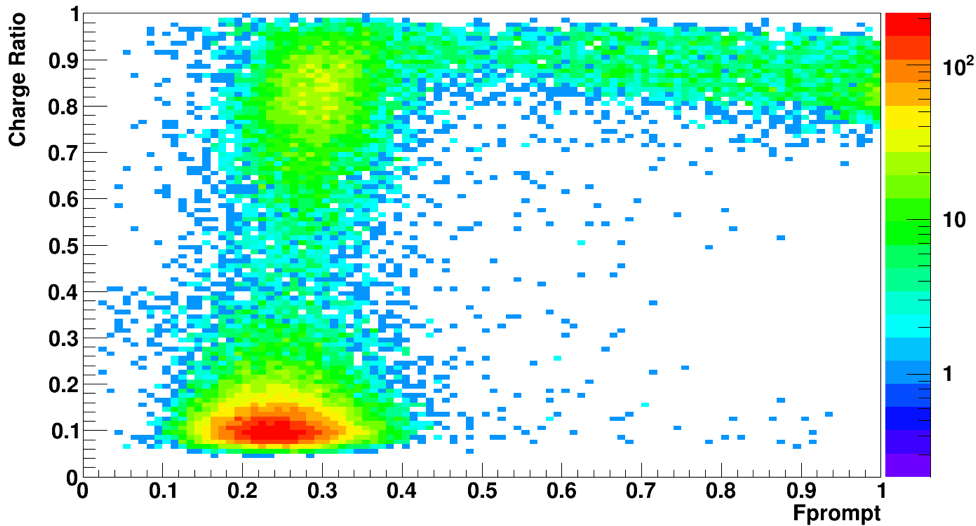}}
\hfill
\caption{ Charge ratio - Fprompt distribution. The timing order starts from (a) to (d), where (a) is earlier pump and purge run and the (d) is latter pump and purge runs.}
\label{fig:fpumppurgeruns}
\end{figure}
\begin{figure}
\hfill
\subfloat[]{\includegraphics[width=7cm]{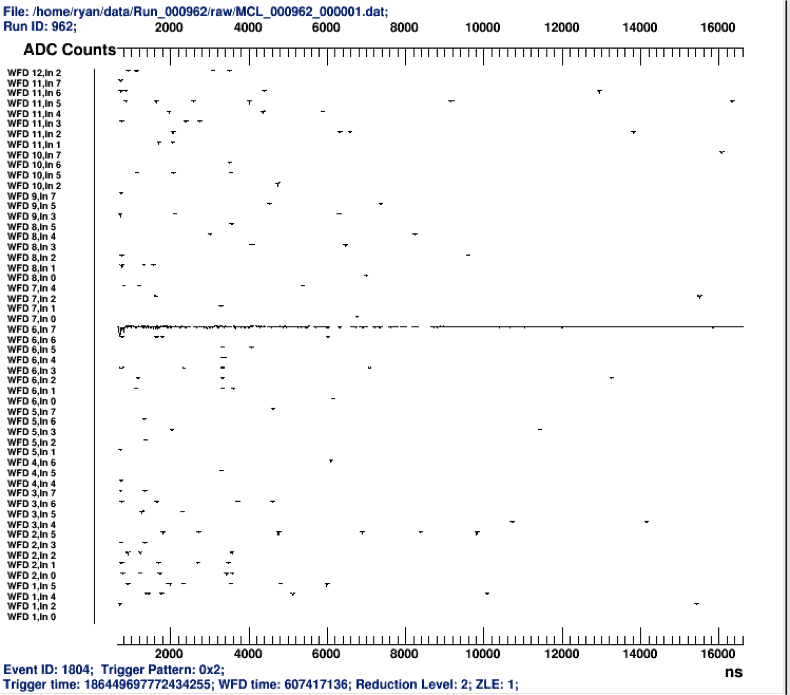}}
\hfill
\subfloat[]{\includegraphics[width=7cm]{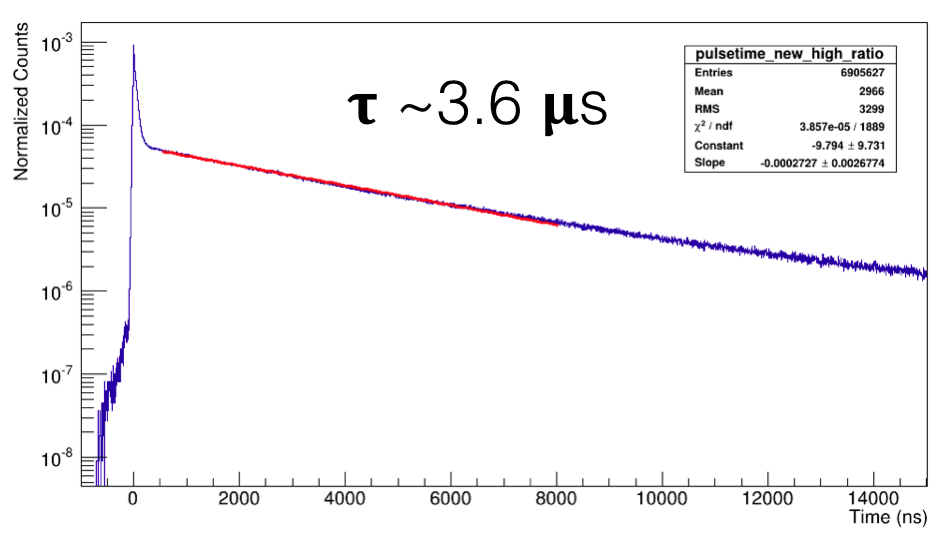}}
\hfill
\caption{(a) PMT discharging event viewed in CLEANViewer. (b) Example fit of the pulse-time distribution of discharging event.}
\label{fig:dischargingexample}
\end{figure}
Plotting the charge distribution for events with Frpompt >0.5 (<0.5) as shown in Fig. \ref{fig:qdis} (a) (Fig \ref{fig:qdis} (b)), the difference become much smaller for events with Fprompt < 0.5. For events with Fprompt > 0.5, mostly contains fast events (e.g. Cherenkov light), they agree in low energy and runs before the leak has more relative high energy events. 
Noticed a ``bump'' shows up in high energy (2000 -3500 PE) for runs after the leak with Fprompt < 0.5. These events are due to the artificial high charge of PMT discharging events. Using the pulse cut to identify the channel with discharging and remove it from the charge distribution for events with Fprompt < 0.5, the ``bump'' disappeared as shown in Fig. \ref{fig:qdisnodis}. The two distribution have similar end point and the runs before the leak has more high energy events. This may attribute to the decreased PMT gain for runs after the leak such that the trigger rate decreased and the contribution from higher energy events decreased.
\begin{figure}
\hfill
\subfloat[]{\includegraphics[width=7cm]{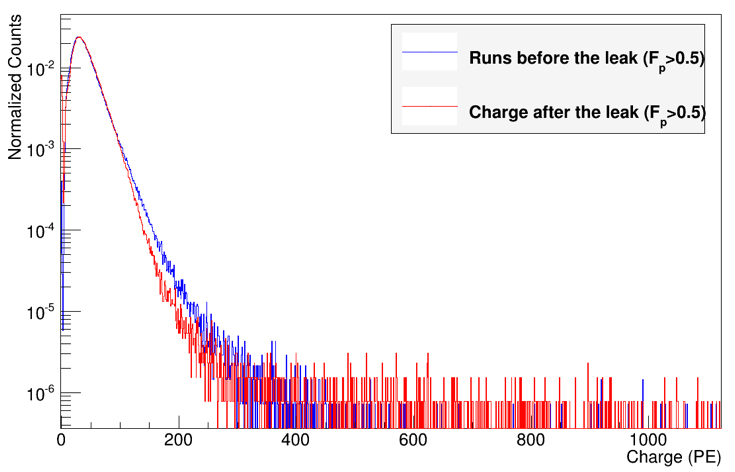}}
\hfill
\subfloat[]{\includegraphics[width=7cm]{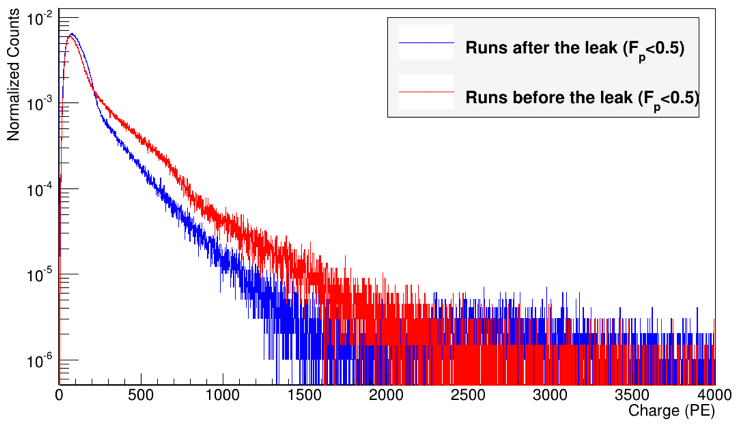}}
\hfill
\caption{Charge distribution for runs before and after the leak. (a) Fprompt > 0.5. (b) Fprompt < 0.5.}
\label{fig:qdis}
\end{figure}
\begin{figure}[htbp]
\centering
\graphicspath{{./fig/Warm_gas/}}
\includegraphics[scale=0.4]{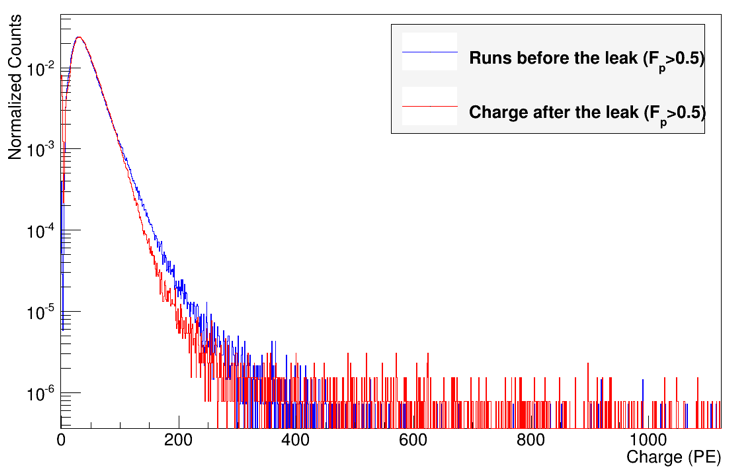}
\caption{  Charge distribution of events with Fp < 0.5. The discharging channel is removed from the charge distribution.}
\label{fig:qdisnodis}
\end{figure}
\section{Post pump and purge run}
The cooling of MiniCLEAN detector continue after the triplet lifetime is restored. Another reduction was found around 420 hours after the beginning of the pump and purge cycle. In order to understand the source of the leak, a series of excise are performed to identify the source. Figure \ref{fig:postpumppurge1} shows the triplet lifetime as a function of time. Noticed that in the zoom-in plot, several excises are performed to test the hypothesis. First, just before the time that the reduction of triplet lifetime happened. The condenser is running to flow LAr into IV. In the same time a unexpected power outage shuts the system off, thus the cryocooler was shut off for 10 hours. This makes the cold finger warm up to over 140 K which might release the oxygen which condensed on the cold finger after the leak into the IV and causing the reduction of triplet lifetime. The other hypothesis is that during the condenser running, the pressure of the condenser soar to unusual high pressure. This may loosen some fitting inside the nitrogen space of the condenser and results in leaking nitrogen or possibly some other impurity into the IV. The results shows the triplet lifetime decreased further when the cold finger warmed up, and stable for the rest of the test. This indicates that the impurity condensed on the cold finger might responsible for the reduction of the triplet lifetime.\par

\begin{figure}[htbp]
\centering
\graphicspath{{./fig/Warm_gas/}}
\includegraphics[scale=0.4]{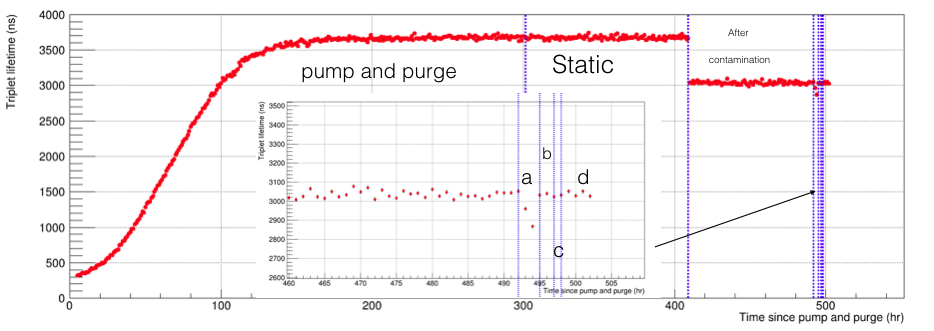}
\caption{ Triplet lifetime as a function of time. In the inset plot : a : cold finger warm up. b : condenser pressurized. c : condenser depressurized. d : condenser running.}
\label{fig:postpumppurge1}
\end{figure}
Furthermore, to rule out the possible temperature/pressure effect , the pressure, temperature and density of the argon gas is plotted against time as shown in Fig. \ref{fig:pump_purge_density_large}. Figure \ref{fig:pump_purge_density_zoom} shows the same quantities for the time zoomed in at right before the cold finger warm up (second dashed line to the right in Fig. \ref{fig:pump_purge_density_large}). In theory, with higher density the triplet lifetime is decreased and vis versa. From the plot, although in Fig. \ref{fig:pump_purge_density_zoom} the triplet lifetime seems correlated with the increased density. However, in Fig. \ref{fig:pump_purge_density_large}, the longer triplet lifetime is attained even with higher density. Thus the density effect should not be the main reason for reduction of triplet lifetime.\par

\begin{figure}[htbp]
\centering
\graphicspath{{./fig/Warm_gas/}}
\includegraphics[scale=0.3]{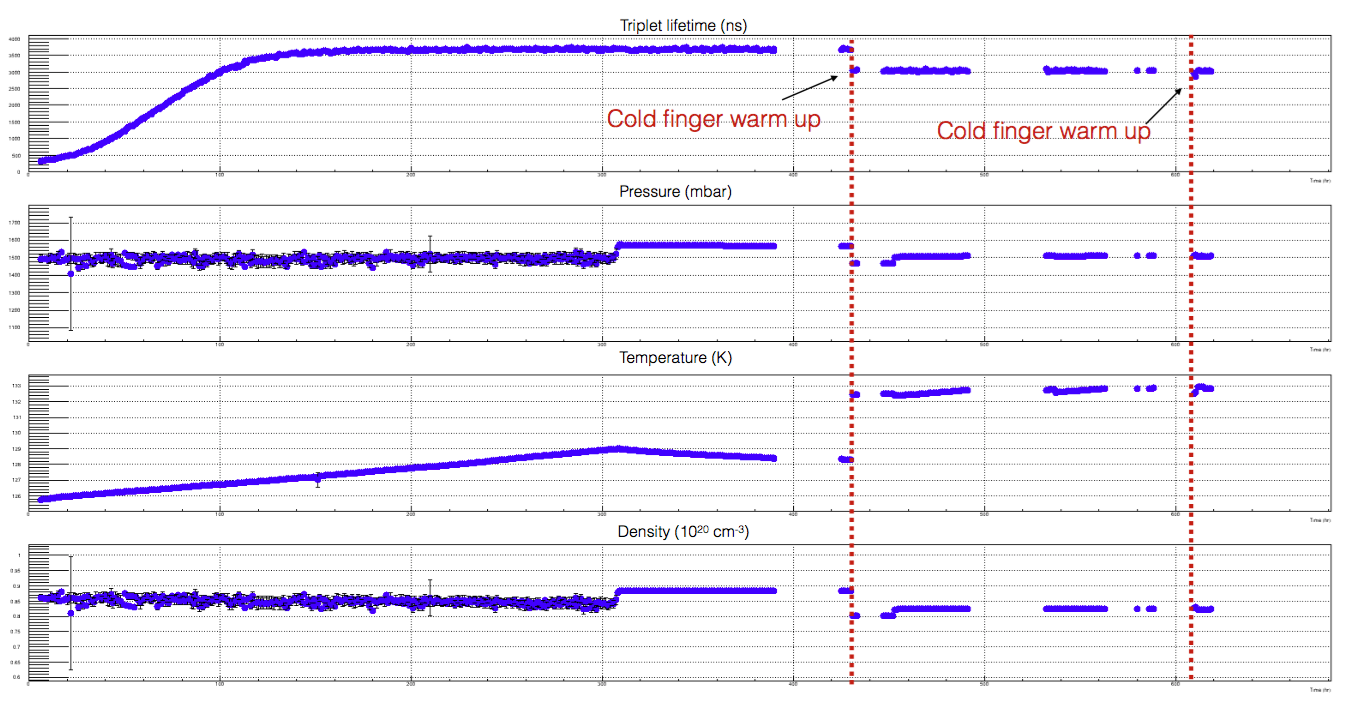}
\caption{ From the top to bottom panel. triplet lifetime,pressure,temperature and density.}
\label{fig:pump_purge_density_large}
\end{figure}

\begin{figure}[htbp]
\centering
\graphicspath{{./fig/Warm_gas/}}
\includegraphics[scale=0.3]{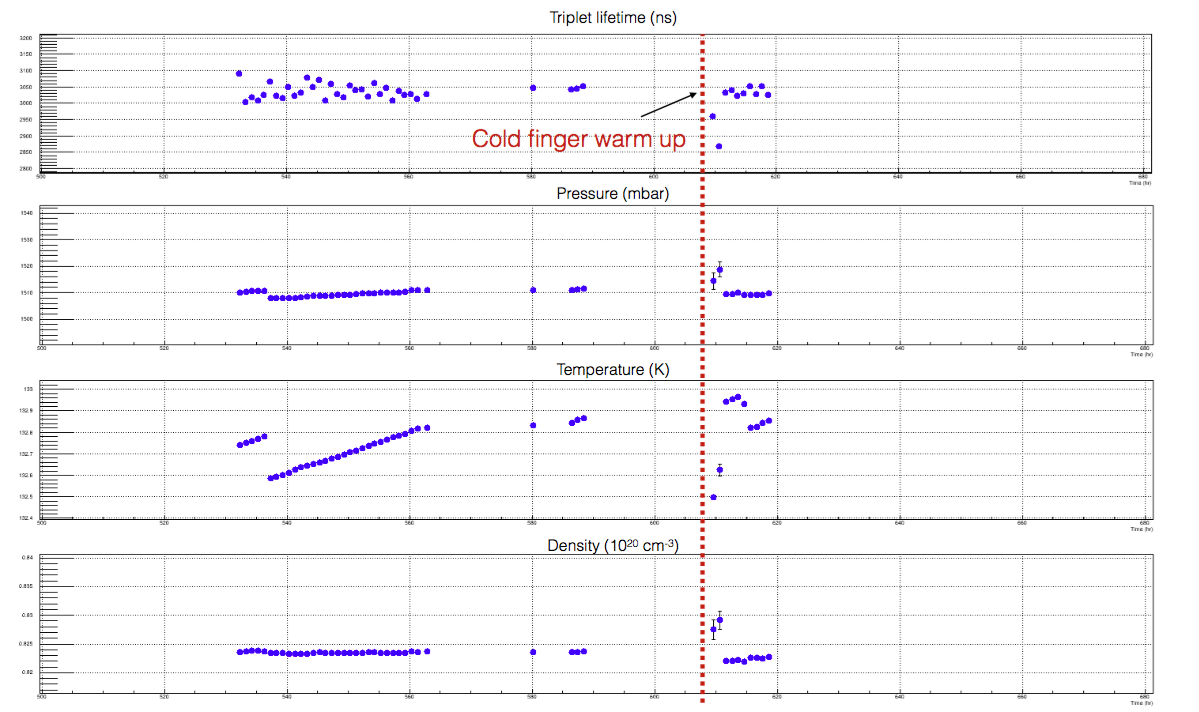}
\caption{ From the top to bottom panel. triplet lifetime,pressure,temperature and density. Zoom in at right before the second red dashed line to the right.}
\label{fig:pump_purge_density_zoom}
\end{figure}
A simple model can be used to estimate how much impurity has been released into the IV. The cold finger is well-aboved boiling point of oxygen for 10 hours which according to the relationship between triplet lifetime and impurity level (Fig. \ref{fig:fimpurity}) introduced 0.12 ppm of the additional impurity. In terms of atoms, 3.36 $\times$10$^{17}$ atoms is released from cold finger over 10 hours. The rate is 0.336 $\times$ 10$^{17}$ atoms/s. For the cold finger warm up excise, the cold finger is warm for 2 hours and the impurity level increased by 0.02 ppm. The calculated rate is 0.285 $\times$ 10$^{17}$ which in agreement with previous result. Considering the impurity level after the leak is 300 ppm, with the surface area of cold finger is around 700 cm$^2$, it takes more than 20 hours to purge the impurity out of IV.\par

Subsequently, the second pump and purge cycle is performed. During the cycle, the same excise to warm up the cold finger is test again to ensure this is main reason for the reduction of triplet lifetime. Figure \ref{fig:second_test} shows that right after the cryocooler warms up, the triplet lifetime drops and recovered  after the cryocooler is turned back on. This indicates the impurity condensed on cold finger is the main reason for the reduction of the triplet lifetime.
\begin{figure}[htbp]
\centering
\graphicspath{{./fig/Warm_gas/}}
\includegraphics[scale=0.3]{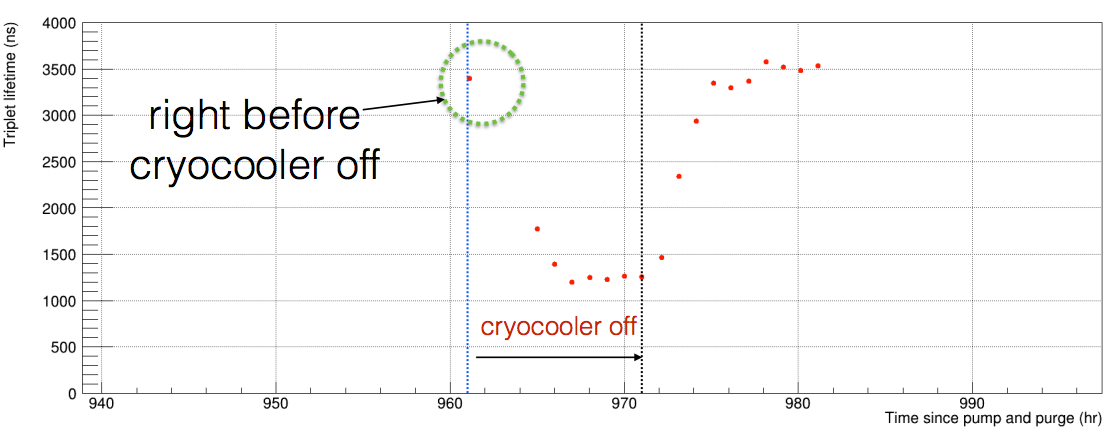}
\caption{ Triplet lifetime as a function of time. Notice that when the cryocooler is off, the triplet lifetime drops and increased when cryocooler is turned back on.}
\label{fig:second_test}
\end{figure}

\chapter{Triplet lifetime measurement in Cold Gas}\label{ch:cold}
Understanding of the relationship between triplet lifetime and impurity level of the gas is important. The limit of RGA sensitivity is around sub-ppm level. However, the required purity of argon is < 100 ppb. Therefore if the relationship of triplet lifetime and impurity level is known, the triplet lifetime can be an indicator to reveal the current impurity level of the gas. Using it to monitor the detector health has been described in detail in Chapter \ref{ch:gasrun}. In this chapter, the detail analysis of finding the relationship of triplet lifetime and impurity level is presented.
\section{Triplet lifetime}\label{sec:1}
\setlength{\parindent}{5ex}
\subsection{Analysis summary}\label{sec:dys}
During different data taking period, different numbers of PMTs are turned on. Normally, 62 PMTs are on except from the beginning of pump and purge (Run 931) until the middle of pump and purge (Run 951), 55 PMTs are activated. The position of PMTs is shown in Fig. \ref{fig:fmap}, numbers in color indicates the PMT is off, the explanation of color index is described in the caption. In the cold gas run, many PMTs are removed from analysis for different reasons :
\begin{itemize}
  \item PMT Channel Number : 7,15,18,19,24,43,66,68,76,88,91, total 11 PMTs with no connection (deactivated).
  \item PMT Channel Number : 14,17,53,54,55,70,72, total 7 PMTs without conformal coating (deactivated).
  \item PMT Channel Number : 1,3,6,12,26,44,63,75,80,89, total 10 PMTs has very low gain.
  \item PMT Channel Number : 35, with high background.
  \item PMT Channel Number : 41, known as noisy PMT.
  \item  PMT Channel Number : 0,2,4,9,10,11,13, total 7 PMTs found to be noisy, turned off during early pump and purging runs (Run 931 - 951). They are included in the later runs (Run 953 - 978).
\end{itemize}
 The existing of PMT discharging events in one channel will bias the centroid reconstructed radius toward the edge of the detector which results in the excessive events near the edge of IV (Figure \ref{fig:fnumber}). Therefore, the additional radius cut is applied to the data and define as $(R/R_{TPB})^3 < 0.7$. Where R is the centroid reconstructed radius, $R_{TPB}$ is the radius of the TPB (wavelength shifter) which define the maximum radius of active volume. The cut efficiency determined by MC simulation decreased to 88.30\%.
\begin{figure}
\hfill
\subfloat[]{\includegraphics[width=7cm]{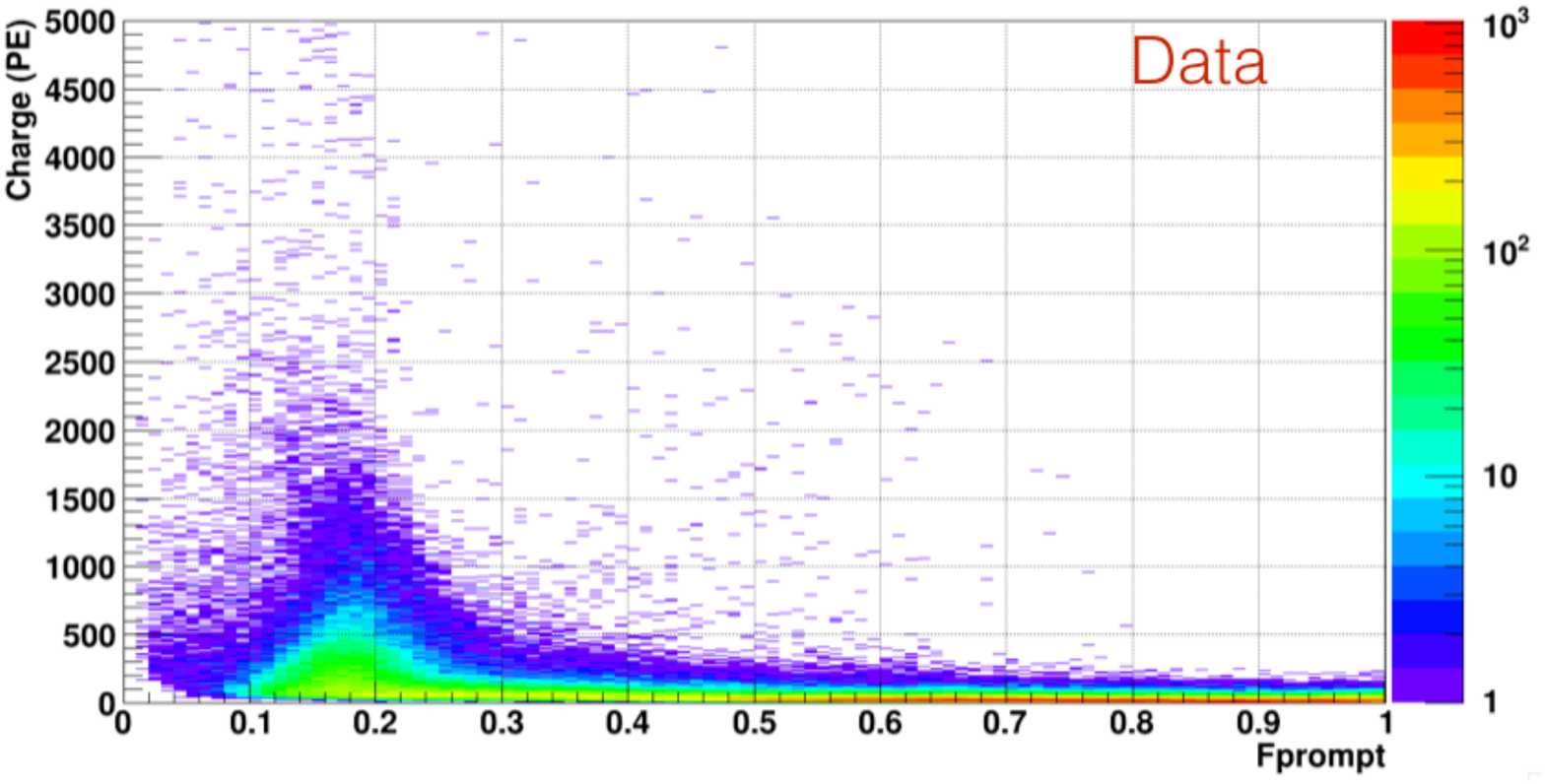}}
\hfill
\subfloat[]{\includegraphics[width=7cm]{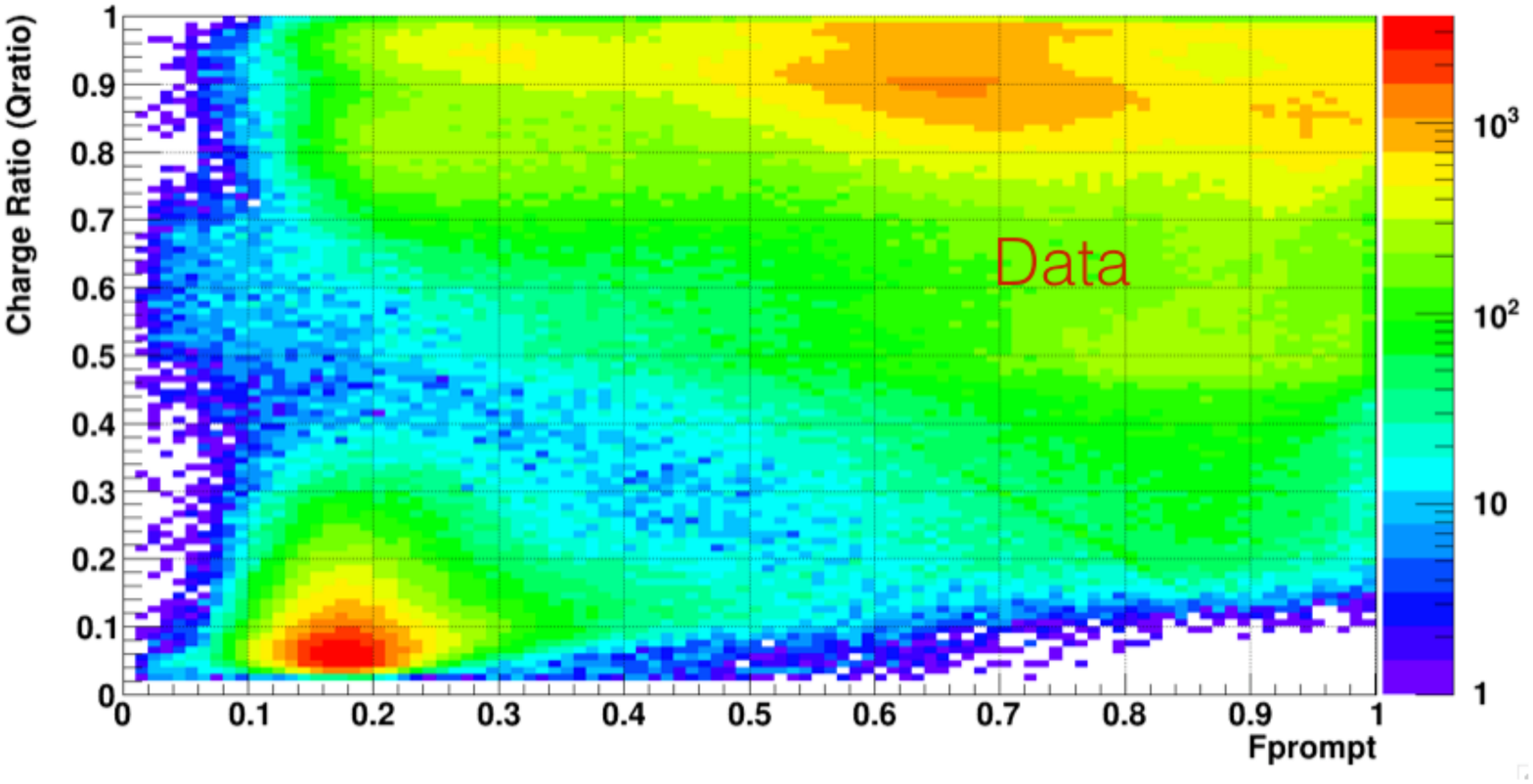}}
\hfill
\subfloat[]{\includegraphics[width=7cm]{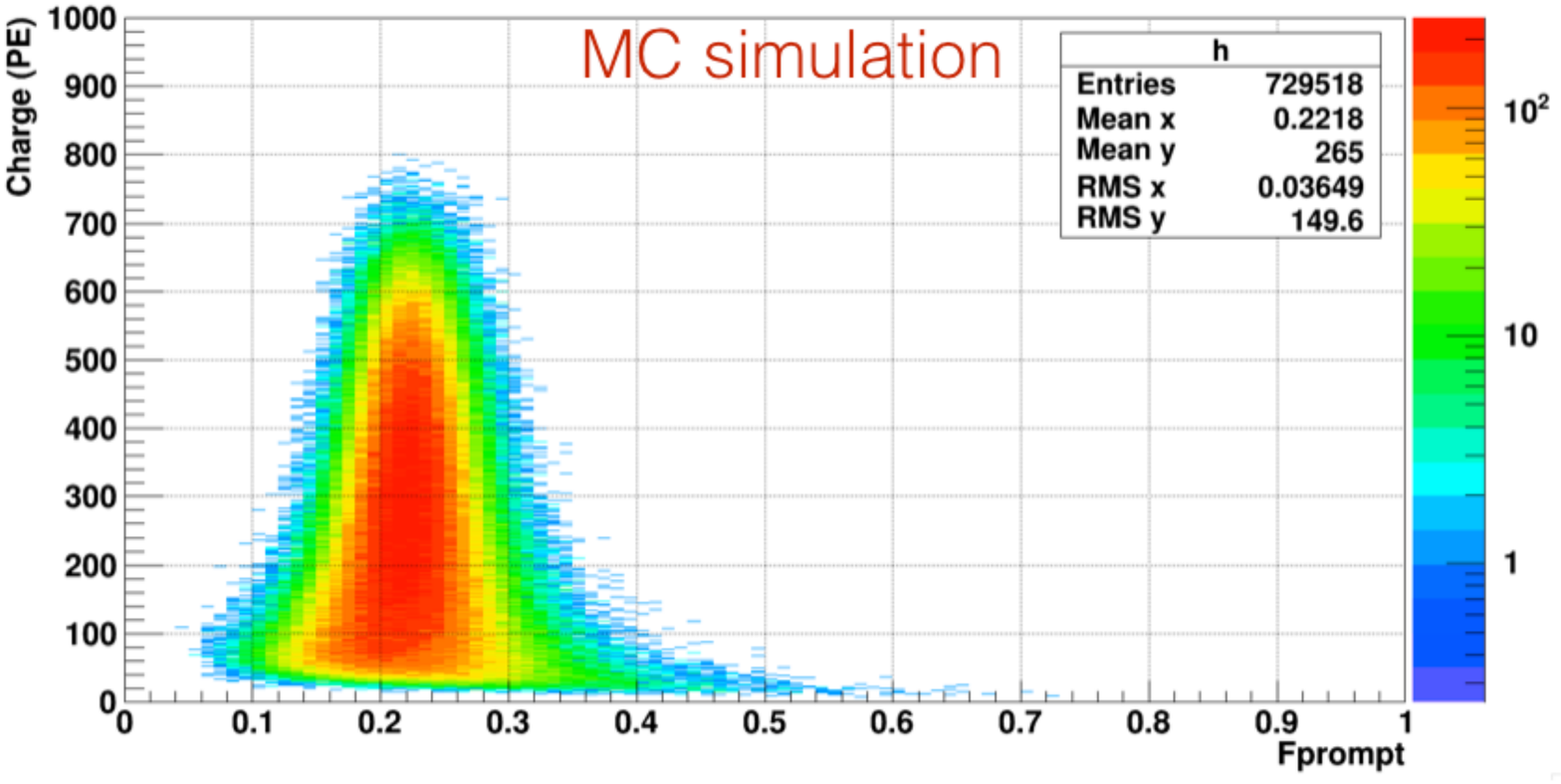}}
\hfill
\subfloat[]{\includegraphics[width=7cm]{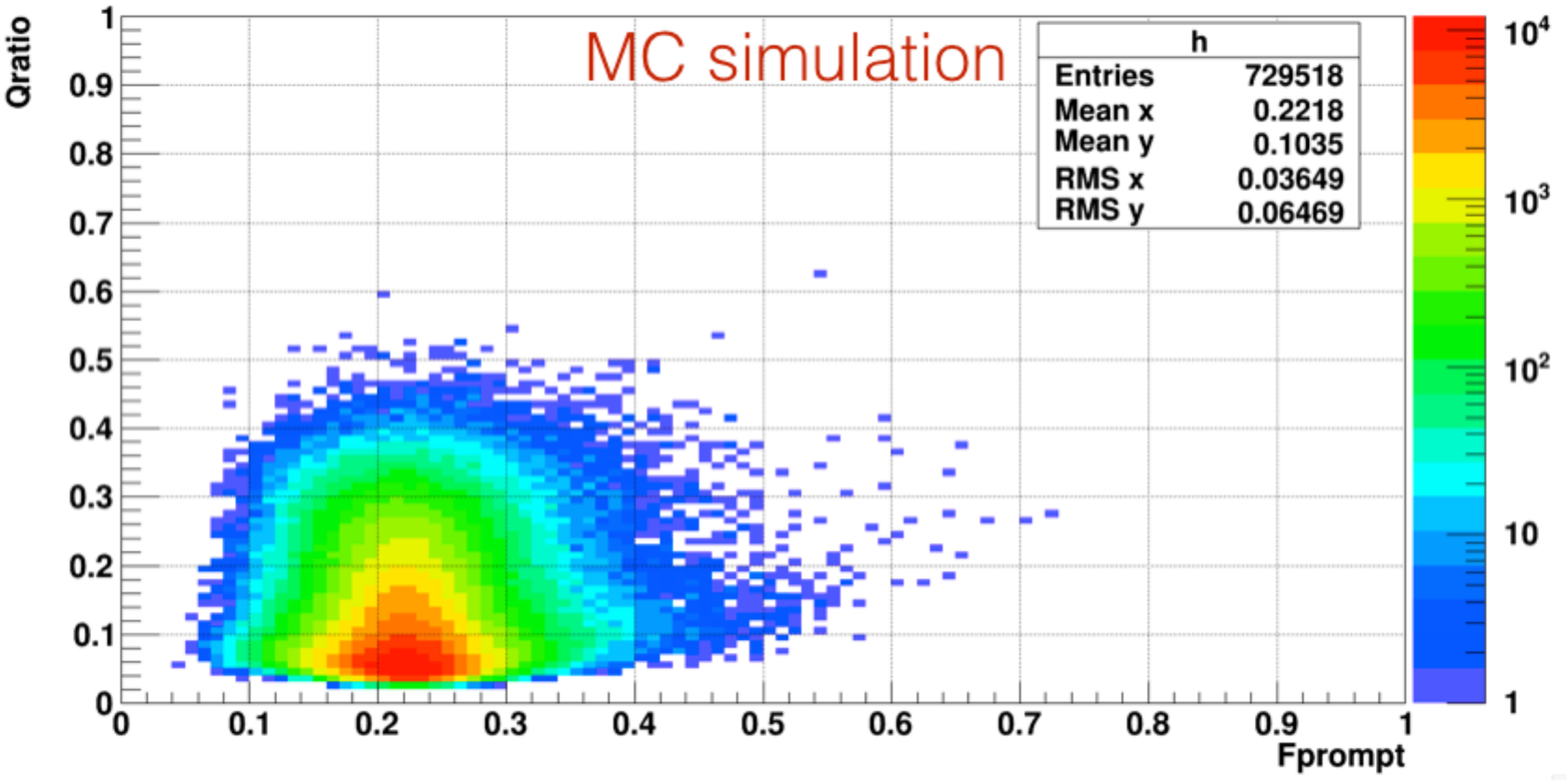}}
\hfill
\caption{(a) Charge vs Fprompt  from the data. The peak at low prompt region is from electronic recoil in gaseous argon from the data. (b) Charge Ratio (Qratio) vs Fprompt  from the data. The group of events at low Qratio and low prompt is from electronic recoil which corresponds to the peak in Charge-Fprompt plot. (c) Charge vs Fprompt from Monte Carlo simulation of $^{39}$Ar events. Noticed the ploting scale on y-axis is different from (a). (d) Charge Ratio(Qratio) vs Fprompt from Monte Carlo simulation of $^{39}$Ar events.}
\label{fig:fcharge}
\end{figure}
\begin{figure}[tbp]
\centering
\graphicspath{{./fig/}}
\includegraphics[scale=0.4]{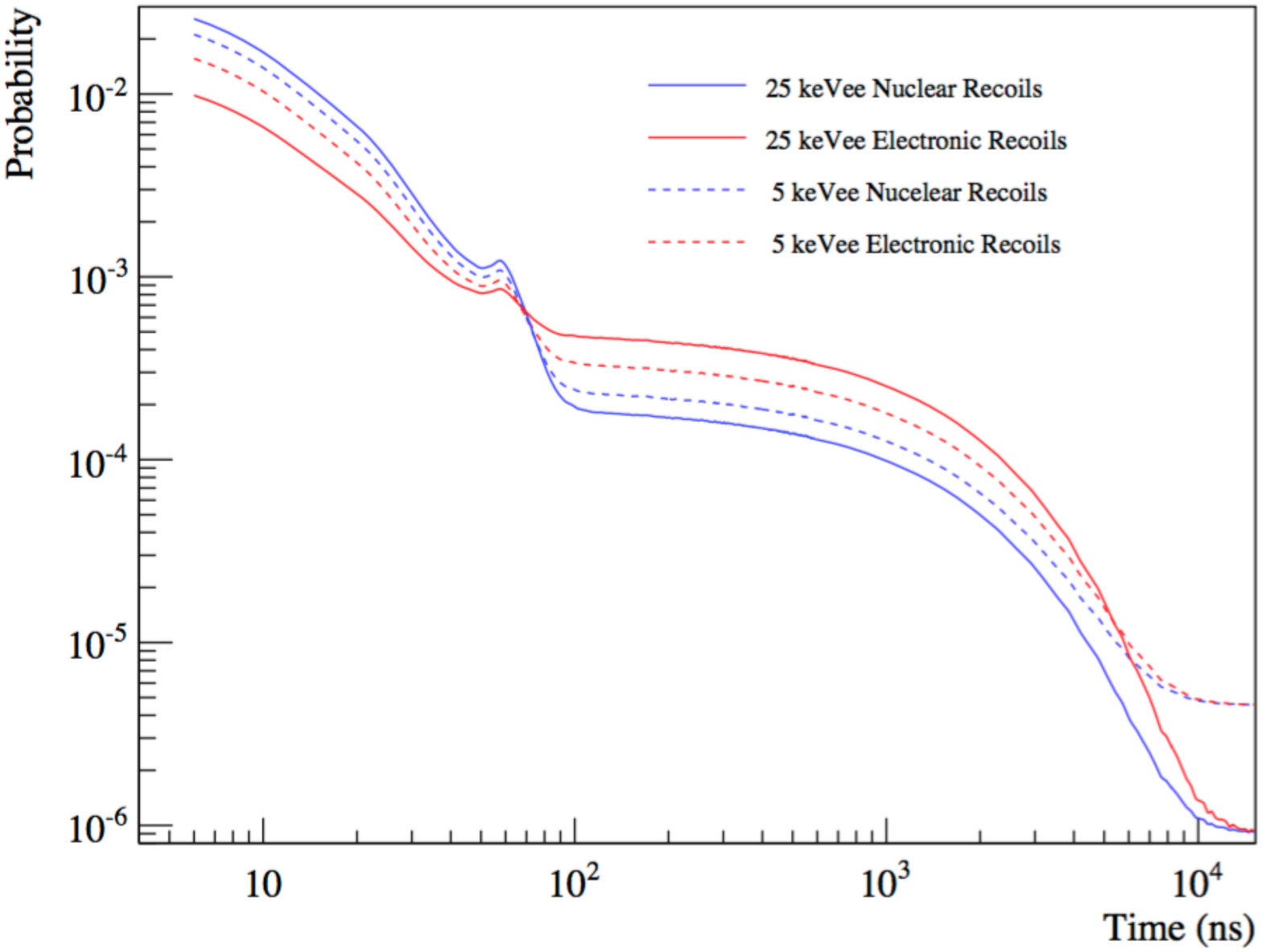}
\caption{Photoelectron detection time PDFs for electronic and nuclear recoils at 5 keVee and 25 keVee energies from MiniCLEAN Monte-Carlo simulation.}
\label{fig:fscin}
\end{figure}
\begin{figure}[tbp]
\centering
\graphicspath{{./fig/}}
\includegraphics[scale=0.4]{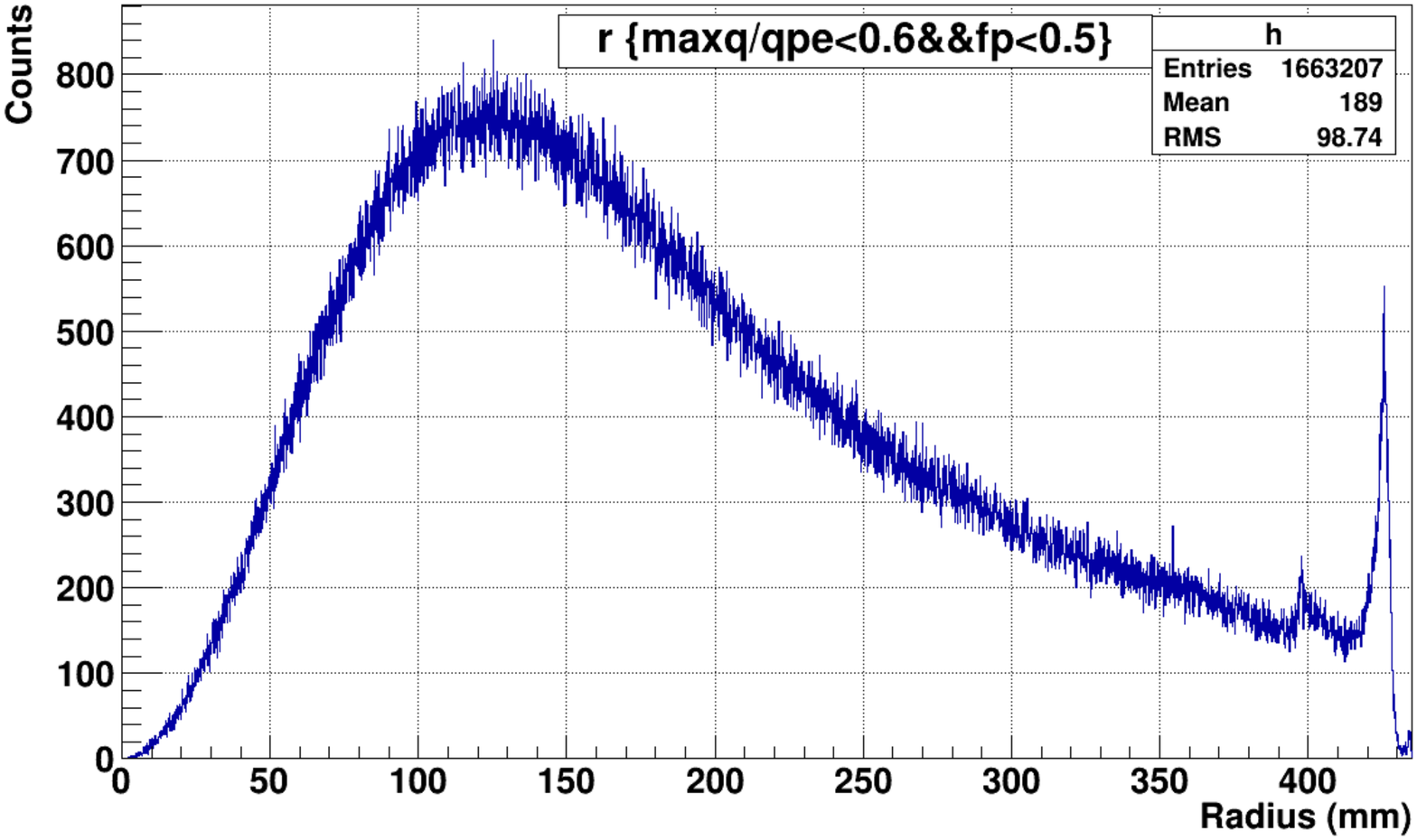}
\caption{ Number of counts vs centroid radius (mm).  }
\label{fig:fnumber}
\end{figure}
The data passed cut are fitted with simple exponential function plus constant background:
\begin{ceqn}\begin{align}\label{eq:2}
F(t) = p_0 \cdot[ (1-p_1) e^{ -t/\tau } +p_1]
\end{align}\end{ceqn}
where $p_{0}$ and $p_{1}$ are fitting parameters, $\tau$ is the fitting parameter which represents the triplet triplet lifetime . A sample fit is shown in Fig \ref{fig:ftrip}(a).
The prompt, intermediate and late components are identified using the three exponential convoluted with Gaussian response function.
\begin{ceqn}\begin{align}\label{eq:3}
f = G(t,\sigma) \otimes [A \cdot e^{-\frac{t}{\tau_{1}}} +B\cdot e^{-\frac{t}{\tau_{2}}}+ (1-A-B) \cdot e^{-\frac{t}{\tau_{3}}}]
\end{align}\end{ceqn}
where G is gaussian response function. The parameters $\tau_{1}$ , $\tau_{2}$ and $\tau_{3}$ are time constant of the fast, intermediate and slow-decaying states respectively. The parameters A and B are the fractions of prompt and intermediate state respectively. In order to determine each component, more precise SPE arrival time is needed. The estimated SPE arrival time using Bayesian techniques developed by MiniCLEAN collaboration is used\cite{AkashiRonquest201540}.
\begin{figure}
\hfill
\subfloat[]{\includegraphics[width=7cm]{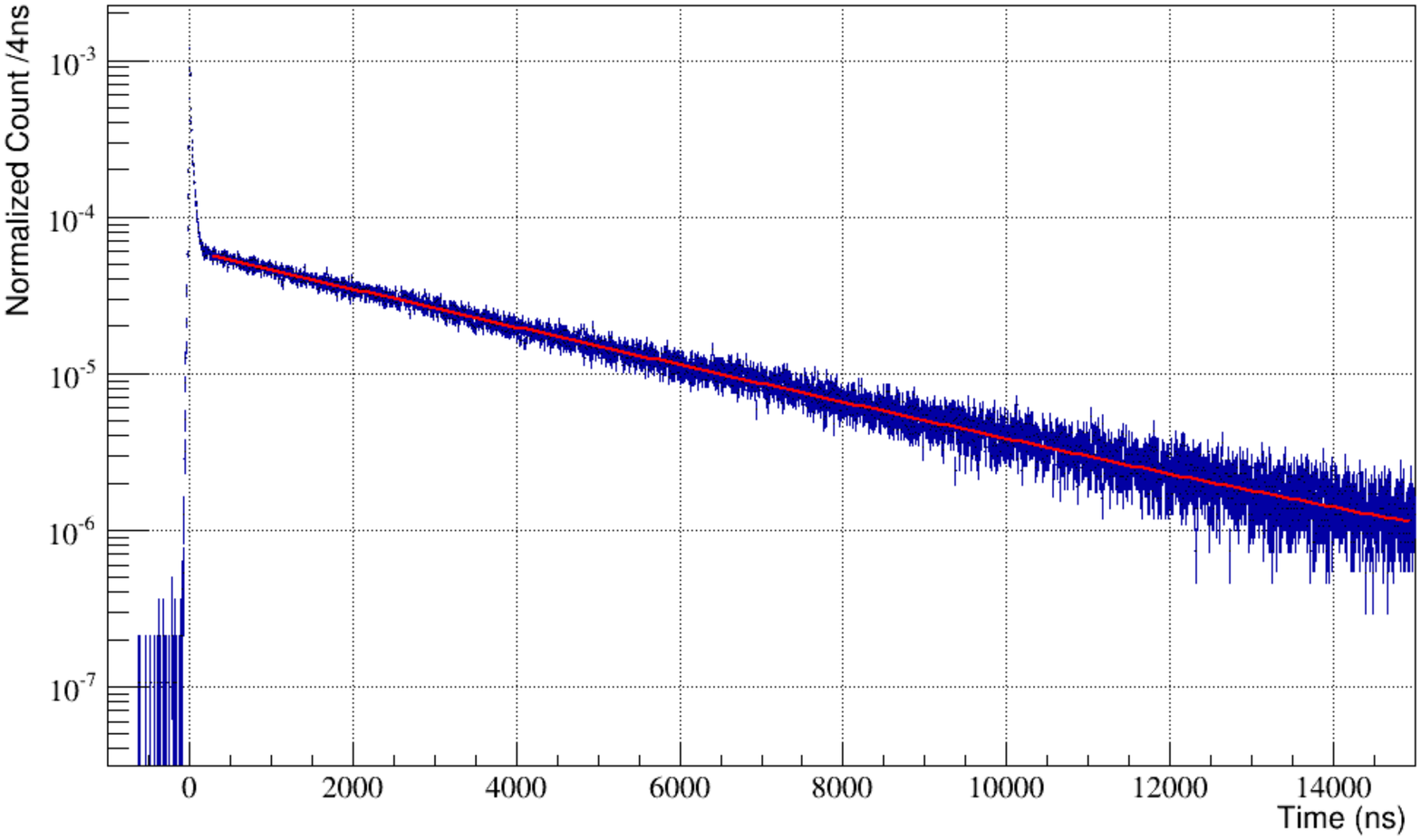}}
\hfill
\subfloat[]{\includegraphics[width=7cm]{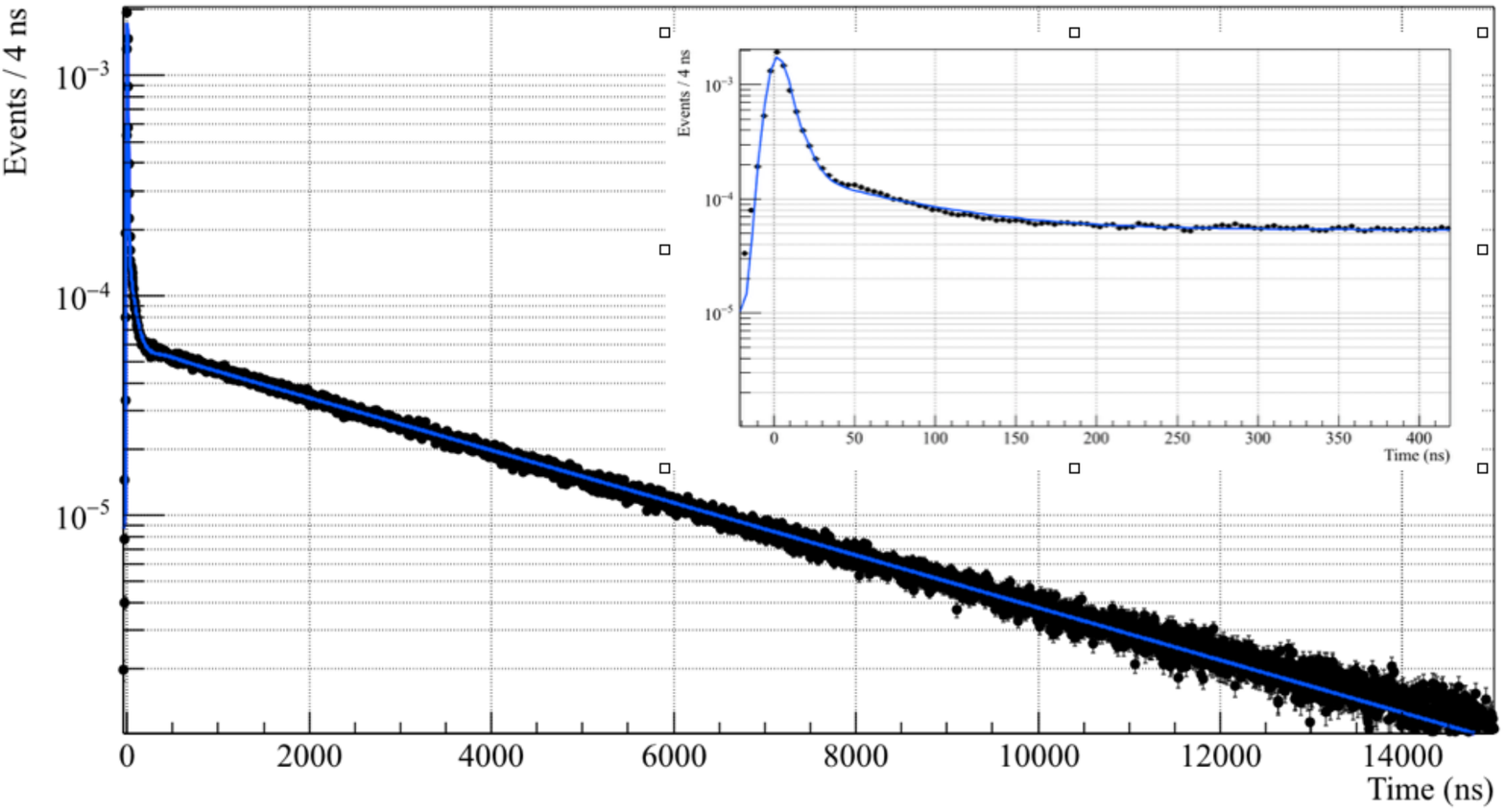}}
\hfill
\caption{(a) Pulse time distribution fit with fit function from Eq. \ref{eq:2}. (b)  Example fit of single photonelectrons arrival time from scintillation events with three exponential convoluted with gaussian resolution function (Eq. \ref{eq:3}) in cold gas. The inset plot shows the first 400 ns.}
\label{fig:ftrip}
\end{figure}
\subsection{First cold gas runs : Oct. 2016  to  Feb. 2017}
The IV continues to be to cooled while data taking.  During this period the fitted triplet lifetime is used to monitor the detector health as well as the dependance of triplet lifetime on pressure/temperature.  Throughout the cooling we must continuously flow argon through the purification system, condenser and into the IV. 
Potential contamination could arise from a leak between fittings or from material outgassing. 
Data is taken every week to ensure the purity of the argon gas and the stability of triplet lifetime indicates the IV is free from contamination.
Figure \ref{fig:fcondenser} shows the stability of triplet lifetime as a function of time that flowing was temporarily stopped (static mode).
Figure \ref{fig:ftriptime} shows the triplet lifetime over a time period in which a leak was discovered.
The triplet lifetime began decreasing after the bottom temperature sensor reached the liquefaction point of argon.  
While the exact timing of the gas contamination is unknown, the leak appears to have been developed sometime around mid-December 2016. 
The source of leak was determined to be exhaust from the IV pump (see Figure \ref{fig:fleak}).
\begin{figure}[tbp]
\centering
\graphicspath{{./fig/}}
\includegraphics[scale=0.4]{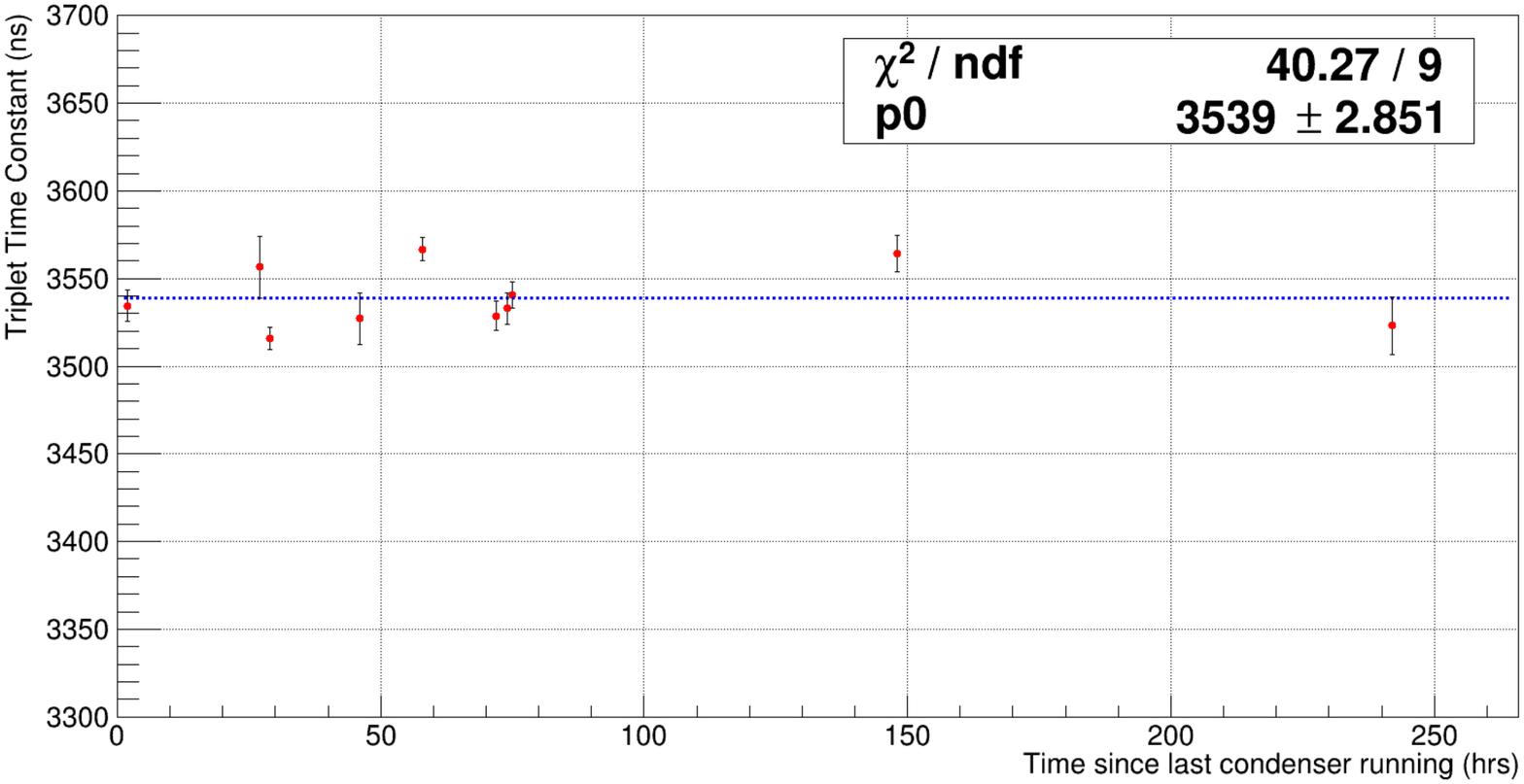}
\caption{Stability of triplet lifetime after the gas sits in static. The zero-th order polynomial fit is present.}
\label{fig:fcondenser}
\end{figure}


\subsection{Pump and purge runs}\label{sec:12}
As soon as the leak is identified and fixed, we begin the pump and purge cycle to purge the impurities from the IV. In the beginning, we are pumping the gas out of IV and purge with pure argon (impurity < 1 ppb). However, it seems this approach didn't pump out the impurity efficiently, probably due to the complicated geometry of the IV leaving trapped gas. We then decided to increase the pressure while purging and lower the pressure when pumping on IV. This method quickly increase the purity of the gas inside the IV. From the point of view of data taking, this procedure starts from Run 931 (3/30/2017) and end in Run 970 (4/13/2017), with total pump and purge of around 434 hours, we successfully restored the triplet lifetime to the good value prior to the leak. Figure \ref{fig:fpumptime}, shows the changes in triplet lifetime as a function of pump and purge time. In the early pump and purge runs, the triplet tail has very few events, so different fitting range is used to get a reasonable $\chi^{2}$/NDF and lifetime.
Figure \ref{fig:fitex} shows the example of fitting in Run 931, noticed that the electronic switching noise is presented around 2600 ns which results in larger $\chi^{2}$/NDF ($\sim$ 2.83) value. The overall $\chi^{2}$/NDF for each hour is shown in Fig. \ref{fig:fchi}.\par
The IV leak gave us the opportunity to investigate the relationship between triplet lifetime and the associated impurity level. Unfortunately the RGA is not sensitive to the impurity level below 1 ppm, thus a model to describe the impurity level decreased with pump and purge cycle is needed. The initial impurity level is well above 10 ppm thus it can be attained directly from measurement of RGA. In the early pump and purge cycle, RGA samples the IV outlet gas approximately every 40 minutes. After RGA background subtraction, the total initial impurity level is estimated as 40.86 $\pm$ 2.63 ppm. The pressure during the pump and purge cycle can be extracted from slow control system\footnote{Slow Control System is builded base on MySQL database system, which record the detector parameter in real time and store in MySQL database\cite{nikkelcontrol}.}as shown in Fig. \ref{fig:fcycle}. With this information, the fraction of gas pump out from IV can be estimated for each cycle. Therefore the behavior of degradation of impurity level can be approximated by this equation
\begin{ceqn}\label{eq:impurity1}
\begin{align}
\frac{dI(t)}{dt} = - \frac{\epsilon f}{T}\cdot I_0
\end{align}\end{ceqn}
where I(t) is the impurity level as a function of time, T is the pump and purging period, $I_0$ is the initial impurity level, $f$ is the fraction of gas pumped out in each cycle and $\epsilon$ is the purge efficiency. For each cycle the temperature changes is very small ($\ll$ 1K), so we will ignore the effect from the temperature variations. In this analysis, we assume the purge efficiency is 1, and use the recursive form (Eq. \ref{eq:impurity1}) to estimate the impurity level at the end of each cycle
\begin{ceqn}\label{eq:impurity2}
\begin{align}
I_{i+1} = I_i\cdot(1-f_i)
\end{align}\end{ceqn}
However, due to the low trigger rate($\sim$ 8 Hz), we need more time to accumulate events in order to fit for the triplet lifetime. Therefore, two different approaches are used to map triplet lifetime to the impurity level. \par
Firstly, the impurity level is changing during the pump and purge cycle, using Eq. \ref{eq:impurity2} and integrated over the period of cycle, one can get the average impurity level across the pump and purge cycle.
\begin{ceqn}\label{eq:impurity3}
\begin{align}
I_{Avg} = I_i\cdot\frac{1-e^{-f_{i+1}}}{f_{i+1}} 
\end{align}\end{ceqn}
This gives the average impurity level ($I_{Avg}$) for $(i+1)$-th cycle. Using the taylor expansion Eq. \ref{eq:impurity3} can be approximated as $I_i\cdot(1-f_{i+1}/2)$. With average fraction of pump out gas is 6.6\%, the average impurity for each cycle is $\approx I_{i}\cdot(0.97)^{i+1}$ for (i+1)-th cycle. The pulse time of scintillation events are populated into histogram during the period of time T for each cycle and map to the average impurity level in the same period of time.\par
Another way to map the impurity level and lifetime is to use the impurity level at the end of pumping, then populate the pulse time of scintillation events in the following time until the beginning of next pump and purge cycle. In this period of time T', the IV is approximately at static, so the impurity level should not change significantly. Therefore we can map the lifetime measured in this period of time to the impurity level which estimated just right before the beginning of IV at static. The cartoon describe both methods is shown in Fig. \ref{fig:fcartoon}.\par

\begin{figure}[htbp]
\centering
\graphicspath{{./fig/Triplet/}}
\includegraphics[scale=0.6]{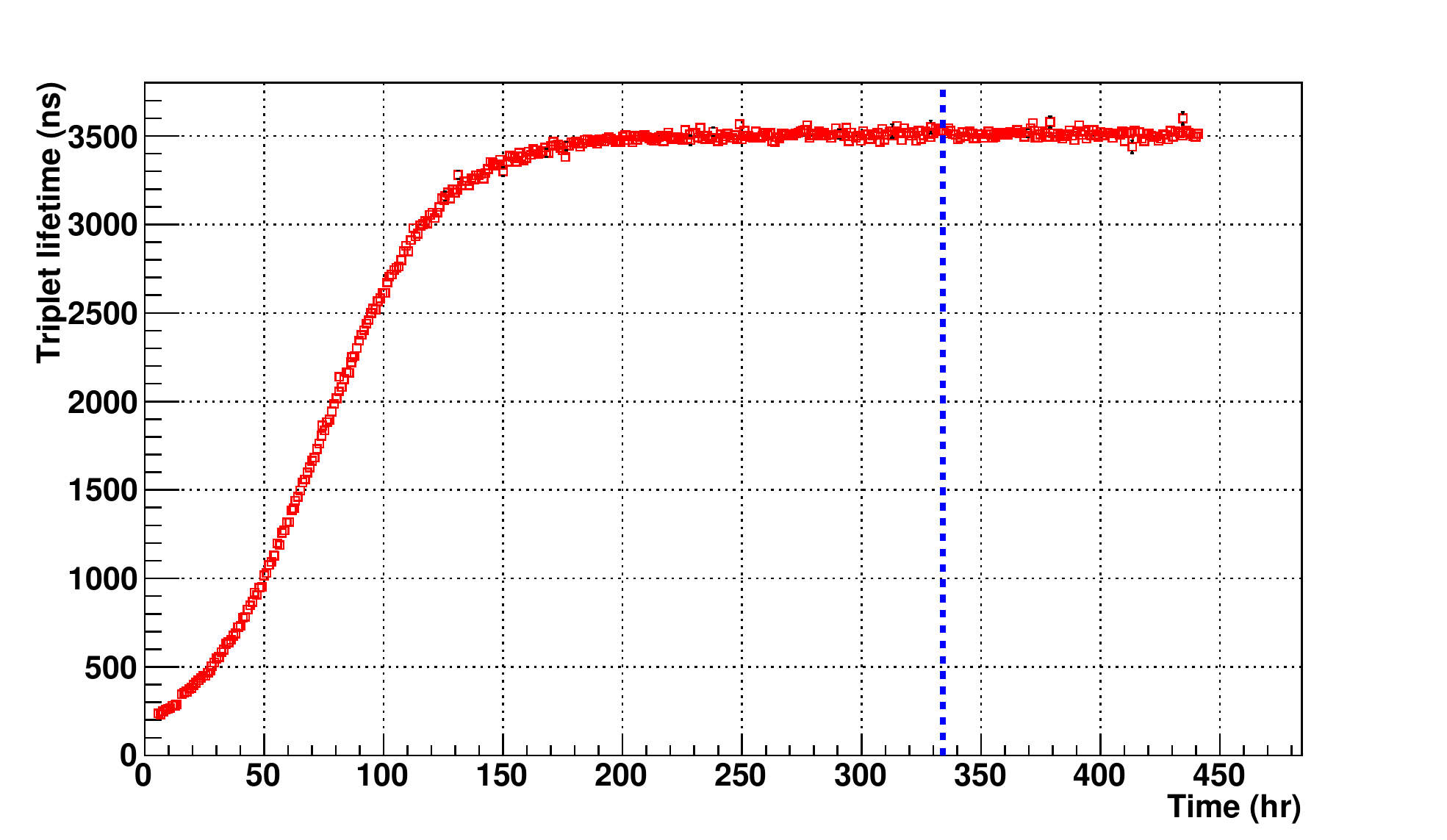}
\caption{ Triplet lifetime vs the time since the beginning of pump and purge, the blue dashed line indicate the time pumping is stopped, IV at static after that. }
\label{fig:fpumptime}
\end{figure}
%

\begin{figure}[htbp]
\centering
\graphicspath{{./fig/Triplet/}}
\includegraphics[scale=0.6]{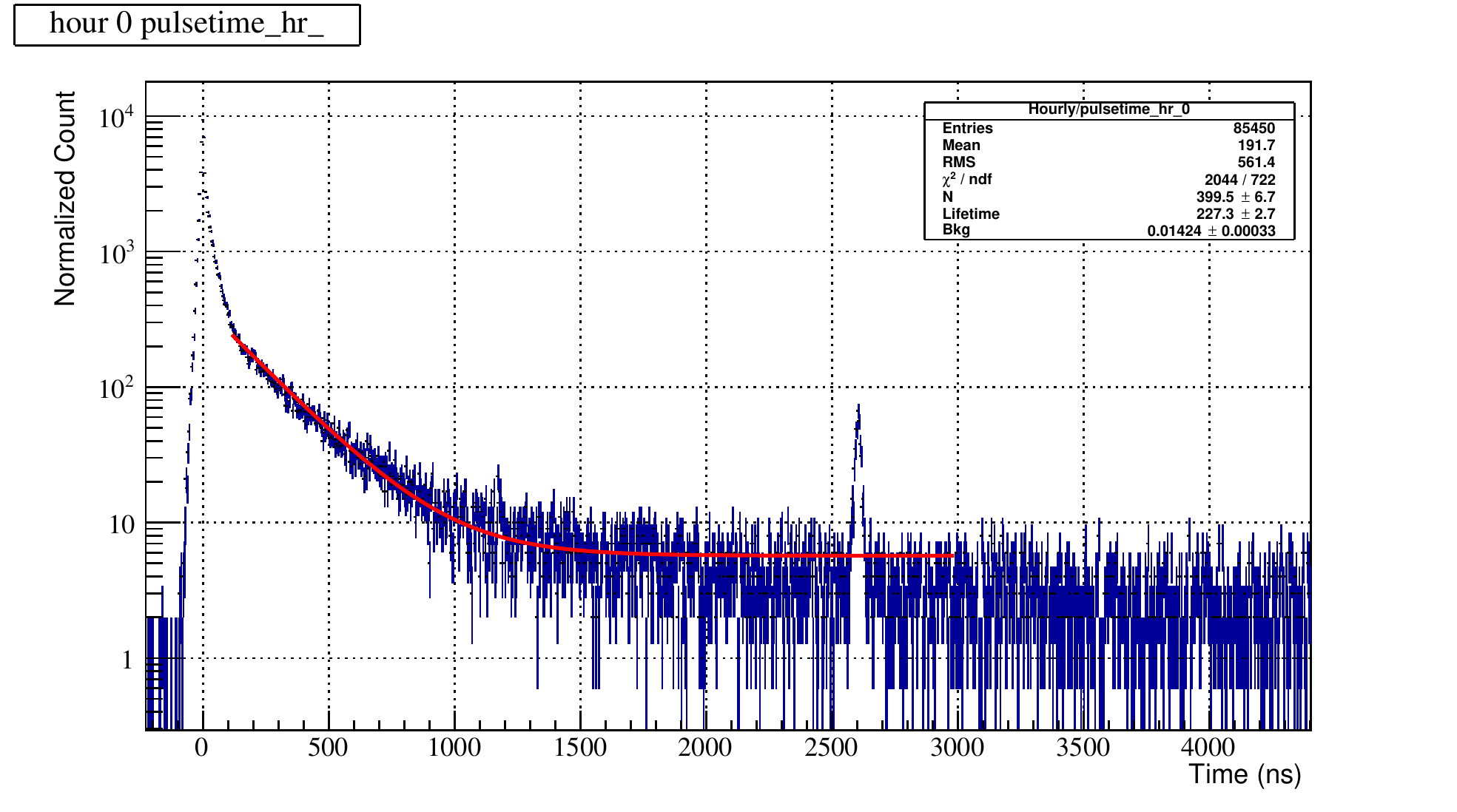}
\caption{ Fitting example in Run 931 which is the earliest run of pump and purging cycle. The peak at $\sim$ 2600 ns is due to the electronic switching noise which increases the $\chi^{2}$. The fitting window is 200 to 3000 ns. }
\label{fig:fitex}
\end{figure}

\begin{figure}[htbp]
\centering
\graphicspath{{./fig/Triplet/}}
\includegraphics[scale=0.5]{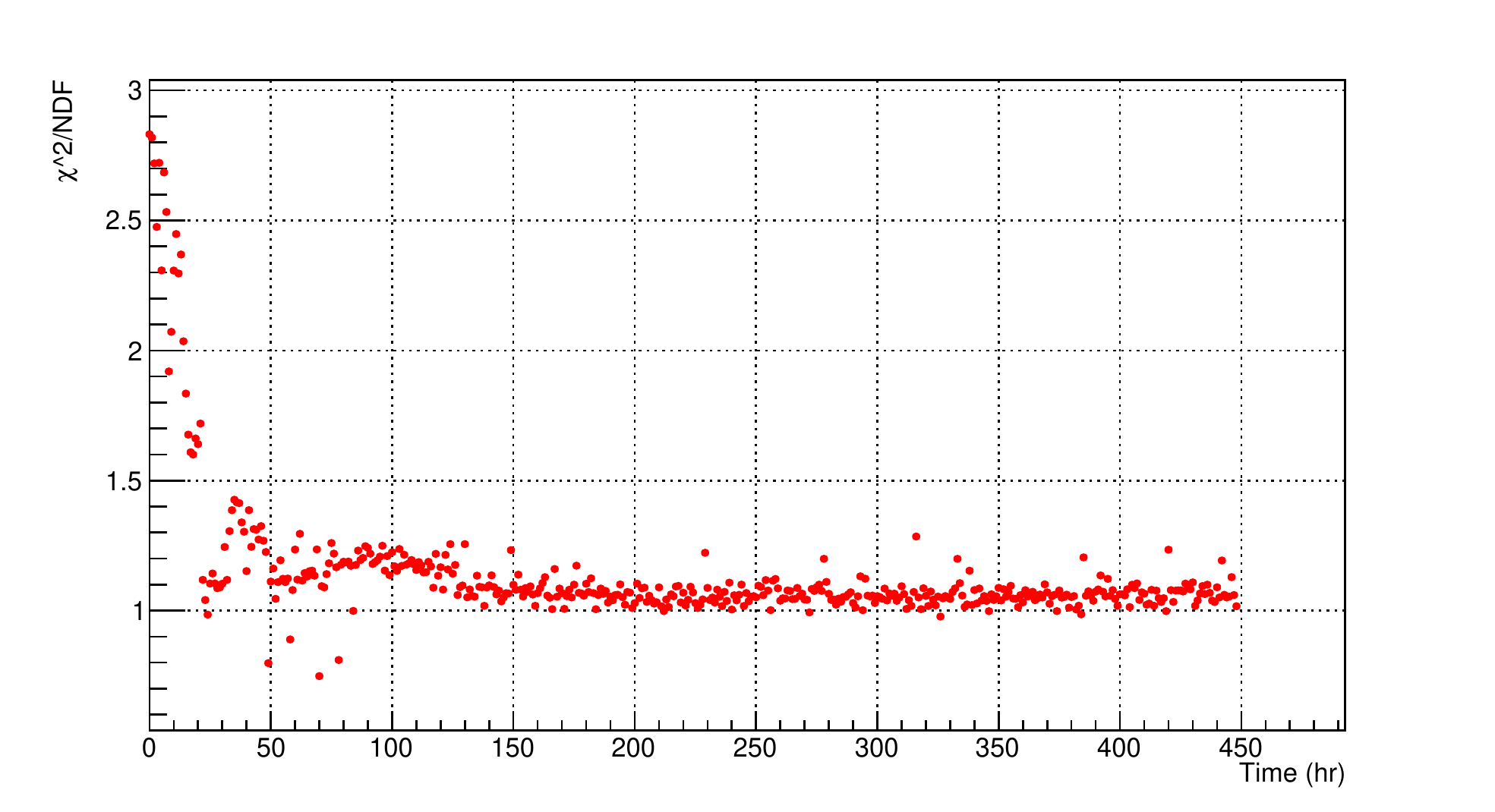}
\caption{ $\chi^{2}$/NDF value for each hour. Due to low statistic and electronic switching noise presented in the data (see Fig. \ref{fig:fitex}) results in larger value of $\chi^2$ in earlier runs of pump and purge. }
\label{fig:fchi}
\end{figure}

\begin{figure}[htbp]
\centering
\graphicspath{{./fig/Triplet/}}
\includegraphics[scale=0.4]{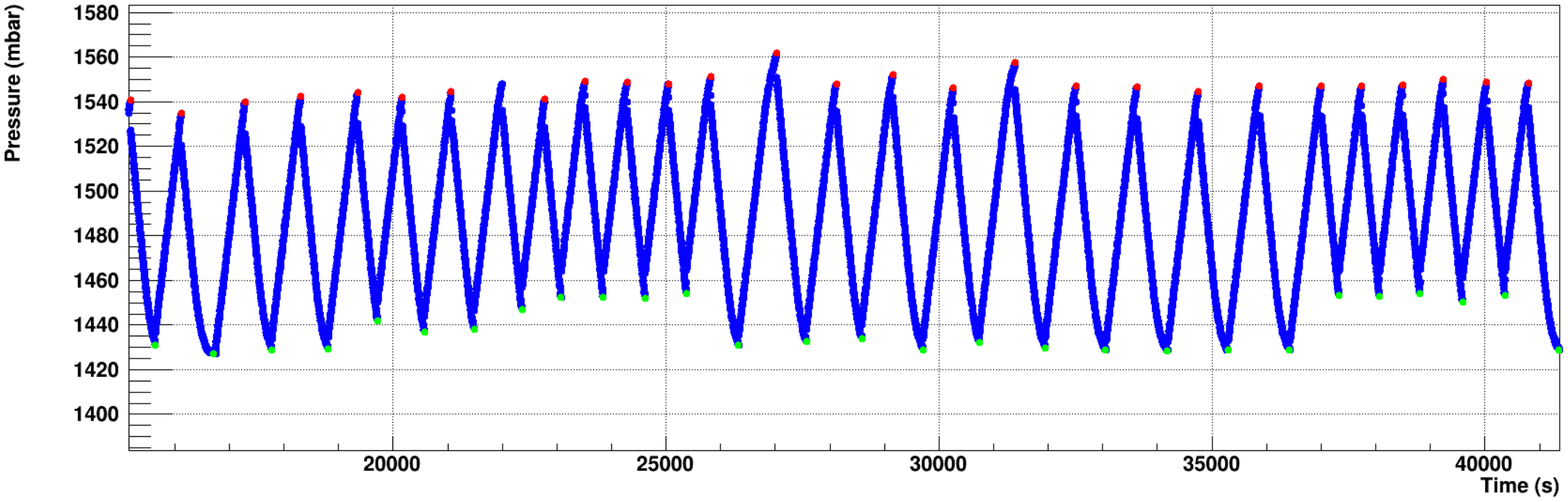}
\caption{ Pressure changes in the pump and purging cycle. The red dot is the local maximum and green dot is local minimum, both are identified by the program.  The average pumped out gas fraction is 6.6 $\pm$ 0.77\%.}
\label{fig:fcycle}
\end{figure}

\begin{figure}[htbp]
\hfill
\subfloat[]{\includegraphics[width=7cm]{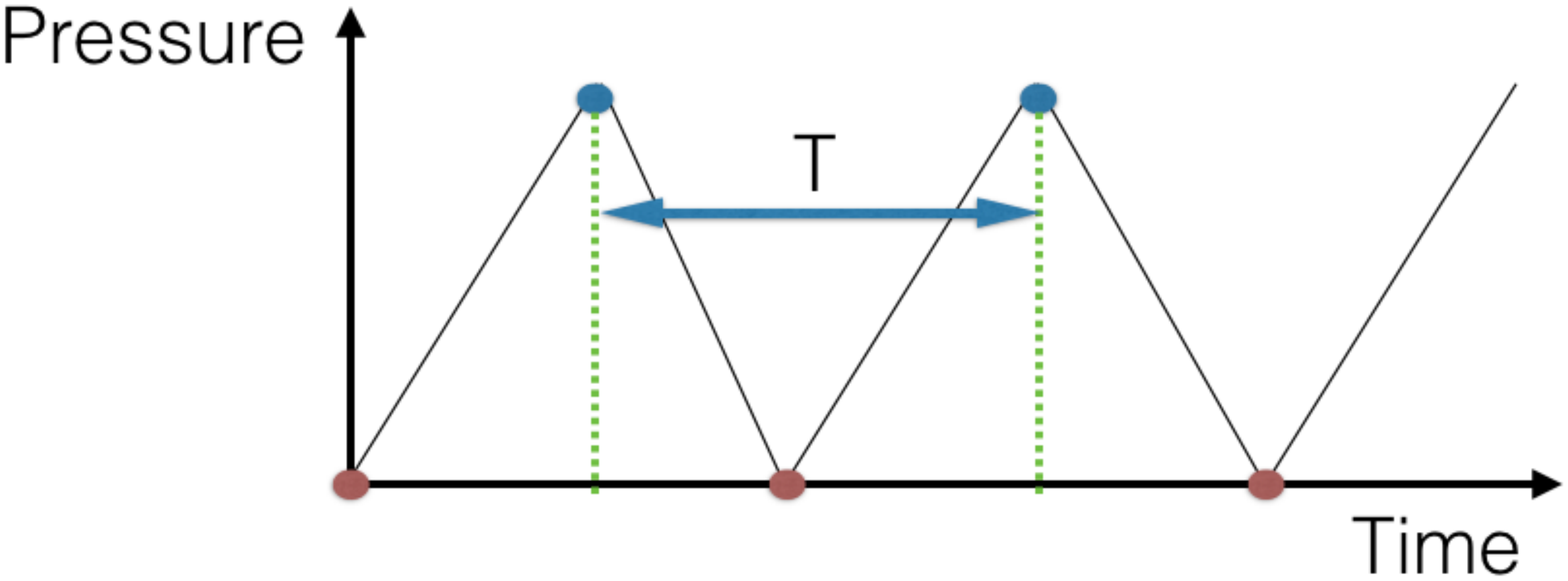}}
\hfill
\subfloat[]{\includegraphics[width=7cm]{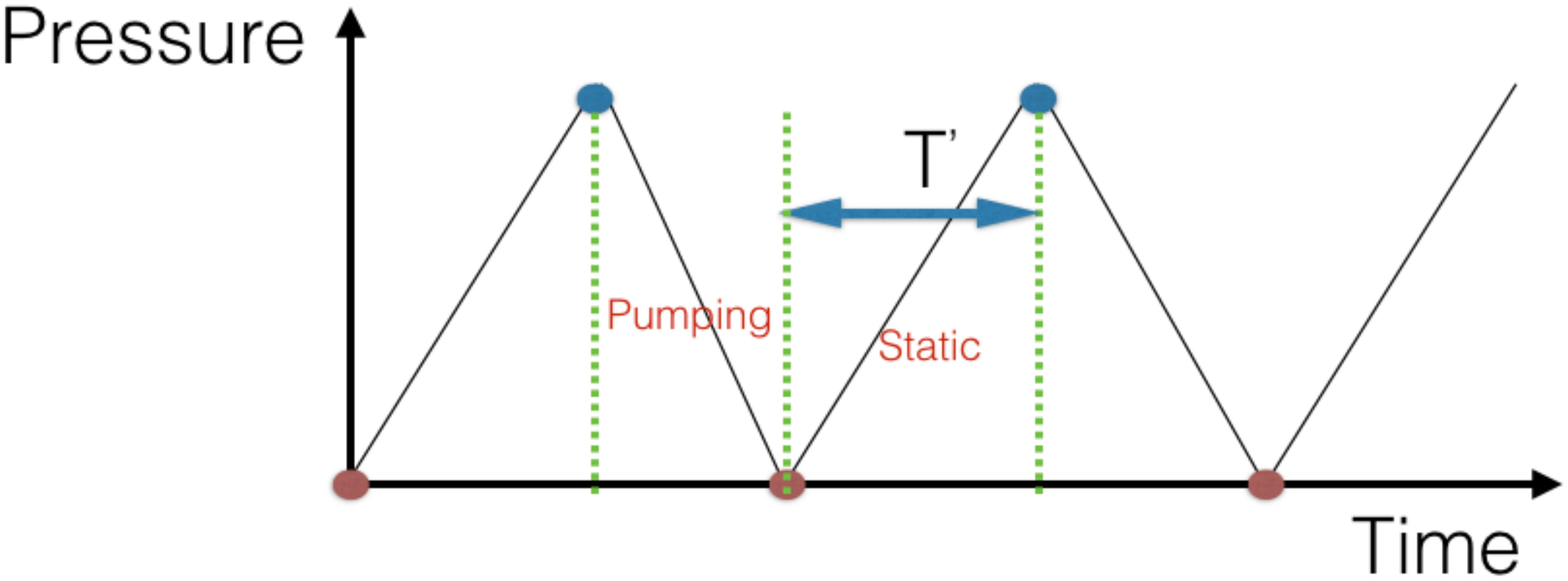}}
\hfill
\caption{A cartoon describes mapping impurity level to fitted triplet lifetime : (a) Determine the average impurity in period T using eq.(4) and populate the pulse-time of  scintillation events into histogram in period T to be fitted for triplet lifetime. (b) Estimate impurity level at the end of each cycle and populate the pulse-time of scintillation events into histogram while IV at static. }
\label{fig:fcartoon}
\end{figure}
\subsection{Results}
Using the average impurity of the period of time T to match the triplet lifetime measured from accumulating events during the period T.  
 The triplet lifetime and the associate impurity level is shown in Figure \ref{fig:fimpurity}. The curve is fitted to Birk's law like function
\begin{ceqn}\begin{align}\label{eq:tripletlifeitme}
\tau_{m} = \frac{\tau_{N}}{1+ k\cdot\eta }
\end{align}\end{ceqn}
where $\tau_{m}$ is the measured triplet lifetime, $\tau_{N}$ is a fit parameter representing the natural triplet lifetime with zero impurity present in gaseous argon, $\eta$ is the total impurity level in ppm and $k$ is the fitting constant. The result is confirmed by using the other method of measuring the impurity as described in previous section, the fitting results is consistent with the results from using the average impurity as shown in \ref{fig:fimpurity2}. We will adopt the average impurity for following analysis. \par
Alternatively, we can convert the lifetime to the decay rate and use the inverse function of Eq. \ref{eq:tripletlifeitme} to fit a line 
\begin{ceqn}\begin{align}\label{eq:7}
R_{m} = R_N\cdot (1+ k\cdot\eta)
\end{align}\end{ceqn}
where $R_m$ and $R_N$ are the inverse lifetimes (decay rates). Figure \ref{fig:fimpurity2} shows the decay rate as a function of impurity level.
The product of $R_N$ and $k$ is the reaction rate per ppm between argon and the impurity molecule. The reaction rate between argon and impurity molecule has been measured by various group\cite{doi:10.1063/1.437923}\cite{doi:10.1063/1.441532}\cite{doi:10.1063/1.436447}. The dominant impurity species in MinCLEAN detector are oxygen and nitrogen at
the operating temperature (<140 K). The quenching effect from nitrogen diminished with impurity level less than 1 ppm\cite{1748-0221-6-08-P08003}, therefore we assume the quenched light yield is mainly due oxygen when impurity 
level is below 1 ppm. Thus the reaction rate obtained from the fit represent the reaction rate between triplet state of argon and oxygen. The result is  $11.7 \times 10^{-10} cm^{3}/s$, which is in rough agreement with the results from the literature $2.6\times10^{-10} cm^{3}/s$\cite{doi:10.1063/1.437923} and the results from K. Mavrokoridis ($35\times10^{-10} cm^{3}/s$)\cite{1748-0221-6-08-P08003}.\par
We note that the $\chi^{2}$ of both fits are large, due to a discrepancy in the high impurity level region. This implies 
either the functional form of does not describe the behavior in high impurity level (> 1ppm) or some additional systematic error occurs in that region. We do not know of any such systematic error in our data.  We speculate that the functional form is altered at large impurity due to nitrogen which is known to strongly quench the triplet state for concentration above 1 ppm\cite{1748-0221-6-08-P08003}. As a cross check we refit the data averaging the lifetime and impurity every four hours and obtained very similar results with satisfactory $\chi^{2}$(Figure \ref{fig:ffour}). This confirms that the fitting results presents here is reliable.\par
In the literature, P. Moutard \cite{doi:10.1063/1.452869} collect the measurement from different experiments and acquired a equation to describe the behavior of triplet decay rate as a function of number density. Figure \ref{fig:fimpurityrate} shows P. Moutard's function in the green dashed line, the black points are older experimental results (prior 2000), the blue points is newer results (after 2000) and the red points are the results from this analysis. This plot shows our results is in agreement with previous experimental results. Table \ref{tab:title} summarize the results from previous measurements.


{\centering
\begin{tabular}{ C{1in} C{1in} *4{C{1in}}}\toprule[1.5pt]
\bf $\tau$ ($\mu$s) & \bf Ref. & \bf Number density ($10^{20}cm^{-3}$) & Estimate impurities level & Induced particle type\\\midrule
2.8 & Thonnard \textit{et al.}\cite{PhysRevA.5.1110} & 0.19 & < 2 ppm&$\beta$\\
2.84$\pm$0.02 & Gleason \textit{et al.}\cite{doi:10.1063/1.434079} & 4 & < 1 ppm&$\beta$\\
2.86 & P. Millet \textit{et al.}\cite{0022-3700-15-17-024} &0.1- 0.26& < 1 ppm&$\alpha$\\
2.88$\pm$0.08&K. Mavrokoridis \textit{et al.}\cite{1748-0221-6-08-P08003} &0.24 & < 1ppb&$\alpha$\\
2.9     &  Carvalho \textit{et al.}\cite{CARVALHO1979487}     &3 & not reported & $\beta$\\
3.0$\pm$0.05& Suemoto \textit{et al.}\cite{Suemoto1977131} &  2.2 & < 10 ppm&$\beta$\\
3.14 $\pm$0.067& C. Amsler  \textit{et al.}\cite{1748-0221-3-02-P02001} & 0.32 & < 9 ppb&$\alpha$\\
3.15$\pm$0.05 & P. Moutard \textit{et al.}\cite{doi:10.1063/1.452869} & 0& < 1 ppm &$ \gamma$ \\
3.2 $\pm$ 0.3& Keto \textit{et al.}\cite{PhysRevLett.33.1365} &  2.6 & < 2 ppm&$\beta$\\
3.22$\pm$0.042& Oka \textit{et al.}\cite{doi:10.1063/1.437923} & 0.32  & < 5 ppm &$\beta$\\
3.24$\pm$0.05&F. Marchal \textit{et al.}\cite{0953-4075-42-1-015201}& 0.2& <1 ppm & $\gamma$\\
2.524 $\pm$ 0.005 & this work & 0.85 & $\sim$1 ppm &$\beta$, $\gamma$ \\  
3.48 $\pm$ 0.01 & this work & 0.85 & \color{blue} < 1 ppb &$\beta$, $\gamma$\\\bottomrule[1.25pt]
\end {tabular}
\captionof{table}{Triplet lifetime in gaseous argon.   The variation of lifetimes is due to both density and (presumably) impurity level.  Only upper limits on impurity are reported. } \label{tab:title} 
}
\begin{figure}[tbp]
\centering
\graphicspath{{./fig/}}
\includegraphics[scale=0.4]{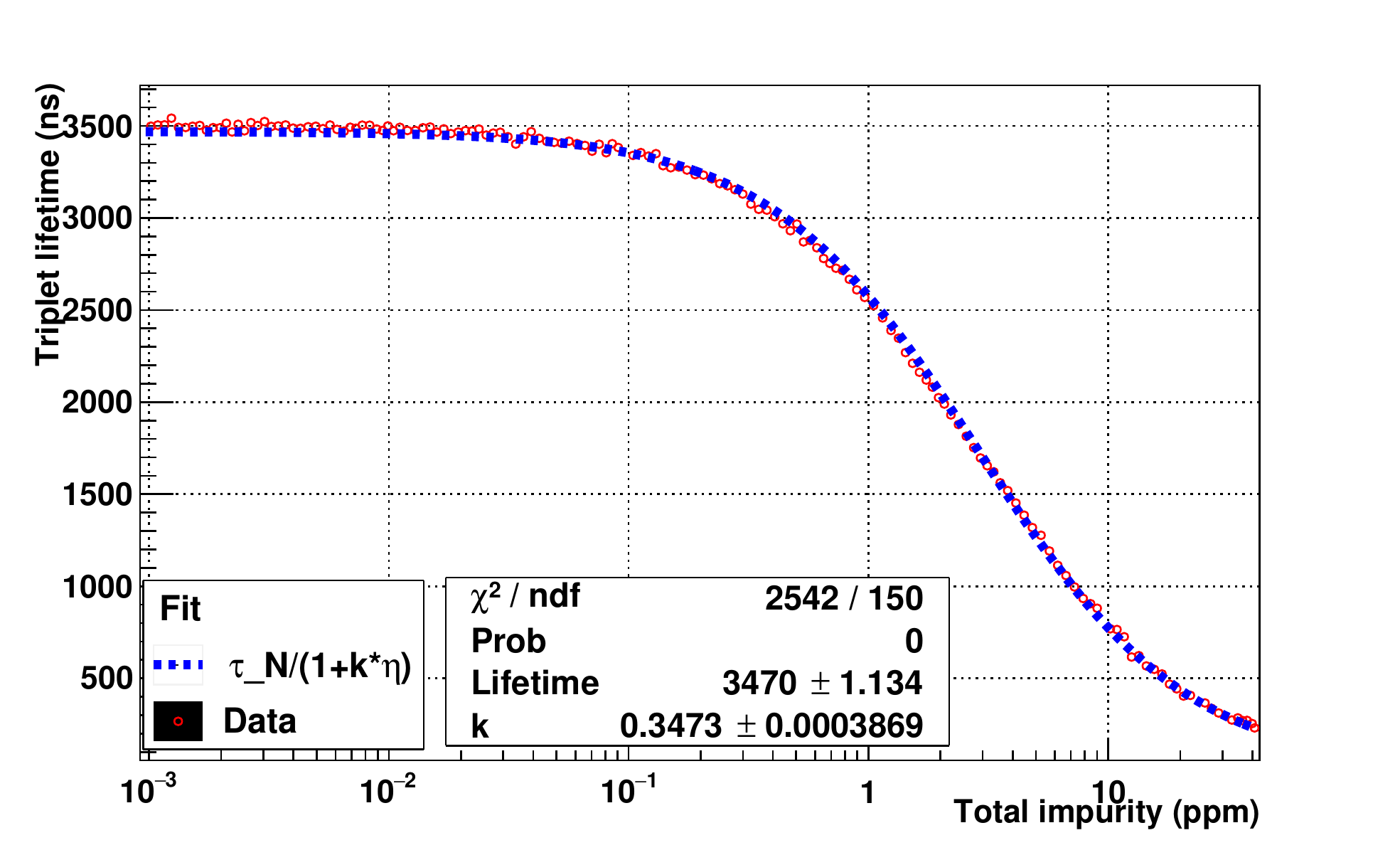}
\caption{Triplet lifetime vs average impurity level. The blue dashed line is the fitting function (Eq. \ref{eq:tripletlifeitme}).}
\label{fig:fimpurity}
\end{figure}

\begin{figure}[tbp]
\centering
\graphicspath{{./fig/}}
\includegraphics[scale=0.4]{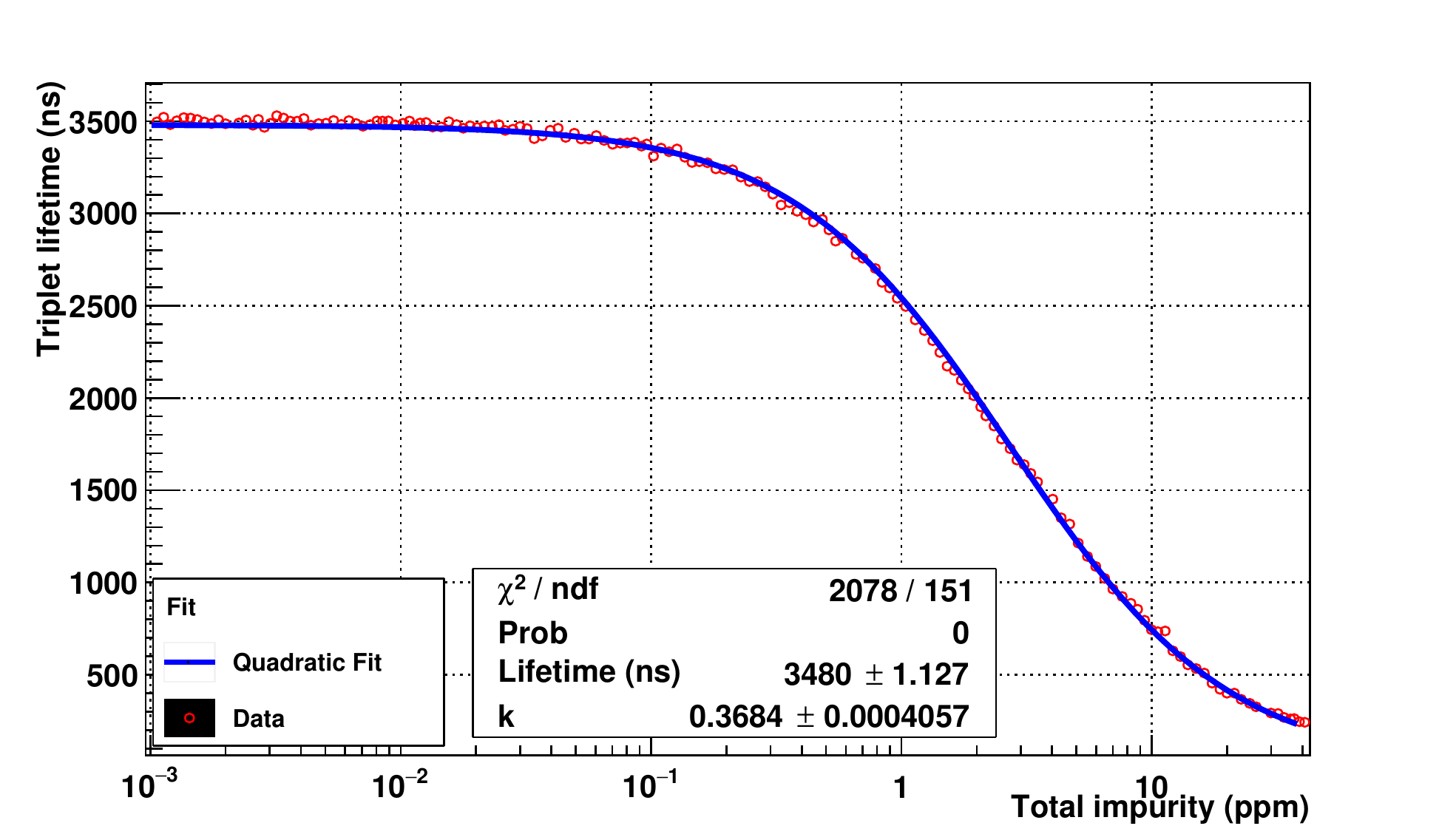}
\caption{Triplet lifetime vs total impurity level calculated at the end of each pumping cycle. The blue dashed line is the fitting function (Eq. \ref{eq:tripletlifeitme}).}
\label{fig:fimpurity2}
\end{figure}

\begin{figure}[tbp]
\centering
\graphicspath{{./fig/}}
\includegraphics[scale=0.4]{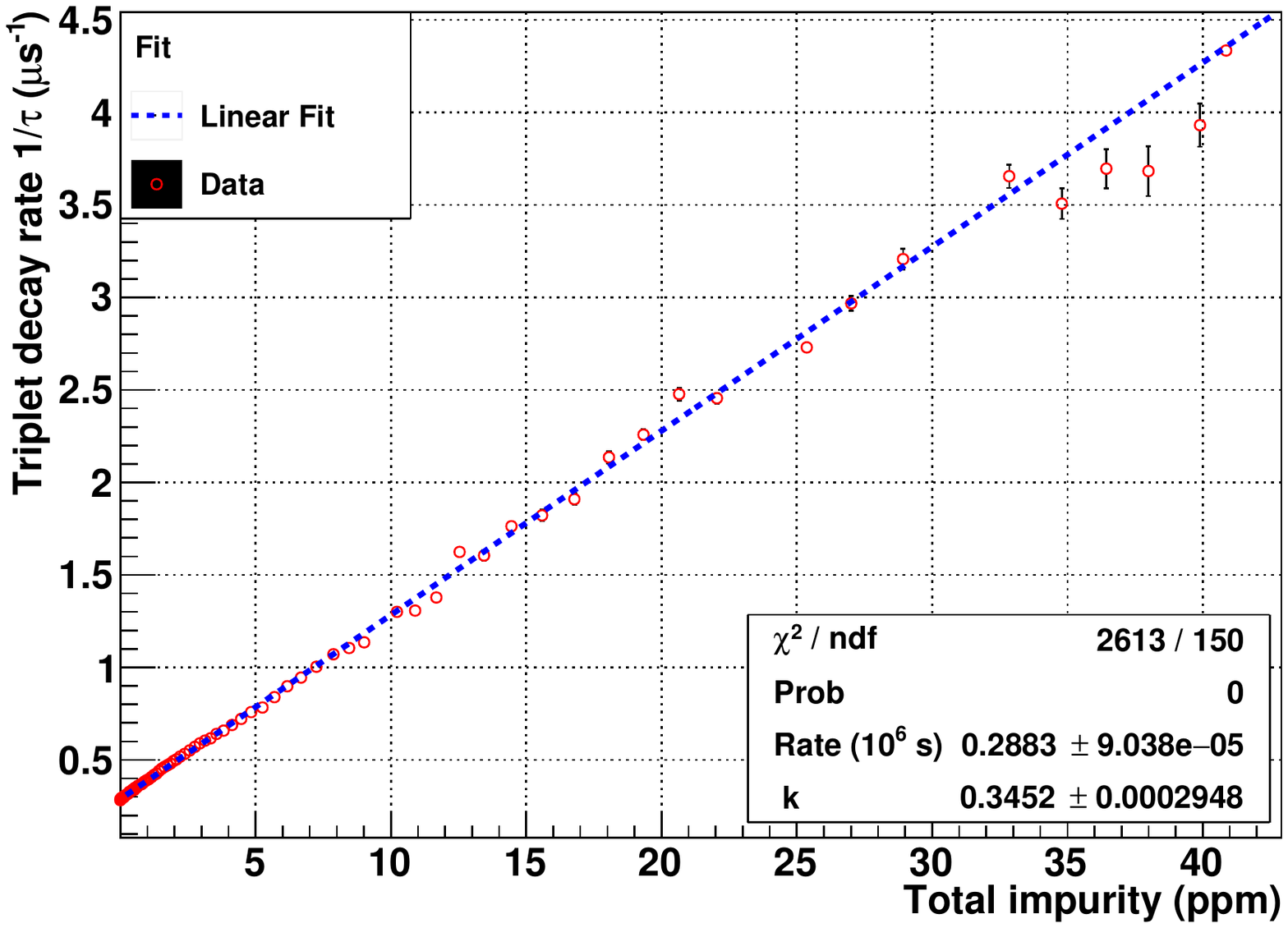}
\caption{Triplet lifetime vs total impurity level. The blue dashed line is the fitting function (Eq. \ref{eq:7}).}
\label{fig:fimpurity2}
\end{figure}
\begin{figure}
\hfill
\subfloat[]{\includegraphics[width=7cm]{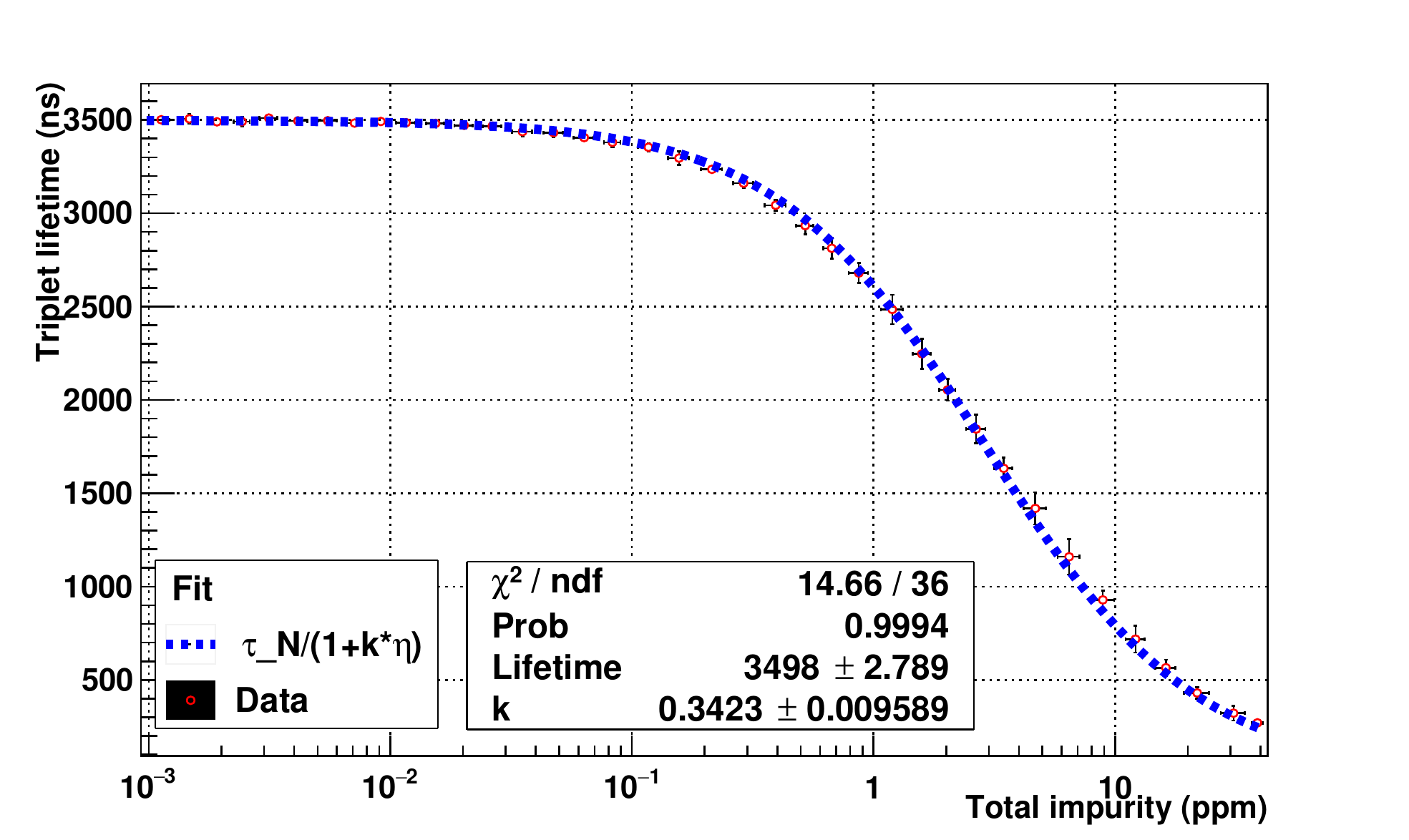}}
\hfill
\subfloat[]{\includegraphics[width=7cm]{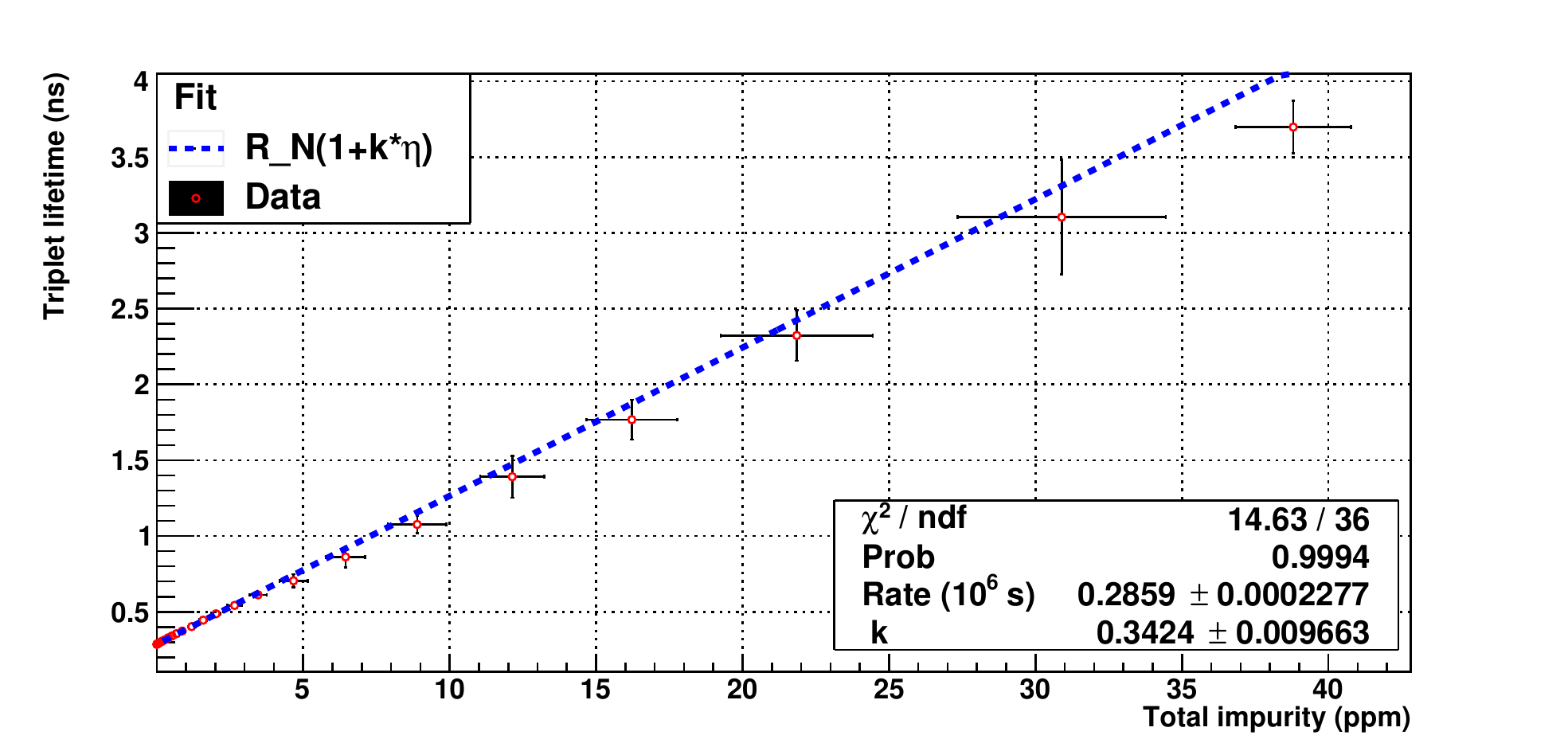}}
\hfill
\caption{(a) Triplet lifetime vs impurity level. Each point are the average value of 4 hrs data. (b) Triplet decay rate vs impurity level. Each point are the average value of 4 hrs data }
\label{fig:ffour}
\end{figure}

\begin{figure}[tbp]
\centering
\graphicspath{{./fig/}}
\includegraphics[scale=0.4]{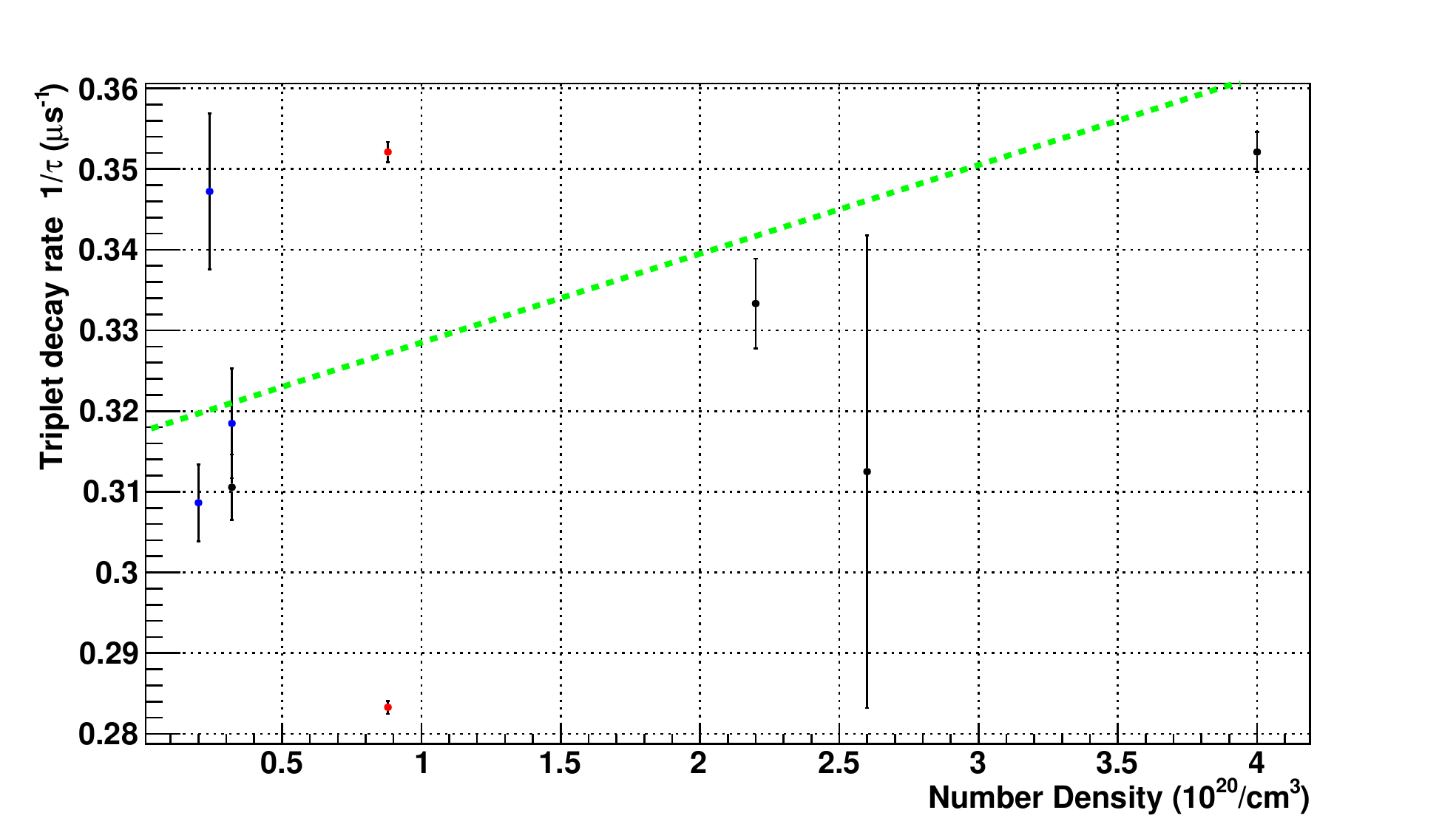}
\caption{Triplet decay rate (inverse lifetime) versus density from Table.  The green line is the fit of rate versus density from Moutard.  The two red points are the  results reported here, with the lower point at the best argon purity and the upper point at a purity of 1 ppm.}
\label{fig:fimpurityrate}
\end{figure}
\section{LY and Late/Prompt Ratio}
To determine the light yield (LY), we use the same cut to select $^{39}$Ar events. In gaseous argon (GAr), the scintillation light is produced in three continuous wavelength bands. In principle, the prompt light is mainly from longest wavelength band (third continuum, 180 - 210 nm) and the intermediate component is from singlet state of the second continuum (128 nm)\cite{Wieser2000233}. The triplet state of second continuum is responsible for the late component. Using the single PE counter and fit it with three exponential convoluted with Gaussian response function, the mean and sigma of these three component can be determined. Figure \ref{fig:fpecounterfit} shows the example of fitting. \par

Light yield (LY) for each component. We choose the prompt component using the mean $\pm$ 3$\sigma$ region to populate
the charge of each events into a histogram and determine the mean of the distribution. 
The starting point to accumulate events for the late component is from 80 ns to the end of data acquisition window. 
Lastly, the time window for intermediate component is from the end of prompt window until the beginning of the late window. The mean light yield for each components is determined from the charge distribution and plotted against triplet lifetime as shown in Figure \ref{fig:fmeanpe}. As can be seen in the figure, the prompt and intermediate component is relatively
flat compared to the late component. It is confirmed that the singlet state of second continuum (intermediate component) is not affected by the impurities while triplet state of second continuum (slow component) is strongly quenched. Moreover, the prompt states which comes from mostly the third continuum are not affect by impurities either.\par

The late/prompt ratio is obtained by directly integrating the summed waveform according to the different windows mentioned in last paragraph. The ratio is determined from summed waveform hourly and calculate the mean and standard deviation in each run.  In the cold gas runs, the electronic switching noise from unknown sources present at around 2600 ns makes the mean of the late integrated charge artificially large especially for runs with low triplet lifetime. 
The additional PMTs are removed from determining the late/prompt ratio to suppress the switching noise. 
This brings the total number of PMTs down to 56 (49) for pump and purge runs (early pump and purge runs). 
In addition, for each run, the flat background is identified by the fitter and subtracted from the data. The final results as a function of the average triplet lifetime run by run as shown in Figure \ref{fig:flpratiosum}. The late/prompt ratio determined by this method is 6.215 $\pm$ 0.007.
The discrepancy between our results and the results from \cite{1748-0221-3-02-P02001} is mainly from the different ionizing particle. 
The value 5.5 $\pm$ 0.6 is obtained using alpha particles known to produces more prompt light than electronic recoils. 
Therefore our value extracted from electronic recoil is in the reasonable range and consistent with the their result.
\begin{figure}[htbp]
\centering
\graphicspath{{./fig/Triplet/}}
\includegraphics[scale=0.4]{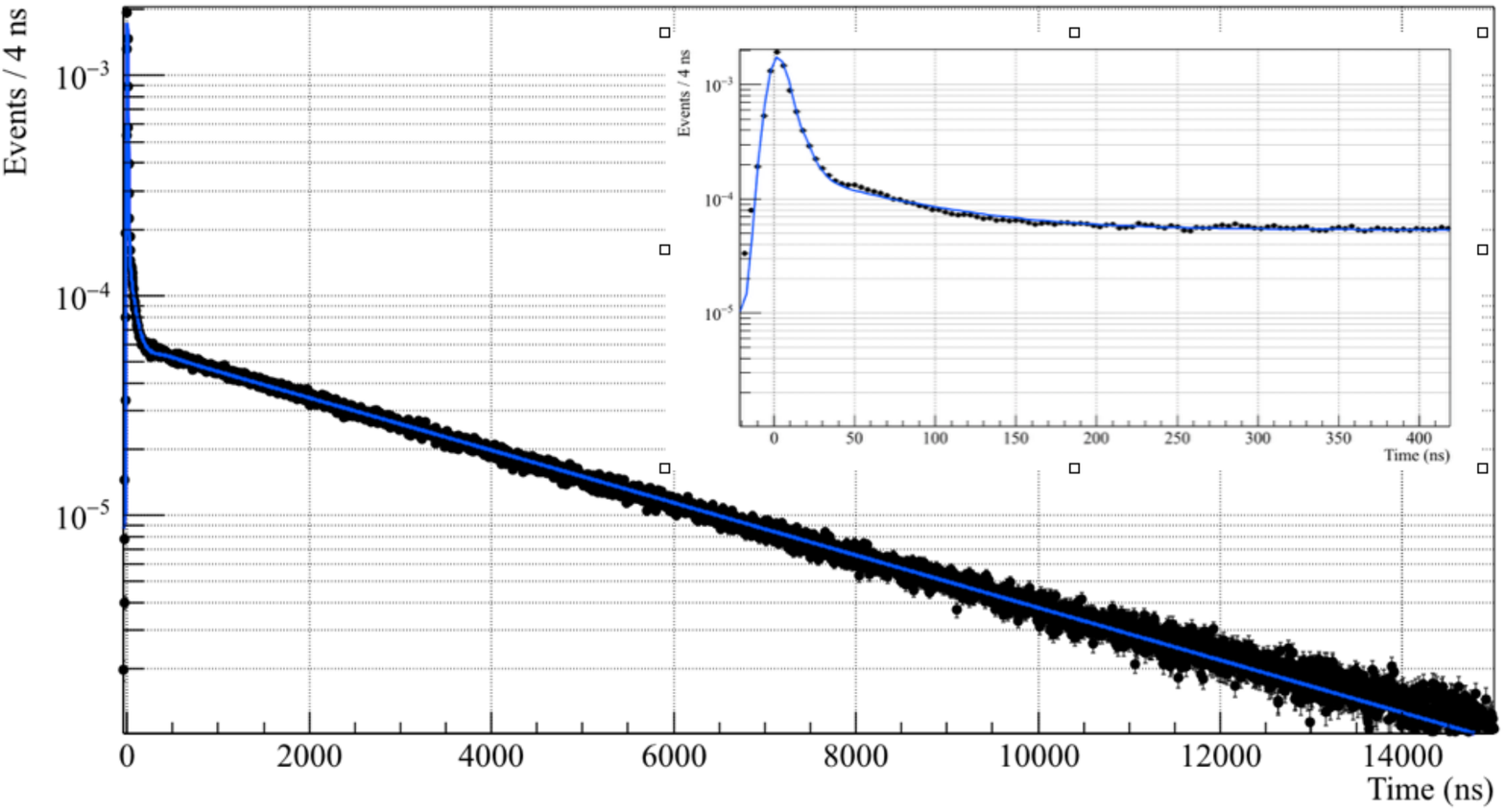}
\caption{ Example fit of single photoelectrons arrival time from scintillation events with three exponential convoluted with Gaussian resolution function in cold gas. The inset plot shows the first 400 ns.}
\label{fig:fpecounterfit}
\end{figure}

\begin{figure}[htbp]
\centering
\graphicspath{{./fig/Triplet/}}
\includegraphics[scale=0.6]{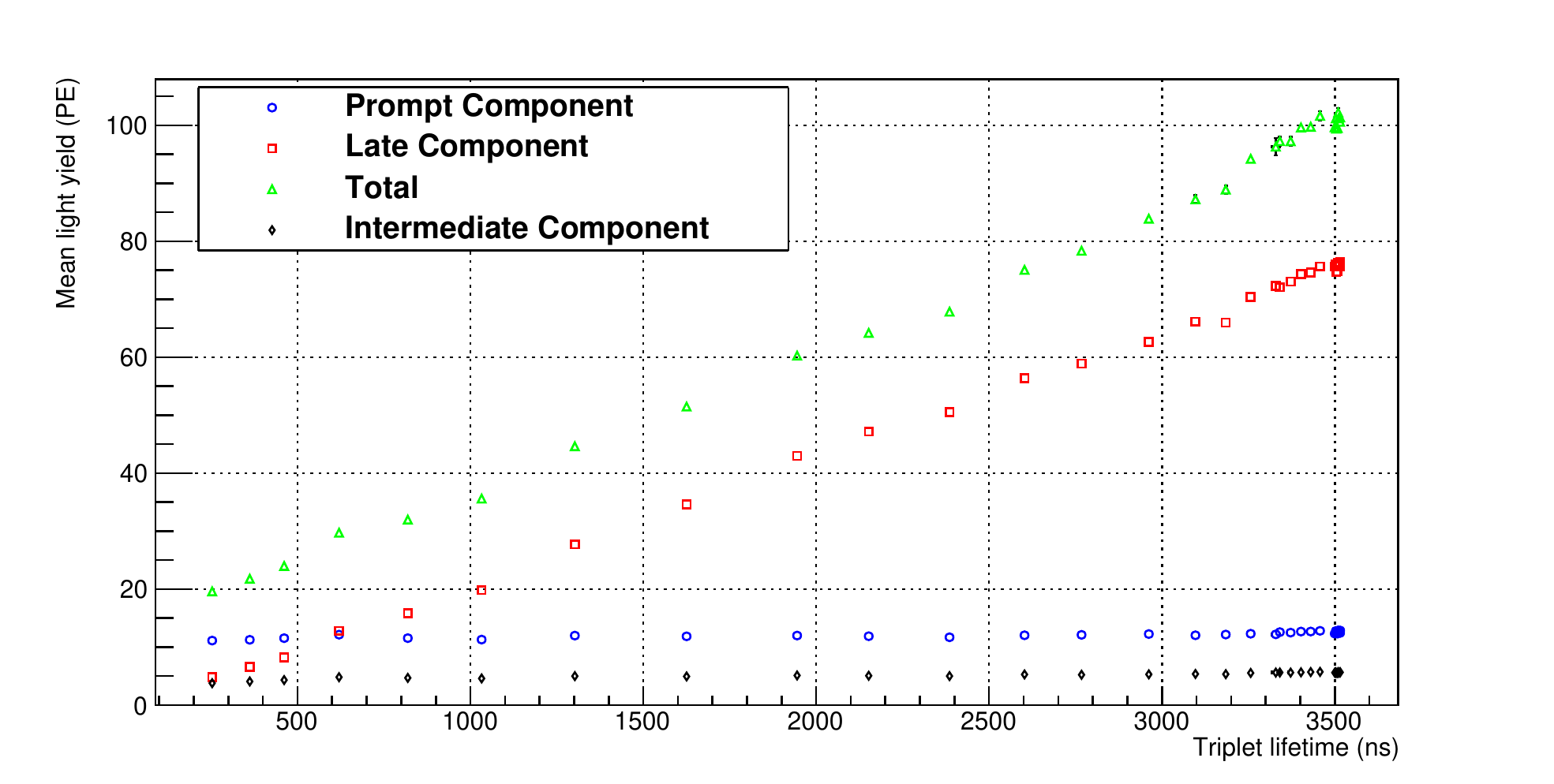}
\caption{ Mean light yield for each component vs triplet lifetime. }
\label{fig:fmeanpe}
\end{figure}
\begin{figure}[tbp]
\centering
\graphicspath{{./fig/}}
\includegraphics[scale=0.6]{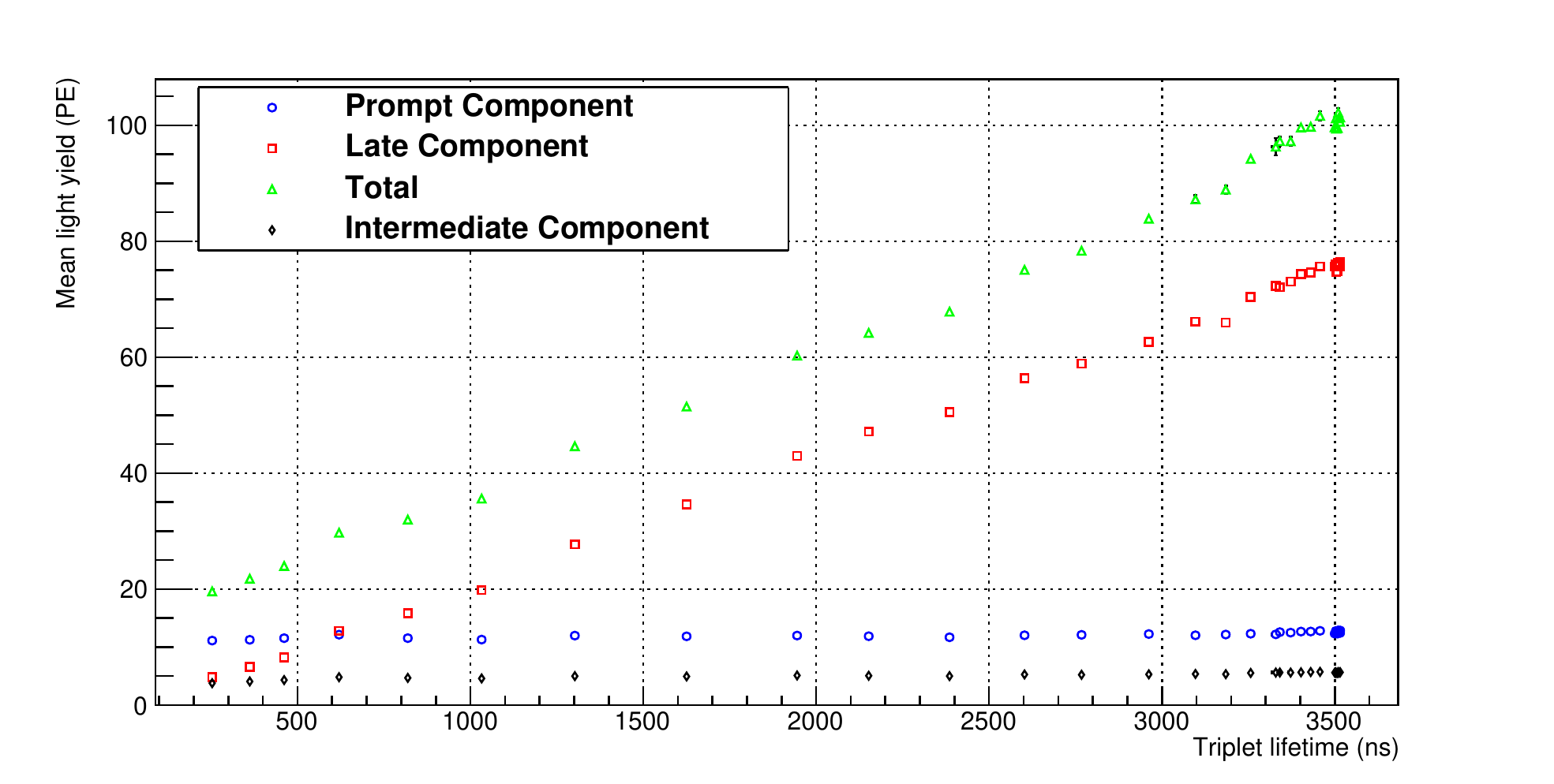}
\caption{ Mean light yield for each component vs triplet lifetime. }
\label{fig:fmeanpe}
\end{figure}


\begin{figure}[htbp]
\centering
\graphicspath{{./fig/Triplet/}}
\includegraphics[scale=0.6]{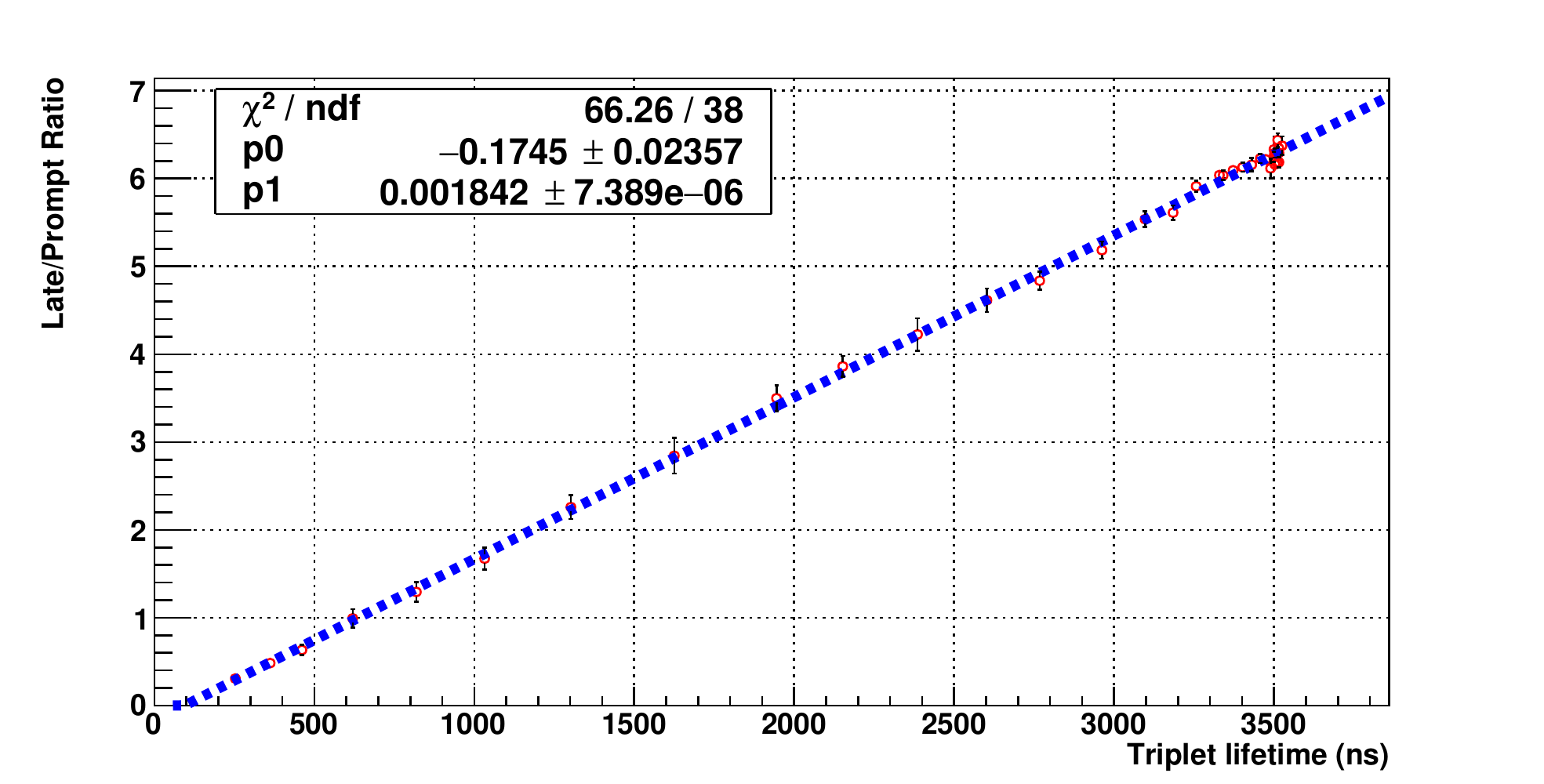}
\caption{ Ratio of late and prompt component determine from sum of waveform vs triplet lifetime.}
\label{fig:flpratiosum}
\end{figure}

\begin{figure}[htbp]
\centering
\graphicspath{{./fig/Triplet/}}
\includegraphics[scale=0.6]{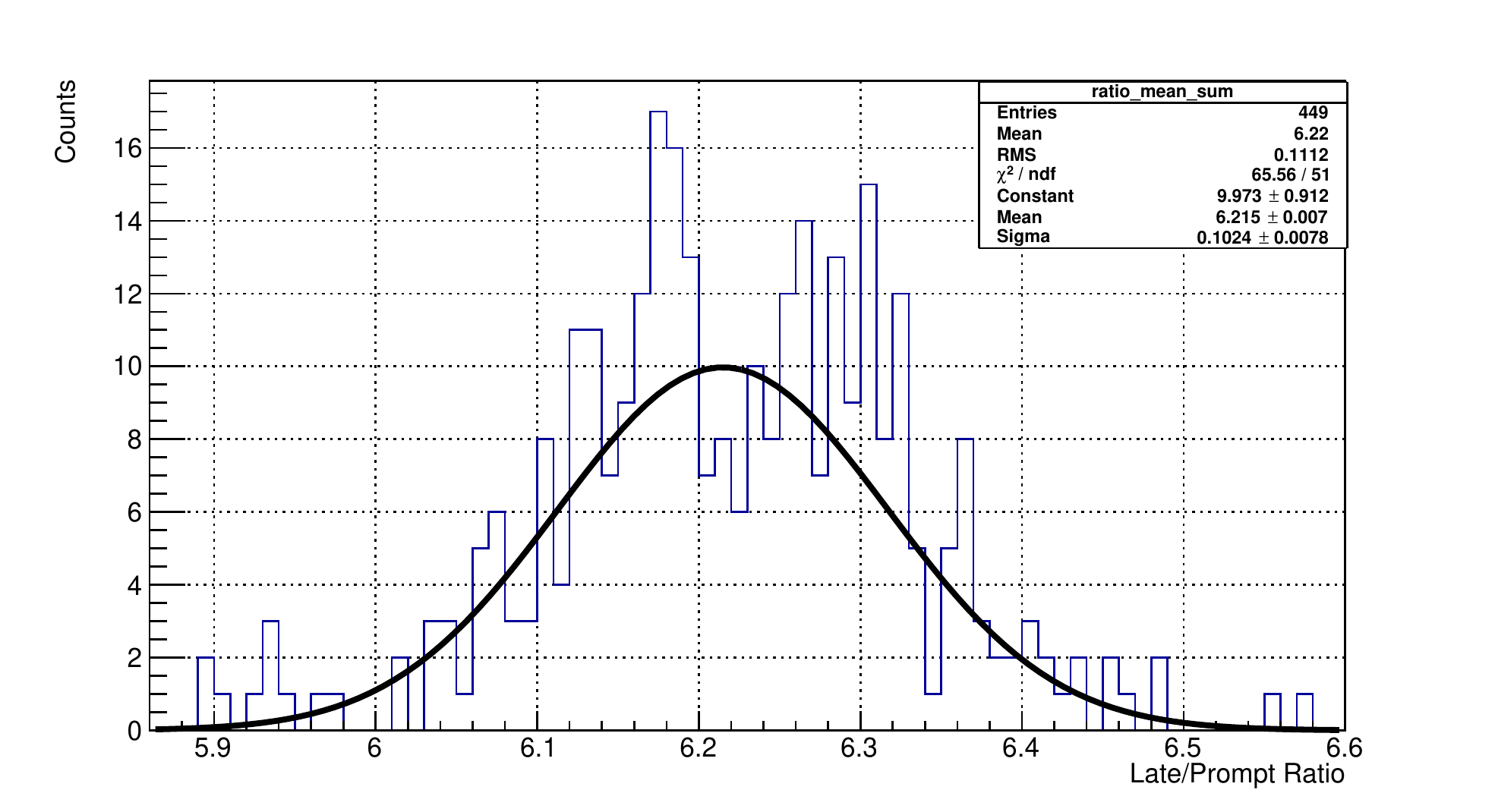}
\caption{ Ratio of late/prompt for runs having triplet lifetime larger than 3400 ns determined from sum of waveform.}
\label{fig:flpratiomeansum}
\end{figure}

\section{Systematic error of triplet lifetime measurement}
The possible sources of systematic error on triplet lifetime measurement are listed below. The detail descriptions of each source of systematic error are described in the following subsections.
\begin{itemize}
  \item Density effect
  \item PMT gain variation
  \item Pulse finding algorithm
  \item Pumping on IV
  \item Variations between PMTs
  \item Radius effect
  \item Systematic error on determining the impurity level
\end{itemize}
The 1-$\sigma$ region or the RMS of distribution will be used to quote the systematic error in the following analysis. 
\subsection{Density effect}
Throughout the pump and purging cycle, the temperature of IV is slowly increased and the pressure is changing. We can use average pressure and temperature in each hour to find out the variation of density during the pump and purge cycle. Using the equation from P. Moutard's \cite{doi:10.1063/1.452869}
(green dashed line in Fig. \ref{fig:fimpurityrate}) to find out the systematic errors causing by density variation.  Figure \ref{fig:fdensity} shows the number of density for each hour in pump and purging cycle populated into histogram. The histogram is fitted to  a Gaussian distribution, and using the mean and 1-$\sigma$ variation, after mapping to the Fig. \ref{fig:fimpurityrate}, we estimate the systematic error from the density effect is 0.07\%.

\begin{figure}[htbp]
\centering
\graphicspath{{./fig/Triplet/}}
\includegraphics[scale=0.6]{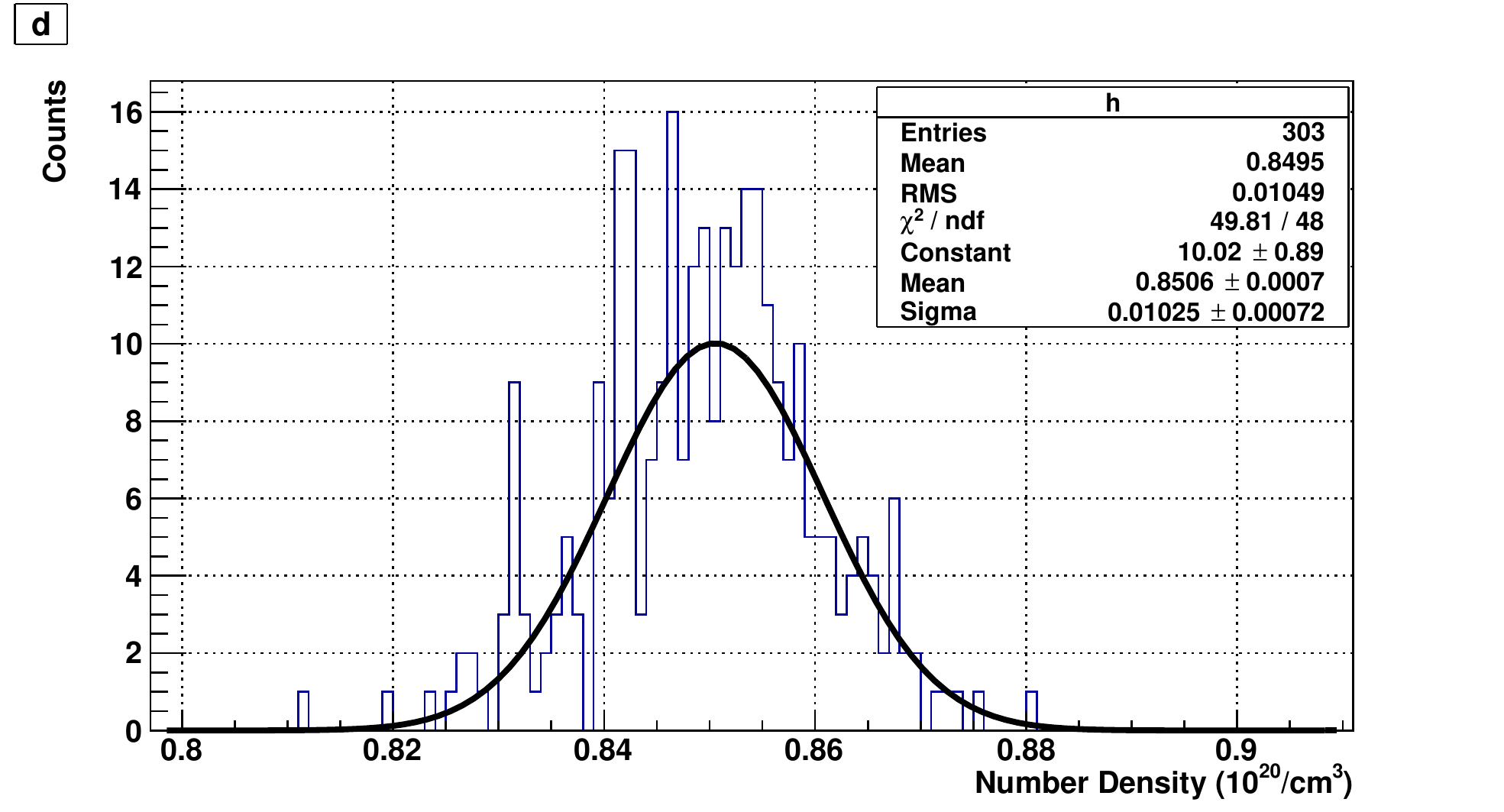}
\caption{Histogram of number of density of each hour during the pump and purge cycle and fitted with Gaussian distribution.  }
\label{fig:fdensity}
\end{figure}
\subsection{PMT gain variation}
In the pump and purge runs, we utilized the late light of scintillation events to determine the SPE value for each PMTs. The SPE value is determined on run to run basis, thus in order to estimate the gain variation hourly, we use the prompt light of scintillation events to estimate the PMT gain variation. The variation could affect the efficiency on pulse finding and results in systematic errors on determining the triplet lifetime. Figure  \ref{fig:fgainpmt} shows the relative gain variation during Run 962 (> 33 hr),  this run is used to estimate the systematic errors due to the PMT gain variations. The relative gain variation is defined as RMS of gain variation in 33 hr divided by mean gain in the 33 hr. Using the relative gain variation in Fig. \ref{fig:fgainpmt}, modify the PMT gain in the pulse finding algorithm. Then fit the modified pulse-time distribution and compare it with unmodified result as shown in Fig. \ref{fig:fpmtgain}(a). We can estimate the systematic error from the histogram of difference between two results divided by unmodified result as shown in \ref{fig:fpmtgain}(b). The estimated systematic error from the RMS of the distribution is $\pm$ 0.26\%.

\begin{figure}[htbp]
\centering
\graphicspath{{./fig/Triplet/}}
\includegraphics[scale=0.5]{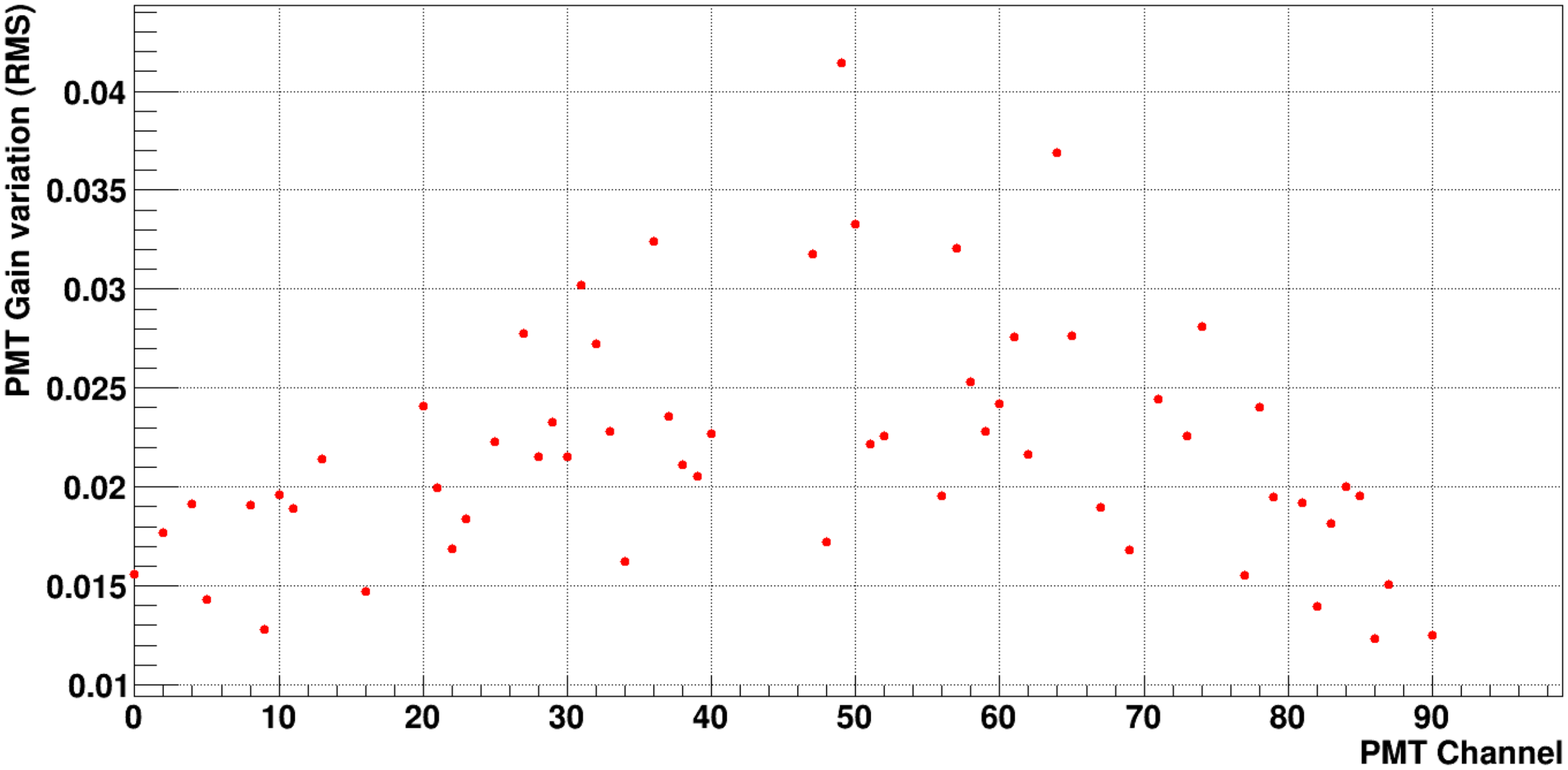}
\caption{ PMT relative RMS gain variation in Run 962.  }
\label{fig:fgainpmt}
\end{figure}
\begin{figure}[htbp]
\hfill
\subfloat[]{\includegraphics[width=7cm]{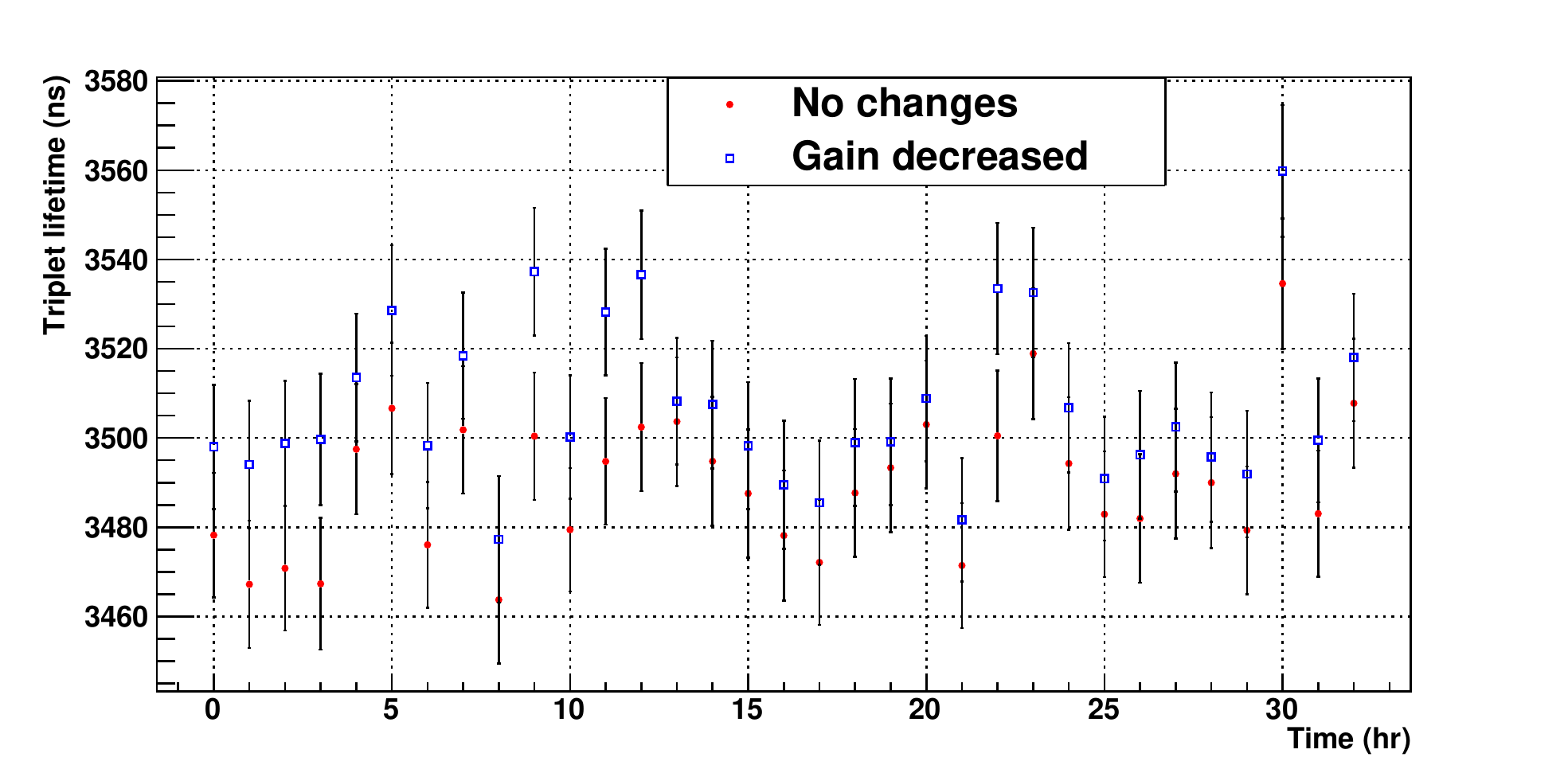}}
\hfill
\subfloat[]{\includegraphics[width=7cm]{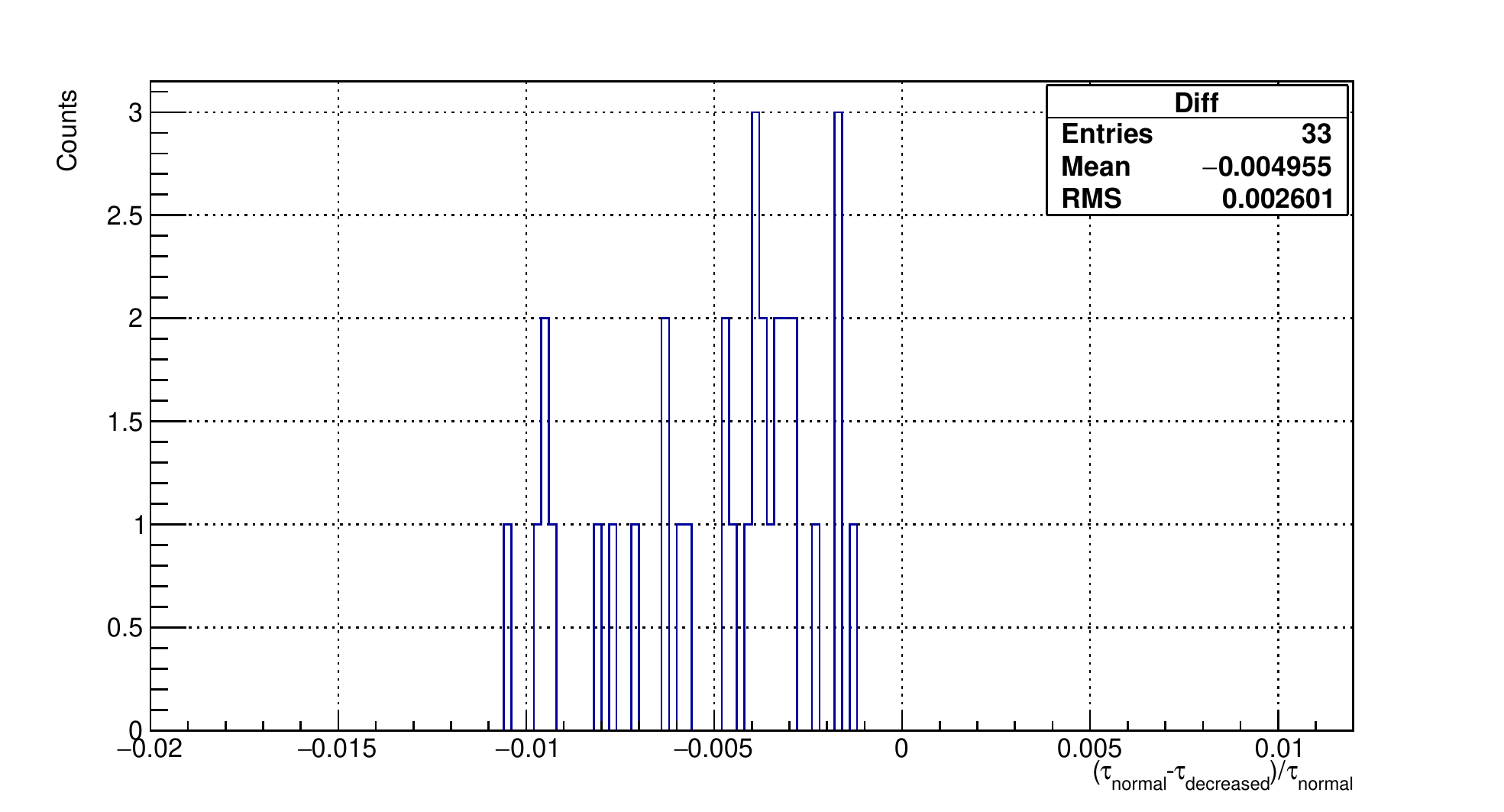}}
\hfill
\caption{(a) Triplet lifetime measured hourly in Run 962. (b) Difference in triplet lifetime for before/after gain changed. }
\label{fig:fpmtgain}
\end{figure}
\subsection{Pulse finding algorithm}
By comparing the fitting results between pulse-time distribution and summed waveform, we can estimate the systematic error results from pulse finding algorithm. Figure \ref{fig:fsum} shows the fitted triplet lifetime from pulse-time distribution and overlap with results from summed waveform and Fig. \ref{fig:fweight} shows the weighted mean for two different methods. The example of fitting for pulse-time distribution and summed waveform is shown in Fig. \ref{fig:fexampletwo}. Due to the larger error bar of the bins in the summed waveform, the fitted result has larger error bar for each hour of the data. The systematic error is estimated from comparing the results of the fit, this gives the estimated systematic error to be 0.39\% as shown in Fig. \ref{fig:fsyspmts}.

\begin{figure}[htbp]
\centering
\graphicspath{{./fig/Triplet/}}
\includegraphics[scale=0.6]{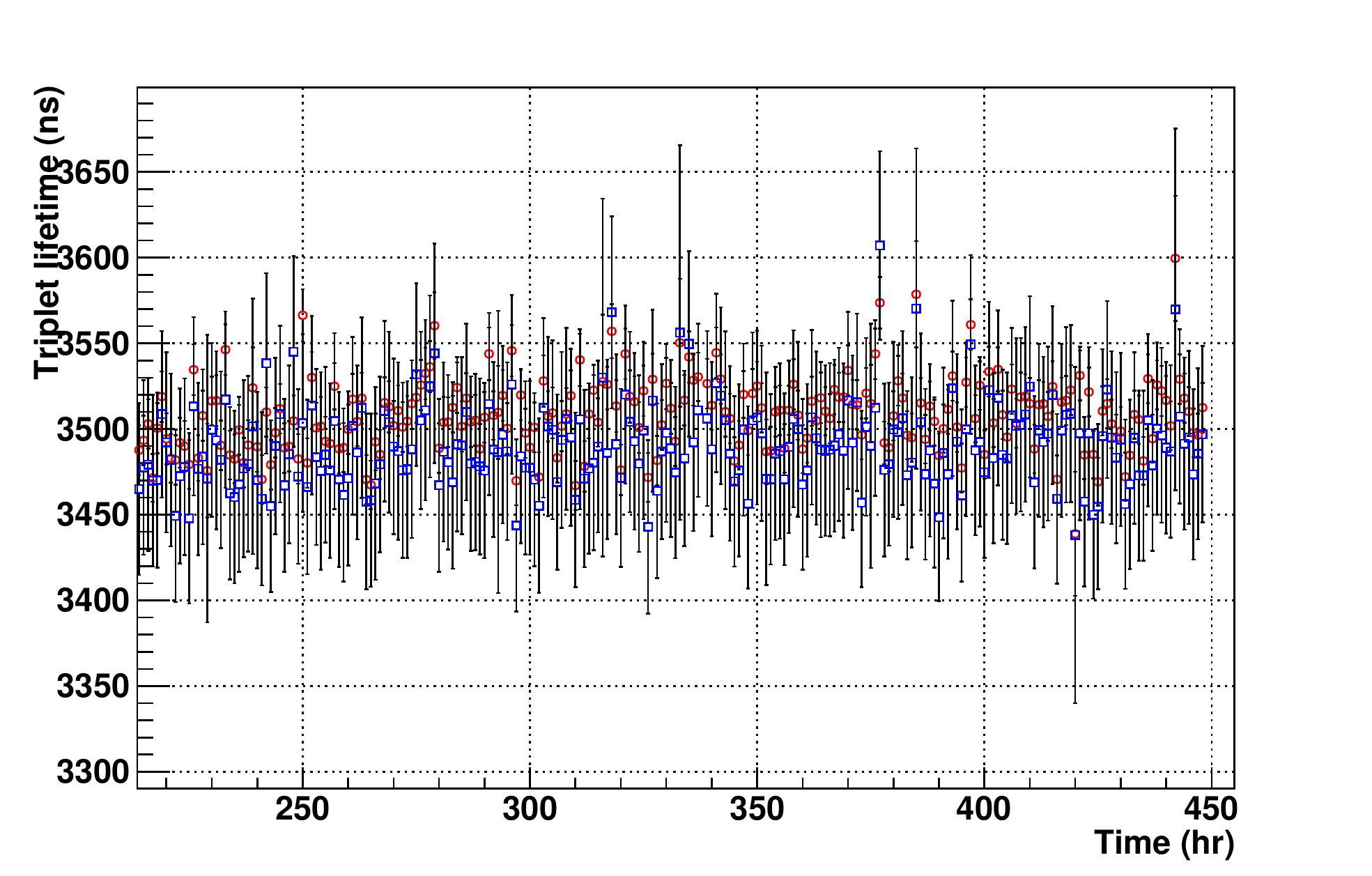}
\caption{ Fitting results from pulse-time distribution (red) and sum of the waveform (blue) overlap in the plot.  }
\label{fig:fsum}
\end{figure}
\begin{figure}[htbp]
\hfill
\subfloat[]{\includegraphics[width=7cm]{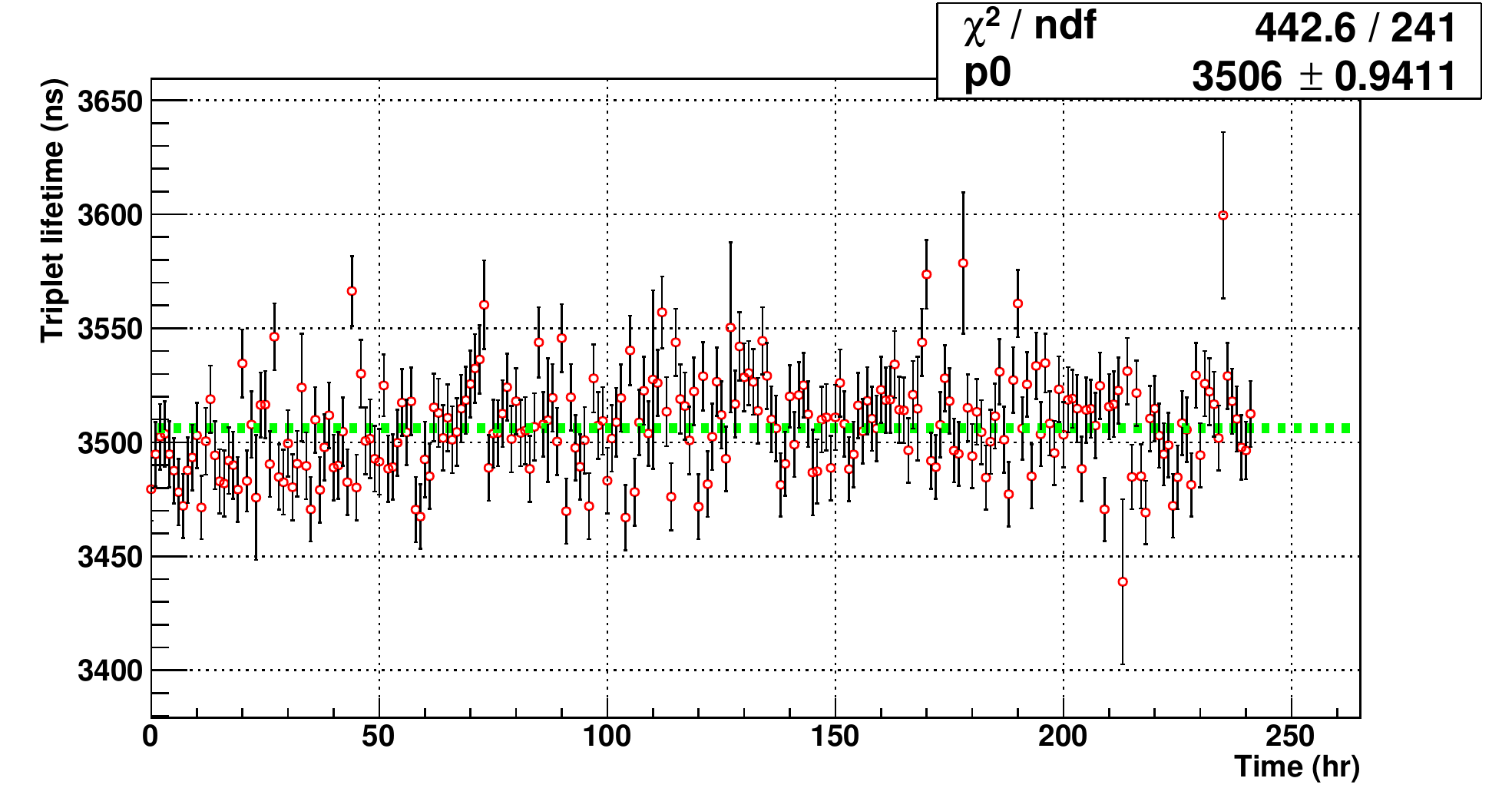}}
\hfill
\subfloat[]{\includegraphics[width=7cm]{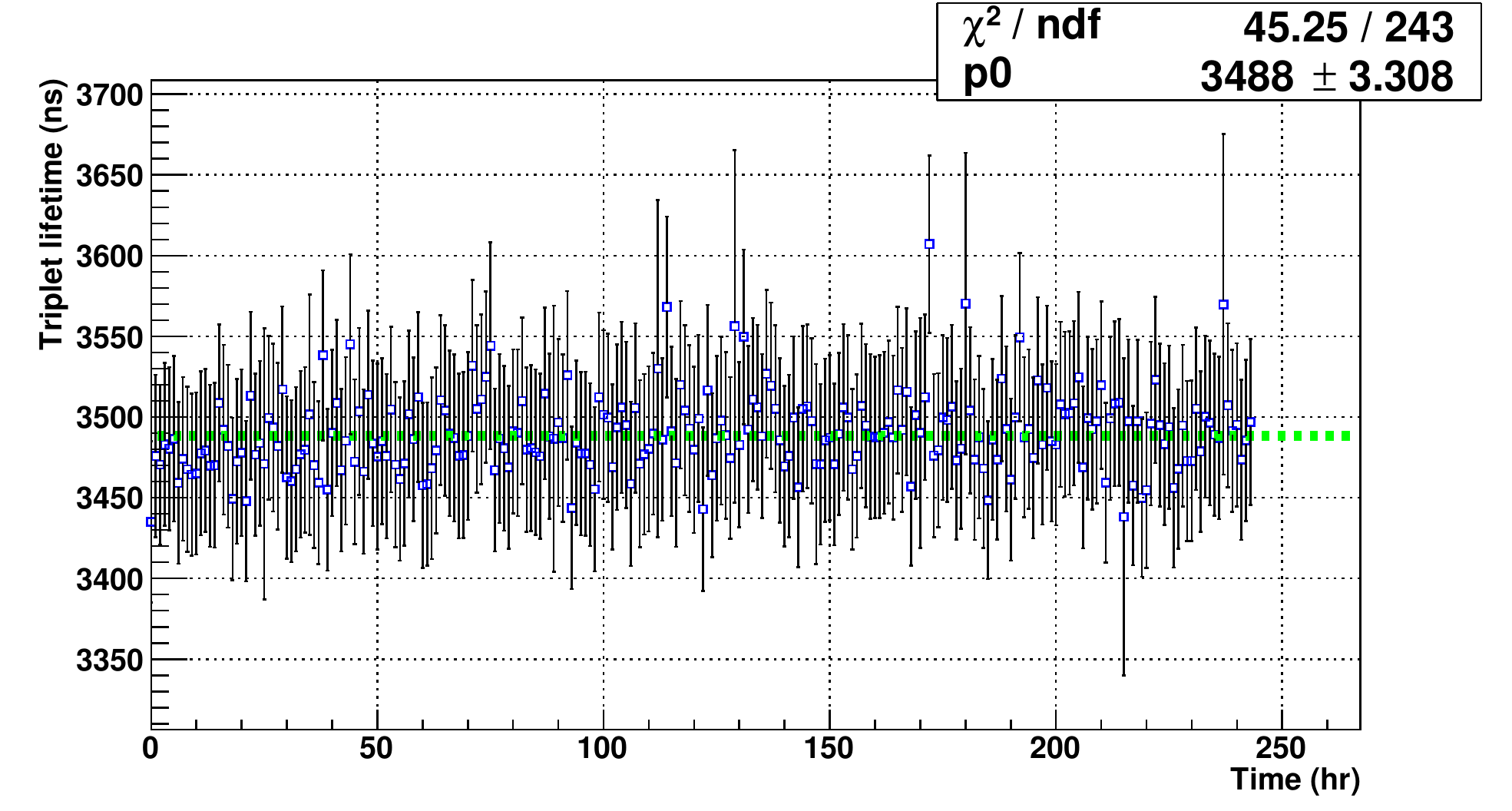}}
\hfill
\caption{(a) Weighted mean of lifetime from pulse-time distribution (b) Weighted mean of lifetime from sum of the waveform. }
\label{fig:fweight}
\end{figure}
\begin{figure}[htbp]
\hfill
\subfloat[]{\includegraphics[width=7cm]{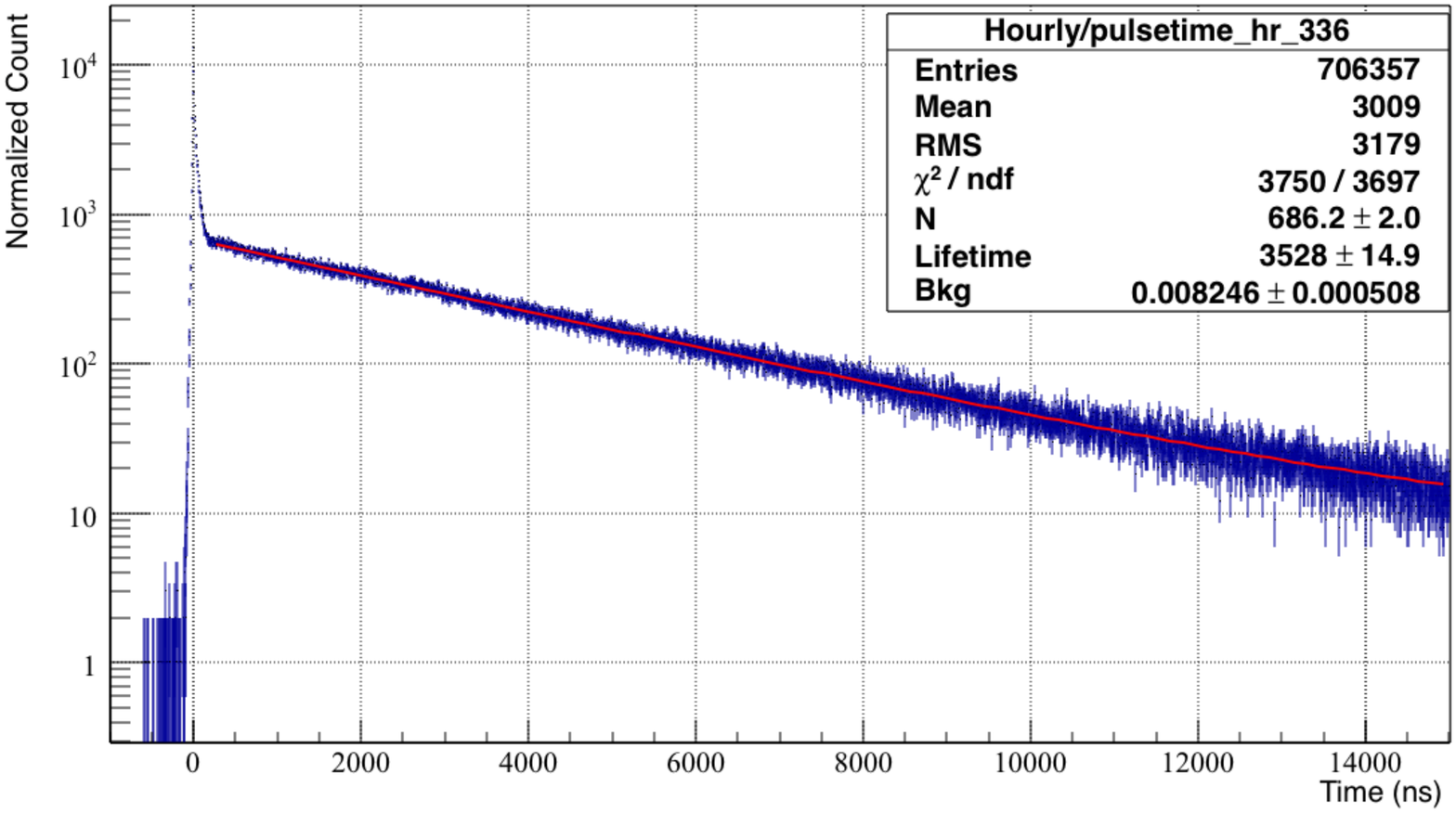}}
\hfill
\subfloat[]{\includegraphics[width=7cm]{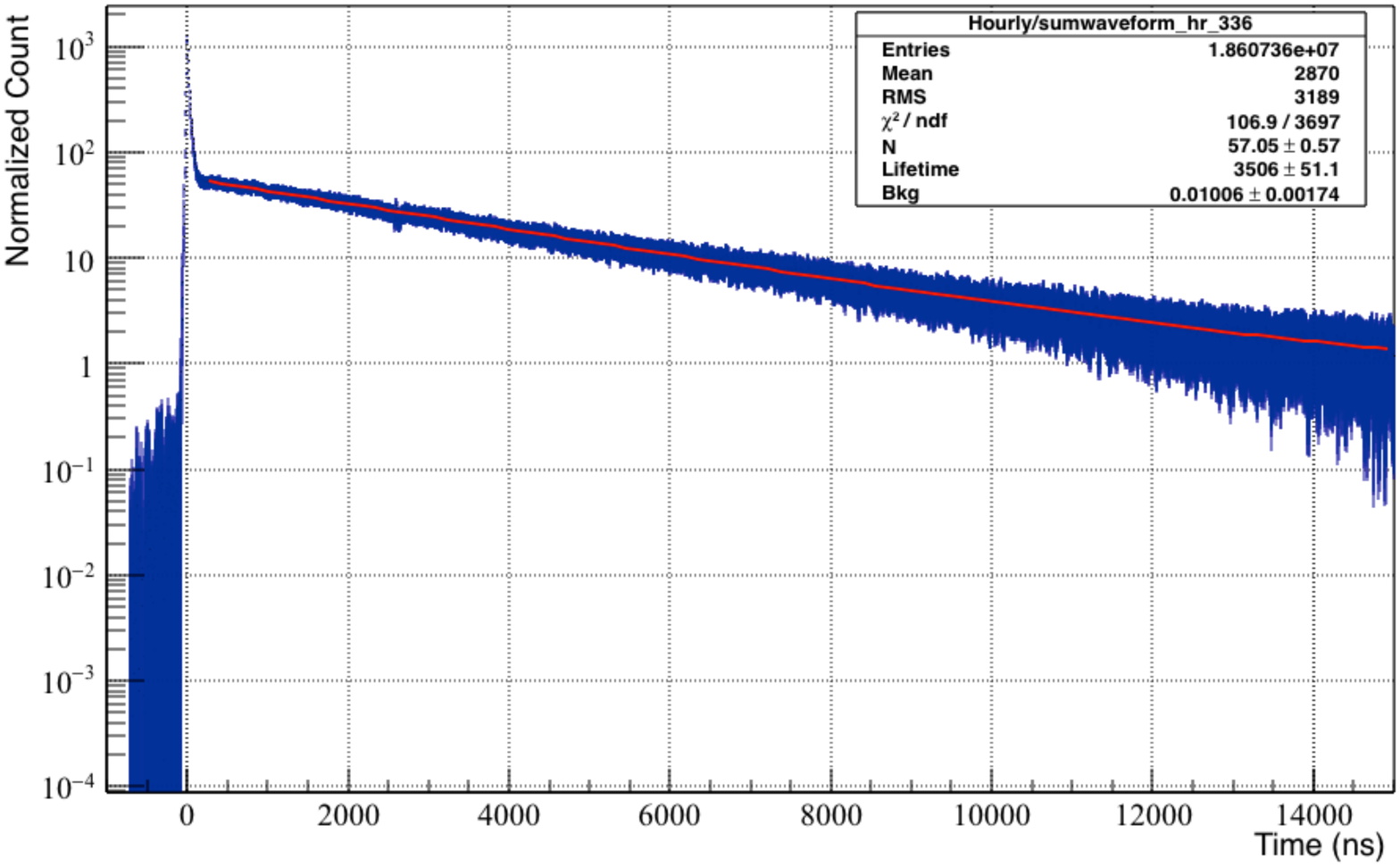}}
\hfill
\caption{Fitting example of the same data set : (a) Example fit of pulse-time distribution (b) Example fit of sum of the waveform. }
\label{fig:fexampletwo}
\end{figure}
\begin{figure}[htbp]
\centering
\graphicspath{{./fig/Triplet/}}
\includegraphics[scale=0.6]{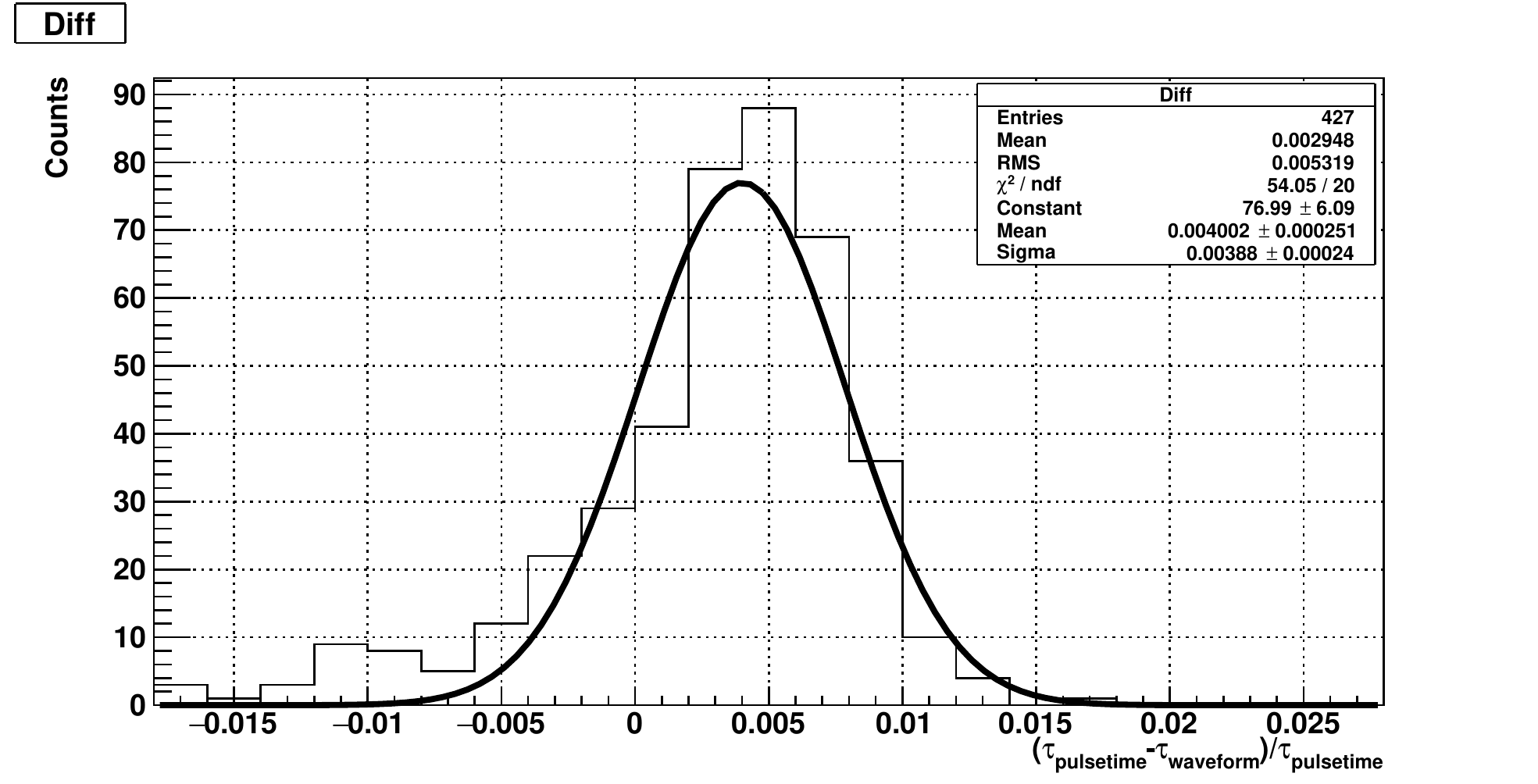}
\caption{ Histogram of difference in triplet lifetime measured from pusle-time and sum of the waveform.  }
\label{fig:fsyspmts}
\end{figure}
\subsection{Pumping on the IV}
Comparing the triplet lifetime between IV being pumped and IV at static can tell us the effect on triplet lifetime from pumping the IV. Figure \ref{fig:fpumpstatic} shows the triplet lifetime during pumping the IV and IV at static. The estimated systematic error results from pumping the detector is $\pm$ 0.73\% as shown in \ref{fig:fpumpstaticsys}.

\begin{figure}[htbp]
\hfill
\subfloat[]{\includegraphics[width=7cm]{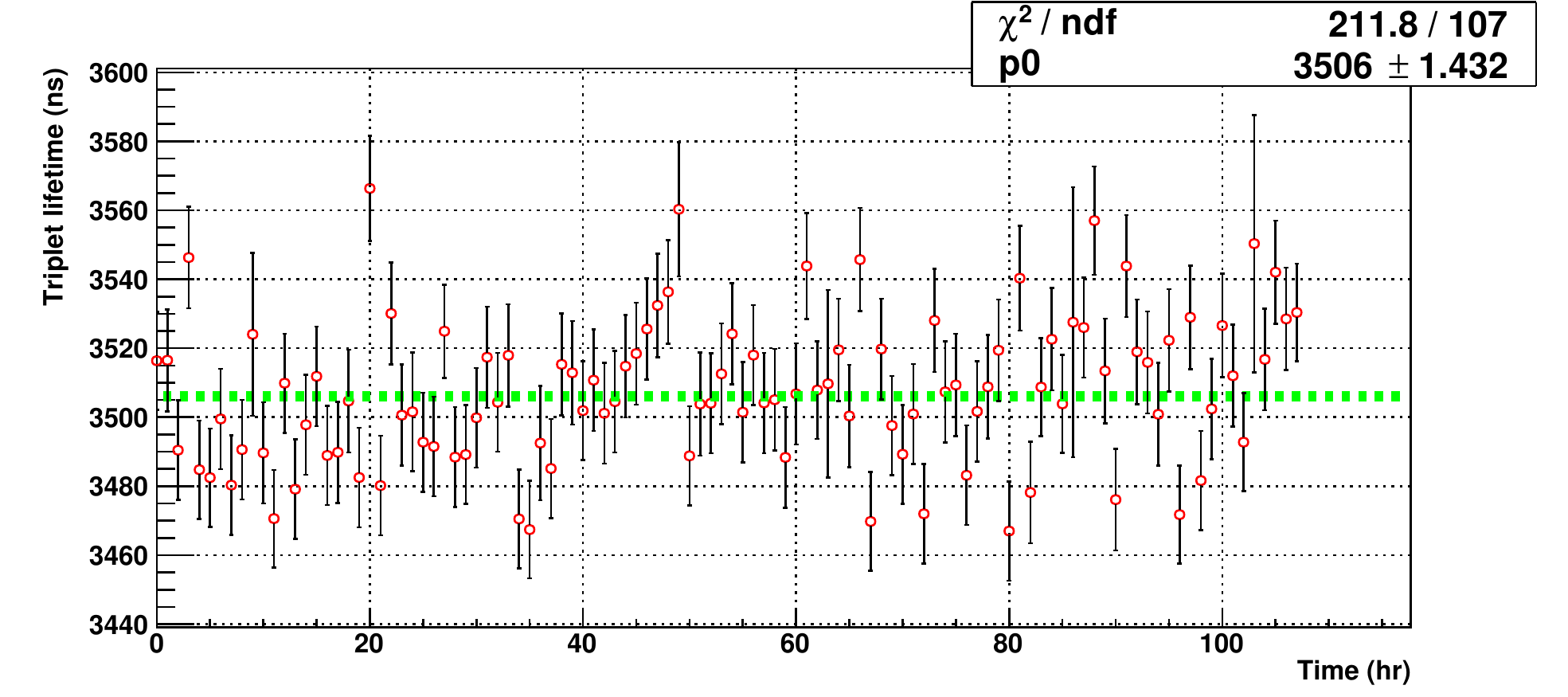}}
\hfill
\subfloat[]{\includegraphics[width=7cm]{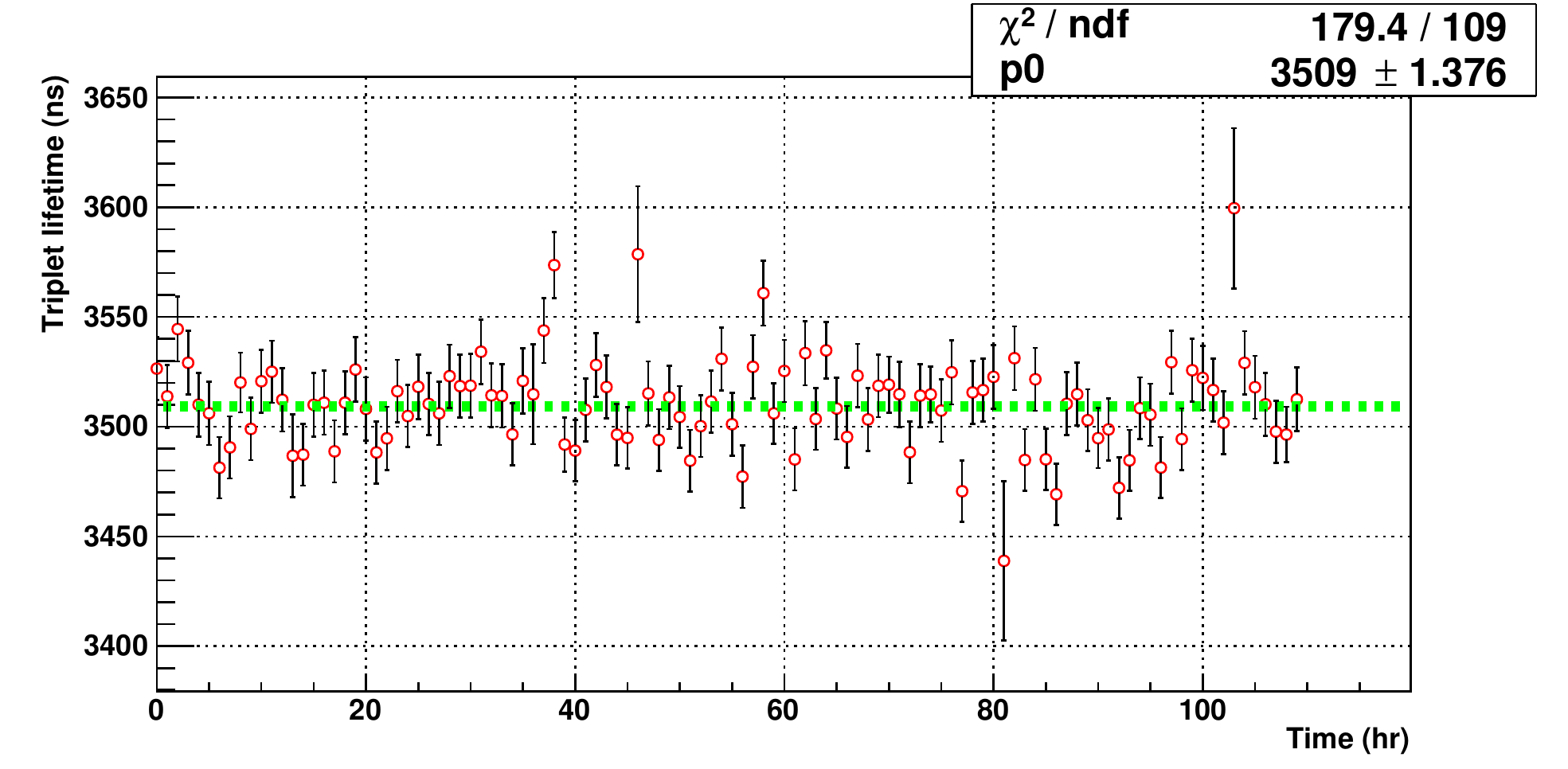}}
\hfill
\caption{(a) Triplet lifetime during the period of pumping the IV. (b) Triplet lifetime during the period of IV at static. }
\label{fig:fpumpstatic}
\end{figure}
\begin{figure}[htbp]
\centering
\graphicspath{{./fig/Triplet/}}
\includegraphics[scale=0.6]{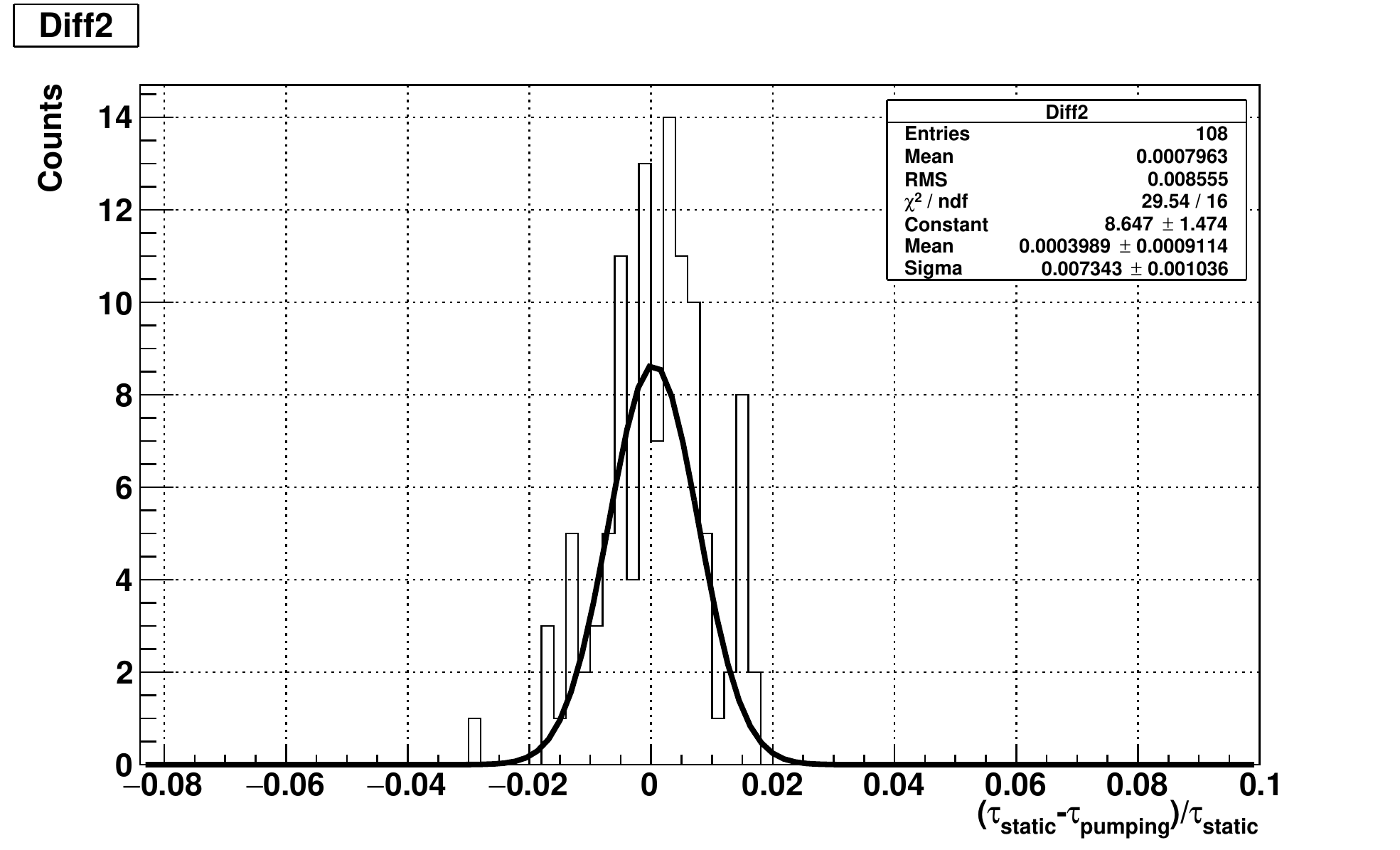}
\caption{ Histogram of the difference of triplet lifetime between pumping and static divided by static.  }
\label{fig:fpumpstaticsys}
\end{figure}
\subsection{Variation between PMTs}
In the earlier results, the variation of triplet lifetime is large. Figure \ref{fig:fbkg} shows the triplet lifetime of PMTs with cut : Qratio <0.6 and Fprompt <0.5 using $\chi^{2}$ fit. It is clear to see that the PMT 35 has larger background than the rest of PMTs, so we decided to remove this PMT from analysis. In this section, the data used to estimate the triplet lifetime variation between PMTs are after the lifetime reaches stable states (see Fig. \ref{fig:fpumptime}). Using the same fitting function (eq. (1)) to fit the pulse-time distribution PMT by PMT. The result is shown in Fig. \ref{fig:fpmtvar}. The possible reason of the variations may due to the fact that the top and bottom PMTs has higher event rates than the others and affects the final fitting results. The estimated systematic error from PMTs variation is 0.96\% as shown in Fig. \ref{fig:feachPMTs}.

\begin{figure}[htbp]
\centering
\graphicspath{{./fig/Triplet/}}
\includegraphics[scale=0.6]{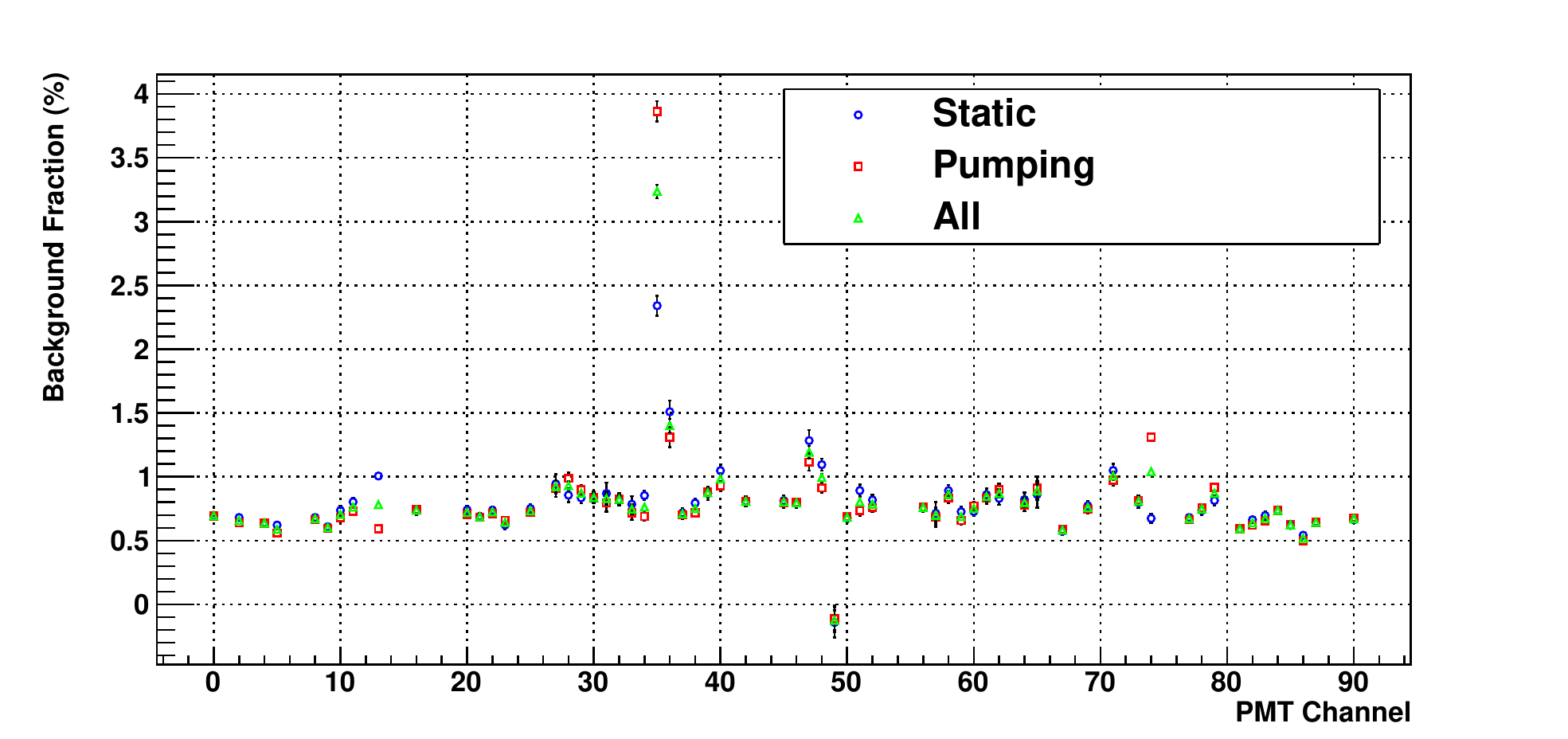}
\caption{ Fitted background fraction for each PMTs.  }
\label{fig:fbkg}
\end{figure}

\begin{figure}[htbp]
\centering
\graphicspath{{./fig/Triplet/}}
\includegraphics[scale=0.6]{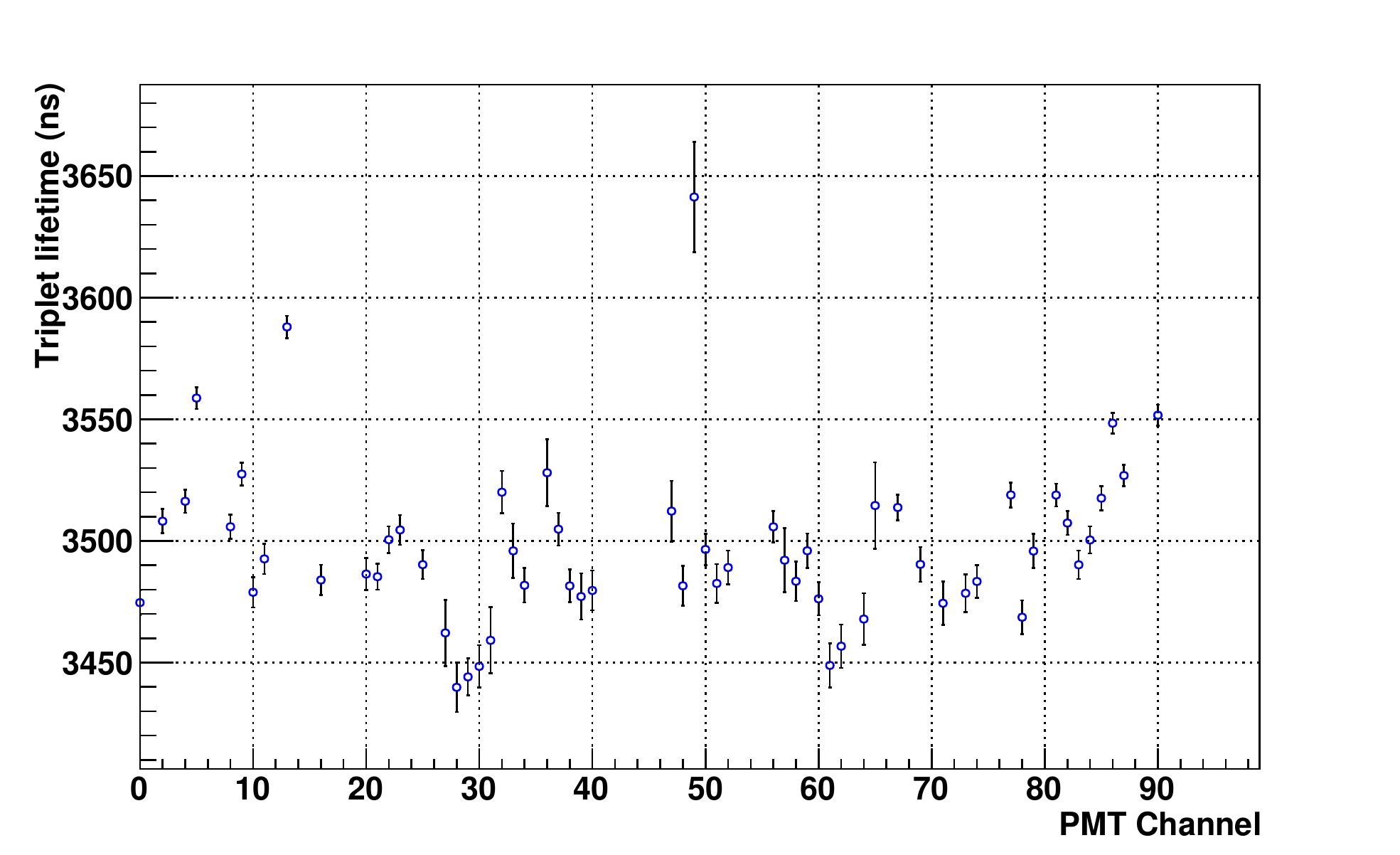}
\caption{ Triplet lifetime for different PMTs  }
\label{fig:fpmtvar}
\end{figure}

\begin{figure}[htbp]
\centering
\graphicspath{{./fig/Triplet/}}
\includegraphics[scale=0.6]{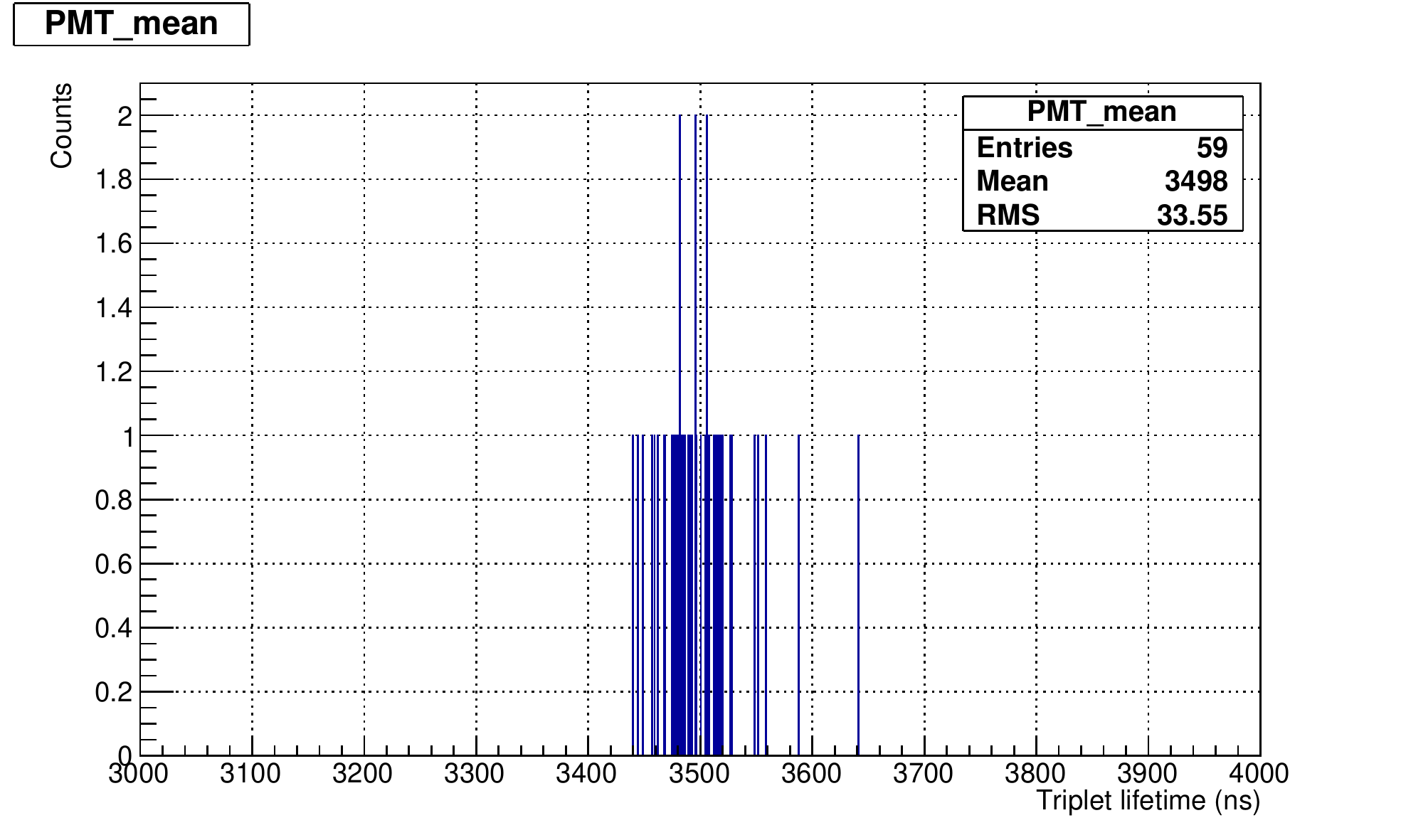}
\caption{ Histogram of triplet lifetime from each PMTs. }
\label{fig:feachPMTs}
\end{figure}

\subsection{Radius}\label{sec:radius}
In the earlier results, the triplet lifetime  increased with increasing radius as shown in Fig. \ref{fig:fradius}. The triplet lifetime seems correlated not only with radius but also with the background fraction as shown in Fig. \ref{fig:frbkg}. In order to understand the radius effect, we decouple the background dependance of the triplet lifetime. We perform a pseudo-experiment to investigate the effect of background on the triplet lifetime. First, using the pulse-cut algorithm (described in \ref{app:pulsecut}) we can identify the background events, which mostly from PMT discharging as shown in Fig. \ref{fig:fdis}. These PMT discharging background will fake long lifetime and pull the centroid reconstructed radius toward the edge of IV. In Fig. \ref{fig:fnumber}, it is clear to see that the peaks close to the edge of IV, therefore the radius cut is added ($(R/R_{TPB})^{3}<0.7$, approximately R $\simeq$ 386 mm. see section \ref{sec:1}) to reduce the effect from the PMT discharging events. After applied the radius cut, the residual background events looks fairly flat as shown in Fig. \ref{fig:fflat}.
We then define a pseudo-experiment to investigate the radius effect. If we assume the true lifetime is some number, then we pick a random number according to the Fig. \ref{fig:fnumber}, we can find the corresponding background fraction in Fig. \ref{fig:fbkgr} which is fitted to 4-th oder of polynomial function. Using the fitting function (eq. (1)) to produce the random number, fix the triplet lifetime and vary the background component. We then use the same function to fit the pseudo data and compare with the fitting results from the data. In each hour of data, the scintillation events comes from different positions inside the IV, the pseudo-experiment takes this effect into account. The difference reveals the radius effect on the triplet lifetime. The results is shown in Fig. \ref{fig:fsysr}. The bias of the histogram is due to the fact that true lifetime is inaccessible. The estimate systematic error from radius effect is $\pm$ 1.31\%.

\begin{figure}[htbp]
\centering
\graphicspath{{./fig/Triplet/}}
\includegraphics[scale=0.6]{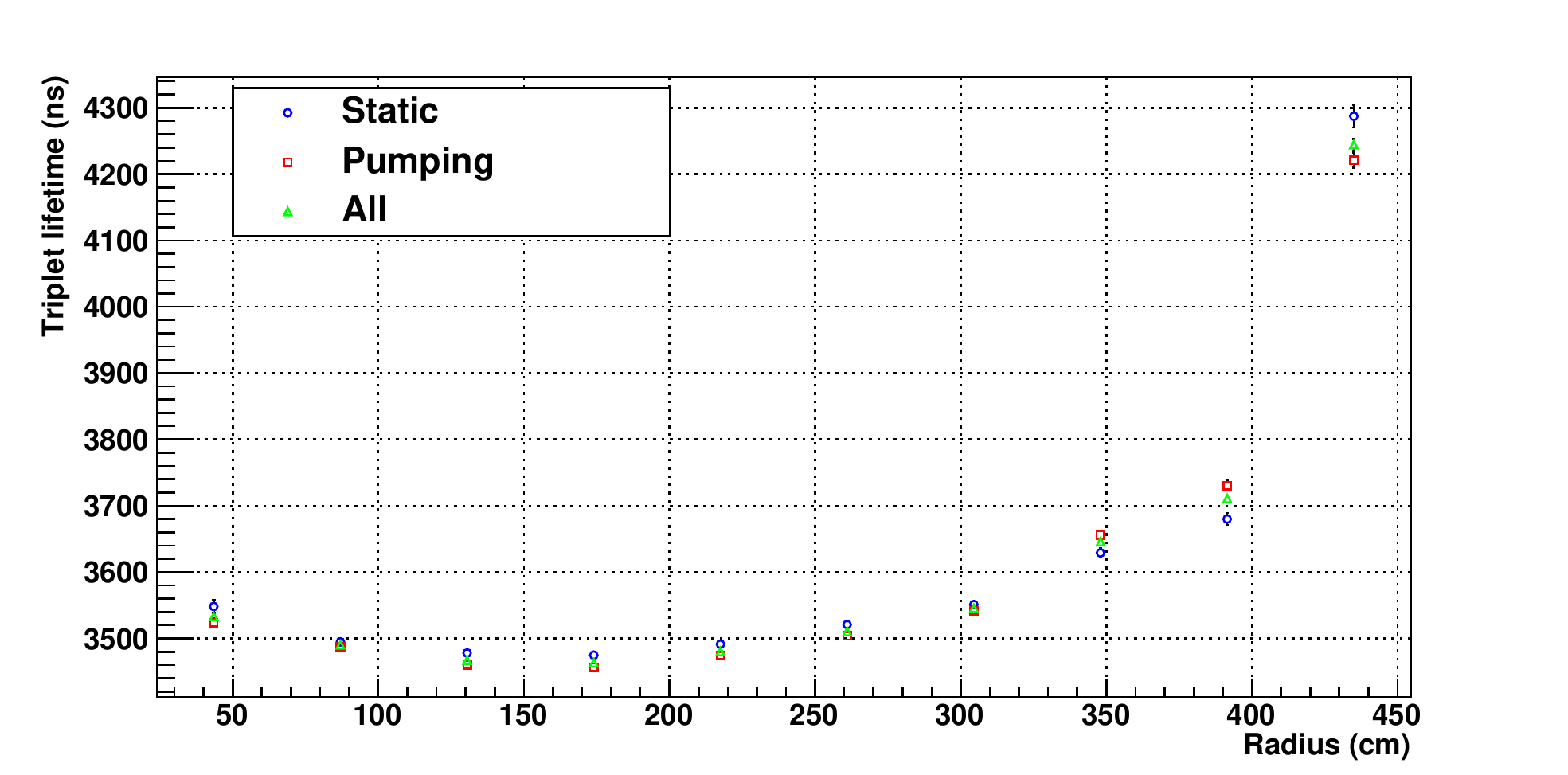}
\caption{ Triplet lifetime vs radius for static and pumping IV states.  }
\label{fig:fradius}
\end{figure}
\begin{figure}[htbp]
\centering
\graphicspath{{./fig/Triplet/}}
\includegraphics[scale=0.6]{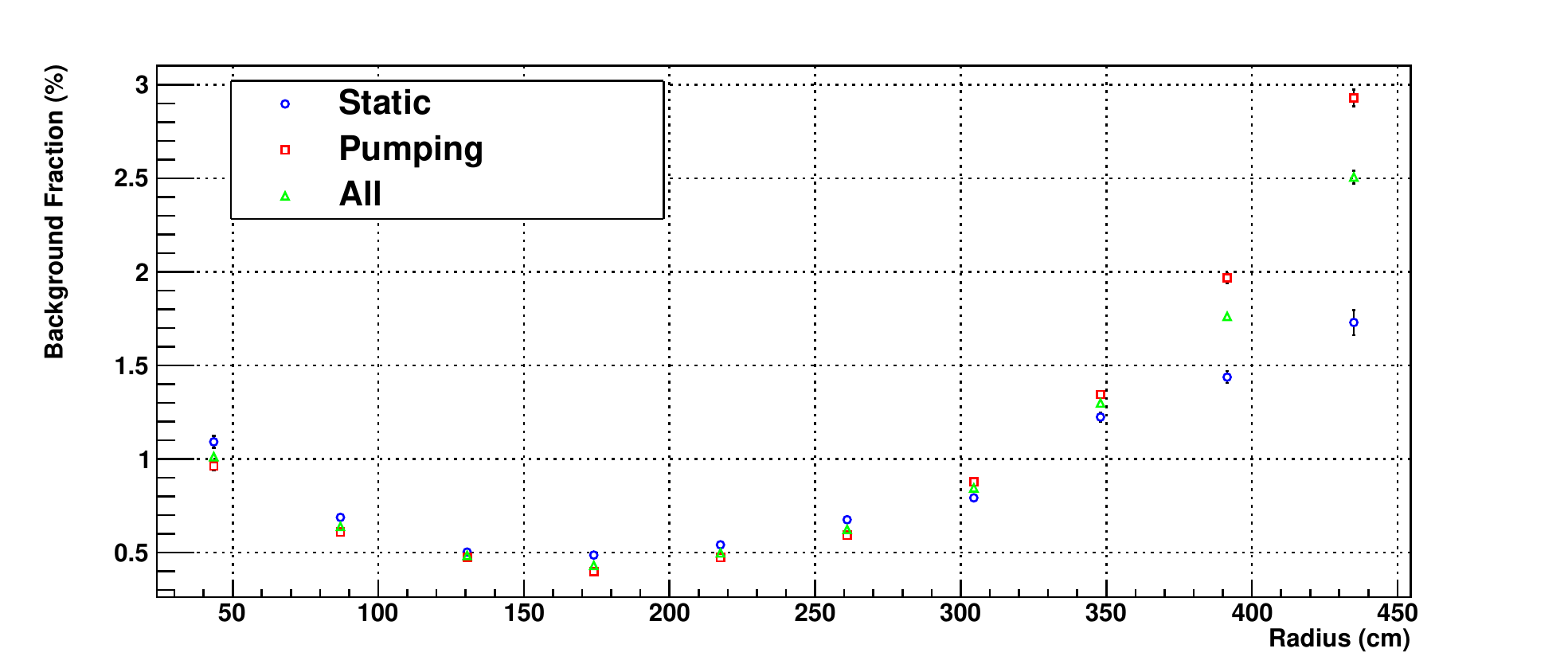}
\caption{ Background vs radius for different stage of IV.  }
\label{fig:frbkg}
\end{figure}

\begin{figure}[htbp]
\centering
\graphicspath{{./fig/Triplet/}}
\includegraphics[scale=0.5]{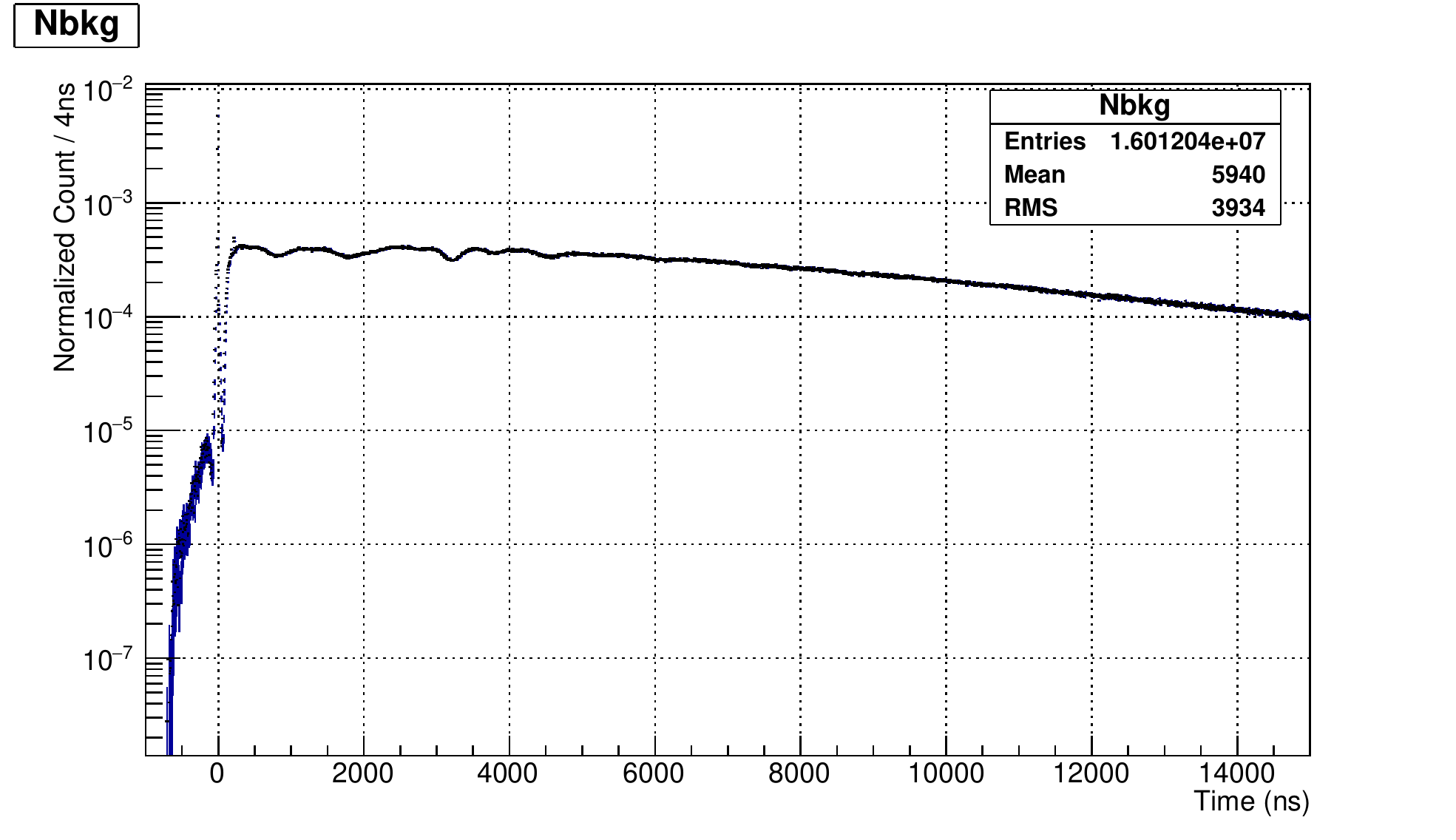}
\caption{ Subtracted background pulse-time distribution using pulse-cut.  }
\label{fig:fdis}
\end{figure}

\begin{figure}[htbp]
\centering
\graphicspath{{./fig/Triplet/}}
\includegraphics[scale=0.4]{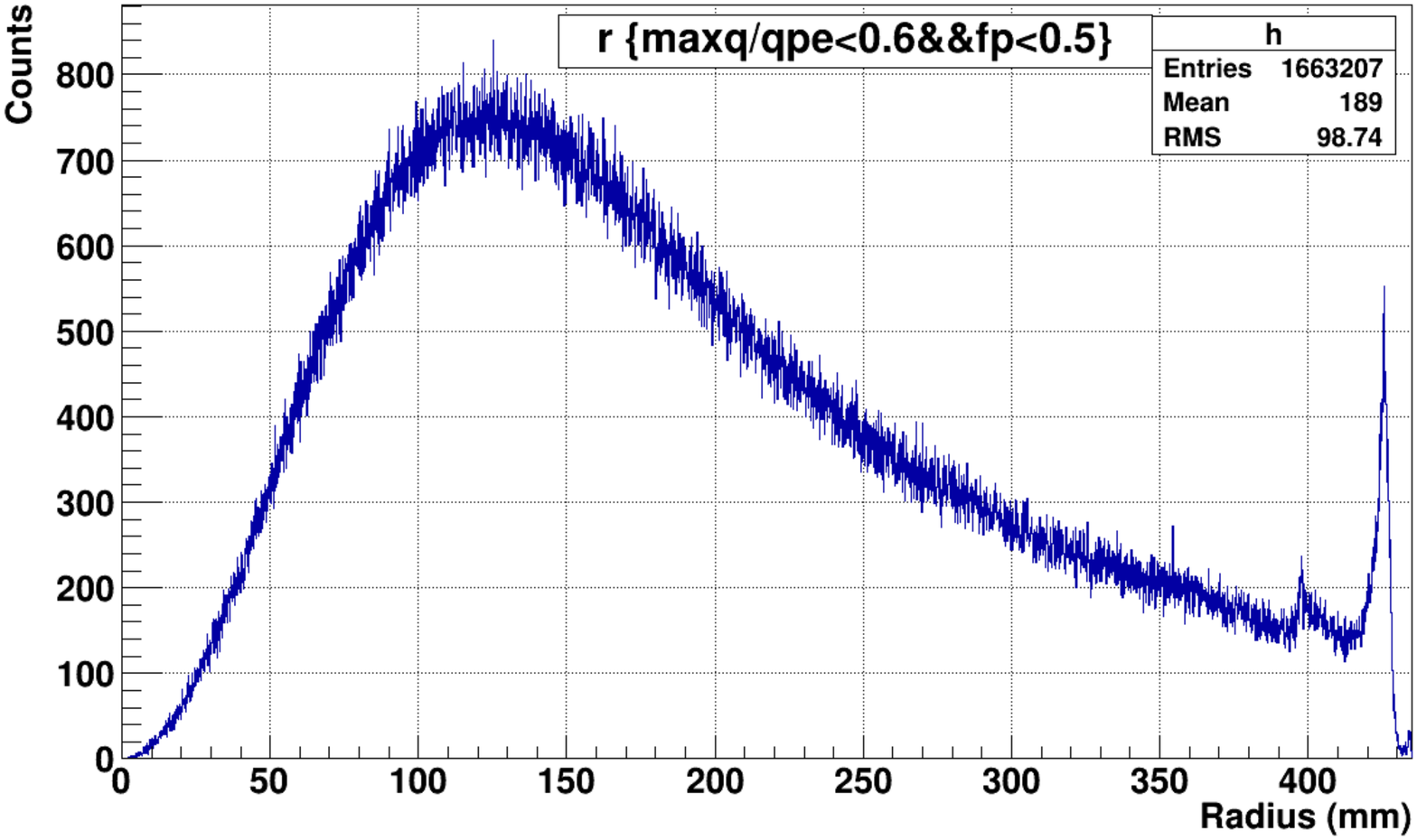}
\caption{ Number of counts vs centroid radius (mm).  }
\label{fig:fnumber}
\end{figure}

\begin{figure}[htbp]
\centering
\graphicspath{{./fig/Triplet/}}
\includegraphics[scale=0.4]{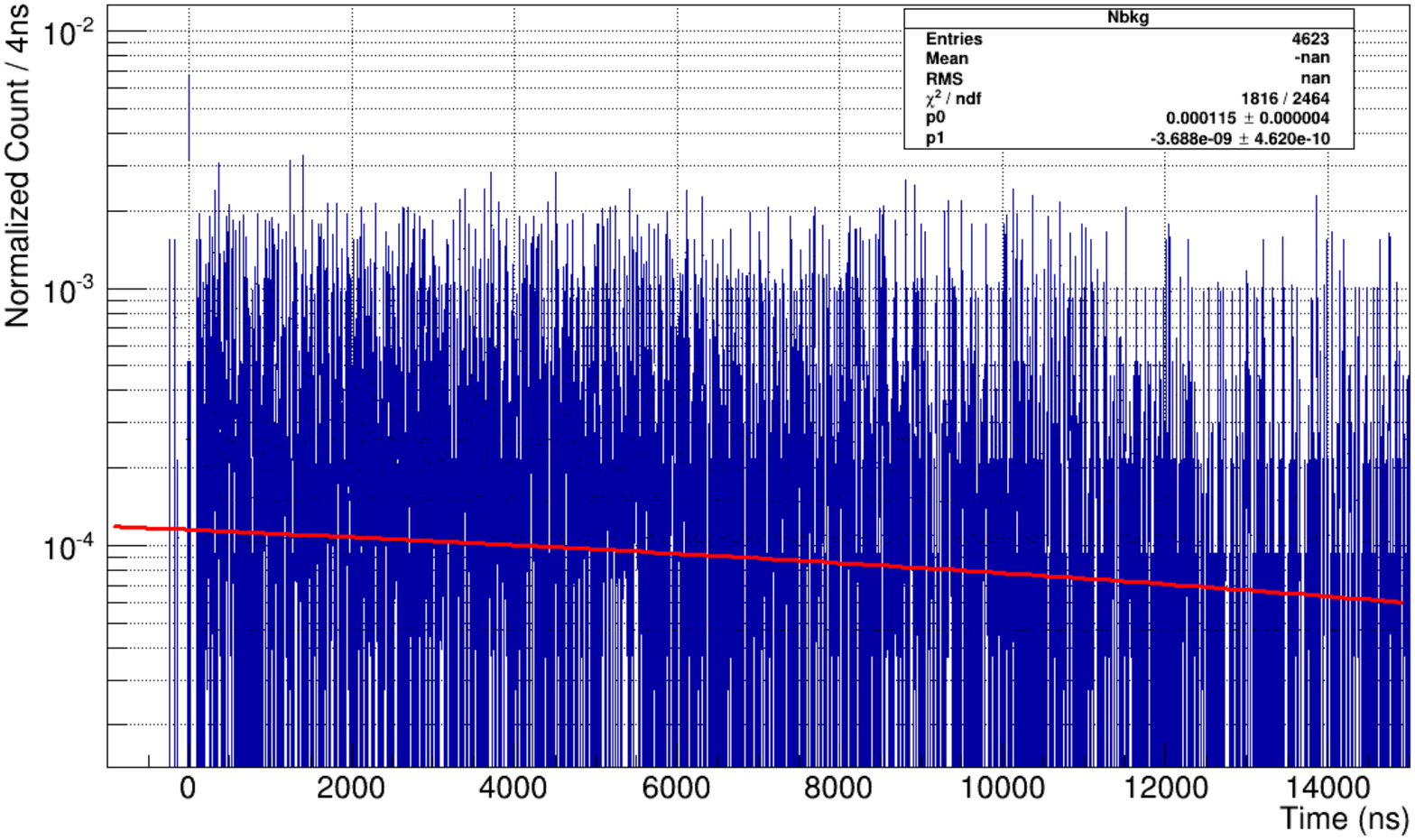}
\caption{ Pulse-time distribution of background events selected after pulse-cut. }
\label{fig:fflat}
\end{figure}

\begin{figure}[htbp]
\centering
\graphicspath{{./fig/Triplet/}}
\includegraphics[scale=0.4]{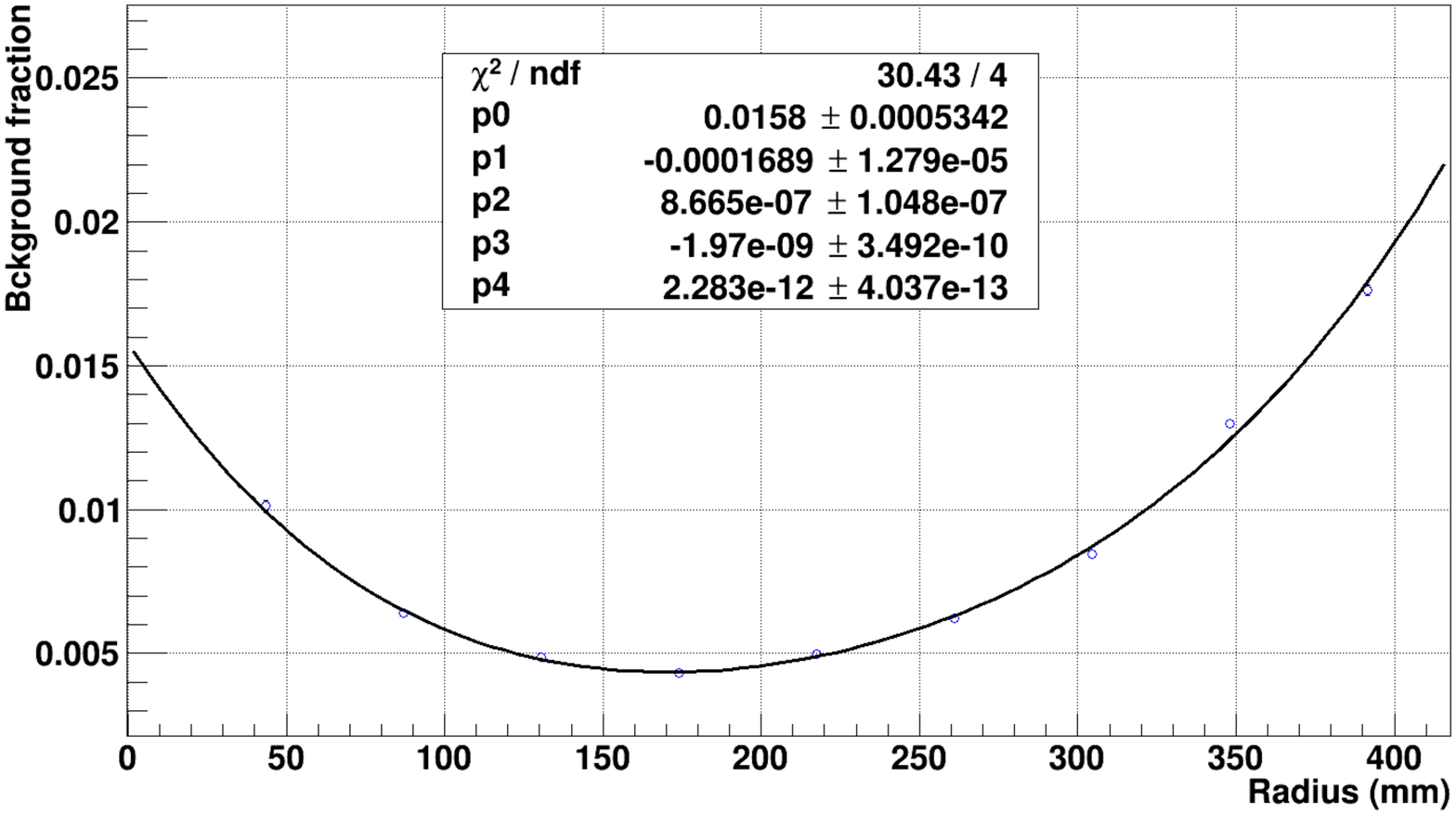}
\caption{ Fitted background fraction vs radius. }
\label{fig:fbkgr}
\end{figure}

\begin{figure}[htbp]
\centering
\graphicspath{{./fig/Triplet/}}
\includegraphics[scale=0.4]{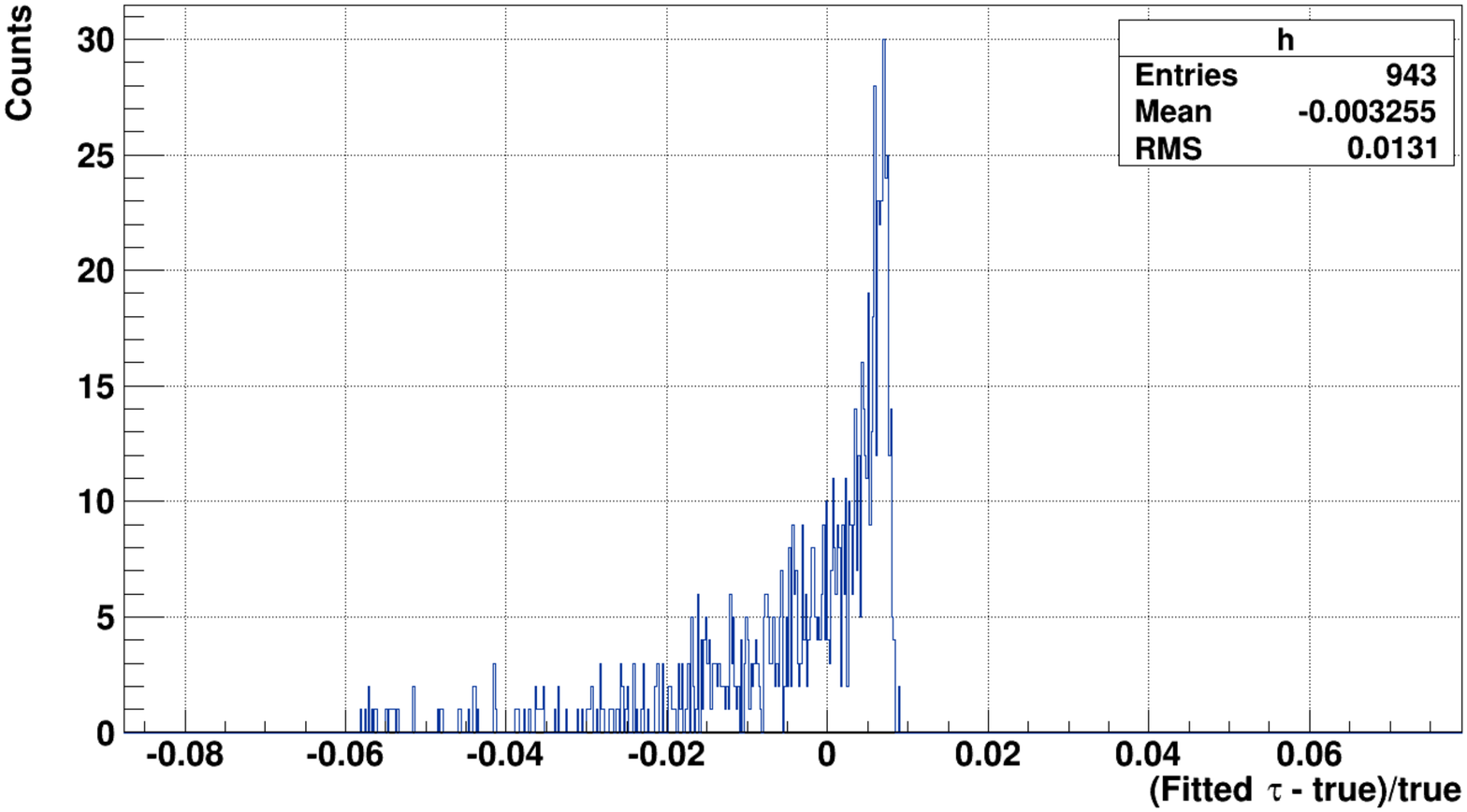}
\caption{ Systematic error of radius effect (see text).  }
\label{fig:fsysr}
\end{figure}

\subsection{Systematic error of impurity level}
The systematic error of impurity level can be studied using results from the two methods of determining the impurity level which described in section \ref{sec:12}. The results in shown in Fig. \ref{fig:fimdiff} , and the estimated systematic error is $\pm$ 7.08\%. The systematic error on determine the initial value is $\pm$ 6.44\%.

\begin{figure}[htbp]
\centering
\graphicspath{{./fig/Triplet/}}
\includegraphics[scale=0.5]{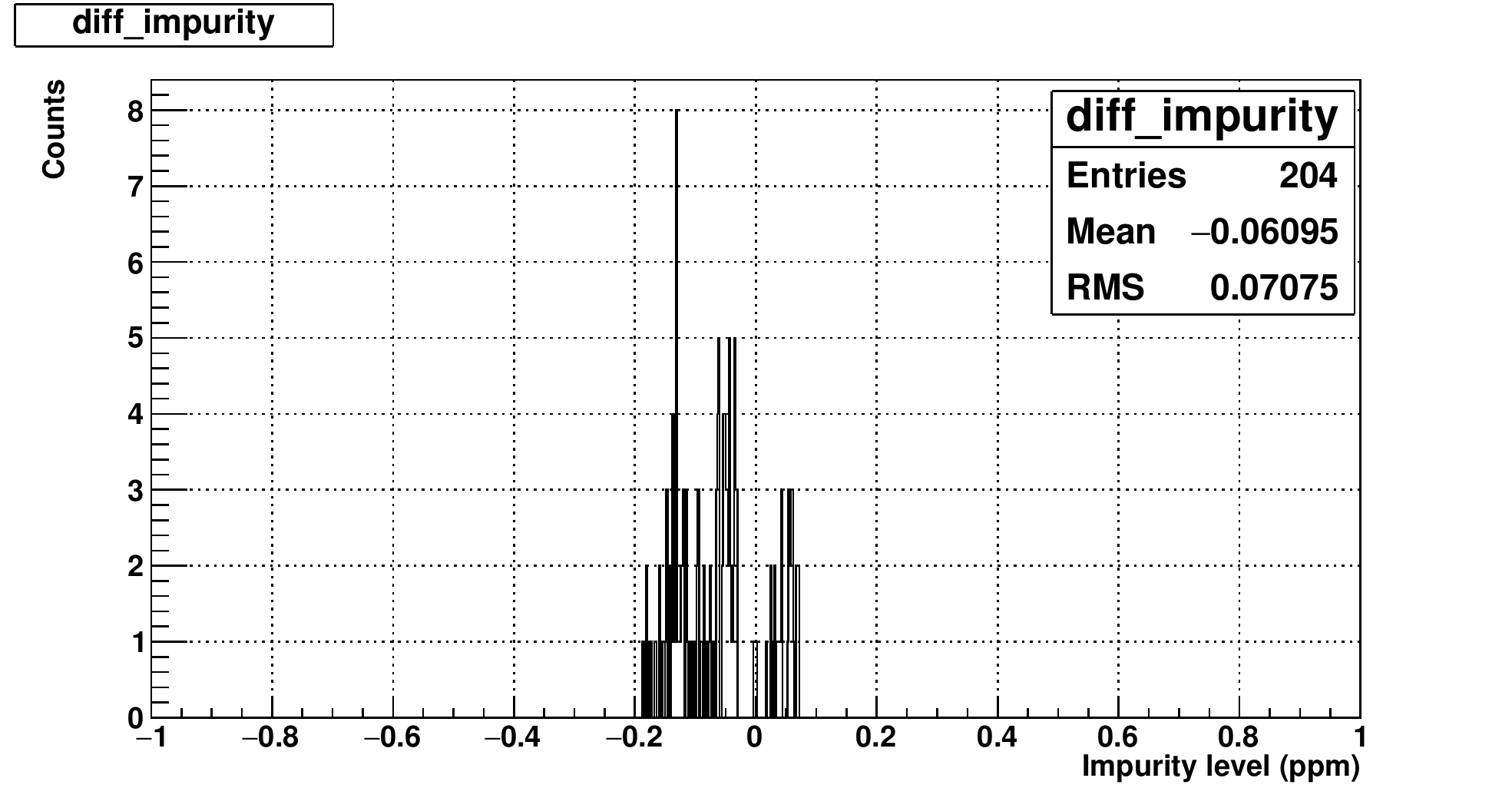}
\caption{ Difference between two methods of get impurity level divided by average impurity level.  }
\label{fig:fimdiff}
\end{figure}
\subsection{Summary on triplet lifetime measurement}
The systematic errors on triplet lifetime measurement is summarized in Table \ref{tab:syslifetime}\\
\\


{\centering
\begin{tabular}{ C{2in} C{2in} *4{C{1in}}}\toprule[1.5pt]
\bf Source & Systematic error \\\midrule
Density & $\pm$ 0.07\%\\\midrule
PMT gain variation & $\pm$ 0.26\%\\\midrule
Pulse finding algorithm & $\pm$ 0.39\%\\\midrule
Pumping on the IV & $\pm$ 0.73\%\\\midrule
Variations between PMT & $\pm$ 0.96\%\\\midrule
Radius & $\pm$ 1.31\%\\\midrule
\bf{Total systematic error}& \bf{$\pm$ 1.84\%}\\\bottomrule[1.25pt]
\end {tabular}
\captionof{table}{The source and associated systematic errors on triplet lifetime measurement. } \label{tab:syslifetime} 
}
\section{Systematic error of LY measurement}
\subsection{Gain variation}
After using SPE constant to calibrate for the energy, we can use the prompt LY to study the gain variation. Figure \ref{fig:fpromptLYPMT} shows the variation of the mean prompt LY vs PMTs. The systematic error from gain variations between PMTs can be determine by populate the mean prompt LY into histogram and fit with the Gaussian distribution as shown in Fig. \ref{fig:fprompthist}. The results on estimated systematic error is $\pm$ 14.58\%.

\begin{figure}[htbp]
\centering
\graphicspath{{./fig/Triplet/}}
\includegraphics[scale=0.5]{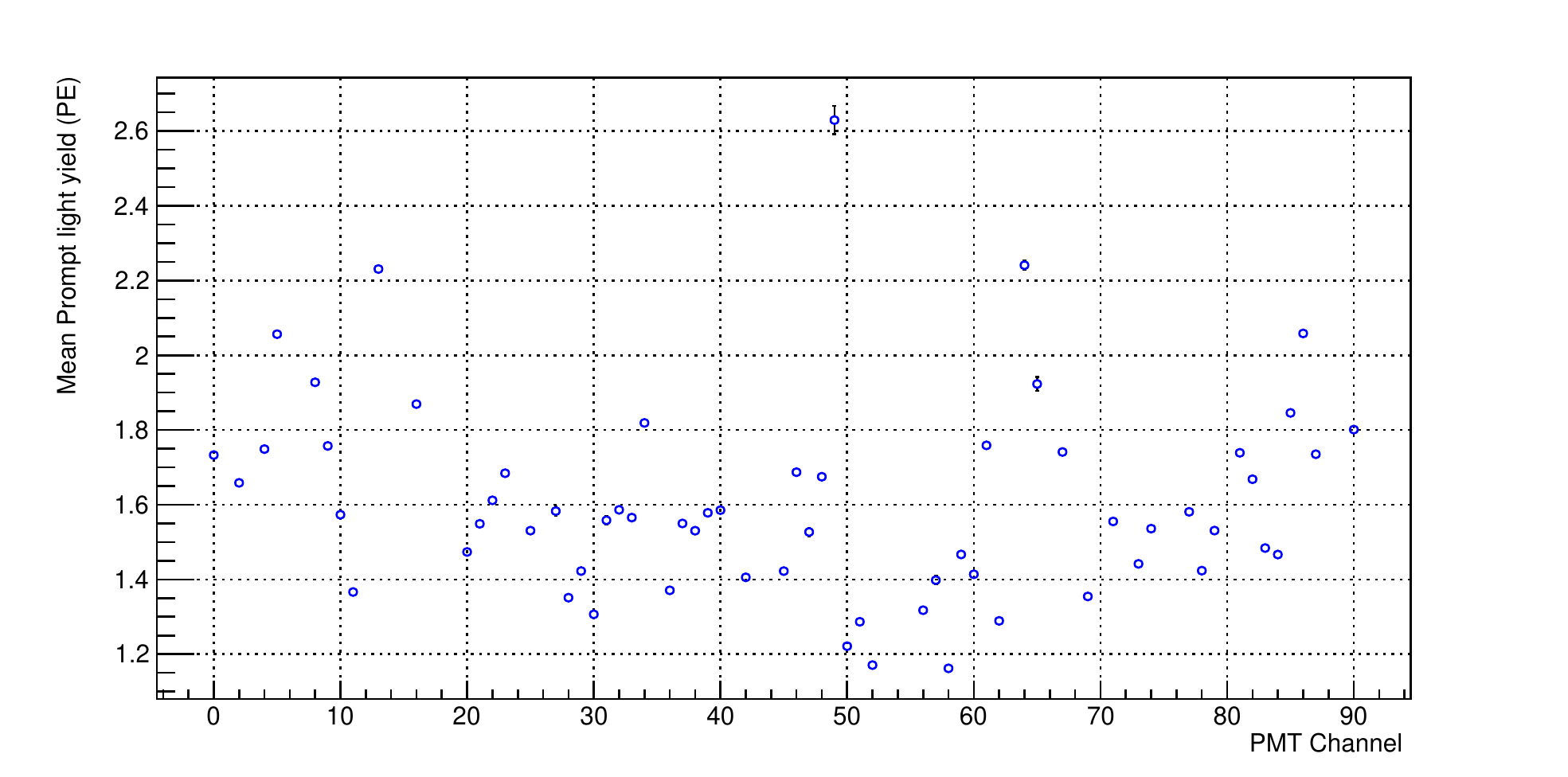}
\caption{ Mean prompt light yield of PMTs. }
\label{fig:fpromptLYPMT}
\end{figure}
\begin{figure}[htbp]
\centering
\graphicspath{{./fig/Triplet/}}
\includegraphics[scale=0.5]{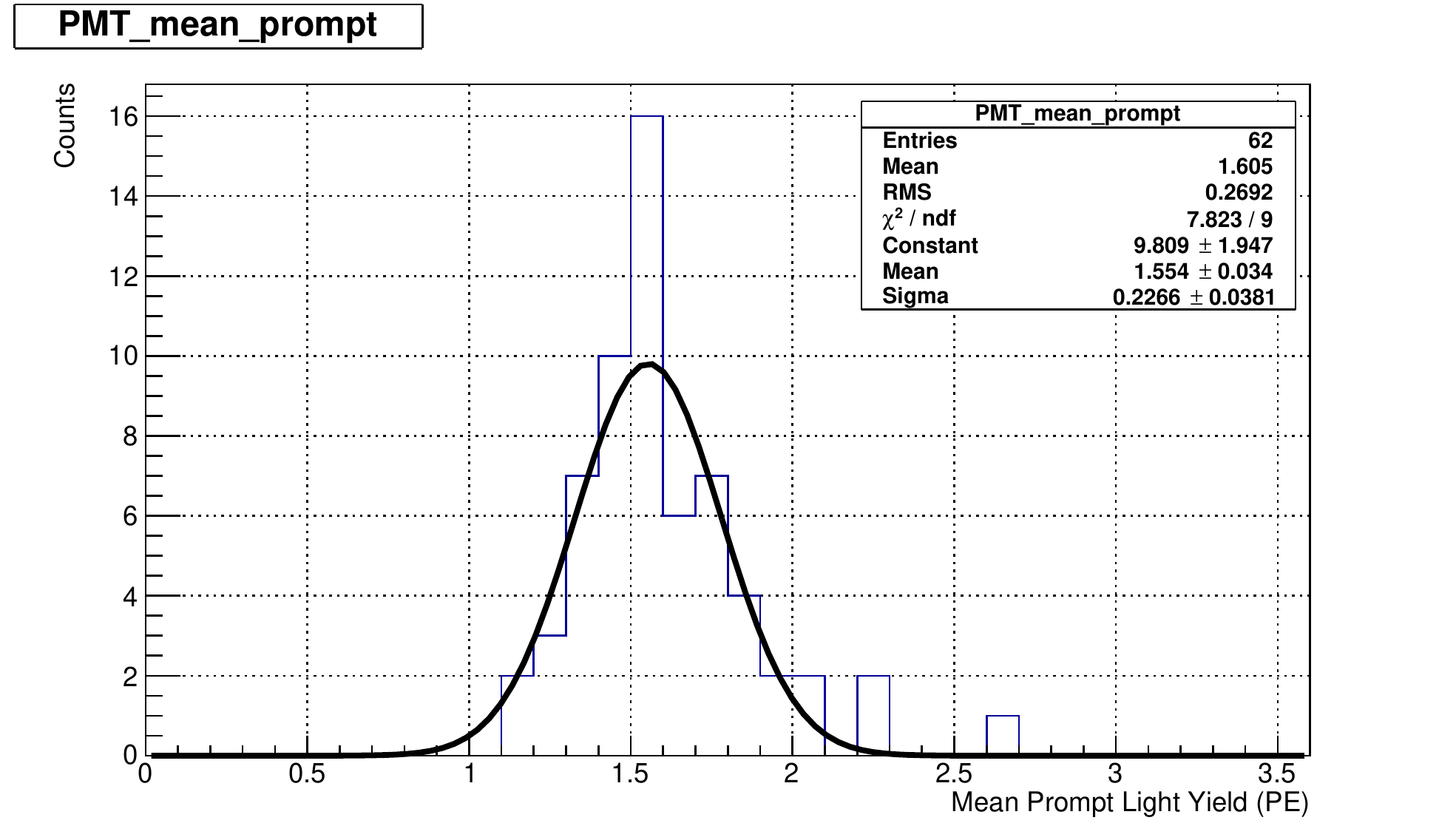}
\caption{ Variation of mean prompt light yield.  }
\label{fig:fprompthist}
\end{figure}
\subsection{PMT variation}
Due to the inhomogeneity of gas distribution in the IV, the collection efficiency would be different for each PMTs. We can use the late/prompt ratio of each PMT to observe the LY variations between PMTs. The mean late/prompt ratio is shown in Fig. \ref{fig:flpratioPMT}. Populating the ratio from each PMTs into histogram and fit with Gaussian distribution, the systematic error can be determined. The results is $\pm$ 12.38\% as shown in Fig. \ref{fig:flpratioPMThist}.
\begin{figure}[htbp]
\centering
\graphicspath{{./fig/Triplet/}}
\includegraphics[scale=0.5]{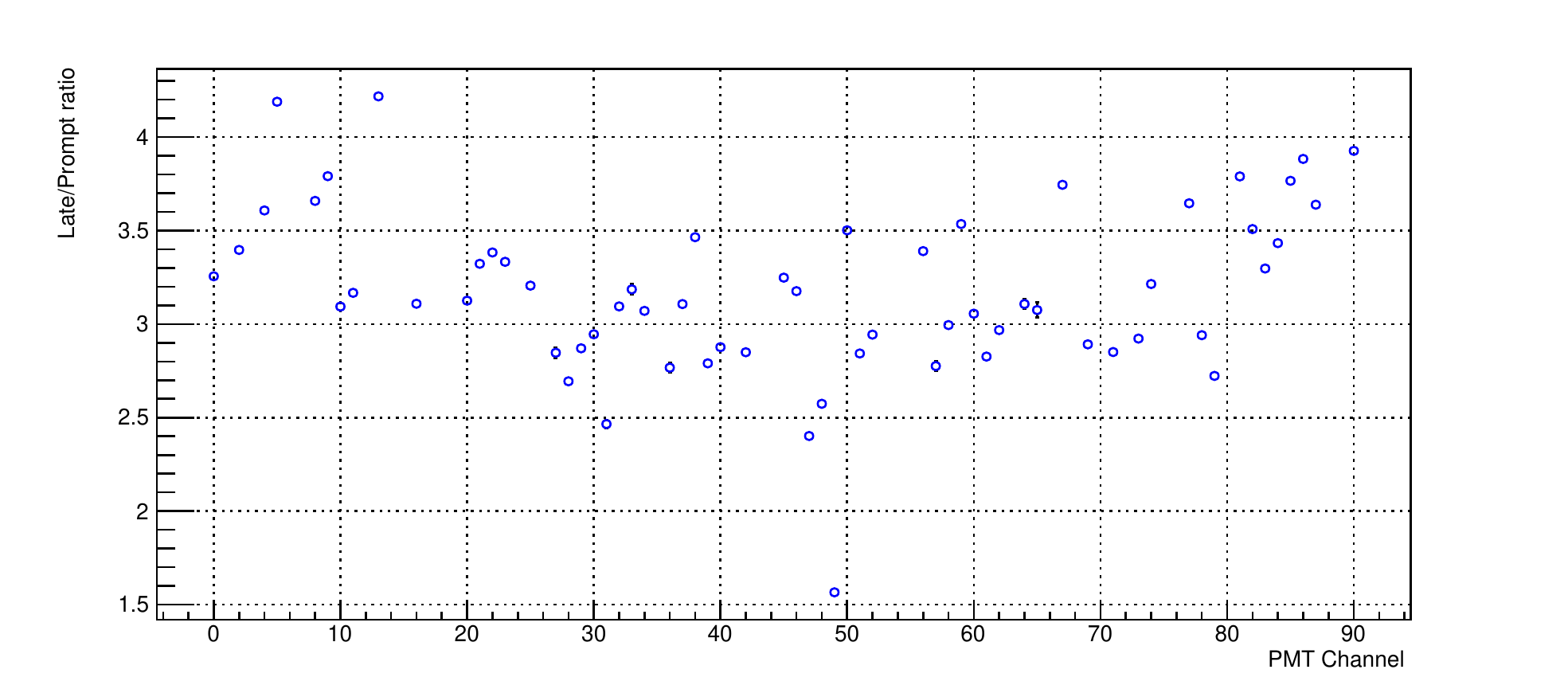}
\caption{ Mean late/prompt ratio of PMTs. }
\label{fig:flpratioPMT}
\end{figure}
\begin{figure}[htbp]
\centering
\graphicspath{{./fig/Triplet/}}
\includegraphics[scale=0.5]{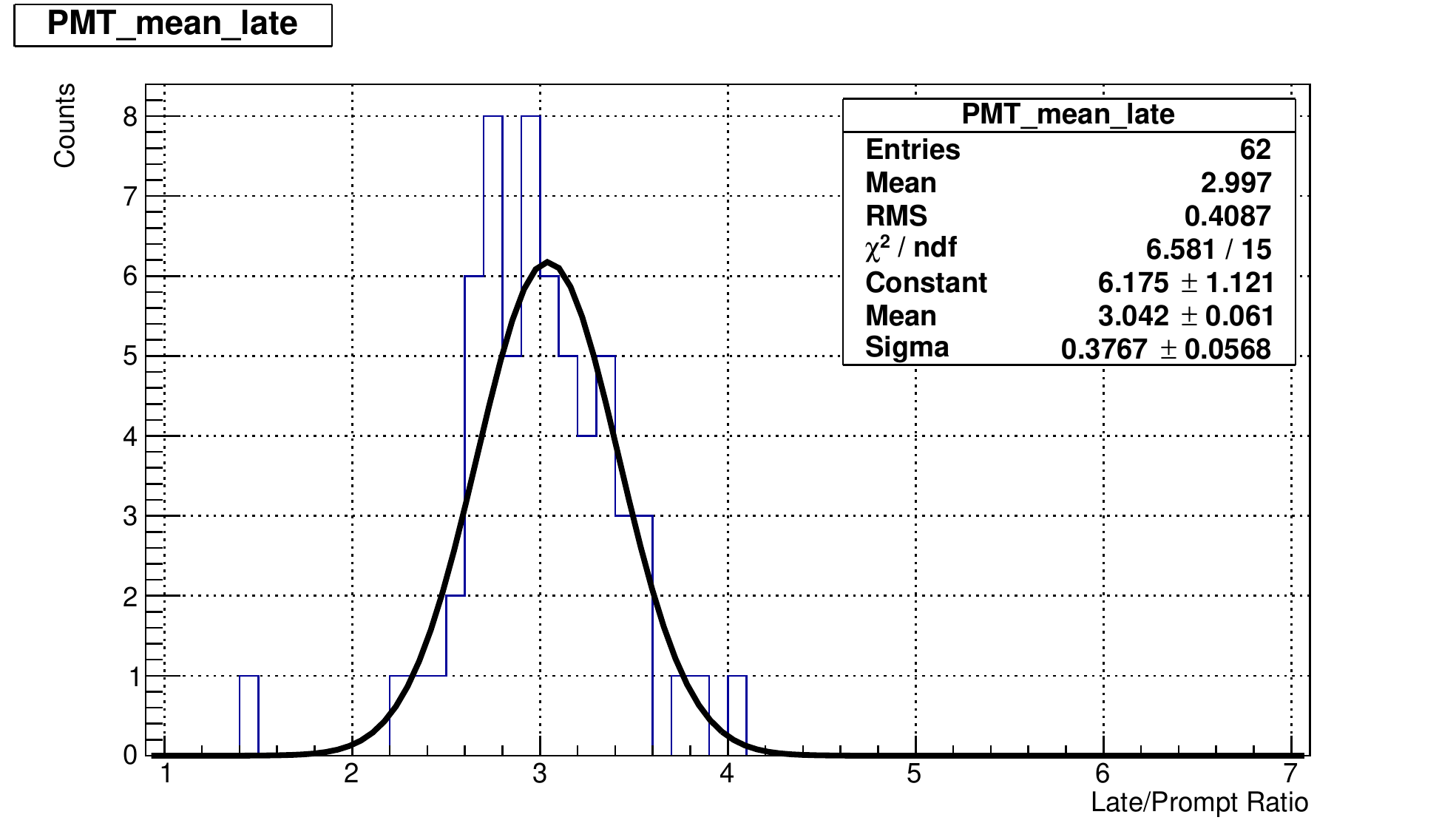}
\caption{ Variation of late/prompt ratio for PMTs.  }
\label{fig:flpratioPMThist}
\end{figure}
\subsection{Background}
Using pulse-cut to identify which PMT channel is discharging for each events, then throw away the charge of identified channel from charge distribution, we can estimate the systematic error from the background. Figure \ref{fig:fbkgLY} shows the charge distribution before and after background subtraction. Estimated systematic error from background is $\pm$ 15.53\%.

\begin{figure}[htbp]
\centering
\graphicspath{{./fig/Triplet/}}
\includegraphics[scale=0.4]{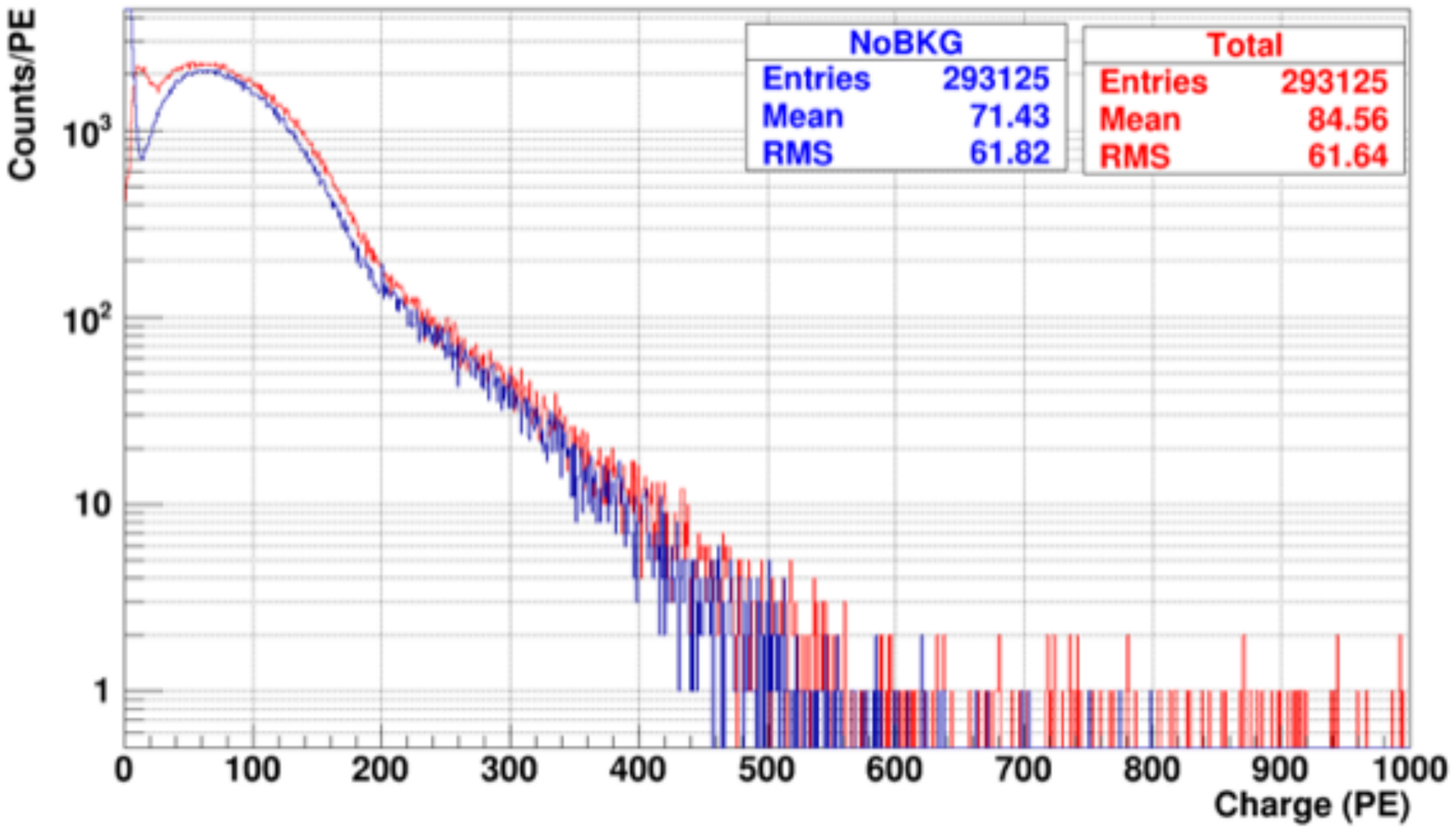}
\caption{ Charge distribution of before(red)/after(blue) background subtraction.  }
\label{fig:fbkgLY}
\end{figure}
\subsection{Summary on systematic error of light yield}
The source of systematic errors are summarized in Table \ref{tab:sysLY}
\\
\\

{\centering
\begin{tabular}{ C{2in} C{2in} *4{C{1in}}}\toprule[1.5pt]
\bf Source & Systematic error \\\midrule
PMT Gain variation & $\pm$ 14.58\%\\\midrule
PMT variation & $\pm$ 12.38\%\\\midrule
Background & $\pm$ 15.53\%\\\midrule
\bf{Total systematic error}& \bf{$\pm$ 24.64\%}\\\bottomrule[1.25pt]
\end {tabular}
\captionof{table}{The source and associated systematic errors on light yield measurement. } \label{tab:sysLY} 
}
\subsection{Systematic errors of Late/Prompt ratio}
\subsubsection{Cut}
The integral window for prompt charge is determine by the mean and sigma from the fitting function (eq. (6)). The default window is mean $\pm$ 3$\sigma$. By varying the integral window, we can access the systematic error from the cut as shown in Fig. \ref{fig:fexample24sigma}.  The estimated systematic error is $\pm$ 11.19\%
\begin{figure}[htbp]
\hfill
\subfloat[]{\includegraphics[width=7cm]{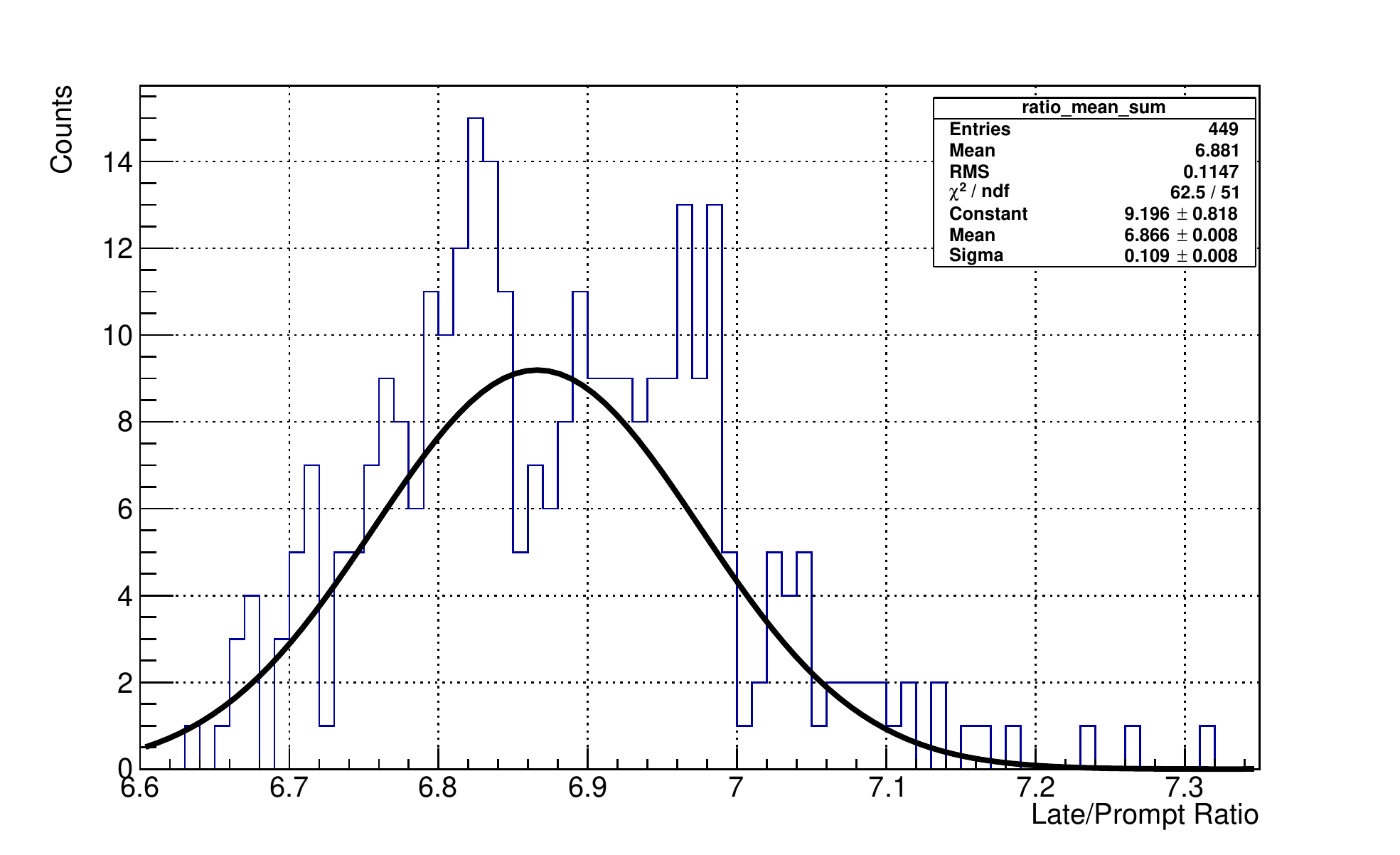}}
\hfill
\subfloat[]{\includegraphics[width=7cm]{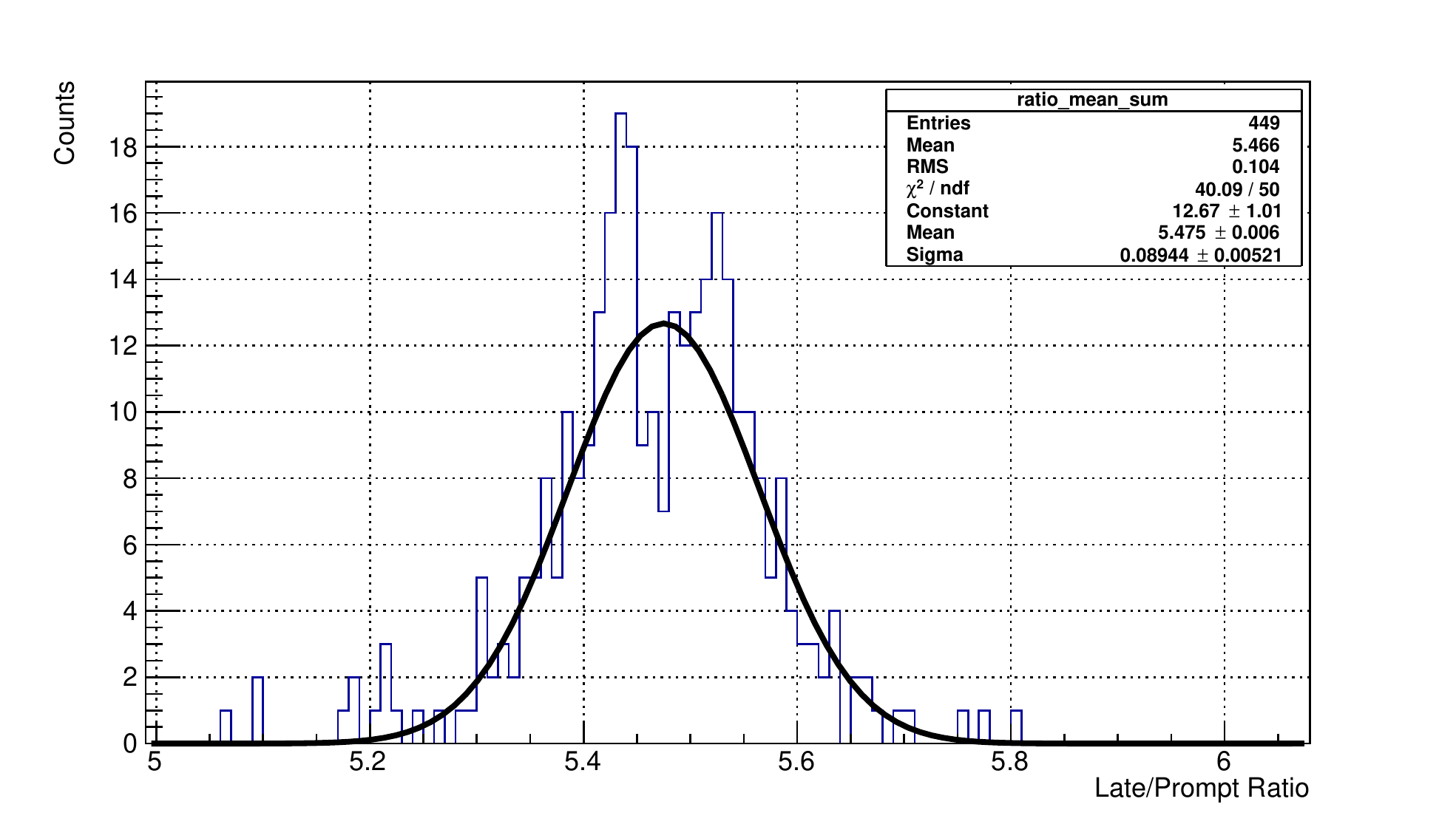}}
\hfill
\caption{Ratio of late and prompt component for (a) taking 2-$\sigma$ region of prompt peak. (b) taking 4-$\sigma$ region of prompt peak. }
\label{fig:fexample24sigma}
\end{figure}

\subsubsection{Background}
By varying the background fraction according to the fitted background error, the systematic error from background can be estimated. Figure \ref{fig:fbkgerror} shows the ratio of late/prompt for increasing/decreasing by background error. The estimated systematic error is $\pm$ 0.43\%
\begin{figure}
\hfill
\subfloat[]{\includegraphics[width=7cm]{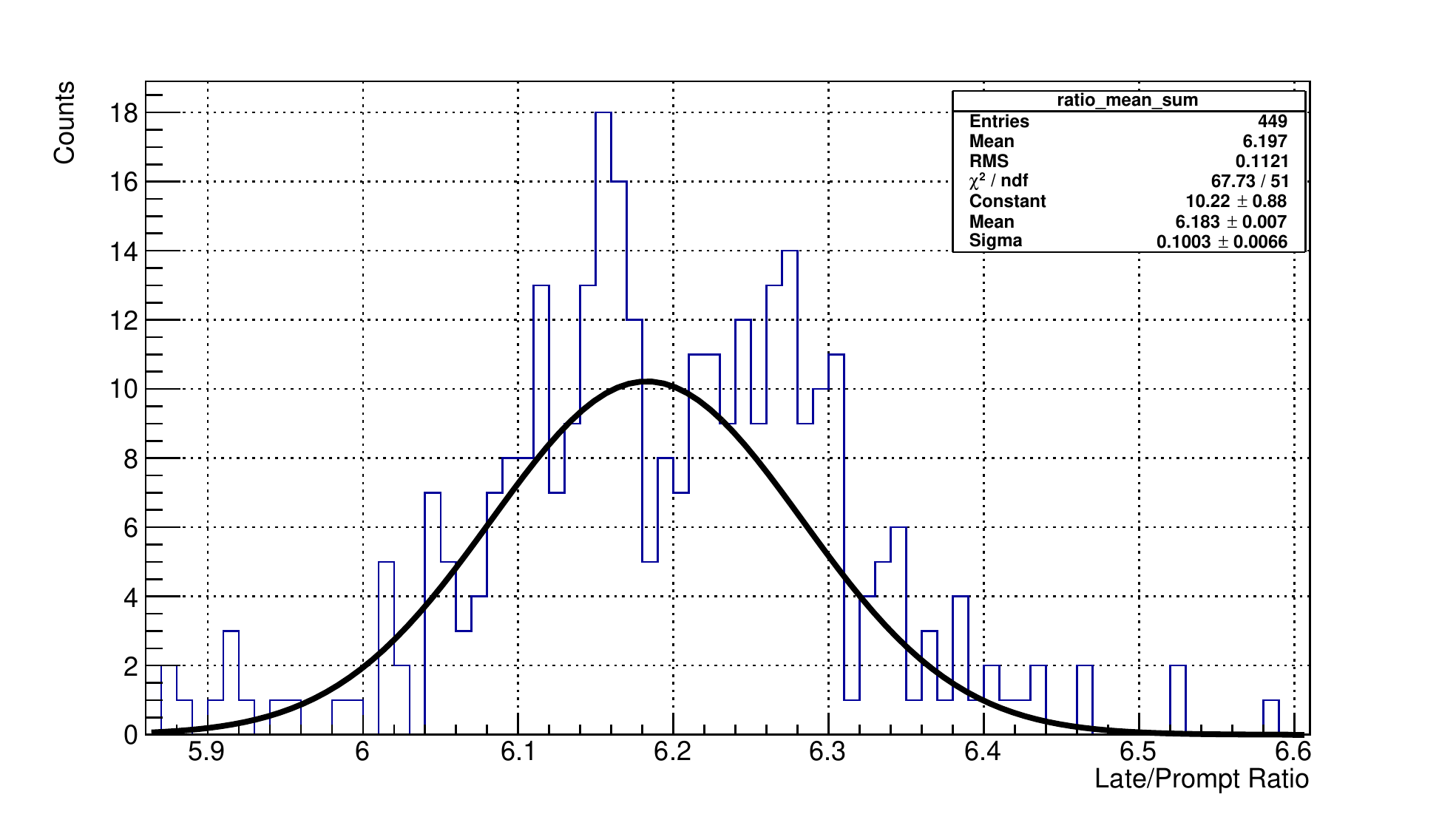}}
\hfill
\subfloat[]{\includegraphics[width=7cm]{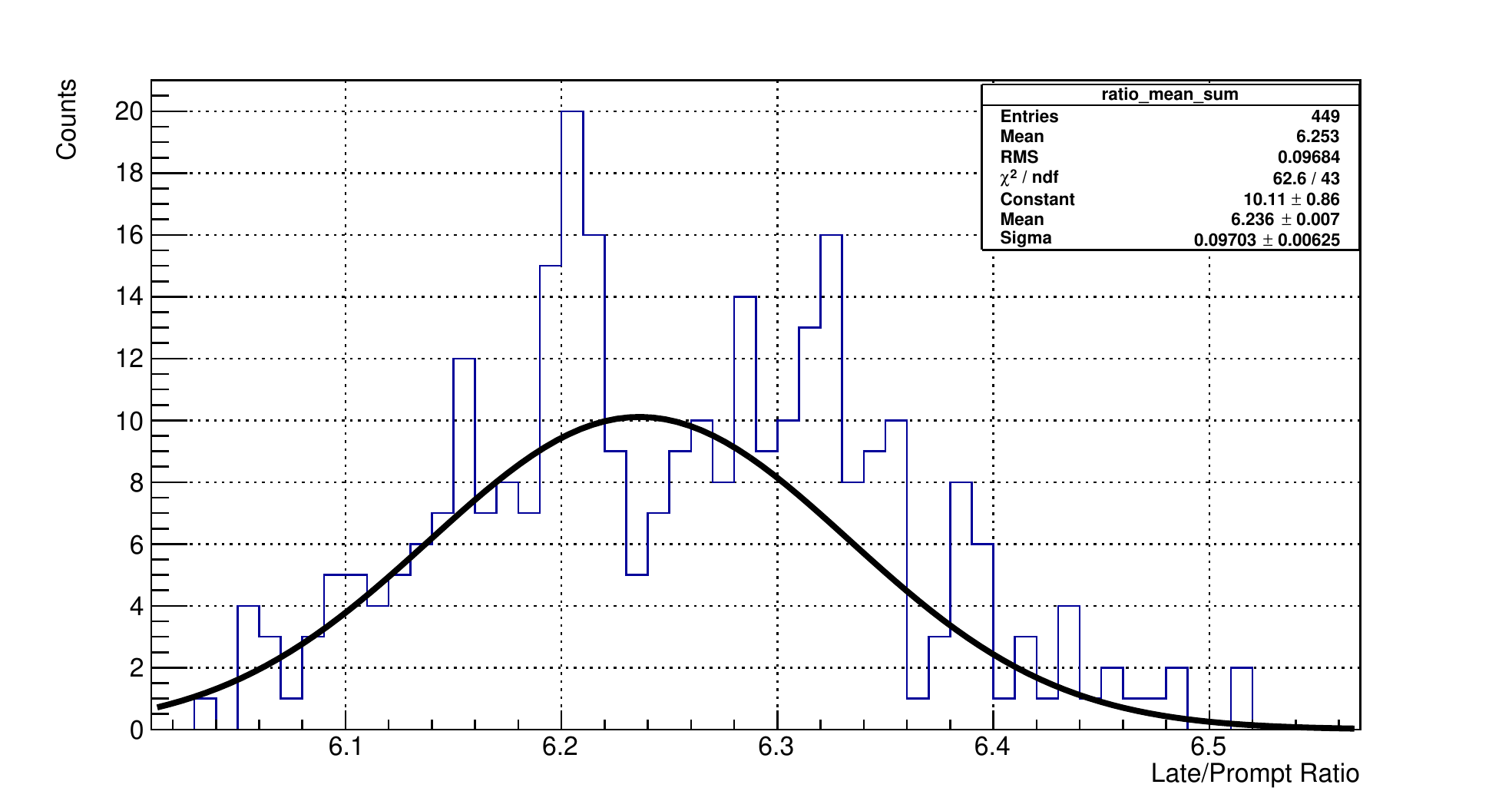}}
\hfill
\caption{Ratio of late and prompt component for (a) Increasing background fraction by 1-$\sigma$ error. (b) Decreasing background fraction by 1-$\sigma$ error. }
\label{fig:fbkgerror}
\end{figure}
\subsection{Summary on systematic error of late/prompt ratio}
The source of systematic errors on late/prompt ratio are summarized in Table \ref{tab:sysLYratio}
\\
\\

{\centering
\begin{tabular}{ C{2in} C{2in} *4{C{1in}}}\toprule[1.5pt]
\bf Source & Systematic error \\\midrule
Cut & $\pm$ 11.19\%\\\midrule
Background & $\pm$ 0.43\%\\\midrule
\bf{Total systematic error}& \bf{$\pm$ 11.20\%}\\\bottomrule[1.25pt]
\end {tabular}
\captionof{table}{The source and associated systematic errors on late/prompt measurement. } \label{tab:sysLYratio} 
}
\section{Summary}
Triplet lifetime is important for doing pulse shape discrimination in gaseous/liquid argon detector. The residual impurities in argon will lead to the degradation of both triplet lifetime and light yield. Subsequently affects the PSD and the energy resolution of the detector. We investigate the detail correlation between triplet lifetime and impurity level which helps to monitor the detector health before and after the commission of the detector. Moreover, with better PMT coverage, we found the triplet lifetime with zero impurity is 3.470 $\pm$ 0.001 $\mu s$ which is the longest lifetime has been measured at 1.5 bar and temperature less than 140 K, to our knowledge. \par
Furthermore, we present the results of relative light yield and the late/prompt ratio. The relative light yield reveals the origin of the quenching effects of impurity molecule. They mainly interact with the triplet state of second continuum and results in loss of light yield. The detail analysis on light yield from each component suggest that the impurity molecule did not affect the prompt (third continuum) and singlet state (second continuum) except at very high impurity level (>100 ppm). This is probably due to the relative fast lifetime of these two components and low reaction rate when impurity level is low. The late/prompt ratio gives a quantitative definition of the scintillation timing structure. The best value of the late/prompt ratio has been measured as 6.215 $\pm$ 0.007 which has better precision than previously measured\cite{1748-0221-3-02-P02001} due to the high purity of our argon as evidenced by our long triplet lifetime. The measured long triplet lifetime in gaseous argon can in principle improve the PSD rejection ability with no extra cost. Also, the low density of gaseous argon reduced the chance of multi-scattering of neutron which is the main background in the region of WIPMs searching. Despite the low event rate in the gaseous argon, the above properties provides better background rejection and could be a new approach of designing the detector for  future dark matter experiment.

\chapter{Energy scale calibration}\label{ch:ar39}
$^{39}$Ar beta decay in the LAr is the major background for LAr detector. However, it is uniformly distribute throughout the volume and with definite energy spectrum. It can be used to calibrate the detector and monitoring the detector health on daily basis. 
\section{$^{39}$Ar beta decay}
$^{39}$Ar is a radioactive nulclide exists in the atmosphere with radioactivity of 1Bq/kg. Argon is produced by the fractional distillation of liquid air which contains the $^{39}$Ar as well. The radioactive $^{39}$Ar decay to $^{39}$K through first forbidden beta decay
\begin{ceqn}\begin{align}
^{39}Ar \longrightarrow ^{39}K + e^- + \bar{\nu_e}
\end{align}\end{ceqn} 
The parent nucleus and the recoil daughter nucleus are assumed at static before and after the reaction. Which means that the electron and the neutrino carries most of the energy.\footnote{Even in the worst possible case(free neutron decay), the largest possible energy carried by proton is about 0.4 keV whcih is just 0.05\% of the reaction Q-value.}. Unlike the alpha decay which the end product ($alpha$) has a distinct energy, the beta decay has a continuous energy distribution due to the momentum conservation. The energy of beta depends on the energy that carried away by the ``invisible'' neutrino.\par
The total spin ($S_{\beta}$) of beta and neutrino can be either 1 or 0 (Fig. \ref{fig:fermi_decay}), both of these can combine with S=1/2 of the neutron for a resultant vector of 1/2. The two different type of beta decay can be defined as Fermi beta decay ($S_{\beta}$ = 0) and Gamow-Teller(GT) beta decay ($S_{\beta}$ = 1) which are named after the people who first described the mode. The Q-value of the reaction can be calculated and the $\beta$-endpoint which is the maximum energy carried by the electron (i.e. the energy of neutrino approaches zero) can be determined. In the beta decay, the decay rate depends on the overlap of the wave functions of the ground state of the parent and the state of the daughter. The decay rate can be calculated using the Fermi's Golden Rule : 
\begin{ceqn}\begin{align}
\lambda = \frac{2\pi}{\hbar} \mid\mel{\Psi_{final}}{V}{\Psi_{initial}}\mid^{2}\rho ,
\end{align}\end{ceqn}
where V is a potential that causes the transition from the initial quantum state $\Psi_{initial}$(parent) to a final quantum state $\Psi_{final}$, that include the wavefunctions of the daughter nucleus, the electron and the neutrino and $\rho$ is the density of the final state. The interaction found by Fermi between the electron and neutrino is called \textit{weak interaction} with a constant g (0.88$\times$ 10$^{-4}$ MeV/fm$^3$) to represent its strength. This is approximately 10$^{-3}$ of the electromagnetic force constant. To calculate the transition probability from Fermi's golden rule for $^{39}$Ar is difficult, several assumptions are made to simplify the calculation. First, the mass of neutrino is assumed to be zero  and the nucleus are treated as a point charge. In the large Z atom, the electron screen correction need to be taken into account. When the negative beta particle escape from the parent nucleus, it will be attracted and slowed down. This effect results in more low energy electrons are produced. In addition, the size of nucleus is finite which affects the statistical shape of the final spectrum. These effects are incorporate by Fermi by using coulomb-distorted wave functions and are contained in a spectrum distortion expression called the Fermi function. Moreover, for decays that have a difference in spin between the parent and daughter nucleus of larger than one angular momentum, the lepton must carry away the access angular momentum. Assuming a decay occurs at the nuclear surface at r = 5 fm, with angular momentum L of 1$\hbar$, the energy can be written
\begin{ceqn}\begin{align}
L = rp \Longrightarrow p=\frac{L}{r} \Longrightarrow E = pc = \frac{Lc}{r},
\end{align}\end{ceqn}
where c is the speed of light. For L= 1$\hbar$ and r = 5 fm
\begin{ceqn}\begin{align}
E = \frac{\hbar c}{r} = \frac{197 MeV\cdot fm}{5 fm} \simeq 20 MeV,
\end{align}\end{ceqn}
This shows that the leptons are emitted preferentially with no orbital angular momentum. The lepton can still emitted with the angular momentum larger than 1. However, the energy available to the leptons is constraint by the Q-value of the reaction. It is possible that the leptons are emitted at large radii, however, the probability of large radius emission is suppressed. As such, the largest beta decay rate happened with L=0, the higher angular momentum emission is possible at the cost of small probability or long lifetime. The conservation of angular momentum impose the following condition on the parity of the initial and the final state
\begin{ceqn}\begin{align}
\Delta\pi = (-1)^L,
\end{align}\end{ceqn}
Table \ref{tab:spin} summarized the different group of beta decays.\par
Using this approximation, the general form for the beta spectra can be expressed as\cite{GK} 
\begin{ceqn}\begin{align}
d\lambda = (W_0 - W)^2p_e W F(\pm Z,W)S(\pm Z,W)dW.
\end{align}\end{ceqn}
where W is the total energy of the electron, $W_0$ represent the end-point energy of the electron, and $p_e$ is the electron momentum. The two terms $F(\pm Z,W)$ and $S(\pm Z,W)$ are the Fermi correction and the correction of forbidden decay, respectively. The $^{39}$Ar beta decay spectrum is calculated and used by RAT to create simulation events as shown in Fig. \ref{fig:ar39_beta_spec}

\begin{figure}
\hfill
\subfloat[]{\includegraphics[width=7cm]{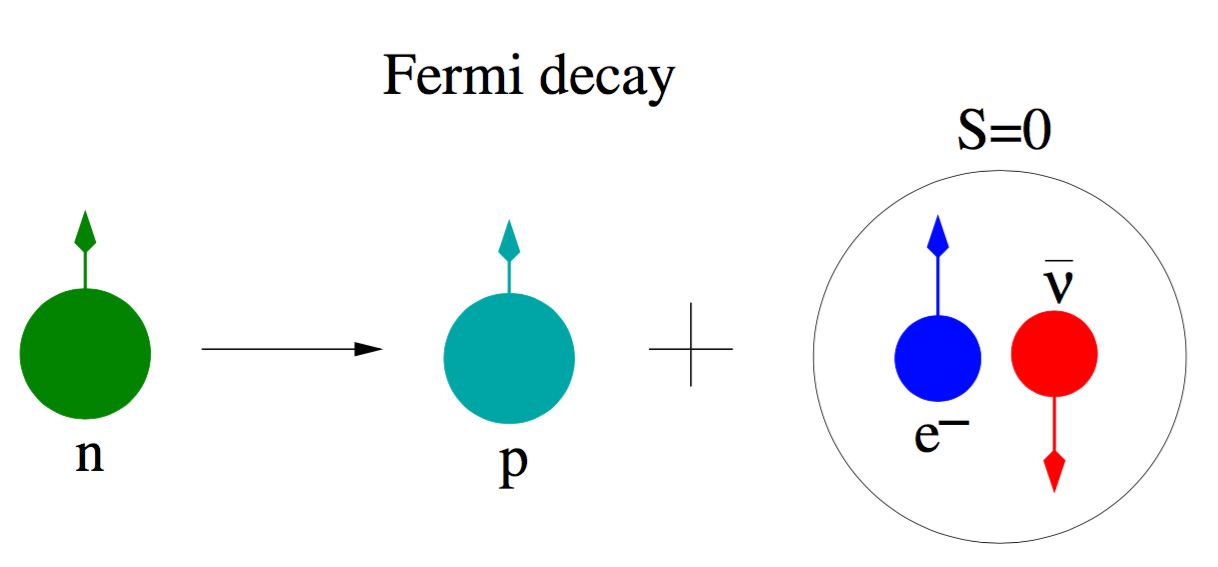}}
\hfill
\subfloat[]{\includegraphics[width=7cm]{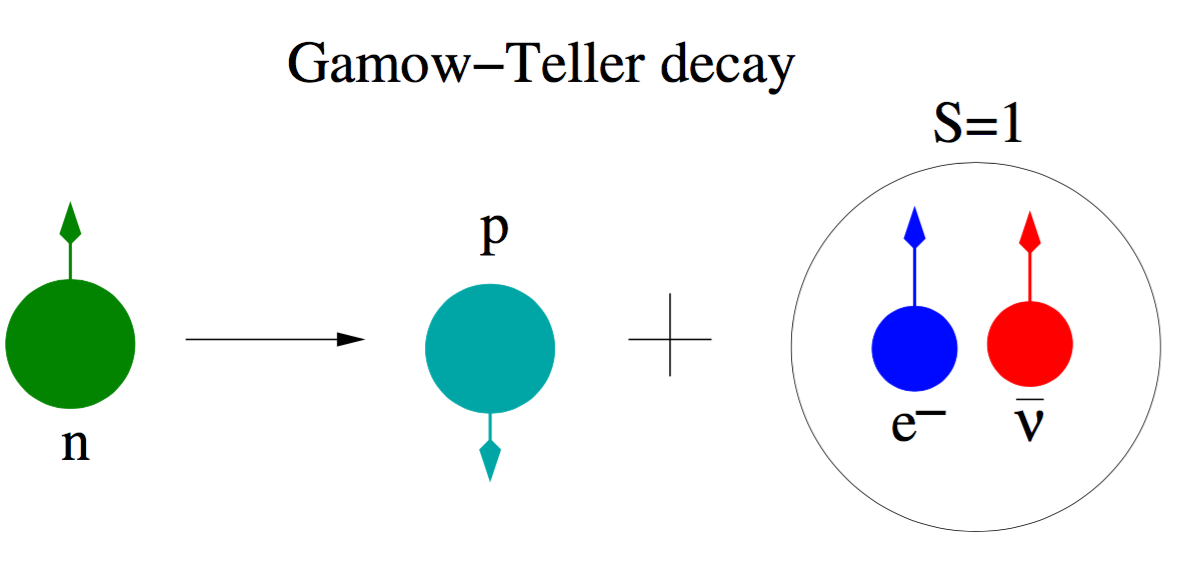}}
\hfill
\caption{(a) Fermi beta decay mode. (b) Gamow-Teller(GT) beta decay mode.}
\label{fig:fermi_decay}
\end{figure}

\begin{table}[htbp]
\centering 
\begin{tabular}{ c c c c c} 
\toprule[1.5pt]
\bf Type & L & $\Delta\pi$ & $\vec{S}=\vec{0}$ Fermi &  $\vec{S}=\vec{1}$ Gam-Tel  \\\midrule
Super-allowed & 0 & + & 0 & 0 \\
Allowed & 0 & + & 0 & 0,1\\
First forbidden & 1 & - & 0,1 & 0,1,2\\
Second forbidden & 2 & + & 1,2 & 1,2,3\\
Third forbidden & 3& - & 2,3 & 2,3,4\\\bottomrule[1.25pt]
\end {tabular}
\captionof{table}{Allowed transition of beta decay. } \label{tab:spin} 
\end{table}
\begin{figure}[htbp]
\centering
\graphicspath{{./fig/Energy/}}
\includegraphics[scale=0.3]{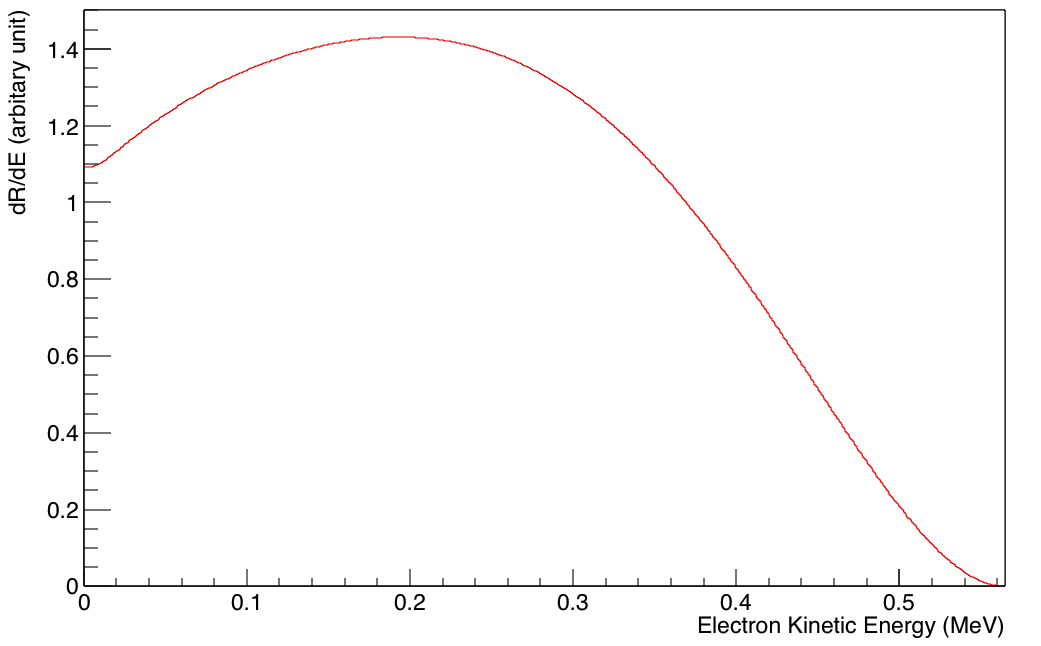}
\caption{ $^{39}$Ar energy spectrum as used in RAT. }
\label{fig:ar39_beta_spec}
\end{figure}
\section{Simulation}
To study the fitting of $^{39}$Ar energy spectrum and extract the energy scale and resolution, 2 million $^{39}$Ar simulation events are produced using RAT.  The trigger efficiency determined by simulation is 98\%, the untriggered event mostly are produced around the edge and escape from the active volume before interaction. Assuming the $f(E) = ((1/N)dN/dE)$, the predicted $\beta$-decay energy spectrum, where N is the total number of electrons and E is the energy of the electrons. The resolution function $R(E,E_i,\sigma_i)$ represents the probability of observing an event with energy in bin $E_i$ when the true energy is E. Assuming the R is Gaussian, with different $\sigma_i$ in each energy bin and an overall energy scale bias $b$. The expected number of events ($\mu_i$) in each bin is :
\begin{ceqn}\begin{align}
\mu_i = N_T \int_{0}^{\infty} f(E)R(E,\bar{E_i},\sigma_i)dE,
\end{align}\end{ceqn}  
where $N_T$ is the total number of observed events (triggered). Since $f$ is finely binned data, thus the integral can be replaced with a sum over all bins
\begin{ceqn}\begin{align}\label{eq:arfit}
\mu_i \approx N_T\delta E\sum\limits_{j} f(E_j)R(E_j,\bar{E_i},\sigma_i),
\end{align}\end{ceqn}
Binning the simulated data, the Eq. \ref{eq:arfit} can be used to fit for the energy spectrum and the associated resolution. Figure \ref{fig:edet_edep} shows the energy dependent bias and resolution. The true deposited energy is taken from MC and the detected energy is from the sum of total PMT reconstructed charge multiplied by the PMT gain (nominally 5 pC/mV) and divided by a nominal photon yield ($Y_N$). The nominal photon yield is calculated using theoretical value and multiply the detector efficiency list in Table \ref{tab:eff}. \par
Figure \ref{fig:flow_RAT} shows the schematic of procedure of the simulation. If the final resolution is Gaussian and the process are independent of each other the total resolution is quadruple sum of the resolution from each process. The energy resolution function is derived from taken to be a Gaussian with mean 
\begin{ceqn}\begin{align}
\mu(E) = E\times(Y_N/Y)
\end{align}\end{ceqn}
and sigma 
\begin{ceqn}\begin{align}\label{eq:resolution}
\sigma(E) = p_1\sqrt{\frac{E}{E_M}} + p_0
\end{align}\end{ceqn}
Where $E_M$ = 220 is the mean of the true energy spectrum and fitted light yield is $Y$ in p.e./keV$_{ee}$. $Y_N$ is the nominal yield which goes into the denominator of the detected energy cancels. The parameters $p_0$ and $p_1$ have dimensions of keV$_{ee}$. The resolution function is integrated over the theoretical spectrum, normalized to the number of events (``norm'') to get the fitting function.  The energy spectrum from simulation is then fit to the four parameters fitting function : normalization, bias, resolution $\sigma$ and yield. The fitting results is shown in Fig. \ref{fig:ar39_fit}, the associated parameters is list in Table \ref{tab:par}.\par
\begin{figure}[htbp]
\centering
\graphicspath{{./fig/Energy/}}
\includegraphics[scale=0.6]{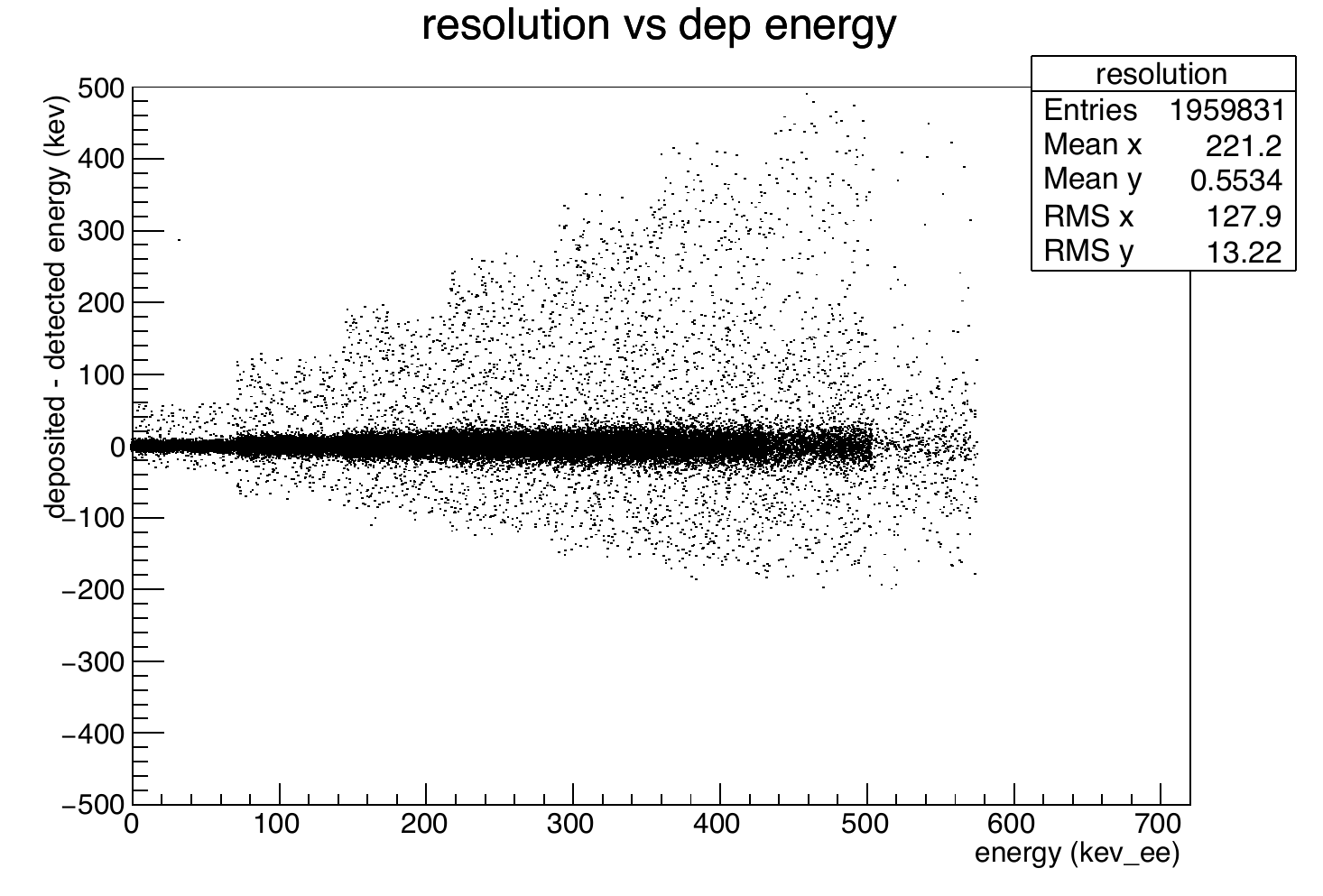}
\caption{ RAT simulated deposited energy - detected energy versus deposited energy. This plot is used to derive the resolution function. }
\label{fig:edet_edep}
\end{figure}

\begin{table}[htbp]
\centering 
\begin{tabular}{ c c c} 
\toprule[1.5pt]
\bf Parameter & Value & p.e./keV$_{ee}$  \\\midrule
yield & 40.0 & 40.0\\
PMT efficiency & 0.192 & 7.674\\
TPB conversion eff. & 0.934 & 7.168\\
TPB re-emit & 1.200 & 8.601\\
TPB efficiency & 0.806 & 6.934\\
Acrylic absorption & 0.112 & 6.150\\
LG absroption & 0.123 & 5.394\\
Fit factor & 1.228 & 6.623\\\bottomrule[1.25pt]
\end {tabular}
\captionof{table}{Efficiency for each component in rat. Right column shows expected p.e./keV$_ee$ with efficiencies applied. Last line is predicted experimental yield including the fitted factor from Fig. \ref{fig:fit_factor} } \label{tab:eff} 
\end{table}
\begin{figure}[htbp]
\centering
\graphicspath{{./fig/Energy/}}
\includegraphics[scale=0.5]{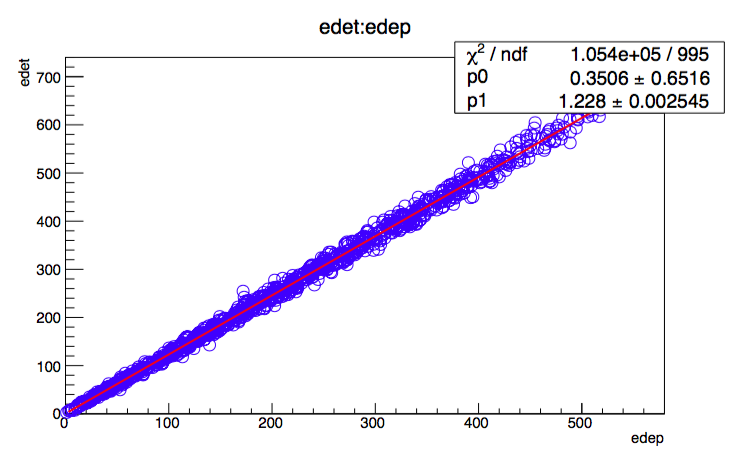}
\caption{ Fit to correction factor applied as last line in Table \ref{tab:eff}. }
\label{fig:fit_factor}
\end{figure}
\begin{figure}[htbp]
\centering
\graphicspath{{./fig/Energy/}}
\includegraphics[scale=0.4]{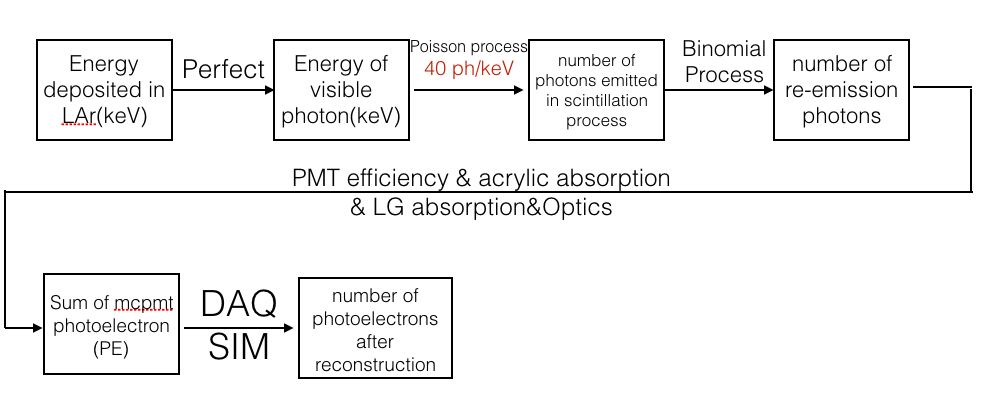}
\caption{Schematic of each process in simulation.}
\label{fig:flow_RAT}
\end{figure}
\begin{figure}[htbp]
\centering
\graphicspath{{./fig/Energy/}}
\includegraphics[scale=0.25]{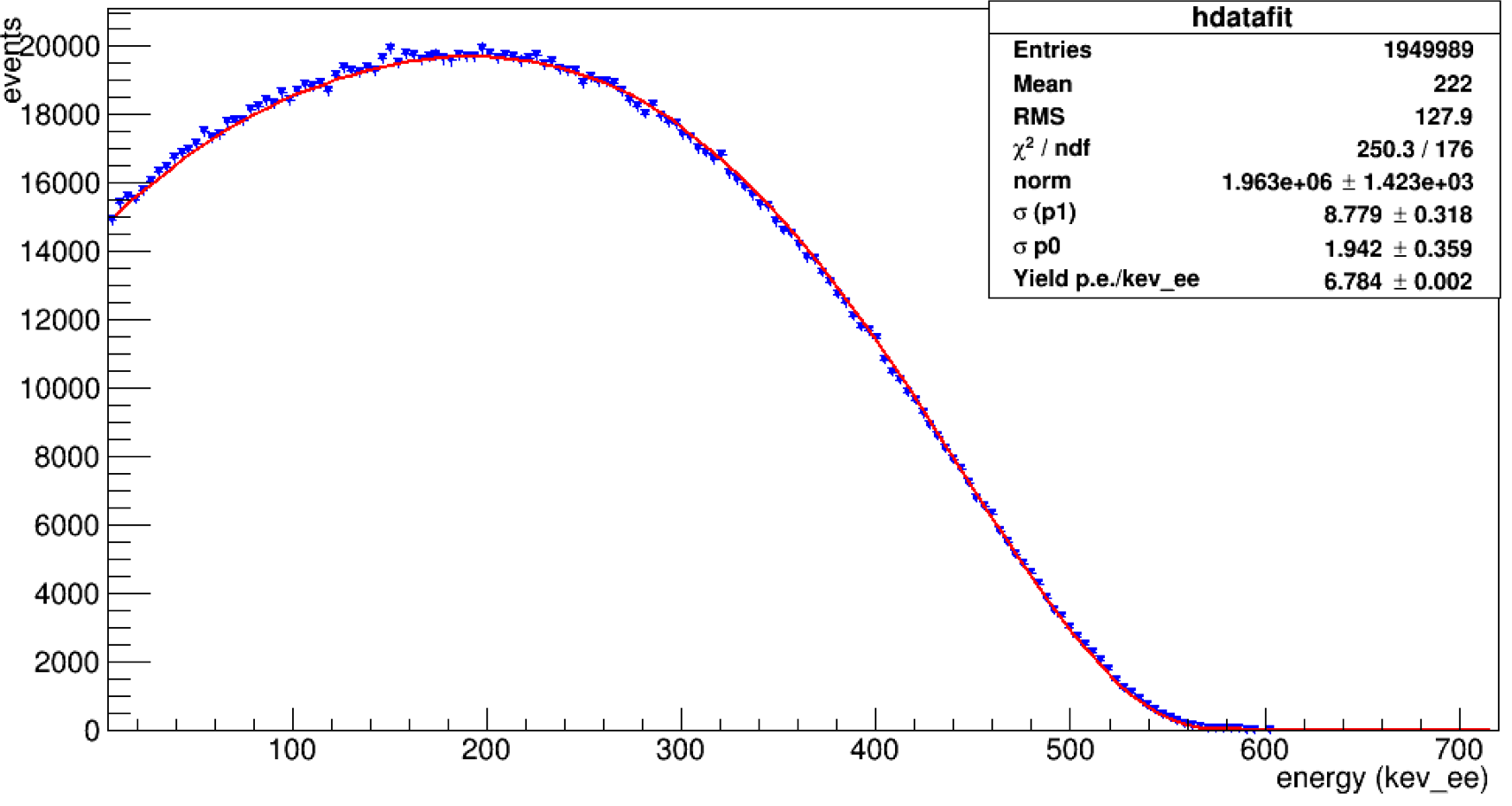}
\caption{ The fit of simulated $^{39}$Ar energy spectrum. }
\label{fig:ar39_fit}
\end{figure}

\begin{table}[htbp]
\centering 
\begin{tabular}{ c c } 
\toprule[1.5pt]
\bf Parameter & Value \\\midrule
$\chi^2$ & 250.3 \\
NDF & 176\\
norm & 1.963$\times$10$^{6}$ $\pm$ 1.423$\times$10$^3$\\
$p_1$ & 8.779 $\pm$0.318 keV$_{ee}$\\
$p_0$ & 1.942 $\pm 0.359$ keV$_{ee}$\\
Yield & 6.784 $\pm$ 0.002 p.e./keV$_{ee}$\\\bottomrule[1.25pt]
\end {tabular}
\captionof{table}{Fitting parameters of Fig. \ref{fig:ar39_fit}.} \label{tab:par} 
\end{table}
In order to understand the fitting, each process has been studied thoroughly (Fig. \ref{fig:flow_RAT}). In the first process, RAT assuming all the energy deposited by the interaction convert to the energy of corresponding photons. The VUV photons emission is following the Poisson distribution with mean value of 40 photons/keV. To understand the behavior in each energy bins, the bias and the energy resolution is plotted for each energy bins (50 keV$_{ee}$/bin). Figure \ref{fig:scintillation_bias} shows the bias (a) and the resolution as a function of energy. The bias is fitted to first order of polynomial function and the resolution is fitted to Eq. \ref{eq:resolution}. The resolution plot seems to fit well to the Eq. \ref{eq:resolution} while the bias seems to behave non-linearly. This may due to the fact that the process follows the Poisson distribution and approaches to Gaussian when the statistics is high as shown in Fig. \ref{fig:scintillation_detail}.  The next process is the re-emission of TPB as shown in the Fig. \ref{fig:reemission_bias}. The Gaussian fit to each energy bin is shown in Fig. \ref{fig:reemission_detail}. The events from long tail in Fig. \ref{fig:reemission_detail} came from the events produced near the baffles. Extracting the events which in the long tail and plot their true position in the simulation. These events clearly from the position near the baffle as shown in Fig. \ref{fig:reemission_angular}. Figure \ref{fig:reemission_radius} shows the plot of the reconstructed radius versus the MC true radius. These plot shows the events near the edge of baffle tend to be mis-reconstructed. The Shellfit is based on the scintillation profile of LAr, for events are too close to the baffle, the reconstruction algorithm will have difficulty to handle the reflection process correctly. Moreover, some photons may be absorbed in the baffle or escape from the active volume, results in mis-reconstructed radius and energy. Figure. \ref{fig:reconstructed_missed_1} shows the radius and angular distribution of events originate outside the fiducial volume with reconstructed radius inside the fiducial volume. This shows that 5.7 $\pm$ 0.04\% of the events leak into the fiducial volume after the reconstruction.\par    
\begin{figure}
\hfill
\subfloat[]{\includegraphics[width=7cm]{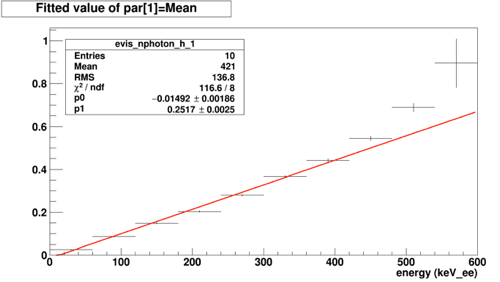}}
\hfill
\subfloat[]{\includegraphics[width=7cm]{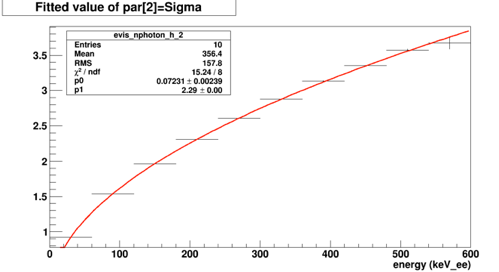}}
\hfill
\caption{(a) Bias versus energy. Unit of Y-axis is  keV$_{ee}$. (b) Resolution versus energy. Unit of Y-axis is  keV$_{ee}$.}
\label{fig:scintillation_bias}
\end{figure}
\begin{figure}[htbp]
\centering
\graphicspath{{./fig/Energy/}}
\includegraphics[scale=0.4]{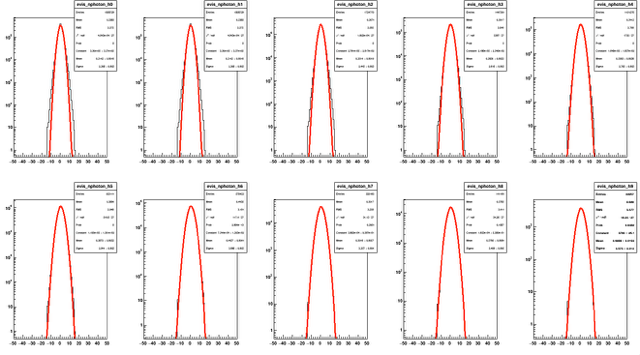}
\caption{ The fitted gaussian function to each energy bin(50 keV$_{ee}$). Starts from upper left is (0-50 keV$_{ee}$) and the bottom right is (550-600 keV$_{ee}$). The X-axis is the counts and the Y-axis is the number of photons from scintillation process minus the number of emitted photons.  }
\label{fig:scintillation_detail}
\end{figure}
\begin{figure}
\hfill
\subfloat[]{\includegraphics[width=7cm]{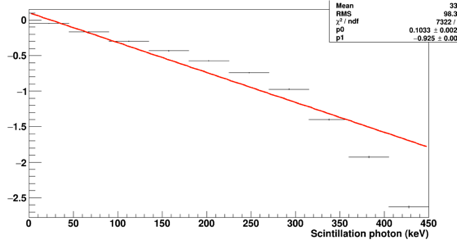}}
\hfill
\subfloat[]{\includegraphics[width=7cm]{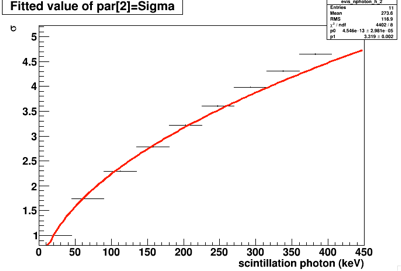}}
\hfill
\caption{(a) Bias versus energy. Unit of Y-axis is  keV$_{ee}$. (b) Resolution versus energy. Unit of Y-axis is  keV$_{ee}$.}
\label{fig:reemission_bias}
\end{figure}
\begin{figure}[htbp]
\centering
\graphicspath{{./fig/Energy/}}
\includegraphics[scale=0.4]{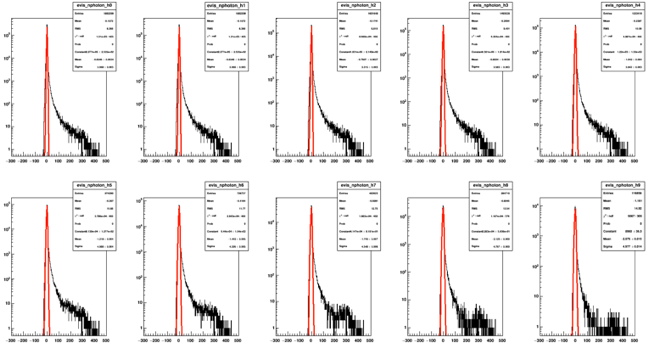}
\caption{ The fitted gaussian function to each energy bin(50 keV$_{ee}$). Starts from upper left is (0-50 keV$_{ee}$) and the bottom right is (550-600 keV$_{ee}$). The X-axis is the counts and the Y-axis is the number of photons from produced minus the number of re-emitted photons of TPB.  }
\label{fig:reemission_detail}
\end{figure}
\begin{figure}
\hfill
\subfloat[]{\includegraphics[width=7cm]{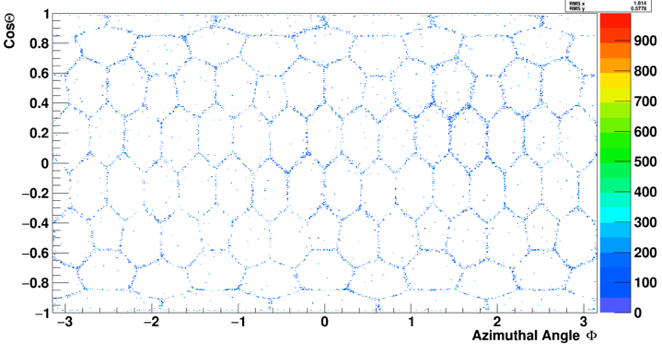}}
\hfill
\subfloat[]{\includegraphics[width=7cm]{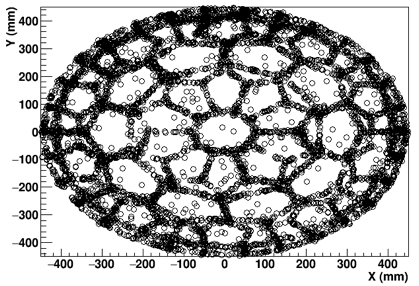}}
\hfill
\caption{(a) MC angular distribution of the mis-constructed events. (b) 3D position of the mis-constructed events.}
\label{fig:reemission_angular}
\end{figure}
\begin{figure}[htbp]
\centering
\graphicspath{{./fig/Energy/}}
\includegraphics[scale=0.4]{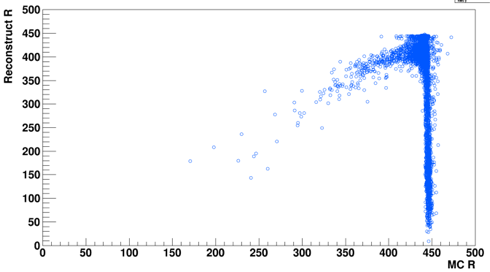}
\caption{ Reconstructed event radius versus MC true radius.  }
\label{fig:reemission_radius}
\end{figure}
\begin{figure}
\hfill
\subfloat[]{\includegraphics[width=7cm]{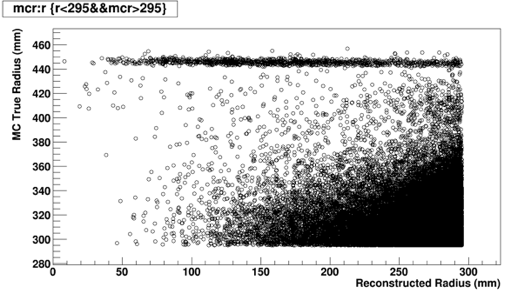}}
\hfill
\subfloat[]{\includegraphics[width=7cm]{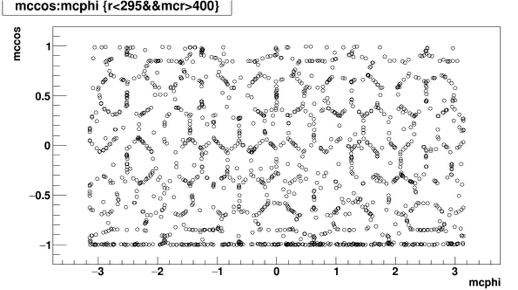}}
\hfill
\caption{(a) MC true radius versus reconstructed radius. (b) Angular distribution for events originate outside the fiducial volume with reconstructed radius inside the fiducial volume.}
\label{fig:reconstructed_missed_1}
\end{figure}
The process of converting MC charge to reconstructed charge is done by simulating the DAQ system. This process should follow the binomial distribution. Figure \ref{fig:MC_EV_bias} shows the bias and resolution versus the energy. The Gaussian fit is shown in Fig. \ref{fig:rMC_EV_detail}. The flat bias at high energy bins might due to the saturation effect of the simulated DAQ system. The last step is to convert the reconstructed charge to the detected photoelectrons. Again, the plot to described this process is shown in Fig. \ref{fig:PE_bias} and Fig. \ref{fig:PE_detail}. Noticed that the negative tail comes from the events reconstructed near some PMTs which has large PMT gain as shown in Fig. \ref{fig:PE_PMT}. For example PMT 74 has gain of 14 pC compare to the nominal gain of 5 pC. This create the excessive energy compare to the input reconstructed energy.\par
\begin{figure}
\hfill
\subfloat[]{\includegraphics[width=7cm]{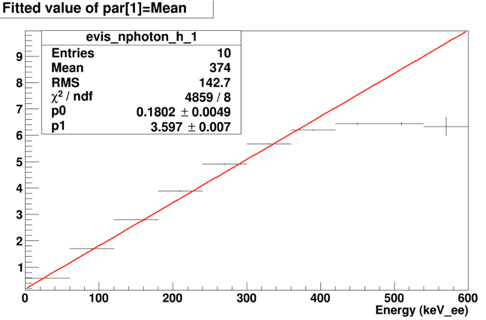}}
\hfill
\subfloat[]{\includegraphics[width=7cm]{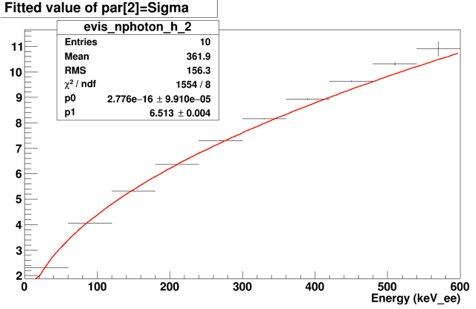}}
\hfill
\caption{(a) Bias versus energy. Unit of Y-axis is  keV$_{ee}$. (b) Resolution versus energy. Unit of Y-axis is  keV$_{ee}$.}
\label{fig:MC_EV_bias}
\end{figure}
\begin{figure}[htbp]
\centering
\graphicspath{{./fig/Energy/}}
\includegraphics[scale=0.4]{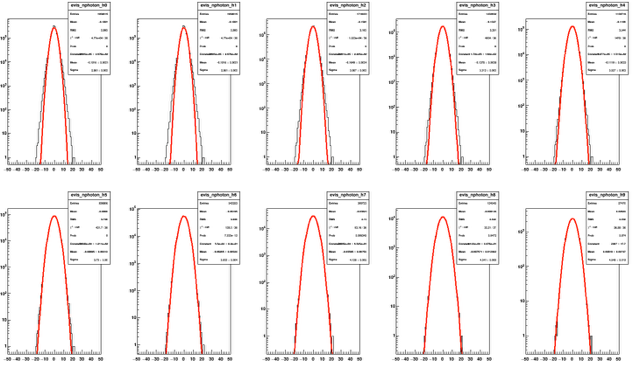}
\caption{ The fitted gaussian function to each energy bin(50 keV$_{ee}$). Starts from upper left is (0-50 keV$_{ee}$) and the bottom right is (550-600 keV$_{ee}$). The Y-axis is the counts and the X-axis is the MC charge minus reconstructed charge.  }
\label{fig:rMC_EV_detail}
\end{figure}
\begin{figure}
\hfill
\subfloat[]{\includegraphics[width=7cm]{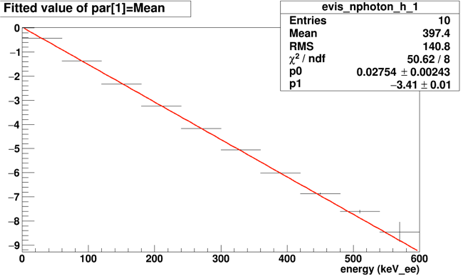}}
\hfill
\subfloat[]{\includegraphics[width=7cm]{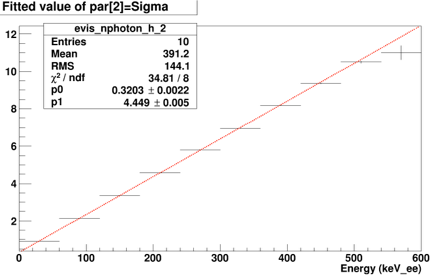}}
\hfill
\caption{(a) Bias versus energy. Unit of Y-axis is keV$_{ee}$. (b) Resolution versus energy. Unit of Y-axis is keV$_{ee}$.}
\label{fig:PE_bias}
\end{figure}
\begin{figure}[htbp]
\centering
\graphicspath{{./fig/Energy/}}
\includegraphics[scale=0.4]{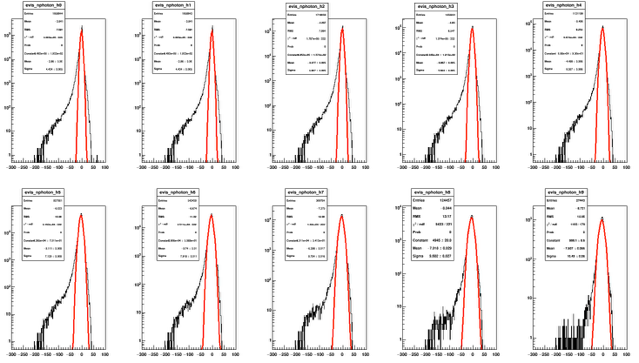}
\caption{ The fitted gaussian function to each energy bin(50 keV$_{ee}$). Starts from upper left is (0-50 keV$_{ee}$) and the bottom right is (550-600 keV$_{ee}$). The Y-axis is the counts and the X-axis is the difference between calibrated and uncalibrated charge.  }
\label{fig:PE_detail}
\end{figure}
\begin{figure}[htbp]
\centering
\graphicspath{{./fig/Energy/}}
\includegraphics[scale=0.5]{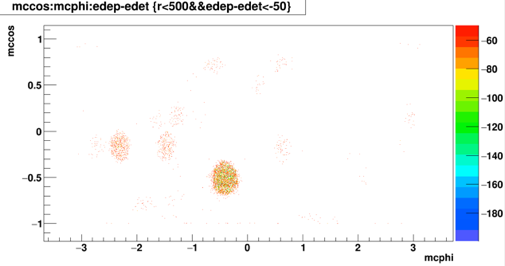}
\caption{ The MC angular distribution for events in the negative tail in Fig. \ref{fig:PE_detail}.  }
\label{fig:PE_PMT}
\end{figure}
The low energy part of spectrum of Fig. \ref{fig:ar39_fit} has slightly excessive events. This may be explained by above discussion. The baffle events tend to be under-estimate, create a excessive events in low energy region of the spectrum. However, the $\chi^2$/NDF (1.42) of $^{39}$Ar spectrum is good, indicates the fit results is reliable.
The final energy resolution as a function of energy is shown in Fig. \ref{fig:final_resolution}.   Alternatively, the energy bias and resolution as a function of reconstructed radius can be investigated. Figure \ref{fig:energy_radius_bias} shows the energy bias and resolution as a function of MC radius. The events originate from edge has larger bias and resolution are harder to be reconstructed correctly and has under-estimated energy. Nevertheless, with real data from LAr, these plots are helpful to monitor the reconstruction algorithm. 
\begin{figure}[htbp]
\centering
\graphicspath{{./fig/Energy/}}
\includegraphics[scale=0.5]{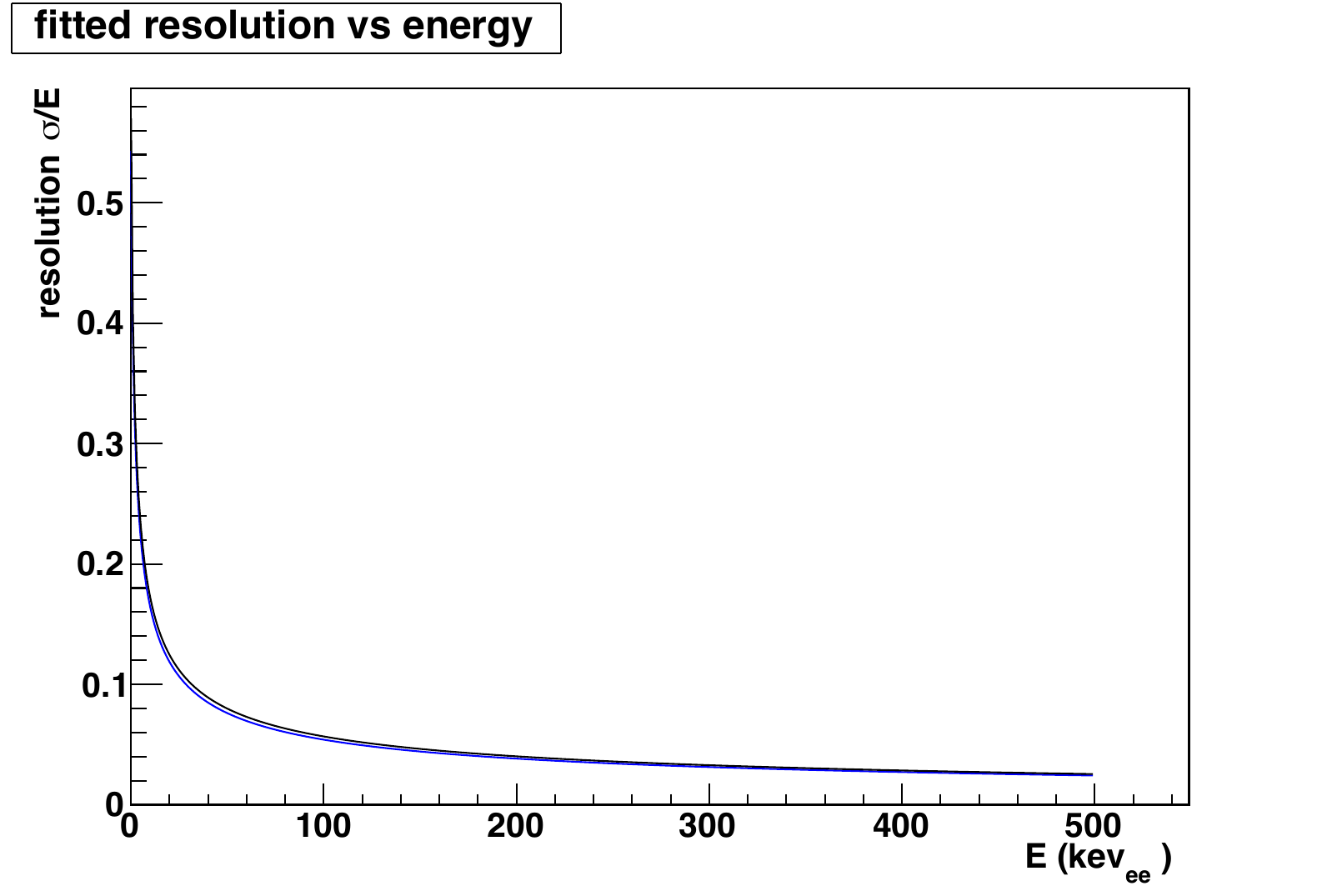}
\caption{ Energy resolution as a function of energy. }
\label{fig:final_resolution}
\end{figure}
\begin{figure}
\hfill
\subfloat[]{\includegraphics[width=7cm]{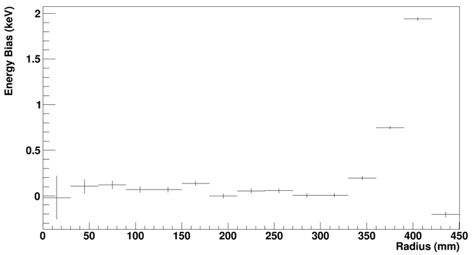}}
\hfill
\subfloat[]{\includegraphics[width=7cm]{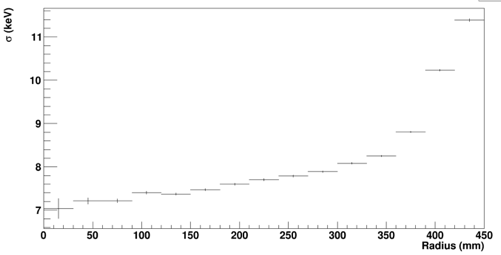}}
\hfill
\caption{(a) Bias versus radius. Unit of Y-axis is keV$_{ee}$. (b) Resolution versus radius. Unit of Y-axis is keV$_{ee}$.}
\label{fig:energy_radius_bias}
\end{figure}
An effort has been made to fit the $^{39}$Ar energy spectrum in cold gas. However, it is difficult to get a good fit due to the following reasons.
\begin{enumerate}
	\item The density of cold gas is around 0.006 g/cm$^2$ which is nearly three order of magnitude less than liquid (1.394 g/cm$^{2}$). The observed events decreased correspondingly.
	\item Due to the low density, the electron emitted by $^{39}$Ar beta decay close to the edge will escape from of active volume results in non-detected events. This distort the energy spectrum.
	\item There are 28 PMTs are deactivated, which reduces the PMT coverage.
	\item The contribution from internal $\beta/\gamma$ background to the electronic recoil is indistinguishable. This will not be a problem in liquid due to high $^{39}$Ar beta decay rates and increased density.
\end{enumerate}
The first point can be simply calculated using the density of gas/liquid argon. In the gaseous argon, the total mass of gas is 2.29 kg which gives the rate of $^{39}$Ar beta decay of 2.29 Hz. However, with the liquid, the rate is around 500 Hz even without the $^{39}$ spike injection. The statistics of the $^{39}$Ar energy spectrum in the cold gas is way smaller than in the liquid, which gives large error on fitting results. As for point two, it can be seen in the simulation that the electron originated near the edge will have large probability to escape without been detected. Figure \ref{fig:Radius_sim_gas} shows the energy of electron as a function of radius ratio cubed.  In the liquid, the triggered events should be uniformly distributed according to the radius cubed, however,  in the gas there are very few high energy events are detected. Moreover, the excessive low energy events are caught at the edge indicating the high energy events which originate closer to the center of the detector just deposit partial energy in the detector. The maximum distance which electrons can travel in the gaseous argon is defined by CSDA (continuous-slowing-down-approzimation) and gives 58 cm. The diameter of the IV active volume is 90 cm, thus agree with the simulation. Figure \ref{fig:Radius_cut_sim} shows cutting the events which close to the edge results in eliminating the low energy events with high energy events stay the same. Moreover, the reducing PMT coverage cost some photons emitted without being detected. In theory this should be modeled correctly by simulation. However, the pressure and the temperature of the IV is changing while taking data. This creates the inhomogeneity in the IV which is hard to be model correctly in the simulation. Figure \ref{fig:angular_cold_gas} shows the angular distribution for both cold gas data and simulation. In the data, a group of events are reconstructed near the top and bottom while in the simulation, the pattern is not seen. Finally the background from the internal $\beta/\gamma$ which originate from the radioactive impurities of detector material will create the same electronic recoil. With reduction of PMT coverage, the shellfit can not reconstructed these events correctly. These indistinguishable events will further distort the energy spectrum of $^{39}$Ar beta decay. However, due to the self shielding in the liquid argon, the fitting of energy spectrum can be done with a proper radius cut. Figure \ref{fig:liquid_cold_gas} shows the simulated beta-deay energy spectrum in liquid and cold gas which displays the distorted energy spectrum of the gas compare to the normal spectrum in the liquid. Figure \ref{fig:gas_data_sim} shows the comparison of charge distribution between real data and the simulation. In the data, more low energy event presented, this may due to the inhomogeneity of the gas such that high energy electron deposit less energy than expected in the simulation. Moreover, low energy gamma (< 40 keV) can also contribute to the charge distribution. The attenuation length for gamma particle with energy of 30 keV is 62 cm, which could only deposit its partial energy in the gaseous argon.

\begin{figure}[htbp]
\centering
\graphicspath{{./fig/Energy/}}
\includegraphics[scale=0.4]{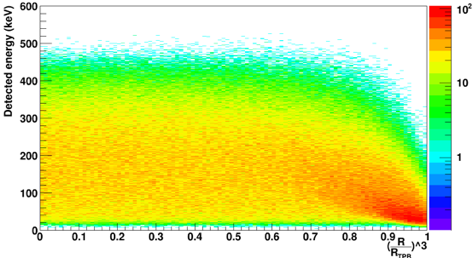}
\caption{ Energy of electrons versus  ratio of radius to radius of TPB cubed in cold gas simulation with 64 PMTs activated (total number of PMT is 92.). }
\label{fig:Radius_sim_gas}
\end{figure}
\begin{figure}[htbp]
\centering
\graphicspath{{./fig/Energy/}}
\includegraphics[scale=0.4]{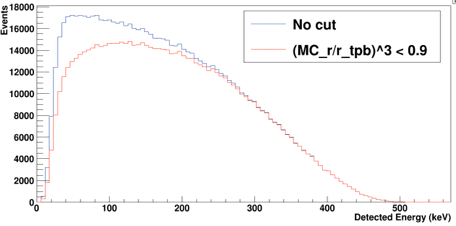}
\caption{ Energy spectrum before and after the cut. Blue curve shows no cut apply, and red curve shows cutting the events originate from the edge. }
\label{fig:Radius_cut_sim}
\end{figure}
\begin{figure}
\hfill
\subfloat[]{\includegraphics[width=7cm]{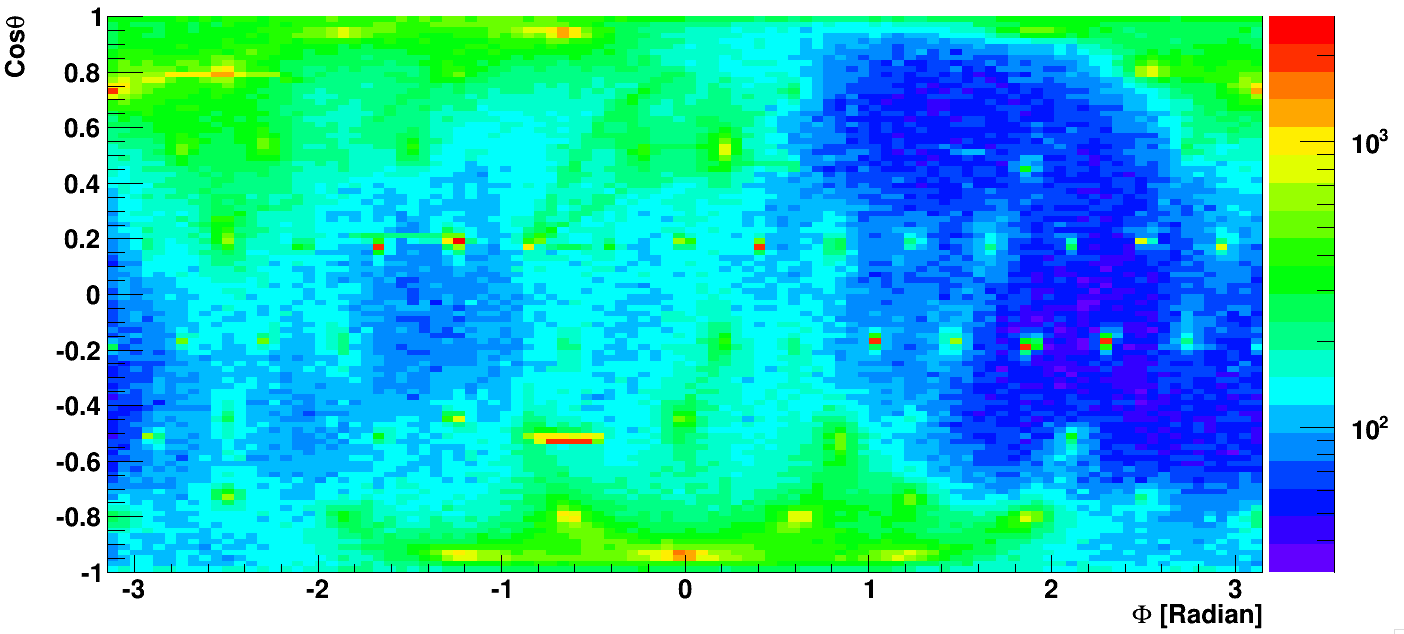}}
\hfill
\subfloat[]{\includegraphics[width=7cm]{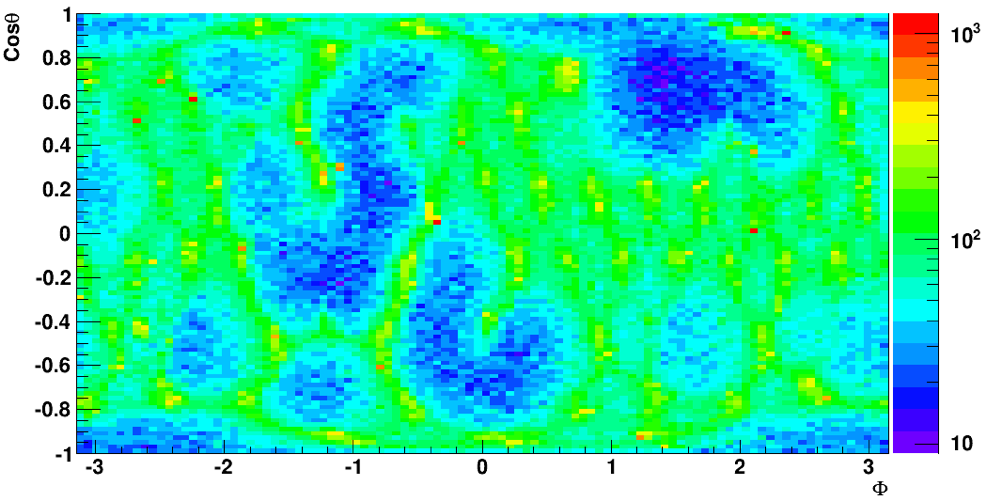}}
\hfill
\caption{Angular distribution of $^{39}$Ar events in (a) cold gas data. (b) cold gas simulation.}
\label{fig:angular_cold_gas}
\end{figure}

\begin{figure}[htbp]
\centering
\graphicspath{{./fig/Energy/}}
\includegraphics[scale=0.25]{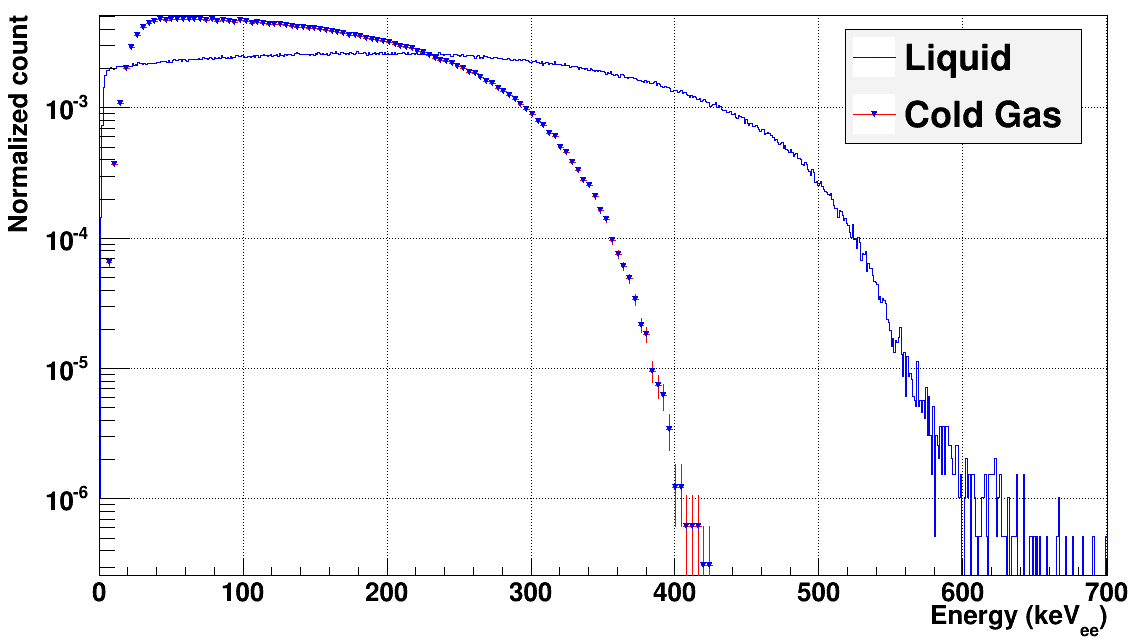}
\caption{ $^{39}$Ar energy spectrum for liquid (blue) and cold gas (with 64 PMTs) (red). }
\label{fig:liquid_cold_gas}
\end{figure}
\begin{figure}[htbp]
\centering
\graphicspath{{./fig/Energy/}}
\includegraphics[scale=0.3]{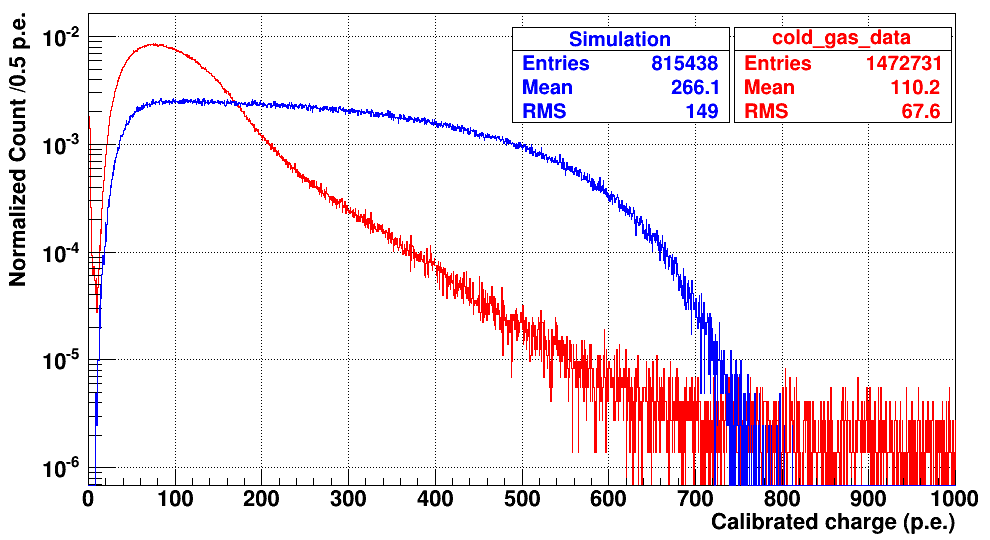}
\caption{ Charge distribution of cold gas data (red) and the simulation (blue).}
\label{fig:gas_data_sim}
\end{figure}

\FloatBarrier

\chapter{Conclusion}
Currently MiniCLEAN detector is in the cooling phase (Aug. 2017) and we expect to start filling the detector with LAr within a month. The preliminary operation in cold gas revealed several dysfunctional PMTs which are summarized in Chapter \ref{sec:dys}. Nonetheless, some PMTs with low gain could be recovered after official commissioning and the PMTs without conformal coating should also perform normally in LAr. For the \textit{In-situ} optical calibration system, several LEDs are not functioning normaly, only three blue LEDs with measurable impedance. We suspect these are the results when the detector warming unexpectedly fast during the incident described in Chapter \ref{sec:coolingphase}.  We infer that the cable were loosened or the electrical connection broken. We perform a MC simulation to evaluate the reduced detector performance compare to the full performance\cite{chrisJreduced}. Ten thousand of $^{39}$Ar events were produced to confirm the light yield and the position reconstruction performance under the reduced PMT coverage. Figure \ref{fig:reduction_pe} shows the reduction of the photoelectrons with only 82 PMTs coverage. The MC simulation indicates the 11.5\% reduction on light yield and 14\% reduction in resolution of position reconstruction as shown in Fig. \ref{fig:reduction_resolution}. After the commissioning, the full performance can be estimate by extrapolate the results from MC simulation.\par
The minimum cross-section limit from MiniCLEAN is estimated to be 9$\times$10$^{-45}$ cm$^2$ at a 94 GeV WIMP mass. It is about 2 order of magnitude larger than the current best results. However, the $^{39}$Ar spike runs allow us to study the limit of PSD and the light yield fitting from $^{39}$Ar energy spectrum. Figure \ref{fig:PSD_ability} shows the  leakage probability of electronic recoil (PSD capability) as a function of energy threshold. This plot shows that with 150 kg LAr in fiducial volume and expected light yield of 6 p.e./keV, and if the PSD rejection ability reaches 10$^{-10}$ level , the MiniCLEAN will have demonstrated what could be achieved in a 1-ton detector.  This extraordinary PSD background rejection ability 
suggests that a single phase LAr detector would be competitive for with other techniques for a dark matter WIMP search.
\begin{figure}[htbp]
\centering
\graphicspath{{./fig/Energy/}}
\includegraphics[scale=0.3]{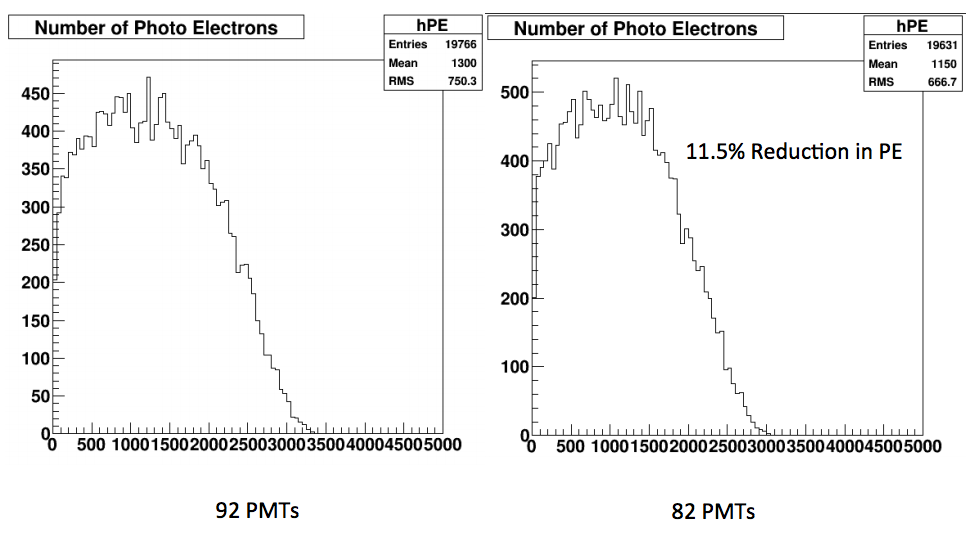}
\caption{ Charge distribution of $^{39}$Ar events. Left : with 92 PMTs (full coverage) Right : with 82 PMTs (reduced coverage). }
\label{fig:reduction_pe}
\end{figure}
\begin{figure}[htbp]
\centering
\graphicspath{{./fig/Energy/}}
\includegraphics[scale=0.3]{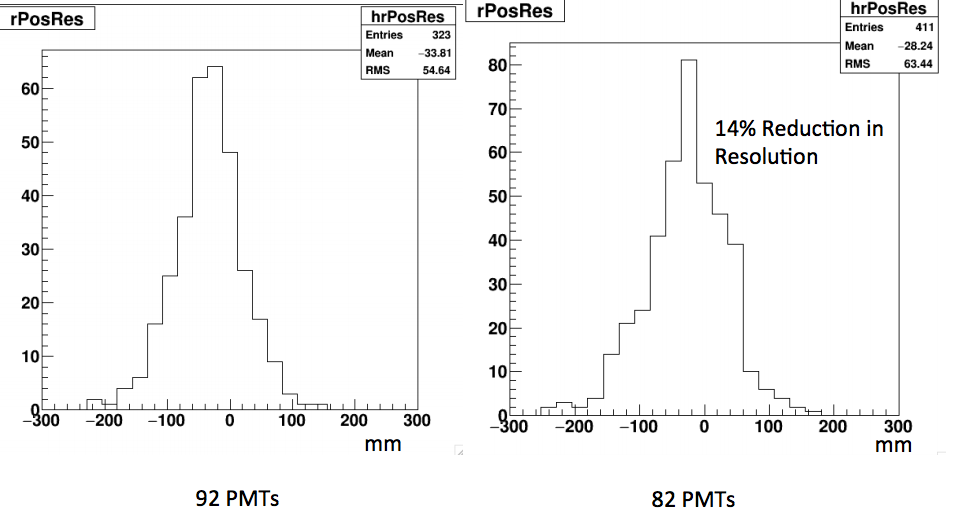}
\caption{Resolution of position reconstruction using ShellFit algorithm for $^{39}$Ar events. Left : with 92 PMTs (full coverage) Right : with 82 PMTs (reduced coverage). }
\label{fig:reduction_resolution}
\end{figure}
\begin{figure}[htbp]
\centering
\graphicspath{{./fig/Energy/}}
\includegraphics[scale=0.3]{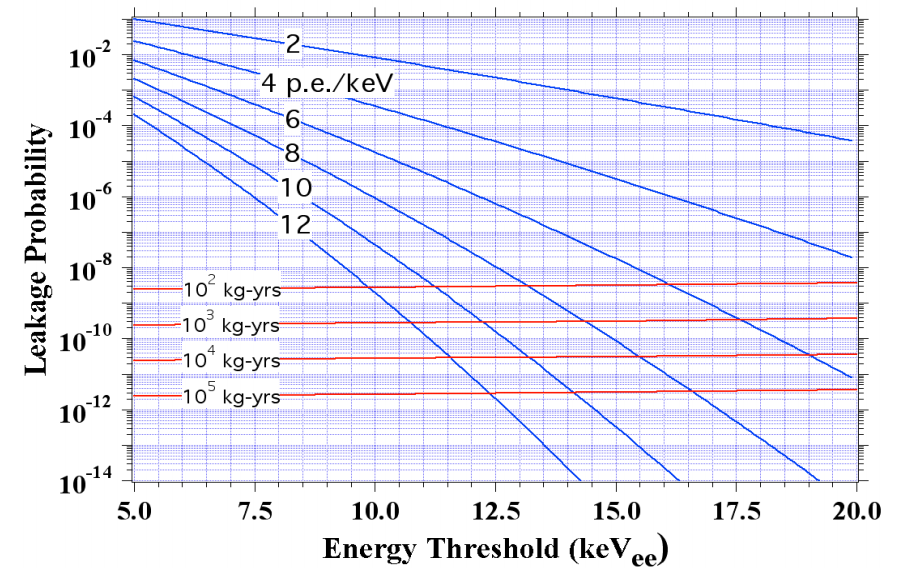}
\caption{ A study of $^{39}$Ar background leakage as a function of detector energy threshold. Leakage is into the nuclear recoil region of interest with 50\% acceptance. The red horizontal lines are indicative of the leakage thresholds required such that dark matter detector (with target mass of 100, 1000, 10,000 kg) have only one background event/year. The blue curve indicates the achieved light yield. Figure from \cite{andrewlight}. }
\label{fig:PSD_ability}
\end{figure}

\chapter*{Appendices}


\appendix
\chapter{Magnetic Compensation Sensors}
\section{Description of device}
The magnetic sensor is designed to measure the magnetic field strength inside the water tank and the results is used to determine the nominal current in magnetic compensation coils.  The other important application of sensors is PMT gain calibration. The PMT gain is affected by the magnetic field. In order to optimize the PMT performance, we need to know the field strength around the PMT and adjust the gain accordingly. \\
The sensor boards were custom made at UNM(Fig. \ref{fig:fmagsensor}). These are based on the HMC1001(1D) and HMC1002 (2D) magnetoresistive sensor ICs. For each channel, 1 magnetoresistive Wheatstone bridge made of ferromagnetic resistors and a coil to set the initial magnetization of the sensitive elements. The sensor board is contained in a PVC case made of an asymmetric T for plumbing, with the 2 wider extremities closed by caps, and a PVC tube connected to the third by way of spigot. The PVC tube is designed to be water tight and has been tested in the water. The wire bundle from the board is a 3M 10 wire ribbon, which exists the case through the spigot, and  runs in a transparent and flexible PVC tube up to water surface.
\begin{figure}[htbp]
\graphicspath{{./fig/appendix/}}
\includegraphics[scale=0.5]{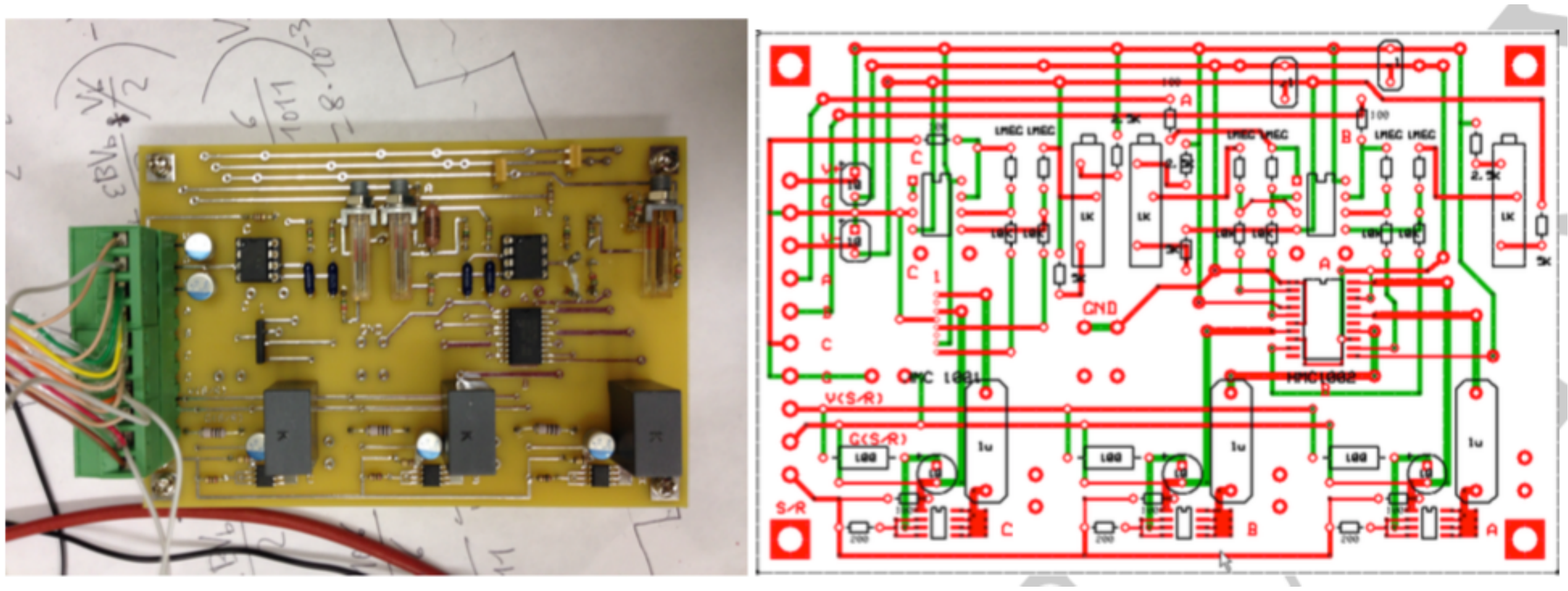}
\caption{ Left: photo of the magnetic sensor board. Right: layout of the magnetic sensor }
\label{fig:fmagsensor}
\end{figure}
\section{sensor position and orientation}
The five sensors are installed onto OV perimeter and the top of OV (Fig. \ref{fig:fmagposition}). For sensors on the OV perimeter,  two clamps are installed on the PVC case along with steel wire which hangs the PVC case onto cable horn (Fig. \ref{fig:ftruepipe}). The sensor on the top of OV was installed the same way onto the 8 inches pipe. The detail orientation and position are follows :
\begin{itemize}
	\item North sensor : Hanging on OV-S,  3\si{\degree} East of North.
	\item East sensor  : Hanging on OV-R, 15\si{\degree} North of East.
	\item South sensor  : Hanging on OV-L, 3\si{\degree} West of South.
	\item West sensor  : Hanging on OV-K, 15\si{\degree} South of West.
	\item Top sensor  : Hanging above the OV on center, pointing North.
\end{itemize}
\begin{figure}[htbp]
\graphicspath{{./fig/appendix/}}
\includegraphics[scale=0.4]{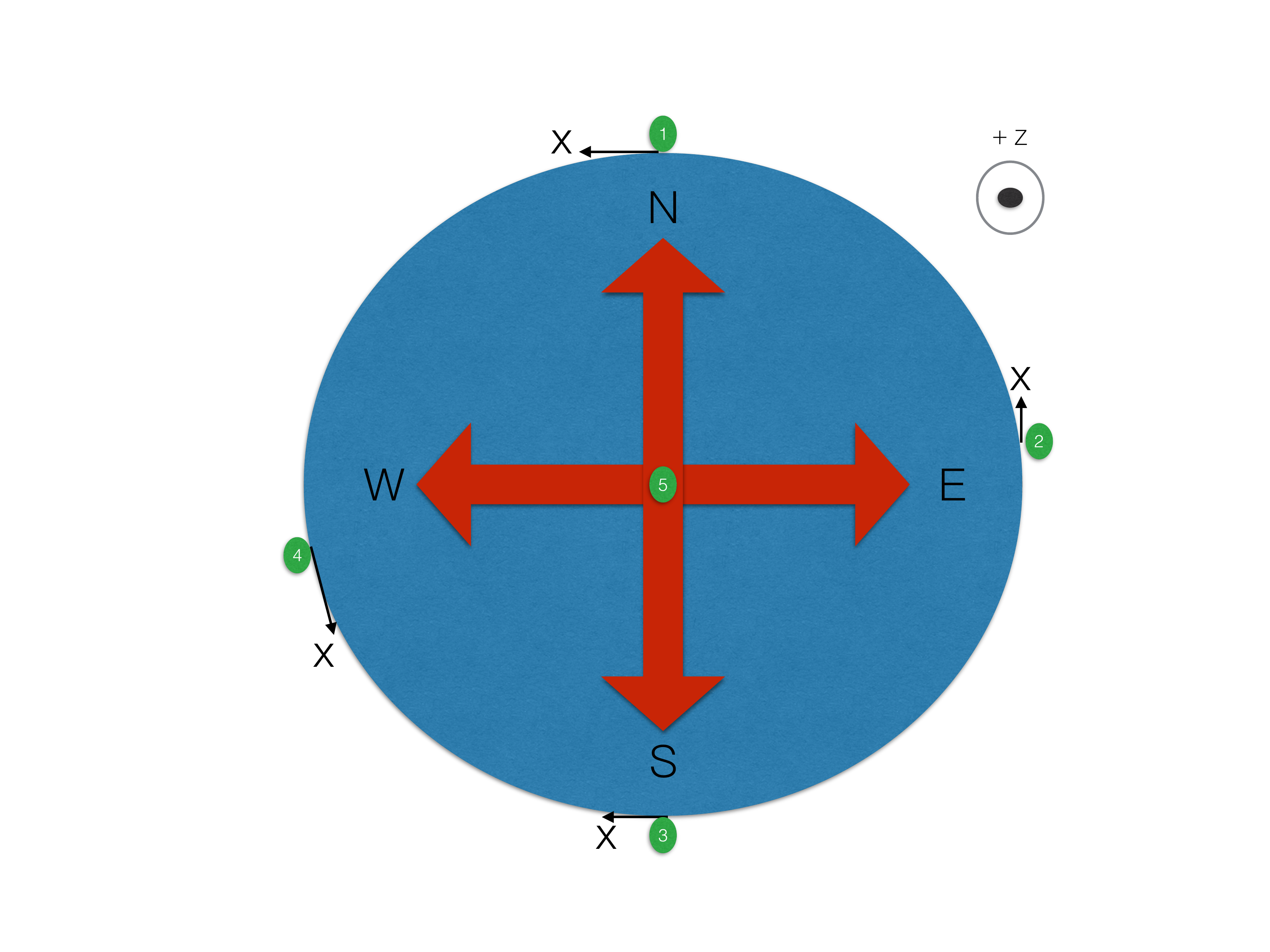}
\caption{ Top down view of position and orientation of sensors. Figure is not drawn to scale.}
\label{fig:fmagposition}
\end{figure}
\begin{figure}[htbp]
\graphicspath{{./fig/appendix/}}
\hfill
\subfloat[]{\includegraphics[width=7cm]{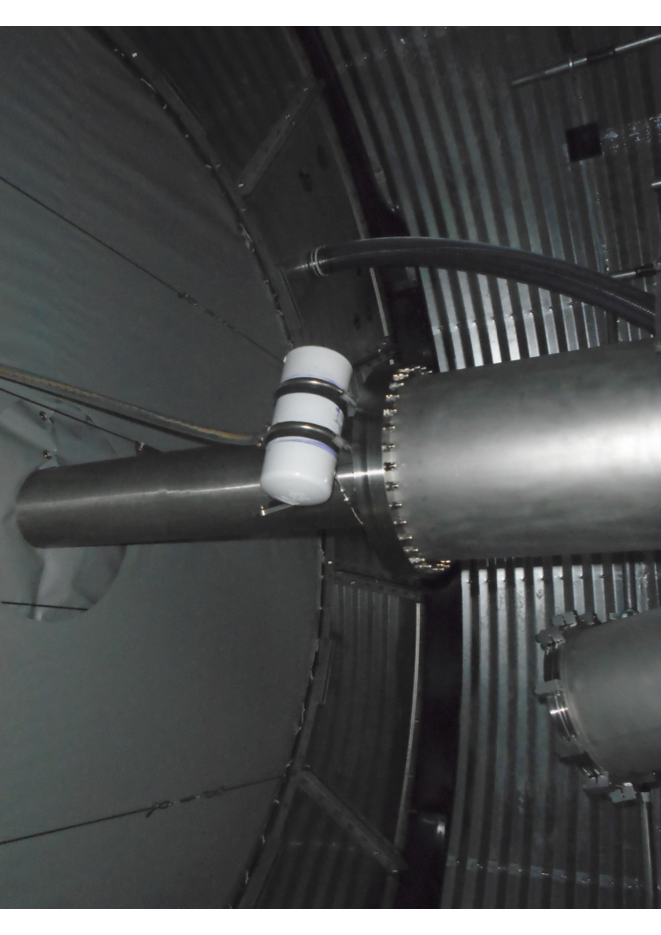}}
\hfill
\subfloat[]{\includegraphics[width=7cm]{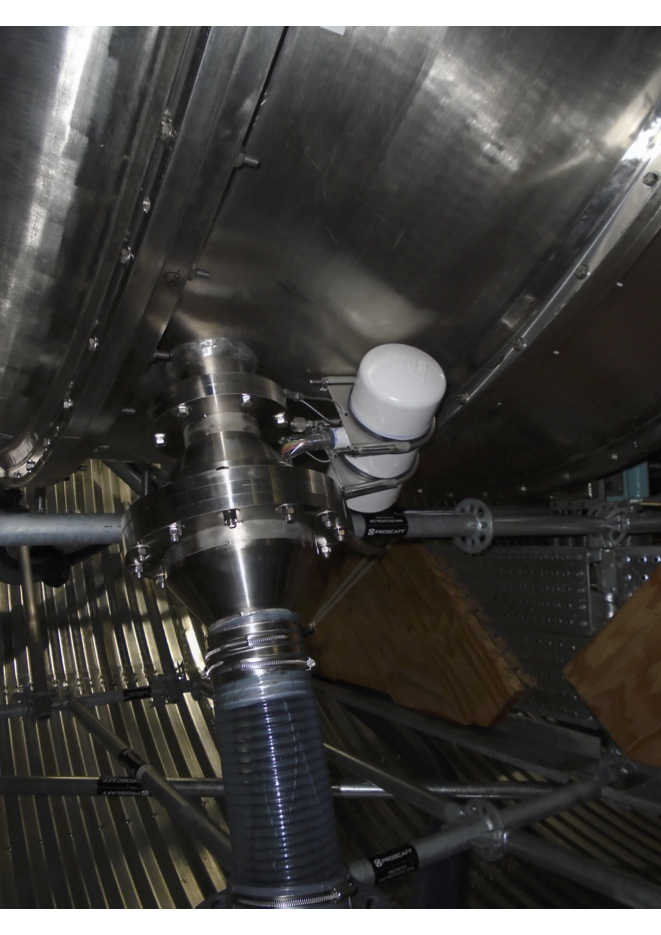}}
\hfill
\caption{Sensor installation}
\label{fig:ftruepipe}
\end{figure}

\section{operation procedure}
\begin{enumerate}
\item Make sure each component is properly connected.
\begin{enumerate}
\item The FPGA should connect to the port on Labjack ADC according to the port map (Fig. \ref{fig:fportmap}) and then gives trigger pulse for sensor.
\item Each sensor has three output wire which should be connected to ADC according to the port map.
\item The DC power supply should feed to both ADC and sensor board as follows.
\begin{enumerate}
\item $\pm$ 5V : The sensors should be feed with pair of same voltage (e.g. $\pm$ 3V, $\pm$ 4V etc), for sensors to output correct signal.
\item LV : 3V to drive ADC board.
\end{enumerate} 
\end{enumerate}
\item Using the FPGA software (Digilent Adept)\footnote{\url{http://store.digilentinc.com/digilent-adept-2-download-only/}} to load the pulse width modulation code onto FPGA through USB connection. 
\item Feed the DC voltage from power supply to sensor board. The typical $V_{cc}$ is 5 V (max 12 V). 
\item Drive the ADC with DC power supply.
\item Using the software from Labjack (LJstremUD)\footnote{\url{https://labjack.com/support/software/applications/ljstreamud}} to collect data. 
\item Repeat the procedure 2-5 for each sensor.
\item Upload the collected data (in .txt format ) to cleanpc06 for further off-line analysis.
\end{enumerate}
\begin{figure}[htbp]
\graphicspath{{./fig/appendix/}}
\includegraphics[scale=0.6]{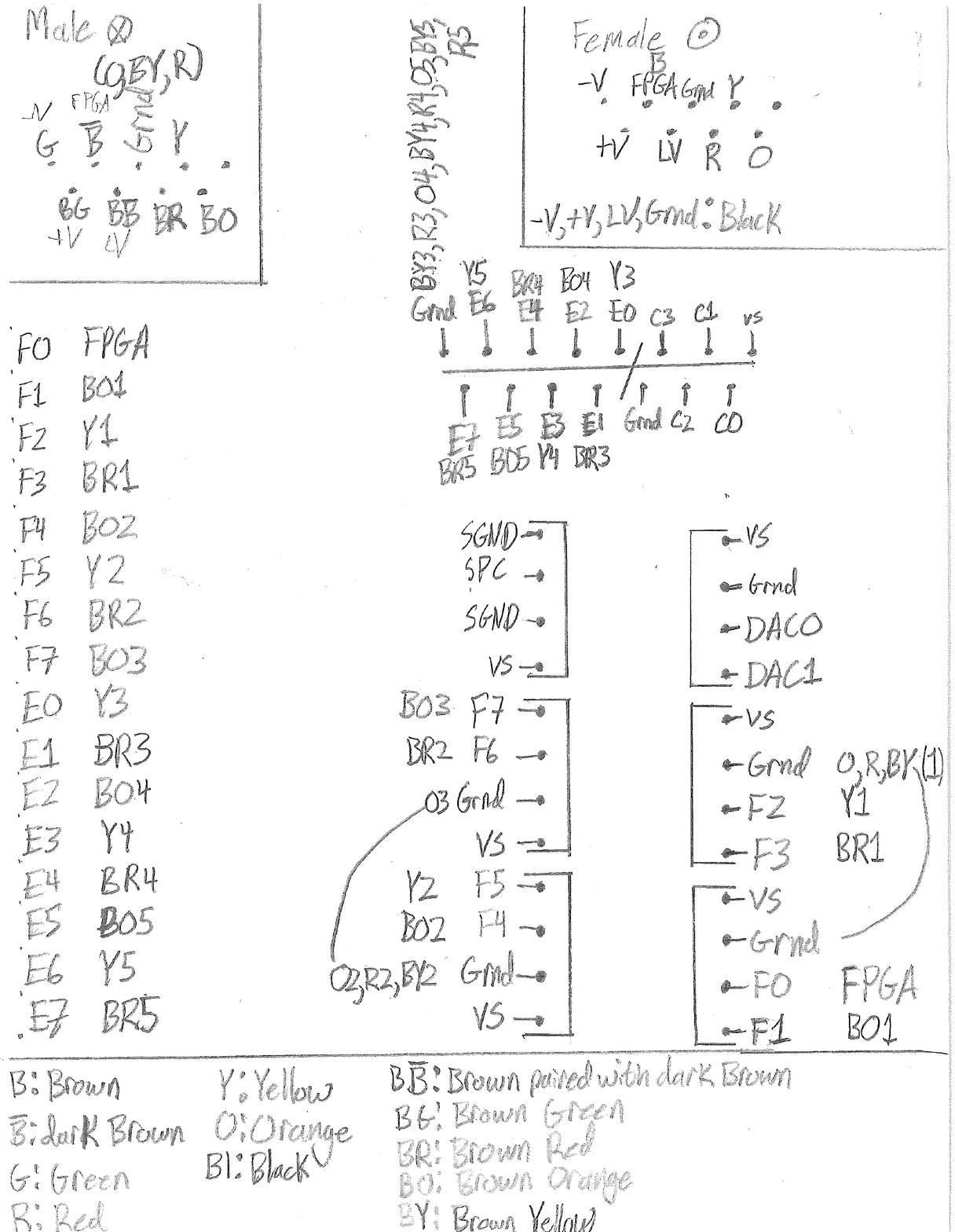}
\caption{ Port map for ADC }
\label{fig:fportmap}
\end{figure}

\chapter{Test Studying Mitigation of Oxygen Deficiency Hazards}
\section{introduction}
MiniCLEAN detector holds $\sim$ 500 kg liquid argon located at Cube Hall in the SNOLab. If there's any unexpected situation e.g. overpressure or seismic events happened such that the insulation system failed. Huge amount of liquid argon will boil-off in very short time, these argon gases are heavier than oxygen, so that at the bottom of Cube Hall the oxygen will be repel and create oxygen deficiency hazards. A series of tests were perform to understand if the mitigation of oxygen-deficiency hazards in the Cube Hall was sufficient to handle the situation of upset conditions in MiniCLEAN. 
\section{Oxygen Sensors}
During the tests, the oxygen sensors used to monitoring the oxygen level in cube hall are from the SNOLab fixed oxygen sensors and a pair of portable logging sensor. The fixed sensors located in the Cube Hall are constantly monitoring the oxygen concentration, the value of sensors are read out by the SNOLAB Building Automation System (BAS). All the BAS and data are achived to the SNOLAB PI server which is read-only copy of the BAS Historian database. Before the test, we go through every sensors and check its functionality. A bag with pure nitrogen or argon covers the sensor until the sensor read less than 1\% oxygen. Subsequently, the bag and gas supply are removed quickly. The rate of decreasing oxygen level were then determined. Based on the order in which the sensors were read out the wiring and addressing was double checked to ensure data integrity. The readout of the fixed sensors is done using a 4-20 mA circuit analog output from the sensors. The time stamps of the measurement are from the BAS system.\par
A long-life zirconium oxide cell was used in the sensor with a range of 0 to 25\%. The device is sensitive to the airflow and be cooled by it results in inaccuracy. The ventilation system in SNOLAB changes the readout value by 0.1\%. The BAS database only readout the value when oxygen level changes, this is inconvenient for doing the data analysis. Therefore it was changed to readout the value every 10 seconds after the first tests. On the other hand, the portable sensors log data every 10 seconds. The readout software is from Dr$\ddot{a}$ger hardware and software. However, there's no a priori guarantee of clock synchronization between the PAC7000 units and the fixed sensors.
\section{Sensor Comparisons}
In oder to compare the sensors properly they need to be exposed to the same atmosphere with a various bumps to the oxygen level. The sensors were placed in a rectangular bucket of approximately 50-L volume and the top was covered except for a small feed through on the opposite side as shown in Fig. \ref{fig:O2sensorinbox}. The rectangular box was covered with a flexible plastic bag and sealed carefully on all sides with aluminum tape. Nitrogen gas was introduced through a long 1/4" tube with a vale and the digital display of the sensors were monitored throughout the test.\par
During the first comparison (Fig. \ref{fig:O2firsttest})the nitrogen gas was introduced quickly into the volume and then the room air allowed to mix back in slowly through the opening (occationally with fanning or tapping the cover). The oxygen concentration decreasing curve follows a near-perfect exponential  curve indicating the mixing in the volume.However, the portable sensors retuened to 20.9\% in a jump from 20.6\% or 20.7\%. Note that the fixed sensor remained at a slightly diminished value of 20.7\%. This diminished value is more credible given how the end of the test was carried out. A similar behavior was seen in the second test, the nitrogen gas was introduced at low concentrations. This time the nitrogen being added slowly and the portable sensors do not show a value below 20.9\% until they have reached 20.3\%. The portable sensors jump from 20.4\% on the recovery side of the test (Fig. \ref{fig:O2secondtest}).\par
To understand the effects comes from the time delay or purely a response function to charges in concentration, the experiment was repeated. This time we add the nitrogen more slowly than previous tests. Figure \ref{fig:O2thirdtest} shows the oxygen concentration as a function of time. Note that the small initial burst was introduced intentionally  to synchronize clocks between sensors. The same behavior was observed, the portable sensor do not show a value below 30.9\% until they have reached 20.3\%. Although the time scale for this test is much longer than the previous test, the jump in portable sensors still occurred at the same concentration.\par
In summary, several characteristic of the sensors response were observed :
\begin{itemize}
	\item The reading from portable PAC7000 and PureAire fixed sensors agree with each other for oxygen concentration belows 20.5\%.
	\item The clocks in the PAC7000 units are not guaranteed to be synced with the fixed units. Thus absolute time between the two PAC7000 units or between a PAC7000 unit and fixed sensor is not to be trusted unless the calibrated during that particular test.
	\item If the reading from portabl PAC700 reads 20.9\%,  it should be interpreted as reporting between 20.3\% and 20.9\%.
\end{itemize}

\begin{figure}[htbp]
\centering
\graphicspath{{./fig/appendix/}}
\includegraphics[scale=0.3]{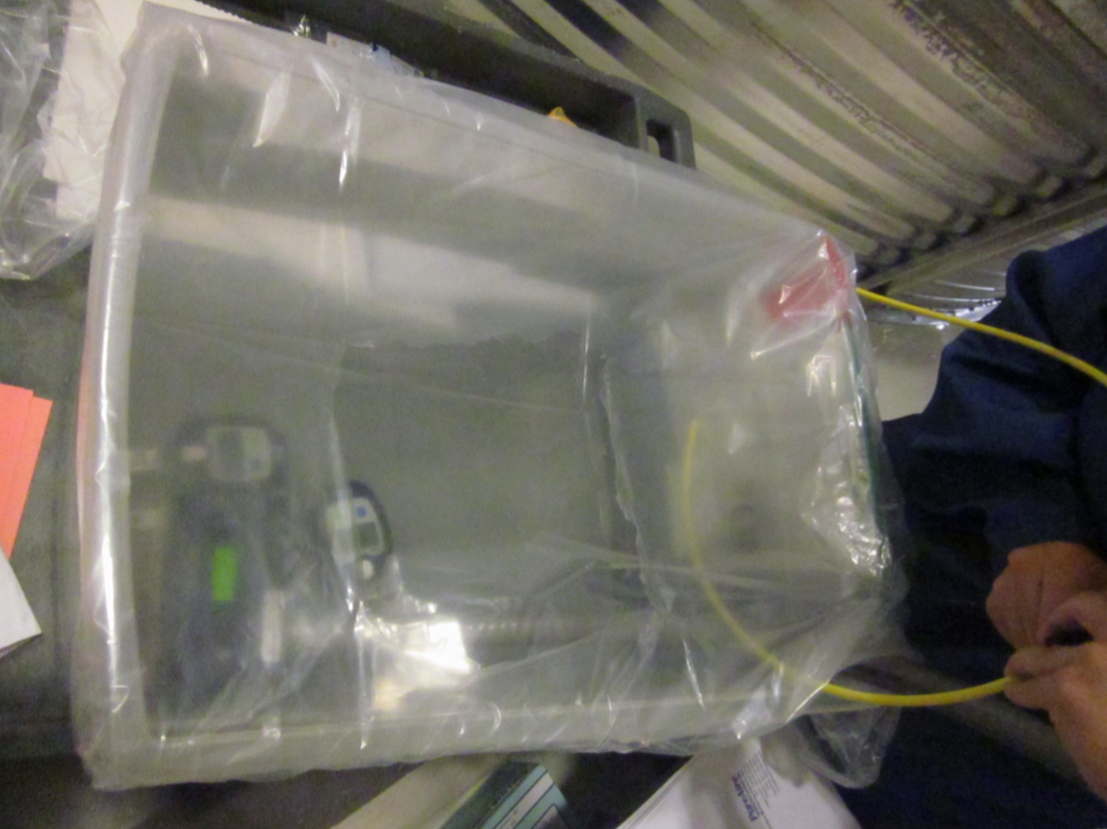}
\caption{ The setup for comparing the response of the oxygen sensors. }
\label{fig:O2sensorinbox}
\end{figure}
\begin{figure}[htbp]
\centering
\graphicspath{{./fig/appendix/}}
\includegraphics[scale=0.4]{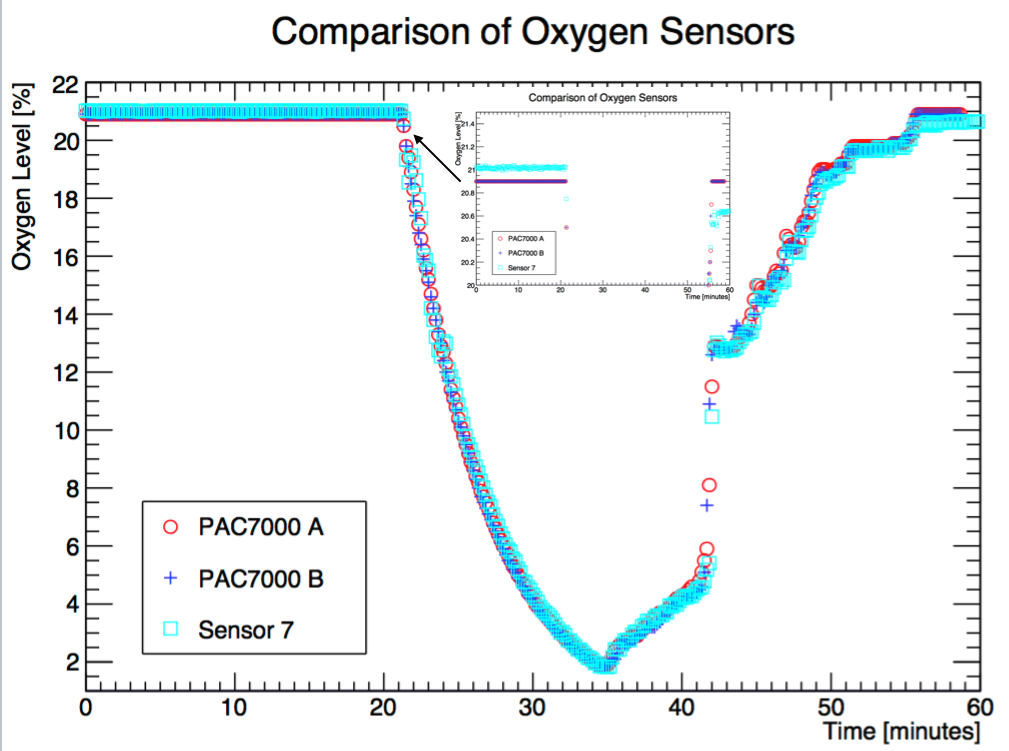}
\caption{ The first comparison of the oxygen sensors. The decline in oxygen concentration is a near-perfect exponential indicating mixing in the volume. The inset plot zoom in the sudden drop described in context. }
\label{fig:O2firsttest}
\end{figure}

\begin{figure}[htbp]
\centering
\graphicspath{{./fig/appendix/}}
\includegraphics[scale=0.3]{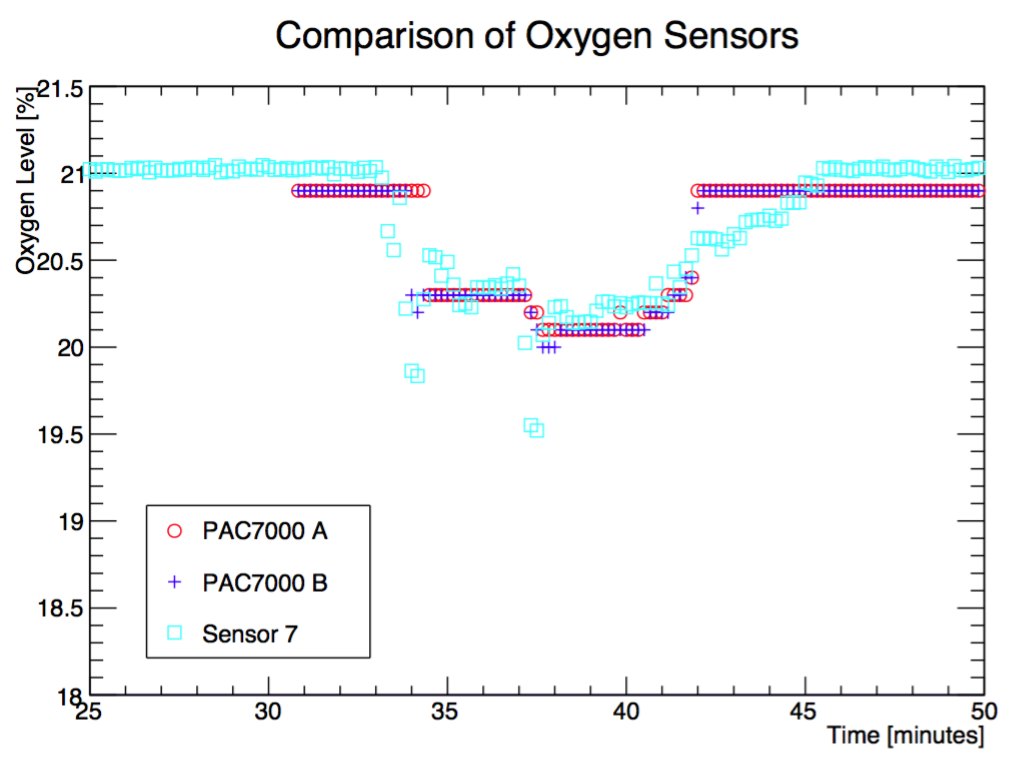}
\caption{ The second comparison of the oxygen sensors. }
\label{fig:O2secondtest}
\end{figure}
\begin{figure}[htbp]
\centering
\graphicspath{{./fig/appendix/}}
\includegraphics[scale=0.3]{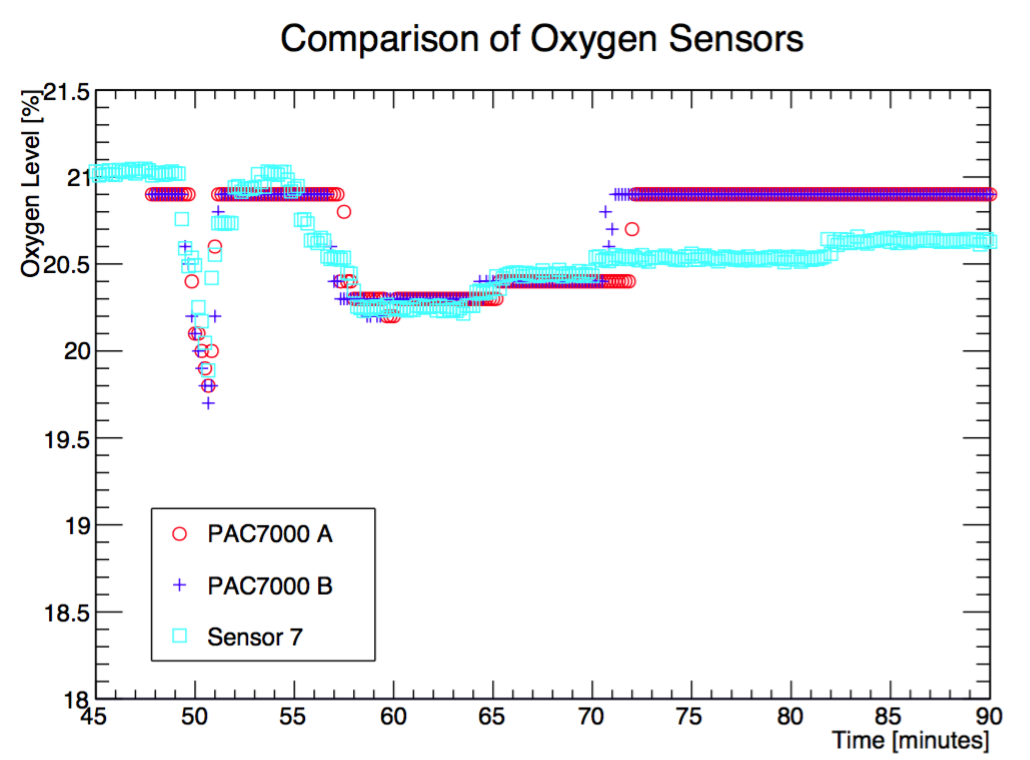}
\caption{ The third comparison of the oxygen sensors. }
\label{fig:O2thirdtest}
\end{figure}
\section{Test in Cube Hall -- Overview}
A series of tests carried out in the different location in Cube Hall to understand the impact of oxygen deficiency hazards.  Figure \ref{fig:cubehallv} shows the ventilation route and the volume of each area in the Cube Hall. First two tests were performed at the staging area which is at the top of cube hall. The third test was carried out on the deck which is on top of MiniCLEAN water tank. The rest of tests were carried out in various location at floor level. Table \ref{tab:O2sum} summarize the tests and Fig. \ref{fig:cubehallb} shows some of the location described above. These area will be defined as Oxygen Deficiency Hazards Area (ODHA) if the oxygen level is less than 19.5\%. However, SNOLab is 6,800 ft underground, the measured atmosphere pressure in SNOLAB is 17.3 psia (1.18 atm or 895 Torr). Thus under such pressure, if the oxygen level drops below 13.5\%, should people starts to have symptoms due to lack of oxygen. During the spill test, all personnel retreat to outside of Cube Hall for safety, and control the test through online system. The primary ventilation is from air handler 7 (at 14,000 cfm $\sim$ 6.608 $m^3/s$). Two additional fans (670 cfm) in staging area and three fans (5700 cfm) on the floor level were installed to improve the air flow. The staging area is a open space that connects to the top of the Cube Hall. No barrier between the staging area and the Cube Hall, therefore the air can mix with Nitrogen/Argon gas freely.
A 1000 cfm fan was installed in the staging area to help the air mixing. The Main Hall Access Drift (MHAD) connects the rest of SNOLAB to the staging/CubeHall. The air space are separated by fire doors and  the MHAD is at lower pressure than the Cube Hall, forcing the air through the unsealed door during the normal operation. This design is to prevent the contamination get into the Cube Hall from the outside environment. The Bottom Access Drift (BAD) connects to both the bottom of the Cube Hall and Cryopit. It is at lower pressure than   
Cube Hall and Cryopit as well thus the air from the Cube Hall/Cryopit moves through the fire doors into the BAD.

\begin{figure}[htbp]
\centering
\graphicspath{{./fig/appendix/}}
\includegraphics[scale=0.3]{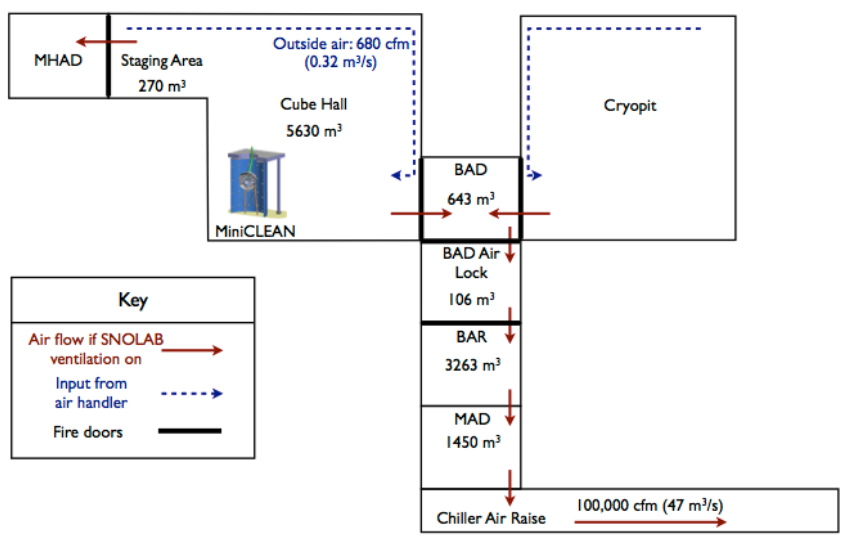}
\caption{ The ventilation path in Cube Hall. }
\label{fig:cubehallv}
\end{figure}


{\centering
\begin{tabular}{ C{0.9in} C{.9in} *4{C{1in}}}\toprule[1.5pt]
\bf Date & Liquid & Location & Ventilation On/Off & Boil-off rate \\\midrule
09/15/2014 & LN$_2$ & Staging area & On & 57 g/s\\\midrule
09/16/2014& LN$_2$ & Staging area  & Off & 70 g/s\\\midrule
09/18/2014 & LN$_2$ & Deck & Off & 65 g/s\\\midrule
09/23/2014 & LN$_2$ & Floor level & Off & < 1767 g/s\\\midrule
09/25/2014 & LN$_2$ & Floor level & Off & 68 g/s\\\midrule
10/02/2014 & LAr & Floor level & Off & 204 g/s\\\midrule
10/06/2014& LAr & Floor level  & Off & < 2185 g/s\\\bottomrule[1.25pt]
\end {tabular}
\captionof{table}{Brief summary of oxygen deficiency tests. } \label{tab:O2sum} 
}
\begin{figure}[htbp]
\centering
\graphicspath{{./fig/appendix/}}
\includegraphics[scale=0.3]{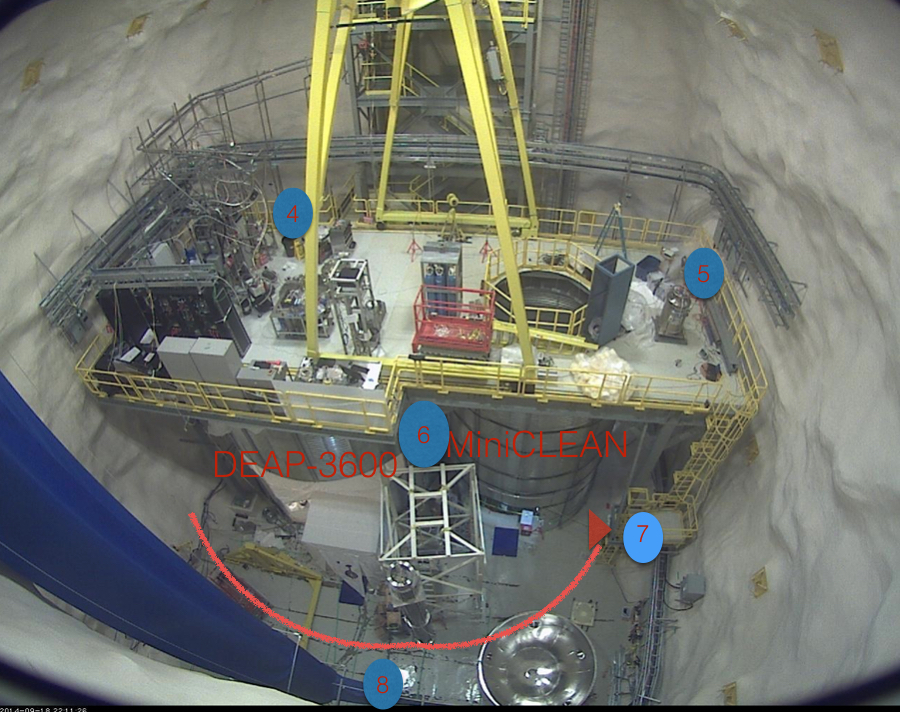}
\caption{ Bird's-eye view of Cube Hall. Numbers in the plot corresponds to the number of the fixed oxygen sensors described in the text. \ref{tab:O2sum}, and the red arrow indicates the direction of air flow. }
\label{fig:cubehallb}
\end{figure}
\section{First  and Second Test in Cube Hall : With Cube Hall Recirculation On}

The first test was performed on Sep. 15 2014. For the first test a spill was effected in the Cube hall staging area. The SNOLAB ventilation was on circulating air within the Cube Hall (drawing from the staging area and distributing along the east wall) at a rate of approximately 14000 cfm. The fresh-air intake into the Cube Hall was turned off for the test. The DEAP clean tent had its ventilation left on for the test. Thus pulled air from the floor of the cube hall behind the tanks near the argon dewar into the clean tent and out the door of the clean tent. The fans were all at floor level.\par
The total amount of nitrogen dispensed from the dewar was approximately 154kg. Of that approximately 50 kg accumulated in the catch basin and did not produce vapour during the test. (The amount that did not produce vapour is imprecise because the catch basin contained both liquid nitrogen and water ice. The total volume was estimated with a dip stick but the ice fraction was visually estimated to be 25\% of the material in the catch basin.) Thus the total boil off was approximately 105 kg. At STP this corresponds to 84 m3. The total volume of the Cube Hall (including the volume of the shielding tanks is approximately 5700 m3. Thus approximately 1.47\% of the air is displaced with pure nitrogen. Given that air is 20.9\% oxygen we expect the oxygen concentration to drop by 0.3\% on average.\par
The second test was carried out on Sep. 16 2014. The same procedure with first test was performed but with slight larger spill rate and the ventilation was off to test extreme situation.
\par
The spill test was carried out near the downstream side of sensor 3. Figure \ref{fig:testonescale} shows the average spill rate was 57 g/s. The portable sensors are brought close to the test site. One was approximately 30 cm off the floor and 2 meters high for the other. Both sensors shows 20.9\% of oxygen level which implies the true level of oxygen is at or above 20.4\%. Figure \ref{fig:ffirst_sensors} shows the reading from fixed oxygen sensors, note that the sensor 3 is the nearest sensor to the test site. The height of sensor 3 is approximately chest height and the spill was below that level. During the first test, the SNOLAB ventilation work as expected and average out the ODH throughout the Cube Hall. Each sensors just dropped between 0.25\% and 0.3 \% and is closed to expectations. The oxygen concentration at any place in the Cube Hall is always above 20\% except within 1 meter from the spill site.\par
In second test, the locations of sensors are the same with the first test. The oxygen concentration in Cube Hall except the staging area reads very similar readings with first test. However, for sensors located at staging area as shown in Fig. \ref{fig:fsecond_sensors} dropped below 20\%. This confirmed that the ventilation system help to push nitrogen to flow to some other space in Cube Hall and mixed with air. Without the ventilation on, the air takes longer time to mix with nitrogen in staging area and nitrogen did not fully mix with the air in the Cube Hall. Nevertheless, the oxygen level still far from the dangerous level (13.8\%), indicating that the personnel still would be safe to work in the area.

\begin{figure}[htbp]
\centering
\graphicspath{{./fig/appendix/}}
\includegraphics[scale=0.3]{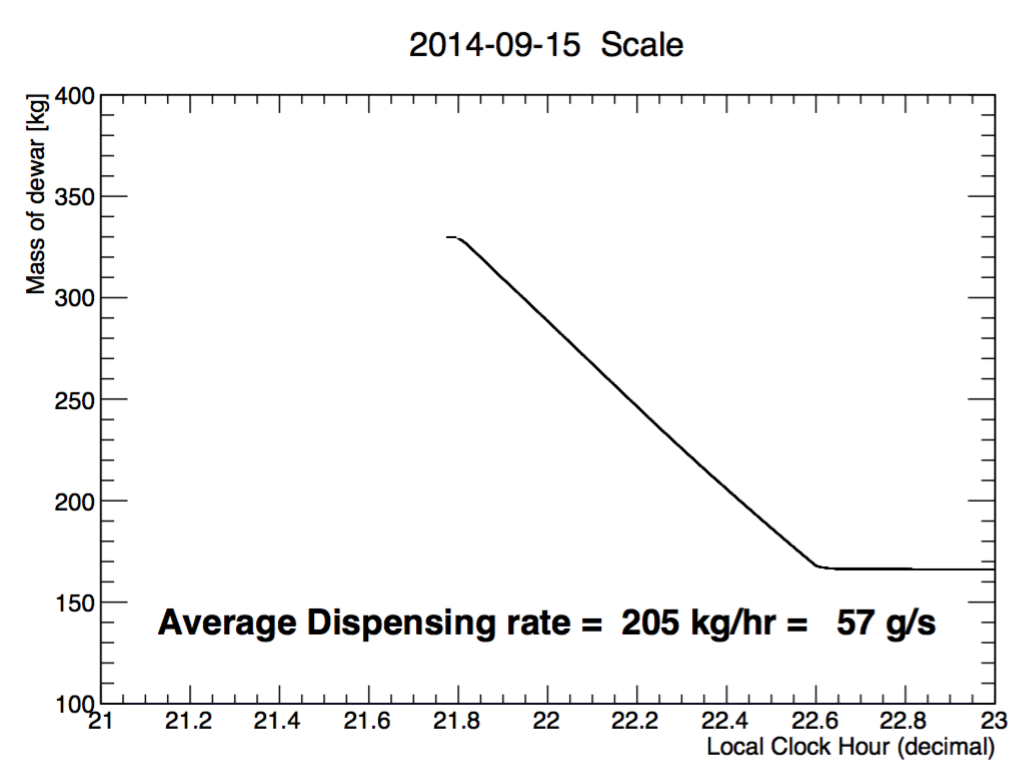}
\caption{The mass of the liquid-nitrogen dewar. The sloped shows a dispensing rate of 57 g/s. the effective vapor generation rate was lower as some liquid accumulated in the spill container as discussed in the text. The average vapor generation rate was approximately 40 g/s. }
\label{fig:testonescale}
\end{figure}

\begin{figure}[htbp]
\hfill
\subfloat[]{\includegraphics[width=7cm]{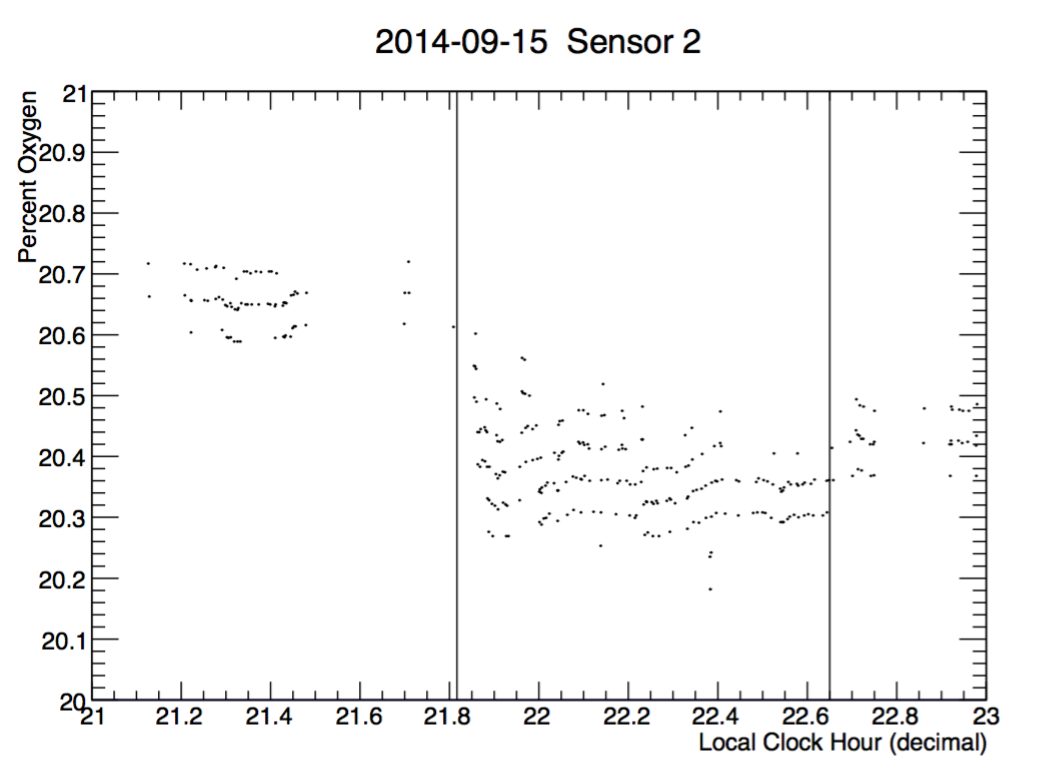}}
\hfill
\subfloat[]{\includegraphics[width=7cm]{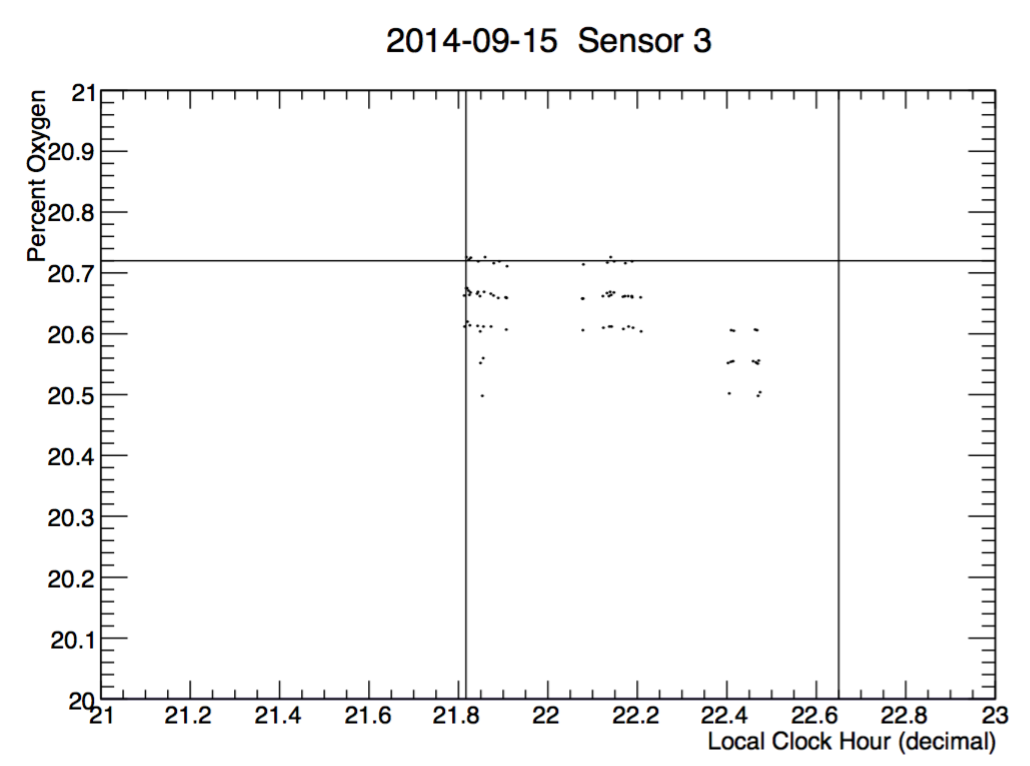}}
\hfill
\subfloat[]{\includegraphics[width=7cm]{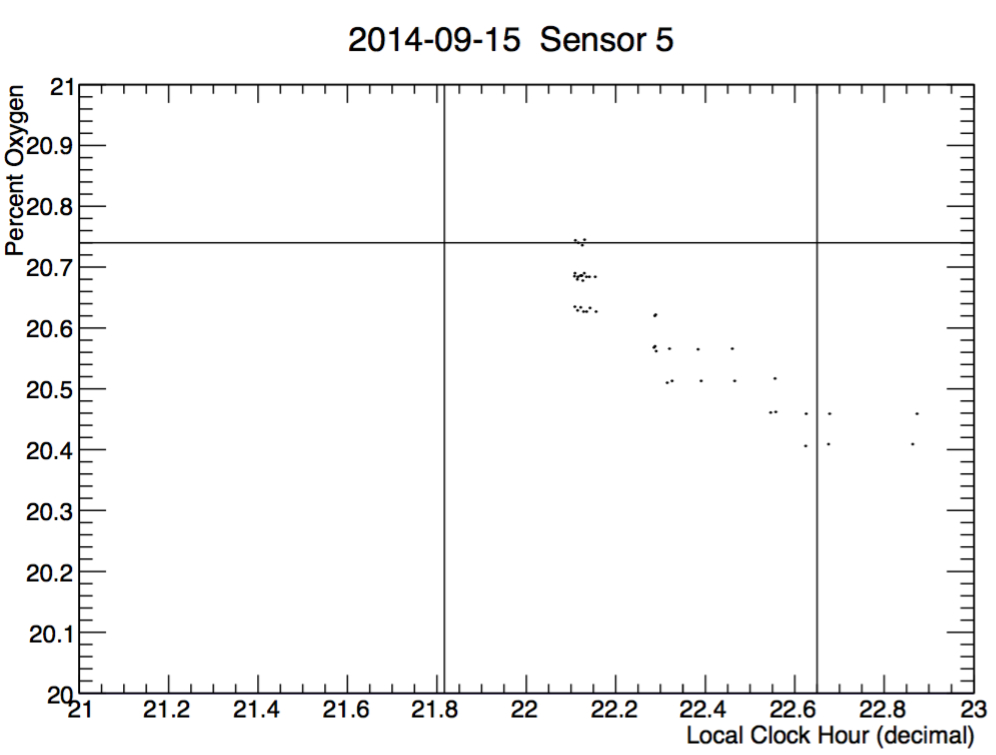}}
\hfill
\subfloat[]{\includegraphics[width=7cm]{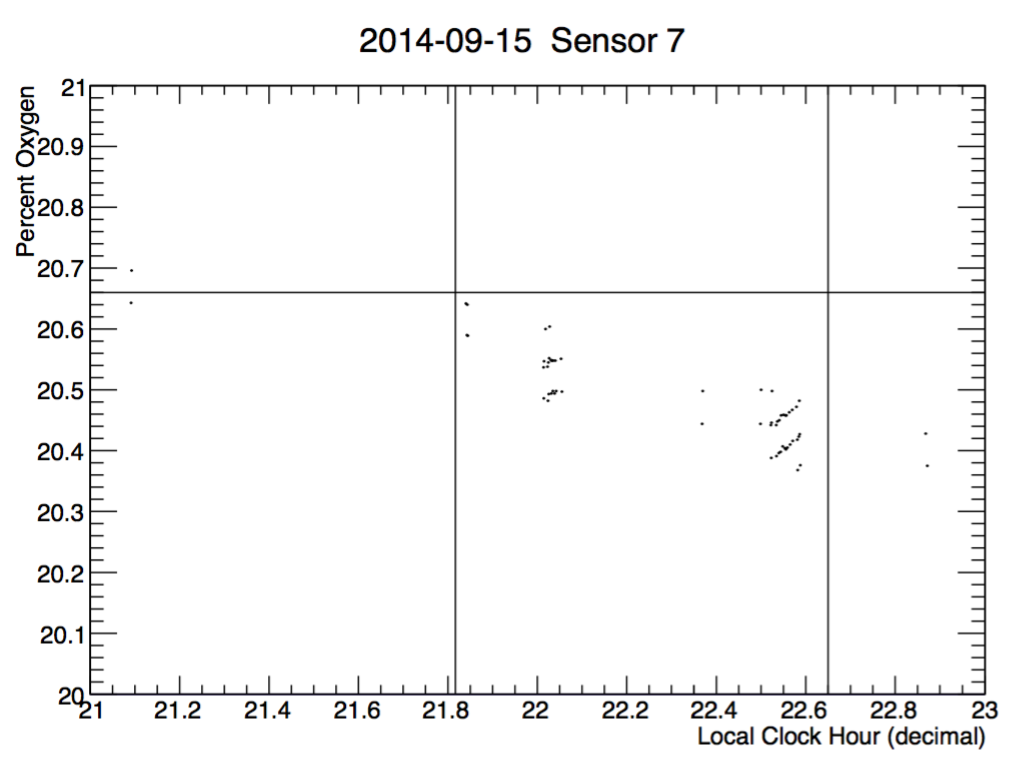}}
\hfill
\caption{Reading from different fixed oxygen sensors in Cube Hall from first test. Note the two vertical line indicates the start and end time of the spill test. }
\label{fig:ffirst_sensors}
\end{figure}
\begin{figure}[htbp]
\hfill
\subfloat[]{\includegraphics[width=7cm]{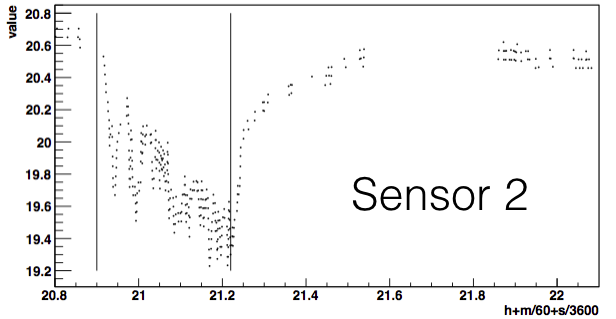}}
\hfill
\subfloat[]{\includegraphics[width=7cm]{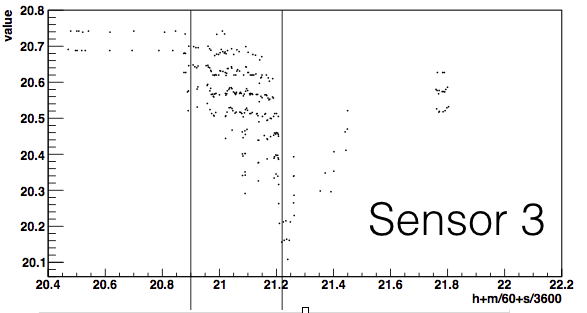}}
\hfill
\caption{Reading from different fixed oxygen sensors in Cube Hall from second test. Note the two vertical line indicates the start and end time of the spill test.}
\label{fig:fsecond_sensors}
\end{figure}
\section{Third Test in Cube Hall : With Cube Hall Recirculation On}

The third test was carried out on the MinCLEAN side of the Cube Hall deck. The SNOLAB ventilation system was off but ODH fans were on except the fan near the MiniCLEAN which is uninstalled at the moment. The liquid nitrogen was decanted from a 230 L dewar as quickly as possible. The portable oxygen sensors were installed in two locations : one approximately 5 feet above the deck and 1 foot above the deck for the other. Both sensors were placed near the spill site with approximately 1.5 meters from it. Soon after the test starts, one irregularity was found. Through the camera which installed for monitoring the test, found that some liquid nitrogen was missing the spill container. The test was halted for a few minutes and a block was put in place to direct the spill into the bucket. To add heat to the nitrogen a peristaltic pump dripped water at a 240 mL/minute onto the stream of nitrogen from the hose. There was no accumulation of nitrogen in the spill bucket.\par

The spill rate is about 70 g/s as shown in Fig. \ref{fig:testtwoscale}. Figure \ref{fig:fthird_sensors} shows the sensors which has large changes. Note that the sensor 5 is the nearest sensor to the spill site but has only a alight reaction to the test. Instead, the first response is from sensor 7 which is located on a post supporting the deck approximately 4 feet above the floor at the corner below the spill site. This may due to the air flow forcing by the ODH fans. During the test, no sensors has dropped below 20 \%.


\begin{figure}[htbp]
\centering
\graphicspath{{./fig/appendix/}}
\includegraphics[scale=0.3]{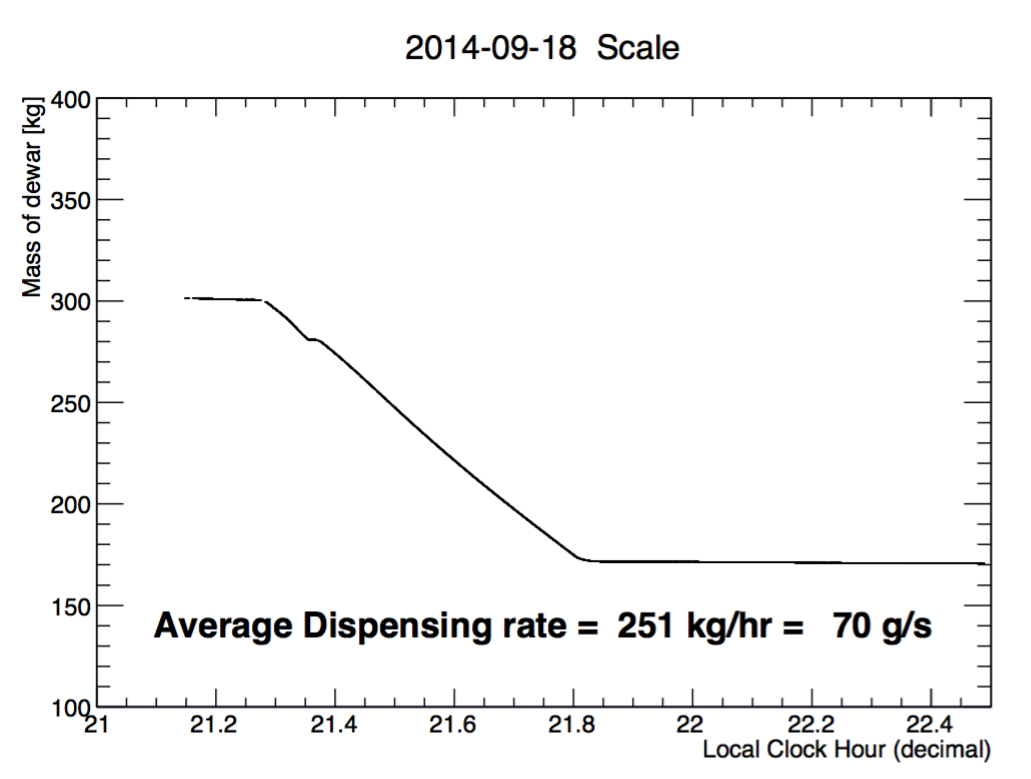}
\caption{The spill profile for this test. Note the plateau just before 21.4 hours was a genuine short pause to fix a problem. The spill rate shown does not include that pause in the calculation. }
\label{fig:testtwoscale}
\end{figure}
\begin{figure}[htbp]
\hfill
\subfloat[]{\includegraphics[width=7cm]{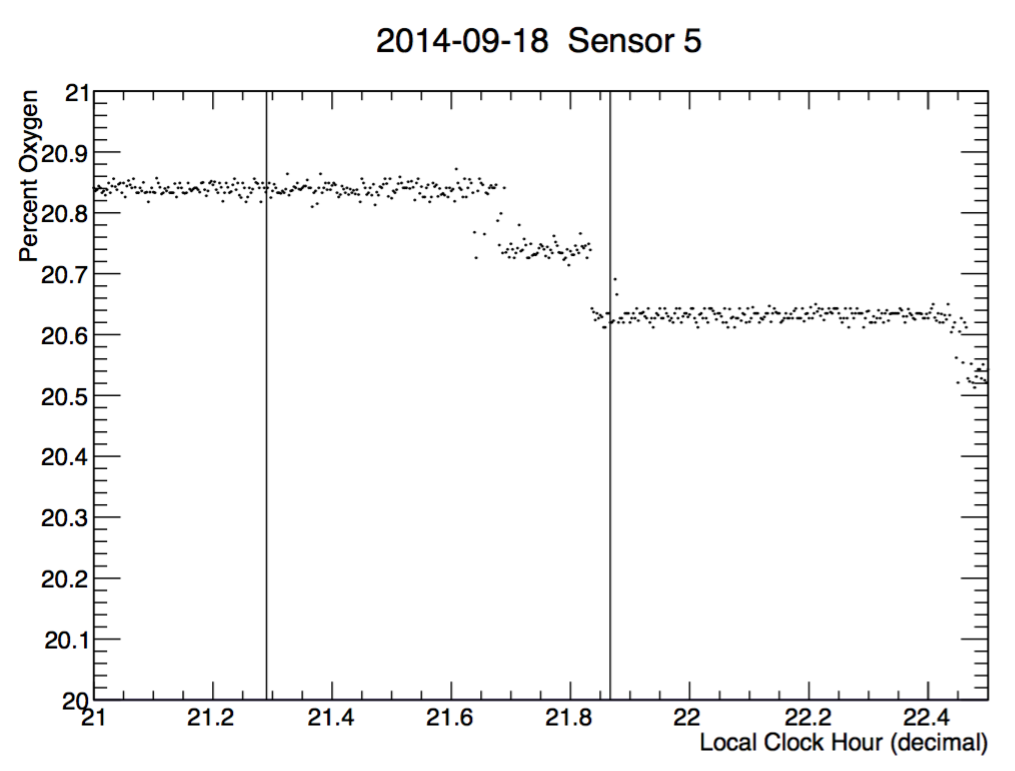}}
\hfill
\subfloat[]{\includegraphics[width=7cm]{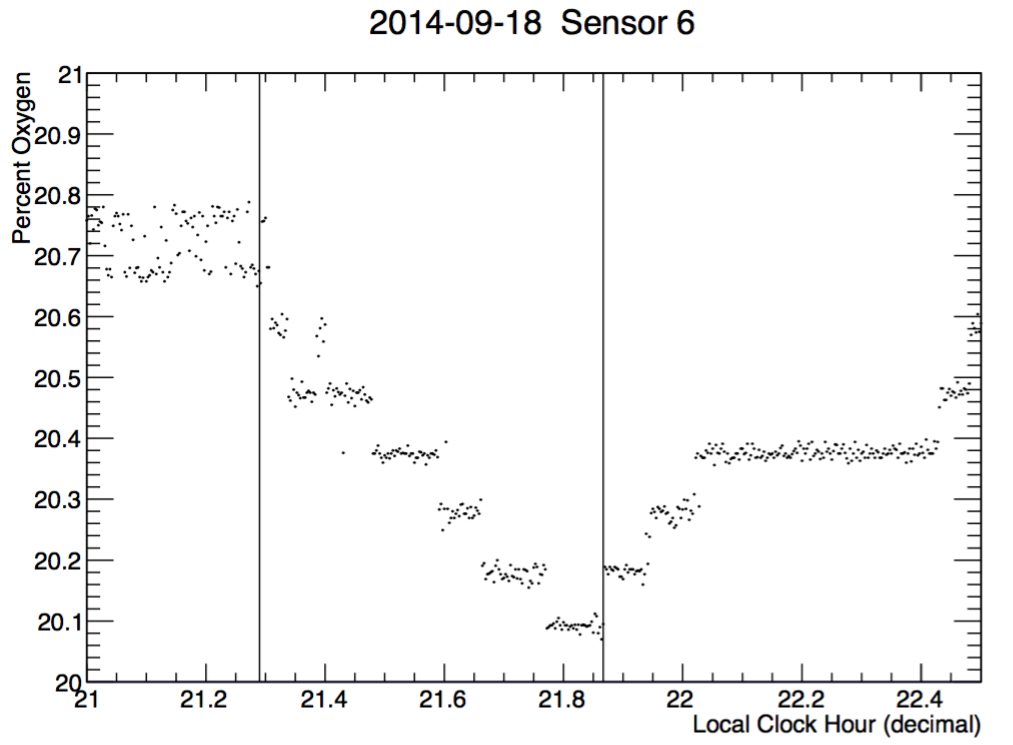}}
\hfill
\subfloat[]{\includegraphics[width=7cm]{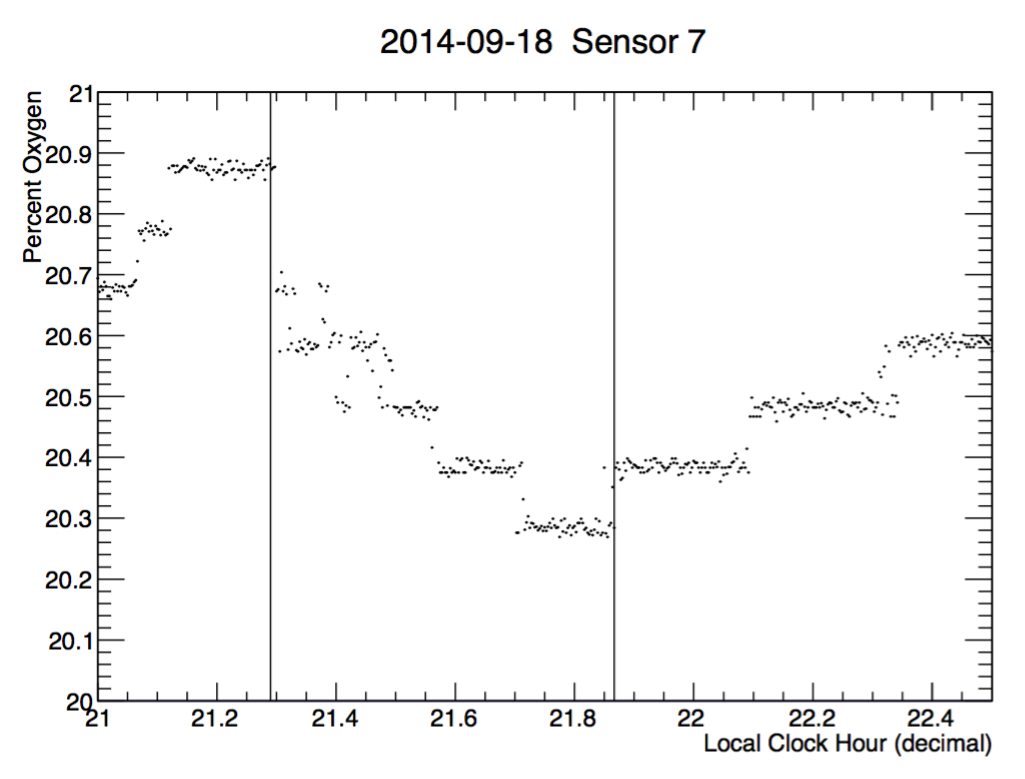}}
\hfill
\subfloat[]{\includegraphics[width=7cm]{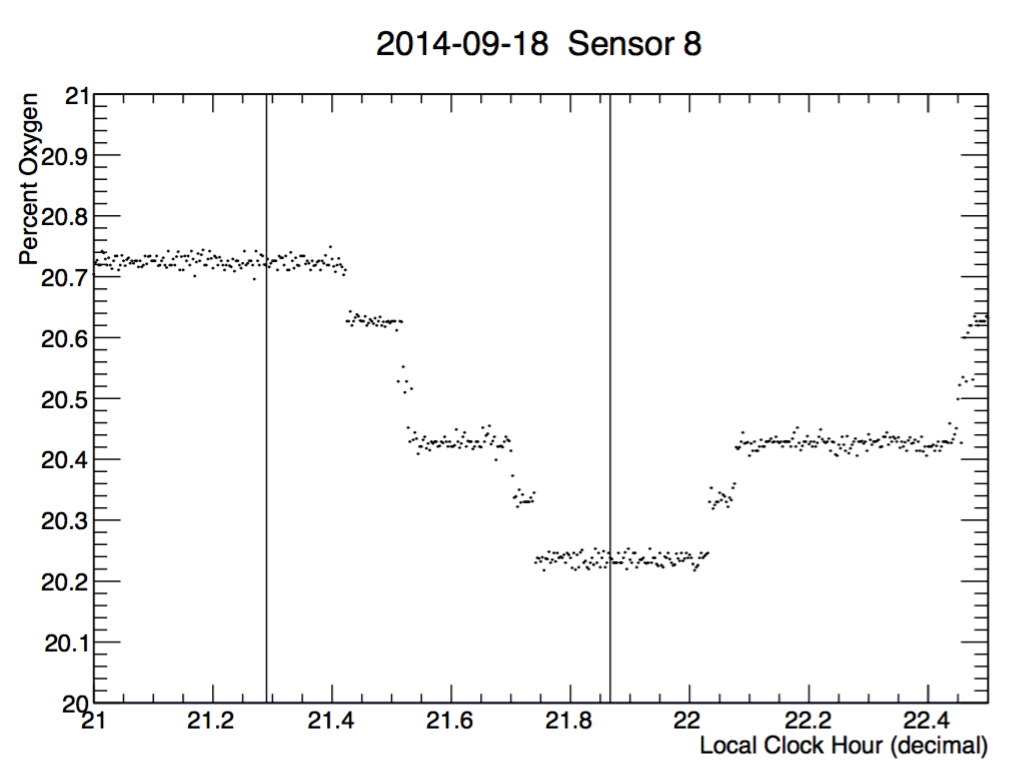}}
\hfill
\caption{Reading from different fixed oxygen sensors in Cube Hall from third test. Note the two vertical line indicates the start and end time of the spill test.}
\label{fig:fthird_sensors}
\end{figure}
\section{Fourth Test in Cube Hall : With Cube Hall Recirculation Off}
In this test, more rapid spill rate was test on the floor level of Cube Hall. The 106 kg of liquid nitrogen was decanted into an open container. A volume of water was placed in a nearby container and pumped into the nitrogen at > 2 liter/s causing the nitrogen to boil-off in less than a minutes. During the test, the SNOLAB ventilation was off. Two of the three large Cube Hall ODH fans were on as were one ODH fan on deck, one at the argon dewar and one at the staging area. Figure \ref{fig:tesfourscale} shows the setup of the test.\par
The three sensors on the Cube Hall floor are 6, 7, and 8. Note that 8 is closest to the spill and shows a large effect. Note that in fact sensor 6 showed a signal before sensor 7 despite the general sense of the air flow suggesting 7 would see the event first. However, the clean tent provided an almost straight-line path for gas to go from the spill site to sensor 6. (The door of the clean tent on the spill side is open and at the back of the clean tent there is a large area of fans with HEPA filters.) The plot of these sensors are shown in Fig \ref{fig:ffourth_sensors}. After approximately 10 minutes, all sensors went back to above 20\% of oxygen concentration.
\begin{figure}[htbp]
\centering
\graphicspath{{./fig/appendix/}}
\includegraphics[scale=0.3]{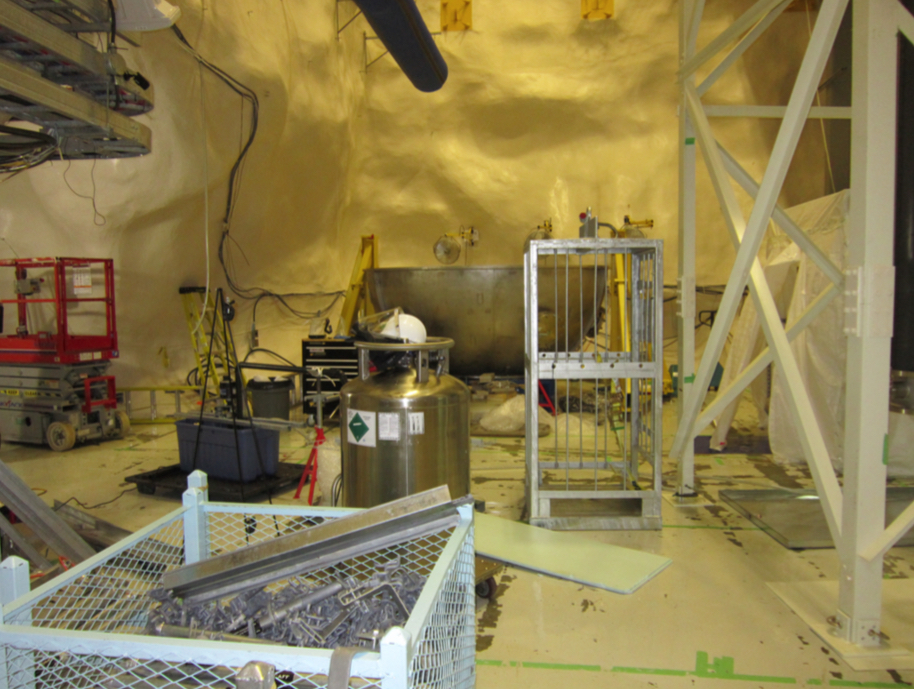}
\caption{The floor of the Cube Hall at the time of the test. Note that the steel shell impeded air flow on the floor. Each of the three fans behind the steel shell on the wall are rated to 5700 cfm. Two were on for this test. The liquid nitrogen was displaced into an insulated bucket behind the liquid-nitrogen dewar. The portable air sensors are mounted on the tripod, one about 1 foot off the floor and the other about 5 feet off the floor. }
\label{fig:tesfourscale}
\end{figure}
\begin{figure}[htbp]
\hfill
\subfloat[]{\includegraphics[width=7cm]{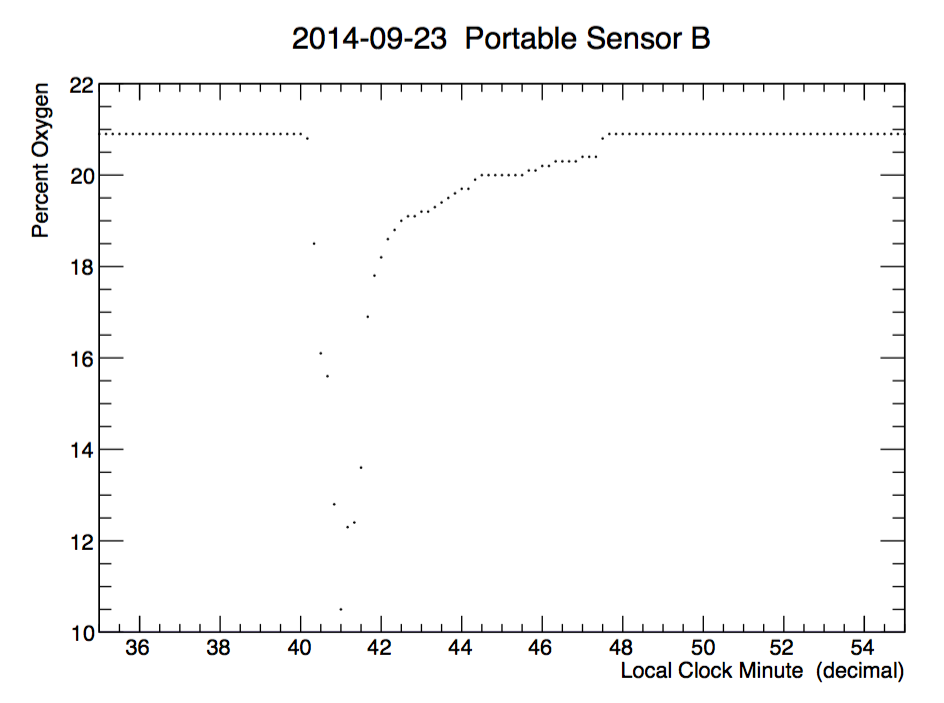}}
\hfill
\subfloat[]{\includegraphics[width=7cm]{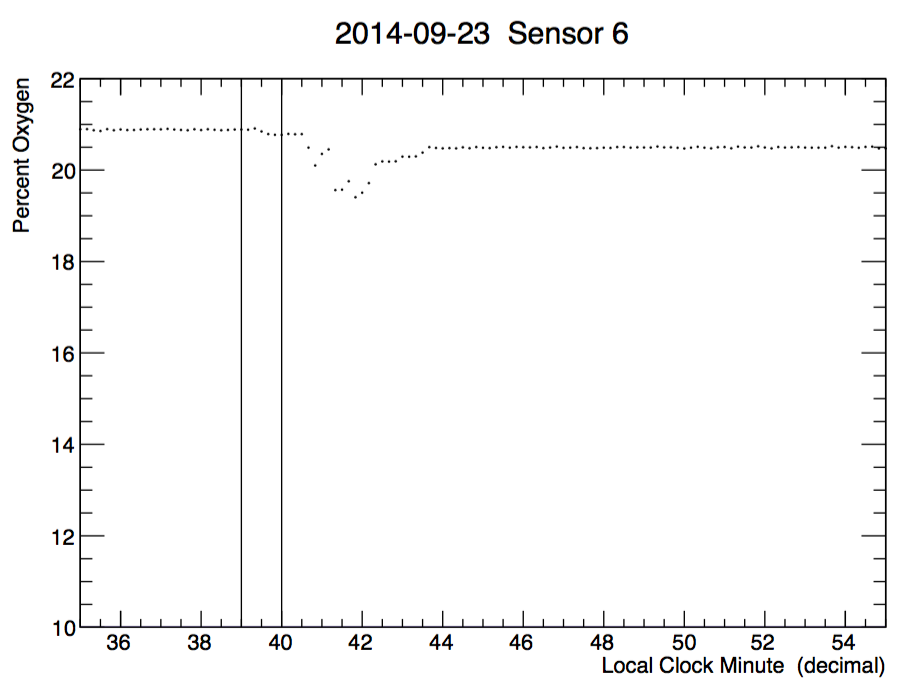}}
\hfill
\subfloat[]{\includegraphics[width=7cm]{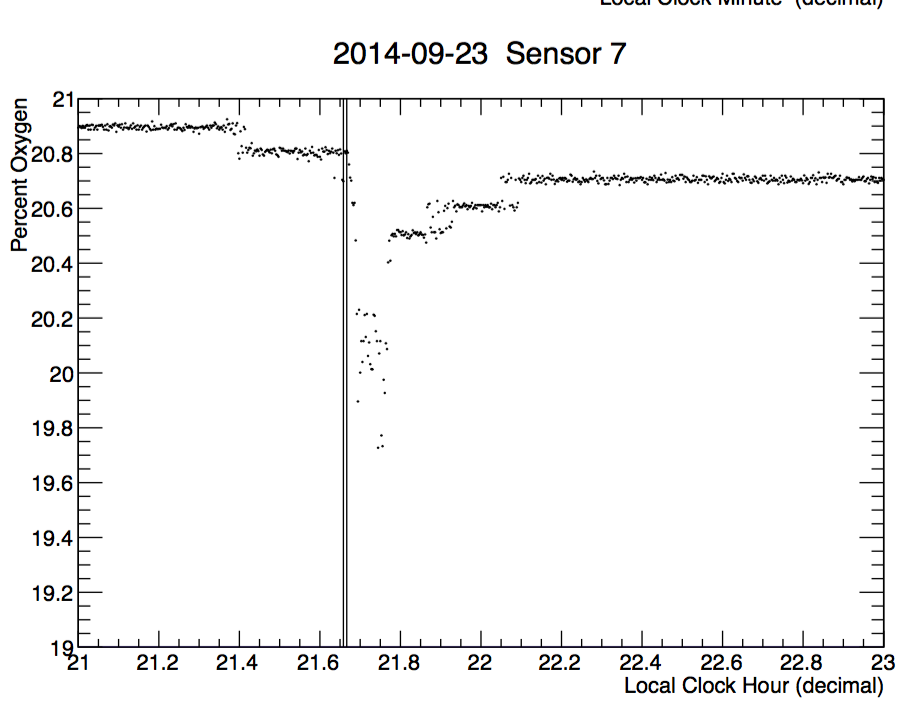}}
\hfill
\subfloat[]{\includegraphics[width=7cm]{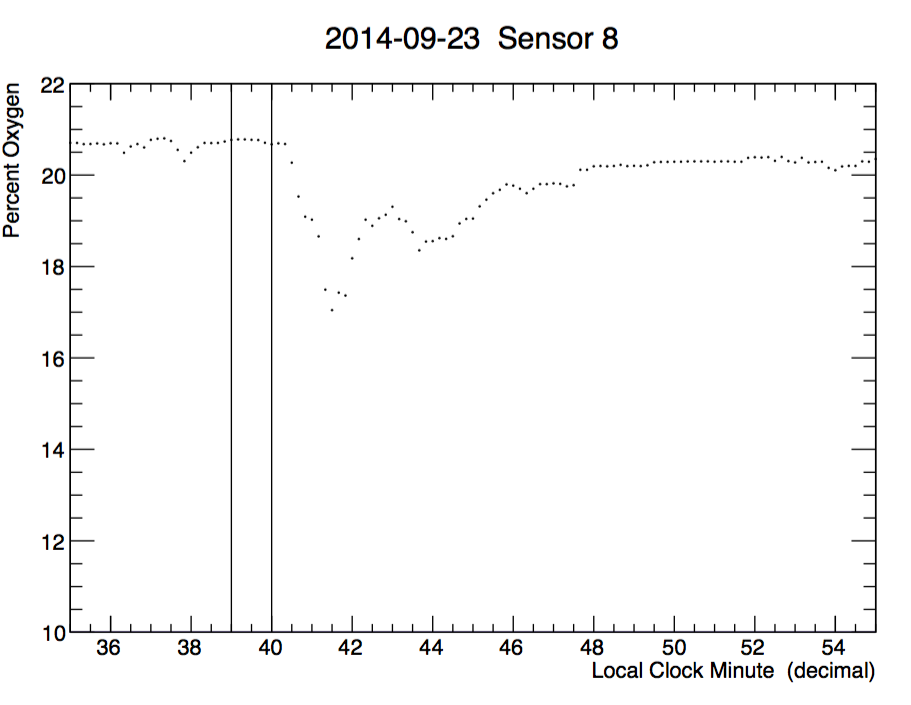}}
\hfill
\caption{The portable air sensor B located approximately 1.5 meters from the spill and approximately 1 foot off the floor. The approximate time of the spill is given by the vertical lines.}
\label{fig:ffourth_sensors}
\end{figure}
\section{Fifth and Sixth Test in Cube Hall : With Cube Hall Recirculation Off}
The fifth test was carried out on Sep 25. 2014. Th location of the test is on floor level and right behind the DEAP water tank as shown in Fig. \ref{fig:tesfifthscale}. The procedure is the same as fourth test. The LN$_2$ spill rate is at 68 g/s. More ODH fans were installed on floor level of Cube Hall to help the air mixing. The oxygen level of sensors are well above 20.5\% throughout the test except the portable sensors just next to the spill site. The readings of fixed sensors is shown in Fig. \ref{fig:fifthtestsensors}. \par
The sixth test is the same with fifth, only this time we were using LAr and spill with faster rate ($\sim$ 204 g/s). The test was carried out on Oct. 2 2014. No sensors dropped below 13.8 \%. However, with faster spill rate, the mixing of the air is slower than previous test. Figure \ref{fig:fsisxth_sensors} shows the reading from various sensors. Note that the vertical mixing didn't take place until the ventilation came back on. Nonetheless, the oxygen level still well above 18 \%.

\begin{figure}[htbp]
\centering
\graphicspath{{./fig/appendix/}}
\includegraphics[scale=0.3]{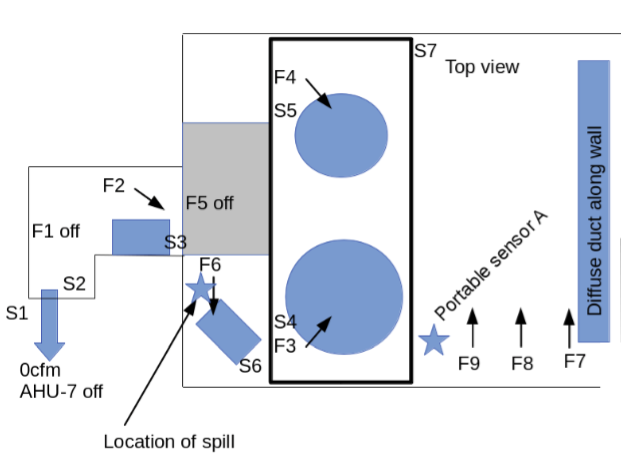}
\caption{The location of fifth and sixth test. }
\label{fig:tesfifthscale}
\end{figure}
\begin{figure}[htbp]
\centering
\graphicspath{{./fig/appendix/}}
\includegraphics[scale=0.3]{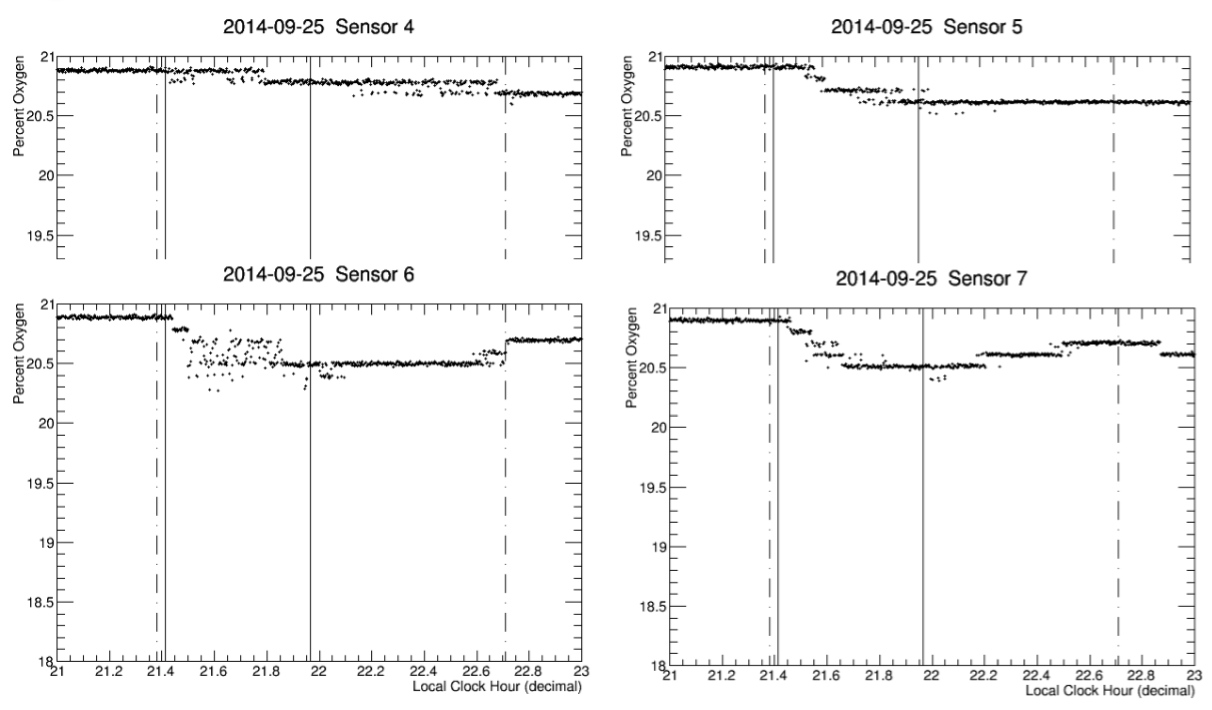}
\caption{Readings from fixed sensors in fifth test. The black line indicates the start and end time of the test. The black dashed line indicates the time the ventilation was turned off and on.}
\label{fig:fifthtestsensors}
\end{figure}
\begin{figure}[htbp]
\hfill
\subfloat[]{\includegraphics[width=7cm]{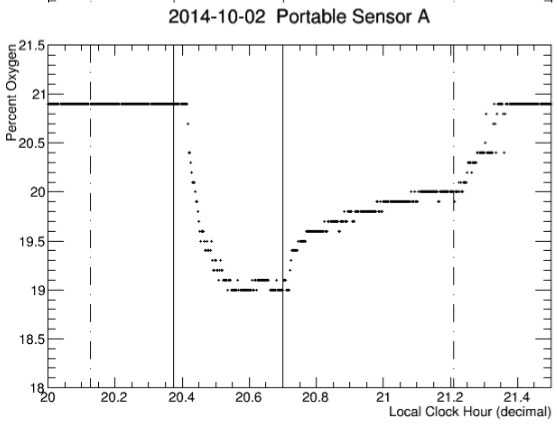}}
\hfill
\subfloat[]{\includegraphics[width=7cm]{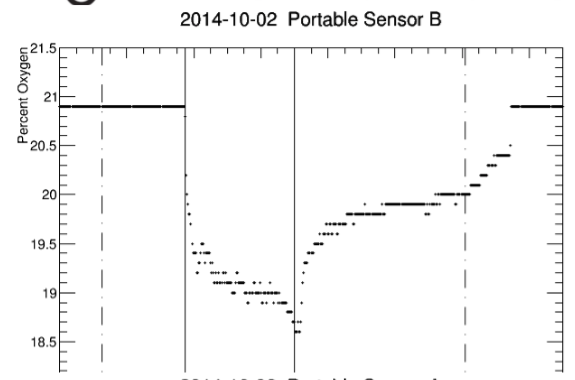}}
\hfill
\subfloat[]{\includegraphics[width=7cm]{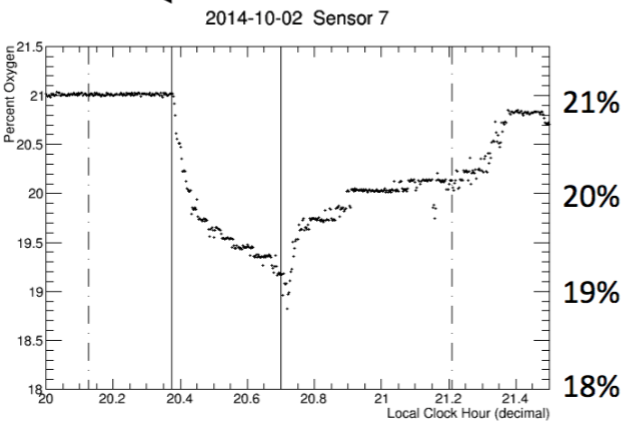}}
\hfill
\subfloat[]{\includegraphics[width=7cm]{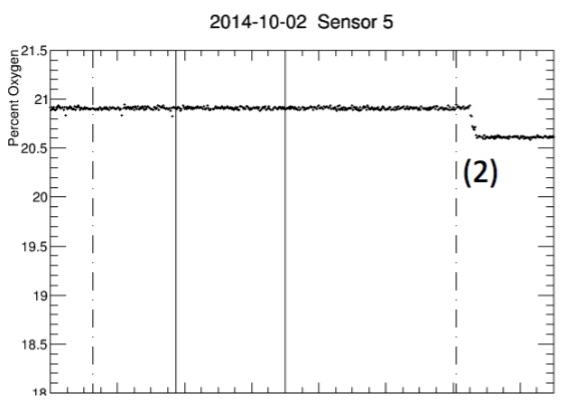}}
\hfill
\caption{Readings from the sensors in sixth test. Note the (2) region on sensor 5 indicates the ventilation is on and the vertical mixing starts.}
\label{fig:fsisxth_sensors}
\end{figure}
\section{Seventh Test in Cube Hall : With Cube Hall Recirculation Off}
The seventh test take place on the floor level and right on front of MiniCLEAN water tank as shown in Fig. \ref{fig:fsevenlocation}. This test is to simulate the worst case scenario when huge amount of LAr boil-off very rapidly ($\sim$ 2185 g/s). Note that the difference between test the the worst case of scenario is that the area and volume of the Cube Hall is larger in the test because the water tank is open. Moreover, the gas was produced at the floor level instead of just below the deck. The sensors on the floor level shows that the initial fast mixing on less than one minute time scale. However, the stratification of argon prevents complete mixing in the Cube Hall thus the hazardous condition will develop over time. The readings of the sensor as a function of time are shown in Fig. \ref{fig:fseven_sensors}.

\begin{figure}[htbp]
\hfill
\subfloat[]{\includegraphics[width=7cm]{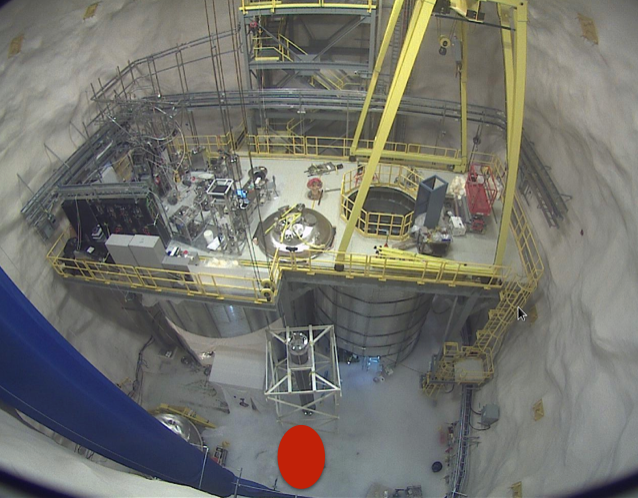}}
\hfill
\subfloat[]{\includegraphics[width=7cm]{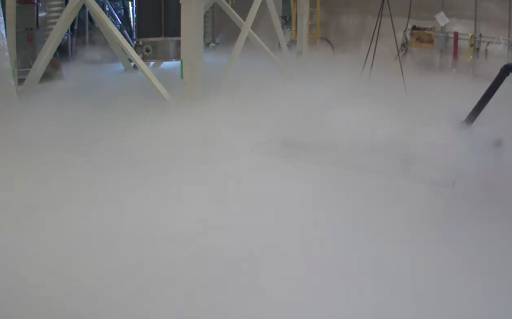}}
\hfill
\caption{(a) Location of seventh test. the red dot showing the exact location the seventh test was carried out. (b) Th picture taken during the test.}
\label{fig:fsevenlocation}
\end{figure}

\begin{figure}[htbp]
\hfill
\subfloat[]{\includegraphics[width=7cm]{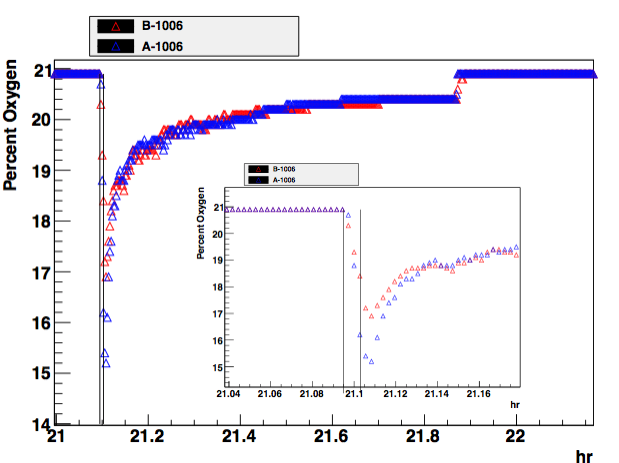}}
\hfill
\subfloat[]{\includegraphics[width=7cm]{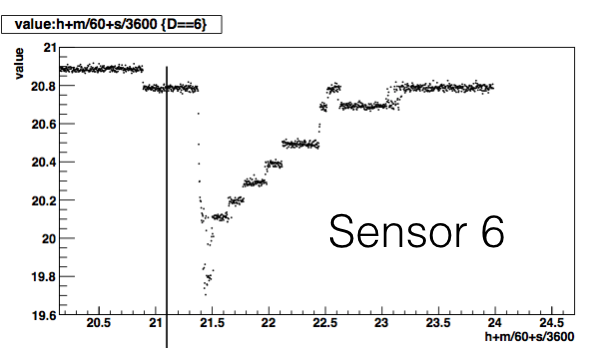}}
\hfill
\subfloat[]{\includegraphics[width=7cm]{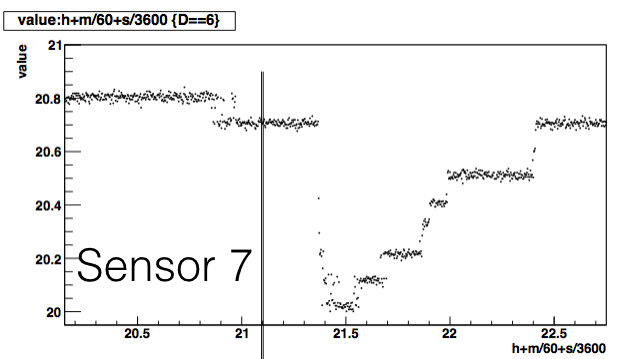}}
\hfill
\subfloat[]{\includegraphics[width=7cm]{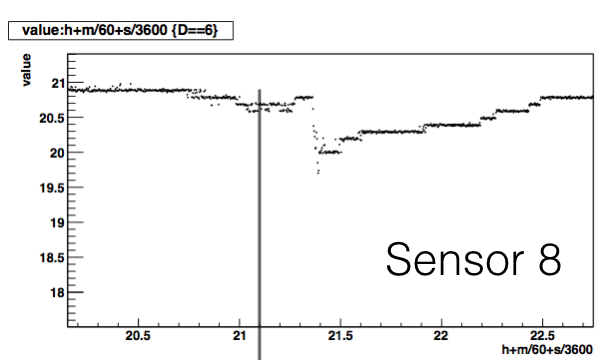}}
\hfill
\caption{Readings from the sensors in seventh test. Note the two vertical black line indicates the start and end time of the spill test and the dashed lines indicates the ventilation On/Off.}
\label{fig:fseven_sensors}
\end{figure}

\section{Conclusion}
After a series of tests, it reveal the time of mixing with air needs to be improved. The ODH fans (5000 cfm) were added in the Cube Hall to help the vertical mixing and mitigate the oxygen deficiency hazards. Therefore in any scenario, the estimated fatalities per hour is less than 10$^{-7}$ with the current setup of Cube Hall. The work rules under ODH condition is work out and meet the safety requirement of SNOLAB. 
\chapter{Waveform Reduction in RAT}\label{app:reduction}
\section{Overview}
In light of dedicated $Ar_{39}$ spike runs, the data reduction level is crucial for DAQ system to handle the event trigger rate over 500 Hz. The two basic data reduction level set by hardware (WFD) are : zero-supressed waveform and full waveform which can just be set on run by run basis. Then the DAQ pass theses events along to data reduction PC for further reduction on event by event basis. In DCDAQ, the DecisionMaker may call a piece of code in RAT called Software Trigger to assign reduction level based on three criteria. 
\begin{enumerate}
\item Total ADC counts in the event 
\item (An estimate of) fprompt
\item Maximum charge seen in a single PMT 
\end{enumerate}
Threshold for each of these quantities can be set in either DAQ.ratdb or during the run configuration in the macro. It is also possible to configure a run with all events set to the same reduction level. In the end of process, each event will has a "data type" integer that identifies its reduction level. Each data reduction level is briefly  described below.\\
\begin{enumerate}
\item Zero-supressed waveform :
	\begin{itemize}
		\item Reduction Level : : \texttt{CHAN\_ZLE\_WAVEFORM} = 2.
		\item Saves full ADC counts within each '"ZLE window".
		\item Data consists of Waveform Blocks (one for each ZLE window).
		\item Each Waveform Block contains : 
			\begin{itemize}
				\item Start time ( \texttt{uint32\_t})
				\item End time (\texttt{uint32\_t})
				\item ADC values (\texttt{uint16\_t})
				\item Example waveform : Fig. \ref{fig:zle} (a)
			\end{itemize}
	\end{itemize}

\item ZLE Integral Reduction Level
	\begin{itemize}
		\item Reduction Level : : \texttt{CHAN\_ZLE\_INTEGRAL} = 3.
		\item Saves full ADC integral for each '"ZLE window".
		\item Data consists of Integral Blocks.
		\item Each Integral Block contains : 
			\begin{itemize}
				\item Start time ( \texttt{uint16\_t})
				\item Width (\texttt{uint8\_t})
				\item Integral (float)
				\item Example waveform : \ref{fig:zle} (b)
			\end{itemize}
	\end{itemize}
	
\item Prompt/Total Reduction Level
	\begin{itemize}
		\item Reduction Level : : \texttt{CHAN\_PROMPT\_TOTAL} = 4.
		\item Saves ADC integral for prompt region and whole event window.
		\item Data consists of PromptTotal Blocks.
		\item Each PromptTotal Block contains : 
			\begin{itemize}
				\item Prompt integral (float)
				\item Total integral (float)
				\item Example waveform : Fig. \ref{fig:zle} (c)
			\end{itemize}
	\end{itemize}

\item Full Waveforms
	\begin{itemize}
		\item This is not a reduction level; it is a WFD setting.
		\item Not done on an event by event basis -- A whole run must be zero-supressed or not.
		\item Data consists of PromptTotal Blocks.
		\item The whole event window will be treated as one big ZLE window by DCDAQ.
		\item Example Waveform : Fig. \ref{fig:zle} (d)
	\end{itemize}
	
\end{enumerate}
\begin{figure}[htbp]
\hfill
\subfloat[]{\includegraphics[width=7cm]{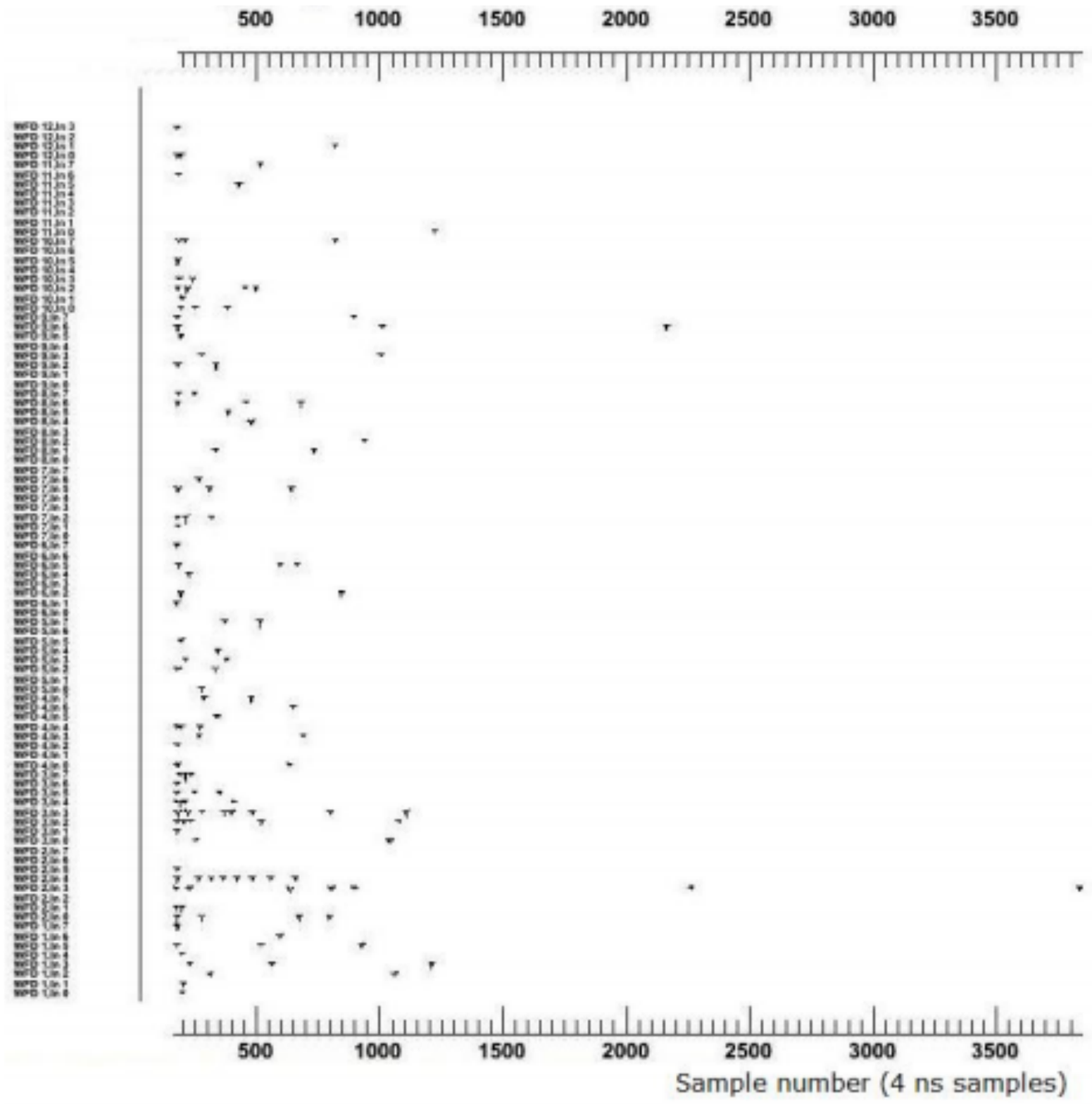}}
\hfill
\subfloat[]{\includegraphics[width=7cm]{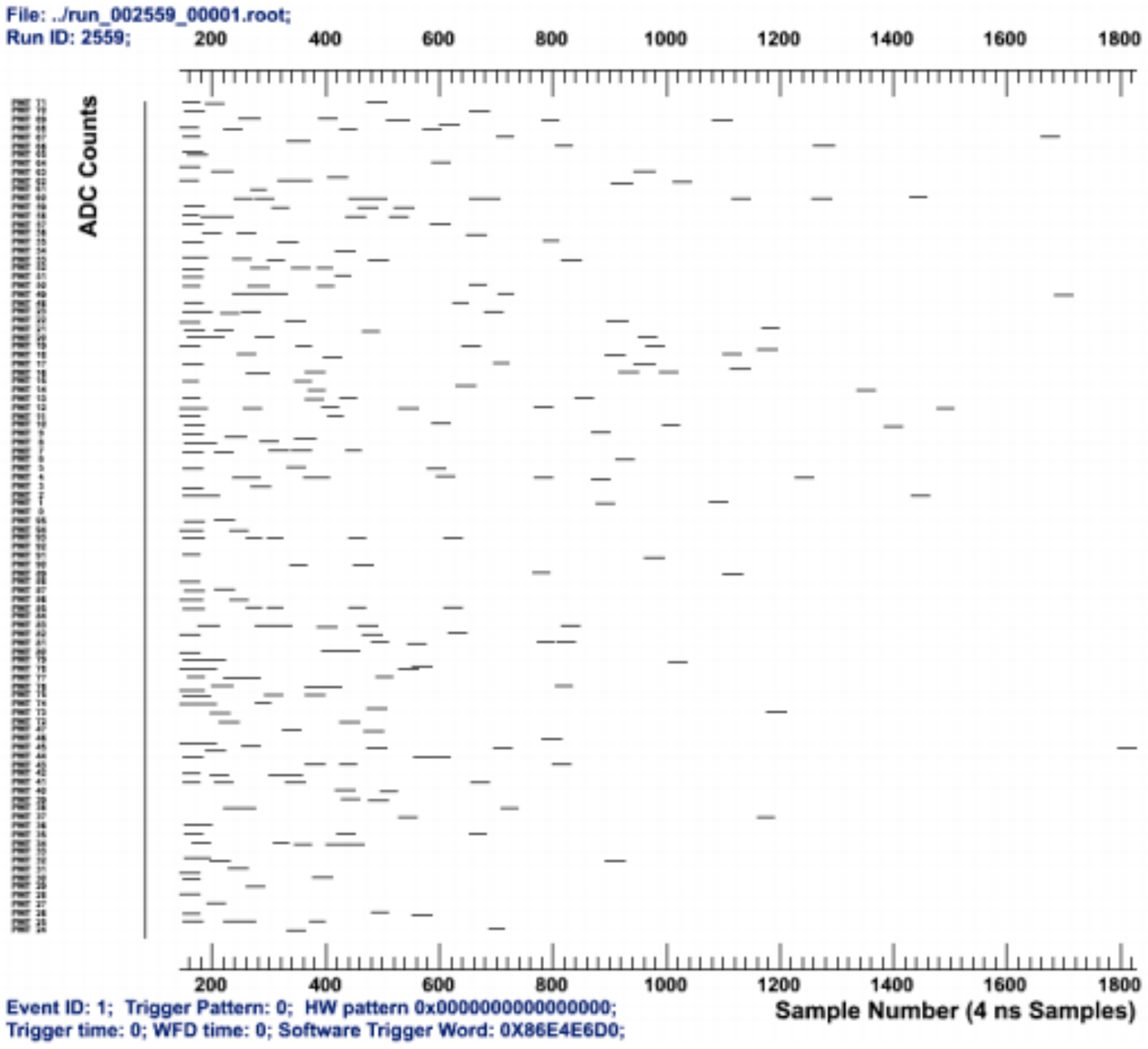}}
\hfill
\subfloat[]{\includegraphics[width=7cm]{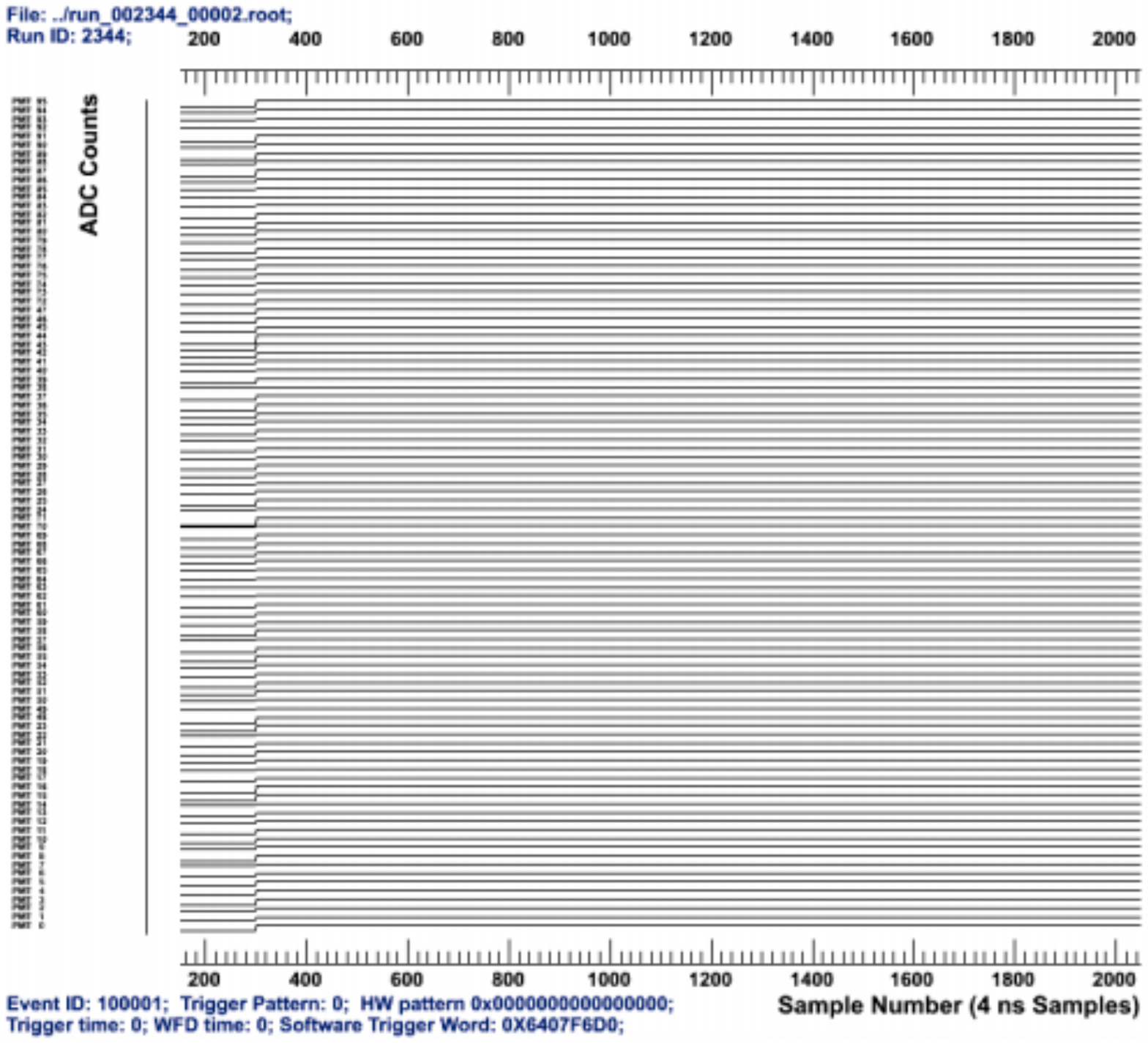}}
\hfill
\subfloat[]{\includegraphics[width=7cm]{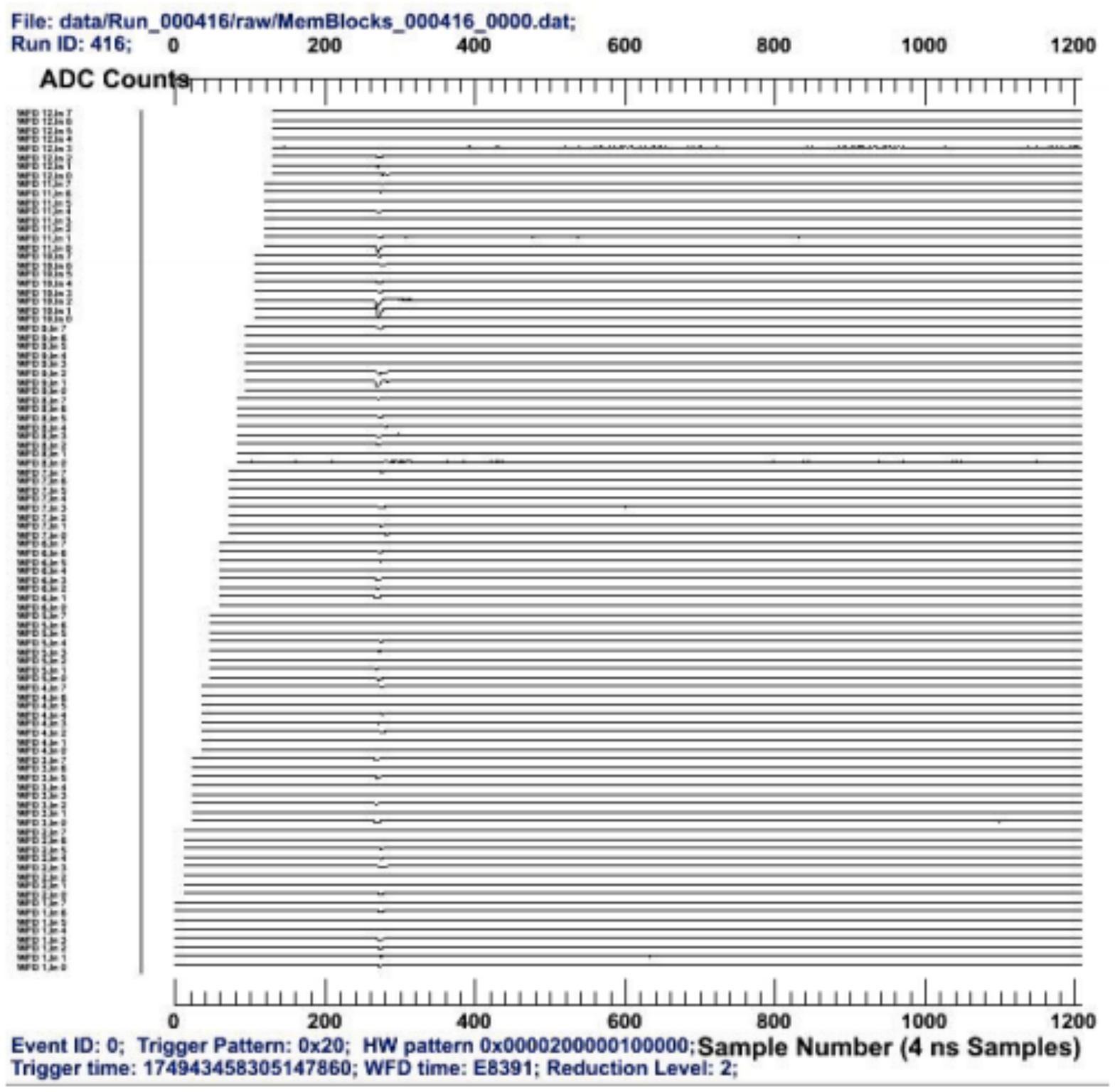}}
\hfill
\caption{ Example of waveform for each reduction level. (a)\texttt{ZLE\_Waveform}. (b) \texttt{ZLE\_Integral}. (c) \texttt{ZLE\_PromptTotal}. (d) \texttt{ZLE\_Fullwaveform}.}
\label{fig:zle}
\end{figure}

\section{Bitwise operation}
The software trigger using bitwise operation to assign the correct reduction level to each event base on the predefined criteria in DAQ.ratdb. In the beginning of the process, all events are assumed as \texttt{ZLE\_PromptTotal} waveform. Then the processor will use the preliminary values on adc counts, fprompt and maxQ to test if this event pass the criteria. This process is implemented using bitwise operation. We use the " trigger " word as our test bit and initialized it to 3. The pass bit is set to power of 2 to avoid possible confliction. For every 100 events we force the processor to assign  Zero-supressed waveform to the given event. The same is true for every 100 events passed charge cut , prompt Q cut and maxQ cut.  The threshold value of each discrimination parameters are in \textbf{DAQ.ratdb}, user can change values as they see fit. Currently the criteria for Sofware tigger to assign reduction level are following :

\begin{enumerate}
	\item \textbf{CHAN\_ZLE\_WAVEFORM}
		\begin{enumerate}
			\item PassPromptQ is true
			\item PassIntQ is true
			\item PassDSRcuts is true
		\end{enumerate}
	\item \textbf{CHAN\_ZLE\_INTEGRAL}
		\begin{enumerate}
			\item Either one of PassPromptQ and PassIntQ is false 
		\end{enumerate}
	\item \textbf{CHAN\_ZLE\_INTEGRAL}
		\begin{enumerate}
			\item Both of PassPromptQ and PassIntQ are false 
		\end{enumerate}
\end{enumerate}

\section{Decision Maker}
As described in previous section, the software trigger processor will sort events into these reduction level base on the parameters. For events which has been identified as \texttt{ZLE\_PROMPT\_TOTAL} or \texttt{ZLE\_INTEGRAL}, the samples in the blocks will be discarded and just keep the information on total charge and the start and end time of the block. The issue happened when the events identified as \texttt{ZLE\_INTEGRAL} waveform\footnote{Reduction Level and waveform sometimes are interchangeable. }, the FPrompt processor will gives unreasonable value on fprompt (Fig. \ref{fig:fpzle}). This happens because for the blocks came in  at boundary of start and end of prompt window, the charge inside the prompt window are miscalculated results in bizarre fprompt value.  To fix this, one can get the charge in each bins in the block, and deduce the portion of the bin that came in inside the prompt window to get prompt charge (Fig. \ref{fig:zlepo}). However, since we throw the samples away, we don't have any means to get exact calibrated time, also the pulse is not evenly distribute throughout the block which results in artificially larger late charge(Fig. \ref{fig:zlehis}). \\
User should aware of this issue but should not worry about it. The threshold value for each parameters to decide the reductional level can be modified in DAQ.ratdb table. This way when doing the data analysis, user can optimal the threshold value of the parameter to dump unwanted events.
\begin{figure}[htbp]
\centering
\graphicspath{{./fig/appendix/}}
\includegraphics[scale=0.3]{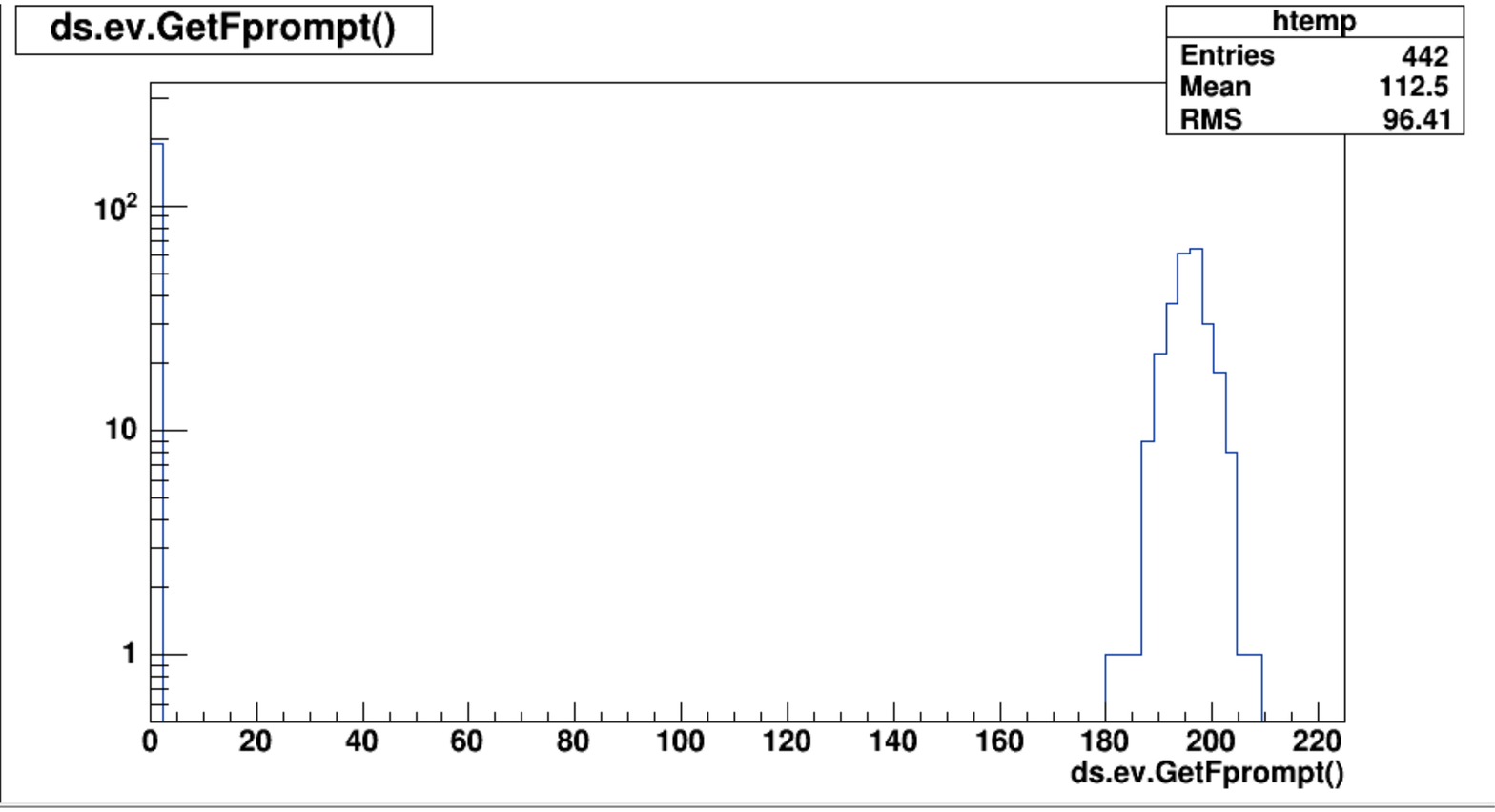}
\caption{ Fprompt value due to bad Fprompt calculation }
\label{fig:fpzle}
\end{figure}
\begin{figure}[htbp]
\centering
\graphicspath{{./fig/appendix/}}
\includegraphics[scale=0.3]{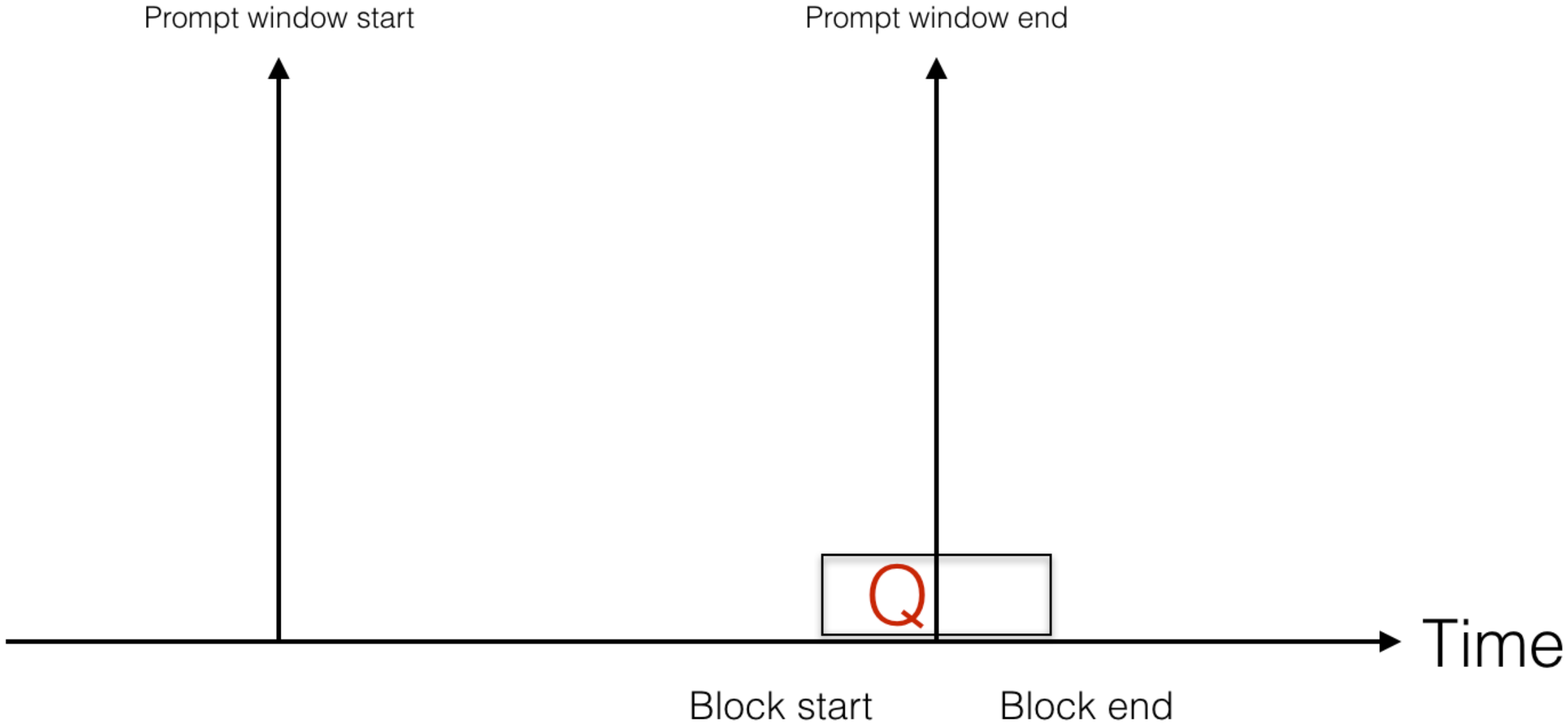}
\caption{Possible situation for Fprompt calculation (see context)}
\label{fig:zlepo}
\end{figure}

\begin{figure}[htbp]
\centering
\graphicspath{{./fig/appendix/}}
\includegraphics[scale=0.3]{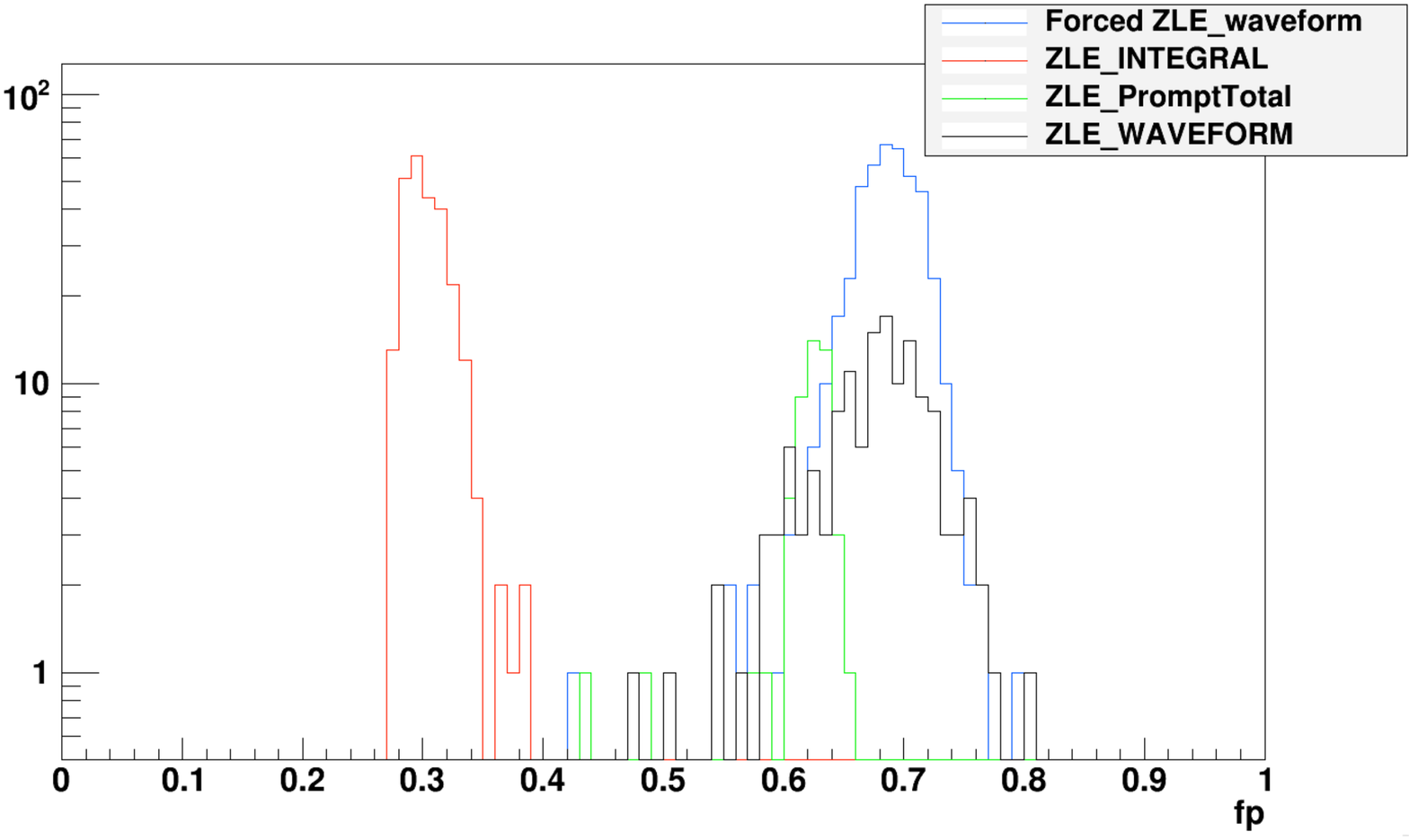}
\caption{ The plot is from two data sets. The blue histogram is from the dataset that has been forced to set reduction level as  \texttt{ZLE\_WAVEFORM} while being reconstructed. The black, red and green histogram represent different reduction level choosen by Software trigger.  }
\label{fig:zlehis}
\end{figure}

\bibliographystyle{unsrt}
\bibliography{./Thesis_ref}

\end{document}